# HOMOMORPHISM PROBLEMS
## in GRAPH DATABASES
## and AUTOMATIC STRUCTURES

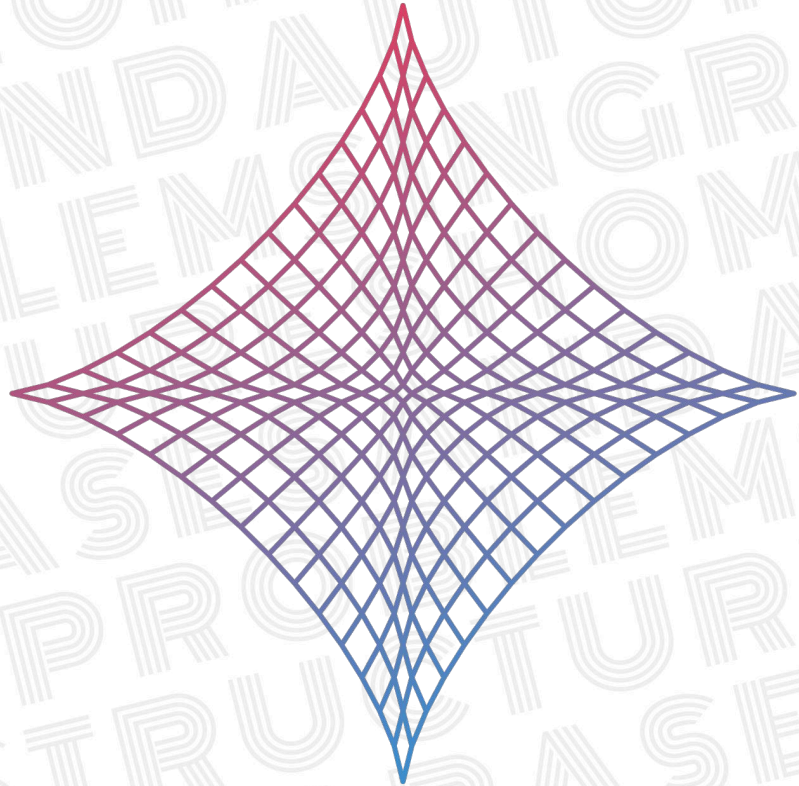

## Rémi Morvan
supervised by Diego Figueira
and Nathanaël Fijalkow

Ph.D. thesis in Computer Science
ED Mathématiques & Informatique
Université de Bordeaux

# Homomorphism Problems in Graph Databases and Automatic Structures

## Problèmes de morphismes sur les bases de données graphes et les structures automatiques

Thèse présentée pour obtenir le grade de docteur de l'Université de Bordeaux

*Rémi Morvan*

# *Preface*

## *Organization*

The chapters of this thesis have been written so that they could mainly be read independently from one another.

The introduction (Chapter I) is targeted at anyone with a reasonable understanding of theoretical computer science—corresponding roughly to a first year of Master's degree. This chapter is not technical and introduces as few definitions as possible. Its goal is to provide an overview of the foundational results of the field, the questions we studied in this thesis, our contributions, and the key questions that remain unsolved. **If you should only read one chapter of this thesis, let it be Chapter I: it provides a complete non-technical overview of the contents of this manuscript.**

The next chapter presents the general preliminaries (Chapter II): it serves no other purpose but to make all notions unambiguous. We suggest that readers skim it initially and return to it as needed.

This thesis is then divided in two independent parts: the first one focuses on database theory (Part 1), and the second one on automatic structures (Part 2). Each part starts with a quick survey of the domain (Chapters III and VII), followed by two chapters presenting our main contributions (Chapters IV and V for database theory, and Chapters VIII and IX for automatic structures). Each part is concluded by a discussion (Chapters VI and X) that highlights open problems and reflects on the techniques we explored, including those that proved unsuccessful. See Figure P.1 for the dependency graph of these chapters.

A succinct global table of contents is presented after this preface and the acknowledgements, and each chapter is preceded by a more detailed one.

## *Knowledge & Knowledge-Clustering*

This thesis was written using Thomas Colcombet's `knowledge package`, allowing one to click on a notion (be it textual or symbolic) to go to its definition: for instance try clicking on 'automatic structure', or on the brackets of $[\![\phi]\!]^{\Sigma^*}$! Most pdf viewers allow you to go back to where you previously were in the document before clicking.

The extensive use of `knowledge` was permitted by `knowledge-clustering`,

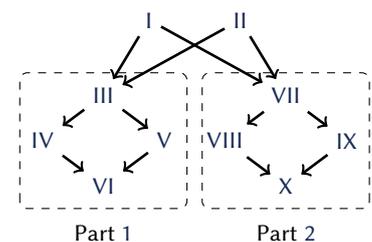

Figure P.1: Dependency graph of the chapters of this thesis.

a command-line tool I developed to help streamline and scale the use of the knowledge package in large LaTeX documents. I would like to thank all the people that provided me with suggestions, feature requests or bug reports, with a special thought for Thomas Colcombet, Aliaume Lopez and Antonio Casares.

### Proofs

For the sake of readability, elementary proofs—which are often the result of elementary set manipulation or applying the previous propositions—are sometimes omitted by a nonchalant "it immediately follows that". Naturally, we reserve this logical blasphemy to statements that are not harder to prove than $1 + 1 = 2$, see [WR10, ∗ 54.43] to [WR12, ∗ 110.643].

### On Black Holes

Many results of this thesis assume the undecidability of the halting problem: we hence assume that the reader lives in our usual universe. Should the reader be in a Malament–Hogarth spacetime [Hog94], we kindly suggest that they momentarily set aside this thesis and focus on resolving the more pressing astrophysical situation, see Figure P.2.

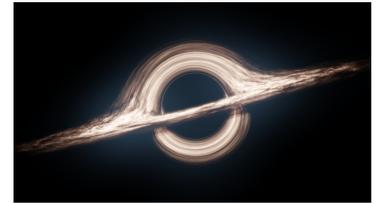

Figure P.2: Computer scientists tend to do badly around black holes. Illustration by 852278-MCS, licensed under CC BY SA 4.0.

# *Acknowledgements · Remerciements*

While this thesis is undoubtedly the proudest achievement of my many years of studies, I must admit that this is by far not a *personal* accomplishment. Beyond the support I received from many people, which I will get to in a moment, the academic journey that led me to this document and these four years in Bordeaux is the result of many influences, most of which I believe have been pivotal into leading me where I am today.

Je voudrais d'abord remercier Marylise Autran, Cyril Germain et Michel Cognet qui ont su tous trois cultiver mon appétence pour les maths par l'incroyable passion qu'ils m'ont transmis dans leurs cours. Si mes années à l'ENS n'ont pas commencé de façon très studieuse, je dois remercier Sylvain Schmitz et Jean Goubault-Larrecq pour m'avoir redonné le goût au fait de rester assis derrière une table grâce à leur pédagogie, leur rigueur et leur engouement (et, dans le cas de Jean, d'une maîtrise sans pareille des parenthèses (oui, même quand elles sont imbriquées!)), et à Dietmar Berwanger pour avoir canalisé et entretenu mon goût pour la théorie des jeux.

My various internships have been a crucial point in my studies, and the reason I started a PhD is unquestionably because these experiences have been fantastic. To Marcin: since my first day you managed to make me feel at ease and valued in our work together. You managed both to teach me many things while giving me the opportunity to explore some more original ideas. To Thomas and Sam, the second part of my thesis is a good witness of the influence you had on me: I went to Bordeaux to do a PhD on databases and ended up with half my thesis on automata and algebraic language theory! My five months with you in Paris have been both rich and interesting, and the pair of you makes a unique blend of rigour and ~~craziness~~ hand-wavyness, allowing the ideas to flow like I've rarely seen.

Nath, dès mon premier jour tu m'as parlé comme à un collègue et non un étudiant. S'il y a bien une constante dans ces quatre années, c'est que tu as toujours tout fait pour me mettre en avant, m'inclure dans notre communauté scientifique et dans divers projets, que ce soit pour me proposer de rejoindre un comité de programme ou créer un cours d'IA.[1] Tu as su cultiver mon autonomie et ma volonté de toucher à tout : si je me sens aujourd'hui aussi à l'aise au sein de notre domaine, tu n'y es pas étranger !

Diego, je peux enfin de te l'avouer : j'avais fait l'impasse sur les bases de données à l'agrégation. Tu as remarquablement su m'y donner goût par ta gentillesse et ta patience. Pendant ces quatre années j'ai profondément admiré tes intuitions et tes idées—même si j'ai essentiellement passé ma première année à démanteler nos résultats de tree-width ! Tu as réussi à supporter mes envies de parler de monades et de limites projectives sans me brider ni m'en décourager, tout en me poussant à avancer sur nos idées plus concrètes. Si cette thèse n'est pas qu'une suite de folles conjectures et contient tout de même quelques résultats, c'est avant tout grâce à toi !

[1] To Andrew Ng: thank you for teaching me what I taught my students very soon after!

Marcin, Sam, Thomas, Nath and Diego, you all are very different people, and most of you are even working in different fields. Yet, you have one thing in common: your kindness and passion. Working with you has been deeply fulfilling, and perhaps the only thing that makes me prouder than saying I was your student is being able to say I am your colleague.

Thejaswini, Pablo, Miguel, Arthur, Pierre V., Antoine C. and Arka, thank you for being such nice co-authors and co-workers. I've learned so many things while working with you. Pablo and Miguel, you've been very welcoming hosts; I will never forget my time in Santiago!

To the members of my jury: first to Wim, for accepting to (very thoroughly) review my thesis. We have only met a few weeks before my defence, but I think it's now quite clear that I found your work inspiring. Mikołaj, I'm quite proud to have such a good-hearted researcher as a reviewer. You've always made me feel welcomed and like I belonged there, wherever we met, be it in Warsaw or in conferences. Then, to Anca, Antoine, Balder, Bartek and Sophie: thank you for accepting to be part of my jury. It was important to me to have people whose work I found inspiring, but also with profound human qualities. With this in mind, I can honestly say that I couldn't have dreamed of a better jury.

Thomas et Antoine, je tiens à encore une fois vous remercier, mais cette fois-ci pour votre engagement écologique au sein de notre communauté, que ce soit à travers de TCS4F, du comité de pilotage de Highlights, ou encore de l'EATCS. Au-delà de rendre service à notre communauté—et à l'avenir de l'espèce humaine—, avoir des chercheurs si reconnus qui prennent de tels engagements m'a profondément inspiré. Notamment, il a été beaucoup plus facile pour moi de refuser de prendre l'avion pour aller en conférence en sachant que vous le faisiez déjà. Ce que vous faites est crucial.

Thomas, encore, je me dois de te remercier pour `knowledge`, qui a eu une influence déraisonnablement grande sur ma production scientifique. Quand la situation actuelle ne te plaît pas, tu préfères mettre les mains dans le cambouis plutôt que de te plaindre, et `knowledge` en est un bon exemple. Aliaume, merci d'avoir partagé mon enthousiasme pour `knowledge` ainsi que de trop nombreuses bouteilles de vin pour oublier les exposés de *sheafification* de la journée. Ton enthousiasme[2] est inspirant : maintenant que nous habitons dans la même ville, j'espère que ce n'est que le début de notre aventure.

More generally, I would like to thank the members of the ANR Delta and

[2] Et tes nombreux projets Github, pas tous initialisés…

GT DAAL, and the Highlights and Finite Model Theory communities for being particularly welcoming. Charles, tu es la seule personne que je connaisse qui puisse faire de la théorie algébrique des langages et à qui on reproche de minorer les applications industrielles de ses projets. Je n'ai retrouvé chez personne d'autre ta capacité à manier la théorie comme la pratique, à écrire des ensembles d'équations profinies entre deux lignes de Rust et une discussion sur le séquençage de l'ADN. Ton travail et ta personnalité sont tout simplement une source d'émerveillement pour moi.

Michaël C., pour un américain, t'es étonnament un type bien ! J'espère bien que nos chemins se recroiseront : merci pour ces belles parties de billard et ces belles découvertes houblonnées. À Laure, Denis, Mikaël, Michaël T., Benjamin, Nathan, Karoliina, Pierre O., Sophie P. et à tous·tes celleux que j'oublie : merci d'avoir égayé mes diverses visites, notre communauté a de la chance de vous avoir. Anton, León and Christoph : the events we've shared would simply not have been the same without you.

Aux membres du LaBRI ensuite, à commencer par Vincent. Tu es, je crois, le premier membre du labo avec qui j'ai interagi : c'était à la réunion DELTA à Paris, alors que je n'étais encore qu'un stagiaire de M2. J'ai tout de suite su que je ne m'ennuyerai pas en venant faire ma thèse au LaBRI ; les faits ne m'ont pas donné tort. Merci à Laurent, pour être un incroyable collègue et chef d'équipe, à Floréal pour avoir été un super responsable de cours, à Marthe pour les découvertes lacto-fermentées et m'avoir mis en contact avec Georgette, à Joanna, Géraud, Guillaume L., Pascal, Marc et Meghyn pour avoir été non seulement des collègues chaleureux, mais aussi pour votre oreille attentive. Mireille, Nicolas B. et Xavier, il me suffisait de vous croiser dans les couloirs et d'échanger un bonjour avec vous pour que ma journée soit illuminée ; votre bonne humeur est terriblement communicative ! Enfin, un grand merci à l'équipe administrative du LaBRI, et tout particulièrement Auriane et Élia, pour votre soutien.

Je souhaite aussi remercier mes co-bureaux de 325, pour ces nombreuses heures partagées : Tobias, Alex et Théo M. ; Françoise, ta présence illumine le bureau ; Gabriel B., pour nos nombreuses discussions passionnantes ; Sarah, pour m'avoir proposé tant d'activités, et Clara, pour être une MJ hors pair, pour nos sorties à la salle et pour m'avoir tant empêché de travailler, mais au passage m'avoir fait découvrir tant de choses.

Ma reconnaissance va ensuite aux membres de l'AFoDIB, et particulièrement celleux qui ont contribué à l'animer pendant ma première année : j'ai une pensée toute particulière pour Aline, Joséphine, Quentin M. et Jonathan.[3] Sans vous, mon arrivée à Bordeaux aurait très probablement été bien plus solitaire. À Claire, ensuite : en plus de cette très belle couronne de lauriers, tu m'as laissé de très grosses chaussures à remplir.

Merci à Rohan, pour les sorties au bar, Timothée, pour l'humour de qualité, Theo DCP., pour les sessions JDR, Manon, pour ta joie rayonnante, Pierre V., pour les nombreux quiz et l'arrachage de cheveux nocturne sur le projet de machine learning, Romain, bébou !, pour avoir égayé mes midis dans l'atrium, Floriane, pour mettre des paillettes dans mes yeux, Quentin C., pour les interminables ragots. Thibault, comme toute une génération de doctorant·e·s, j'ai pensé à me nourrir grâce à toi. Corto et Antonio, ce fut un plaisir de vous connaître, au travail comme en dehors. J'espère de tout cœur que l'avenir vous donnera ce que vous souhaitez et méritez. Vincent M. : merci pour ce savant mélange de monoïdes, de musique, de catégories, et de choses plus personnelles qui forment folles discussions.

[3] Preuve de leur implication hors du commun : encore aujourd'hui, il arrive à des membres de l'AFoDIB de crier « Aline ! Aline ! » pour qu'elle revienne...

Yanis, il est difficile de décrire ces trois dernières années, avec ses pauses café, nos[4] observations naturalistes ou nos verres le soir. J'espère que, quand ça sera ton tour d'avoir fini de rédiger, nos aventures pourront reprendre ! Arnaud : merci d'avoir partagé ces nombreuses découvertes, notamment—mais pas exclusivement—celle de ton lieu de travail. Mathilde, merci non seulement pour nos verres, mais aussi pour ta patience avec mon manque de disponibilité. Aymeric, merci pour nos belles soirées passées ensemble.

Ensuite, je souhaiterais particulièrement remercier Émile, pour avoir changé mon rapport au mobilier urbain, Edgar, pour avoir partagé cette folle aventure de débuter l'apprentissage d'un instrument à cordes frottées, Timothé, pour ton enthousiasme sans pareil pour faire de la musique, organiser des randos, ou encore faire des raclettes, et Taïssir, pour ton soutien émotionnel sans faille, notamment pendant les dernières semaines de rédaction qui ne furent pas aisées.

Je suis tout particulièrement reconnaissant à tous·tes mes amis·es qui, malgré la distance, réchauffent mon existence. Je pense particulièrement à Quitterie, Agathe, Nathalie, Thomas, Guillaume, Esther, Julie, Margot, Julien, Emma et Juliette. Alice, voici dix ans que nous sommes rencontrés et que tu illumines ma vie au grand dam de mon portefeuille et de mon taux de cholestérol : pour tous ces moments que nous avons partagés et ceux encore à venir, je ne peux que me montrer que reconnaissant. Merci à Roland, pour ta bonne humeur à toute épreuve, à Nicolas, pour avoir partagé avec moi l'apprentissage de l'arithmétique ordinale parisienne tout comme les bars à tapas espagnols, à Arthur, pour ton enthousiasme pédestre, et enfin à tous les membres de Montautre pour me faire passer, chaque année, un si beau week-end.

Merci enfin à ma famille : naturellement à mes parents et à mon frère, Sylvain, pour ces nombreuses années passées ensemble, mais aussi tout particulièrement à Jean, Dominique, Hélène et Olivier, pour votre accueil indéfectible.

Finalement, je tiens à exprimer ma profonde gratitude à Hélène Largillier : le violon m'a fourni une nécessaire bouffée d'oxygène pendant cette éprouvante aventure que fut cette thèse. Ton savant mélange de bienveillance et d'exigence a non seulement su entretenir ma motivation à apprendre la musique malgré la difficulté de l'exercice,[5] mais est aussi un modèle pour moi lorsque j'enseigne.

# Contents

# Résumé étendu en français

## Les deux facettes du problème du morphisme

Cette thèse s'intéresse à des variations du *problème du morphisme*, qui est un concept central en informatique théorique. La notion de *morphisme* formalise l'idée de fonction préservant de l'information relationnelle d'une structure à une autre. Ces objets, ainsi que le problème de décision associé, apparaissent naturellement dans divers domaines : citons par exemple en *logique*, en tant que caractérisation du *model-checking* pour les formules primitives positives ; en *théorie des bases de données*, comme la sémantique des requêtes conjonctives ; et en *programmation logique*, comme formalisation du fait qu'une structure satisfait un ensemble de contraintes.

Dans cette section, nous commençons par formaliser la notion de morphisme, avant de donner deux exemples représentatifs de l'utilisation des problèmes de morphismes en informatique fondamentale : l'évaluation de requêtes SQL et la résolution de Sudoku. Ces deux exemples révèlent une *dualité* sur la façon dont les problèmes sont encodés : dans certains cas, la requête—ou formule logique—apparaît à gauche du morphisme, alors que dans d'autres cas elle apparaît à droite. Le premier type d'encodage correspond à ce que nous appelons des *problèmes existentiels*, qui forment les fondements de la *théorie des bases de données*, que nous présenterons dans la seconde section. Quant au second type, il encode des *problèmes universels*, étudiés dans le domaine des *problèmes de satisfaction de contraintes*, présenté dans la dernière partie de ce résumé. Cette dualité intrinsèque du problème du morphisme—ou plutôt de l'encodage de problèmes naturels dans le problème du morphisme—guidera cette thèse jusque dans sa structure, que l'on peut notamment observer par sa division en deux parties.

La structure mathématique la plus simple est peut-être celle d'un graphe fini (orienté) : elle consiste en un ensemble fini $V$ de sommets (également appelé *domaine*), ainsi qu'en un ensemble d'*arêtes* $\mathcal{E} \subseteq V \times V$. Un *morphisme* d'un graphe vers un autre est alors une fonction $f$ définie sur leurs sommets qui préserve les arêtes, au sens où, si $\langle u, v \rangle$ est une arête, alors $\langle f(u), f(v) \rangle$ doit également en être une : nous représentons un exemple de morphisme de graphes en Figure R.1. La notion plus riche de $\sigma$-structure, non explicitée dans

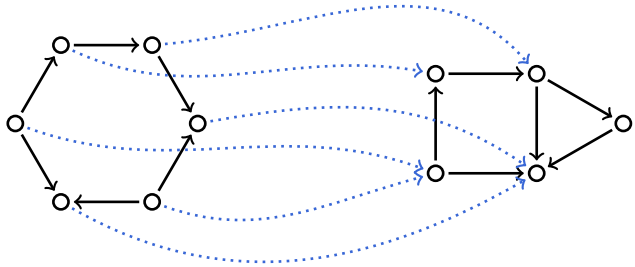

ce résumé, généralise la notion de graphe en permettant d'avoir plusieurs types de relations.

> Le problème de morphisme sur $\sigma$
> *Input* :   Deux $\sigma$-structures **A** et **B**.
> *Question* :   Existe-t-il un morphisme de **A** vers **B** ?

Au-delà d'un simple problème de décision, qui se trouve être NP-complet, le problème de morphisme sur $\sigma$ peut en fait être vu comme un cadre dans lequel reformuler des problèmes naturels.

*Bases de données.*    Considérons une requête SQL, disons la requête $\gamma$ suivante, qui retourne les titres de films ainsi que leur horaire de projection.

```
1   SELECT title, time
2   FROM Movies, Projections
3   WHERE Projections.movie_id = Movies.id;
```

Une base de données **D** peut être décrite comme une $\sigma$-structure—où la signature $\sigma$ décrit le *schéma* de la base—, notée également **D**. Il est possible de construire[6] de manière assez triviale—notamment en temps linéaire—une $\sigma$-structure $\mathbf{Q}_{\gamma}$, contenant notamment deux sommets title et time, tels que pour toute paire d'éléments $\langle x, y \rangle$ de la base de données, il existe un morphisme de $\mathbf{Q}_{\gamma}$ vers **D** envoyant title vers $x$ et time vers $y$ si, et seulement si, la paire $\langle x, y \rangle$ est une sortie valide de la requête $\gamma$ lorsqu'elle est évaluée sur **D**, c'est-à-dire que le film $x$ est projeté à l'horaire $y$. Autrement dit, ce problème de requêtage peut être formulé comme un problème de morphisme.[7]

*Sudoku.*    Donnons maintenant un second exemple, dans un domaine tout à fait différent. Nous considérons une grille de Sudoku $G$, telle que celle Figure R.2. Cette grille peut être encodée comme une $\sigma$-structure $\mathbf{S}_G$, ayant un sommet par case de la grille.[8] On peut alors définir une $\sigma$-structure **T** dont les sommets sont les entiers $1, 2, \ldots, 9$, de sorte que les morphismes $f$ de $\mathbf{S}_G$ vers **T** décrivent exactement les complétions de la grille de Sudoku. Notamment, étant donné un tel morphisme $f$ et une position $\langle i, j \rangle$, la valeur $f(i, j)$ est la valeur rentrée dans la grille en position $\langle i, j \rangle$ lorsque complétée selon $f$.

[6] Voir Exemple I.1.2 pour plus de détails.

[7] *Stricto sensu* le problème de morphisme mentionné dans ce paragraphe ne rentre pas dans le cadre décrit précédemment, puisqu'il doit gérer des constantes (title et time). En fait, il est possible de gérer ces constantes en les faisant passer dans la signature, et donc ce problème s'inscrit bien dans le cadre étudié.

[8] Voir Exemple I.1.3 pour les détails de cet encodage.

Fig. R.2 : Une grille de Sudoku préremplie. (Reproduction de Figure I.6.)

Ces deux exemples illustrent que le PROBLÈME DE MORPHISME est versatile et peut encoder des problèmes naturels d'une grande diversité. Notons cependant une distinction assez notable entre les deux exemples décrits :

- dans le cas des bases de données, les données sont encodées à droite du morphisme, et la logique à gauche : on demande s'il y a un morphisme de $\mathbf{Q}_\gamma$ (la structure encodant la requête) vers $\mathbf{D}$ (la base de données) ;
- au contraire, pour le Sudoku, on encode les données à gauche du morphisme, et la logique à droite : on demande s'il y a un morphisme de $\mathbf{S}_G$ (la structure représentant la grille) vers $\mathbf{T}$ (la structure représentant les contraintes du Sudoku).

Cette asymétrie entre ces deux encodages a pour conséquence que, bien que ces deux exemples partagent le même cadre, nous allons généralement étudier des questions différentes sur le PROBLÈME DE MORPHISME selon qu'on soit dans le premier cas ou le second. Il est généralement pertinent d'étudier ce qu'il se passe lorsque l'on fixe la logique et que les données sont variables : dans le cas des bases de données, cela revient à fixer le côté gauche du morphisme, mais dans le cas du Sudoku, cela correspond à fixer le côté droit. Il en résulte notamment des propriétés de monotonie et de complexité assez différentes. Nous séparerons le reste de cette thèse, et de ce résumé, en deux parties.

### Requêtes SQL & bases de données

L'exemple donné précédemment se généralise aisément : on peut montrer que toute requête SQL de la forme

<p align="center">SELECT ... FROM ... WHERE</p>

satisfait la même propriété, c'est-à-dire que l'évaluation d'une requête $\gamma$ de ce type sur une base de données $\mathbf{D}$ revient à résoudre un PROBLÈME DE MORPHISME où $\gamma$ est encodée à gauche et $\mathbf{D}$ à droite. L'algorithme naïf pour résoudre le PROBLÈME DE MORPHISME consiste essentiellement à énumérer toutes les

$$|\text{structure de droite}|^{|\text{structure de gauche}|}$$

fonctions allant de la structure de gauche à celle de droite, puis à vérifier pour chacune si elle constitue un morphisme. Dans ce cas, la structure de droite étant une base de données, et sachant que celles-ci peuvent atteindre plusieurs téraoctets en pratique, une requête de taille moyenne peut rendre cet algorithme totalement inefficace.

*Optimisation de requêtes conjonctives.* Une question centrale en théorie des bases de données consiste donc à optimiser la taille des requêtes : c'est-à-dire, étant donnée une requête, peut-on en trouver une équivalente—c'est-à-dire qui retourne les mêmes résultats sur toute base de données—mais plus suc-cincte ? Ce problème, *a priori* difficile, est particulièrement bien compris pour les requêtes du fragment <span style="color:red">SELECT-FROM-WHERE</span> de SQL, qui correspond, de manière plus abstraite, au fragment existentiel positif du premier ordre, ou encore aux *requêtes conjonctives*. Cette optimisation est rendue possible grâce au prisme offert par le PROBLÈME DE MORPHISME et à l'encodage d'une telle requête comme une $\sigma$-structure. Par cette méthode, pour chaque requête conjonctive $\gamma$, nous pouvons construire une requête $\widehat{\gamma}$, sémantiquement équivalente—c'est-à-dire fournissant les mêmes réponses sur toute base de données—et par ailleurs plus petite. De manière remarquable, ici, « plus petite » peut être compris au sens de « utilisant moins de variables », mais également de manière plus générale : toute mesure de taille raisonnable,[9] est optimisée par cette requête $\widehat{\gamma}$.

[9] Formellement, par raisonnable on entend « fermée par sous-structure ».

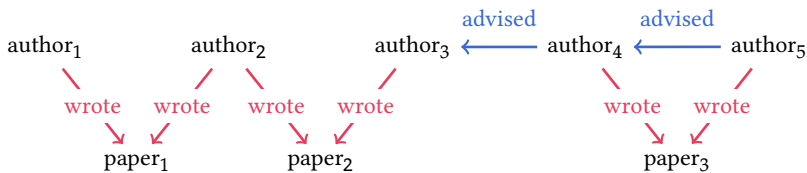

FIG. R.3 : Une base de données graphe avec deux types d'arêtes. (Reproduction de Figure III.22.)

*Au-delà des requêtes conjonctives.* Bien que les requêtes conjonctives soient particulièrement faciles à optimiser, leur expressivité reste limitée. Ceci est particulièrement vrai pour les bases de données graphes,[10] qui sont utiles pour stocker, par exemple, des données bibliographiques, biologiques ou de classification. Une telle base de données est illustrée en Figure R.3. Elle contient deux types d'arêtes : $x \xrightarrow{\text{wrote}} y$ indique qu'une personne $x$ a écrit un article $y$, alors que $x \xrightarrow{\text{advised}} y$ indique qu'une personne $x$ a supervisé la thèse de $y$. Il est alors impossible d'exprimer la requête « l'ensemble des paires $y, z$ telles que $z$ est l'ancêtre scientifique de $y$ » avec une requête conjonctive, ou, de manière équivalente, avec une requête du fragment <span style="color:red">SELECT-FROM-WHERE</span> de SQL. Pour pallier cette limitation, les *requêtes conjonctives à chemins réguliers*— « conjunctive regular path queries », ou « CRPQ » en anglais—introduisent un mécanisme supplémentaire dans le langage de requête : il devient possible d'exiger l'existence d'un chemin reliant un point $x$ à un point $y$, dont la suite d'étiquettes appartient à un langage régulier. Par exemple, la CRPQ

[10] Appelées aussi « bases de données orientées graphe ».

$$\gamma(x,y) \triangleq x \xrightarrow{\text{wrote}} z \land z' \xrightarrow{\text{wrote}} z \land y \xrightarrow{(\text{advised})^r} z',$$

demande toutes les paires $\langle x, y \rangle$ vérifiant les contraintes suivantes :
- $x$ a écrit $z$,
- $z'$ a écrit $z$, et
- il existe un chemin de $y$ à $z'$ étiqueté par un mot du langage $(\text{advised})^*$.

Autrement dit, cette requête identifie toutes les paires de personnes $\langle x, y \rangle$ telles que $x$ a co-écrit un article avec un descendant scientifique de $y$.

Les CRPQ modélisent le principal mécanisme de requêtage des langages utilisés pour interroger des bases de données graphes, comme SPARQL. L'op-timisation de ces requêtes est donc un problème fondamental en théorie des bases de données. Alors que les CRPQ étendent les requêtes conjonctives, les bonnes propriétés d'optimisation de ces dernières ne se généralisent pas, et leur optimisation reste un problème complexe, qui constitue le cœur de la première moitié de cette thèse.

> CONTRIBUTION. Étant donnée une requête conjonctive à chemins réguliers et un entier $k$, nous pouvons décider s'il existe une requête équivalente utilisant au plus $k$ contraintes/atomes, et produire la requête correspondante le cas échéant.

Le nombre de contraintes, ou *atomes*—c'est-à-dire le nombre de termes de la forme $x \xrightarrow{L} y$ où $L$ est un langage régulier—est une mesure pertinente de la complexité d'évaluation d'une CRPQ. Nous avons également montré, de manière surprenante, qu'il existe des requêtes minimales dans ce sens, c'est-à-dire non équivalentes à une CRPQ avec strictement moins d'atomes, mais équivalentes à une *union finie* de CRPQ utilisant chacune strictement moins d'atomes. L'évaluation d'une telle union étant facilement parallélisable, sa complexité dépend surtout du nombre maximal d'atomes, même si le nombre de CRPQ dans l'union est élevé.

> CONTRIBUTION. Étant donnée une requête conjonctive à chemins réguliers et un entier $k$, nous pouvons décider s'il existe une *union finie* de CRPQ équivalente, dont chaque CRPQ utilise au plus $k$ atomes, et produire cette union si elle existe.

Lorsque la réponse à ces problèmes est « oui », l'algorithme fournit un témoin explicite. Lorsque la réponse est « non », il est plus difficile de justifier pourquoi aucune requête plus petite n'existe.

> CONTRIBUTION. Nous donnons une condition suffisante pour qu'une requête soit difficile à exprimer, c'est-à-dire qu'elle ne soit pas équivalente à une requête plus petite.

Le tour de force de ce résultat réside dans le fait que « plus petit » peut être interprété de manière large, et ne se limite pas au nombre d'atomes. Cette condition est utile pour démontrer des bornes inférieures de complexité. Cependant, elle est seulement suffisante, et non nécessaire. Trouver une condition à la fois nécessaire et suffisante reste le principal problème ouvert, et diverses pistes sont discutées en conclusion. Une telle condition, si elle était effective, pourrait améliorer significativement les algorithmes existants.

Nous nous intéressons ensuite à la *largeur arborescente*, une mesure plus fine de la complexité d'une CRPQ que le simple nombre d'atomes. Barceló,

Romero et Vardi ont proposé en 2016 un algorithme permettant de décider si une CRPQ est équivalente à une union finie de CRPQ en forme d'arbre, c'est-à-dire de largeur arborescente 1. Le cas général pour une largeur arborescente $k$ (entier arbitraire) restait cependant ouvert.

> CONTRIBUTION. Nous donnons un algorithme qui prend en entrée une CRPQ et un entier $k$, et qui décide s'il existe une union finie de CRPQ de largeur arborescente au plus $k$, et produit cette union si elle existe.

Les algorithmes précédents, ainsi que ceux optimisant le nombre d'atomes, sont peu efficaces et utilisent de l'espace exponentiel, voire doublement exponentiel. Nous avons également démontré que ces bornes sont intrinsèques : le problème est fondamentalement difficile. Nous considérons donc un fragment du langage, en limitant les langages réguliers autorisés dans les atomes. Une étude empirique montre que 75 % des atomes pratiques sont de cette forme. Dans ce fragment, nous fournissons des algorithmes bien plus efficaces.

> CONTRIBUTION. Nous montrons que la minimisation du nombre d'atomes (resp. de la largeur arborescente) d'une CRPQ n'utilisant que des expressions régulières simples est située dans la hiérarchie polynomiale.

En résumé, les problèmes de morphismes où la logique est encodée à gauche constituent un socle solide pour la théorie des bases de données et pour l'optimisation des requêtes conjonctives. Dans la première partie de cette thèse, nous avons étendu ces résultats aux *requêtes conjonctives à chemins réguliers* (CRPQ).

## Problèmes de satisfaction de contraintes

La deuxième partie de ce travail est consacrée aux problèmes de satisfaction de contraintes : essentiellement, cela revient à étudier des PROBLÈMES DE MORPHISMES où la structure de droite est fixée. Cette structure décrit les contraintes du problème : une certaine structure correspond au Sudoku, une autre à la $k$-coloration de graphes (pour $k$ un entier), et une autre encore aux contraintes de 3-SAT, c'est-à-dire la satisfaction de formules booléennes en forme normale 3-conjonctive.

Une question majeure de ce domaine consiste à déterminer, étant donné un ensemble de contraintes décrites par une structure, la complexité du problème consistant à savoir si des données vérifient ces contraintes, c'est-à-dire s'il existe un morphisme des données vers ces contraintes. Ce problème ne peut jamais être plus dur que NP, et est en fait NP-complet pour les trois exemples précédents. Cependant, il existe des contraintes pour lesquelles le PROBLÈME

DE MORPHISME devient polynomial, voire de complexité encore plus faible. En 2017, Bulatov et Zhuk ont tous deux indépendamment démontré le résultat majeur du domaine, conjecturé vingt ans plus tôt par Feder et Vardi : pour tout ensemble de contraintes, son PROBLÈME DE MORPHISME est soit dans P, soit NP-complet. Autrement dit, ces problèmes ne peuvent avoir de complexité NP-intermédiaire.

La majorité des résultats de ce domaine, comme celui-ci, étudient la satisfaction de contraintes sur un ensemble *fini* de données. Il est cependant courant en informatique de manipuler des structures infinies. Deux exemples viennent en tête : d'une part en théorie des modèles, lorsque l'on considère des modèles mathématiques ; par exemple, étudier l'arithmétique revient à étudier la structure des nombres entiers munis de l'addition et de la multiplication. D'autre part, pour une machine, bien que celle-ci soit finie, son ensemble de configurations ne l'est pas. Une machine de Turing, ou même un automate fini, définit un graphe de configurations dont les sommets sont les configurations et les arêtes décrivent les transitions élémentaires, et ce graphe est quasi systématiquement infini. Ce dernier exemple présente toutefois une propriété remarquable : bien qu'infini, il peut être décrit finement, et il est pertinent de poser des questions algorithmiques à son sujet.

La notion de *structure automatique* a précisément été introduite pour généraliser cette idée : ce sont des structures infinies mais décrites finiment, à l'aide d'automates. Les graphes de configurations des machines de Turing sont en particulier des structures automatiques. Cette notion est profondément naturelle et canonique : pour preuve, elle a été (ré)introduite au moins cinq fois indépendamment entre les années 1970 et la fin du siècle. Nous nous intéressons alors au problème de satisfaction de contraintes sur cette classe de structures.

Il se trouve que, bien qu'extrêmement naturelle, cette question a été remarquablement peu étudiée. Le seul résultat connu, démontré par Köcher en 2014, est que, étant donné un graphe automatique, il est indécidable de savoir s'il est 2-coloriable. La notion de 2-coloriage, ou plus généralement de morphisme d'une structure automatique vers une structure finie, présente toutefois une asymétrie curieuse : le domaine est automatique, c'est-à-dire potentiellement infini mais forcément finiment descriptible. En revanche, un tel morphisme n'est pas forcément finiment descriptible : il faut spécifier, pour chaque élément du domaine, sa valeur... Nous introduisons donc la notion de *morphismes réguliers*, correspondant exactement aux morphismes finiment descriptibles par des automates.

Notre première contribution est de montrer que le problème de 2-coloriage régulier, c'est-à-dire, étant donné un graphe automatique, déterminer si l'on peut séparer ses sommets en deux moitiés par un automate fini, de sorte que deux sommets adjacents ne puissent appartenir à la même moitié, est lui aussi indécidable. Cela laisse penser que de nombreux problèmes de satisfaction de contraintes sont en fait indécidables. Remarquablement, nous parvenons à généraliser ce résultat à tous les problèmes de morphismes.

> CONTRIBUTION. Nous prouvons une dichotomie pour les structures automatiques : pour tout ensemble fini de contraintes, le fait de savoir si une structure automatique satisfait cet ensemble de contraintes est soit indécidable, soit décidable en NL. La propriété s'applique lorsque l'on restreint les morphismes aux morphismes réguliers. Par ailleurs, les ensembles de contraintes pour lesquels le problème est décidable coïncident dans les deux cas : ce sont précisément ceux qui ont « dualité finie », ou, de façon équivalente, qui sont descriptibles par la logique du premier ordre.

En se rappelant que le graphe de configuration d'une machine de Turing est toujours automatique, on peut voir ce résultat comme une variation (toutefois incomparable) du théorème de Rice : toute propriété non triviale—dans le sens de « indescriptible au premier ordre »—sur le graphe de configuration d'une machine de Turing est indécidable, sous réserve que cette propriété soit un problème de satisfaction (régulière ou non) de contraintes.

*Propriété des relations automatiques*   Le (méta-)résultat précédent décrit une très large classe de problèmes sur les structures automatiques comme étant entièrement indécidable. La suite de cette thèse cherche donc à identifier un îlot de tractabilité dans cet océan d'indécidabilité.

Notre résultat s'inscrit dans le domaine de la théorie algébrique des langages. En bref, tous les langages réguliers ne se valent pas : certains sont intrinsèquement plus compliqués à décrire que d'autres. Ici, le terme « plus compliqué » doit être compris non pas tant au sens de « avoir plus d'états », mais « nécessite une forme d'automate plus complexe ». Formellement, cette notion de complexité est notamment liée à la complexité de circuits à la compilation.[11] Pour comprendre quels langages sont « simples »—formellement, lesquels appartiennent à une certaine classe de langages satisfaisant de bonnes propriétés de clôture—, la théorie algébrique des langages a été développée depuis les années 1960. Essentiellement, cette théorie associe à chaque langage une structure algébrique—un *monoïde*—qui reflète assez clairement les propriétés dudit langage. Les questions sur les langages se traduisent alors en questions sur les monoïdes, qui se révèlent être des objets plus aisés à manipuler.

Cette question fondamentale—un langage est-il « facile » à décrire ?—s'étend naturellement aux structures automatiques, puisque celles-ci sont décrites par des automates appelés *automates synchrones*. Cependant, la théorie algébrique classique ne s'applique pas telle quelle à ces automates : la raison principale est que ces automates synchrones ignorent certaines de leurs entrées, ou plutôt que le comportement de l'automate sur certaines entrées n'a aucune influence sur la structure automatique définie.[12] Malheureusement, la structure algébrique de monoïde est définie en prenant en compte le comportement de l'automate sur tous les mots.

[11] En termes techniques, la notion de simplicité se traduit par « le langage appartient-il à une pseudovariété de langages fixée ».

[12] On peut faire un parallèle avec la programmation : peu importe le comportement d'une fonction `pgcd` sur des chaînes de caractères si elle n'est appelée que sur des entiers !

Notre première contribution sur ce sujet est d'introduire une structure algébrique, appelée *algèbre synchrone*, qui traduit le comportement des automates synchrones uniquement sur les entrées pertinentes.

> CONTRIBUTION. Nous étendons la théorie algébrique des langages—ses objets et ses théorèmes principaux—aux relations reconnues par des automates synchrones.

Dans un second temps, nous nous appuyons sur cette théorie pour résoudre la question d'origine. En général, il n'existe pas d'algorithme pour déterminer l'appartenance à chaque classe de langage. En revanche, nous montrons que dès que c'est possible, ce théorème peut être relevé aux automates synchrones.

> CONTRIBUTION. Nous montrons que dès que l'on peut décider si un langage appartient à une telle classe—ce qui revient essentiellement à dire que ce langage peut être décrit par un automate de « forme simple »—alors on peut toujours décider si un automate synchrone est équivalent à un autre automate synchrone de « forme simple ».

Notons que ce résultat est en fait un méta-théorème : ce n'est pas une caractérisation d'une classe, mais une façon de produire, à partir d'un algorithme caractérisant des langages, un autre algorithme qui caractérise la sémantique des automates synchrones.

En conclusion, cette seconde partie de thèse, qui étudie d'une part les problèmes de satisfaction de contraintes sur des structures infinies et d'autre part la complexité des automates décrivant ces structures, présente deux (méta-)théorèmes diamétralement opposés : le premier caractérise exactement quels problèmes de satisfaction de contraintes sont décidables sur les structures automatiques, et ils s'avèrent en fait particulièrement rares, tandis que le second montre que la construction des automates synchrones préserve la décidabilité, permettant souvent de déterminer si une relation décrite par un automate synchrone peut être représentée par un automate synchrone de « forme simple ».

# Chapter I

# *Introduction*

### Abstract

This chapter serves both as an introduction and as an extended abstract of the thesis. It provides an accessible overview of the broader research area, the specific questions addressed, and the contributions made in this work. Homomorphism problems lie at the heart of many foundational questions in logic, database theory and programming. After exposing the foundational results of these domains in the framework of homomorphisms, this chapter then summarizes the key contributions of this thesis.

While this chapter is intended for readers with a background in theoretical computer science, it avoids formal definitions in favour of intuition and high-level explanations. Formal definitions can be found in Chapters II, III and VII. When needed, terms and symbols can be clicked to navigate directly to their definitions elsewhere in the document.

### Contents

## I.1  The Two Sides of the Homomorphism Problem

This thesis is devoted to studying variations of the HOMOMORPHISM PROBLEM, a central concept in theoretical computer science. Homomorphisms capture the idea of maps preserving relational information from a mathematical structure to another. They arise naturally across a variety of domains: in logic, as the mechanism behind model-checking of primitive-positive formulas; in database theory, as the semantics of conjunctive query evaluation; and in constraint satisfaction, as the formalization of whether a structure satisfies a given set of rules. This thesis approaches the HOMOMORPHISM PROBLEM as a unifying framework, by emphasizing its complementary role in data querying and constraint solving.

To introduce this problem, we present a range of examples: automata acceptance, SQL query evaluation, and Sudoku solving that can all be encoded as HOMOMORPHISM PROBLEMS.[1] Interestingly, they reveal a dichotomy on how problems are encoded: in some, the query appears on the left-hand side of the homomorphism, while on others it appears on the right-hand side. The first type of encoding correspond to what we call *existential problems*, and are rooted in the field of *database theory*: we introduce this domain in Section I.2. On the other hand, the dual encoding deals with *universal problems* studied in the domain of *constraint satisfaction*: it is the focus of Section I.3. This dual perspective on the HOMOMORPHISM PROBLEM forms the basis for the two-part division of this thesis.

*Homomorphisms.*  The simplest mathematical structure is perhaps that of a (directed) finite graph: it consists of a finite set $V$ of vertices (also called *domain*), together with a set of edges $\mathcal{E} \subseteq V \times V$. Then, a homomorphism from a graph to another is a function $f$ between their vertices that preserves the edges, in the sense that if $\langle u, v \rangle$ is an edge, then $\langle f(u), f(v) \rangle$ must also be an edge: we depict an example of graph homomorphism in Figure I.1.

[1] We will see even further examples—3-colourability and SAT-solving—in the next sections.

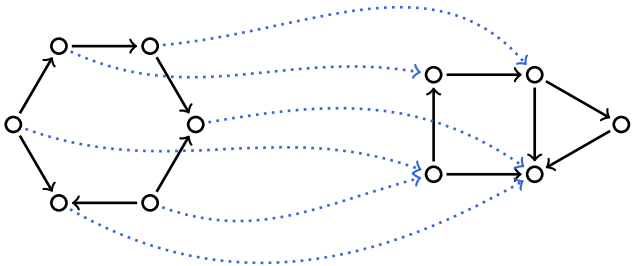

Figure I.1:  Two graphs (in black) and a homomorphism (blue dotted arrows) from the graph on the left-hand side to the right one.

To enrich the structure—but also, perhaps more enjoyably, to add a splash of colour to this thesis—we will consider more complex structures by allowing for multiple edge relations, or even relations of higher arity linking the vertices. The types and arities of relations allowed in a structure are specified by its signature $\sigma$. These richer structures are known as $\sigma$-structures or relational structures—see Figure I.2—, and homomorphisms between $\sigma$-structures are asked to preserve *all* relations in the signature $\sigma$.

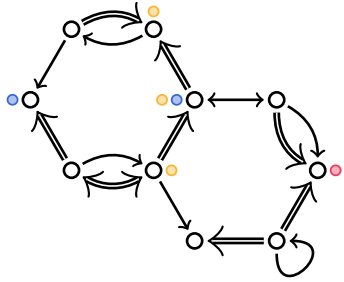

Figure I.2: A relational structure with two kinds of binary relations (represented by simple and double arrows) and three kinds of unary relations (represented by small red, yellow and blue circles next to the vertices).

> The HOMOMORPHISM PROBLEM over $\sigma$
>
> *Input*: Two finite $\sigma$-structures **A** and **B**.
> *Question*: Is there a homomorphism from **A** to **B**?

In the problem above, we refer to **A** as the *source structure* and to **B** as the *target structure*, and we denote by **A** $\xrightarrow{\text{hom}}$ **B** the existence of a homomorphism from **A** to **B**.

More than a mere decision problem—which is easily seen to lie in NP—, the HOMOMORPHISM PROBLEM should rather be understood as a *framework* or *language* to formalize many problems arising in computer science.

**Example I.1.1** (Non-deterministic automata). A non-deterministic automaton **A** can be seen as a relational structure on the signature with two unary predicates (one for describing initial states, one for final states), and one binary predicate for each letter of the alphabet $\Sigma$ describing the transitions. As expected, its vertices are its states, and each predicate is naturally interpreted. Any finite word $u \in \Sigma^*$ can in turn be seen as a relational structure $\mathbf{W}_u$ with $[\![0, |u|]\!]$ as its domain, where 0 is initial, $|u|$ is final, and for each $i \in [\![1, |u|]\!]$, there is an edge from $i - 1$ to $i$ whose type is given by the $i$-th letter of $u$, see Figure I.3. Then, there is a homomorphism from $\mathbf{W}_u$ to **A** if, and only if, the automaton **A** accepts $u$. $\diamond$

Note that in Example I.1.1, not only is the existence of a homomorphism equivalent to the existence of a solution, but the set of homomorphisms naturally corresponds to the set of solutions: homomorphisms from $\mathbf{W}_u$ to **A** exactly correspond to accepting runs of the automaton over $u$.

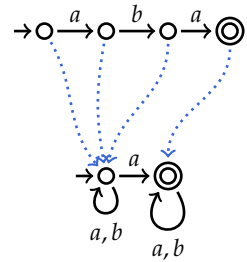

Figure I.3: Automata acceptance as a HOMOMORPHISM PROBLEM: structure representing the finite word *aba* (above), structure representing the minimal automaton of $(a + b)^* a (a + b)^*$ (below) and a homomorphism from the former to the latter (blue dotted arrows). Vertices with a double circle (resp. incoming dangling arrow) represent final (resp. initial) states.

| MOVIES | | | |
|---|---|---|---|
| id | title | duration | director |
| 197 | Eyes Wide Shut | 159 | Stanley Kubrick |
| 205 | J'ai tué ma mère | 96 | Xavier Dolan |
| 304 | Amadeus | 161 | Miloš Forman |
| 321 | 120 Battements par minute | 143 | Robin Campillo |

| ROOMS | |
|---|---|
| id | capacity |
| 1 | 108 |
| 2 | 124 |
| 3 | 96 |
| 4 | 102 |

| PROJECTIONS | | |
|---|---|---|
| movie_id | room_id | time |
| 197 | 2 | 2025-03-28 14:00 |
| 205 | 3 | 2025-03-28 14:30 |
| 321 | 4 | 2025-03-28 14:30 |
| 197 | 1 | 2025-03-28 17:00 |

Table I.4: A relational database consisting of three tables, representing data stored by a cinema.

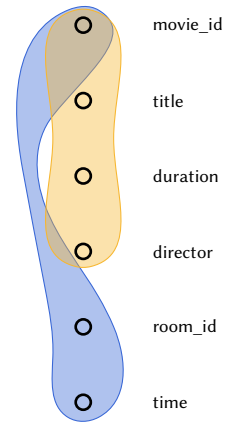

Figure I.5: A SQL query seen as a relational structure. The yellow potato represents the single tuple of the MOVIES relation, and the blue potato surrounds the only tuple that belongs to the PROJECTIONS relation.

**Example I.1.2** (SQL queries). A relational database, such as the one depicted on Table I.4, can easily be seen as a relational structure whose domain is the set of elements occurring somewhere in a table, with one relation for each table.[2]

Consider the following SQL query, which outputs all pairs of movie titles together with their projection time.

```
1  SELECT title, time
2  FROM Movies, Projections
3  WHERE Projections.movie_id = Movies.id;
```

This query $\gamma$ can in fact be seen itself as a relational structure $\mathbf{Q}_\gamma$: its domain has six elements, corresponding to the attributes of the MOVIES and PROJECTIONS table, merged on the attribute Projections.movie_id = Movies.id. Both relations MOVIES and PROJECTIONS consist of a single tuple, and the relation ROOMS is empty, as depicted in Figure I.5.

Then, the set of pairs $\langle x, y \rangle$ such that there is a homomorphism from $\mathbf{Q}_\gamma$ to the relational database, sending title to $x$ and time to $y$ is exactly the output set of the SQL query on the database. ◇

We now provide a last example: Sudoku grids. While it is also encoded as a HOMOMORPHISM PROBLEM, we will see in fact it is of a different nature than the reductions of Examples I.1.1 and I.1.2.

**Example I.1.3** (Sudoku grids). We represent an empty Sudoku grid as the relational structure whose domain is $[\![1, 9]\!] \times [\![1, 9]\!]$—corresponding to coordinates in the grid—with three kinds of binary predicates: $\mathcal{R}$, $\mathcal{C}$ and $\mathcal{S}$, that describe when two coordinates are on the same row, column or 3∗3-square, respectively. To deal with prefilled grids, we add nine unary predicates $P_k$ ($k \in [\![1, 9]\!]$), and if coordinate $\langle i, j \rangle$ is prefilled with number $k$, then predicate $P_k$ must hold on the element $\langle i, j \rangle$. Given a prefilled grid $G$, we denote by $\mathbf{S}_G$ the associated relational structure.

We then define the target structure $\mathbf{T}$ to have $[\![1, 9]\!]$ as its domain, with each vertex representing a possible value filling a cell. The binary relations

---

[2] In fact the only difference between relational databases and relational structures precisely lies in the fact that in the case of the former, the domain is implicit, while for the latter it is explicit. While this difference alters the theory, the difference is mostly negligible for the query languages we will study, see Chapter III.

Figure I.6: A prefilled Sudoku grid.

$\mathscr{R}$, $\mathscr{C}$ and $\mathscr{S}$ all three consist of all pairs $\langle x, y \rangle$ s.t. $x \neq y$. This translates the fact that, if two cells are on the same row, column, or $3*3$ grid, then they must have different values. Moreover, each unary relation $P_k$ ($k \in [\![1, 9]\!]$) is defined to hold only on $\{i\}$.

Then, a prefilled grid $G$ can be completed if, and only if,[3] $\mathbf{S}_G$ has a homomorphism to $\mathbf{T}$. More precisely, homomorphisms $f : \mathbf{S}_G \to \mathbf{T}$ exactly correspond to complete Sudoku grids that extend $G$, with $f(\langle i, j \rangle)$ giving the number contained in cell $\langle i, j \rangle$. ◇

> FOUNDATION. The HOMOMORPHISM PROBLEM is a natural framework in which we can express many logical questions, ranging from database evaluation to Sudoku solving.

All three problems we presented can actually be seen as instances of model-checking: part of the input represents some *data*—or *model*—and part of it represents a *query*—or *logical question/specification*. However, as depicted in Table I.7, the encodings of these problems into HOMOMORPHISM PROBLEMS can be of two types:

- the *query* is encoded in the source structure, and the *data* in the target structure: we denote this family of problems by 'query $\xrightarrow{\text{hom}}$? data';
- in the other cases, the *data* is encoded in the source structure and the *query* is the target structure: we denote these problems by 'data $\xrightarrow{\text{hom}}$? query'.

|  | data | query | hom problem |
|---|---|---|---|
| Ex I.1.1 | automata | is $u$ accepted? | query $\xrightarrow{\text{hom}}$? data |
| Ex I.1.2 | database | SQL query | query $\xrightarrow{\text{hom}}$? data |
| Ex I.1.3 | Sudoku grid | is solvable? | data $\xrightarrow{\text{hom}}$? query |
| Ex I.3.1 | graph | is $k$-colourable? | data $\xrightarrow{\text{hom}}$? query |
| Ex I.3.2 | SAT formula | is satisfiable? | data $\xrightarrow{\text{hom}}$? query |

The two situations are far from symmetric: in model-checking, the size of the data is usually much larger than the query: for instance, when analysing the documents in the *Panama Papers* scandal, the data represented 2.9 TB, while the queries were a few lines long [Neo]. Hence, we often study problems where the query is fixed: accordingly, not only does this fundamental asymmetry between data and query roles motivates the structure of this thesis in two independent parts, it also underlies distinct research domains.

> FOUNDATION. Encodings of model-checking problems as HOMOMORPHISM PROBLEMS are dual by nature, leading to two schools that developed different techniques to tackle them: *database theory* and *constraint satisfaction problems*.

[3] In fact, for this example we could use only one binary predicate instead of three. Note that this encoding is actually quite close to graph colouring (Example I.3.1) with an extra trick to force some values. This trick—formally called marked structure—will actually prove crucial in Chapter VIII.

Table I.7: Summary of the encodings of some natural model-checking problem into HOMOMORPHISM PROBLEMS. The last column indicates whether the structure encoding the data (resp. the query) is on the left-hand side (*i.e.* acts as the source structure) or the right-hand hand (the target structure) of the HOMOMORPHISM PROBLEM. The last two examples will be described in Section I.3.

## I.2 *Existentialism is a Database Theory*

### I.2.1 *Conjunctive Queries*

We now turn to the query $\xrightarrow{\text{hom}}$? data side of the homomorphism problem. This perspective captures classical database evaluation and highlights how such queries naturally express existential, monotonic properties. For instance, if **G** is the graph with two nodes $u$ and $v$ and a single edge from $u$ to $v$, then asking if there is a homomorphism from **G** to a graph **H** amounts to asking if there exists at least one edge in **H**.

As we would expect for any existential problem, they are monotonic: if a solution exists, and we add more data, then a solution still exists. More formally, for any structure **A**, **B** and **B′**, if **B** is a substructure of **B′** and $\mathbf{A} \xrightarrow{\text{hom}} \mathbf{B}$, then $\mathbf{A} \xrightarrow{\text{hom}} \mathbf{B'}$.

SQL queries (Example I.1.2) actually represent more than a mere example: every HOMOMORPHISM PROBLEM $\mathbf{A} \xrightarrow{\text{hom}} \mathbf{B}$ can be seen as a SQL query evaluation problem where **A** encodes a query in the SELECT-FROM-WHERE fragment of SQL and **B** encodes a relational database. This fragment can also be characterized as the fragment of first-order logic where neither universal quantification, nor negation, nor union is allowed. For instance, the SQL query of Example I.1.2 can be expressed by the formula

$$\phi(\text{title}, \text{time}) \triangleq \exists \text{movie\_id}. \exists \text{duration}. \exists \text{director}. \exists \text{room\_id}.$$
$$\text{Movies}(\text{movie\_id}, \text{title}, \text{duration}, \text{director})$$
$$\wedge \text{Projections}(\text{movie\_id}, \text{room\_id}, \text{time}).$$

Yet another characterization of these queries is as conjunctive queries, that consist of a relational structure together with a tuple of vertices, called "output"—this tuple plays the same role as the SELECT statement in SQL. For instance, the previous query can be expressed as the conjunctive query of Figure I.9. The semantics of such a query $\gamma = \langle \mathbf{G}, \bar{x} \rangle$ is defined as follows: given a relational database, seen as a relational structure **D**, it returns every possible tuple $\bar{d}$ of elements of **D** such that there exists a homomorphism from **G** to **D** that sends $\bar{x}$ to $\bar{d}$. Notice how this formalism uses the connection with homomorphisms described in Example I.1.2 to define a semantics. Overall, these characterizations show this query language to be quite robust.

> FOUNDATION. Problems of the form query $\xrightarrow{\text{hom}}$? data exactly correspond to the evaluation of conjunctive queries. Finite unions of conjunctive queries can also be described as the SELECT-FROM-WHERE fragment of SQL, or as the primitive-positive fragment of first-order logic.

We now turn to the complexity of evaluating these queries. As we have seen,

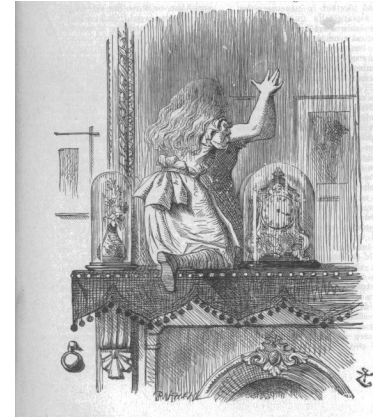

Figure I.8: *Looking glass room*, by John Tenniel.

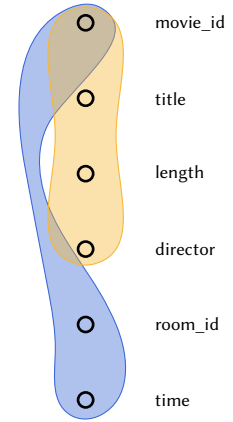

output: title, time

Figure I.9: A conjunctive query.

CONJUNCTIVE QUERY EVALUATION boils down to HOMOMORPHISM PROBLEMS of the form query $\xrightarrow{\text{hom}}$? data. Assuming that the query is fixed, the naive algorithm to solve the HOMOMORPHISM PROBLEM—consisting in enumerating every possible function from $A$ to $B$ and checking whether some of them define a homomorphism—works in polynomial time, as there are only $|B|^{|A|}$ such functions. In fact, it is straightforward to devise an algorithm in uniform-$AC^0$—actually, the depth of the circuit does not even depend on $\mathbf{A}$.[4]

While $|B|^{|A|}$ is indeed polynomial when $\mathbf{A}$ is fixed, recall that $\mathbf{B}$ represents a database: even though most theoreticians pretend that polynomial-time is tractable, a polynomial-time algorithm of degree seven, run over a 2.9 TB database, will likely execute more instructions than there are atoms in the observable universe. This leads to two natural directions:

- optimizing the size of the exponent, *i.e.* replacing the query with a semantically equivalent one of smaller size,
- studying the parameterized complexity of evaluating SQL queries, when parameterized by the size of the query. This provides a finer complexity notion than the classical $NP/AC^0$ approach; our previous remark shows the result to be slicewise polynomial (XP), which is not as well-behaved in practice as fixed-parameter tractable (FPT) problems.

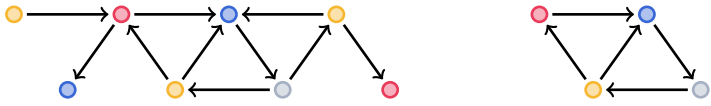

*Query minimization.* The problem of optimizing the SELECT-FROM-WHERE fragment of SQL is well-understood, precisely by using the framework of conjunctive queries and relational structures. This problem amounts to, given a finite $\sigma$-structure $\mathbf{A}$, deciding if there exists a strictly smaller $\sigma$-structure $\mathbf{A}'$ *s.t.*,[5] for any finite $\sigma$-structure $\mathbf{B}$, then

$$\mathbf{A} \xrightarrow{\text{hom}} \mathbf{B} \quad \textit{iff} \quad \mathbf{A}' \xrightarrow{\text{hom}} \mathbf{B}.$$

The property above is in fact equivalent to

$$\mathbf{A} \xrightarrow{\text{hom}} \mathbf{A}' \quad \text{and} \quad \mathbf{A}' \xrightarrow{\text{hom}} \mathbf{A} \tag{I.1}$$

and is hence decidable. The optimization procedure then goes as follows: starting from $\mathbf{A}$, we check for every possible vertex $a \in A$ if $\mathbf{A} \smallsetminus \{a\}$ is equivalent to $\mathbf{A}$ in the sense of Equation (I.1). If some $a$ satisfy the property, we let $\mathbf{A} \smallsetminus \{a\}$ be our new query and start the process again. Otherwise, we get a minimal query, called core of $\mathbf{A}$. This core is unique—which is not obvious since we defined it with a greedy procedure—and is, by construction, a substructure of $\mathbf{A}$, see Figure I.10. In particular, it implies that the core does not only minimize the number of vertices of $\mathbf{A}$ while being semantically equivalent: it also minimizes any parameter that is closed under taking substructures, such as *e.g.* the number of edges or the tree-width![6] Therefore, this notion of core, together with seeing SELECT-FROM-WHERE queries as relational structures/conjunctive queries, provides a remarkably robust tool for solving most optimization problem on these queries.

[4] There is roughly one layer simulating the existential quantifiers, and another one simulating the conjunctions.

[5] Here, our size measure is simply the number of vertices of the structure.

Figure I.10: On the left-hand side a graph, and its core on the right. The colours are not part of the structure, but are used to describe a homomorphism from the original structure to its core.

> FOUNDATION. Conjunctive queries can be minimized by computing their core. This process minimizes the number of variables/vertices, but also many other parameters, such as the number of edges, the path-width, the tree-width, etc.

*FPT algorithms.* In short, the field of parameterized complexity studies the computational complexity of decision problems in a finer way than classical complexity theory: each problem is associated to a parameter—which is smaller than the size of the whole instance—and the goal is to understand the influence of the size of this parameter on the complexity of the problem. Let us mention the parameterized classes FPT and W[1], which are roughly the parameterized equivalent of P and NP. The problem of whether a graph contains a $k$-clique, when parameterized by $k$, is known to be W[1]-complete: in some sense, it means that this problem is hard, and that this parameter $k$ plays a crucial role in the hardness of the problem.

The problem of whether a graph contains a $k$-clique easily reduces to a HOMOMORPHISM PROBLEM where the source structure is fixed—and equal to the $k$-clique. It follows that the HOMOMORPHISM PROBLEM, parameterized in the size of the source structure, is W[1]-hard. Unfortunately, assuming that W[1] ≠ FPT,[7] it follows that there cannot be an FPT (*i.e.* "efficient") algorithm for evaluating conjunctive queries. Hence, a quest emerged to find classes of conjunctive queries with an FPT evaluation.

First, one can notice that if said query is tree-shaped, such as the conjunctive query of Figure I.11, then the naive bottom-up evaluation algorithm works in time that is polynomial both in the size of the query and in the size of the database. Now, assume that $\gamma$ is a conjunctive query that is not tree-shaped, but that is equivalent to a tree-shaped query. This is the same as saying that the core of $\gamma$ is tree-shaped. Hence, to evaluate $\gamma$, one can instead:

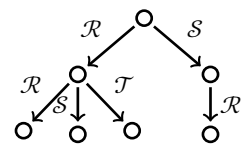

• first compute its core,[8]

• then evaluate this core on the database.

The interest of this approach is that, while databases are big and ever-changing, queries are short and fixed. Hence, spending effort optimizing them can be beneficial, since it might lead to performance gains for *every* evaluation of the query: this is why studying the complexity of the evaluation problem parameterized by the size of the query is relevant. Formally, the previous procedure yields an algorithm that works in time

Figure I.11: A tree-shaped conjunctive query over a signature with three binary relations denoted by $a$, $b$ and $c$.

$$\mathcal{O}(f(|\text{query}|) \cdot \text{poly}(|\text{core}|, |\text{database}|)).$$

This precisely means that evaluating conjunctive queries that are semantically equivalent to tree-shaped queries is FPT.

In fact, for this reasoning to work, the notion of "tree-shaped" need not be as restrictive as what is shown in Figure I.9: for instance, edges could be reversed. More generally, if the query has tree-width at most $k$, then we still get a polynomial-time evaluation algorithm—where the order of the polynomial depends on $k$. In turns, it means that for every $k \in \mathbb{N}_{>0}$, evaluating conjunctive queries that are semantically equivalent to a query of tree-width at most $k$ is FPT.[9]

Remarkably, this is *exactly* the limit of tractability for these queries: Grohe showed that a class of conjunctive queries has FPT evaluation when parameterized in the size of the query *iff* it has bounded "semantic tree-width"—meaning that there exists $k \in \mathbb{N}_{>0}$ *s.t.* every query in the class is semantically equivalent to a query of tree-width at most $k$ [Gro07].[10,11]

> FOUNDATION. Conjunctive queries of bounded semantic tree-width are exactly the classes of conjunctive queries with tractable evaluation.

### I.2.2 Adding Regular Path Predicates

Overall, the previous result proves that the theory of *conjunctive queries* is well-understood. However, even when considering other features from SQL, such as aggregate functions—COUNT, SUM, etc.—, the query language of conjunctive queries faces a big limitation: it is *intrinsically local*. Consider two structures $\mathbf{A}$ and $\mathbf{B}$, and two elements $a$ and $a'$ of $A$. For any homomorphism $f$ from $\mathbf{A}$ to $\mathbf{B}$, the distance from $f(a)$ to $f(a')$ in $\mathbf{B}$ is at most[12] the distance from $a$ to $a'$ in $\mathbf{A}$. Now assume that $\mathbf{B}$ is a graph, whose vertices are humans, and whose edges represent the 'is a child of' relation. For any $k \in \mathbb{N}$, it is easy to build a conjunctive query $\langle \mathbf{A}_k, \langle a, a' \rangle \rangle$ outputting all pairs $\langle b, b' \rangle$ *s.t.* $b'$ is a depth-$k$ descendant of $b$—see Figure I.12 for $k = 2$. However, since homomorphisms contract distances, there is no conjunctive query $\langle \mathbf{A}_*, \langle a, a' \rangle \rangle$ outputting all pairs $\langle b, b' \rangle$ *s.t.* $b'$ is a descendant, at *any* depth, of $b$. In other words, conjunctive queries are not closed under transitive closure.

More generally, human-centered data does not usually go well with relational databases, as they are not designed to allow graph traversal. To face this issue, graph databases have been introduced: they can essentially be modelled as relational structures whose relations are all binary. In other words, they correspond to edge-labelled graphs, see Figure I.14.[13] To illustrate this point, we consider Wikidata, which is a graph database containing more than one hundred million vertices, whose data is used amongst others on Wikipedia.

We would like to obtain all literary works published before 1990 and whose French title contains the string "exist". This can be done using the SPARQL query language, which is roughly the equivalent of SQL for knowledge bases:

[9] In fact, they can even be evaluated in polynomial time, but the argument is more involved [CR00].

[10] Technical point: actually, for the equivalence to hold, the class of queries needs to be recursively enumerable.

[11] The same equivalence holds for polynomial-time evaluation.

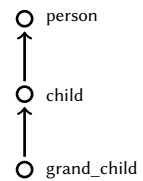

output: person, grand_child

Figure I.12: A conjunctive query outputting all pairs of people with their grandchildren.

[12] Indeed, this follows from the definition of a homomorphism, by using a trivial induction on the distance.

[13] If the reader is familiar with *knowledge graphs*, everything we are saying about graph databases also applies to knowledges graphs. While the two notions are distinct in practice, their fundamental concept—and hence our theoretical modelling of them—is identical: storing data as a graph.

```
1   SELECT ?workLabel ?authorLabel
2   WHERE
3   {
4     ?work wdt:P31/wdt:P279* wd:Q7725634;
5           rdfs:label ?workLabel;
6           wdt:P577 ?date;
7           wdt:P50 ?author.
8     ?author rdfs:label ?authorLabel.
9     FILTER(LANG(?workLabel) = "fr" && LANG(?authorLabel) = "fr").
10    FILTER(CONTAINS(?workLabel, "exist")).
11    FILTER(YEAR(?date) <= 1990).
12  }
13  LIMIT 7
```

Figure I.13: A SPARQL query (above), together with a human-friendly translation (below).
[☞ Run the query.]

```
1   SELECT ?workLabel ?authorLabel
2   WHERE
3   {
4     ?work 'is instance of'/'is sublcass of'* 'literary work';
5           'has label' ?workLabel;
6           'was published on' ?date;
7           'was authored by' ?author.
8     ?author 'has label' ?authorLabel.
9     FILTER(LANG(?workLabel) = "fr" && LANG(?authorLabel) = "fr").
10    FILTER(CONTAINS(?workLabel, "exist")).
11    FILTER(YEAR(?date) <= 1990).
12  }
13  LIMIT 7
```

the query is represented in Figure I.13. The central notion in knowledge graphs and SPARQL is the notion of triplets: x R y. refers to an edge of the R-relation going from x to y. Then x R y; S z. is an abbreviation for the conjunction x R y. x S z. Hence, the central part (ll. 4–8) of the SPARQL query of Figure I.13 should be understood as follows: we are looking for variables ?work, ?typeOfWork (implicit), ?workLabel, ?date and ?authorLabel such that:

- there is a path from ?work to ?typeOfWork obtained by taking an edge 'is instance of', and then an arbitrary number of edges of type 'is subclass of',
- ?typeOfWork should exactly correspond to the vertex called 'type of work',
- ?work and ?author have label ?workLabel and ?authorLabel, respectively,
- ?work was published on ?date, and
- ?work was authored by ?author.

The key feature of query languages for graph databases is illustrated by the wdt:P31/wdt:P279* instruction: this expression does not refer to a single edge of the knowledge graph, but rather to a regular expression formed from these edges. These regular expressions are in fact precisely what allows for easy graph traversal! An example of a match of this expression is provided in red in Figure I.14. The output of the query of Figure I.13 is provided in Table I.15.

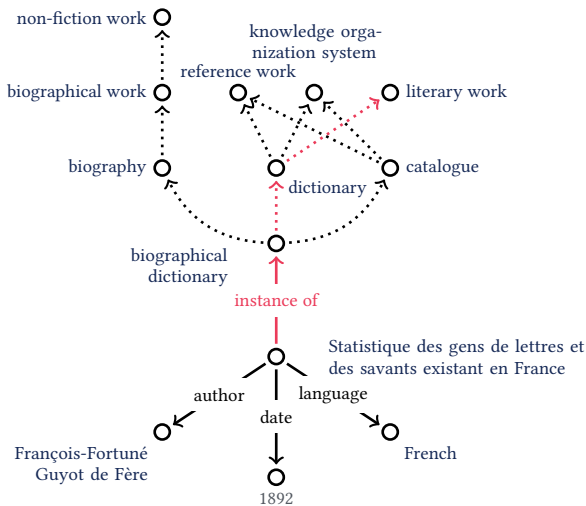

Figure I.14: Part of the Wikidata graph database. Dotted arrows represent the relation 'subclass of'. The red path matches the expression wdt : P31 / wdt : P279 *. For readability, labels are written next to the vertices rather than as separate vertices linked with a 'has label' relation.

| workLabel | authorLabel |
|---|---|
| Statistique des gens de lettres et des savants existant en France | François-Fortuné Guyot de Fère |
| Le Chevalier inexistant | Italo Calvino |
| L'existentialisme est un humanisme | Jean-Paul Sartre |
| Ennui existentiel | Anton Tchekhov |
| Les Ennuis de l'existence | Anton Tchekhov |
| La tentation d'exister | Emil Cioran |
| Inexistence | David Zindell |

Table I.15: Output of the SPARQL query of Figure I.13.

We formalize the fundamental features of SPARQL as conjunctive regular path queries: they consist of conjunctive queries, except that their atoms are no longer of the form $x \xrightarrow{\mathcal{R}} y$ (for some binary relation $\mathcal{R} \in \sigma$), but can be more generally of the form

$$x \xrightarrow{L} y \quad \text{for some regular language } L \text{ over } \sigma.$$

For instance, ll. 1–8 of the SPARQL query of Figure I.13 can be modelled by the conjunctive regular path query of Figure I.16: notice the regular expression in red.

> FOUNDATION. Graph databases/knowledge graphs store information as edge-labelled graphs. To allow for graph traversal, we extend conjunctive queries to conjunctive regular path queries by adding regular expressions.

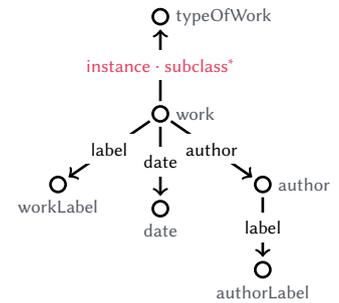

output: workLabel, authorLabel

Figure I.16: A conjunctive regular path query modelling the core part of Figure I.13.

## I.2.3 Minimization and Structure of Conjunctive Regular Path Queries

While conjunctive regular path queries share some enjoyable properties of conjunctive queries—for instance the decidability of semantical equivalence, in contrast to *e.g.* first-order logic—their semantics is more complex: graph-like phenomena (homomorphisms) intertwine with regular languages. Not only does this lead to a complexity blow-up—semantical equivalence is NP-complete for conjunctive queries but ExpSpace-complete for conjunctive regular path queries—, it also breaks the nice theory of cores.

As a consequence, optimizing conjunctive regular path queries poses a significant challenge to untwist graph properties from automata-theoretic ones. This first part of this thesis is dedicated to this problem. After exposing the basic theory of conjunctive regular path queries in Chapter III, we study the MINIMIZATION PROBLEM in Chapter IV: given such a query and $k \in \mathbb{N}_{>0}$, we can decide if it is equivalent to a query of size at most $k$, and if so we can produce it.

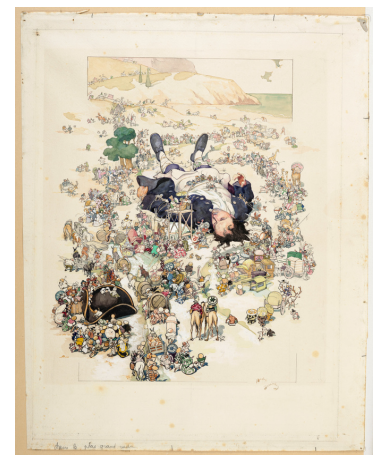

Figure I.17: *Gulliver à Lilliput : Le réveil sur la plage*, by André Devambez.

> CONTRIBUTION. Whether a conjunctive regular path query can be minimized is decidable, and minimization is effective.

We notice that, somewhat unexpectedly, there are some conjunctive regular path queries that are minimal in the sense above, but that are equivalent to a *finite union*—in the semantical sense—of strictly smaller conjunctive regular path queries.[14] We argue that measuring a union of such queries by the maximal size of a query in the union is a sensible thing to do—because the complexity of evaluating such a union depends mostly on this parameter—, and prove that given a conjunctive regular path query and $k \in \mathbb{N}$, we can decide if it is equivalent to a finite union of queries which are all of size at most $k$.

[14] This contrasts with the case of conjunctive queries, where the notion of core and the order-theoretic properties of relational structures precisely prevents this phenomenon from appearing. In other words, this phenomenon precisely emerges by interlacing the graph structure and the automata of the query.

> CONTRIBUTION. Whether a conjunctive regular path query can be minimized as a union of strictly smaller queries is decidable, and minimization is effective.

Both algorithms are essentially brute-force, and the main technical difficulty lies in proving that there are finitely many candidates to test, which is not trivial because we do not ask for any bound on the size of the regular languages.[15] The ideas behind the two algorithms are in fact surprisingly different:

- in the first case—when union is not allowed—, we prove that if a query is equivalent to another one with few atoms (but potentially big languages), then it must be equivalent to a query with few atoms *and* small languages. This property is proved by understanding the subtle interactions between languages and the graph structure;
- in the second case—when union is allowed—, we build a canonical finite union of queries, corresponding to the maximal under-approximation by a finite union of small queries: it is the *best* under-approximation—in the sense that it logically entails the query—and so, if the original query is equivalent to a finite union of small ones, then it must be equivalent to this maximal under-approximation. The difficulty there lies in proving the existence of maximal under-approximations, or rather that it can be expressed by a *finite* union. This construction can essentially be seen as a smart brute-forcing, obtained by agglomerating all possible smaller queries.

One reason we resolve to using a brute-force algorithm is that it is remarkably hard to understand when a query cannot be minimized. The case of conjunctive queries is much simpler: if the core of the query has $k$ edges (resp. tree-width $k$), then any conjunctive query semantically equivalent to it must use at least $k$ edges (resp. have tree-width at least $k$).

Another of our contributions is to identify a sufficient condition on a query so that *any* query that is semantically equivalent to it must contain a "complex pattern". The strength of this theorem lies in its general applicability, as the notion of "complex pattern" is formalized as a "minor-closed class of graphs"—examples include the class of all graphs with at most $k$ atoms, or the class of all graphs of tree-width at most $k$.

> CONTRIBUTION. We introduce the semantical structure theorem, that provides a way to prove lower bounds on the number of atoms, or tree-width, or in fact any minor-closed property, that is necessary to express a query.

This tool proves useful to show minimality of specific examples—*i.e.* for

proofs—, and to prove complexity lower bounds for the decision problems. However, this only provides a sufficient condition that is often not necessary. In particular, it fails at providing a *simple* algorithm to test minimality, justifying why we resolved to using brute-force algorithms.

Then, in Chapter V, we turn to the question of tree-width. Similarly to conjunctive queries, finite unions of conjunctive regular path queries of bounded tree-width can be evaluated in polynomial time. It leads to the question of deciding when a query is actually equivalent to a query of small tree-width.

> FOUNDATION. Barceló, Romero and Vardi [BRV16] devised an algorithm to test if a conjunctive regular path query is equivalent to a finite union of "acyclic"—meaning of tree-width 1—queries.

The general question for tree-width $k$ is left open in their paper as the authors did not know how to extend their technique to this more general setting. We extend their result, relying again on the notion of maximal under-approximation:[16] we prove the existence and computability of the maximal under-approximation by finite unions of queries of tree-width at most $k$ of a given conjunctive regular path query.

> CONTRIBUTION. Given $k \in \mathbb{N}_{>0}$ and a conjunctive regular path query, we can decide if the latter is semantically equivalent to a finite union of queries of tree-width (resp. path-width) at most $k$.

The proof of existence of this maximal under-approximation is much harder than in the case of minimizing the number of atoms. It needs to deal with two kinds of information: the structure of the query, *i.e.* its underlying graph, and its languages, and so the proof precisely massages the query to preserve information, at the same time, about the tree decomposition—serving as a witness of the small tree-width of the query—and about the semantics of the query.

Amusingly, we originally thought that our proof was not able to capture the case $k = 1$ that was already handled, and that the constructions of Barceló, Romero and Vardi and ours were orthogonal. While writing the journal version of this paper—that was originally published at ICDT '23—, we wanted to extend the results to path-width,[17] but part of our construction broke. Introducing the technical tool to fix it[18] actually leads to a unified solution, that handles both the case of tree-width $k$ (including $k = 1$) and path-width.[19]

Lastly, all these algorithms rely on testing the equivalence of conjunctive regular path queries, which is ExpSpace-complete. It leads to resource-hungry algorithms—although it has to be noted that it is worth running exponential algorithms on smallish queries in order to optimize their evaluation on huge

[16] The paper of Barceló, Romero and Vardi also relies on maximal under-approximations, and this notion already existed for conjunctive queries.

[17] The main motivation behind this is that the evaluation of queries of bounded path-width is not only polynomial but even NL!

[18] See the notions of contractions and contracted path-width.

[19] The order of presentation of these results does not follow the timeline of their discovery: our work on semantic tree-width was done in 2022–23 and published at ICDT '23, while the one on minimization was done in 2024–25 and published at PODS '25.

databases!—which leads to a natural quest for identifying subclasses of queries that admit more efficient algorithms.

As witnessed by the example of Figure I.13, the regular expressions used in practice are often much simpler than the ones required to obtain the ExpSpace-hardness of the equivalence problem. Fortunately, conjunctive regular path queries over simple regular expressions, where the regular languages allowed are restricted to be concatenations of edge labels and reflexive-transitive closure, *a.k.a.* Kleene star, of edge labels, were already known to have a more efficient algorithm for testing semantical equivalence.

> CONTRIBUTION. We prove that the problem of minimizing the number of atoms (resp. tree-width) of conjunctive regular path queries over simple regular expressions lies in the polynomial hierarchy.

A consequence of our work on tree-width is that, given a conjunctive regular path query that is promised to be equivalent to a query of tree-width $k$, we can first compute said equivalent query of small tree-width by using our algorithm, and then evaluate it on any database. This proves that the evaluation problem for conjunctive regular path queries of bounded semantic tree-width is FPT when parameterized by the size of the query.[20]

Whether the converse holds remains a mystery: many attempts have been tried to extend Grohe's proof for conjunctive queries to this setting, but all failed, precisely because of the difficulty posed by the intertwining of the graph structure and the automata. We conclude this part of the thesis by a discussion of this problem, as well as whimsical ideas related to conjunctive regular path queries in Chapter VI.

> OPEN PROBLEM. Characterize the classes of CRPQs with FPT evaluation when parameterized by the size of the query.

To summarize, the query $\xrightarrow{\text{hom}}?$ data formulation of the HOMOMORPHISM PROBLEM provides a robust foundation for classical database theory. The first part of this thesis extends this framework to a richer context: graph databases and queries extended with regular path predicates.

## I.3 Everyone Who Wants to Do Constraint Satisfaction Always Ends Up in Universal Problems

### I.3.1 Constraint Satisfaction Problems

This second part explores the complexity of the HOMOMORPHISM PROBLEM when the data is fixed and the query varies, focusing on constraint satisfaction problems and automatic structures.

[20] This result was in fact already known—but proven differently, with an incomparable complexity (better preprocessing but worse polynomial exponent)—by Romero, Barceló and Vardi [RBV17].

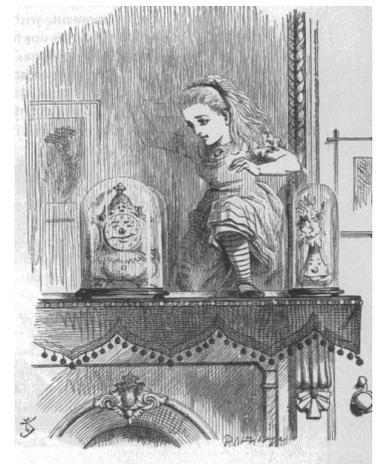

Figure I.18: *Looking glass room*, by John Tenniel.

*Constraint Satisfaction Problems.* Going to the other side, encodings of model-checking problems as HOMOMORPHISM PROBLEMS of the form data $\xrightarrow{\text{hom}}$? query can be thought of as "universal problems"—here "universal" does not refer to some form of completeness, but simply to universal quantification. Notice *e.g.* that they are anti-monotonic with respect to the data: for all structures **A**, **A**′ and **B**, if **A** $\xrightarrow{\text{hom}}$ **B** and **A**′ is a substructure of **A** then **A**′ $\xrightarrow{\text{hom}}$ **B**. Moreover, while problems of the form query $\xrightarrow{\text{hom}}$? data can be solved locally—whether a vertex of the data is part of a solution (a homomorphism) only depends on vertices at a bounded distance—, problems of type data $\xrightarrow{\text{hom}}$? query cannot.

**Example I.3.1** (Graph colouring). Let $k \in \mathbb{N}_{>0}$. We let the *k-clique*, denoted by $\mathbf{K}_k$, to be the graph whose vertices are $[\![1, k]\!]$, and with an edge from $i$ to $j$ (with $i, j \in [\![1, k]\!]$) *iff* $i \neq j$, see Figure I.19. The classical graph-theoretical notion of *k-colouring* of a graph **G** consists of a map from vertices of **G** to $[\![1, k]\!]$ *s.t.* no two adjacent vertices are sent on the same colour/number. We then say that a graph is *k-colourable* when it admits at least one *k-colouring*. In other words, a *k-colouring* corresponds precisely to a homomorphism from **G** to $\mathbf{K}_k$, where colours correspond to the vertices of the clique, see *e.g.* Figure I.20. Hence, a graph is *k-colourable* if, and only if, there is a homomorphism from **G** to $\mathbf{K}_k$. ◇

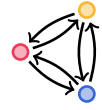

Figure I.19: The 3-clique $\mathbf{K}_3$.

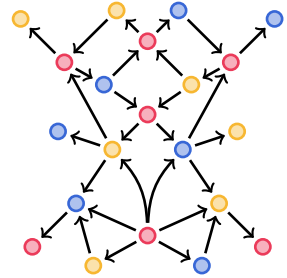

Figure I.20: A 3-colouring of some beetle-shaped graph.

For instance, 3-COLOURABILITY is a global property of a graph and cannot be solved by gluing local solutions, or with greedy algorithms. In particular, this implies that fixing the query does not necessarily result in a drop in the complexity: the 3-COLOURABILITY PROBLEM—which takes as input a finite graph and asks whether it is 3-colourable—is already NP-complete! The next example shows that, even when the target structure is fixed, the HOMOMORPHISM PROBLEM provides a flexible framework to encode problems.

**Example I.3.2** (SAT solving). We consider a 3-SAT instance, namely a finite conjunction of disjunctions of three literals, say

$$\phi \triangleq \bigwedge_{i=1}^{n} \ell_{i,1} \vee \ell_{i,2} \vee \ell_{i,3},$$

where each $\ell_{i,j}$ is either a variable, or the negation of a variable. We assume *w.l.o.g.* that in each clause, positive variables appear before negative ones: this of course can be achieved by a simple syntactical rewriting of each clause.[21] We let **B** be the structure whose domain has two elements $\{0, 1\}$, equipped with four ternary relations $\mathcal{R}_0$ through $\mathcal{R}_3$, where $\mathcal{R}_j$ encodes clauses with exactly $j$ negated literals. They are formally defined as

$$\mathcal{R}_0 \triangleq \{0,1\}^3 \smallsetminus \{\langle 0,0,0 \rangle\}, \qquad \mathcal{R}_1 \triangleq \{0,1\}^3 \smallsetminus \{\langle 1,0,0 \rangle\},$$

$$\mathcal{R}_2 \triangleq \{0,1\}^3 \smallsetminus \{\langle 1,1,0 \rangle\} \quad \text{and} \quad \mathcal{R}_3 \triangleq \{0,1\}^3 \smallsetminus \{\langle 1,1,1 \rangle\}.$$

We then encode $\phi$ into the relational structure $\mathbf{F}_\phi$ whose domain is the set of variables of $\phi$, and for every $i \in [\![1, n]\!]$, $\mathcal{R}_j$ ($j \in [\![0, 3]\!]$) consists of all triplets

[21] Meaning *e.g.* that $x \vee \neg y \vee z$ is not allowed, contrary to $x \vee z \vee \neg y$.

of variables $\langle x, y, z \rangle$ *s.t.* there is a clause of $\phi$ containing variables $x$, $y$ and $z$ (with multiplicity), and exactly $j$ of these variables occur negatively. For instance, $\langle x, y, \neg x \rangle \in \mathcal{R}_1$, and $\langle \neg x, \neg y, \neg z \rangle \in \mathcal{R}_3$. A function $f$ from the domain of $\mathbf{F}_\phi$ to the domain of $\mathbf{B}$ amounts to picking a Boolean valuation of the variables occurring in $\phi$. Observe that, by definition of the relations $\mathcal{R}_j$, given a clause $\psi$ containing variables $x, y, z$, $f$ is a homomorphism from $\mathbf{F}_\psi$ to $\mathbf{B}$ *iff* $f$, seen as a valuation,[22] satisfies $\psi$. By taking conjunction, the conclusion follows: homomorphisms from $\mathbf{F}_\phi$ to $\mathbf{B}$ exactly correspond to valuations satisfying $\phi$. In particular, there is a homomorphism from $\mathbf{F}_\phi$ to $\mathbf{B}$ *iff* $\phi$ is satisfiable.[23] ◇

[22] For instance, if all variables are positive, then all valuations except the one putting all variables to false satisfy the formula. This is why $\mathcal{R}_0$ is defined on $\mathbf{B}$ as $\{0, 1\}^3 \smallsetminus \{\langle 0, 0, 0 \rangle\}$.

[23] Note that this example can be easily generalized to $k$-SAT for any $k \in \mathbb{N}_{>0}$. However, the signature depends on $k$.

However, contrary to 3-COLOURABILITY and SAT solving, not all of these problems are NP-hard. For example, 2-colourability is not only polynomial-time, but can be solved using a greedy algorithm. This begs the question of understanding what makes a relational structure hard for the HOMOMORPHISM PROBLEM when it is used as the target structure. This question is not only motivated by theory: constraint logic programming has emerged in the 1980s with Prolog II and III; and modern programming languages such as answer-set programming provide an efficient way of doing constraint solving.

```
1   x(1..9). % abscissa
2   y(1..9). % ordinate
3   n(1..9). % value
4
5   {sudoku(X,Y,N): n(N)}=1 :- x(X), y(Y).
6
7   subgrid(X,Y,A,B) :- x(X), x(A), y(Y), y(B), (X-1)/3 == (A-1)/3, (Y-1)/3 == (B-1)/3.
8
9   :- sudoku(X,Y,N), sudoku(A,Y,N), X != A.
10  :- sudoku(X,Y,N), sudoku(X,B,N), Y != B.
11  :- sudoku(X,Y,V), sudoku(A,B,V), subgrid(X,Y,A,B), X != A, Y != B.
12
13  #show sudoku/3.
```

Figure I.21: A Clingo program (answer-set programming) to solve Sudoku grids. Written by Enrico Höschler [source]. Try running it on https://potassco.org/clingo/run/!

Answer-set programming can be thought of, albeit caricaturally, as a human-readable SAT-solver. It deals with variables, relations between these variables, and logical rules between these variables. These rules take the form 'A :- B', which can be understood as 'if $B$, then $A$'. The right-hand side is parsed conjunctively while the left-hand side is parsed disjunctively: 'A, B :- C, D' should be understood as 'if $C$ and $D$, then $A$ or $B$'. Figure I.21 provides an example of a Clingo program for solving Sudoku grids.

- it starts by declaring three types of variables: abscissa $x$, ordinates $y$ and values $n$ (representing a value in the grid), as well as their range;
- it introduces a sudoku ternary relation, where sudoku$(x, y, n)$ represents the fact that entry $(x, y)$ of the grid has value $n$, and it says that there should be exactly one value per entry;
- it introduces a subgrid relation, saying when two entries belong to the

same 3∗3-square;
- finally, it says that any two values on the same column, row or subgrid must be different.[24]

To solve a specific grid using the program of Figure I.21, it suffices to add declarations of the form 'sudoku(4,1,5).', where 'A.' is a shorthand for 'A :-.'. This specifies that the cell at position $(4,1)$ has value 5.

Contrary to more classical programming languages, this paradigm does not explain *how* things should be computed, but *what constraints* the memory/solution should satisfy. In HOMOMORPHISM PROBLEMS, the target structure precisely plays this role of encoding constraints. For instance, the only constraint for a graph 3-colouring is that adjacent vertices must be mapped to distinct colours: this constraint is reflected in $\mathbf{K}_3$ by the fact that the edges of $\mathbf{K}_3$ are exactly the pairs of distinct colours.

The field of *constraint satisfaction problems* precisely aims at classifying the structures $\mathbf{B}$ *w.r.t.* to the complexity of the HOMOMORPHISM PROBLEM when the target structure is $\mathbf{B}$. One of the earliest and most impactful result of the domain was found by Schaefer [Sch78], who proved that such problems are either in P or NP-complete when the domain of $\mathbf{B}$ has two elements—this already captures the example of SAT-solving (Example I.3.2) from earlier. A decade later, Hell and Nešetřil [HN90] proved a similar result for undirected graphs. Moreover, in both cases, these dichotomies are effective: given a structure, we can decide if its HOMOMORPHISM PROBLEM is in P or is NP-complete. These results, together with the importance of *constraint satisfaction* in computer science, led Feder and Vardi at the end of the 1990s to state their celebrated *dichotomy conjecture*: "for any *finite structure* $\mathbf{B}$, the HOMOMORPHISM PROBLEM with target structure $\mathbf{B}$ is either P or NP-complete" [FV98]. Despite receiving lots of attention, the conjecture remained wide open for two decades, until Bulatov [Bul17] and Zhuk [Zhu20] independently showed the conjecture to be true.[25]

However not all problems in P are complete for this class: some are even simpler and are complete for NL or L, or even FO^fin—i.e. when it is a first-order definable class of finite structures. One result that will be of major importance in this thesis is a dichotomy theorem by Larose and Tesson separating structures in FO^fin from those that are L-hard [LT09].

> FOUNDATION. The field of constraint satisfaction problems classifies target structures depending on the complexity of their HOMOMORPHISM PROBLEM.

### I.3.2   *Automatic Structures: The Dream Is Not Over Yet*

The second part of this thesis is dedicated to pushing these results to their limit. The structures handled by the HOMOMORPHISM PROBLEM, like most problems in computer science, are usually assumed to be finite. We discuss in this section the generalization of the problem to infinite structures. This is motivated by two facts: not only do infinite structures naturally arise from computational models—the runs of a machine usually form an infinite structure—, they can also model 'mathematical universes'.

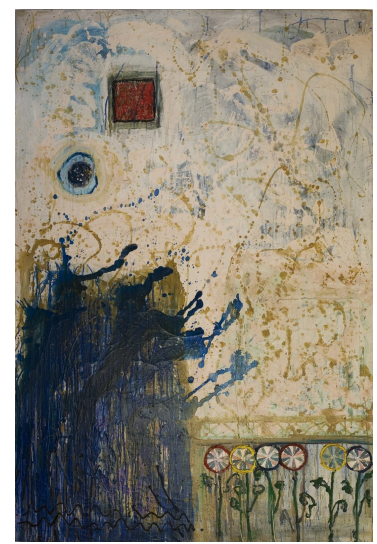

Figure I.22: *Flowers for Algernon*, by Marshall P Baron. Licensed under CC BY SA 4.0.

Von Neumann would have said "It's all over"[26] after hearing Gödel expose his famous incompleteness theorem in 1930: any *effective* (recursively enumerable) consistent theory that is expressive enough to express arithmetic is incomplete—*i.e.* it contains statements that are neither valid (true in all models), nor unsatisfiable (false in all models). In other words, there is no reasonable set of axioms to capture all mathematics: some statements must necessarily fall outside the scope of the theory.

Completeness is perhaps best understood as follows: if a theory (a set of axioms) is consistent—*i.e.* it has at least one model—and complete, then pick any of its models. A statement is then valid in the theory if and only if it is true in this model. Dually, if you are given a theory for which there exists a model with this property (a statement is valid in the theory *iff* it is true in the model), then the theory is complete. In essence, not only does Gödel's result put a dent in Hilbert's dream ("Wir müssen wissen. Wir werden wissen."[27]) of building solid foundations for mathematics: in fact, it shatters the Platonistic idea of an unequivocal and universal mathematical world, or at the very least of one that can be captured by axioms.

Ironically, what makes Gödel's incompleteness theorem a proper abomination for computer scientists is probably another theorem of Gödel, known as Gödel's completeness theorem and that he proved only a year earlier in his Ph.D. dissertation: first-order logic admits a complete proof system. Or, said differently, what is valid—that is, true on every model satisfying the axioms—is exactly what can be proven. Hence, the Gödel of 1929 could have dreamt of a complete theory for mathematics. If a such theory existed, to determine if a statement $\phi$ was valid, it would suffice to enumerate in parallel all possible proofs of $\phi$ and of $\neg\phi$. By completeness, this procedure would always stop, and either conclude that $\phi$ is valid, or that $\neg\phi$ is valid. Are there cardinals strictly between $\aleph_0$ and the continuum? Start Turing's nifty device—invented in 1936—, and you would (eventually) get an answer! In this strange world, automatic theorem proving would be a reality, and this thesis would probably look very different.

Sadly for the Gödel of 1929, the Gödel of 1930 came, and so... "It's all over"! Since then, theories—such has Zermelo–Fraenkel set theory plus the axiom of choice—have been developed, and while not being complete, they manage to capture most of the parts of mathematics we are interested in.[28]

Yet, after half a century of efforts to build solid foundations, the incompleteness of this consolation prize is actually frustrating, and mathematicians still often resort to denial.

"On foundations we believe in the reality of mathematics, but of course when philosophers attack us with their paradoxes we rush to hide behind formalism

[26] This quote is mentioned as historically accurate by [DPPDP09], however this claim seems undocumented [Man11].

[27] "We must know. We will know."

"Another case of men devoting their lives to studying more and more about less and less." —Daniel Keyes, *Flowers for Algernon*.

[28] For the sake of sanity, we assume throughout this introduction that the pronoun 'we' excludes set theorists.

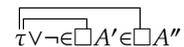

Figure I.23: The first-order sentence $\exists x. (x \notin A') \lor (x \in A'')$ as written by Bourbaki in [Bou06, § 1].

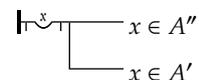

Figure I.24: The first-order sentence of Figure I.23 written using Frege's notations (1879).

and say: "Mathematics is just a combination of meaningless symbols," and then we bring out Chapters 1 and 2 on set theory. Finally we are left in peace to go back to our mathematics and do it as we have always done, with the feeling each mathematician has that he is working with something real. This sensation is probably an illusion, but is very convenient. That is Bourbaki's attitude towards foundations." —Jean Dieudonné [Die70].

For the reader intrigued by what could possibly frighten philosophers in Bourbaki's first volume, we refer them to the formula of Figure I.23—interestingly, this is not the most frightening way to write formulas, see Figure I.24!

However, not all hope is lost: while the Platonistic mathematical world might not be understood, some of it restrictions might be axiomatized. In 1929, Presburger proved that a natural set of axioms for doing arithmetic with only addition is also both complete and decidable—the formulas that are valid in this theory exactly correspond to the sentences that are true on $\langle \mathbb{N}, + \rangle$. Around the same time, Tarski formalized Euclidean geometry as a first-order theory, and proved that is was complete and decidable, see [TG99].

Hence, logic is not completely useless at capturing complex infinite structures! Interestingly, generalizing the idea behind the decidability of Presburger's arithmetic, mathematicians and computer scientists kept rediscovering the notion of automatic structures during the second half of the XXth century.[29] This notion captures the idea on why $\langle \mathbb{N}, + \rangle$ can be simply axiomatized: in this sense automatic structures salvage the shards of mathematicians' shattered dreams. Given an automatic structure and a first-order sentence, we can decide whether it holds on this structure. These structures can be infinite, but are, by definition, describable by finite-state automata—which is what makes decidability possible.

> FOUNDATION. The first-order theory of every automatic structure is decidable.

Unsurprisingly, the foremost example of an automatic structure is $\langle \mathbb{N}, + \rangle$.[30] On the other hand, it should be noted that Peano's arithmetic, namely natural number with addition and multiplication $\langle \mathbb{N}, +, \cdot \rangle$ is not automatic, as it is already undecidable. While automatic structures cannot express "mathematical universes" serving as foundations for a universal mathematical theory, they are surprisingly adequate to express infinite structures arising from computational models. For instance, the graph of runs of a finite-state automaton is automatic, since the unfolding of any finite graph is automatic. Even more generally, the configuration graph of any Turing machine is automatic...[31]

Hence, in Chapter VIII, we naturally study the HOMOMORPHISM PROBLEM when the source structure is allowed to be any automatic structure. Surprisingly, very little was known about this problem: the only known result by Köcher states that whether an automatic graph is 2-colourable is undecidable [Köc14]. Said otherwise, it is undecidable whether an automatic graph admits a homomorphism to the 2-clique. This led us to conjecture that actually most HOMOMORPHISM PROBLEMS on automatic structures should be undecidable, since $\mathbf{K}_2$ is actually somewhat "simple".

[29] It should be noted that while the most common way of proving the decidability of Presburger's arithmetic today is by using automata, this is not Presburger's original proof, who relies on quantifier elimination.

[30] See Example VII.3.1.

[31] See Example VII.3.13.

While the HOMOMORPHISM PROBLEM seems quite natural at first glance, a homomorphism $f$ from an automatic structure $\mathbf{A}$ to a finite one $\mathbf{B}$ does not live in the same world as $\mathbf{A}$ and $\mathbf{B}$, in the sense that it might not be finitely presentable—its domain is infinite and so, *a priori* require infinite information to be described. We introduce the notion of regular homomorphisms, that corresponds to homomorphisms that are finitely presentable in the same fashion as automatic structures, and show that this notion differs from the notion of homomorphism, see Figure VIII.15 and Example VIII.3.9.

Our first contribution is to show that whether a graph admits a regular 2-colouring is undecidable. We then notice that a particular type of HOMOMORPHISM PROBLEM is decidable: for instance, if the target structure is a transitive tournament. This is best understood on an example: consider the 3-transitive tournament depicted in Figure I.25. A homomorphism $f$ from a graph $\mathbf{G}$ to $\mathbf{T}_3$ amounts to a function from the set of vertices of $\mathbf{G}$ to $[\![0, 3]\!]$ *s.t.* for any vertices $u$ and $v$, if there is an edge from $u$ to $v$, then $f(u) < f(v)$. It is clear that the existence of such a mapping is in fact equivalent to asking that there is no path of length 4 in $\mathbf{G}$. In turn, this property can be expressed by a first-order formula, and is hence decidable on automatic graphs. More generally, this property can be extended to any target structure $\mathbf{B}$ with the property that the class of (finite or infinite) structures $\mathbf{A}$ that admit a homomorphism to $\mathbf{B}$ is first-order definable. Luckily for us, this class has been well-studied, and is known as the class of structures with finite duality [Ats08]. In particular, let us cite the result of Larose and Tesson who proved that any HOMOMORPHISM PROBLEM whose target structure does not have finite duality must be L-hard [LT09].

The HOMOMORPHISM PROBLEM on automatic structures is undecidable when the target structure is a finite clique. Yet, it becomes decidable when the target is a finite transitive tournament. This contrast leads to conjecture that finite duality represents the frontier of decidability for automatic structures. In Chapter VIII, we manage to prove this result, and extend it to regular homomorphisms.

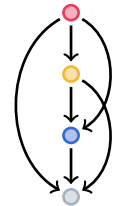

Figure I.25: The 3-transitive tournament $\mathbf{T}_3$.

> CONTRIBUTION. We provide a dichotomy theorem for automatic structures: for any finite structure $\mathbf{B}$, the HOMOMORPHISM PROBLEM with target structure $\mathbf{B}$ is either in NL or is undecidable. The same holds for regular homomorphisms. Moreover, in both cases, these two problems are decidable precisely when $\mathbf{B}$ has finite duality.

Part of the proof, namely the 'undecidability' part, are proven by generalizing Larose and Tesson's reduction, although proving that the 'base problem' of the reduction is undecidable is non-trivial. For proving the decidabil-ity of regular homomorphisms when **B** has finite duality, we provide two alternative proofs: a logical one—which is short but quite abstract—and a graph-theoretical one—the algorithm is much more concrete, but the proof of correctness is quite long. This last proof actually provides new hindsights on an existing algorithm from the literature, known as the hyperedge consistency algorithm.

Since the configuration graph of any Turing machine is an automatic graph, it follows that this dichotomy theorem[32] can be understood as a variation on Rice's theorem, that states that any non-trivial semantical property of a Turing machine is undecidable. Our dichotomy theorem hence implies the following result.

[32] See Theorem VIII.4.1 for a formal statement.

> CONTRIBUTION. Any non-trivial—in the sense of "non-first-order definable"— property on the configuration graph of a Turing machine is undecidable, provided that this property can be expressed as a HO-MOMORPHISM PROBLEM.

One of our motivations for studying this problem was actually originating in the AUT/REC-SEPARABILITY PROBLEM, which takes as input two automatic relations—namely the binary relations over finite words described by synchronous automata—and asks if they can be separated by a recognizable relation, *i.e.* a finite union $\bigcup_i K_i \times L_i$ of Cartesian products of regular languages. We prove this problem to be equivalent to the one taking an automatic graph and asking whether it has a finite regular colouring, which amounts to testing if there exists some $k \in \mathbb{N}_{>0}$ for which the graph admits a regular homomorphism to $\mathbf{K}_k$. We still don't know whether this problem is decidable, even if the results of Chapter VIII hint at its undecidability. Our undecidability result for regular homomorphisms actually yields, when translated back to the vocabulary of separability, that it is undecidable whether two automatic relations are separable by a recognizable relation that can be written as a finite union of $k$ Cartesian products, whenever $k \geq 2$ is *fixed*.

### I.3.3 Language-Theoretic Properties of Automatic Structures

When dealing with regular languages, separability problems are quite common: given a class $\mathcal{C}$ of regular languages, understanding when two regular languages can be separated by a language from $\mathcal{C}$ usually requires a much deeper understanding of the class $\mathcal{C}$ than solving the membership problem for $\mathcal{C}$. In some sense, solving this latter problem only requires a qualitative understanding of $\mathcal{C}$, while the separability problem requires quantitative knowledge on the class. A remarkably efficient tool to prove them decidable is algebraic language theory: this theory associates to every language a canonical algebra, called syntactic monoid, with the property that it is finite if, and only if, the language is regular. Moreover, the language-theoretic and logical properties of the language can be translated to algebraic properties of this monoid: more formally, there is a natural bijection between classes of finite monoids and classes of regular languages under mild closure assumptions.

> FOUNDATION. Algebraic properties of finite monoids correspond to language-theoretic and logical properties of regular languages.

In Chapter IX, motivated by the AUT/REC-SEPARABILITY PROBLEM, we introduce an algebraic theory for automatic relations: these algebras are called synchronous algebras. We prove that each finite-word relation[33] admits a syntactic synchronous algebra, and that this algebra is finite if, and only if, the relation is automatic.

[33] A finite-word relation is simply a subset $\mathcal{R} \subseteq \Sigma^* \times \Sigma^*$ for some finite alphabet $\Sigma$.

We then prove that classes formed of these algebras are in bijection with the classes of automatic relations, under some mild closure assumptions.

> CONTRIBUTION. We extend algebraic language theory to handle relations of finite words rather than only languages of finite words.

Furthermore, we show that this algebraic theory is relevant, in the following sense. A synchronous automaton encodes a pair (from a binary relation) as a finite word. This encoding is injective, but not surjective: not all finite words correspond to encodings of pairs. Hence, the semantics of a synchronous automaton can be precisely seen as the semantics of a classical automaton, together with the promise that it will be only fed inputs that correspond to valid encodings. In other words, the behaviour of such an automaton on words that do not correspond to a valid encoding plays no role whatsoever in its semantics, see Figure I.26.

This approach is ubiquitous in mathematics, and especially in logics: for instance, first-order logic over finite structures is precisely defined as first-order logic over all structures, restricted to finite structures! While being natural and ubiquitous, this construction fails to preserve most properties of the logics: for instance, first-order logic over all structures admits a complete proof system, but does not when restricted to finite structures. The model-checking problem is coRE-complete over all structures, but goes to RE-complete—an incomparable complexity class—for finite structures. Proving a meta-theorem on such a restriction that explains how some property behaves in the restricted universe simply by knowing how it behaves on the larger universe is hence somewhat unexpected but very welcomed!

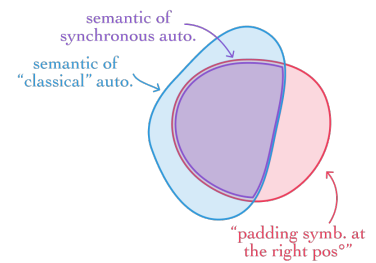

Figure I.26: Semantics of a synchronous automaton.

> CONTRIBUTION. We prove that, for any class of regular languages satisfying mild closure properties, assuming we can decide if a language belongs to this class, then we can decide if an automatic relation can be written as the restriction of a regular language in this class to the set of all valid encodings of pairs of words.

Let us point out that actually, to arrive at this result the notion of synchronous algebras is somewhat intricate. While a more naive definition exists and makes sense, we show that such a result cannot be proven using this simpler notion.

This algebraic theory could provide an interesting framework to study the AUT/REC-SEPARABILITY PROBLEM. While the class of recognizable relations has some desirable closure properties we need, it lacks others: unfortunately, it implies that proving the decidability of the AUT/REC-SEPARABILITY PROBLEM *via* this framework would be highly non-trivial.

> OPEN PROBLEM. Can we decide, given two automatic relations, if they are separable by a recognizable relation?

In summary, this thesis explores two fundamental perspectives on HOMOMORPHISM PROBLEMS: the first extends the theory of conjunctive queries in database theory by adding regular path predicates, to capture natural query languages for human-centered graph-shaped data. It focuses on the problem of query minimization, both in terms of its total number of atoms, or its tree-width, which is a relevant parameter to capture the complexity of its evaluation. The second part focuses on the complexity of problems related to constraint satisfaction over automatic structures. We show that most *structural* problems, probing the shape of the infinite object at hand, are undecidable. In contrast, *language-theoretic* problems, dealing with how these infinite structures are represented—or rather how easy it is to represent them—, remain decidable.

# CHAPTER II

# *Prolegomena*

## ABSTRACT

We introduce the basic definitions and notations used throughout this thesis. Rather than reading it linearly, we recommand to skim it to get an idea of what it contains, and to only go back to this chapter only when needed, using the numerous internal hyperlinks.

## CONTENTS

*Notations.* We try to use notations that syntactically reflect their type: for instance, given a set $X$, we use Roman lowercase letters ($x, y, z, \dots$) to denote elements of $X$, Roman uppercase letters for its subsets ($A, B, C, \dots$), and cursive letters ($\mathcal{X}, \dots$) for sets of subsets of $X$. Functions are denoted by $f, g, h$, etc., and relations by uppercase cursive letters ($\mathcal{R}, \mathcal{S}, \dots$).

Tuples $\langle x_1, \dots, x_k \rangle$ are sometimes denoted by $\bar{x}$. Machines (Turing machines, automata, etc.) are also denoted with uppercase cursive letters ($\mathcal{T}, \mathcal{A}, \dots$). On the other hand, Greek letters are used to either denote queries ($\gamma, \delta, \dots$), formulas ($\phi, \chi, \psi, \dots$) or monoid morphisms ($\phi, \chi, \psi, \eta$). When possible, we try to use the letter to recall the semantics of the object: for instance in Chapter V, $\gamma$ will be used to denote a base query, $\alpha$ for one of its **a**pproximations, $\rho$ for a **r**efinements and $\chi$ for an e**x**pansion. We reserve boldface letters for 'complex objects' (*e.g.* a relational structure $\mathbf{A}$ or a monad $\mathbf{S}$), and blackboard bold for canonical objects (*e.g.* the natural numbers $\mathbb{N}$) or pseudovarieties.

We will use $\mathbb{A}, \mathbb{B}, \dots$ to denote alphabets in Part 1 and $\Sigma, \Gamma, \dots$ to denote them in Part 2.[1] Lastly, decision problems are typesetted in small caps (*e.g.* FINITE REGULAR COLOURABILITY), complexity classes and categories in sans-serif (coNP, ExpSpace, Set, Pos).

## II.1 Set and Functions

We denote by $\mathbb{N}$, $\mathbb{N}_{>0}$ and $\mathbb{Z}$ the sets of natural numbers—that naturally contains zero—, of strictly positive natural numbers, and of integers, respectively. For any $n \in \mathbb{N}_{>0}$, $\mathbb{Z}/k\mathbb{Z}$ denotes the set of integers modulo $k$, and for any $i, j \in \mathbb{Z}$, $[\![i, j]\!]$ is the set of integers from $i$ to $j$, bounds included.

The powerset—*i.e.* the set of all subsets—of a set $X$ is denoted by $\mathfrak{P}(X)$, and $\mathfrak{P}_+(X)$ is defined analogously except that subsets are required to be non-empty. The disjoint union of two sets is denoted by $\sqcup$.

While a function $f$ from set $X$ to set $Y$ is denoted by $f \colon X \to Y$, we reserve $\rightharpoonup$ and $\twoheadrightarrow$ for partial functions and surjections, respectively. The restriction of a function $f$ to a subset $A$ of its domain is denoted by $f|_A$, and the identity function $x \mapsto x$ over a set $X$ is denoted by $\mathrm{id}_X$. The domain of a function, be it either total or partial, is denoted by $\mathrm{dom}(f)$. Given a function $f \colon X \to Y$ and a subset $A \subseteq X$, we denote by $f[A]$ the direct image of $A$ by $f$.[2] Similarly, given a subset $B \subseteq Y$, $f^{-1}[B]$ denotes the inverse image of $B$ by $f$, and when $B$ is a singleton $\{b\}$, we will write $f^{-1}[b]$ instead.

A binary relation $\mathcal{R}$ over sets $X$ and $Y$—*i.e.* a subset $\mathcal{R} \subseteq X \times Y$— is said to be *functional* when for every $x$, there exists at most one $y$ *s.t.* $\langle x, y \rangle \in \mathcal{R}$. When moreover $X = Y$, we say that it is:

- *reflexive* when $\langle x, x \rangle \in \mathcal{R}$ for all $x$;
- *symmetric* when $\langle x, y \rangle \in \mathcal{R}$ *iff* $\langle y, x \rangle \in \mathcal{R}$ for all $x, y$;
- *transitive* when $\langle x, y \rangle \in \mathcal{R}$ and $\langle y, z \rangle \in \mathcal{R}$ imply $\langle x, z \rangle \in \mathcal{R}$ for all $x, y, z$.

An equivalence relation over $X$ is any binary relation satisfying the three previous axioms. Given an equivalence class $\sim$ over a set $X$ and an element $x \in X$, we denote by $[x]^{\sim}$ the equivalence class of $x$ under $\sim$. Lastly, $\hat{=}$ denotes the definition symbol: $x \hat{=} y$ means that $x$ is defined as $y$.

## II.2 Relational Structures

### II.2.1 Basic Notions on Structures

Relational structures generalize the concept of graphs by allowing (1) multiple kinds of relations and (2) relations of higher arity. This data is made explicit in the signature—also called *vocabulary* or even *schema*. We start by defining a *purely relational signature*, which consists in a set of elements, called *predicates*,[3] together with, for each of these elements, a strictly positive natural number, called *arity*. We denote by $\mathcal{R}_{(k)} \in \sigma$ the fact that predicate $\mathcal{R}$, of arity $k$, is part of signature $\sigma$. Then, a *relational signature*, or *signature* for short, consists of a purely relational signature together with a set of constant symbols.[4]

Then, given a signature $\sigma$, a *$\sigma$-structure* **A** consists of:

- a set $A$, called *domain*,
- for each predicate $\mathcal{R}_{(k)} \in \sigma$, a $k$-ary relation $\mathcal{R}(\mathbf{A}) \subseteq A^k$, and
- for each constant $c \in \sigma$, an element $c(\mathbf{A}) \in A$.

We call $\mathcal{R}(\mathbf{A})$ (resp. $c(\mathbf{A})$) the *interpretation* of predicate $\mathcal{R}_{(k)}$ (resp. constant $c$) in **A**. By analogy with graphs, elements of the domain are sometimes referred to as *vertices*. See Figure I.2 for an example of $\sigma$-structure.

The *graph signature* is a purely relational signature consisting of a single binary predicate, either written $\mathcal{E}$ in prefix notation or $\rightarrow$ in infix notation. Then, the $\sigma$-structures over this signature exactly consists of *directed graphs*.

An element of $\mathcal{R}_{(k)}(\mathbf{A})$ is called an *$\mathcal{R}$-tuple* of structure **A**. We also use the terminology *edge* in place of tuple for binary predicates. An *hyperedge* of **A** will designate any of its $\mathcal{R}$-tuples, indifferently of the predicate $\mathcal{R}$.

A $\sigma$-structure **A** is said to be *finite* when its both its domain and its set of hyperedges are finite. In particular, note that this last condition amounts to asking that (1) for every predicate $\mathcal{R}_{(k)} \in \sigma$, the relation $\mathcal{R}_{(k)}(\mathbf{A})$ is finite, and (2) there exists only finitely many predicates $\mathcal{R}_{(k)} \in \sigma$ *s.t.* $\mathcal{R}_{(k)}(\mathbf{A})$ is non-empty.

A *substructure* of a $\sigma$-structure **A** is another $\sigma$-structure **A′** such that:

- the domain $A'$ of **A′** is a subset of $A$,
- each interpretation of a predicate in **A′** is a subset of its interpretation in **A**, and
- every constant of **A** belongs to $A'$, and the interpretation of the constants in both structures coincide.

A *proper substructure* is a substructure for which at least one of the inclusions in the first two points above is strict: in other words, such a substructure should miss at least one element, or one hyperedge. Given a subset $X$ of the domain $A$ of a $\sigma$-structure **A**, the *substructure of* **A** *induced by $X$* is:

[3] We do *not* assume this set to be finite.

[4] Usually the notion of signature also allows for function symbols beyond the degenerate case of constants. However all the signatures considered in the thesis will be relational, justifying the abuse of terminology.

- undefined if not all constants of **A** belong to $X$,
- otherwise, it is obtained from **A** by restricting its domain to $X$, and by intersecting its $k$-ary relations with $X^k$.

The roles played by constants and predicates are obviously not symmetric. For this reason we will often rather deal with pointed structures. Formally, given a purely relational signature $\sigma$, a *pointed $\sigma$-structures* consists of a $\sigma$-structure together with a *tuple* of elements of its domain. Note that pointed $\sigma$-structures with tuples of size $k \in \mathbb{N}$ are in natural bijection with the set of $\sigma'$-structures, where $\sigma'$ is obtained from $\sigma$ by adding a set of $k$ constants.

### II.2.2 Constructions on Structures

The *disjoint union* of two structures over a purely relational signature is obtained by taking the disjoint union of their domain and of their predicate interpretations, and is denoted by $\uplus$. Over other signatures, we apply the same construction but then identify the constants in both structures.[5]

Let again $\sigma$ be any signature. Given two $\sigma$-structures **A** and **B**, their *Cartesian product* $\mathbf{A} \times \mathbf{B}$ is defined by taking the Cartesian product of their domain, of their predicate interpretations and of their interpretations of constants. Their *block product* has the same domain and constants, but now the interpretation of predicate $\mathcal{R}_{(k)}$ in the block product consists of all tuples

$$\langle\langle a_1, b_1\rangle, \dots, \langle a_k, b_k\rangle\rangle \quad s.t.$$

- either $\langle a_1, \dots, a_k\rangle \in \mathcal{R}(\mathbf{A})$ and $b_1 = \dots = b_k$, or
- $a_1 = \dots = a_k$ and $\langle b_1, \dots, b_k\rangle \in \mathcal{R}(\mathbf{B})$.

See Figures II.1 and II.2 for examples. The $k$-fold iteration of **A** is denoted by $\mathbf{A}^k$ and is defined as

$$\underbrace{\mathbf{A} \times \dots \times \mathbf{A}}_{k \text{ times}}.$$

Note that by construction, the disjoint union, Cartesian product and block product of two structures over a signature yields a structure over the *same* signature.[6]

### II.2.3 Adjacencies

Given a $\sigma$-structure **A** and $a \in A$, we define the *adjacency*[7] of $a$ in **A** to be the tuple of sets

$$\mathcal{A}dj_{\mathbf{A}}^{\mathcal{R},i}(a) \triangleq \Big\{\langle a_1, \dots, a_{i-1}, a_{i+1}, \dots, a_k\rangle \in A^{k-1}$$
$$\mid \langle a_1, \dots, a_{i-1}, a, a_{i+1}, \dots, a_k\rangle \in \mathcal{R}(\mathbf{A})\Big\},$$

when $\mathcal{R}$ ranges over predicate of arity $k$ of $\sigma$ and $i \in [\![1, k]\!]$. For graphs, the adjacency of a vertex corresponds to its set of predecessors and its set of successors.

[5] Hence this union is not entirely disjoint, however we keep the name for uniformity. As we will see in Section III.1.3, this construction actually satisfies the universal property of coproducts.

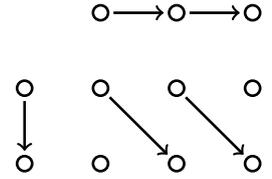

Figure II.1: Two graphs (above and left) and their Cartesian product (below right).

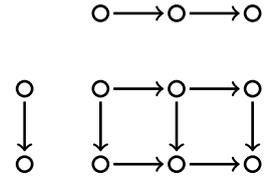

Figure II.2: Two graphs (above and left) and their block product (below right).

[6] This is what motivated the somewhat strange definition of disjoint union for structures over signatures that are not purely relational.

[7] We do not use the terminology *neighbourhood* since it usually refers to a set of elements, namely the set of elements occurring in the adjacency.

### II.2.4  Undirected Paths

An *undirected path* in a $\sigma$-structure **A** consists of a sequence

$$\langle a_0, \bar{h}_0, a_1, \ldots, \bar{h}_{n-1}, a_n \rangle, \text{ with } n \in \mathbb{N},$$

where $a_i \in A$ and each $\bar{h}_i$ is a hyperedge of **A** *s.t.* both $a_i$ and $a_{i+1}$ occur in $\bar{h}_i$. When such an undirected path exists, we say that there is an undirected path between $a_0$ and $a_n$, or equivalently that $a_0$ and $a_n$ are *connected*.[8] A connected component of **A** consists of an equivalence class under this relation.

An *undirected graph* consists of a domain, together with a set of (unordered) pairs of elements of the domain. The *incidence graph* of a $\sigma$-structure, for some signature $\sigma$, is the following undirected graph:

- its domain is the disjoint union of $A$ and the hyperedges of **A**;
- there is an edge between two vertices *iff* one of them is a vertex $a$ of **A**, and the other is an hyperedge $\bar{h}$ of **A**, with the property that $a \in \bar{h}$.

The *distance* between two vertices of a $\sigma$-structure is defined as half of their distance in the incidence graph.[9] The *diameter* of a structure is the maximum over $u$ and $v$ of the distance between vertices $u$ and $v$.

The ball $\mathcal{B}_{\mathbf{A}}^m(a)$ centred at vertex $a \in A$ and of radius $r \in \mathbb{N}$ of a $\sigma$-structure **A** is the substructure of **A** induced by all vertices at distance at most $r$. A structure is said to be *locally finite* when every ball of finite radius is finite.

A *simple path* in a $\sigma$-structure is a simple path in its incidence graph—in other words, such a path should alternate between vertices and hyperedges, and visit each of them at most once.

### II.2.5  Graphs

A *directed cycle* in a graph consists of a non-empty directed path from a vertex to itself. A *directed acyclic graph*, or *DAG* for short, is any graph that contains no directed cycle.

A *directed tree* is a non-empty graph *s.t.* there exists a vertex $r$ with the property that for every vertex $v$, there exists exactly one directed path from $r$ to $v$: see Figure II.3.

The *chromatic number* of a graph is the least cardinal $k$ *s.t.* it is $k$-colourable. We say that a graph is *finitely colourable* when its chromatic number is finite.

### II.2.6  Homomorphisms

A *homomorphism* from a $\sigma$-structure **A** to to a $\sigma$-structure **B** consists of a function $f$ from $A$ to $B$, *s.t.* for every $\mathcal{R}_{(k)} \in \sigma$, for every $\langle a_1, \ldots, a_k \rangle \in \mathcal{R}(\mathbf{A})$, we have $\langle f(a_1), \ldots, f(a_k) \rangle \in \mathcal{R}(\mathbf{B})$. Moreover, for every constant $c \in \sigma$, we must have $f(c(\mathbf{A})) = c(\mathbf{B})$.

An *embedding* is an injective homomorphism, whereas a *strong onto homomorphism* is a homomorphism that is both surjective on the domain and on the hyperedges. The existence of a strong onto homomorphism from a structure **A** to a structure **B** is denoted by $\mathbf{A} \xrightarrow{\text{hom}} \mathbf{B}$, and we say that **B** is a *homomorphic image* of **A**.

[8] Note that this defines an equivalence relation.

[9] By construction the incidence graph is bipartite and hence the distance between two vertices of the structure is even.

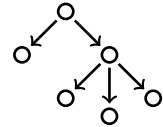

Figure II.3: A directed tree.

An *isomorphism* from **A** to **B** is a homomorphism $f\colon \mathbf{A} \to \mathbf{B}$ *s.t.* there exists another homomorphism $g\colon \mathbf{B} \to \mathbf{A}$ with the property that $g \circ f = \text{id}_A$ and $f \circ g = \text{id}_B$. Equivalently, it is a strong onto homomorphism that is also an embedding. Two structures are *isomorphic*, denoted by $\cong$, if there is an isomorphism between them, and an *automorphism* is an isomorphism from a structure to itself.

Given structures $\mathbf{A}_1, \dots, \mathbf{A}_k$ sharing the same signature, the $i$-th projection from the Cartesian product $\mathbf{A}_1 \times \dots \times \mathbf{A}_k$ to $\mathbf{A}_i$, defined by $\langle a_1, \dots, a_k \rangle \mapsto a_i$ ($i \in [\![1, k]\!]$), is a homomorphism, and is denoted by $\pi_i$.

Given a $\sigma$-structure **A**, a *congruence* on **A** is an equivalence class $\sim$ of $A$ *s.t.* for every $\mathcal{R}_{(k)} \in \sigma$, for every $\langle a_1, \dots, a_k \rangle \in \mathcal{R}(\mathbf{A})$, for any tuple $\langle a'_1, \dots, a'_k \rangle \in A^k$ *s.t.* $a_i \sim a'_i$ for each $i \in [\![1, k]\!]$, then $\langle a'_1, \dots, a'_k \rangle \in \mathcal{R}(\mathbf{A})$. The *quotient structure* defined by a congruence $\sim$ on a structure **A** has the equivalence classes of $\sim$ as its domain, and the natural interpretation of the predicates and constants.

We will often implicitly use Noether's first isomorphism theorem: the substructure of **B** induced by the image $f[A]$ of $A$ is isomorphic to the quotient of **A** by ker $f$.

### II.2.7 Cores

Two $\sigma$-structures **A** and **B** are *homomorphically equivalent* if $\mathbf{A} \xrightarrow{\text{hom}} \mathbf{B}$ and $\mathbf{B} \xrightarrow{\text{hom}} \mathbf{A}$.

A retraction of **A** is a substructure **A′** of **A** together with a homomorphism from **A** to **A′** with the property that any vertex of $A'$ is sent on itself.

**Proposition II.2.1.** Any finite $\sigma$-structure **A** admits a unique[10] minimal—in the number of vertices—retraction.

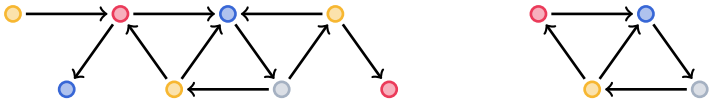

*Proof.* The existence is trivial. For the uniqueness, consider two retractions $f_1\colon \mathbf{A} \xrightarrow{\text{hom}} \mathbf{B}_1$ and $f_2\colon \mathbf{A} \xrightarrow{\text{hom}} \mathbf{B}_2$. We want to prove that $\mathbf{B}_1$ and $\mathbf{B}_2$ are isomorphic. Since $\mathbf{B}_1$ is a substructure of **A**, consider $f_2|_{B_1}\colon \mathbf{B}_1 \xrightarrow{\text{hom}} \mathbf{B}_2$. By minimality of $\mathbf{B}_2$, this homomorphism must be surjective. By symmetry, $f_1|_{B_2}\colon \mathbf{B}_2 \xrightarrow{\text{hom}} \mathbf{B}_1$ is also a surjective homomorphism. By composition, we obtain a surjective homomorphism from $\mathbf{B}_1$ to itself and from $\mathbf{B}_2$ to itself. By finiteness, these surjective homomorphisms must actually be automorphisms. Hence, it follows that $f_2|_{B_1}$ and $f_1|_{B_2}$ are isomorphisms, and hence $\mathbf{B}_1$ is isomorphic to $\mathbf{B}_2$. □

This unique minimal retraction of **A** is called *core* of **A** and is denoted by $\check{\mathbf{A}}$. By construction, the core of $\mathbf{A}$ is a substructure of $\mathbf{A}$ to which it is homomorphically equivalent, see Figure II.5. In general, a *core* is any $\sigma$-structure such that is the core of some structure—or equivalently of itself.

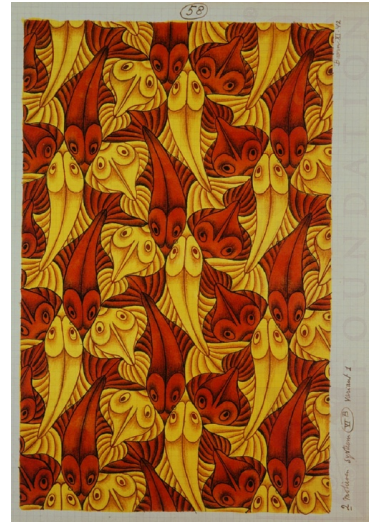

Figure II.4: *2 motifs system VI(b) variant 1*, M. C. Escher, © The M.C. Escher Company.

[10] Throughout this thesis, by "unique" we mean "unique up to isomorphism".

Figure II.5: On the left-hand side a graph, and its core on the right. The colours are not part of the structure, but are used to describe the retraction of the original structure onto its core. (Replica of Figure I.10.)

**Proposition II.2.2.** A finite $\sigma$-structure $\mathbf{C}$ is a core if, and only if, every homomorphism from $\mathbf{C}$ to itself is an automorphism.

*Proof.* For the left-to-right implication, we let $f\colon \mathbf{C} \to \mathbf{C}$ be a homomorphism. Then $f[\mathbf{C}]$ must be isomorphic to $\mathbf{C}$, otherwise we would obtain a strictly smaller retraction. Hence, $f$ is a strong onto homomorphism from $\mathbf{C}$ to itself, and hence is an automorphism.

Conversely, assuming that any homomorphism from $\mathbf{C}$ to itself is an automorphism we get in particular that any retraction must be an automorphism, and hence that $\mathbf{C}$ is isomorphic to $\check{\mathbf{C}}$. $\square$

**Proposition II.2.3.** Two finite structures are homomorphically equivalent if, and only if, their cores are isomorphic.

*Proof.* The right-to-left implication is trivial. For the converse one, denote the two structures by $\mathbf{A}_1$ and $\mathbf{A}_2$, and suppose that $\mathbf{A}_1$ is homomorphically equivalent to $\mathbf{A}_2$. Using the homomorphical equivalence of $\mathbf{A}_1$ and $\mathbf{A}_2$, we get retractions of $\mathbf{A}_2$ onto $\check{\mathbf{A}}_1$ and of $\mathbf{A}_1$ onto $\check{\mathbf{A}}_2$. It follows that we have surjective homomorphisms from $\check{\mathbf{A}}_1$ to $\check{\mathbf{A}}_2$ and conversely. Hence, $\check{\mathbf{A}}_1$ and $\check{\mathbf{A}}_2$ are isomorphic. $\square$

**Proposition II.2.4.** Given a $\sigma$-structure $\mathbf{B}$, if $\mathbf{B}$ is a core, then two elements $b_1$ and $b_2$ of $\mathbf{B}$ have the same adjacency *iff* $b_1 = b_2$.

*Proof.* The right-to-left implication is trivial. For the converse one, consider the homomorphism from $\mathbf{B}$ to itself which maps $b_2$ to $b_1$, and all elements of $B \smallsetminus \{b_2\}$ to themselves. Since we assumed that $b_1$ and $b_2$ have the same adjacency, this is indeed a homomorphism, which is clearly not bijective, and $\mathbf{B}$ is not a core. $\square$

## II.3  Logic Related Notions

### II.3.1  First-Order Logic and Beyond

We fix a purely relational signature $\sigma$. A *Boolean (semantical) query* is any subclass of the class of all $\sigma$-structures. More generally, a $k$-ary *query* is a function that maps any $\sigma$-structure to a (potentially empty) set of $k$-tuples of vertices. We now turn to more syntactical definitions of queries.

We assume that we are given a countable infinite set of variables. A *first-order formula* is any formula generated by the grammar

$$\phi ::= \mathcal{R}(x_1, \ldots, x_k) \mid \neg\phi \mid \phi \vee \phi \mid \phi \wedge \phi \mid \exists x.\phi \mid \forall x.\phi,$$

where the $x_i$'s range over the set of variables and $\mathcal{R}_{(k)}$ over the signature $\sigma$.

> HYPOTHESIS. We assume that, even when not mentioned explicitly, the signature contains a binary predicate = that is interpreted over all structures as equality.

A *first-order sentence* is any first-order formula with no free variable. Given a first-order formula $\phi$ with free variables $\bar{x}$, denoted by $\phi(\bar{x})$, and a pointed $\sigma$-structure $\langle \mathbf{A}, \bar{a} \rangle$ where the arity of $\bar{a}$ coincides with the one of $\bar{x}$, we denote by $\langle \mathbf{A}, \bar{a} \rangle \vDash \phi(\bar{x})$ the fact that $\langle \mathbf{A}, \bar{a} \rangle$ is a model of $\phi(\bar{x})$: this can be defined by a trivial induction on the formula, by interpreting:

- the free variable $x_i$ as $a_i$,
- the predicates $\mathcal{R}$ as the relation $\mathcal{R}(\mathbf{A})$,
- $\neg$, $\vee$, $\wedge$, $\exists$ and $\forall$ as the Boolean operators of negation, disjunction and conjunction, and as the existential and universal quantifiers, respectively.

Given a $\sigma$-structure $\mathbf{A}$, we then denote by $[\![\phi(\bar{x})]\!]^{\mathbf{A}}$ the set of tuples $\bar{a}$ *s.t.* $\langle \mathbf{A}, \bar{a} \rangle \vDash \phi(\bar{x})$.

**Remark II.3.1.** Strictly speaking, the data $\phi(\bar{x})$ is richer than just a formula, however we will abusively still call this pair a *formula*. $\diamond$

We now define the classes $\Pi_n$ and $\Sigma_n$ ($n \in \mathbb{N}$) of first-order formulas as least fixpoints, with the property that $\Sigma_0 \subseteq \Sigma_1 \subseteq \ldots$ and dually for $\Pi_n$. We let $\Sigma_0 = \Pi_0$ be the set of quantifier-free formulas. Then for all $n \in \mathbb{N}$, we consider the following rules:

$$\frac{\phi \in \Sigma_n}{\neg \phi \in \Pi_n} \qquad \text{and} \qquad \frac{\phi \in \Pi_n}{\neg \phi \in \Sigma_n},$$

moreover $\Pi_n \subseteq \Sigma_{n+1}$ and $\Sigma_{n+1}$ is closed under disjunction, conjunction and existential quantification, and dually $\Sigma_n \subseteq \Pi_{n+1}$ and $\Pi_{n+1}$ is closed under disjunction, conjunction and universal quantification. Formally, the hierarchies

$$\Sigma_0 \subseteq \Sigma_1 \subseteq \cdots \qquad \text{and} \qquad \Pi_0 \subseteq \Pi_1 \subseteq \cdots$$

are defined as the smallest sets of formulas satisfying these rules. Then, the *quantifier alternation rank* of a formula $\phi(\bar{x})$ is the least $n \in \mathbb{N}$ *s.t.* $\phi(\bar{x})$ belongs to either $\Sigma_n$ or $\Pi_n$. Formulas from $\Sigma_1$ are called *existential formulas*, and we define a few important fragments of this logic in Table II.6 by restricting which operators are allowed to build formulas.

| Fragment of existential FO | Operators allowed |
|---|---|
| *existential-positive formulas* | $\exists$, $\wedge$, $\vee$ |
| *primitive-positive formulas* | $\exists$, $\wedge$ |
| *positive quantifier-free formulas* | $\wedge$ |

Table II.6: A few fragments of existential formulas defined by restricting the operators that are allowed.

For instance, $\exists x. \exists y. \neg \mathcal{R}(x, y)$ is an existential formula but is not existential-positive. On the other hand, $\exists x. \exists y. \mathcal{R}(x, y)$ is primitive-positive, and hence also existential-positive.

A relation over a $\sigma$-structure $\mathbf{A}$ is said to be *first-order definable* when it can be written as $[\![\phi(\bar{x})]\!]^{\mathbf{A}}$ for some first-order formula $\phi(\bar{x})$. Moreover, a class of $\sigma$-structures is said to be *first-order definable* when there exists a first-order sentence $\phi$ *s.t.* the class of structures $\mathbf{A}$ *s.t.* $\mathbf{A} \vDash \phi$ is precisely the class itself.

*First-order logic* simply consists of the syntax of first-order formulas together with their semantics. Lastly, *monadic second-order logic* (resp. *second-order logic*) is obtained from first-order logic by also allowing quantifications over subsets of the structure (resp. relations of arbitrary arity over the structure).

### II.3.2 Automata Theory

We assume the reader to be familiar with the basic notions of automata theory. Given a set $X$, we denote by $X^*$ and $X^+$ the set of finite words over $X$, and of non-empty finite words over $X$, respectively. The empty word is denoted by $\varepsilon$. An *alphabet* is nothing else but a finite set, and we denote by $2$ the binary alphabet $\{0, 1\}$.

In an automaton $\mathcal{A}$, we denote by $p \xrightarrow{a} q \in \mathcal{A}$ the fact that there is an $a$-labelled transition from $p$ to $q$. A *regular language* is any language—*i.e.* subset of $\Sigma^*$ for some alphabet $\Sigma$—that can be recognized by a finite-state automaton.

The signature of words over an alphabet $\Sigma$ has a binary predicate $\leq$ as well as a unary predicate $a$ for each $a \in \Sigma$. A word $w_0 \cdots w_{n-1}$ of length $n$ can be seen as a structure over the signature of $\Sigma$-words by taking $[\![0, n-1]\!]$ as its domain, interpreting $\leq$ naturally, and interpreting $a \in \Sigma$ as the set of $i \in [\![0, n-1]\!]$ *s.t.* $w_i = a$. It is well-known[11] that a language is regular if, and only if, it is definable in monadic second-order logic. When $\Sigma = 2$, the signature of $\Sigma$-words is also called the *signature of binary strings*.

[11] See Section VII.1.1 for details.

### II.3.3 Monoids

We refer the reader to Pin's seminal lecture notes [Pin22] for an introduction to algebraic language theory.

A *monoid* $\mathbf{M} = \langle M, \cdot, 1 \rangle$ is a set $M$ together with an associative binary operator $\cdot$ called *product*, and an element $1 \in M$, called *unit*, *s.t.* $x \cdot 1 = x = 1 \cdot x$ for all $x \in M$. A *monoid morphism* is a function between monoids that preserve the product and unit.

Monoids—or rather monoid morphisms—can be used to recognize languages as $\Sigma^*$ is itself a monoid under concatenation—actually, it is *the free monoid* over $\Sigma$. A language $L$ is regular if, and only if, there exists a finite monoid $\mathbf{M}$, and subset $\mathrm{Acc} \subseteq M$ (called *accepting elements*), and a monoid morphism $\phi \colon \Sigma^* \to M$ *s.t.* $L = \phi^{-1}[\mathrm{Acc}]$. Another way of thinking of the pair $\langle \phi, M \rangle$ is as follows: a deterministic complete semiautomaton can be described as a set $Q$ together with a monoid right action of $\Sigma^*$ over $Q$. On the other hand, any monoid morphism $\phi \colon \Sigma^* \to M$ yields both a monoid left action and a monoid right action of $\Sigma^*$ over a set $M$, with the properties that $u \cdot (x \cdot v) = (u \cdot x) \cdot v$. In other words, while automata states represent some information on the word that is updatable by appending letters on the right, monoid elements represent an information on the word that is updatable both by appending letters on the left or the right.

Unsurprisingly, there is a notion of "minimal information" required to recognize a monoid, giving rise to the notions of *syntactic monoids* and *syntactic morphisms*, see *e.g.* [Boj20, Theorem 1.7].

Submonoids and quotient structures are defined analogously to substructures and quotient structures for relational structures. We say that a monoid *divides* another one if it is a submonoid of one of its quotients. A *pseudovariety of monoids* $\mathbb{V}$ consists in a set of finite monoids closed under finite Cartesian products and monoid division.[12] For more details on these, see [Pin22, § XI.1, p. 189] under the name "variety". On the other hand, a *\*-pseudovariety of regular languages* consists of *stream*, *i.e.* a function $\mathcal{V}$ from alphabets to languages over this alphabet—for which the image of an alphabet $\Sigma$ is denoted by $\mathcal{V}_\Sigma$ rather than $\mathcal{V}(\Sigma)$—, and with the following properties:[13]

- it is closed under Boolean operators;
- it is closed under preimages by monoid morphisms, in the sense that for every monoid morphism $\phi \colon \Gamma^* \to \Sigma^*$, for any $L \in \mathcal{V}_\Sigma$, then $\phi^{-1}[L] \in \mathcal{V}_\Gamma$, and
- it is closed under residuals, in the sense that for any $L \in \mathcal{V}_\Sigma$, for any $u \in \Sigma^*$, then $u^{-1}L \triangleq \{v \in \Sigma^* \mid uv \in L\}$ and $Lu^{-1} \triangleq \{v \in \Sigma^* \mid vu \in L\}$ both belong to $\mathcal{V}_\Sigma$.

A seminal result by Eilenberg shows that, mapping a pseudovariety of monoids $\mathbb{V}$ to the stream associating to $\Sigma$ the set of languages over $\Sigma$ that are recognized by a monoid from $\mathbb{V}$ yields a *-pseudovariety of regular languages, and moreover, this operation is a bijection! This result generalizes Schützenberger's famous theorem, showing that star-free languages are exactly those recognized by aperiodic monoids: for this reason, this bijection is called the *Eilenberg-Schützenberger correspondence*.[14]

Lastly, given a stream of regular languages $\mathcal{V}$, the $\mathcal{V}$-MEMBERSHIP PROBLEM takes as input an alphabet $\Sigma$ and a regular language $L \subseteq \Sigma^*$, and asks if $L \in \mathcal{V}_\Sigma$. When $\mathcal{V}$ is a pseudovariety of regular languages, a powerful technique to solve this problem is to prove that the membership problem of the corresponding pseudovariety of monoids is decidable.

## II.4  Computability and Complexity

### II.4.1  Turing Machines

[15] We assume that the reader is familiar with Turing machines, see *e.g.* [AB09, § 1]. Unless stated otherwise, a *Turing machine* is assumed to have a single tape, which is bounded on the left but not on the right. The *configuration*

[12] "Pseudovariety of *foo*" and "variety of finite *foo*" are used interchangeably in the literature.

[13] See also [Pin22, § XIII.3, p. 226].

[14] See [Pin22, Theorem XIII.4.10, p. 228] for more details.

[15] "Mais dans ce sandwich absurde entre mon initialisation et ma terminaison, que m'aura-t-il vraiment manqué ? Peut-être un corps… un corps qui ressent et tutti quanti !" —Gilles Dowek, Laurence Devillers & Serge Abiteboul, *Qui a hacké Garoutzia*.

of a Turing machine consists of the content of the tape at a given point, the position of the machine's head, at well as its current state. It can be summarized by a triple $\langle u, q, v \rangle$, where $q$ denotes the current state, $u$ is the word written strictly on the left of the head, and $v$ is the word written to the right of the head (head included). The *initial configuration* of a Turing machine $\mathcal{M}$ refers to the configuration $\langle \varepsilon, q_0, \varepsilon \rangle$ where $q_0$ is the initial state of $\mathcal{M}$. A configuration is *reachable* whenever it can be obtained from the initial configurations by applying a finite sequence of transitions of the machine.

## II.4.2 Elements of Complexity Theory

We assume familiarity with basic complexity classes, see *e.g.* [AB09, §§ 2–5]. We say that two decision problems are *computationally equivalent* when there are many-one reductions between them.

*Complexity Classes.* Here are the most important classes that appear in this document—the other ones will all point to their respective entry in the Complexity Zoo:

- *FO*$^{\mathrm{fin}}$ refers to first-order definable classes, see Section VII.2.2 for the definition;
- *L* and *NL* refer to problems solvable in deterministic and non-deterministic logarithmic space; for hardness we usually consider first-order reductions;
- *P* (resp. *PSpace*) refers to problems solvable in deterministic polynomial time (resp. polyomial space); hardness is usually considered under logarithmic-space reductions;
- *NP* and *coNP* refer to the first level of the polynomial-time hierarchy [AB09, § 2], and $\Sigma_2^p, \Pi_2^p$ to its second level, see *e.g.* [AB09, § 5]; hardness is usually considered under logarithmic-space reductions;
- *k-ExpTime* and *k-ExpSpace* are the problems solvable in time and space

$$n \mapsto \underbrace{2^{2^{\cdot^{\cdot^{2^{\mathrm{poly}(n)}}}}}}_{k \text{ times}},$$

respectively; hardness is considered under polynomial-time reductions;
- *Tower* is the class of problems that can be solved in time $n \mapsto \mathrm{tower}(f(n))$ for some elementary function $f$,[16] where $\mathrm{tower}(n) \triangleq t(n, n)$ and $t$ is the function defined recursively by $t(p, q) \triangleq 2^{t(p-1, q)}$ and $t(0, q) \triangleq q$: in other words $\mathrm{tower}(n)$ it is a tower of exponentials whose height is given by the input; hardness is defined under elementary reductions, see [Sch16] for more details on this class.

[16] Recall that an function is said to be elementary when it is bounded by a tower of exponentials of fixed height.

*Computability Classes.* We now turn to undecidable classes: all hardness results are considered up to many-one reductions, *i.e.* computable functions preserving both membership and non-membership to the problem. We denote by *RE* and *coRE* the classes of recursively enumerable and co-recursively enu-merable problems, respectively. The next levels $\Sigma_n^0$ and $\Pi_n^0$ of the *arithmetical hierarchy* can be defined as the classes of sets that are definable by a first-order formula from the fragments $\Sigma_n$ and $\Pi_n$, respectively, in the structure $\langle \mathbb{N}, +, \cdot \rangle$. It can be shown that $\Sigma_0^0 = \Pi_0^0$ corresponds to the class of all decidable problems, and $\Sigma_1^0 = \mathsf{RE}$ and $\Pi_1^0 = \mathsf{coRE}$. The only other levels that will be of interest to us will be $\Sigma_2^0$ and $\Pi_2^0$. The *analytical hierarchy* is defined analogously, by replacing first-order logic with second-order logic.

*Decision Problems.* CONNECTIVITY IN FINITE GRAPHS is the decision problem that takes as input a graph and two vertices $s$ and $t$, and asks if they are connected. REACHABILITY IN FINITE GRAPHS is defined analogously, but asks rather if there is a directed path from $s$ to $t$. Surprisingly, despite their resemblance, these two problems have a distinct complexity: REACHABILITY IN FINITE GRAPHS is $\mathsf{NL}$-complete, see *e.g.* [AB09, Theorem 4.18], while CONNECTIVITY IN FINITE GRAPHS is only in $\mathsf{L}$ by Reingold's theorem, see *e.g.* [AB09, Theorem 21.21]. In fact, the problem is $\mathsf{L}$-complete: Etessami even proved that the problem was $\mathsf{L}$-hard under first-order reductions even if the graph is restricted to be a directed path [Ete97, Theorem 4.2]!

### II.4.3 *Parametrized Complexity*

A parameterized language is any subset $L \subseteq \Sigma^* \times \mathbb{N}$: the first component represents an instance, and the second a parameter. We define *FPT* as the class of *fixed-parameter tractable* problems, for which there is a Turing machine $\mathcal{M}$ *s.t.*

$$\mathcal{M} \text{ accepts } \langle x, k \rangle \quad \textit{iff} \quad \underbrace{\langle x, k \rangle}_{} \in L,$$

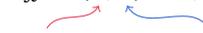 input 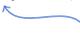 parameter

and, moreover, $\mathcal{M}$ runs in time $\mathcal{O}(f(k) \cdot |x|^c)$, for constant $c$ that does not depend on $x$ nor $k$, and where $f \colon \mathbb{N} \to \mathbb{N}$ is a computable function.

*XP* is the class of slicewise polynomial-time problems $L \subseteq \Sigma^* \times \mathbb{N}$: *s.t.*, for every bounded value of the parameter, then the resulting language must be in $\mathsf{P}$. Formally, for each $k \in \mathbb{N}$, we ask that the (classical) problem

$$\{u \in \Sigma^* \mid \langle u, k \rangle \in L\}$$

belongs to $P$.

Note that $\mathsf{FPT}$ is, by construction, included in $\mathsf{XP}$. However, an algorithm that runs in time $\mathcal{O}(|x|^k)$ is a typical example of algorithm that is $\mathsf{XP}$ but not $\mathsf{FPT}$.

The class *W[1]*, which we will not formally define here, can roughly be seen as the equivalent of $\mathsf{NP}$. See [Cyg+15] for a formal definition. Some problems in $\mathsf{XP}$ belong to $\mathsf{FPT}$, and are considered to be tractable, *i.e.* efficiently solvable in practice. On the other hand, others problems, such as the $k$-clique problem, when parameterized by $k$, are in XP but are W[1]-hard, and hence conjectured not to be in FPT. We think of these problems as non-tractable.

# Querying Graph Databases

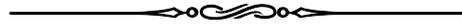

CHAPTER **III**

# *Query Languages for Relational and Graph Databases*

ABSTRACT

This preliminary chapter briefly surveys the literature on the notion of *conjunctive queries*, *conjunctive regular path queries* and related notions. We start by presenting the notion of duality for conjunctive queries, and how it can be used for the *static analysis* of this language. We then turn to the more complex language of conjunctive regular path queries designed for graph databases, which will be the focus of the rest of this part of the thesis.

ACKNOWLEDGEMENTS

Parts of Section III.2 come from [FM25; FMR25]—mostly from the introductions and preliminaries.

# Contents

## III.1 Relational Databases

We assume familiarity with basic SQL and the concept of tables—however a very high-level understanding of these notions should be more than enough to follow this chapter.

### III.1.1 SQL and First-Order Logic

The most common model of databases is by far that of relational databases, in which data is stored in *tables*: an example is depicted in Table III.1.

MOVIES

| id | title | duration | director |
|----|-------|----------|----------|
| 197 | Eyes Wide Shut | 159 | Stanley Kubrick |
| 205 | J'ai tué ma mère | 96 | Xavier Dolan |
| 304 | Amadeus | 161 | Miloš Forman |
| 321 | 120 Battements par minute | 143 | Robin Campillo |

| ROOMS | |
|----|----------|
| id | capacity |
| 1 | 108 |
| 2 | 124 |
| 3 | 96 |
| 4 | 102 |

| PROJECTIONS | | |
|----------|---------|------|
| movie_id | room_id | time |
| 197 | 2 | 2025-03-28 14:00 |
| 205 | 3 | 2025-03-28 14:30 |
| 321 | 4 | 2025-03-28 14:30 |
| 197 | 1 | 2025-03-28 17:00 |

Table III.1: A relational database consisting of three tables, representing data stored by a cinema. (Replica of Table I.4.)

Formally, a *pointed relational database* $\langle \mathbf{D}, \bar{d} \rangle$ over a purely relational signature $\sigma$ consists of, for each predicate $\mathcal{R}$ of arity $k$ in $\sigma$, a finite $k$-ary relation $\mathcal{R}_{(k)}(\mathbf{D})$, as well as a tuple $\bar{d}$ of elements, called *constants*. A *relational database* is a pointed relational database whose tuple is empty.[1] The data consisting of a tuple occurring in some relation $\mathcal{R}_{(k)}(\mathbf{D})$, together with its predicate $\mathcal{R}_{(k)}$, is called a *fact*. From a practical perspective, relations model the tables, while each fact corresponds to some row: for instance the database of Table III.1 has three relations, twelve facts and no constants.[2]

Naturally, each pointed relational database yields a $\sigma$-structure whose domain is the set of elements occurring in some fact of the database or as a constant. This structure has the property that each vertex either belongs to some hyperedge or is a constant: in other words, it has no *isolated vertices*. This mapping is in fact one-to-one: pointed relational databases over $\sigma$ are in bijection with pointed $\sigma$-structures with no isolated vertices.

> HYPOTHESIS/NOTATION. We identify (pointed) relational databases with (pointed) relational structures with no isolated vertices.

As mentioned in Chapter I, from a theoretical perspective, a very natural way of querying these structures is *via* first-order logic. Remarkably, it is

[1] It is often asked that at least one $\mathcal{R}_{(k)}(\mathbf{D})$ is non-empty. Whether this condition is imposed does change the theory—for instance the first-order sentences that are valid over all non-empty structures is a strict supset of those valid over all (possibly empty) structures. However, this condition is mostly required for historical reasons, and whether the databases are allowed to be non-empty will be of little importance for the query languages we will consider.

[2] Many variations on the definition above exists: for instance the columns of the table are often given a name (called *attribute*), see *e.g.* [AHV95, §§ 3.1–3.2]. This is usually done to make the syntax of relational algebra easier on the eye, however this is only syntactic sugar [AHV95, Proposition 5.1.2].

exactly as expressive as the fragment of $k$-ary SQL queries generated by

$$\phi_{(k)} ::= \text{SELECT DISTINCT attribute}_1, \dots, \text{attribute}_k$$
$$\qquad \text{FROM } \phi_{(i_1)}, \dots, \phi_{(i_n)}$$
$$\qquad \text{WHERE some condition} \qquad\qquad\qquad\text{(III.1)}$$
$$\qquad | \text{ some table of arity } k$$
$$\qquad | \phi_{(k)} \text{ UNION } \phi_{(k)} | \phi_{(k)} \text{ EXCEPT } \phi_{(k)} | \phi_{(k)} \text{ INTERSECT } \phi_{(k)},$$

where $i_1 + \dots + i_n = k$ and the condition after WHERE is a conjunction of equalities between attributes and/or constants. Dealing with the arity $k \in \mathbb{N}$ is required to ensure *e.g.* that the union is homogeneous—*i.e.* that all facts in the union have the same arity.[3]

**Proposition III.1.1** (Codd's theorem [Cod72, § 4]).[4] First-order logic over $\sigma$-structures with no isolated vertices is equally expressive to the SQL fragment defined in (III.1). Moreover, this equivalence is effective.

*Proof sketch. Warning:* dealing with all the subtleties of the proof is actually somewhat tedious. Hence, we provide an *informal* proof, which prioritizes intuition over formalism.

❧ *From SQL to FO.* Expressing a SQL query as a first-order formula is intuitive, by generalizing the idea given in Section I.2.1:

- UNION, EXCEPT AND INTERSECT are interpreted as the union, set difference (*i.e.* $- \wedge (\neg -)$) and intersection;
- a table of arity $k$, modelled as a relation $\mathcal{R}_{(k)}$, is encoded as the atomic formula

$$\phi_{(k)}(x_1, \dots, x_k) \mathrel{\hat{=}} \mathcal{R}_{(k)}(x_1, \dots, x_k)$$

with $k$ free variables which are all fresh;

- a query

$$\text{SELECT DISTINCT attribute}_1, \dots, \text{attribute}_k$$
$$\text{FROM } \phi^1_{(i_1)}, \dots, \phi^n_{(i_n)}$$
$$\text{WHERE } \theta$$

is encoded as

$$\psi(y_1, \dots, y_k) \mathrel{\hat{=}} \exists \bar{x}. \, \tilde{\phi}^1_{(i_1)}(x_1^1, \dots, x_{i_1}^1) \wedge \dots \wedge \tilde{\phi}^n_{(i_n)}(x_1^n, \dots, x_{i_n}^n) \wedge \theta,$$

with $y_1, \dots, y_k$ being the variables associated to the attributes—*e.g.* if attribute$_1$ is the third attribute of $\phi^2$, then $y_1 \mathrel{\hat{=}} x_3^2$—and $\bar{x}$ is the tuple of all variables of the form $x_j^i$ that are distinct from the $y_i$'s. Moreover, each $\tilde{\phi}^i$ and $\tilde{\theta}$ denotes the encodings of $\phi^i$ and $\theta$, which can be defined recursively for $\phi^i$ and trivially for $\theta$.

❧ *From FO to SQL.* The converse encoding, *i.e.* from first-order logic to SQL queries, is a little more tricky:

[3] Also, the DISTINCT keyword is only necessary because SQL has a multi-set semantics rather than a set-based semantics…

[4] Codd's theorem actually deals with *relational algebra* rather than SQL. The equivalence between *relational algebra* and the fragment (III.1) of SQL is however straightforward. See also [AHV95, Theorem 5.4.6].

- union is encoded with UNION, intersection and existential quantification with SELECT DISTINCT-FROM-WHERE;
- we encode negation using EXCEPT and a SQL query that outputs every $k$-tuple of vertices of the structure—this query can be written as a big union of SELECT DISTINCT-FROM queries;
- lastly, universal quantification can then be obtained using the tautology

$$\forall x. \ \phi(x) \equiv \neg \exists x. \ \neg \phi(x). \qquad \square$$

A crucial ingredient that ensures the correctness of these encodings is actually the fact that relational databases are encoded as relational structures *with no isolated variables*.[5] Take for instance the first-order formula $\neg\text{ROOM}(x, y)$: we translated it to the SQL query

All pairs EXCEPT  (SELECT DISTINCT $x, y$ FROM  ROOM)

where 'All pairs' is a query outputting all pairs in the database. This latter query can actually be expressed as a union of SELECT DISTINCT-FROM queries precisely thanks to the lack of isolated vertices in the structure: for more details, we refer the reader to the term "active domain" in [AHV95].[6]

The expressiveness of this fragment of SQL however comes to the cost of computational efficiency.

**Proposition III.1.2** (Folklore)**.** Given a first-order formula $\phi(\bar{x})$, a relational structure $\mathbf{D}$ with no isolated vertices, and a tuple $\bar{d}$, deciding if $\bar{d} \in [\![\phi(\bar{x})]\!]^{\mathbf{D}}$ is PSpace-complete.

*Proof sketch.* The upper bound can be proven by considering the naïve algorithm that recurses on the formula. The lower bound follows from a trivial reduction from the quantified Boolean formula problem. $\qquad \square$

Even worse, when turning to the *static analysis* of these queries, the problems become undecidable. Given two (semantical) queries $\phi$ and $\psi$, we say that they are *semantically equivalent*—implicitly over finite relational databases—when for every finite relational databases $\mathbf{D}$, we have $\mathbf{D} \in \phi$ *iff* $\mathbf{D} \in \phi'$. This fact is denoted by $\phi \equiv \phi'$.

**Proposition III.1.3** (Trakhtenbrot, see *e.g.* [AHV95, Theorem 6.3.1 & Corollary 6.3.2])**.**[7] It is undecidable whether a first-order formula is satisfiable over finite relational databases. In turn, validity and semantical equivalence are also undecidable.

In turn, it means that there is no hope to optimize a SQL query from the fragment (III.1), in the sense of Section I.2.

**Corollary III.1.4** (Folklore (but not often mentionned), see *e.g.* [AW12, Remark 5.3])**.** Given a first-order formula $\phi$ and $k \in \mathbb{N}$, it is undecidable whether $\phi$ is semantically equivalent to a formula with at most $k$ variables.

[5] The restriction of first-order logic to relational databases is usually called *relational calculus*.

[6] The query $\neg\text{ROOM}(x, y)$ is actually a good example of why there is no built-in negation in SQL, even though it would not change its expressiveness: it is actually very hard to imagine a situation where knowing that 'Xavier Dolan' is not the id of a room with capacity '2025-03-28 14:00' would be useful...

[7] *Stricto sensu* Trakhtenbrot's theorem deals with all relational structures, and the notion of semantical equivalence differs when considered over relational databases or all relational structures. But the proof of undecidability in the two settings do not differ substantially.

### III.1.2  Conjunctive Queries to the Rescue

The undecidability results of Proposition III.1.3 and Corollary III.1.4, together with the fact that queries occurring in practice—see Section I.2.1—are much simpler than the formulas occurring in the undecidability proofs, motivate the study of well-behaved query fragments. We will focus on conjunctive queries, which arise from the grammar

$$\phi_{(k)} ::= \text{SELECT DISTINCT } \text{attribute}_1, \ldots, \text{attribute}_k$$
$$\text{FROM } \phi_{(i_1)}, \ldots, \phi_{(i_n)} \tag{III.2}$$
$$\text{WHERE some condition}$$
$$| \text{ some table of arity } k$$

of SQL, where, once again, $i_1 + \ldots + i_n = k$ and the condition after WHERE is a conjunction of equalities between attributes. From the proof of Proposition III.1.1 it actually follows that this fragment is exactly as expressive as the fragment of first-order logic, restricted to relational databases, generated by

$$\phi ::= \mathcal{R}_{(k)}(x_1, \ldots, x_k) \mid \phi \wedge \phi \mid \exists x.\ \phi, \tag{III.3}$$

where $\mathcal{R}_{(k)}$ ranges over the signature.

Now observe that, when dealing with the fragment (III.3), the first-order formula $(\exists x.\ \phi(x, \bar{y})) \wedge \psi(\bar{z})$ is equivalent to $\exists x'.\ (\phi(x', \bar{y}) \wedge \psi(\bar{z}))$ where $x'$ is any variable that occurs neither in $\bar{y}$ nor in $\bar{z}$. For instance, we have

$$(\exists x.\ \mathcal{P}(x)) \wedge (\exists x.\ \exists y.\ \mathcal{R}(x, y)) \equiv \exists z.\ \exists x.\ \exists y.\ \mathcal{P}(z) \wedge \mathcal{R}(x, y).$$

This leads to a simple rewriting system that puts every formula from (III.3) in so-called prenex form.

**Proposition III.1.5** (*Prenex form*)**.** Every formula from (III.3) can be written in the form

$$\exists \bar{x}.\ \bigwedge_{i=1}^{n} \mathcal{R}^i_{(k_i)}(y^i_1, \ldots, y^i_{k_i}). \tag{III.4}$$

Translating back the formulas in prenex form to SQL queries, it implies that the fragment (III.2) is no more expressive than its induction-free fragment

$$\phi_{(k)} ::= \text{SELECT DISTINCT } \text{attribute}_1, \ldots, \text{attribute}_k$$
$$\text{FROM } \text{table}_{(i_1)}, \ldots, \text{table}_{(i_n)} \tag{III.5}$$
$$\text{WHERE some condition.}$$

Hence, we define a conjunctive query to be any first-order formula in prenex form—see Equation (III.4). Recall that, as mentioned in Remark II.3.1, we assume that formulas come with a tuple of variables, containing all free variables. Hence, a $k$-ary *conjunctive query* over a purely relational signature $\sigma$, or *CQ* for short, amounts to a finite set of atomic formulas of the form $\mathcal{R}_{(k)}(x_1, \dots, x_k)$, called *atoms*, with $\mathcal{R}_{(k)} \in \sigma$, together with tuple of $k$ variables $\bar{x}$, called *free variables* or *output variables*.[8] The set of atoms is denoted conjunctively. So, an example of conjunctive query is

$$\gamma(\text{title}, \text{time}) \triangleq \text{MOVIES}(\text{movie\_id}, \text{title}, \text{length}, \text{director})$$
$$\wedge \text{PROJECTIONS}(\text{movie\_id}, \text{room\_id}, \text{time}).$$

Semantically, it is interpreted as the first-order formula in prenex form in which every variable that is not an output variable is quantified existentially, giving in our case

$$\exists \text{movie\_id}. \, \exists \text{length}. \, \exists \text{director}. \, \exists \text{room\_id}.$$
$$\text{MOVIES}(\text{movie\_id}, \text{title}, \text{length}, \text{director})$$
$$\wedge \text{PROJECTIONS}(\text{movie\_id}, \text{room\_id}, \text{time}).$$

Interestingly, the semantics of conjunctive queries can be described using homomorphisms, via the theory of *duality*: this was first noticed by Chandra and Merlin in their seminal 1977 paper [CM77].

**Definition III.1.6.** The *canonical database* associated to a conjunctive query

$$\gamma(\bar{x}) = \bigwedge_{i=1}^{n} \mathcal{R}^i_{(k_i)}(y^i_1, \dots, y^i_{k_i})$$

over the purely relational signature $\sigma$ is the pointed relational database over $\sigma$ with tuple $\bar{x}$ and whose facts are $\mathcal{R}^i_{(k_i)}(y^i_1, \dots, y^i_{k_i})$ for $i \in [\![1, n]\!]$.

Since it is somewhat impractical to graphically depict $k$-ary relations for $k \geq 3$, and since signatures that consist only of unary predicates are a degenerate case, we will often consider examples in which all relations are binary. We will denote these relations by $\xrightarrow{a}$, where $a$ ranges over some alphabet $\mathbb{A}$: moreover, we will often not make the effort to specify which alphabet we are using, as any alphabet containing all letters occuring in the example will do. For instance, the canonical database of the conjunctive query

$$\gamma() \triangleq x \xrightarrow{a} x' \wedge x \xrightarrow{b} y \wedge y \xrightarrow{a} y'$$

is the pointed relational database of Figure III.2.

Naturally, Boolean queries are denoted by $\gamma()$. In this case, its canonical database is not only a pointed relational database but in fact a relational database. We will sometimes denote some queries (either Boolean or not) by $\gamma$ (with no brackets): this is simply to ease the notations when no ambiguity can arise—we will often explicitly mention the tuple of output variables when introducing the query, and then simply the notation in a second time.

**Proposition III.1.7** (*Duality*)**.** Let $\gamma(\bar{x})$ be a conjunctive query, and let $\langle \mathbf{G}, \bar{x} \rangle$ denote its canonical database.[9] For any pointed relational database $\langle \mathbf{D}, \bar{d} \rangle$,

[8] A $k$-ary CQ over a purely relational signature is equivalent to a Boolean CQ over a relational signature with $k$ constants.

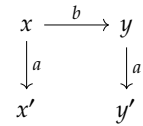

Figure III.2: The canonical database of $\gamma() \triangleq x \xrightarrow{a} x' \wedge x \xrightarrow{b} y \wedge y \xrightarrow{a} y'$.

[9] By convention, we denote the canonical database using the Roman uppercase associated with the Greek letter used to denote the query.

we have

$$\bar{d} \in [\![\gamma(\bar{x})]\!]^{\mathbf{D}} \qquad \textit{iff} \qquad \langle \mathbf{G}, \bar{x} \rangle \xrightarrow{\text{hom}} \langle \mathbf{D}, \bar{d} \rangle.$$

*Proof.* This follows from the definition of homomorphisms and the semantics of first-order logic. ☐

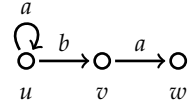

Figure III.3: A relational database satisfying the conjunctive query of Figure III.2.

For instance, consider the relational database of Figure III.3. It satisfies the Boolean query

$$\gamma() \mathrel{\hat{=}} x \xrightarrow{a} x' \wedge x \xrightarrow{b} y \wedge y \xrightarrow{a} y'$$

asking if there is a $b$-edge ($x \xrightarrow{b} y$) whose extremities have outgoing $a$-edges ($x \xrightarrow{a} x'$ and $y \xrightarrow{a} y'$): indeed, there is a $b$-edge from $u$ to $v$, and both $u$ and $v$ have outgoing $a$-edges. In terms of duality, this is witnessed by the homomorphism from the canonical database of Figure III.2 to the database of Figure III.3, that sends $x$ and $x'$ onto $u$, $y$ onto $v$ and $y'$ onto $w$. As witnessed by this example, the homomorphism $\mathbf{G} \to \mathbf{D}$ actually has a natural meaning when thinking about model checking. For this reason, such a homomorphism is also called an evaluation map—or even abusively *homomorphism*—from $\gamma$ to $\mathbf{D}$.

Duality has *many* consequences. The first one is that the following evaluation problem lies in NP, as it can be encoded into a HOMOMORPHISM PROBLEM.

> **CONJUNCTIVE QUERY EVALUATION**
> *Input*: A purely relational signature $\sigma$, a conjunctive query $\gamma(\bar{x})$ over $\sigma$, and a pointed relational $\sigma$-database $\langle \mathbf{D}, \bar{d} \rangle$.
> *Question*: Does $\bar{d} \in [\![\gamma(\bar{x})]\!]^{\mathbf{D}}$?

For Boolean conjunctive queries, both tuples are empty, and so this amounts to asking whether the query is true on the database.[10] It is fact NP-complete [CM77, Theorem 7]. More importantly, duality has the consequence that semantical equivalence, as well as the finer notion of containment, are decidable for CQs.

Given two queries $\gamma(\bar{x})$ and $\gamma'(\bar{x}')$, we say that $\gamma(\bar{x})$ is *contained* in $\gamma'(\bar{x}')$ whenever (1) $\bar{x}$ and $\bar{x}'$ have the same arity, and (2) $[\![\gamma(\bar{x})]\!]^{\mathbf{D}} \subseteq [\![\gamma'(\bar{x}')]\!]^{\mathbf{D}}$ for every finite relational database $\mathbf{D}$. This is denoted by $\gamma(\bar{x}) \sqsubseteq \gamma'(\bar{x}')$; the notion of containment is also known as *entailment* or *logical implication*. Clearly, semantical equivalence can be obtained as the symmetric closure of containment.

**Proposition III.1.8.**[11] Let $\gamma(\bar{x})$ and $\gamma'(\bar{x}')$ be two conjunctive queries. The following are equivalent:

1. $\gamma(\bar{x}) \sqsubseteq \gamma'(\bar{x}')$;
2. $\langle \mathbf{G}, \bar{x} \rangle \vDash \gamma'(\bar{x}')$;
3. $\langle \mathbf{G}', \bar{x}' \rangle \xrightarrow{\text{hom}} \langle \mathbf{G}, \bar{x} \rangle$,

where $\langle \mathbf{G}, \bar{x} \rangle$ and $\langle \mathbf{G}', \bar{x}' \rangle$ are the canonical databases of $\gamma(\bar{x})$ and $\gamma'(\bar{x}')$, respectively.

[10] Indeed, for Boolean CQs, $[\![\gamma()]\!]^{\mathbf{D}}$ is either empty (that we interpret as *false*), or equal to the singleton consisting of the empty tuple (that we interpret as *true*).

[11] In fact, another consequence of duality is that the quantification over finite relational databases in the definition of containment can be equivalently replaced by a quantification over finite structures, or even over all structures!

*Proof.* ❧ *(1) ⟹ (2).* By duality, $\langle \mathbf{G}, \bar{x} \rangle$ is a model of $\gamma(\bar{x})$, and so, since this query is contained in $\gamma'(\bar{x}')$, it follows that $\langle \mathbf{G}, \bar{x} \rangle \vDash \gamma'(\bar{x}')$.

❧ *(2) ⟹ (3).* By duality.

❧ *(3) ⟹ (1).* Assume that $\langle \mathbf{G}', \bar{x}' \rangle \xrightarrow{\text{hom}} \langle \mathbf{G}, \bar{x} \rangle$, and let us prove that $\gamma(\bar{x}) \subseteq \gamma'(\bar{x}')$. Let $\langle \mathbf{D}, \bar{d} \rangle$ be a pointed relational database, and assume that $\langle \mathbf{D}, \bar{d} \rangle \vDash \gamma(\bar{x})$. By duality, we get $\langle \mathbf{G}, \bar{x} \rangle \xrightarrow{\text{hom}} \langle \mathbf{D}, \bar{d} \rangle$, and by precomposing with any homomorphism witnessing that $\langle \mathbf{G}', \bar{x}' \rangle \xrightarrow{\text{hom}} \langle \mathbf{G}, \bar{x} \rangle$, we get that $\langle \mathbf{G}', \bar{x} \rangle \xrightarrow{\text{hom}} \langle \mathbf{D}, \bar{d} \rangle$. Once again, by duality, this amounts to $\langle \mathbf{D}, \bar{d} \rangle \vDash \gamma'(\bar{x}')$, which concludes the proof that $\gamma(\bar{x}) \subseteq \gamma'(\bar{x}')$. □

**Corollary III.1.9.** Containment (and hence semantical equivalence) of conjunctive queries is decidable, and in fact is NP-complete.

### III.1.3  The Preordered Set of Conjunctive Queries

Duality takes its name from the fact that Proposition III.1.8 can be simply rephrased as "the preordered set of conjunctive queries over $\sigma$ under containment is *dually isomorphic* to the preordered set of relational databases over $\sigma$ under the homomorphism ordering." Symbolically, letting $\mathrm{RelDb}_\sigma$ denote the set of all relational databases over $\sigma$, we have

$$\langle \mathrm{CQ}_\sigma, \subseteq \rangle \cong \langle \mathrm{RelDb}_\sigma, \xleftarrow{\text{hom}} \rangle.$$

Naturally, to go from relational databases to conjunctive queries, we associate to any pointed relational database $\langle \mathbf{G}, \bar{g} \rangle$ a *canonical conjunctive query* $\gamma(\bar{g})$ with one atom $\mathcal{R}_{(k)}(x_1, \dots, x_k)$ for every hyperedge $\langle x_1, \dots, x_k \rangle \in \mathcal{R}_{(k)}(\mathbf{G})$. This map is precisely the inverse of the construction defining canonical databases.

This dual isomorphism has many consequences: essentially every theory that deals with relational databases can be applied to study conjunctive queries!

**Corollary III.1.10** (of duality and Proposition II.2.3)**.** Two conjunctive queries are semantically equivalent *iff* the core of their canonical database are isomorphic.

*Graphical depiction of the preordered set of relational databases.* Note that for each conjunctive query $\gamma(\bar{x})$, the class of pointed relational databases $\langle \mathbf{D}, \bar{d} \rangle$ satisfying the query is *closed under homomorphisms*, i.e.

if  $\langle \mathbf{D}, \bar{d} \rangle \vDash \gamma(\bar{x})$  and  $\langle \mathbf{D}, \bar{d} \rangle \xrightarrow{\text{hom}} \langle \mathbf{D}', \bar{d}' \rangle$  then  $\langle \mathbf{D}', \bar{d}' \rangle \vDash \gamma(\bar{x})$.

We represent the preordered set of relational databases ordered by $\xrightarrow{\text{hom}}$ as follows: each equivalence class of homomorphically equivalent relational databases is represented by a single point. In other words, points are in one-to-one correspondence with cores. Then, we represent a point $\check{\mathbf{G}}$ below another point $\check{\mathbf{D}}$ whenever $\check{\mathbf{G}} \xrightarrow{\text{hom}} \check{\mathbf{D}}$. For Boolean queries, this ordering[12] admits a unique minimal element, which is the empty database. For non-Boolean queries, there is also a minimal relational database with no facts, but that has constants. Similarly, there is always a unique maximal element: the database with a unique vertex $x$ and such that $\mathcal{R}_{(k)}(x, \ldots, x)$ holds for every $\mathcal{R}_{(k)} \in \sigma$, and all constants are interpreted as $x$. We now prove that this poset has a non-trivial structure, provided that the signature is itself non-trivial.

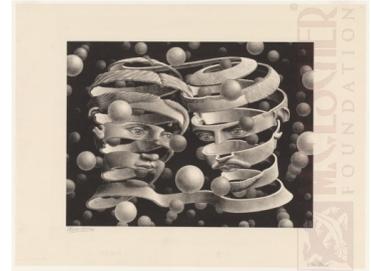

Figure III.4: *Bond of Union*, M. C. Escher, © The M.C. Escher Company.

[12] Formally, from the preordering over relational databases we obtained a partial order over the quotient of relational databases by the equivalence class induced by $\xrightarrow{\text{hom}}$, which happens to be the poset of cores. Hence, we will interchangeably use to terms *preordering* and *(partial) ordering*.

**Proposition III.1.11.** Assume that $\sigma$ contains at least one symbol of arity at least 2. The poset of relational databases admits infinite chains, infinite co-chains and infinite antichains.

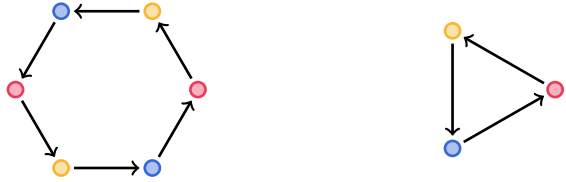

Figure III.5: The graphs $\mathbf{C}_6$ (left) and $\mathbf{C}_3$ (right) and a homomorphism from the former to the latter, described by colour coding.

*Proof.* For the sake of simplicity, we assume that we actually have a binary predicate: this assumption is *w.l.o.g.* since we can encode the binary relation used in these constructions into any $k$-ary relation provided that $k \geq 2$ by encoding $\mathcal{E}(x, y)$ as $\mathcal{R}_{(k)}(x, y, \ldots, y)$.

Clearly, directed paths provide an infinite chain

$$\mathbf{P}_1 \xrightarrow{\text{hom}} \mathbf{P}_2 \xrightarrow{\text{hom}} \cdots \xrightarrow{\text{hom}} \mathbf{P}_n \xrightarrow{\text{hom}} \mathbf{P}_{n+1} \xrightarrow{\text{hom}} \cdots .$$

We now let $\mathbf{C}_n$ ($n \in \mathbb{N}_{>0}$) denote the directed cycle with domain $\mathbb{Z}/n\mathbb{Z}$ and with an edge from $i$ to $j$ *iff* $i + 1 = j$, see Figure III.5. It is then routine to check that for $n, m \in \mathbb{N}_{>0}$, we have $\mathbf{C}_n \xrightarrow{\text{hom}} \mathbf{C}_m$ *iff* $n$ is a multiple of $m$. In particular,[13] we have

$$\mathbf{C}_1 \xleftarrow{\text{hom}} \mathbf{C}_2 \xleftarrow{\text{hom}} \mathbf{C}_4 \xleftarrow{\text{hom}} \cdots \xleftarrow{\text{hom}} \mathbf{C}_{2^n} \xleftarrow{\text{hom}} \mathbf{C}_{2^{n+1}} \xleftarrow{\text{hom}} \cdots .$$

Finally, $\langle \mathbf{C}_p \rangle_{p \text{ prime}}$ is an infinite antichain. $\square$

[13] In fact, we obtain a *projective system*. We will discuss projective limits in Chapter VI.

Based on Proposition III.1.11, we provide an illustration of the ordered set of relational databases in Figure III.6. Notice that, since the semantics of conjunctive queries is closed under homomorphisms, they are represented by upper-closed sets. Moreover, this set has a unique minimum, corresponding to its canonical database. Notice how this representation naturally represents the concept of duality:

- points of the poset (*i.e.* relational databases) are in natural bijection with upper-closed sets that admit a minimum (*i.e.* conjunctive queries);
- a conjunctive query is contained in another *iff* the canonical database of the first is below that of the second in Figure III.6.

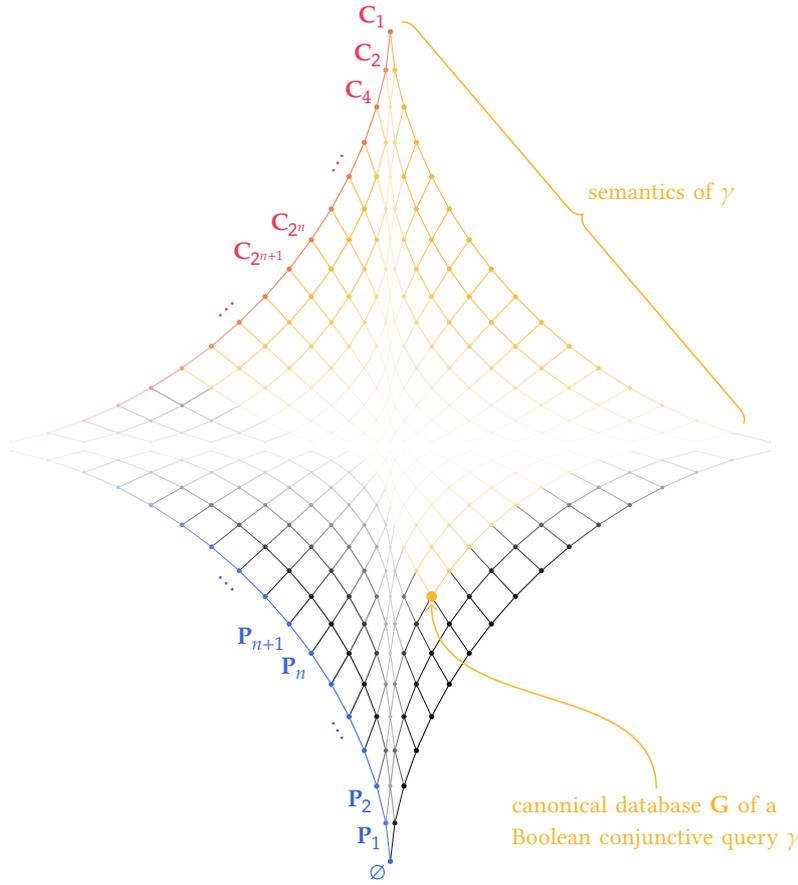

For the sake of simplicity—and compilation time of this document!—, we abstract the picture of Figure III.6 into Figure III.7.

**Remark III.1.12.** À propos relational databases *vs.* relational structures, these posets are actually isomorphic since a relational structure is always homomorphically equivalent to the structure in which we removed isolated vertices. ◇

*The Distributive Lattice of Relational Databases.* Proposition III.1.11 shows that the poset of relational databases has a somewhat complex structure, in the sense that it has infinite height, co-height and width. However, we next show that it has a rich algebraic structure.

In light of Remark III.1.12, the Cartesian product of two relational databases $\langle \mathbf{D}_1, \bar{d}_1 \rangle$ and $\langle \mathbf{D}_2, \bar{d}_2 \rangle$ whose tuples have the same arity is well-defined: we consider their product $\langle \mathbf{D}_1, \bar{d}_1 \rangle \times \langle \mathbf{D}_2, \bar{d}_2 \rangle$ as relational structures, and remove all isolated vertices. We slightly abuse the notation and still denote this product by ×. It is routine to check that this Cartesian product in indeed a Cartesian product in the categorical sense, *i.e.* that it satisfies the universal property that it has homomorphisms $\pi_1$ and $\pi_2$ to both $\langle \mathbf{D}_1, \bar{d}_1 \rangle$ and $\langle \mathbf{D}_2, \bar{d}_2 \rangle$, and that moreover it is the smallest object satisfying this property, in the sense that for every relational database $\langle \mathbf{U}, \bar{u} \rangle$ with homomorphisms $f_1$ and $f_2$ to $\langle \mathbf{D}_1, \bar{d}_1 \rangle$ and $\langle \mathbf{D}_2, \bar{d}_2 \rangle$, then there exists a unique homomorphism $\mathbf{G}$ from $\langle \mathbf{U}, \bar{u} \rangle$ to $\langle \mathbf{D}_1, \bar{d}_1 \rangle \times \langle \mathbf{D}_2, \bar{d}_2 \rangle$ such that the diagram of Figure III.8 commutes: this homomorphism $\mathbf{G}$ is actually $f_1 \times f_2 \colon u \mapsto \langle f_1(u), f_2(u) \rangle$. Going back to the poset structure, this implies that any pair of points must have an infimum.

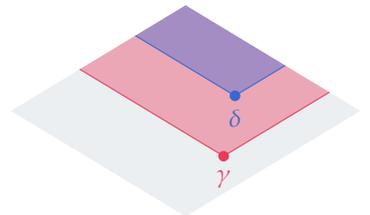

Figure III.7: A more abstract view of the poset of relational databases: we represent two conjunctive queries $\gamma$ (in red) and $\delta$ (in blue). The semantics of each query is represented by a filled diamond, and its canonical database by a large dot.

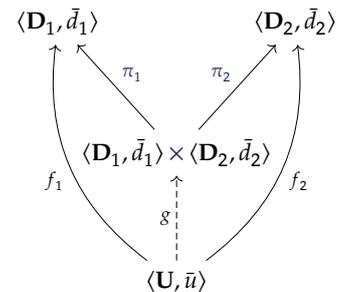

Figure III.8: Universal property satisfied by the Cartesian product.

**Fact III.1.13.** Given two finite relational databases, their Cartesian product is their greatest lower bound in the poset of relational databases ordered by $\xrightarrow{\text{hom}}$.

Similarly, the disjoint union satisfies the property dual to Figure III.8.

**Fact III.1.14.** Given two finite relational databases, their disjoint union is their least upper bound in the poset of relational databases ordered by $\xrightarrow{\text{hom}}$.

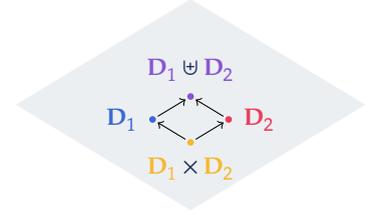

Figure III.9: The distributive lattice structure of relational databases: we represent two structures, as well as their least upper bound and greatest lower bound.

Figure III.10: Two databases (left and right), their Cartesian product (below) and disjoint union (above).

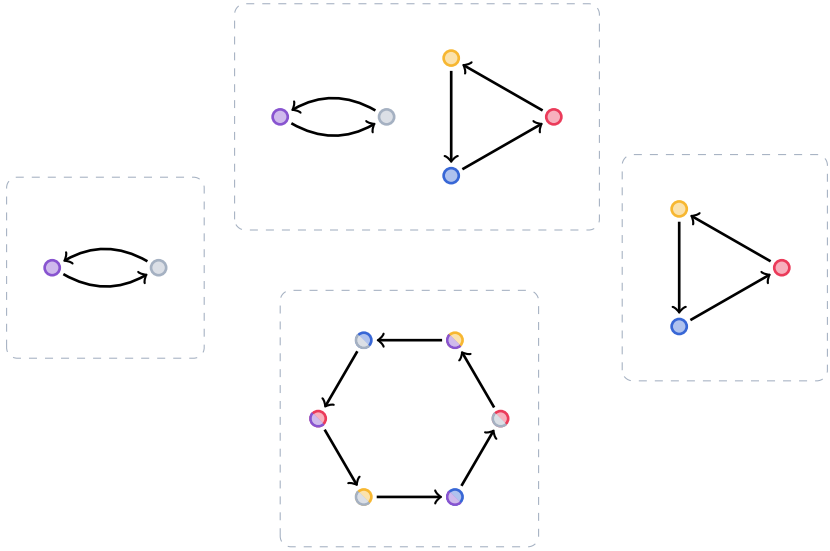

Put together, these facts imply that the poset of relational databases is actually a bounded lattice, as depicted in Figures III.9 and III.10. It is moreover distributive since the isomorphism (and hence homomorphic equivalence)

$$\mathbf{A} \times (\mathbf{B} \uplus \mathbf{C}) \cong (\mathbf{A} \times \mathbf{B}) \uplus (\mathbf{A} \times \mathbf{C})$$

holds for any databases $\mathbf{A}$, $\mathbf{B}$ and $\mathbf{C}$. Then, by duality, we get that the poset of conjunctive queries under containment is also a distributive lattice! We shall see that the greatest lower bound and least upper bound have a natural logical interpretation, and that moreover this structure of distributive lattice will help us deal with CQs: and in particular solve the SYNTHESIS PROBLEM.

*The Distributive Lattice of Conjunctive Queries.* Given two CQs $\gamma(\bar{x})$ and $\delta(\bar{y})$, where $\bar{x}$ and $\bar{y}$ have the same arity, we define their *disjoint conjunction*, denoted by $\gamma(\bar{x}) \, \text{\textcircled{$\wedge$}} \, \delta(\bar{y})$, to be the CQs obtained by taking the disjoint union of atoms of the CQs and then identifying the elements of $\bar{x}$ and $\bar{y}$ pointwise. For instance, letting $\gamma(x) \mathrel{\hat{=}} x \xrightarrow{a} y$ be the query asking for all elements with an outgoing $a$-edge, and $\delta(x) \mathrel{\hat{=}} x \xrightarrow{b} y$ be the query asking for all elements with an outgoing $b$-edge, then their disjoint conjunction $\gamma(x) \varowedge \delta(x)$ is

$$\alpha(x) \triangleq x \xrightarrow{a} y \wedge x \xrightarrow{b} y',$$

which outputs all elements with both an outgoing $a$-edge and an outgoing $b$-edge.[14]

**Fact III.1.15.** The canonical database of the disjoint conjunction equals the disjoint union of the canonical databases. By duality, it follows that the disjoint conjunction is the greatest lower bound of two CQs.[15]

The other operator (the dual of Cartesian product) does not have such a nice intuitive interpretation—however it does not mean that it will be less useful, on the contrary! While conjunctive queries are not closed under semantical union—we will see this in Section III.1.6—, this operator acts as the best approximation of it.

**Definition III.1.16.** We define the *weak union* $\varovee$ of two CQs with the same number of output variables as the canonical CQ of the Cartesian product of their canonical databases.

More explicitly, the weak union of two CQs $\gamma(x_1, \ldots, \bar{x}_n)$ and $\delta(x_1, \ldots, x_n)$ is defined as follows:

- its variables are the pairs of variables of $\gamma(\bar{x})$ and of $\delta(\bar{x})$,
- for any $\mathcal{R}_{(k)} \in \sigma$, there is an atom $\mathcal{R}(\langle y_1, z_1 \rangle, \ldots, \langle y_k, z_k \rangle)$ iff $\mathcal{R}(y_1, \ldots, y_k)$ and $\mathcal{R}(z_1, \ldots, z_k)$ are atoms of $\gamma(\bar{x})$ and of $\delta(\bar{x})$, respectively, and
- its output variables are $\langle x_1, x_1 \rangle, \langle x_2, x_2 \rangle, \ldots, \langle x_n, x_n \rangle$—which we rename as $x_1, x_2, \ldots, x_n$ to keep the same signature.

By construction, it is their greatest upper bound, in the sense that $\gamma(\bar{x}) \subseteq \gamma(\bar{x}) \varovee \delta(\bar{y})$, $\delta(\bar{y}) \subseteq \gamma(\bar{x}) \varovee \delta(\bar{y})$, and $\gamma(\bar{x}) \varovee \delta(\bar{y})$ is the smallest CQ satisfying this property.

For instance, letting $\gamma() \triangleq x \xrightarrow{a} y$ be the CQ asking for the existence of an $a$-edge, and letting $\delta() \triangleq x \xrightarrow{b} y$ be the one asking for a $b$-edge, then $\gamma() \varovee \delta()$ is (semantically equivalent to) the empty CQ, *i.e.* the CQ that always true: knowing that a database contains either an $a$-edge or a $b$-edge is as good as knowing nothing in terms of CQ expressivity. On the other hand, letting $\gamma(x) \triangleq x \xrightarrow{a} y \xrightarrow{b} z$ be the query outputting the source of all $ab$-paths and $\delta(x) \triangleq x \xrightarrow{a} y \xrightarrow{c} z$ be the CQ outputting the source of all $ac$-paths, then $\gamma(x) \varovee \delta(x)$ is homomorphically equivalent to $v(x) \triangleq x \xrightarrow{a} y$ which outputs all sources of $a$-edges. The query $\gamma(x) \varovee \delta(x)$ is depicted in Figure III.11.

The distributive lattice structure of conjunctive queries is depicted in Figure III.12. Observe how this choice of representation makes obvious the fact that (1) the disjoint conjunction of two CQs is actually its conjunction in the semantical sense, but (2) their weak union is strictly bigger than their semantical union, unless we are in a degenerate case.

We summarize the properties of the two distributive lattices in Table III.13. Note that all notions occurring in this table, be it the orders or the binary operators, require the CQs to have the same number of output variables, and the (pointed) relational databases to have tuples of the same size, and so in fact we obtain two lattices for every possible arity/tuple size.

[14] As for disjoint union, the name disjoint conjunction is slightly abusive but justified by its universal property.

[15] Careful: duality is precisely a *dual* isomorphism, *i.e.* it reverses the order. Hence, a least upper bound (disjoint union) becomes a greatest lower bound (disjoint conjunction).

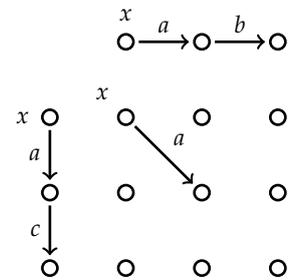

Figure III.11: Two CQs (above and left) $\gamma(x) \triangleq x \xrightarrow{a} y \xrightarrow{b} z$ and $\delta(x) \triangleq x \xrightarrow{a} y \xrightarrow{c} z$ and their weak union $\gamma(x) \varovee \delta(x)$ (below right).

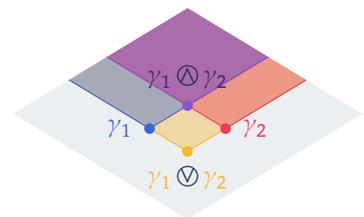

Figure III.12: The distributive lattice structure of Boolean conjunctive queries: we represent two structures, as well as their least upper bound and greatest lower bound, and the natural homomorphism they come equipped with (projections and canonical embeddings).

| | Conjunctive Queries | Relational Databases |
|---|---|---|
| preorder | containment $\subseteq$ | existence of homomorphisms $\xrightarrow{\text{hom}}$ |
| least upper bound | weak union $\lor$ | disjoint union $\uplus$ |
| greatest lower bound | disjoint conjunction $\land$ | Cartesian product $\times$ |
| greatest element | "true" (empty CQ) | single vertex with all hyperedges |
| least element | single variable with all hyperedges | empty database |

Table III.13: The distributive lattices of conjunctive queries and relational databases. By duality, one can go from one lattice to the *opposite* of the other by taking canonical databases or canonical conjunctive queries.

Amongst bounded distributive lattices, the better behaved are the Boolean algebras, in which every element has a complement. In our case, this would be an operator $\neg$ *s.t.* for every database $\mathbf{D}$, then $\mathbf{D} \times \neg\mathbf{D}$ would be homomorphically equivalent to the empty database and $\mathbf{D} \uplus \neg\mathbf{D}$ to the greatest database, consisting of a single vertex and all possible hyperedges over it.

**Proposition III.1.17.** The distributive lattice of databases (and hence of conjunctive queries) does not admit complementation.

*Proof.* Assume that the signature is non-trivial and let $\mathbf{D}$ be a non-trivial database, *i.e.* neither the least nor greatest element of the lattice. To ensure that $\mathbf{D} \uplus \neg\mathbf{D}$ is homomorphically equivalent to the greatest database, $\neg\mathbf{D}$ must actually be homomorphically equivalent to the greatest database itself, but then $\mathbf{D} \times \neg\mathbf{D}$ would be non-empty. Hence, $\neg\mathbf{D}$ cannot exist. $\square$

However, conjunctive queries still have a bit more structure: we shall see that they form a Heyting algebra, namely that it is a bounded distributive lattice with the extra property that for every $\gamma(\bar{x})$ and $\delta(\bar{y})$, where $\bar{x}$ and $\bar{y}$ have the same arity, then there exists a greatest element $\chi(\bar{z})$ *s.t.*

$$\gamma(\bar{x}) \land \chi(\bar{y}) \subseteq \delta(\bar{z}).$$

This greatest element $\chi(\bar{z})$ is denoted by $\gamma(\bar{x}) \Rightarrow \delta(\bar{y})$, and is called *implication*.[16]

**Proposition III.1.18.** The bounded distributive lattice of conjunctive queries is actually a Heyting algebra.

*Proof.* We use duality to prove this and show that relational databases form a "co-Heyting algebra", in the sense that for any database $\langle \mathbf{G}, \bar{g} \rangle$ and $\langle \mathbf{D}, \bar{d} \rangle$, there exists a least element $\langle \mathbf{X}, \bar{x} \rangle$ *s.t.*

$$\langle \mathbf{G}, \bar{g} \rangle \uplus \langle \mathbf{X}, \bar{x} \rangle \xleftarrow{\text{hom}} \langle \mathbf{D}, \bar{d} \rangle :$$

[16] Heyting algebras were introduced to model intuitionistic logic. It is not hard to see that the implication $\phi \Rightarrow \psi$ does satisfy the property that it is the greatest formula $\chi$ *s.t.* $\phi \land \chi$ entails $\psi$.

⟨**X**, $\bar{x}$⟩ is actually obtained by taking the disjoint union of all connected components of ⟨**D**, $\bar{d}$⟩ that cannot be mapped homomorphically to ⟨**G**, $\bar{g}$⟩. The operator ⟹ can then be explicitly constructed by duality. □

By construction of ⟹, it comes naturally equipped with some form of Currying, in the sense that

$$\langle \mathbf{A}, \bar{a} \rangle \vDash \gamma(\bar{x}) \Rightarrow \delta(\bar{y}) \quad iff \quad \langle \mathbf{A}, \bar{a} \rangle \uplus \langle \mathbf{G}, \bar{x} \rangle \vDash \delta(\bar{y}),$$

for any CQs $\gamma(\bar{x})$ and $\delta(\bar{y})$ and any relational database ⟨**A**, $\bar{a}$⟩, where ⟨**G**, $\bar{x}$⟩ denotes the canonical database of $\gamma(\bar{x})$. Similarly, as expected, we have

$$\gamma(\bar{x}) \mathbin{\textcircled{\land}} (\gamma(\bar{x}) \Rightarrow \delta(\bar{y})) \equiv \delta(\bar{y}).$$

In Sections III.1.4 and III.1.6, we will apply some of the theory we developed here to better understand the expressivity of *conjunctive queries*.

### III.1.4   Static Analysis of Conjunctive Queries

*Minimization.*   As we have already seen, the first consequence of duality is that containment and semantical equivalence are both decidable, see Corollary III.1.9. We shall now see that the rich theory of relational structures, and in particular the concept of cores, trivializes the question of minimization.

Fix a relational signature $\sigma$. A *subquery* of $\gamma(\bar{x})$ is any CQ over $\sigma$ obtained by removing variables and/or atoms from $\gamma$: to ensure that we still obtain a CQ over $\sigma$, output variables cannot be removed. In terms of duality, this amounts to taking the canonical conjunctive query of a substructure of the canonical database.[17] We then say that a class $\mathcal{C}$ of conjunctive queries over $\sigma$ is *monotone* if for any CQ $\gamma(\bar{x}) \in \mathcal{C}$, for any subquery $\gamma'(\bar{x})$ of $\gamma(\bar{x})$, we must have $\gamma'(\bar{x}) \in \mathcal{C}$. Monotone classes of CQs naturally model the notion of "simplicity", in the sense that a CQ is "simple" *w.r.t.* $\mathcal{C}$ whenever it belongs to $\mathcal{C}$. The monotonicity assumption precisely ensures that the class formalizes an idea of "simplicity" and not "the CQ has exactly fourteen atoms, two ternary hyperedges, three legs, a moustache and a mustard watch." Typical examples of monotone classes of CQs include:

- CQs with at most $k \in \mathbb{N}$ atoms,
- CQs with at most $k$ variables,
- CQs of tree-width at most $k$ (defined in Section III.1.5),
- CQs of path-width at most $k$,
- CQs in which all cliques are of size at most $k$, etc.

We define the *core* of a conjunctive query, still denoted by $\breve{\ }$, to be the canonical conjunctive query of the core of its canonical database.

**Proposition III.1.19.**   Let $\gamma(\bar{x})$ be a conjunctive query and $\mathcal{C}$ be a monotone class of CQs. Then $\gamma(\bar{x})$ is semantically equivalent to a CQ in $\mathcal{C}$ if, and only if, its core $\breve{\gamma}(\bar{x})$ belongs to $\mathcal{C}$.

[17] Careful: if $\gamma'$ is a subquery of $\gamma$, then $\gamma \subseteq \gamma'$! The fewer constraints there are, the easier it is to satisfy them...

*Proof.* This follows from Proposition II.2.3 and the fact that the core of a structure is always a substructure of it. □

Not only does this imply that we can solve the $\mathcal{C}$-minimization problem, but actually that all these problems can be solved simultaneously, in the sense that if a solution exists to each problem, then a common solution exists!

> **CQ MINIMIZATION PROBLEM OVER $\mathcal{C}$**
> *Input*: A conjunctive query.
> *Question*: Is it semantically equivalent to a CQ of $\mathcal{C}$?

**Corollary III.1.20.** For each monotone class of CQs $\mathcal{C}$, assumed to be described as an oracle, the CQ MINIMIZATION PROBLEM OVER $\mathcal{C}$ is NP. Furthermore, it is NP-hard for some classes $\mathcal{C}$.

*Proof.* ❧ *Upper bound.* A naïve algorithm would be to first compute the core, and then to use the oracle to test if it belongs to $\mathcal{C}$. However, computing the core is not in NP: intuitively, we first need to guess a substructure that is homomorphically equivalent to the whole—which is NP—, and then check that no strictly smaller structure satisfies the property—which is coNP.[18] So, let us not be naïve! We start with a conjunctive query $\gamma(\bar{x})$, guess a subquery $\gamma'(\bar{x})$, and then check if (1)[19] $\gamma'(\bar{x}) \equiv \gamma(\bar{x})$ and if (2) $\gamma'(\bar{x}) \in \mathcal{C}$. The algorithm is clearly in NP, and moreover it is correct by Proposition II.2.3 and monotonicity.

❧ *Lower bound.* We prove that the minimization problem is NP-hard for the class $\mathcal{C}$ of conjunctive queries with at most 3 variables. In fact, by duality, we rather prove that the problem of, given a structure, deciding if it is homomorphically equivalent to a structure with at most 3 vertices is NP-hard. We reduce 3-COLOURABILITY to this latter problem. Given an instance **G** of 3-COLOURABILITY, we reduce it to the instance:

- **G** ⊎ **K**$_3$ if **G** does not contain any self-loop;
- any negative instance otherwise, *e.g.* **K**$_4$.

If **G** is 3-colourable, then it has no self-loop and moreover **G** ⊎ **K**$_3$ $\xrightarrow{\text{hom}}$ **K**$_3$ and hence **G** ⊎ **K**$_3$ $\xrightarrow{\text{hom}}$ **K**$_3$, from which we get that **G** ⊎ **K**$_3$ ≡ **K**$_3$, and hence **G** ⊎ **K**$_3$ is semantically equivalent to a structure with at most 3 vertices. Conversely, assume that the right-hand side of the reduction is semantically equivalent to a structure with at most 3 vertices. Then **G** does not contain any self-loop and moreover, we get that **G** ⊎ **K**$_3$ is homomorphically equivalent to a structure with 3 vertices. In particular, we get a homomorphism from **G** to a structure with 3 vertices, and from the assumption that **G** does not contain any self-loop, we actually get a homomorphism from **G** to **K**$_3$. This concludes the reduction.[20] □

We now turn to the SYNTHESIS PROBLEM, also called *passive learning*, or *reverse engineering*. The idea behind the problem is to infer a query from examples: these examples are models, taking the form of pointed relational databases, together with the knowledge of whether this model should be part of the query (*positive examples*) or not (*negative examples*), see Figure III.14.[21]

[18] In fact, deciding if a structure is a core is actually coNP-complete, even for undirected graphs, [HN92, Theorem 7].

[19] In fact $\gamma \subseteq \gamma'$ always holds so it suffices to check that $\gamma' \subseteq \gamma$.

[20] The trick used for the lower bound, namely that of reducing a homomorphism/containment problem to a minimization problem *via* a disjoint union, will be used in a much trickier way in Chapter IV to prove ExpSpace lower bounds on the MINIMIZATION PROBLEM FOR CRPQs, see Sections IV.A and IV.5.

---

**Synthesis Problem for CQs**

*Input*: Integers $n, p \in \mathbb{N}_{>0}$, and pointed relational databases $\mathbf{D}_1^+, \dots, \mathbf{D}_p^+, \mathbf{D}_1^-, \dots, \mathbf{D}_n^-$.

*Question*: Is there a conjunctive query $\gamma$ s.t. $\mathbf{D}_i^+ \vDash \gamma$ for all $i \in [\![1, p]\!]$ and $\mathbf{D}_j^- \nvDash \gamma$ for all $j \in [\![1, n]\!]$?

---

**Proposition III.1.21.** The synthesis problem for CQs is coNExp.

*Proof.* We denote by $\delta_1^+, \dots, \delta_p^+$ the canonical conjunctive queries of $\mathbf{D}_1^+, \dots, \mathbf{D}_p^+$. We now consider their weak union

$$\upsilon \mathrel{\hat{=}} \delta_1^+ \ovee \cdots \ovee \delta_p^+.$$

We then test for each $j \in [\![1, n]\!]$ if $\mathbf{D}_j^- \nvDash \upsilon$: the conjunction of these tests will be the output of our algorithm!

First, each test $\mathbf{D}_j^- \nvDash \upsilon$ can be done in coNP in the size of $\mathbf{D}_j^-$ and of $\upsilon$. Since the size of $\upsilon$ is the product of the sizes of the $\delta_i^+$'s, it is exponential, and hence we obtain a coNExp algorithm. Then, we prove—or rather notice—its correctness: by construction, the weak union is the least upper bound of CQs, see Figure III.15! *Voilà.* □

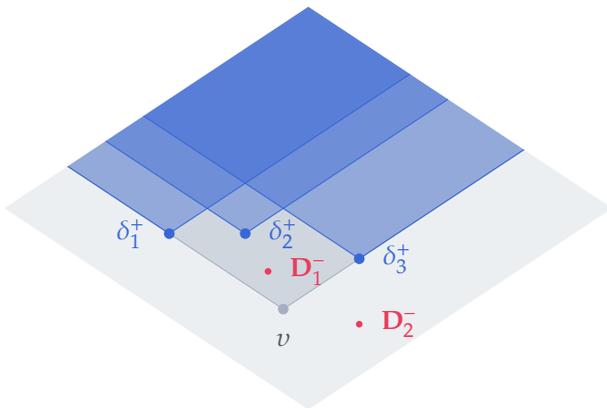

In fact there is a matching lower bound, even if the signature is fixed, but the proof is far from trivial. Willard proved the coNExp-hardness in [Wil10, Theorem 3] when the signature is part of the input[22], and ten Cate and Dalmau proved the same lower bound for some fixed signature [CD15, Theorem 2].[23] In both cases, the reduction is from an exponential tiling problem.

*Active learning.* Active learning, *a.k.a.* Angluin-style learning is a setting where two players interact (*a priori* adversarially): *Student* tries to lean a query that only *Teacher* knows. To achieve this, they can ask two kinds of questions.[24]

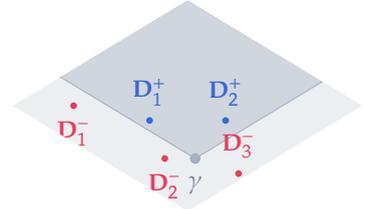

Figure III.14: An instance of the synthesis problem for CQs, as well as one valid solution for this instance.

Figure III.15: A negative instance of the synthesis problem and the construction done in the proof of Proposition III.1.21.

[22] The result is actually phrased for constraint satisfaction problems.

[23] They refer to the synthesis problem as the "CQ-definability problem".

[24] "Questions" are usually called "queries" but we avoid this terminology here for obvious reasons.

- Membership questions: they provide a model and asks if it satisfies the query;

- Equivalence questions: they provide a query and ask Teacher if it is semantically equivalent to theirs; if yes, the game stops, otherwise Teacher provides a counter-example in the form of a model satisfying one query but not the other.

Of course, Student always has a winning strategy: enumerate all queries and ask for equivalence... The difficult and hence interesting question is that of *efficient* learning.

Ten Cate, Dalmau and Kolaitis proved that the class of all conjunctive queries could be learned with polynomially many questions [CDK13, Theorem A], but not with polynomially many questions of a given kind provided that the signature is non-trivial [CDK13, Theorem B]—*i.e.* both kinds of questions are necessary to be able to learn efficiently! By relying on the distributive lattice structure of relational databases, ten Cate and Dalmau then exhibited a subclass of conjunctive queries, known as "*c*-acylic", that can be learned with polynomially many membership queries and no equivalence query [CD21, Theorem 5.2].

### III.1.5 *Conjunctive Queries of Small Tree-Width*

Recall that, by duality, CONJUNCTIVE QUERY EVALUATION is NP-complete: this begs the question of finding classes of CQs with faster evaluation. We will see that queries whose underlying structure looks like a tree—formally, queries of bounded tree-width—can be evaluated in polynomial time.

*Tree-Width.* Tree-width is a measure of how much a graph differs from a tree—the notion was introduced and rediscovered numerous times in the 1970 and 1980s; however the first paper that seems to make a substantial connection between tree-width and tractability seems to be the work of Arnborg and Proskurowski [AP89]. For a gentle but thorough introduction to tree-width, we also refer the reader to [NM12, § 3.6].

Formally, a *tree decomposition* of an undirected graph $\mathbf{G}$ is a pair $\langle \mathbf{T}, \mathbf{v} \rangle$ where $\mathbf{T}$ is a tree and $\mathbf{v} : T \to \mathfrak{P}(G)$ is a function that associates to each node of $\mathbf{T}$, called *bag*, a set of vertices of $\mathbf{G}$. When $x \in \mathbf{v}(b)$ we shall say that the bag $b \in T$ contains vertex $v$. Further, it must satisfy the following three properties:

- each vertex $v$ of $\mathbf{G}$ is contained in at least one bag of $\mathbf{T}$,

- for each edge $\{u, v\}$ of $\mathbf{G}$, there is at least one bag of $\mathbf{T}$ that contains both $u$ and $v$, and

- for every vertex $v$ of $\mathbf{G}$, the set of bags of $\mathbf{T}$ containing $v$ is a connected subset of $T$.

We give an example of tree decomposition in Figure III.16:

- In Figure III.16a, we give the "full" representation of the decomposition: we draw $\mathbf{T}$, and inside each of the four bags $b$ of $\mathbf{T}$ we represent a copy of $\mathbf{G}$.

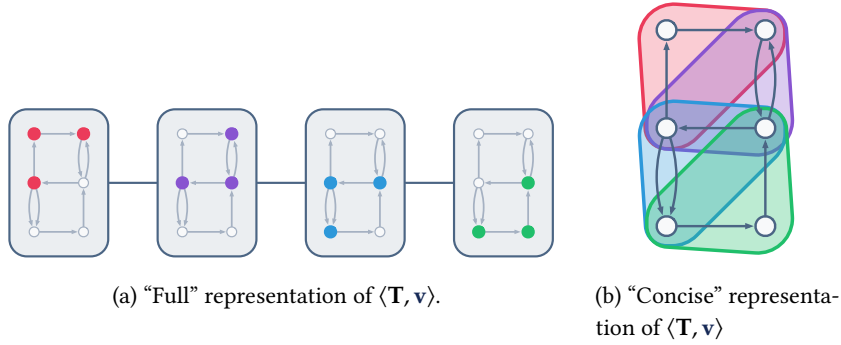

(a) "Full" representation of $\langle \mathbf{T}, \mathbf{v} \rangle$.

(b) "Concise" representation of $\langle \mathbf{T}, \mathbf{v} \rangle$

Figure III.16: Two different representations of the same tree decomposition $\langle \mathbf{T}, \mathbf{v} \rangle$ of a directed graph $\mathbf{G}$ with six vertices. The underlying tree is a path with four nodes and each bag contains 3 vertices—hence the decomposition has width 2.

Nodes of $\mathbf{G}$ belonging to $b$ are highlighted, while the others are dimmed. Sometimes, we will only write the name of the nodes contained in the bag, instead of drawing the graph.

- In Figure III.16b, we give a "concise" representation: we draw over $\mathbf{G}$ a coloured shape for each bag of $\mathbf{T}$. This representation is ambiguous—the structure of $\mathbf{T}$ is not made explicit—and will only be used for the most simple cases.

The *width* of a tree decomposition $\langle \mathbf{T}, \mathbf{v} \rangle$ is the maximum size of a bag minus one, *i.e.* max $\{|\mathbf{v}(b)| - 1 \mid b \in T\}$. The *tree-width* of $\mathbf{G}$ is the minimum of the width of all tree decompositions of $\mathbf{G}$. The notion can be directly generalized to arbitrary relational structures: instead of asking that every edge $\{u, v\}$ is contained in some bag, we require that for every hyperedge $\langle x_1, \dots, x_k \rangle$, there exists a bag that contains all $x_i$'s. The *tree-width* of a CQ is naturally defined as the tree-width of its canonical database. The undirected graphs of tree-width at most 1 are exactly the *forests*, *i.e.* the disjoint unions of trees. On the other hand, the graph of Figure III.16 has tree-width 2. The $k$-clique has tree-width exactly $k - 1$ since any tree decomposition of $\mathbf{K}_k$ must actually contain a bag that contains all vertices of $\mathbf{K}_k$.

**Proposition III.1.22** ([CR00, Theorem 3]).[25],[26] For any fixed signature $\sigma$, for every $k \in \mathbb{N}$, CONJUNCTIVE QUERY EVALUATION can be solved in polynomial time when restricted to CQs of tree-width at most $k$.

Proposition III.1.22, as most algorithms on structures of bounded tree-width actually relies on the ability to compute a tree decomposition.

**Proposition III.1.23** (Bodlaender's algorithm [Bod96, Theorem 1.1]).[27] For every *fixed* $k \in \mathbb{N}_{>0}$, there is a linear-time algorithm which takes as input a finite structure and decides if it has tree-width at most $k$, in which case it also outputs a witness of the form of a tree decomposition of width at most $k$.

*Proof sketch of Proposition III.1.22.* We are given as input a CQ $\gamma(\bar{x})$ of tree-width at most $k$, and a pointed relational database $\langle \mathbf{D}, \bar{d} \rangle$, and need to decide if $\langle \mathbf{D}, \bar{d} \rangle \vDash \gamma(\bar{x})$. *W.l.o.g.*, using Bodlaender's algorithm (Proposition III.1.23), we assume that we also have a tree decomposition $\langle \mathbf{T}, \mathbf{v} \rangle$ of the canonical database $\langle \mathbf{G}, \bar{x} \rangle$ of $\gamma(\bar{x})$.

We do a bottom-up algorithm on the tree $\mathbf{T}$, which maintains a set $\mathcal{H}$

[25] Theorem 3 of [CR00] talks about query containment of CQs, which is in fact equivalent to the EVALUATION PROBLEM for CQs. Moreover, the theorem deals with "query width", but this parameter is equivalent up to a multiplicative constant to the tree-width [CR00, Lemma 2] assuming that the signature is fixed.

[26] An equivalent result was in fact proven a decade ealier by Freuder [Fre90, Theorem 3] using the vocabulary of constraint satisfaction problems.

[27] The fact that $k$ is fixed is crucial: if it is also part of the input, the problem becomes NP-complete [ACP87, Theorem 3.3]

of partial homomorphisms from $\langle \mathbf{G}, \bar{x} \rangle$ to $\langle \mathbf{D}, \bar{d} \rangle$. In light of Figure III.16a, the idea is to compute for each bag where the variables it contains could be mapped on the databases. Formally, we want this procedure to satisfy the following invariant: when dealing with bag $b$, a partial homomorphism $f \colon \langle \mathbf{G}, \bar{x} \rangle \rightharpoonup \langle \mathbf{D}, \bar{d} \rangle$ belongs to $\mathcal{H}_b$ if, and only if,

- the domain of $f$ is $\mathbf{v}(b)$, and
- $f$ can be extended into a partial homomorphism $\tilde{f} \colon \langle \mathbf{G}, \bar{x} \rangle \rightharpoonup \langle \mathbf{D}, \bar{d} \rangle$ defined exactly on the union of $\mathbf{v}(b')$ where $b'$ ranges over vertices that are below $b$ in the tree $\mathbf{T}$.

For leaves $b$, the procedure is easy: we enumerate every possible map from $\mathbf{v}(b)$ to $\mathbf{D}$, and only keep those that define a partial homomorphism $\langle \mathbf{G}, \bar{x} \rangle \rightharpoonup \langle \mathbf{D}, \bar{d} \rangle$: this yields a set $\mathcal{H}_b$ of partial homomorphisms. Then, when dealing with a node $b$ whose children are $b_1, \dots, b_k$, again we enumerate every possible map $f$ from $\mathbf{v}(b)$ to $\mathbf{D}$, but then only keep those that (1) define a partial homomorphism $\langle \mathbf{G}, \bar{x} \rangle \rightharpoonup \langle \mathbf{D}, \bar{d} \rangle$, and (2) agree with the children bags, in the sense that there must exist partial homomorphisms $f_1 \in \mathcal{H}_{b_1}, \dots, f_k \in \mathcal{H}_{b_k}$ s.t. $f$ and $f_i$ agree on their common vertices, i.e. $f|_{\mathbf{v}(b) \cap \mathbf{v}(b_i)} = f_i|_{\mathbf{v}(b) \cap \mathbf{v}(b_i)}$ for all $i \in [\![1, k]\!]$.

The correctness of this procedure follows from the assumption that, in a tree decomposition, the set of bags containing any node must be a connected subset: in some sense this is what allows us to make consistent choices. Then, to decide if $\langle \mathbf{D}, \bar{d} \rangle \vDash \gamma(\bar{x})$, we check if the set of partial homomorphisms associated to the root is non-empty. If so, since every vertex must appear in some bag, the invariant yields the existence of a homomorphism from $\langle \mathbf{G}, \bar{x} \rangle$ to $\langle \mathbf{D}, \bar{d} \rangle$. Otherwise, there is no such homomorphism. Correctness follows by duality.

Lastly, concerning the complexity, observe that the tree decomposition has width at most $k$, so there at most $|D|^{k+1}$ maps from any fixed bag to $\mathbf{D}$: we can enumerate them all in polynomial time since $k$ is fixed! To compute the partial homomorphisms $f$ that agree with some partial homomorphisms from $\mathcal{H}_{b_1}, \dots, \mathcal{H}_{b_k}$, we can first sort each table $\mathcal{H}_{b_i}$ according to their value on $\mathbf{v}(b)$, and then use a binary search. Sorting can be done in $\mathcal{O}(|D|^{k+1} \log(|D|^{k+1})) = \mathcal{O}(|D|^{k+1} \log(|D|))$, and the binary search—one for each partial homomorphism that is a candidate for $\mathcal{H}_b$—runs in $\mathcal{O}(\log(|D|^{k+1})) = \mathcal{O}(\log(|D|))$. Overall, we get an algorithm that runs in time

$$\mathcal{O}(|T| \cdot |D|^{k+1}) = \mathcal{O}(|\mathrm{vars}(\gamma)| \cdot |D|^{k+1}),$$

up to logarithmic factors. $\qquad\qquad\square$

Note also that, given a tree decomposition of a structure $\mathbf{A}$, and a substructure $\mathbf{B}$ of $\mathbf{A}$, by intersecting each bag with $B$, we obtain a tree decomposition of $\mathbf{B}$. It follows that the tree-width of a substructure is always upper bounded by the tree-width of the full structure. In other words, conjunctive queries of tree-width at most $k \in \mathbb{N}$ form a monotone class of CQs.

Grohe, Schwentick and Segoufin proved that there are no graph property other than tree-width that ensures polynomial time evaluation [GSS01, Corollary 19]. Importantly, this statement deals with *graph properties*, *i.e.* with classes of CQs defined by restricting their underlying graph.[28] We will see next that actually there are other classes of CQs with tractable evaluation.

*Bounded Semantic Tree-Width.* As mentioned in Section I.2, Proposition III.1.22 together with the notion of core actually yields a tractability result for a larger class of queries than those of bounded tree-width.

**Proposition III.1.24.** For any $k \in \mathbb{N}$, the CQ EVALUATION PROBLEM, restricted to conjunctive queries that are semantically equivalent to a CQ of tree-width at most $k$, is fixed-parameter tractable when parameterized by the size of the query.

*Proof.* The algorithm goes as follows: we start by computing the core $\check{\gamma}(\bar{x})$ of the CQ $\gamma(\bar{x})$, and then instead of evaluating $\gamma(\bar{x})$ on the database, we instead evaluate $\check{\gamma}(\bar{x})$, using the polynomial-time algorithm of Proposition III.1.22. Overall, the algorithm runs in time

$$\mathcal{O}(f(\|\gamma\|) \cdot |\text{vars}(\gamma)| \cdot |D|^{k+1}),$$

up to logarithmic factors, and where $f(\|\gamma\|)$ is the time required to compute the core of $\gamma$.[29] Hence, the problem is FPT when parameterized by $\|\gamma\|$. □

In fact, Dalmau, Kolaitis and Vardi improved this result and proved that this class could actually be evaluated in polynomial time [DKV02, Corollary 5]—importantly, this is a *promise* problem: we work under the assumption that the input is equivalent to one of bounded tree-width, but we do not need to verify this claim. This is far from trivial, since computing the core of a structure is hard in general [HN92, Theorem 7], and so one cannot naively first compute the core of the query, and then evaluate it, to achieve polynomial time.[30]

Remarkably, Grohe then proved the converse implication: these are the only classes of CQs that are tractable!

**Proposition III.1.25** (*Grohe's theorem* [Gro07, Theorem 1.1]). Let $\sigma$ be a fixed signature. Assuming that W[1] ≠ FPT, for any recursively enumerable class of CQs over $\sigma$, the following are equivalent:

1. there exists $k \in \mathbb{N}$ *s.t.* the tree-width of the cores of the queries in the class is bounded by $k$;
2. its evaluation problem is fixed-parameter tractable when parameterized by the size of the query;
3. its evaluation problem can be solved in polynomial time.

The difficult implication in this theorem is (3) ⇒ (1). Grohe proves it by contraposition, by generalizing the ideas of [GSS01]: in short, given a class whose cores have unbounded tree-width, using the Excluded Minor Theorem

[28] Formally, each relational structure is associated to an undirected graph with the same vertices, and with an edge between two variables if they appear in the same hyperedge.

[29] This can be done in exponential time.

[30] Dalmau, Kolaitis and Vardi's result in fact imply that, for any $k \in \mathbb{N}$, we can compute in polynomial-time the core of a CQ $\gamma(\bar{x})$ under the promise that this core has tree-width at most $k$. Indeed, it suffices for each subquery $\gamma'(\bar{x})$ of $\gamma(\bar{x})$ with exactly one less variable, to let $\langle \mathbf{G}', \bar{x} \rangle$ denote its canonical database, and check if $\langle \mathbf{G}', \bar{x} \rangle \vDash \gamma(\bar{x})$. The latter can be done in polynomial time using [DKV02, Corollary 5]. If the procedure fails for each $\gamma'(\bar{x})$, then $\gamma(\bar{x})$ must be a core. Otherwise, if it succeeds for some $\gamma'(\bar{x})$ we iterate the procedure until we obtain a core. We thank Antoine Amarilli for pointing this out to us.

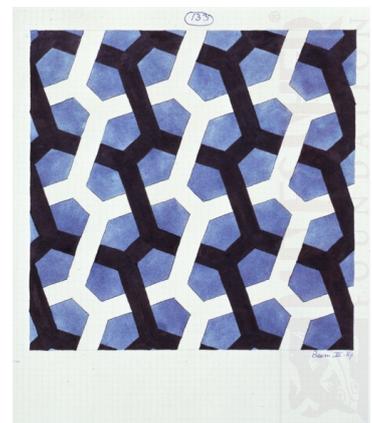

Figure III.17: *E 133*, M. C. Escher, © The M.C. Escher Company.

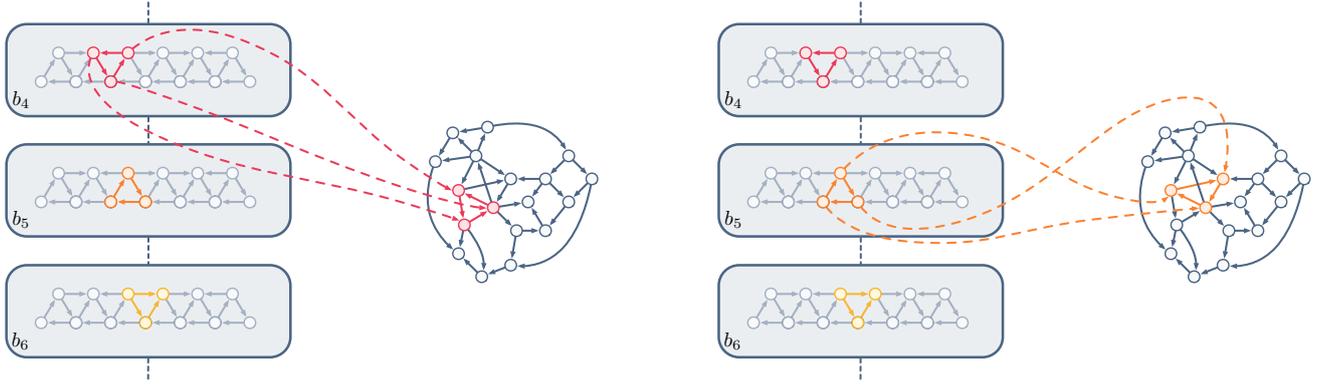

(a) Partial homomorphism computed at the fourth step of the algorithm.

(b) Partial homomorphism computed at the fifth step of the algorithm.

Figure III.18: Algorithm to evaluate a CQ of path-width at most $k$ (here $k = 2$) in NL. Each subfigure represents a path decomposition of the CQ (on the left-hand side) and a relational database (on the right-hand side), together with a partial homomorphism from the first to the second.

[RS86, ∗ (1.5)] one can find an query in the class whose core contains an arbitrarily large grid. In turn, the clique problem, which is W[1]-hard, is then reduced to the evaluation problem for these queries with big grids.

Grohe's theorem deals with fixed signatures: it was later generalized [CGLP20, Theorem 1] for characterizing FPT evaluation when the signature is also part of the input. This is done by replacing the notion of tree-width with that of "submodular width", introduced by Marx in [Mar13].

*Bounded Path-Width.* We now focus on path-width, which is a parameter upper bounded by the tree-width. So, if a class of CQs has bounded path-width, then it has bounded tree-width and so by Proposition III.1.22 it can be evaluated in polynomial time. We shall see that CQs of small path-width can in fact be evaluated even more efficiently, namely in NL! We define a *path decomposition* to be a tree decomposition $\langle \mathbf{T}, \mathbf{v} \rangle$ in which $\mathbf{T}$ is a path, such as in Figure III.16. The *path-width* of a structure is the minimum of the width of all of its path decompositions.

**Lemma III.1.26** ([FM25, Lemma 8.10]).[31],[32]  For each $k \geq 1$, the CQ EVALUATION PROBLEM, restricted to CQs of path-width at most $k$, is NL-complete.

*Proof.* ↬ *Lower bound.* We reduce the problem of REACHABILITY IN FINITE GRAPHS which is NL-hard. Given an instance $\langle \mathbf{G}, s, t \rangle$ of this problem, we reduce it to

$$\langle \mathbf{G}', s, t \rangle \vDash^? \rho(x_1, x_n) \mathrel{\hat{=}} x_1 \to x_2 \wedge x_2 \to x_3 \wedge ... \wedge x_{n-1} \to x_n,$$

where $n = |G|$ and $\mathbf{G}'$ is the graph obtained from $\mathbf{G}$ by adding a self-loop on $t$. Clearly, there is a path from $s$ to $t$ in $\mathbf{G}$ *iff* there is a path from $s$ to $t$ of length at most $n$ in $\mathbf{G}$—by the pigeon-hole principle—, which in turn is equivalent to asking for a path from $s$ to $t$ of length *exactly* $n$ in $\mathbf{G}'$ thanks to the extra self-loop. To conclude, note that $\rho$ has path-width one.

↬ *Upper bound, first part: with the path decomposition.* First, we assume that a path decomposition of width at most $k$ of the query of is also provided as part of the input. So, we are given as input:

- a database $\langle \mathbf{D}, \bar{d} \rangle$,
- a CQ $\gamma(\bar{x})$, and
- a path decomposition $\langle T, \mathbf{v} \rangle$ of width at most $k$ of $\gamma(\bar{x})$.

The algorithm, illustrated in Figure III.18, maintains a partial homomorphism $f : \mathbf{G} \to \mathbf{D}$. We scan the bags of the decomposition from top to bottom.

- Initially—before even scanning the first bag—$f$ is the map with empty domain.
- Then, when scanning the $i$-th bag $b_i$, we start by restricting $f$ to variables of $\mathrm{dom}(f) \cap \mathbf{v}(b_i)$. Then, we extend $f$ so that it is defined on the whole bag $\mathbf{v}(b_i)$. For every variable $y$ in $\mathbf{v}(b_i) \smallsetminus \mathrm{dom}(f)$:
  - if it belongs to $\bar{x}$, say $y = x_i$, we let $f(x_i) \triangleq u_i$;
  - otherwise, we non-deterministically guess the value of $f(y)$.

  We then check, for every atom $\mathcal{R}_{(k)}(x_1, \ldots, x_k)$ of $\gamma(\bar{x})$ s.t. all $x_j$'s occur in $\mathbf{v}(b_i)$ if $\langle f(x_1), \ldots, f(x_k) \rangle \in \mathcal{R}_{(k)}(\mathbf{D})$. If not, we reject.

If the algorithm manages to scan the whole bag decomposition without rejecting, it accepts.

Completeness of the algorithm is trivial. Correctness follows from the fact that if a variable occurs in bags $b_i$ and $b_k$ with $i \leq k$, then it must also belong to every bag $b_j$ for $j \in [\![i, k]\!]$. As a consequence, a variable $x$ is assigned exactly one value $f(x)$ during the whole process.

Concerning the space complexity, by construction, at the $i$-th step of the algorithm, $f$ is defined exactly on $b_i$, so on at most $k + 1$ variables. And so, $f$ can be stored in space $(k + 1) \log(|D|)$. We also need a counter with $\log(|T|)$ bits to scan through the path decomposition. Overall, the algorithm runs in non-deterministic space $\mathcal{O}(k \log(|D|) + \log(|T|)) = \mathcal{O}(\log(|D|) + \log(|T|))$, which is logarithmic in the size of the input.

↪ *Upper bound, second part: without the path decomposition.* Then, we claim that the original problem—when the tree decomposition is not part of the input—also lies in NL. This is because one can compute, from $\gamma$, a path decomposition in (deterministic) logarithmic space by[33] [KM10, Theorem 1.3, p. 2]. The conclusion follows since functions computable in non-deterministic logarithmic space are closed under composition [AB09, Lemma 4.17, p. 88].

□

[33] This result is an adaption of a similar statement for tree-width [EJT10, Theorem I.1, p. 143]. Note that the promise that the query has bounded path-width—in fact bounded tree-width suffices—is a crucial assumption of [EJT10, Theorem I.1, p. 143].

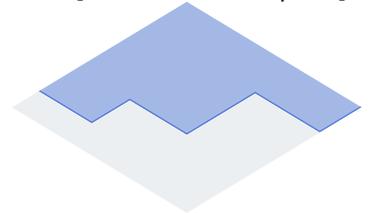

Figure III.19: Semantics of a union of conjunctive queries (in blue) in the distributive lattice of relational databases.

### III.1.6 Unions of Conjunctive Queries

We now show that the desirable properties of conjunctive queries can be lifted to finite unions of such queries. As mentioned in Section III.1.3, conjunctive queries are not closed under semantical union: for instance, the query "the database contains either an $a$-edge or a $b$-edge" cannot be expressed by a conjunctive query. In fact, our graphical depiction of the distributive lattice of relational databases precisely helps us understand this.

Formally, we define a *union of conjunctive queries*, or *UCQ* for short, as a finite set of conjunctive queries that all have the same number of output variables. This set is denoted disjunctively. Its semantics is defined as the union of the semantics of the conjunctive queries it contains.[34] For instance, if $\gamma(x) = x \xrightarrow{a} y$ outputs all vertices with an outgoing $a$-edge and $\delta(x') = x' \xrightarrow{b} y' \xrightarrow{c} z'$ outputs all vertices with an outgoing $bc$-path, then $\gamma(x) \lor \delta(x')$ is the UCQ asking for all nodes that are either the source of an $a$-edge or of a $bc$-path. We denote UCQs with capital Greek letters, and we call *disjunct* of $\Gamma$ any conjunctive query belonging to a UCQ $\Gamma$.

Notice that union of conjunctive queries are also closed under homomorphisms. Graphically, we represent them as... unions of conjunctive queries: instead of being diamond-shaped, UCQs will hence be depicted as inverted mountain ranges, see Figure III.19. By the proof of Proposition III.1.1, we can notice that unions of conjunctive queries are as expressive as existential-positive formulas, *i.e.* first-order formulas built out of $\exists$, $\land$ and $\lor$.

*Hom-Minimality.* We say that a graph database $\langle \mathbf{G}, \bar{u} \rangle$ satisfying a semantical query $\gamma(\bar{x})$ is *hom-minimal* if for every other graph database $\langle \mathbf{G}', \bar{v} \rangle$, if $\langle \mathbf{G}', \bar{v} \rangle$ also satisfies $\gamma(\bar{x})$ and if $\langle \mathbf{G}', \bar{v} \rangle \xrightarrow{\text{hom}} \langle \mathbf{G}, \bar{u} \rangle$ then $\langle \mathbf{G}', \bar{v} \rangle \equiv \langle \mathbf{G}, \bar{u} \rangle$.[35] Graphically, this literally corresponds to the minimal elements of a set in Figure III.9. In the definition, the quantification over all "$\langle \mathbf{G}', \bar{v} \rangle$ that satisfy $\gamma(\bar{x})$" can actually be replaced, for UCQs, by a quantification over all canonical databases.

**Proposition III.1.27.** Let $\Gamma$ be a UCQ. The following are equivalent:

1. $\Gamma$ is semantically equivalent to a conjunctive query;
2. the semantics of $\Gamma$ contains a unique hom-minimal element;
3. some disjunct of $\Gamma$ contains all disjuncts;
4. $\Gamma$ is semantically equivalent to one of its disjuncts;
5. $\Gamma$ is semantically equivalent to the weak union of its disjuncts;

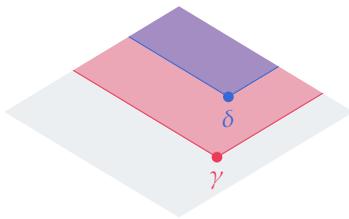

(a) Two CQs $\gamma$ and $\delta$ whose semantical union is equivalent to a CQ.

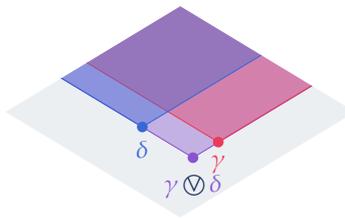

(b) Two CQs $\gamma$ and $\delta$ whose semantical union is not equivalent to a CQ. We also represented their weak union.

*Proof.* All the intuitions are provided in Figure III.21.

☙ *(1) ⟹ (2).* If $\Gamma$ is semantically equivalent to a CQ $\delta$, then the canonical database $\mathbf{D}$ of $\delta$ is the unique minimal element in the semantics of $\Gamma$.

☙ *(2) ⟹ (3).* If the semantics of $\Gamma$ contains a unique minimal element, say $\mathbf{D}$, then since $\mathbf{D} \vDash \Gamma$, there exists a disjunct $\gamma$ of $\Gamma$ *s.t.* $\mathbf{D} \vDash \gamma$. By minimality of

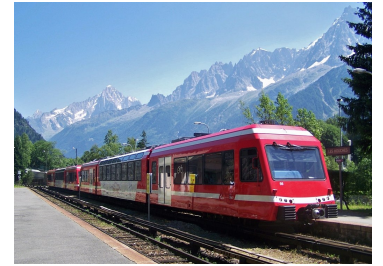

Figure III.20: The study of (co-)UCQs is intriguingly popular amongst finite model theorists. *TER en Gare des Houches*, by Florian Pépellin, licensed under CC BY SA 3.0.

Figure III.21: When is a union of conjunctive queries actually equivalent to a conjunctive query?

**D**, it follows that $\gamma$ is actually the canonical CQ of **D**, and again by minimality of **D** it follows that all disjuncts are contained in $\gamma$.

☙ *(3) ⟹ (4).* If $\Gamma = \gamma_1 \vee ... \vee \gamma_k$ has a disjunct, say $\gamma_i$, that contains all other disjuncts, then

$$\Gamma = \gamma_1 \vee ... \vee \gamma_k \subseteq \gamma_i \vee ... \vee \gamma_i \equiv \gamma_i \subseteq \Gamma$$

and so $\Gamma$ is semantically equivalent to its disjunct $\gamma_i$.

☙ *(4) ⟹ (5).* If $\Gamma = \gamma_1 \vee ... \vee \gamma_k$ is semantically equivalent to one of its disjuncts, say $\gamma_i$, then for each $j$ we have $\gamma_j \subseteq \Gamma \equiv \gamma_i$, and so by definition of the weak union as the least upper bound, it follows that $\gamma_1 \ovee ... \ovee \gamma_k \equiv \gamma_i$.

☙ *(5) ⟹ (1).* If $\Gamma$ is semantically equivalent to the weak union of its disjuncts, then since this query is a CQ, it is equivalent to a CQ... □

**Proposition III.1.28.** Given two UCQs $\Gamma$ and $\Delta$, we have $\Gamma \subseteq \Delta$ if, and only if, for every disjunct $\gamma \in \Gamma$, there exists a disjunct $\delta \in \Delta$ *s.t.* $\gamma \subseteq \delta$.

*Proof.* The right-to-left implication is trivial. For the converse one, assume that $\Gamma \subseteq \Delta$, and let $\gamma$ be a disjunct of $\Gamma$. Letting **G** be its canonical database, we have by duality that $\mathbf{G} \vDash \Gamma$ and since $\Gamma \subseteq \Delta$, it follows that $\mathbf{G} \vDash \Delta$, and hence $\mathbf{G} \vDash \delta$ for some disjunct $\delta \in \Delta$, and so, by duality, $\gamma \subseteq \delta$. □

**Corollary III.1.29.** Containment and semantical equivalence of unions of conjunctive queries are NP-complete.

Essentially, these results are permitted, at least in part, by the fact that semantical union is well-behaved with respect to duality: this is mostly because their semantics is closed under homomorphisms. Hopefully, one could hope to find larger query languages satisfying this property, and then deduce decidability results via duality. Both unfortunately and expectedly, it turns out that there are no other first-order queries which are closed under homomorphisms.

**Proposition III.1.30** (*Rossman's theorem*, [Ros08, Theorem 1.7]).[36],[37] The semantics over finite relational structures of a first-order sentence is closed under homomorphisms if, and only if, it is equivalent to a union of conjunctive queries.

We leave to the reader the care to show that the other properties of CQs can be lifted to UCQs, for instance:

- starting from a UCQ $\gamma_1 \vee ... \vee \gamma_k$, the UCQ obtained by (1) considering the cores $\check{\gamma}_1, ... , \check{\gamma}_k$, (2) putting aside those that are contained in another core, and (3) taking their union yields a UCQ that is semantically equivalent to the original one, and minimal under most reasonable metrics;
- the evaluation of UCQs of bounded tree-width can be done in polynomial time—we naturally extend the notion of tree-width to unions by letting it be the maximum of the tree-width of its disjuncts.

[36] Actually the full statement of the theorem is slightly more precise and relates the quantifier-rank of the formulas. The original construction was far from optimal, and it was recently shown by the same author that this rank can actually be preserved by the construction [Ros25, Theorem 1.4].

[37] Both references mention first-order sentences; however we do not see any reason why it would not apply to all first-order formulas.

In the next section, we will deal with a class of queries closed under homomorphisms that strictly extends unions of conjunctive queries, namely unions of conjunctive regular path queries. Naturally, in light of Rossman's theorem, it will be incomparable with first-order logic, but the fact that it is closed under homomorphisms will help retain a few helpful properties that UCQs have.

## III.2  Graph Databases

For a detailed introduction to CRPQs, the reader can also see [Fig21a]. For a more general introduction to different query languages for graph databases, including CRPQs, see [Bar13], and for a more practical approach, see [Ang+17].

### III.2.1  Conjunctive Regular Path Queries

Section III.1 shows that conjunctive queries—and unions thereof—is a reasonable query language: it is expressive enough to capture many of the queries that are written in practice in SQL, but small enough to be amenable to static analysis thanks to duality. However, conjunctive queries, and in fact the larger language of first-order logic, or *relational calculus*, is remarkably powerless against "graph-traversal queries".

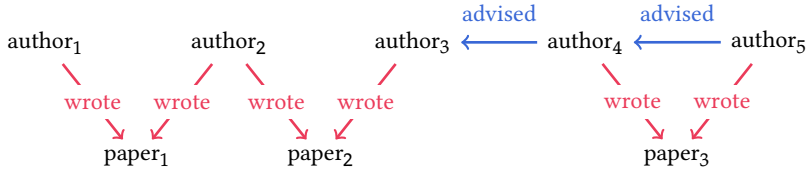

Figure III.22: A relational database with two binary predicates.

We illustrate this idea on a simplified example—to see a real example, see Figures I.13, I.14 and I.16 and Table I.15. We depict in Figure III.22 a relational database over the purely relational signature with two binary predicates $\xrightarrow{\text{wrote}}$ and $\xrightarrow{\text{advised}}$. Vertices of the structure represent people, and an edge $x \xrightarrow{\text{wrote}} y$ indicate that the person $x$ wrote the paper $y$, while edges $x \xrightarrow{\text{advised}} y$ indicate that person $x$ was the Ph.D. advisor of person $y$. While conjunctive queries can express properties like

$$\gamma_1(x, y) = x \xrightarrow{\text{wrote}} z \wedge y \xrightarrow{\text{wrote}} z,$$

that asks for all pairs of co-authors, it cannot express transitive-based properties; in fact, even the full first-order logic cannot express such properties.

**Proposition III.2.1.** There is not first-order formula $\gamma_2(x, y)$ that ouputs all pairs *s.t.* $x$ is a scientific ancestor of $y$—*i.e. s.t.* there is a non-empty path from $x$ to $y$ that only consists of advised edges.

*Proof.* Assume by contradiction that this would be the case. Clearly, $\gamma_2(x, y)$ is closed under homomorphisms, and so by Rossman's theorem, it would be equivalent to a UCQ, say $\delta_1(x, y) \vee \ldots \vee \delta_k(x, y)$. We let $n$ be strictly greater than the diameter of any $\delta_i$, and define $\langle \mathbf{P}_n, p_0, p_n \rangle$ to be the pointed relational database consisting the facts

$$p_0 \xrightarrow{\text{wrote}} p_1 \xrightarrow{\text{wrote}} \ldots \xrightarrow{\text{wrote}} p_n,$$

which is actually a shorthand for

$$\{p_0 \xrightarrow{\text{wrote}} p_1, \ p_1 \xrightarrow{\text{wrote}} p_2, \ \ldots, \ p_{n-1} \xrightarrow{\text{wrote}} p_n\}.$$

Then clearly, $\langle \mathbf{P}_n, p_0, p_n \rangle \vDash \gamma_2(x, y)$ and so $\langle \mathbf{P}_n, p_0, p_n \rangle \vDash \delta_i(x, y)$ for some $i \in [\![1, k]\!]$. By duality, there exists a homomorphism from the canonical database $\langle \mathbf{D}_i, x, y \rangle$ of $\delta_i(x, y)$ to $\langle \mathbf{P}_n, p_0, p_n \rangle$. However, the distance from $x$ to $y$ in $\mathbf{G}_i$ is strictly upper bounded by $n$, since $n$ was chosen to be strictly greater than the diameter of $\mathbf{D}_i$, and the distance from $x$ to $y$ is exactly $n$. Since homomorphisms contract distances, we have reached a contradiction. And hence, the property of Proposition III.2.1 is not first-order definable.[38]  □

Queries such as the one of Proposition III.2.1 do not arise over all kinds of databases: for instance, transitive closures of tables—or of projections of tables— for Table III.1 would not make any sense. On the other hand, it is a very natural construction for databases such as the ones of Figures I.14 and III.22. Transitive closures naturally arise in databases whose nodes are homogeneous, and in particular for human-centered data—see for instance [Neo]. Rather than encoding this kind of data into relational databases, by inferring from the examples of Figures I.14 and III.22, we model graph databases as edge-labelled directed graphs, or equivalently as relational structures over a finite signature consisting only of binary predicates. Formally, given an alphabet $\mathbb{A}$, a *graph database* over $\mathbb{A}$ consists of a pair $\mathbf{G} = \langle G, \mathcal{E}(G) \rangle$, where $\mathcal{E}(G) \subseteq G \times \mathbb{A} \times G$ is a set of labelled edges—sometimes also called atoms.

This definition was motivated by the need to have database models centred around the notion of paths. We then naturally extend the language of conjunctive queries to allow for path navigation: this is done by allowing these queries to ask for the existence of a *path* between two variables labelled by a given regular language. For instance, $x \xrightarrow{(ab)^*} y$ asks that there is a path from $x$ to $y$ that alternates between $\xrightarrow{a}$ and $\xrightarrow{b}$ edges.

**Definition III.2.2** (CRPQ: syntax). A *conjunctive regular path query*, or *CRPQ* for short, $\gamma(\bar{x})$ over $\mathbb{A}$ consists of a set of triples $\langle y, L, z \rangle$ together with a tuple of variables $\bar{x}$, where $L$ is a regular language over $\mathbb{A}$, and $y$ and $z$ are variables.[39] We denote by $x \xrightarrow{L} y$ the triple $\langle x, L, y \rangle$, called *atom*. The set Atoms$(\gamma)$ of atoms is denoted conjunctively. We denote by vars$(\gamma)$ the set of variables occurring in $\gamma$. As for CQs, variables of $\bar{x}$ are called *output variables*.

An example of conjunctive regular path query is

$$\gamma_3(x, y) \mathrel{\hat{=}} x \xrightarrow{\text{wrote}} z \wedge z' \xrightarrow{\text{wrote}} z \wedge y \xrightarrow{(\text{advised})^*} z',$$

[38] A much more common way of proving Proposition III.2.1 is by using the notion of Ehrenfeucht–Fraïssé games, see *e.g.* [Kol07, Proposition 2.3.28]. or Gaifman/Hanf-locality, see *e.g.* [Bar09]. In all cases, the proof relies on an idea that first-order queries are local. In our proof, the notion of locality—formalized by the fact that homomorphisms contract the distances—is extremely strong, but is only valid over the first-order queries amenable to duality, *i.e.* UCQs. This results in a very simple proof, which however relies on a highly non-trivial result, namely Rossman's theorem. On the other hand, a proof based on Ehrenfeucht–Fraïssé games or Gaifman/Hanf-locality does not rely on any deep theorem, but the notion of locality it manipulates is much more complex, as it captures all first-order queries.

[39] In practice, $L$ is specified as a regular expression. For all results of this thesis, however, we assume that it is represented as an NFA. This makes little difference as both representations are equivalent up to a polynomial blow-up.

which intuitively asks for, in a given graph database, all pairs of people $\langle x, y \rangle$ such that

- $x$ has written some paper $z$, and some person $z'$ also wrote $z$,
- this $z'$ can be reached from $y$ by taking an arbitrary sequence of $\xrightarrow{\text{advised}}$ edges.

Reformulating the constraints, we get that $x$ and $z'$ should be co-authors, and that $z'$ should be a scientific descendant—in the broad sense—of $y$. Hence, the query asks for all pairs $\langle x, y \rangle$ *s.t.* $x$ is a co-author of a "scientific descendant" of $y$.

By definition, conjunctive regular path queries over $\mathbb{A}$ can be seen as finite pointed structures over the infinite signature $\sigma_{\text{Reg}(\mathbb{A}^*)}$ that has a binary predicate $\xrightarrow{L}$ for each regular language $L \subseteq \mathbb{A}^*$: this leads to a natural depiction of these queries, see Figure III.23.

**Definition III.2.3** (CRPQ: semantics). Given a CRPQ $\gamma(\bar{x})$ and a pointed graph database $\langle \mathbf{G}, \bar{u} \rangle$, an *evaluation map* from $\gamma(\bar{x})$ to $\langle \mathbf{G}, \bar{u} \rangle$ is a function from $\text{vars}(\gamma)$ to $G$ that sends $\bar{x}$ to $\bar{u}$, and such that for every atom $x \xrightarrow{L} y$ of $\gamma$, there exists a path from $f(x)$ to $f(y)$ in $\mathbf{G}$ labelled by some word of $L$. See Figure III.24 for an example.

When such an evaluation map exists, we say that $\langle \mathbf{G}, \bar{u} \rangle$ *satisfies* $\gamma(\bar{x})$, which we still denote by $\langle \mathbf{G}, \bar{u} \rangle \vDash \gamma(\bar{x})$ or $\bar{u} \in [\![\gamma(\bar{x})]\!]^{\mathbf{G}}$. The set $[\![\gamma(\bar{x})]\!]^{\mathbf{G}}$ is called the *evaluation* of $\gamma(\bar{x})$ over $\mathbf{G}$.

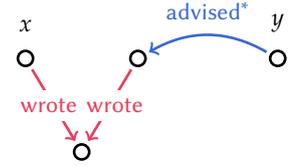

Figure III.23: Graphical representation of the CRPQ $\gamma_3(x, y)$.

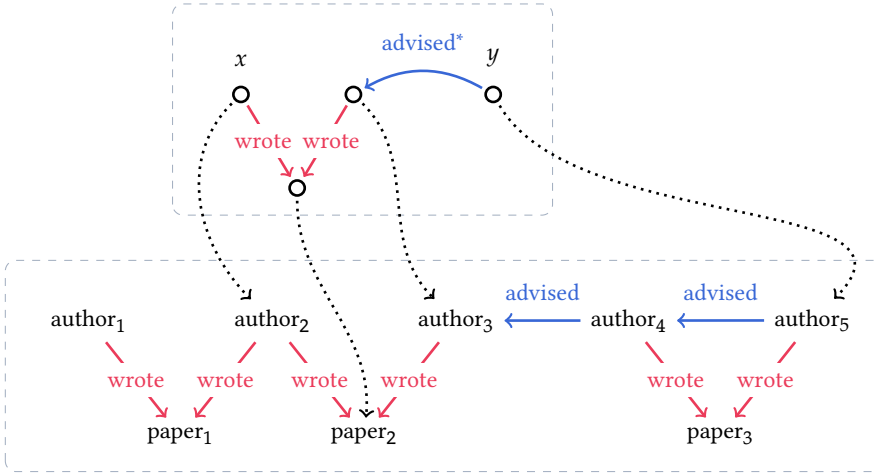

Figure III.24: A conjunctive regular path query (above), a graph database (below), and an evaluation map from the former to the latter.

Evaluation maps are, as Figure III.24 hints at, well-behaved *w.r.t.* homomorphisms: formally, if $f \colon \gamma(\bar{x}) \to \langle \mathbf{H}, \bar{v} \rangle$ is an evaluation map, and $g \colon \langle \mathbf{H}, \bar{v} \rangle \to \langle \mathbf{H}', \bar{v}' \rangle$ is a homomorphism, then their composition $g \circ f \colon \gamma(\bar{x}) \to \langle \mathbf{H}', \bar{v}' \rangle$ is still an evaluation map.

**Fact III.2.4.** In particular, the semantics of CRPQs is closed under homomorphisms.

When $\gamma(\bar{x})$ is a CQ, an evaluation map $\gamma(\bar{x}) \to \langle \mathbf{H}, \bar{v} \rangle$ exactly corresponds to homomorphisms $\langle \mathbf{G}, \bar{x} \rangle \to \langle \mathbf{H}, \bar{v} \rangle$, where $\langle \mathbf{G}, \bar{x} \rangle$ denotes the canonical database of $\gamma(\bar{x})$.[40]

Naturally, any CQ can be seen as a CRPQ by identifying $\xrightarrow{a}$ with $\xrightarrow{[a]}$. Hence, by Proposition III.2.1, it follows that CRPQs are strictly more expressive than CQs. An alternative but equivalent definition for the semantics of CRPQs is to encode them as primitive-positive formulas over the infinite signature $\sigma_{\mathrm{Reg}(A^*)}$. Exactly like for conjunctive queries, we encode *e.g.*

$$\gamma_3(x, y) = x \xrightarrow{\text{wrote}} z \wedge z' \xrightarrow{\text{wrote}} z \wedge y \xrightarrow{(\text{advised})^*} z'$$

as

$$\bar{\gamma}_3(x, y) \mathrel{\hat{=}} \exists z.\ \exists z'.\ x \xrightarrow{\text{wrote}} z \wedge z' \xrightarrow{\text{wrote}} z \wedge y \xrightarrow{(\text{advised})^*} z'.$$

To evaluate a primitive-positive formula $\bar{\gamma}(\bar{x})$ over $\sigma_{\mathrm{Reg}(A^*)}$, we need to build a finite $\sigma_{\mathrm{Reg}(A^*)}$-structure out of a graph database $\mathbf{G}$: we let $\bar{\mathbf{G}}$ share the same vertices as $\mathbf{G}$ and, for every regular language $L$ s.t. $\xrightarrow{L}$ occurs in $\phi$, we interpret the predicate $\xrightarrow{L}$ as the set of pairs $\langle u, v \rangle$ of vertices s.t. there is a path from $u$ to $v$ labelled by a word of $L$.[41] It is then routine to check that the first-order semantics $\llbracket \bar{\gamma}(\bar{x}) \rrbracket^{\bar{\mathbf{G}}}$ coincides with the CRPQ semantics $\llbracket \gamma(\bar{x}) \rrbracket^{\mathbf{G}}$.[42]

Note however that this does not mean that the theory of CRPQs boils down to the theory of CQs: one reason is first that the construction $\mathbf{G} \mapsto \bar{\mathbf{G}}$ actually depends on the query—we need to restrict our construction to languages occurring in the query in order to obtain a *finite* structure. Even more importantly, the construction $\mathbf{G} \mapsto \bar{\mathbf{G}}$ actually fails to preserve any interesting property: notice *e.g.* that if $\mathbf{G}$ is a directed path

$$u_0 \xrightarrow{a} u_1 \xrightarrow{a} ... \xrightarrow{a} u_n,$$

and if only $\xrightarrow{a^*}$ occurs in the query, then $\bar{\mathbf{G}}$ is the $k$-transitive tournament, see Figure III.25. From a very simple structure of path-width 1, we can obtain a structure of unbounded tree-width! Hence, CRPQs deserve to be studied for their own sake.

> CONJUNCTIVE REGULAR PATH QUERIES EVALUATION
> *Input*: An alphabet $\mathbf{A}$, a conjunctive regular path query $\gamma(\bar{x})$ over $\mathbf{A}$, and a pointed graph database $\langle \mathbf{G}, \bar{u} \rangle$.
> *Question*: Does $\bar{u} \in \llbracket \gamma(\bar{x}) \rrbracket^{\mathbf{G}}$?

When dealing with CRPQs, we will assume in our proofs that the regular languages are specified by non-deterministic finite automata; however we will use regular expressions in every example for the sake of readability. Given a CRPQ $\gamma$, we denote its number of variables and atoms by $\|\gamma\|_{\mathrm{var}}$ and $\|\gamma\|_{\mathrm{at}}$, respectively. On the other hand $\|\gamma\|$ denotes the size of any reasonable representation of the query: we agree to take the sum of its number of atoms with the sum of the size of automata (NFAs) used to describe $\gamma$.

**Proposition III.2.5.** The CONJUNCTIVE REGULAR PATH QUERIES EVALUATION PROBLEM is NP-complete.

[40] Because of this, the terminology is sometimes abused in the literature, where evaluations maps for CRPQs can be referred to as "homomorphisms"... We refuse this etymological blasphemy: conjunctive regular path queries and graph databases are not of the same type—formally, these structures do not share the same signature—so they cannot be homomorphic. Partially out of defiance we actually considered calling evaluation maps "heteromorphisms" since they are constraint-preserving maps between objects having a different structure.

[41] This can be computed in NL *e.g.* by doing a Cartesian product between $\mathbf{G}$ and an automaton for $L$, followed by a suitable projection.

[42] Hence, our reuse of $\vDash$ and $\llbracket - \rrbracket^-$ is actually a very reasonable abuse of notation!

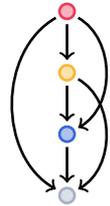

Figure III.25: The 3-transitive tournament $\mathbf{T}_3$. (Replica of Figure I.25.)

*Proof.* The lower bound already holds for CQs by Proposition III.1.7. The upper bound can be proven by (1) guessing a function from the variables of the query to the vertices of the database, and (2) checking that it is an evaluation map. For each atom, each check consists in verifying if there is a path from some vertex to another one that is labelled by a word of some language, which can be done in NL *e.g.* by an easy adaptation of the NL algorithm for REACHABILITY IN FINITE GRAPHS. □

Overall, the algorithms for CQ EVALUATION extend effortlessly to CRPQs. Let us point out that we defined graph databases as a subclass of relational structures and not as relational databases: the motivation behind this definition is that, in light *e.g.* of the Wikidata example of Section I.2—or of the Panama Papers [Neo]!—the vertices of graph databases are first-class citizens!

*Variations on Conjunctive Regular Path Queries.* We mention here a few sub/subclasses of CRPQs that will play an important role. *Regular path queries* (*RPQ*) consist of CRPQs with a single atom: the best example of RPQ is perhaps that of $\gamma(x, y) = x \xrightarrow{\mathbb{A}^*} y$ that described a simple reachability constraint.

In the other direction, the languages used in CRPQs can be extended to navigate edges in both directions. Formally, given a graph database $\mathbf{G}$, we consider the expanded database $\mathbf{G}^{\pm}$ obtained from $\mathbf{G}$ by adding, for every edge $x \xrightarrow{a} y$ in $\mathbf{G}$, an extra edge $y \xrightarrow{a^-} x$. We obtain a graph database on the alphabet $\mathbb{A}^{\pm} = \mathbb{A} \cup \mathbb{A}^-$ where $\mathbb{A}^- \triangleq \{a^- \mid a \in \mathbb{A}\}$. We then define the syntax of a *CRPQ with two-way navigation*, or *C2RPQ*, as a CRPQ on the alphabet $\mathbb{A}^{\pm}$. Its *evaluation* is defined as the evaluation of the CRPQ on $\mathbf{G}^{\pm}$. For instance, the evaluation of the C2RPQ

$$\gamma_4(x, y) \triangleq x \xrightarrow{(\text{wrote} \cdot \text{wrote}^-)^*} y$$

on the graph database of Figure III.22 returns all pairs of individuals linked by a chain of co-authorship. It includes $(\text{author}_1, \text{author}_3)$ or $(\text{author}_1, \text{author}_1)$ but not $(\text{author}_1, \text{author}_4)$.

*Unions of conjunctive regular path queries* (*UCRPQs*) and *unions of conjunctive two-way regular path queries* (*UC2RPQs*) are defined analogously to unions of conjunctive queries. *Infinitary unions* are defined similarly, except that we allow for potentially infinite unions. We often use a set notation to denote the union, especially for infinitary unions.

**Fact III.2.6.** Infinitary unions of CQs, of CRPQs, or even of UC2RPQs all have the same expressivity, which exactly correspond to the semantical queries that are closed under homomorphisms.

*Proof.* Indeed, it is easy to see that each of these query languages are closed under homomorphisms. Conversely, given a query $\phi$ that is closed under homomorphisms, then $\phi$ is semantically equivalent to the infinitary union of

CQs

$$\bigvee \{\gamma \mid \gamma \text{ canonical CQ of some } \mathbf{G} \in \phi\}. \qquad \square$$

These infinite objects will mostly appear as intermediate objects in our constructions and proofs, on our way to proving that they are actually equivalent to a *finite* union.

### III.2.2 Deciding Equivalence of Conjunctive Regular Path Queries

CRPQs form the core navigational mechanism of the new ISO standard Graph Query Language (GQL) [ISO24] and the SQL extension for querying graph-structured data SQL/PGQ [ISO23] (see also [Fra+23a; Fra+23b]). Hence, the static analysis of conjunctive regular path queries is of the foremost importance: we prove in this subsection that semantical equivalence of CRPQs is decidable. Notice however that the containment of $\gamma_1(x, y) \stackrel{\circ}{=} x \xrightarrow{K} y$ into $\gamma_2(x, y) \stackrel{\circ}{=} x \xrightarrow{L} y$ is equivalent to asking for $K \subseteq L$. Since language inclusion for non-deterministic automata is PSpace-hard, it follows that there is no hope for containment to be as easy for CRPQs as it is for conjunctive queries...

Static analysis of conjunctive queries relied on duality theory, which associated to each CQ a canonical database. We show here that for CRPQs, or even for UC2RPQs, we can associate to each query an *infinite* set of canonical databases, which will help us decide containment, and hence semantical equivalence.

*Homomorphisms.* CRPQs and C2RPQs over an alphabet $\mathbb{A}$ can be seen as relational structures over $\sigma_{\mathsf{Reg}(\mathbb{A}^*)}$ and $\sigma_{\mathsf{Reg}(\mathbb{A}^{\pm *})}$, respectively. As such, they come equipped with a notion of homomorphism between them, see *e.g.* Figure III.27b. These homomorphisms behave nicely *w.r.t.* to evaluation maps,[43] in the sense that for every homomorphism of CRPQs $f \colon \gamma'(\bar{x}') \to \gamma(\bar{x})$, for every evaluation map $g \colon \gamma(\bar{x}) \to \langle \mathbf{H}, \bar{v} \rangle$, then their composition $g \circ f \colon \gamma'(\bar{x}') \to \langle \mathbf{H}, \bar{v} \rangle$ is still an evaluation map.

**Fact III.2.7.** Hence, for any CRPQs (resp. C2RPQs) $\gamma(\bar{x})$ and $\gamma'(\bar{x}')$, if $\gamma(\bar{x}) \xrightarrow{\mathsf{hom}} \gamma'(\bar{x}')$ then $\gamma'(\bar{x}') \subseteq \gamma(\bar{x})$.

However, contrary to the case of conjunctive queries (Proposition III.1.8), the converse implication does *not* hold: letting $\gamma_1() \stackrel{\circ}{=} x \xrightarrow{ab} z$ and $\gamma_2() \stackrel{\circ}{=} x \xrightarrow{a} y \xrightarrow{b} z$, then $\gamma_1() \subseteq \gamma_2()$ (in fact these queries are equivalent) but there is no homomorphism from $\gamma_2()$ to $\gamma_1()$—and neither from $\gamma_1()$ to $\gamma_1()$. This lack of converse implication for Fact III.2.7 is precisely what makes the static analysis of CRPQs hard—or interesting, depending on the point of view! The previous example, while important, can actually seem somewhat degenerate: the only problems seems to occur from the fact that an atom $\xrightarrow{ab}$ essentially has access to an internal variable that the homomorphism cannot manipulate. While this is indeed an issue, it only illustrates part of the difficulty posed by CRPQs, and so we turn to a slightly more involved example.

**Example III.2.8** (Simplification of [FLS98, Figure 3].). Consider the Boolean

CRPQs

$$\gamma() \triangleq w \xrightarrow{a} x \xrightarrow{a+b} y \xrightarrow{b} z \quad \text{and} \quad \delta() \triangleq x \xrightarrow{a} y \xrightarrow{b} z.$$

We claim that $\gamma() \subseteq \delta()$: indeed, if **G** is a database that satisfies $\gamma()$ then it contains a path that is either labelled by *aab* or by *abb*, and so in both cases it contains an *ab*-path. However, there is no homomorphism from $\delta()$ to $\gamma()$. Note that here the issue does not arise from an "internal variable", but rather from the fact that to handle one case (the case of *aab*-paths), we would need to map $\langle x, y, z \rangle$ to $\langle w, x, y \rangle$, while in the other case (the case of *abb*-paths) we would need to map $\langle x, y, z \rangle$ (of $\delta$) to $\langle x, y, z \rangle$ (of $\gamma$) to obtain a homomorphism.[44]                                                                          ◇

The idea behind canonical databases for C(2)RPQs is as follows: starting from a CRPQ $\gamma(\bar{x}) = \bigwedge_{i=1}^{n} y_i \xrightarrow{L_i} z_i$, we look at every possible choice of words $\langle u_i \in L_i \rangle_{i \in [\![1,n]\!]}$. Replacing $\xrightarrow{L_i}$ by a directed path labelled by $u_i$ yields a pointed graph database that satisfies $\gamma(\bar{x})$ by construction. These will precisely be the canonical databases of our CRPQ! We illustrate the notion on Figure III.26.

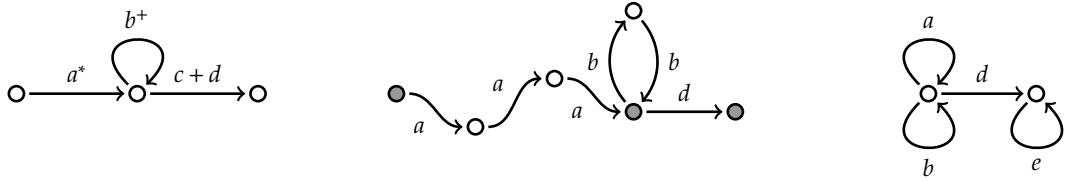

Dealing formally with the case that $u_i = \varepsilon$ is actually somewhat tricky since, as a result, the two ends of this atom needs to be merged. So, to properly define canonical databases for CRPQs, we first need to introduce a few technical tools, that will also prove useful throughout this part of the thesis to manipulate CRPQs.

*Equality Atoms.* C2RPQs with *equality atoms* are queries of the form $\gamma(\bar{x}) = \delta \wedge I$, where $\delta$ is a C2RPQ (without equality atoms) and $I$ is a conjunction of equality atoms of the form $x = y$. Again, we denote by $\text{vars}(\gamma)$ the set of variables appearing in the (equality and non-equality) atoms of $\gamma$. We define the binary relation $=_\gamma$ over $\text{vars}(\gamma)$ to be the reflexive-symmetric-transitive closure of the binary relation $\{\langle x, y \rangle \mid x = y \text{ is an equality atom in } \gamma\}$. In other words, we have $x =_\gamma y$ if the equality $x = y$ is forced by the equality atoms of $\gamma$. Note that every C2RPQ with equality atoms $\gamma(\bar{x}) = \delta \wedge I$ is equivalent to a C2RPQ without equality atoms $\gamma^\approx$, which is obtained from $\gamma$ by collapsing each equivalence class of the relation $=_\gamma$ into a single variable. This transformation gives us a *canonical* renaming from $\text{vars}(\gamma)$ to $\text{vars}(\gamma^\approx)$. For instance, $\gamma(x, y) \triangleq x \xrightarrow{K} y \wedge y \xrightarrow{L} z \wedge x = y$ collapses to $\gamma^\approx(x, x) \triangleq x \xrightarrow{K} x \wedge x \xrightarrow{L} z$.

*Refinements.* For an NFA $\mathcal{A}$ and two states $q, q'$ thereof, we denote by $\mathcal{A}[q, q']$ the *sublanguage* of $\mathcal{A}$ recognized when considering $\{q\}$ as the set of initial

---

[44] Actually this idea of having maps that covers only *some* cases but not all can be formalized, and leads to an algorithm for deciding containment: see the notion of "canonical database *covered by a query mapping*" in [FLS98].

Figure III.26: A Boolean CRPQ (left-hand side), one of its canonical databases (middle), and a graph database that satisfies the query but is not one of its canonical databases.

states and $\{q'\}$ as the set of final states. An *atom m-refinement* of a C2RPQ atom $\gamma(x, y) = x \xrightarrow{L} y$ where $L$ is given by the NFA $\mathcal{A}_L$ is any C2RPQ of the form

$$\rho(x, y) = x \xrightarrow{L_1} t_1 \xrightarrow{L_2} \dots \xrightarrow{L_{n-1}} t_{n-1} \xrightarrow{L_n} y \tag{III.6}$$

where $1 \le n \le m$, $t_1, \dots, t_{n-1}$ are fresh (existentially quantified) variables, and $L_1, \dots, L_n$ are such that there exists a sequence $(q_0, \dots, q_n)$ of states of $\mathcal{A}_L$ such that $q_0$ is initial, $q_n$ is final, and for each $i$, $L_i$ is either of the form

i. $\mathcal{A}_L[q_i, q_{i+1}]$,

ii. $\{a\}$ if the letter $a \in \mathbb{A}$ belongs to $\mathcal{A}_L[q_i, q_{i+1}]$, or

iii. $\{a^-\}$ if $a^- \in \mathbb{A}^-$ belongs to $\mathcal{A}_L[q_i, q_{i+1}]$.

Additionally, if $\varepsilon \in L$, the equality atom "$x = y$" is also an atom $m$-refinement. Thus, an *atom m-refinement* can be either of the form (III.6) or "$x = y$".

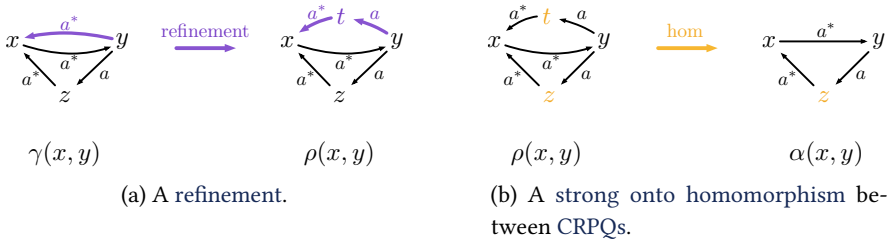

Figure III.27: Homomorphisms and refinements between CRPQs.

$\gamma(x, y)$     $\rho(x, y)$

(a) A refinement.

$\rho(x, y)$     $\alpha(x, y)$

(b) A strong onto homomorphism between CRPQs.

By convention, $t \xrightarrow{a^-} t'$ is a shorthand for $t' \xrightarrow{a} t$. As a consequence, the underlying graph of an atom $m$-refinement of the form (III.6) is not necessarily a directed path. By definition, note that $L_1 \cdots L_n \subseteq L$ and hence $\rho \subseteq \gamma$ for any atom $m$-refinement $\rho$ of $\gamma$. An *atom refinement* is an atom $m$-refinement for some $m$. An example is provided in Figure III.27a.

**Definition III.2.9.** Given an atom refinement $\rho = x \xrightarrow{L_1} t_1 \xrightarrow{L_2} \dots \xrightarrow{L_{n-1}} t_{n-1} \xrightarrow{L_n} y$ of $\gamma = x \xrightarrow{L} y$ as in (III.6), define a *condensation* of $\rho$ between $t_i$ and $t_j$, where $0 \le i, j \le n$ and $j > i + 1$, as any C2RPQ of the form:

$$\rho' = x \xrightarrow{L_1} t_1 \xrightarrow{L_2} \dots \xrightarrow{L_i} t_i \xrightarrow{K} t_j \xrightarrow{L_{j+1}} \dots \xrightarrow{L_{n-1}} t_{n-1} \xrightarrow{L_n} y$$

such that $K = \mathcal{A}_L[q_i, q_j]$.

**Fact III.2.10.** Every condensation $\rho'$ of $\rho$ is a refinement of $\gamma$, and $\rho \subseteq \rho' \subseteq \gamma$.

Informally, we will abuse the notation and write $[L_i \cdots L_j]$ to denote the language $K$—even if this language does not only depend on $L_i \cdots L_j$.

**Example III.2.11.** Let $\gamma(x, y) = x \xrightarrow{(aa^-)^*} y$ be a C2RPQ atom, where $(aa^-)^*$ is implicitly represented by its minimal automaton. Then $\rho(x, y)$ is a refinement of refinement length seven of $\gamma(x, y)$ and $\rho'(x, y)$ is a condensation of $\rho(x, y)$, where:

$$\rho(x, y) = x \xrightarrow{a} t_1 \xrightarrow{(a^-a)^*} t_2 \xrightarrow{(a^-a)^*} t_3 \xleftarrow{a} t_4 \xrightarrow{(aa^-)^*} t_5 \xrightarrow{(aa^-)^*a} t_6 \xleftarrow{a} y,$$

$$\rho'(x, y) = x \xrightarrow{a} t_1 \xrightarrow{(a^-a)^*} t_2 \xrightarrow{(a^-a)^*} t_3 \xleftarrow{a} t_4 \xrightarrow{(aa^-)^*} y.$$

On the other hand, $\rho''(x, y) = x \xrightarrow{a} t_1 \xleftarrow{a} y$ is not a condensation of $\rho(x, y)$.

$\diamond$

Given a natural number $m$, an *m-refinement* of a C2RPQ $\gamma(\bar{x}) = \bigwedge_i x_i \xrightarrow{L_i} y_i$ is any query resulting from: (1) replacing every atom by one of its $m$-refinements, and (2) should some $m$-refinements have equality atoms, collapsing the variables. A *refinement* is an $m$-refinement for some $m$. Note that any atom $m$-refinement is, by definition, also an atom $m'$-refinement when $m \leq m'$: as a consequence, in the refinement of a C2RPQ the atom refinements need not have the same length. For instance, both $\rho(x, x) = x \xrightarrow{c} x$ and $\rho'(x, y) = x \xrightarrow{a} t_1 \xrightarrow{a} y \xleftarrow{c} y$ are refinements of $\gamma(x, y) = x \xrightarrow{a^\star} y \xleftarrow{c} x$.

For a given C2RPQ $\gamma$, let $\mathrm{Ref}^{\leq m}(\gamma)$ be the set of all $m$-refinements of $\gamma$, and $\mathrm{Ref}(\gamma)$ be the set of all its refinements. Given a refinement $\rho(\bar{x})$ of $\gamma(\bar{x})$, its *refinement length* is the least natural number $m$ such that $\rho(\bar{x}) \in \mathrm{Ref}^{\leq m}(\gamma)$. Note that if the automaton representing a language $L$ has more than one final state, for instance the minimal automaton for $L = a^+ + b^+$, then $x \xrightarrow{L} y$ is not a refinement of itself. However, it will always be equivalent to a union of refinements: in this example, $x \xrightarrow{a^+ + b^+} y$ is equivalent to the union of $x \xrightarrow{a^+} y$ and $x \xrightarrow{b^+} y$, which are both refinements of the original C2RPQ.

*Expansions.* Remember that a C2RPQ whose languages are of the form $\{a\}$ or $\{a^-\}$ for $a \in \mathbb{A}$ is in effect a CQ. The *expansions* of a C2RPQ $\gamma$ is the set $\mathrm{Exp}(\gamma)$ of all CQs which are refinements of $\gamma$. In other words, an expansion of $\gamma$ is any CQ obtained from $\gamma$ by replacing each atom $x \xrightarrow{L} y$ by a path $x \xrightarrow{w} y$ for some word $w \in L$. For instance, $\xi(x, y) = x \xrightarrow{a} t_1 \xleftarrow{a} t_2 \xrightarrow{a} t_3 \xleftarrow{a} y$ is an expansion of $\rho(x, y) = x \xrightarrow{(aa^-)^\cdot} y$.

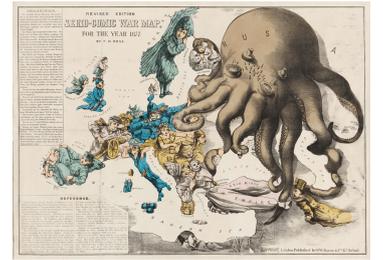

Figure III.28: *Serio-comic war map for the year 1877*, by F. W. Rose.

*Canonical databases.* We define the *canonical databases* of a C2RPQ as the canonical databases of the expansions of the query. We denote by $\langle \mathbf{G}, \bar{u} \rangle \vDash^\star \gamma(\bar{x})$ the fact that $\langle \mathbf{G}, \bar{u} \rangle$ is a canonical database of $\gamma(\bar{x})$. We extend the notions of expansions and of canonical databases to UC(2)RPQs by taking the union.

Notice that any UC2RPQ is equivalent to the infinitary union of its expansions. In light of this, the semantics for UC2RPQ can be rephrased as follows. Given a UC2RPQ $\Gamma(\bar{x})$ and a graph database $\mathbf{G}$, the evaluation of $\Gamma(\bar{x})$ over $\mathbf{H}$, is the set of tuples $\bar{v}$ of nodes for which:

- there is $\xi \in \mathrm{Exp}(\Gamma)$ such that there is an evaluation map from $\xi$ to $\mathbf{H}$ that sends $\bar{x}$ onto $\bar{v}$, or equivalently
- there exists $\langle \mathbf{G}, \bar{u} \rangle \vDash^\star \Gamma(\bar{x})$ s.t. $\langle \mathbf{G}, \bar{u} \rangle \xrightarrow{\mathrm{hom}} \langle \mathbf{H}, \bar{v} \rangle$.

In this specific case, since $\xi$ is a CQ, the notion of evaluation map coincides with that of homomorphism: duality justifies that we use these two notions interchangeably, which we will happily do from now on.

Similarly, containment of UC2RPQs can also be characterized in terms of expansions.

**Proposition III.2.12** (Folklore).[45],[46] Let $\Gamma_1$ and $\Gamma_2$ be UC2RPQs. The following are equivalent:

[45] The proof is elementary, see *e.g.* [FLS98, Proposition 3.2] or [CDLV00, Theorem 2].

[46] The proposition works for Boolean as well as non-Boolean queries: we dropped the tuples of variables for the sake of readability.

- $\Gamma_1 \subseteq \Gamma_2$;
- for every $\xi_1 \in \mathrm{Exp}(\Gamma_1)$, we have $\xi_1 \subseteq \Gamma_2$;
- for every $\xi_1 \in \mathrm{Exp}(\Gamma_1)$ there exists $\xi_2 \in \mathrm{Exp}(\Gamma_2)$ such that $\xi_2 \xrightarrow{\mathrm{hom}} \xi_1$;
- for every $\mathbf{G}_1 \vDash^\star \Gamma_1$, we have $\mathbf{G}_1 \vDash \Gamma_2$;
- for every $\mathbf{G}_1 \vDash^\star \Gamma_1$, there exists $\mathbf{G}_2 \vDash^\star \Gamma_2$ *s.t.* $\mathbf{G}_2 \xrightarrow{\mathrm{hom}} \mathbf{G}_1$;

As an example, we consider the queries

$$\gamma() \triangleq w \xrightarrow{a} x \xrightarrow{a+b} y \xrightarrow{b} z \quad \text{and} \quad \delta() \triangleq x \xrightarrow{a} y \xrightarrow{b} z$$

from Example III.2.8. Then $\delta()$ has a unique canonical database that we denote by $\mathbf{D}$, and $\gamma()$ has two canonical databases $\mathbf{G}_a$ and $\mathbf{G}_b$. We represent them in Figures III.29 and III.30, together with witnesses that $\gamma() \subseteq \delta()$.

Proposition III.2.12 by itself is not enough to conclude to the decidability of containment since the set of expansions or canonical databases of a UC2RPQ, or even simply CRPQs, is infinite. To get decidability, the easiest way is by proving a small model property.

**Proposition III.2.13** (Small model property for C2RPQs, folklore)**.** Given two C2RPQs $\gamma_1(\bar{x}_1)$ and $\gamma_2(\bar{x}_2)$, if $\gamma_1(\bar{x}_1) \nsubseteq \gamma_2(\bar{x}_2)$, then there exists $\langle \mathbf{G}_1, \bar{u}_1 \rangle \vDash^\star \gamma_1(\bar{x}_1)$ of size at most $f(\|\gamma_1\| + \|\gamma_2\|)$ *s.t.* $\langle \mathbf{G}_1, \bar{u}_1 \rangle \nvDash \gamma_2(\bar{x}_2)$, where $f$ is a doubly-exponential function.

*Proof sketch.* We start with some $\langle \mathbf{G}_1, \bar{u}_1 \rangle \vDash^\star \gamma_1(\bar{x}_1)$ of arbitrary size such that $\langle \mathbf{G}_1, \bar{u}_1 \rangle \nvDash \gamma_2(\bar{x}_2)$. We then look at how $\gamma_2$ maps into $\mathbf{G}_1$ in the following sense: in an atom refinement in $\mathbf{G}_1$

$$x_0 \xrightarrow{a_1} x_1 \xrightarrow{a_2} \dots \xrightarrow{a_n} x_n$$

we associate to every sequence $a_1 a_2 \cdots a_i$ the function which takes a tuple of set of automaton states (one set for each automaton of $\gamma_2$), and returns the tuple of set of states they can reach after reading $a_i a_{i+1} \cdots a_j$. By the pigeon-hole principle, if $n$ has at least double-exponential size, then there must exist two indices $i < j$ that have the same behaviour. Then, shrinking

$$x_0 \xrightarrow{a_1} x_1 \xrightarrow{a_2} \dots \xrightarrow{a_n} x_n \quad \text{into} \quad x_0 \xrightarrow{a_1} x_1 \xrightarrow{a_2} \dots \xrightarrow{a_i} x_i = x_j \xrightarrow{a_{j+1}} \dots \xrightarrow{a_n} x_n$$

yields a strictly smaller canonical database $\mathbf{G}_1'$ of $\gamma_1$ that behaves like $\mathbf{G}_1$ *w.r.t.* to the automata of $\gamma_2$, and so since we had $\langle \mathbf{G}_1, \bar{u}_1 \rangle \nvDash \gamma_2(\bar{x}_2)$, we still have $\langle \mathbf{G}_1', \bar{u}_1 \rangle \nvDash \gamma_2(\bar{x}_2)$. □

In particular Proposition III.2.13 implies that containment is decidable for UC2RPQs. However, this does not give an optimal algorithm.

**Proposition III.2.14.**[47] The CONTAINMENT PROBLEM for UC2RPQs is ExpSpace-complete. The lower bound already holds for Boolean CRPQs.

Overtime, this proof has been simplified to eventually climax to the following result.

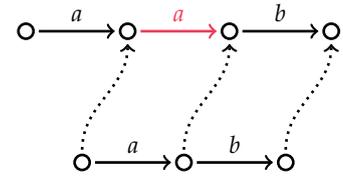

Figure III.29: Homomorphism from $\mathbf{D}$ to $\mathbf{G}_a$.

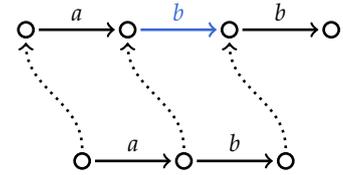

Figure III.30: Homomorphism from $\mathbf{D}$ to $\mathbf{G}_b$.

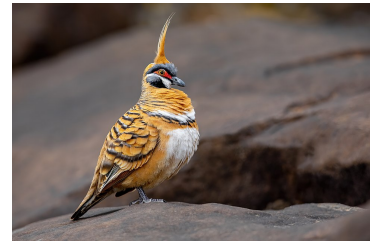

Figure III.31: Fancy pigeon. *Spinifex Pigeon*, by JJ Harrison, licensed under CC BY SA 3.0.

[47] Proven independently in [CDLV00, Theorem 5] for C2RPQs and in [FLS98, § after Theorem 4.8] for CRPQs without inverses but with an infinite alphabet.

**Proposition III.2.15** ([Fig20, Lemma 8]). There is a fixed alphabet over which the CONTAINMENT PROBLEM for Boolean CRPQs is already ExpSpace-hard when restricted to instances of the form

$$\gamma_1() = \bullet \xrightarrow{K} \bullet \qquad \subseteq^? \qquad \bullet \overset{L_1}{\underset{L_p}{\overset{\vdots}{\rightrightarrows}}} \bullet = \gamma_2().$$

Observe that, while the containment

$$(\gamma_1(x,y) \triangleq x \xrightarrow{K} y) \subseteq^? (\gamma_2(x,y) \triangleq x \xrightarrow{L} y)$$

is equivalent to $K \subseteq^? L$, we have that

$$(\gamma_1() \triangleq x \xrightarrow{K} y) \subseteq^? (\gamma_2() \triangleq x \xrightarrow{L} y)$$

is rather equivalent to $K \subseteq^? \mathbb{A}^* L \mathbb{A}^*$, and so the problem of Proposition III.2.15 can be reformulated as

$$K \subseteq^? \mathbb{A}^* \Big( \bigcap_{i=1}^{p} L_i \Big) \mathbb{A}^*.$$

The proof of Proposition III.2.15 is by reduction from the exponential-space tiling problem, and relies on encoding counters with exponential value in polynomial space.

### III.2.3 Queries Over Simple Languages

The high complexity of Propositions III.2.14 and III.2.15 motivates the quest for fragments of conjunctive regular path queries with better complexity. We will present a fragment UCRPQ(SRE) that has a CONTAINMENT PROBLEM that is much better behaved than for general UCRPQs, and moreover is widely used in practice.

A *simple regular expression*, or *SRE*, is a regular expression of the form $a^*$ for some letter $a \in \mathbb{A}$ or of the form $a_1 + \cdots + a_m$ for some $a_1, \ldots, a_m \in \mathbb{A}$. Let UCRPQ(SRE) be the set of all UCRPQ whose languages are expressed via simple regular expressions. Observe that UCRPQ(SRE) is semantically equivalent to the class of UCRPQs over the closure under concatenation of simple regular expressions since $\gamma(x,y) = x \xrightarrow{e_1 \cdot e_2} y$ is equivalent to $\gamma'(x,y) = x \xrightarrow{e_1} z \wedge z \xrightarrow{e_2} y$. Moreover, UCRPQ(SRE) also corresponds to UC2RPQs whose languages are expressed via SREs; in other words adding two-wayness does not increase the expressivity of the class.

**Proposition III.2.16** ([Fig+20, Corollary 5.2]). Containment of UCRPQ(SRE) is $\Pi_2^p$-complete.

In other words, this problem is just one level up the polynomial hierarchy compared to the CQ CONTAINMENT PROBLEM, which sharply contrasts with the costly ExpSpace-completeness result for the full class of UC2RPQs! Moreover, recent studies on SPARQL query logs on Wikidata, DBpedia and other sources show that this kind of regular expressions cover a majority of the queries investigated, *e.g.*, 75% of the "property paths" (C2RPQ atoms) of the corpus of
1.5M queries of Bonifati, Martens and Timm [BMT20, Table 15].

In Chapter IV we will actually need a marginally smaller subclass of these:
we define *positive simple regular expressions* analogously to simple regular
expressions but by replacing $a^*$ with $a^+$.

### III.2.4 Static Analysis

Static optimization for CRPQs has received considerable attention. Beyond
the basic study of containment and equivalence problems for CRPQs that we
already mentioned, let us highlight that these problems have also been inves-
tigated under different scenarios: restrictions on the shape of queries [Fig20],
restrictions on their regular languages [Fig+20], alternative semantics [FR23],
or under schema information [GGIM22; GGIM24]. This has enabled the study
of more advanced static analysis problems motivated by the following general
question: *Can a given query be equivalently rewritten as one from a target
fragment (which enjoys desirable properties)?* In the literature the problem
has been studied where the target fragment are queries which either (i) avoid
having infinite languages, or (ii) have a tree-like structure. This gives rise
to the so-called (i) *boundedness problem* for CRPQs (*i.e.*, whether a CRPQ is
equivalent to a UCQ) [BFR19; FKMP24], and (ii) *semantic treewidth problem*
for CRPQs (*i.e.*, whether a CRPQ is equivalent to one that is tree-shaped)
[BRV16].

In the next chapters, we focus on the problems of minimizing the number
of atoms (Chapter IV) and tree-width (Chapter IV) necessary to express a
(U)C(2)RPQ. In both cases, we will prove decidability of the problem by
providing an algorithm in $k$-ExpSpace for some $k$, and will provide a much
more efficient algorithm—*i.e.* in the polynomial hierarchy—for queries over
simple regular expressions.

CHAPTER IV

# Minimization of Conjunctive Regular Path Queries

ABSTRACT

We study the *minimization problem* for Conjunctive Regular Path Queries (CRPQs) and unions of CRPQs (UCRPQs). This is the problem of checking, given a query and a number $k$, whether the query is equivalent to one of size at most $k$. For CRPQs we consider the size to be the number of atoms, and for UCRPQs the maximum number of atoms in a CRPQ therein, motivated by the fact that the number of atoms has a leading influence on the cost of query evaluation.

We show that the minimization problem is decidable, both for CRPQs and UCRPQs. We provide a 2ExpSpace upper bound for CRPQ minimization, based on a brute-force enumeration algorithm, and an ExpSpace lower bound. For UCRPQs, we show that the problem is ExpSpace-complete, having thus the same complexity as the classical containment problem. The upper bound is obtained by defining and computing a notion of maximal under-approximation. Moreover, we show that for UCRPQs using the so-called *simple regular expressions* consisting of concatenations of expressions of the form $a^+$ or $a_1 + \cdots + a_k$, the minimization problem becomes $\Pi_2^p$-complete, again matching the complexity of containment.

ACKNOWLEDGEMENTS

This chapter is mostly a reproduction of the eponymous paper that was published in PODS '25 [FMR25], that received the *Distinguished Paper Award*. Proofs were incorporated in the body; the preliminaries and part of the introduction was moved to Chapter III. It is a joint work with Diego Figueira and Miguel Romero.

# Contents

## IV.1   Introduction

*Minimization of queries.*   Minimization – that is, the problem of transforming a query into a strictly smaller equivalent query – is perhaps the most fundamental query optimization question. For CQs (and UCQs), minimization is well understood, and there exists a canonical unique minimal query, namely the core. The mechanism for obtaining such a minimal query is simple: eliminate any atom from the query that results in an equivalent query (*i.e.*, any atom which is 'redundant' in the sense of equivalence). In contrast, minimization of CRPQs is poorly understood from a theoretical perspective. In this case, the situation is more challenging: there is no natural notion of 'core', and it is not clear whether a notion of 'canonical' smallest query may even be possible. In particular, eliminating redundant atoms of a CRPQ, as done for CQs, in general results in a query which is neither minimal nor canonical.

In this chapter we study the MINIMIZATION PROBLEM for CRPQs and UCRPQs. In the case of CRPQs, we aim at minimizing the number of atoms of a CRPQ, and hence we formulate the problem as follows ($\equiv$ denotes query equivalence, *i.e.*, the fact that the queries output the same answer for all databases):[1]

---
**MINIMIZATION PROBLEM FOR CRPQs**

*Input*:   A finite alphabet $\mathbb{A}$, a CRPQ $\gamma$ over $\mathbb{A}$ and $k \in \mathbb{N}$.

*Question*:   Is there a CRPQ $\delta$ over $\mathbb{A}$ with at most $k$ atoms such that $\gamma \equiv \delta$?

---

On the other hand, in the case of UCRPQs, we minimize the maximum number of atoms of the CRPQs participating in a UCRPQ:

---
**MINIMIZATION PROBLEM FOR UCRPQs**

*Input*:   A finite alphabet $\mathbb{A}$, a UCRPQ $\Gamma$ over $\mathbb{A}$ and $k \in \mathbb{N}$.

*Question*:   Is there a UCRPQ $\Delta$ over $\mathbb{A}$ whose every CRPQ has at most $k$ atoms *s.t.* $\Gamma \equiv \Delta$?

---

Observe that the minimization problem for CRPQs and UCRPQs are two different problems: an algorithm for the minimization problem for UCRPQs (where the equivalent query may have unions) in principle does not imply any bound on the minimization for CRPQs.

*Contributions.*   We investigate the minimization problem for CRPQs and UCRPQs, and present several fundamental results. More concretely:

- We show that the minimization problem for CRPQs and UCRPQs are both decidable. As explained before, these are different problems and we give two very different algorithms. Contrary to what happens for CQs, minimizing a CRPQ by unions of CRPQs may result in smaller queries, hence in a sense UCRPQ minimization may be seen as a strictly more powerful approach (Proposition IV.4.1).

[1] Note that we can always assume $k$ to be smaller than the number of atoms of the input query, since otherwise the instance of the MINIMIZATION PROBLEM is trivially solvable by answering 'yes'. So, whether $k$ is given in unary or binary does not affect the size of the input.

- For the MINIMIZATION OF CRPQs, the algorithm is essentially by brute-force. By carefully bounding the sizes of the automata involved, we show that the algorithm can be implemented in 2ExpSpace in Theorem IV.3.1. We also show an ExpSpace-hard lower bound in Theorem IV.5.1, leaving an exponential gap.

- For the MINIMIZATION OF UCRPQs we can apply a more elegant solution, and in fact we show how to compute 'maximal under-approximations' of a query by UCRPQs of a given size (Lemma IV.4.6). The minimization then follows by testing whether the given query is equivalent to its approximation of size $k$, yielding an ExpSpace upper bound (Corollary IV.4.7), which is tight with the lower bound (Corollary IV.5.12).

- We consider subclasses of UCRPQs restricted to some commonly used regular expressions as observed in practice, namely, the so-called positive simple regular expressions,[2] We show that minimization of UCRPQs having such simple regular expressions is $\Pi_2^p$-complete (Theorem IV.4.8).

- We explore some necessary and sufficient conditions for minimality. In particular, we show that non-redundancy (*i.e.*, the fact that removing any atom results in a non-equivalent query) is necessary but not sufficient for minimality (also known to be the case for tree patterns [CMNP18]). We also investigate a notion of 'strong minimality' which implies minimality (Corollary IV.2.10), and can be used as a theoretical tool to prove minimality of queries. This result is based on Theorem IV.2.9, which may be of independent interest, providing a tool to extract lower bounds on the number of atoms (and more generally properties on the underlying structure of queries, such as tree-width, path-width, etc.) that is necessary to express a UCRPQ.

- We also discuss an alternative definition of size, where instead of the number of atoms we count the number of variables: we obtain upper bounds for the VARIABLE-MINIMIZATION PROBLEM of CRPQs and UCRPQs in Section IV.6.

*On the chosen size measure.* A naïve algorithm for the evaluation of a union of $t$ CRPQs with $k$ atoms on a graph database $G$ gives a rough bound of $O\big(tk(|G||\mathcal{E}(G)|r) + t|G|^{2k}\big)$, where $r$ is the maximum size of the regular expressions it contains, and $\mathcal{E}(G)$, $G$ are the set of edges and vertices of $G$, respectively.[3] As we see, the most costly dependence is on $k$, since $G$ is the largest object (*i.e.*, the database, several orders of magnitude larger than the remaining parameters in practice). The size of regular expressions and the number of unions have a less predominant multiplicative influence on the cost. Further, unions can be executed in parallel, which justifies the choice of taking the maximum size of the number of atoms of the CRPQs therein. However, other measures may also be reasonable. For example, taking the size to be the number of variables instead of the number of atoms is explored in Section IV.6.1. More complex measures including the size of regular expressions and the number of unions would need to take into account the drastically different roles of the parameters in the evaluation in view of the previous discussion (*e.g.*, a simple sum of the parameters would not be a reasonable choice).

---

[2] The positivity restriction is needed in Lemma IV.5.11.

[3] This is obtained by first materializing a table with the answers to each RPQ atom $x \xrightarrow{L} y$ of the query. For each vertex $u \in V(G)$, we can compute the answers to $u \xrightarrow{L} y$, by a BFS traversal on the product of $G$ and the NFA $\mathcal{A}_L$ for the regular language $L$, taking roughly $\mathcal{O}(|\mathcal{E}(G)|r)$. Then we can evaluate each CRPQ as if it were a conjunctive query on the computed tables (each table having size at most $|G|^2$), in $\mathcal{O}((|G|^2)^k) = \mathcal{O}(|G|^{2k})$.

Our size measure of number of atoms is also natural from a practical perspective. In practice, systems typically evaluate CRPQs by combining on-the-fly "materialization" of CRPQ atoms with relational database techniques, in particular using join algorithms (see *e.g.* [Vrg+24; KBHSN24; CRV23]). The number of atoms (or joins) plays an important role in these algorithms.

*Related work.* Minimization has also been studied for the class of *tree patterns* [FFM08; KS08; CMNP18]. Tree patterns are simple yet widely used tree-like queries for tree-like databases such as XML. These queries allow mild recursion in the form of descendent edges, that is, atoms of the form $x \xrightarrow{a^+} y$, where $x$ is the parent of $y$. Minimization of tree patterns is now well-understood [CMNP18]: it is known that non-redundancy is not the same as minimality, and that the MINIMIZATION PROBLEM is $\Sigma_2^p$-complete, the lower bound being highly non-trivial.

## IV.2  Necessary & Sufficient Conditions for Minimality

This section explores some necessary and sufficient conditions for a query to be minimal. We start with some necessary definitions. We say that a (U)CRPQ is *minimal* if it is not equivalent to any (U)CRPQ having strictly fewer atoms. An *internal variable* of a CRPQ $\gamma$ is any non-output variable with both in-degree and out-degree 1. A non-internal variable is called *external*. A *one-way internal path*,[4] or *internal path* for short, from $x_0$ to $x_n$ of a CRPQ $\gamma$ is a *simple*[5] path

$$x_0 \xrightarrow{L_1} x_1 \xrightarrow{L_2} \cdots \xrightarrow{L_n} x_n \tag{IV.1}$$

in $\gamma$ where $n > 0$ and every $x_i$ is internal with $i \in [\![1, n-1]\!]$. A *segment* of a CRPQ $\gamma$ is a maximal internal path in $\gamma$, where "maximal" means that it cannot be extended on the left or on the right. We say that a segment is *cyclic* if $x_0 = x_n$. We identify two cyclic segments if they are equal up to circular permutation.[6] We say that a segment as in (IV.1) is *incident* to a variable $y$ if $y = x_i$ for some $i$.

**Fact IV.2.1.** The segments—seen as sets of atoms—of $\gamma$ form a partition on its set of atoms.

See Figure IV.1a for an example of a decomposition into segments. Note that each segment is incident to either zero external variable (isolated cycles), to one (non-isolated cycles) or to two (non-cycles). We denote by $\|\gamma\|_{\mathrm{seg}}$ the

[4] This definition comes from [FM25, § 7], under the name 'one-way internal path'; there is also an equivalent notion for C2RPQs.

[5] Meaning that all nodes are pairwise disjoint, except that potentially $x_0 = x_n$.

[6] The endpoint of a cyclic segment can either be internal or not, see Figure IV.1a.

Figure IV.1: Segments and segment graph of a CRPQ $\gamma$.

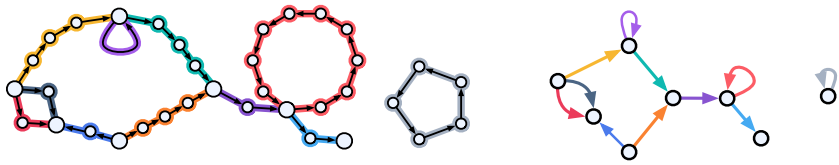

(a) The segments of $\gamma$—labels are omitted. Each segment has a different colour. Internal variables are the smaller circles.

(b) The segment graph of $\gamma$.

number of segments of a CRPQ $\gamma$, and we extend this notation to UCRPQs by letting $\|\Gamma\|_{\text{seg}} \triangleq \max_{\gamma \in \Gamma} \|\gamma\|_{\text{seg}}$. By Fact IV.2.1, $\|\Gamma\|_{\text{seg}} \leq \|\Gamma\|_{\text{at}}$ always holds.

### IV.2.1 Necessary Conditions: Contractions and Redundancy

*Contractions.* One simple (and tractable) way to make a query smaller is to 'contract' any two consecutive atoms sharing an internal variable, and replacing them with a single atom having the concatenation of the languages: we call this new query *contraction*. A CRPQ $\gamma$ is *fully contracted* if it cannot be contracted. In other words, $\gamma$ is fully contracted *iff* $\|\gamma\|_{\text{seg}} = \|\gamma\|_{\text{at}}$. A contraction of a UCRPQ is a contraction of a CRPQ therein (obtaining the UCRPQ where one CRPQ $\gamma$ was replaced by a contraction of $\gamma$). A UCRPQ is then *fully contracted* if each CRPQ therein is fully contracted.

**Fact IV.2.2.** Contractions preserve semantical equivalence. Further, from a UCRPQ $\Gamma$ one can produce, in polynomial time, an equivalent one that is fully contracted with $\|\Gamma\|_{\text{seg}}$ atoms.

In particular, if a UCRPQ $\Gamma$ is minimal then $\Gamma$ is fully contracted and $\|\Gamma\|_{\text{seg}} = \|\Gamma\|_{\text{at}}$;[7] in other words, $\|\Gamma\|_{\text{seg}}$ is an upper bound on the number of atoms of the minimal equivalent query.

[7] As for $\|\Gamma\|_{\text{seg}}$, we define $\|\Gamma\|_{\text{at}}$ to be the maximum of the number of atoms of a disjunct of $\Gamma$.

*Redundancy.* Another way to reduce the number of atoms of a query is to remove any *redundant atom*, that is, any atom whose removal results in an equivalent query. When there are no such redundant atoms, we say that the query is *non-redundant*. However, this is a more difficult problem, since it involves testing for query equivalence, an ExpSpace-complete problem.

**Proposition IV.2.3.** Testing whether a (U)CRPQ is non-redundant is ExpSpace-complete.

In order to prove this result, we first need to mention a variation of Proposition III.2.15.

**Proposition IV.2.4** (Variation of Proposition III.2.15)**.** There is a fixed alphabet over which the CONTAINMENT PROBLEM for Boolean CRPQs is ExpSpace-hard restricted to instances of the form

$$\gamma_1() = \bullet \xrightarrow{\ K\ } \bullet \qquad \subseteq? \qquad \bullet \overset{L_1}{\underset{L_p}{\vdots}} \bullet = \gamma_2(),$$

where moreover:

1. no language among $K$ or the $L_i$'s is empty or contains the empty word $\varepsilon$, and

2. there is no $i$ such that $\mathbb{A}^* L_i \mathbb{A}^* = \mathbb{A}^* \big(\bigcap_j L_j\big) \mathbb{A}^*$.

*Proof.* By inspecting [Fig20, Proof of Lemma 8, pp. 15–17], it can be noticed that actually the first condition is satisfied by Figueira's reduction—using his notation, neither $E$ nor $G_i \cup F_C \cup F_H$ with $i \in [\![0, n]\!]$ are empty.

Moreover, we claim that the reduction can be made so that the second condition also holds. Notice first that the $2^n$-tiling problem is still ExpSpace-complete if we restrict it to instances with $n > 1$ and such that all instances admit one tiling which is "locally valid" but not valid—namely a tiling which satisfies all vertical and horizontal constraints, but not the initial and final tiles conditions. This can be achieved *e.g.* by adding a new tile $t$ *s.t.* $(t, t)$ is both a valid horizontal and vertical configuration, but $t$ cannot be adjacent to any other tile. Then, the second condition amounts to showing that there is no $i$ *s.t.*

$$\mathbb{A}^*(G_i \cup F_C \cup F_H)\mathbb{A}^* = \mathbb{A}^*\Big(\bigcap_{0 \le j \le n} (G_j \cup F_C \cup F_H)\Big)\mathbb{A}^*$$

which is equivalent, by elementary manipulations, to saying that for all $i \in [\![0, n]\!]$

$$\mathbb{A}^* G_i \mathbb{A}^* \not\subseteq \mathbb{A}^*\Big(\bigcap_{0 \le j \le n} G_j\Big)\mathbb{A}^* \cup \mathbb{A}^*(F_C \cup F_H)\mathbb{A}^*.$$

For $i = 0$, this holds because we can consider a valid encoding of a tiling which respects all constraints except that one vertical constraint is violated. For $i \in [\![1, n]\!]$, we consider the encoding of a tiling which is locally valid. Then, it has no vertical error, no horizontal error, and no encoding error, so it does not belong to the right-hand side. However, it belongs to $\mathbb{A}^* G_i \mathbb{A}^*$ for any $i \in [\![1, n]\!]$ since it contains a subword encoding two cells separated by exactly one row. Hence, the second condition also holds. □

*Proof of Proposition IV.2.3.* The upper bound is trivial: for every atom we remove it and check equivalence.

For the lower bound, we use the construction of Proposition IV.2.4 for containment. Let $\delta() \mathrel{\hat=} x' \xrightarrow{\ K\ } y' \wedge \bigwedge_i x \xrightarrow{\ L_i\ } y$ be the disjoint conjunction of $\gamma_1()$ and $\gamma_2()$ as defined in Proposition IV.2.4. We first strengthen the construction to ensure the following two properties:

1. $K$ **cannot be mapped inside any** $L_i$: There is no word of $K$ which appears as factor of a word from some $L_i$. For this, it suffices to add a special letter at the beginning and the end of every word of $K$ which is not in any of the $L_i$'s. That is, we can define a new $K^{new} \mathrel{\hat=} \# \cdot K^{old} \cdot \#$ for a new symbol $\#$.

2. **For every $j$ there is $w_j \in L_j$ such that $w_j \notin L_i$ for every $i \neq j$:** it suffices to add a special word (*e.g.* using a new alphabet letter) to each $L_i$. For example, we can define $L_i^{new} \triangleq L_i^{old} \cup \{@_i\}$, where $@_i$ is a fresh alphabet letter.

It is easy to see that these modifications preserve all the properties needed for the CONTAINMENT PROBLEM to still be ExpSpace-hard.

We show that $\delta()$ is non-redundant iff $x' \xrightarrow{K} y' \subseteq \bigwedge_i x \xrightarrow{L_i} y$.

↬ *$K$ cannot be removed.* We first show that removing the atom $x' \xrightarrow{K} y'$ from $\delta()$ results in a non-equivalent query $\delta'()$. Indeed, if it is removed then for any expansion of $\delta'()$ there will not be any word from $K$ that can be used to map into the expansion due to the first point above.

↬ *If some $L_j$ is redundant, then containment holds.* Consider the result $\delta'()$ of removing an atom $x \xrightarrow{L_j} y$ from $\delta()$. Consider all the expansions of $\delta'()$ that choose $w_i$—defined in the second point above—as the atom expansion for $L_i$ for every $i \neq j$. It follows that for any such expansion there must be an expansion of $\delta()$ that maps necessarily $x \xrightarrow{L_j} y$ to the atom expansion of $x' \xrightarrow{K} y'$ in $\delta'()$. Otherwise, we would be mapping some word of $L_j$ to some $w_i$ with $i \neq j$, which we know it is not possible due to the second point above. This means that $x' \xrightarrow{K} y' \subseteq \bigwedge_i x \xrightarrow{L_i} y$.

↬ *If containment holds, then all $L_i$'s redundant.* Finally, observe that if $x' \xrightarrow{K} y' \subseteq \bigwedge_i x \xrightarrow{L_i} y$ then the query is equivalent to $x' \xrightarrow{K} y'$.

Overall, we obtained that the following are equivalent:

i. $\delta()$ is redundant,

ii. an atom $x \xrightarrow{L_i} y$ of $\delta()$ is redundant,

iii. the containment $x' \xrightarrow{K} y' \subseteq \bigwedge_i x \xrightarrow{L_i} y$ holds,

iv. all atoms $x \xrightarrow{L_i} y$ of $\delta()$ are redundant. □

While in the case of conjunctive queries non-redundancy is the same as minimality, for CRPQs and UCRPQs this is not the case, even if the query is fully contracted.

**Proposition IV.2.5.** There are fully contracted non-redundant CRPQs which are not minimal.

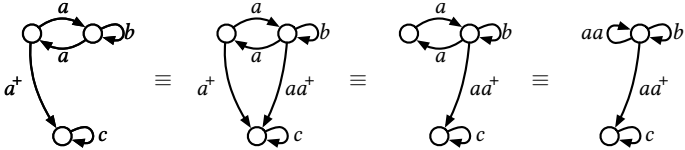

Figure IV.2: Equivalent CRPQs. The leftmost query is fully contracted, non-redundant and not minimal, since it is equivalent to the rightmost query.

*Proof.* Figure IV.2 contains a simple witnessing example. It is trivial to see that the CRPQs in the figure are all equivalent and that the leftmost query is non-redundant and fully contracted. However, it is not minimal since it is equivalent to the rightmost CRPQ. □

### IV.2.2 A Sufficient Condition: Strong Minimality

We have seen some sound ways to reduce the size of queries. But how can we ensure that a query is actually minimal? Here we give a theoretical tool which can ensure this by means of finding some expansion which is a witness for the query to have "many atoms". We call this strong minimality. A UCRPQ $\Gamma$ is *strongly minimal* if it has a hom-minimal[8] expansion $\xi \in \mathrm{Exp}(\Gamma)$ s.t. $\|\check{\xi}\|_{\mathrm{seg}} = \|\Gamma\|_{\mathrm{at}}$. We will next show that this is a sufficient condition for minimality.

We denote by $G_\gamma$ the *underlying graph* of a CRPQ $\gamma$, namely the directed graph whose vertices are variables of $\gamma$, and with an edge from $x$ to $y \in \gamma$ whenever there is some atom from $x$ to $y$ in $\gamma$.

Informally, the segment graph $\mathcal{SG}(\gamma)$ of a CRPQ $\gamma$ is the directed multigraph obtained by first taking the underlying graph, and then replacing segments of $\gamma$ with edges, as illustrated in Figure IV.1b. The main motivation behind the notion of segments is that it is essentially dual to the notion of atom refinements.

**Definition IV.2.6.** Given a CRPQ $\gamma$, we define its *segment graph $\mathcal{SG}(\gamma)$* to be the directed multigraph defined by:

- every external variable of $\gamma$ is a vertex of $\mathcal{SG}(\gamma)$, and moreover, for every cyclic segment $\sigma$ that is not incident to any external variable, we create a new variable $x_\sigma$;
- for every cyclic segment $\sigma$, we have a self-loop around $x_\sigma$, and for any external variable $x$ and $y$ of $\gamma$, we have an edge from $x$ to $y$ in $\mathcal{SG}(\gamma)$ for any segment starting at $x$ and ending at $y$.

**Fact IV.2.7.** The segment graph of $\gamma$ can always be obtained from its underlying graph by contracting internal variables.

A *minor* of a graph is any graph that can be obtained by removing edges, removing vertices—and their adjacent edges—, and *contracting edges*—meaning that we identify the two endpoints of the edge and remove the edge from the graph.[9] Moreover, we say that a class of CRPQs is *minor-closed* when, for every query is the class, any other query whose underlying graph is a minor of the underlying graph of the first query must also belong to the class.

[8] Naturally, we say that an expansion is hom-minimal whenever its canonical database is hom-minimal.

[9] This definition is a trivial generalization of the notion of minors for undirected graphs.

Figure IV.3: Illustration of Proposition IV.2.8: Figure IV.3c can be obtained as a minor of Figure IV.3a.

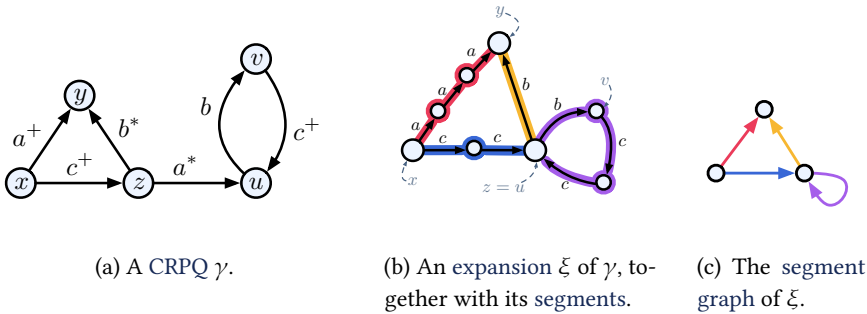

(a) A CRPQ $\gamma$.  (b) An expansion $\xi$ of $\gamma$, together with its segments.  (c) The segment graph of $\xi$.

**Proposition IV.2.8.** Let $\gamma$ be a CRPQ and $\xi$ be an expansion of $\gamma$. Then $\mathscr{SG}(\xi)$ is a minor of the underlying graph of $\gamma$.[10]

In particular, $\|\xi\|_{\mathrm{seg}} \leq \|\gamma\|_{\mathrm{at}}$. See Figure IV.3 for an example.

*Proof.* The underlying graph of $\xi$ is obtained from the underlying graph of $\gamma$ by:

1. contracting some edges—corresponding to atom refinements for $\varepsilon$,
2. potentially removing isolated vertices, and
3. refining some edges—corresponding to atom refinements for words of length at least 2.

Let $G'$ be the underlying graph of $\gamma$ to which we applied all operations of type (1) and (2). By construction, $G'$ is a minor of $G_\gamma$. Notice then that if $H$ is a graph obtained by refining one edge of $G'$, then $\mathscr{SG}(H)$ and $\mathscr{SG}(G')$ are isomorphic. By trivial induction, it follows that $\mathscr{SG}(\xi)$ is isomorphic to $\mathscr{SG}(G')$. In turn, by Fact IV.2.7, $\mathscr{SG}(G')$ is an edge contraction of $G'$, which concludes the proof since the latter is a minor of $G_\gamma$. $\square$

The next proposition provides a helpful tool to prove lower bounds on the number of atoms—but also on the structure—required to express a query.

**Theorem IV.2.9** (*Semantical Structure*)**.** Let $\Gamma$ be a UCRPQ. Let $\xi$ be a hom-minimal expansion of $\Gamma$, and $\Delta$ be any UCRPQ equivalent to $\Gamma$. Then there exists some $\delta \in \Delta$ s.t. the segment graph $\mathscr{SG}(\check{\xi})$ of the core of $\xi$ is a minor of the underlying graph of $\delta$.

*Proof.* Let $\Gamma$ be a fixed UCRPQ, and let $\Delta$ be a equivalent UCRPQ. Let $\xi_\Gamma$ be a hom-minimal expansion of $\Gamma$. Since $\Gamma \subseteq \Delta$, there exists an expansion $\xi_\Delta$ of $\Delta$ s.t. $\xi_\Delta \xrightarrow{\mathrm{hom}} \xi_\Gamma$. Likewise, since $\Delta \subseteq \Gamma$, there exists an expansion $\xi'_\Gamma \in \mathrm{Exp}(\Gamma)$ s.t. $\xi'_\Gamma \xrightarrow{\mathrm{hom}} \xi_\Delta$. Overall, we have $\xi'_\Gamma \xrightarrow{\mathrm{hom}} \xi_\Gamma$ and so, by hom-minimality of $\xi_\Gamma$, it is hom-equivalent to $\xi'_\Gamma$. In turn, this implies that $\xi_\Delta$ is hom-equivalent to $\xi_\Gamma$, and thus there exists an embedding of $\check{\xi}_\Gamma$ into $\xi_\Delta$. Note moreover that such an embedding must send variables of in-degree 0 (resp. out-degree 0) to nodes of in-degree 0 (resp. out-degree 0) and so by Corollary IV.2.13, there is an embedding from $\mathscr{SG}(\check{\xi}_\Gamma)$ into $\mathscr{SG}(\xi_\Delta)$. Letting $\delta$ be the disjunct of $\Delta$ of which $\xi_\Delta$ is an expansion, Proposition IV.2.8 implies that $\mathscr{SG}(\xi_\Delta)$ is a minor of the underlying graph of $\delta$. Hence, $\mathscr{SG}(\check{\xi}_\Gamma)$ is a subgraph of a minor, and hence a minor, of the underlying graph of $\delta$. $\square$

**Corollary IV.2.10** (of Theorem IV.2.9)**.** Every strongly minimal UCRPQ is minimal.

*Proof.* The number of edges of $\mathscr{SG}(\check{\xi})$ equals $\|\check{\xi}\|_{\mathrm{seg}}$, and a minor can only decrease the number of edges. $\square$

In fact, it can be seen that the assumption that $\xi$ is hom-minimal in Theorem IV.2.9 is necessary as otherwise the statement would be false (see Remark IV.2.15 for details). Also, note in particular that Theorem IV.2.9 implies

[10] In general, $\mathscr{SG}(\xi)$ can be obtained by contracting edges, but there are some degenerate cases where we must remove some vertices, for instance when considering the expansion of $x \xrightarrow{a^*} x$ associated with the empty word. This is an artefact of our choice to disallow isolated variables in CRPQs.

$\|\check{\xi}\|_{\mathrm{seg}} \leq \|\Delta\|_{\mathrm{at}}$. But it can also be used to obtain lower bounds on, for instance, the tree-width of $\Delta$, and hence the one-way semantic tree-width[11] of $\Gamma$, or more generally to prove that $\Gamma$ cannot be equivalent to a UCRPQ whose underlying graphs all belong to a minor-closed class of graphs.

**Proposition IV.2.11.** Let $\gamma$ be a CRPQ. The set of atoms of any path $x_0 \xrightarrow{a_1} \cdots \xrightarrow{a_n} x_n$ of $\gamma$ where either $x_0 = x_n$ or where both $x_0$ and $x_n$ are external is a finite union of segments of $\gamma$.

*Proof.* The statement deals with the set of atoms of the path—and not the path itself—, so *w.l.o.g.*, up to a circular permutation of the path, we assume that (⑤) if $x_0 = x_n$ then either $x_0$ and $x_n$ are external, or all $x_i$'s ($i \in [\![1, n-1]\!]$) are internal.

We prove the statement by induction on the length of the path. We identify three cases:

1. each $x_i$ ($i \in [\![1, n-1]\!]$) is both internal and distinct from all $x_j$'s ($j \in [\![0, n]\!]$);
2. $x_0$ and $x_n$ are external and there exists $i \in [\![1, n-1]\!]$ *s.t.* $x_i$ is external;
3. $x_0 = x_n$ and there exists $k \in [\![1, n-1]\!]$ *s.t.* $x_k = x_0$ $(= x_n)$.

Next, we show that this covers all possible cases.

If we are not in the first case, then either some $x_i$ ($i \in [\![1, n-1]\!]$) is external or is equal to some $x_j$ ($j \in [\![0, n]\!]$). For the former, by (⑤) we get that $x_0$ and $x_n$ are necessarily external, and we fall in case 2.

For the former, we know that all $x_i$'s ($i \in [\![1, n-1]\!]$) are internal, and there exists $i \in [\![1, n-1]\!]$ and $j \in [\![0, n]\!]$ *s.t.* $x_i = x_j$. Up to renaming the variables, we get that $i < j$ and $(i, j) \neq (0, n)$. Recall that all $x_k$'s ($k \in [\![1, n-1]\!]$) are internal, and so from $x_i = x_j$ we get by trivial induction that $x_0 = x_{j-i}$. In particular, $x_0$ must be internal and so $x_0 = x_n$. Letting $k \mathrel{\hat{=}} j - i$ we have $k \in [\![1, n-1]\!]$ *s.t.* $x_k = x_0 = x_n$, and so we are in case 3.

We can now proceed with the induction.

1. For the base case, we have a path where each $x_i$ ($i \in [\![1, n-1]\!]$) is both internal and distinct from all $x_j$'s ($j \in [\![0, n]\!]$). By definition this path is a segment.

2. In the second case, $x_0$ and $x_n$ are external, and there exists some $i \in [\![1, n-1]\!]$ *s.t.* $x_i$ is external. We use the induction hypothesis on the paths $x_0 \xrightarrow{a_1} \cdots \xrightarrow{a_i} x_i$ and $x_i \xrightarrow{a_{i+1}} \cdots \xrightarrow{a_n} x_n$ and the conclusion follows.

3. In the last case, there exists $k \in [\![1, n-1]\!]$ *s.t.* $x_0 = x_k = x_n$, and the conclusion follows from applying the induction on $x_0 \xrightarrow{a_1} \cdots \xrightarrow{a_k} x_k$ and $x_k \xrightarrow{a_{k+1}} \cdots \xrightarrow{a_n} x_n$.

This concludes the induction and the proof. $\qquad\square$

**Lemma IV.2.12.** Let $\gamma, \delta$ be CRPQs. If $f \colon \delta \to \gamma$ is a homomorphism that sends external variables of $\delta$ on external variables of $\gamma$, then there is a function from nodes of $\mathscr{SG}(\delta)$ to nodes of $\mathscr{SG}(\gamma)$ that sends an edge of $\mathscr{SG}(\delta)$ to a path of $\mathscr{SG}(\gamma)$.

*Proof.* By Proposition IV.2.11, the image $f[\sigma]$ by $f$ of any segment $\sigma$ of $\delta$ is a union of segments of $\gamma$. $\qquad\square$

**Corollary IV.2.13.** Let $\gamma, \delta$ be two CRPQs such that there is an embedding from $\delta$ to $\gamma$ *s.t.* every node of in-degree 0 (resp. out-degree 0) is sent on a node of in-degree 0 (resp. out-degree 0). Then $\mathscr{SG}(\delta)$ is a minor of $\mathscr{SG}(\gamma)$.

*Proof.* Let $f\colon \delta \to \gamma$ be such an embedding.

*Claim* IV.2.14. If $x \in \delta$ is external, then $f(x)$ is external.

Indeed, if $x$ has in-degree at least 2, or out-degree at least 2, so does $f(x)$ since $f$ is an embedding. Otherwise, either $x$ has either in-degree 0 or out-degree 0, and so does $f(x)$ using the assumption we made on $f$.

We then use Lemma IV.2.12. Note that since $f$ is an embedding, the segments of $\gamma$ occurring in $f[\sigma]$ must be distinct from the segments occurring in $f[\sigma']$ for any segment $\sigma' \neq \sigma$. Overall, we have an injective map from nodes of $\mathscr{SG}(\delta)$ to nodes of $\mathscr{SG}(\gamma)$, which sends an edge to a path, and moreover these paths are pairwise disjoint. This shows that $\mathscr{SG}(\delta)$ is a minor of $\mathscr{SG}(\gamma)$. $\qquad\square$

**Remark IV.2.15.** Note that the assumption that $\xi$ is hom-minimal in Theorem IV.2.9 is necessary: consider the CRPQ $\gamma(x,y) \mathrel{\hat{=}} x \xrightarrow{a^+} y \wedge x \xrightarrow{(aa)^+} y$. For $n, m \in \mathbb{N}_{>0}$, let $\xi_{n,m}(x,y) \mathrel{\hat{=}} x \xrightarrow{a^n} y \wedge x \xrightarrow{a^{2m}} y$. There are two cases:

1. If $n \neq 2m$, then $\lVert \check{\xi}_{n,m} \rVert_{\mathrm{seg}} = 2$ but $\xi_{n,m}$ is not hom-minimal since $\xi_{2m,m} \xrightarrow{\mathrm{hom}} \xi_{n,m}$ but $\xi_{n,m} \xcancel{\xrightarrow{\mathrm{hom}}} \xi_{2m,m}$.

2. If $n = 2m$, then $\lVert \check{\xi}_{n,m} \rVert_{\mathrm{seg}} = 1$ and $\xi_{n,m}$ is hom-minimal.

Hence, using Theorem IV.2.9, we can only get a lower bound of one atom (and not two) on the size of any UCRPQ equivalent to $\gamma(x,y)$, which is consistent with the fact that $\gamma(x,y) \equiv \gamma'(x,y)$ where $\gamma'(x,y) \mathrel{\hat{=}} x \xrightarrow{(aa)^+} y$. If Theorem IV.2.9 would allow for non hom-minimal queries we would obtain, by Item 1 a lower bound of 2 atoms, which is false. $\qquad\Diamond$

However, this is a sound but unsurprisingly not a complete characterization of minimality.

**Proposition IV.2.16.** There are minimal (U)CRPQs which are not strongly minimal.

*Proof.* The Boolean CRPQ $\gamma() = x \xrightarrow{a^+} x$. Since it has one atom, it must be minimal, but it has no hom-minimal expansions. $\qquad\square$

Finally, we show that checking strong minimality is at least as hard as checking containment.

**Proposition IV.2.17.** Testing whether a (U)CRPQ is strongly minimal is ExpSpace-hard.

*Proof.* We use the same reduction as in Proposition IV.2.3. If $\delta$ is strongly minimal, then it is minimal, and so it is non-redundant, and hence by the proof of Proposition IV.2.3, $x' \xrightarrow{K} y' \not\subseteq \bigwedge_i x \xrightarrow{L_i} y$.

We prove the converse implication: assume that $x' \xrightarrow{K} y' \not\subseteq \bigwedge_i x \xrightarrow{L_i} y$. Then there exists $\#u\# \in K$ s.t. $x' \xrightarrow{\#u\#} y'$ does not satisfy $\bigwedge_i x \xrightarrow{L_i} y$. Consider the expansion $\xi$ obtained by replacing $K$ with $\#u\#$, and replacing $L_i$ with $@_i$ for each $i$. Then $\xi$ is a core because of the fresh letters $\#$ and $@_i$. Moreover, we claim that it has to be hom-minimal. Indeed, assume that some other expansion $\xi'$ is *s.t.* $\xi' \xrightarrow{\text{hom}} \xi$. Then because of $\#$, the atom expansion of $x' \xrightarrow{K} y'$ in $\xi'$ must be mapped on $x' \xrightarrow{\#u\#} y'$ in $\xi$. Then, by definition of $u$, the atom expansions of $\bigwedge_i x \xrightarrow{L_i} y$ cannot be mapped on $x' \xrightarrow{\#u\#} y'$, and so they must be mapped on $\bigwedge_i x \xrightarrow{@_i} y$, and hence $\xi' = \xi$. Hence, this shows that $\delta$ is strongly minimal. Overall, we showed that $x' \xrightarrow{K} y' \not\subseteq \bigwedge_i x \xrightarrow{L_i} y$ *iff* $\delta$ is strongly minimal, which concludes our reduction. □

## IV.3 An Upper Bound for Minimization of CRPQs

In this section we show that the MINIMIZATION PROBLEM for CRPQs is decidable, in particular, it belongs to the class 2ExpSpace.

**Theorem IV.3.1.** The MINIMIZATION PROBLEM for CRPQs is in 2ExpSpace.

The proof of Theorem IV.3.1 is based on a key lemma stated below. Intuitively, this lemma tells us that if a CRPQ $\gamma$ is equivalent to another CRPQ $\alpha$, then $\gamma$ is also equivalent to a CRPQ $\beta$, where $\beta$ has the same "shape" than $\alpha$ but the sizes of the NFAs appearing in $\beta$ are bounded by the size of $\gamma$. In particular, if $\gamma$ is equivalent to a CRPQ with at most $k$ atoms, then it is equivalent to a CRPQ with at most $k$ atoms and NFAs whose size is bounded by a function of $k$, and hence the search space in the minimization problem becomes finite. A careful analysis of this idea yields our 2ExpSpace upper bound.

**Lemma IV.3.2.** Let $\gamma$ and $\alpha$ be two CRPQs such that $\alpha \subseteq \gamma$. Then there exists a CRPQ $\beta$ satisfying $\alpha \subseteq \beta \subseteq \gamma$ such that:

1. The underlying graphs of $\alpha$ and $\beta$ coincide.
2. The size of each NFA appearing in $\beta$ is bounded by $f(\|\gamma\|)$, where $f$ is a double-exponential function.

We start by giving a proof sketch of this lemma before giving the full detailed proof.

*Proof sketch.* The idea is to define the CRPQ $\beta$ as the CRPQ obtained from $\alpha$ by replacing each regular language $L$ by a suitable regular language $\tilde{L}$, which depends also on $\gamma$. Consider the following equivalence relation on $\mathbb{A}^*$, where $\mathbb{A}$ is the underlying alphabet. Given $u, v \in \mathbb{A}^*$, we write $u \sim_\gamma v$ if for every NFA $\mathcal{A}$ appearing on $\gamma$, and pair of states $p, q$ of $\mathcal{A}$:

$$u \text{ is accepted by } \mathcal{A}[p, q] \iff v \text{ is accepted by } \mathcal{A}[p, q].$$

Recall that $\mathcal{A}[p, q]$ denote the sublanguage of $\mathcal{A}$ recognized when considering $\{p\}$ as the set of initial states and $\{q\}$ as the set of final states. For $u \in \mathbb{A}^*$, we define its *γ-type* to be:

$$\text{type}_\gamma(u) \triangleq \{([u_1]_{\sim_\gamma}, \dots, [u_\ell]_{\sim_\gamma}) \mid \ell \le (\|\gamma\|_{\text{var}} + 1),$$

$$\text{and } u_1, \dots, u_\ell \in \mathbb{A}^* \text{ s.t. } u = u_1 \cdots u_\ell\}.$$

The idea is that $\text{type}_\gamma(u)$ encodes all the possible ways $u$ can be broken into $\ell \le (\|\gamma\|_{\text{var}} + 1)$ subwords: we are not interested in the particular subwords $u_i$, but in their equivalence classes $[u_i]_{\sim_\gamma}$ with respect to $\sim_\gamma$.

We define the sought CRPQ $\beta$ to be the CRPQ obtained from $\alpha$ by replacing each regular language $L$ by $\tilde{L}$, where:

$$\tilde{L} \triangleq \bigcup_{u \in L} \{z \in \mathbb{A}^* \mid \text{type}_\gamma(u) \subseteq \text{type}_\gamma(z)\}.$$

By definition, the underlying graphs of $\alpha$ and $\beta$ are the same, and hence condition (1) holds. It remains to verify $\alpha \subseteq \beta \subseteq \gamma$ and condition (2). Note that $u \in \{z \in \mathbb{A}^* \mid \text{type}_\gamma(u) \subseteq \text{type}_\gamma(z)\}$ always holds, and hence $L \subseteq \tilde{L}$. It follows that $\alpha \subseteq \beta$.

Showing $\beta \subseteq \gamma$ is more involved. By Proposition III.2.12, we need to prove that for every expansion $\xi_\beta \in \text{Exp}(\beta)$, there exists an expansion $\eta_\beta \in \text{Exp}(\gamma)$, such that $\eta_\beta \xrightarrow{\text{hom}} \xi_\beta$. Assume that $\beta$ and $\gamma$ are of the form $\bigwedge_{j=1}^m x_j \xrightarrow{L_j} y_j$ and $\bigwedge_{i=1}^r t_i \xrightarrow{M_i} s_i$, respectively. In particular, $\alpha$ must be of the form $\bigwedge_{j=1}^m x_j \xrightarrow{L_j} y_j$. Suppose $\xi_\beta$ is of the form $\bigwedge_{j=1}^m x_j \xrightarrow{z_j} y_j$, where $z_j \in \tilde{L}_j$. By construction, $\xi_\beta$ has a corresponding expansion $\xi_\alpha$ of $\alpha$: since each $z_j \in \tilde{L}_j$, there must be a $u_j \in L_j$ such that $\text{type}_\gamma(u_j) \subseteq \text{type}_\gamma(z_j)$, and then we can take $\xi_\alpha$ as $\bigwedge_{j=1}^m x_j \xrightarrow{u_j} y_j$. In turn, since $\alpha \subseteq \gamma$, there exists an expansion $\eta_\alpha \in \text{Exp}(\gamma)$ such that $\eta_\alpha \xrightarrow{\text{hom}} \xi_\alpha$ via a homomorphism $f$.

The idea is to modify $\eta_\alpha$ into another expansion $\eta_\beta$ with $\eta_\beta \to \xi_\beta$, as desired. Note that $f$ maps the external variables of $\eta_\alpha$ to external or internal variables in $\xi_\alpha$. This determines a subdivision for each path $x_j \xrightarrow{u_j} y_j$ of $\xi_\alpha$ into smaller or 'basic paths', whose endpoints correspond to external variables of $\xi_\alpha$ or images of the external variables of $\eta_\alpha$ via $f$. The number of these paths is hence bounded by $\|\gamma\|_{\text{var}} + 1$. Since $\text{type}_\gamma(u_j) \subseteq \text{type}_\gamma(z_j)$, then each path $x_j \xrightarrow{z_j} y_j$ in $\xi_\beta$ can also be subdivided in an equivalent way than $x_j \xrightarrow{u_j} y_j$. Overall, the decomposition of $\xi_\alpha$ into basic paths can be 'simulated' in $\xi_\beta$. This gives us a homomorphism from $\eta_\beta$ to $\xi_\beta$, where $\eta_\beta$ is obtained from $\eta_\alpha$ in the following way: for each path $t_i \xrightarrow{w_i} s_i$ in $\eta_\alpha$, where $w_i \in M_i$, the image of the path via $f$ induces a subdivision of $t_i \xrightarrow{w_i} s_i$ into basic paths of $\xi_\alpha$. We can replace each of these basic paths by its equivalent basic path in $\xi_\beta$. As the label of these paths are equivalent w.r.t the relation $\sim_\gamma$, membership in $M_i$ is maintained after this transformation. Hence $\eta_\beta$ is indeed an expansion of $\gamma$.

For condition (2), it is easy to see that every equivalence class $C$ of the relation $\sim_\gamma$ can be accepted by an NFA $\mathcal{A}_C$ of single-exponential size on $\|\gamma\|$. Also, for each word $u \in \mathbb{A}^*$, and each tuple $\bar{c} = (C_1, \dots, C_\ell) \in \text{type}_\gamma(u)$, there is a single-exponential size NFA $\mathcal{A}_{\bar{c}}$ accepting the language $\{z \in \mathbb{A}^* \mid \bar{c} \in \text{type}_\gamma(z)\}$. By intersecting these NFAs, we obtain an NFA $\mathcal{A}_u$ accepting the language

$$\{z \in \mathbb{A}^* \mid \text{type}_\gamma(u) \subseteq \text{type}_\gamma(z)\}.$$

It is easy to see that the number of tuples of the form $(C_1, \dots, C_\ell)$ is at most single-exponential, and then the size of $\mathcal{A}_u$ is at most double-exponential. It follows that the number of possible $\text{type}_\gamma(u)$ is at most double-exponential, and hence the languages $\bar{L}$ in $\beta$ can be described by a union of double-exponential many NFAs, each of size at most double-exponential. Overall, each $\bar{L}$ can be described by an NFA of at most double-exponential size on $\|\gamma\|$. □

*Proof details.* ❧ *Construction of $\beta$.* Recall that $\beta$ is defined as the CRPQ obtained from $\alpha$ by replacing each regular language $L$ by $\bar{L}$, where:

$$\bar{L} \triangleq \bigcup_{u \in L} \{z \in \mathbb{A}^* \mid \text{type}_\gamma(u) \subseteq \text{type}_\gamma(z)\}.$$

❧ *Correctness of the construction.* By definition, the underlying graphs of $\alpha$ and $\beta$ are the same, and hence condition (1) holds. It remains to verify $\alpha \subseteq \beta \subseteq \gamma$ and condition (2).

❧❧ $\alpha \subseteq \beta \subseteq \gamma$. Note that $u \in \{z \in \mathbb{A}^* \mid \text{type}_\gamma(u) \subseteq \text{type}_\gamma(z)\}$ always holds, and hence $L \subseteq \bar{L}$. It follows that $\alpha \subseteq \beta$.

We check $\beta \subseteq \gamma$ using Proposition III.2.12. Assume $\beta$ is of the form $\bigwedge_{j=1}^m x_j \xrightarrow{L_j} y_j$. Consider an expansion $\xi_\beta \in \text{Exp}(\beta)$ defined by replacing each atom $x_j \xrightarrow{L_j} y_j$ by the path $x_j \xrightarrow{z_j} y_j$, for $z_j \in \bar{L}_j$. There must exist $u_j \in L_j$ such that $\text{type}_\gamma(u_j) \subseteq \text{type}_\gamma(z_j)$. Let $\xi_\alpha \in \text{Exp}(\alpha)$ be the expansion of $\alpha = \bigwedge_{j=1}^m x_j \xrightarrow{L_j} y_j$ obtained from replacing each atom $x_j \xrightarrow{L_j} y_j$ by the path $x_j \xrightarrow{u_j} y_j$. As $\alpha \subseteq \gamma$, then there exists an expansion $\eta_\alpha \in \text{Exp}(\gamma)$ such that $\eta_\alpha \xrightarrow{\text{hom}} \xi_\alpha$. Assume $\gamma$ is of the form $\bigwedge_{i=1}^r t_i \xrightarrow{M_i} s_i$ and $\eta_\alpha$ is of the form $\bigwedge_{i=1}^r t_i \xrightarrow{w_i} s_i$, where $w_i \in M_i$. Let $f$ be a homomorphism from $\eta_\alpha$ to $\xi_\alpha$. We can decompose each path $x_j \xrightarrow{u_j} y_j$ in $\xi_\alpha$ into

$$x_j \xrightarrow{u_{j,1}} h_{j,1} \xrightarrow{u_{j,2}} \cdots \xrightarrow{u_{j,\ell-1}} h_{j,\ell-1} \xrightarrow{u_{j,\ell}} y_j$$

where each $h_{j,p}$ is the image via $f$ of some variable in $\text{vars}(\gamma) = \{s_1, t_1, \dots, s_r, t_r\}$ and each path $x \xrightarrow{u_{j,p}} x'$ satisfies that all of its internal variables are not images via $f$ of variables in $\text{vars}(\gamma)$. Note that $\ell \le (\|\gamma\|_{\text{var}} + 1)$. We say that the paths of the form $x \xrightarrow{u_{j,p}} x'$ are the *basic paths* of $\xi_\alpha$ with respect to $f$. Since $\text{type}_\gamma(u_j) \subseteq \text{type}_\gamma(z_j)$, each path $x_j \xrightarrow{z_j} y_j$ in $\xi_\beta$ can be decomposed into

$$x_j \xrightarrow{z_{j,1}} g_{j,1} \xrightarrow{z_{j,2}} \cdots \xrightarrow{z_{j,\ell-1}} g_{j,\ell-1} \xrightarrow{z_{j,\ell}} y_j$$

where $z_{j,p} \sim_\gamma u_{j,p}$. Again, we say that the paths $x \xrightarrow{z_{j,p}} x'$ are the basic paths of $\xi_\beta$ with respect to $f$. We conclude by showing that there is $\eta_\beta \in \text{Exp}(\gamma)$ such that $\eta_\beta \xrightarrow{\text{hom}} \xi_\beta$. Intuitively, $\xi_\alpha$ and $\xi_\beta$ have "equivalent" decompositions in terms of basic paths. Hence, it is possible to turn $\eta_\alpha$ into another expansion $\eta_\beta \in \text{Exp}(\gamma)$ by replacing basic paths of $\xi_\alpha$ by their corresponding basic paths in $\xi_\beta$. By doing so, we indeed obtain an expansion $\eta_\beta$ of $\gamma$ such that $\eta_\beta \xrightarrow{\text{hom}} \xi_\beta$.

Formally, observe that each path $t_i \xrightarrow{w_i} s_i$ of expansion $\eta_\alpha$ is mapped via $f$ to path in $\xi_\alpha$ that can be decomposed into

$$k_0 \xrightarrow{w_{i,1}} k_1 \xrightarrow{w_{i,2}} \cdots \xrightarrow{w_{i,n-1}} k_{n-1} \xrightarrow{w_{i,n}} k_n$$

where $w_i = w_{i,1} \cdots w_{i,n}$, $k_0 = f(t_i)$, $k_n = f(s_i)$ and each path $k_{q-1} \xrightarrow{w_{i,q}} k_q$ is a basic path of $\xi_\alpha$ w.r.t. $f$. Each path $k_{q-1} \xrightarrow{w_{i,q}} k_q$ has a corresponding basic path $k'_{q-1} \xrightarrow{w_{i,q}'} k_q'$ of $\xi_\beta$ w.r.t. $f$ defined in the natural way: if $k_o \in \text{vars}(\alpha) = \{x_1, y_1, \ldots, x_m, y_m\}$, then $k_o' = k_o$; if $k_o = h_{j,p}$, then $k_o' = g_{j,p}$; and if $w_{i,q} = u_{j,p}$, then $w_{i,q}' = z_{j,p}$. In particular, we have that $w_{i,q} \sim_\gamma w_{i,q}'$ and that the following path belongs to $\xi_\beta$:

$$k_0' \xrightarrow{w_{i,1}'} k_1' \xrightarrow{w_{i,2}'} \cdots \xrightarrow{w_{i,n-1}'} k_{n-1}' \xrightarrow{w_{i,n}'} k_n'$$

It follows that the CQ $\eta_\beta$ obtained from $\gamma$ by replacing each atom $t_i \xrightarrow{M_i} s_i$ by the path $t_i \xrightarrow{w_i'} s_i$, where $w'_i = w_{i,1}' \cdots w_{i,n}'$ satisfies that $\eta_\beta \xrightarrow{\text{hom}} \xi_\beta$. It remains to show that $\eta_\beta$ is actually an expansion of $\gamma$, that is, each $w'_i \in M_i$. Suppose $M_i$ is represented by the NFA $\mathcal{A}_i$. Since $w_i \in M_i$ and $w_i = w_{i,1} \cdots w_{i,n}$, there is a sequence $e_0, \ldots, e_n$ of states of $\mathcal{A}_i$ such that $e_0$ is initial, $e_n$ is final, and $w_{i,q}$ is accepted by $\mathcal{A}_i[e_{q-1}, e_q]$. As $w_{i,q} \sim_\gamma w_{i,q}$, we have that $w_{i,q}'$ is also accepted by $\mathcal{A}_i[e_{q-1}, e_q]$, and hence $w'_i = w_{i,1}' \cdots w_{i,n}'$ is accepted by $\mathcal{A}_i$.

☙☙ *Bounding the size of NFAs in $\beta$.* First note that for any equivalence class $C$ of $\sim_\gamma$ the language $\{z \in \mathbb{A}^* \mid [z]_{\sim_\gamma} = C\}$ is accepted by an NFA $\mathcal{A}_C$ of single-exponential size. Indeed, the class $C$ can be described by a set of triples

$$\mathcal{T}_C \subseteq \mathcal{T} = \{(\mathcal{A}, p, q) \mid \mathcal{A} \text{ is an NFA in } \gamma, \text{ and } p, q \text{ are states of } \mathcal{A}\}$$

in such a way that $v \in C$ iff $v$ is accepted by $\mathcal{A}[p, q]$, for every $(\mathcal{A}, p, q) \in \mathcal{T}_C$, and $v$ is accepted by $\mathcal{A}[p, q]^\complement$, for every $(\mathcal{A}, p, q) \notin \mathcal{T}_C$, where $\mathcal{A}^\complement$ denotes the complement NFA of $\mathcal{A}$. Hence $\mathcal{A}_C$ can be written as the following intersection of NFAs:

$$\mathcal{A}_C = \bigcap_{(\mathcal{A}, p, q) \in \mathcal{T}_C} \mathcal{A}[p, q] \cap \bigcap_{(\mathcal{A}, p, q) \notin \mathcal{T}_C} \mathcal{A}[p, q]^\complement$$

The number of states in $\mathcal{A}[p, q]^\complement$ is at most $2^r$ and $|\mathcal{T}| \leq r^2 \|\gamma\|_{\text{at}}$, where $r$ is the maximum number of states in an NFA of $\gamma$. Hence the number of states of $\mathcal{A}_C$ is at most $2^{r^3 \|\gamma\|_{\text{at}}}$, that is, single-exponential on $\gamma$.

Fix $u \in \mathbb{A}^*$. We claim that $\{z \in \mathbb{A}^* \mid \text{type}_\gamma(u) \subseteq \text{type}_\gamma(z)\}$ can be accepted by an NFA $\mathcal{A}_u$ of at most double-exponential size on $\gamma$. It is easy to see that for every tuple $\bar{c} = (C_1, \ldots, C_\ell) \in \text{type}_\gamma(u)$, the language $\{z \in \mathbb{A}^* \mid \bar{c} \subseteq \text{type}_\gamma(z)\}$ can be described by an NFA $\mathcal{A}_{\bar{c}}$ of single-exponential size: we guess a decomposition $z = z_1 \cdots z_\ell$ of the input word $z$ and check that each $z_i$ is accepted by $\mathcal{A}_{C_i}$. The number of states of $\mathcal{A}_{\bar{c}}$ is $\mathcal{O}(\sum_{i=1}^{\ell} s_i)$, where $s_i$ is the number of states of $\mathcal{A}_{C_i}$. Since $\ell \leq (\|\gamma\|_{\mathrm{var}} + 1)$, this is $\mathcal{O}(\|\gamma\|_{\mathrm{var}} 2^{r^3\|\gamma\|_{\mathrm{at}}})$. Now, the NFA $\mathcal{A}_u$ is simply the intersection of all the NFAs in $\{\mathcal{A}_{\bar{c}} \mid \bar{c} \in \mathrm{type}_\gamma(u)\}$. The number of possible equivalence classes of $\sim_\gamma$ is at most $2^{|\mathcal{T}|} \leq 2^{r^2\|\gamma\|_{\mathrm{at}}}$, and then the number of possible tuples of the form $\bar{c} = (C_1, \dots, C_\ell)$ is $N = \mathcal{O}(2^{r^2\|\gamma\|_{\mathrm{at}}(\|\gamma\|_{\mathrm{var}}+1)})$. It follows that the number of states of $\mathcal{A}_u$ is at most $\mathcal{O}(\|\gamma\|_{\mathrm{var}} 2^{r^3\|\gamma\|_{\mathrm{at}}} N)$, i.e., double-exponential in $\gamma$.

Finally, note that the number of possible $\mathrm{type}_\gamma(u)$ is at most double-exponential in $\gamma$, more precisely, $2^N$. We conclude that every language $\bar{L}$ in $\beta$ can be represented by an NFA $\mathcal{A}_{\bar{L}}$ which is the union of at most double-exponential many NFAs of the form $\mathcal{A}_u$, each of these having at most double-exponential many states. We conclude that the size of $\mathcal{A}_{\bar{L}}$ can be bounded by a double-exponential function $f(\|\gamma\|)$. □

Using this lemma, we can then easily deduce the desired 2ExpSpace upper bound.

*Proof of Theorem IV.3.1.* The 2ExpSpace algorithm proceeds as follows. Let $\gamma$ be a CRPQ over $\mathbb{A}$ and $k \in \mathbb{N}$. Let $f$ be the double-exponential function from Lemma IV.3.2. We enumerate all possible directed multigraphs with at most $k$ edges. For each of these graphs, we enumerate all the possible CRPQs $\beta$ obtained by labelling each edge with an NFA of size bounded by $f(\|\gamma\|)$. If for some of these $\beta$, we have that $\beta \equiv \gamma$ then we accept the instance, otherwise we reject it. Lemma IV.3.2 ensures that this algorithm is correct.

It remains to show that $\beta \equiv \gamma$ can be carried out in 2ExpSpace. We use Proposition V.3.11, which states that containment $\Gamma \subseteq \Delta$, for UCRPQs $\Gamma$ and $\Delta$, can be solved in non-deterministic space $\mathcal{O}(\|\Gamma\| + \|\Delta\|^{c \cdot \|\Delta\|_{\mathrm{at}}})$, where $c$ is a constant. This implies that $\beta \subseteq \gamma$ can be checked in space $\mathcal{O}(\|\beta\| + \|\gamma\|^{c \cdot \|\gamma\|_{\mathrm{at}}})$, and then within 2ExpSpace. The containment $\gamma \subseteq \beta$ can be solved in space $\mathcal{O}(\|\gamma\| + \|\beta\|^{c \cdot k})$, and hence also in 2ExpSpace. □

## IV.4 Minimization of UCRPQs via Approximations

We now focus on minimizations by *finite unions* of CRPQs. This is a different problem than the one seen in the previous section, *i.e.*, there is no obvious reduction in either direction between the MINIMIZATION PROBLEM for CRPQs and the MINIMIZATION PROBLEM for UCRPQs, and indeed we will solve this problem using an altogether different approach.

### IV.4.1 Unions Allow Further Minimization

As it turns out, having unions may help in minimizing the (maximum) number of atoms of a query as the next proposition shows.

**Proposition IV.4.1.** There exist CRPQs which are *minimal* among CRPQs but not among UCRPQs.

*Proof.* The following example is inspired from [FM25, Example 1.2] (*a.k.a.* Example V.1.2). Consider the following CRPQs:

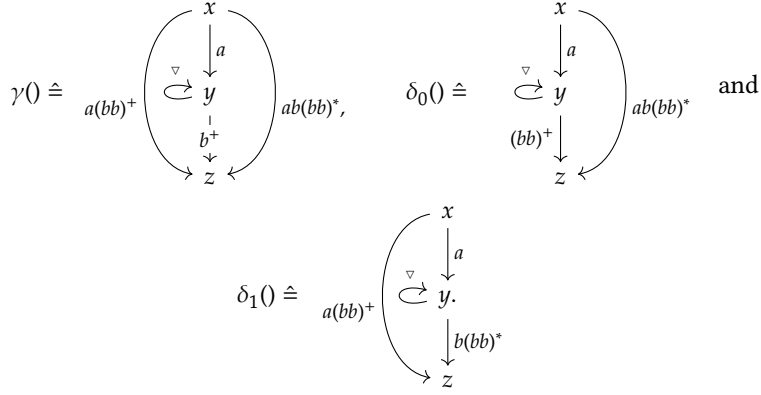

It is easy to see that $\gamma \equiv \delta_0 \vee \delta_1$ by doing a case disjunction on the parity of the path between $y$ and $z$—the even case is handled by $\delta_0$, the odd case by $\delta_1$. Hence, $\gamma$ is not minimal among UCRPQs.

We want to show that $\gamma$ is minimal among CRPQs. Let $\zeta()$ be a CRPQ that is equivalent to $\gamma$, and assume by contradiction that it has at most four atoms. Given natural numbers $l, m, r \in \mathbb{N}_{>0}$ where $l$ is even and $r$ is odd, consider the expansion

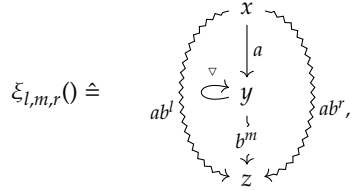

where squiggly arrows represent paths of atoms.

*Claim* IV.4.2. The expansion $\xi_{l,m,r}$ is hom-minimal *iff* $l = m$ or $m = r$.

Indeed, if $l \neq m$ and $m \neq r$, then if $m$ is even then $\xi_{m,m,r} \xrightarrow{\text{hom}} \xi_{l,m,r}$ but $\xi_{l,m,r} \xrightarrow{\text{hom}} \xi_{m,m,r}$ (since $l \neq m$) and dually if $m$ is odd then $\xi_{l,m,m} \xrightarrow{\text{hom}} \xi_{l,m,r}$ but $\xi_{l,m,r} \xrightarrow{\text{hom}} \xi_{l,m,m}$ (since $m \neq r$). In both cases, $\xi_{l,m,r}$ is not hom-minimal.

Conversely, if $l = m$ or $m = r$, assume *w.l.o.g.*, by symmetry, that $l = m$. Let $l', m', r' \in \mathbb{N}_{>0}$ *s.t.* $l'$ is even and $r'$ is odd, and assume that there is a homomorphism $f \colon \xi_{l',m',r'} \to \xi_{l,m,r}$. Because of the $\triangledown$-self-loop, $f(y) = y$ and hence $f(x) = x$. It then follows that $f(z) = z$ because we must have both an $a(bb)^+$- and an $ab(bb)^*$-path from $f(x)$ and $f(z)$. Moreover, we must have $m = m'$, and since $l'$ is even, we must have $l' = l = m$, and dually, since $r'$ is odd, we must have $r' = r$. It follows that $\langle l', m', r' \rangle = \langle l, m, r \rangle$, and hence $\xi_{l,m,r}$ is hom-minimal.

*Claim* IV.4.3. If $m = r$, then $\xi_{l,m,r}^{\smallsmile}$ equals

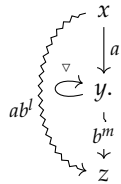

Putting Claims IV.4.2 and IV.4.3 and Theorem IV.2.9 together, we get that $\mathcal{SG}(\xi_{2,1,1}^{\check{\phantom{x}}})$ is a minor of $\zeta$. Since $\zeta$ was assumed to have at most four atoms, it follows that $\zeta()$ must be exactly of the form

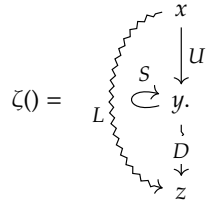

$$\zeta() = \quad L \left\{ \begin{array}{c} x \\ \overset{S}{\underset{\rightleftarrows}{\leadsto}} \; y. \\ \overset{D}{\underset{\scriptscriptstyle z}{\leadsto}} \end{array} \right.$$

Using the containment $\zeta \subseteq \gamma$, since the only directed simple cycle in an expansion of $\gamma$ is labelled by $\triangledown$, we get $S = \{\triangledown\}$. Similarly, using again that $\zeta \subseteq \gamma$ and that any expansion of $\gamma$ must have both an $a(bb)^+$- and an $ab(bb)^*$-path, it can be shown that $U = \{a\}$.

Now let $l, m, r \in \mathbb{N}_{>0}$ with $l$ even and $r$ odd. From $\gamma \subseteq \zeta$, we get that each $\xi_{l,m,r}$ satisfies $\zeta$ and so there is an evaluation map $f \colon \zeta \to \xi_{l,m,r}$. Clearly $f(y) = y$ and then using the $a(bb)^+$- and $ab(bb)^*$-paths, we get that $f(x) = x$ and $f(z) = z$. It then follows that

- $b^m \in D$, and
- either $ab^l \in L$, or $ab^r \in L$, or $a\triangledown^k b^m$ for some $k \in \mathbb{N}$

From the first point we get $b^+ \subseteq D$, and in fact using $\zeta \subseteq \gamma$, $D = b^+$.

Now let $w \in L$ and $m \in \mathbb{N}$, and consider the expansion

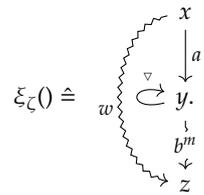

$$\xi_\zeta() \; \hat{=} \quad w \left\{ \begin{array}{c} x \\ \overset{\triangledown}{\underset{\rightleftarrows}{\leadsto}} \; y. \\ \overset{b^m}{\underset{\scriptscriptstyle z}{\leadsto}} \end{array} \right.$$

of $\zeta$. Since $\zeta \subseteq \gamma$, there is an evaluation map $f \colon \gamma \to \xi_\zeta$. Taking $m$ to be any even value, we get that $w$ must be in $ab(bb)^*$, say $w = ab^r$ for some odd $r \in \mathbb{N}_{>0}$. Taking $m$ to be any odd value, we get dually that $w$ must be in $a(bb)^+$, say $w = ab^l$ for some even $l \in \mathbb{N}_{>0}$. Hence, we get $\{ab^l, ab^r\} \subseteq L$.

We can now pick $m$ to be even and $w = ab^l$, and then $\xi_\zeta$ should be a model of $\gamma$. But $\xi_\zeta$ cannot satisfy the atom $x \xrightarrow{ab(bb)^*} z$. Contradiction. Hence, $\gamma$ cannot be equivalent to a single CRPQ with at most four atoms. □

### IV.4.2 Maximal Under-Approximations

Our approach exploits having unions at our disposal, enabling the possibility of defining and computing maximal under-approximations for UCRPQs having some given underlying shape. This will lead to an ExpSpace upper bound for UCRPQ minimization.

For a graph class $\mathcal{C}$ (remember that these are directed multigraphs) we denote by $\mathrm{CQ}(\mathcal{C})$, $\mathrm{CRPQ}(\mathcal{C})$, $\mathrm{UCRPQ}(\mathcal{C})$, $\mathrm{UCRPQ}^\infty(\mathcal{C})$, the class of CQs, of CRPQs, of finite unions of CRPQs and of infinite unions of CRPQs $\gamma$ such that $G_\gamma \in \mathcal{C}$. Given a graph class $\mathcal{C}$ and a UCRPQ $\Gamma$, we define $\mathrm{App}_\mathcal{C}^\infty(\Gamma)$ to be the (infinite) set of all CRPQ($\mathcal{C}$) queries which are contractions of CQs contained in $\Gamma$. More formally:[12]

$$\text{App}_{\mathcal{C}}^{\infty}(\Gamma) \doteq \{\alpha \in \text{CRPQ}(\mathcal{C}) \mid \exists \xi \in \text{Exp}(\Gamma), \exists \eta \in \text{CQ } s.t. \ \xi \xrightarrow{\text{hom}} \eta$$
$$\text{and } \alpha \text{ is a contraction of } \eta\}.$$

**Proposition IV.4.4.** Let $\Gamma$ be a UCRPQ. If $\mathcal{C}$ is a graph class closed under taking minors, then $\text{App}_{\mathcal{C}}^{\infty}(\Gamma)$ is a maximal under-approximation of $\Gamma$ by UCRPQ$^{\infty}$($\mathcal{C}$) queries, in the sense that:

1. $\text{App}_{\mathcal{C}}^{\infty}(\Gamma) \in \text{UCRPQ}^{\infty}(\mathcal{C})$,
2. $\text{App}_{\mathcal{C}}^{\infty}(\Gamma) \subseteq \Gamma$, and
3. for any $\Delta \in \text{UCRPQ}^{\infty}(\mathcal{C})$, if $\Delta \subseteq \Gamma$, then $\Delta \subseteq \text{App}_{\mathcal{C}}^{\infty}(\Gamma)$.

*Proof.* The first point is trivial. For the second one, observe that if $\alpha \in \text{CRPQ}(\mathcal{C})$ is *s.t.* there exist $\xi \in \text{Exp}(\Gamma)$ and $\eta \in \text{CQ } s.t. \ \xi \xrightarrow{\text{hom}} \eta$ and $\alpha$ is a contraction of $\eta$, then $\eta \subseteq \xi$ and $\eta$ is semantically equivalent to $\alpha$ by Fact IV.2.2, and hence $\alpha \subseteq \xi$, from which it follows that $\text{App}_{\mathcal{C}}^{\infty}(\Gamma) \subseteq \Gamma$.

For the third point, let $\Delta \in \text{UCRPQ}^{\infty}(\mathcal{C})$ *s.t.* $\Delta \subseteq \Gamma$. Let $\delta \in \Delta$ and $\xi_{\delta}$ be an expansion of $\delta$. Since $\Delta \subseteq \Gamma$, there exist $\gamma \in \Gamma$ and $\xi_{\gamma} \in \text{Exp}(\gamma)$ *s.t.* $\xi_{\gamma} \xrightarrow{\text{hom}} \xi_{\delta}$. From the fact that $G_{\delta} \in \mathcal{C}$ and that $\xi_{\delta} \in \text{Exp}(\delta)$, it follows that there exists a contraction $\alpha_{\delta}$ of $\xi_{\delta}$ *s.t.* $G_{\alpha_{\delta}}$ is a minor of $G_{\delta}$, and thus belongs to $\mathcal{C}$. Hence, it follows that $\alpha_{\delta} \in \text{App}_{\mathcal{C}}^{\infty}(\Gamma)$. So $\xi_{\delta} \subseteq \text{App}_{\mathcal{C}}^{\infty}(\Gamma)$ and thus $\Delta \subseteq \text{App}_{\mathcal{C}}^{\infty}(\Gamma)$. $\square$

**Remark IV.4.5.** If $\mathcal{C}$ is closed under taking subgraphs and expansions, then

$$\text{App}_{\mathcal{C}}^{\infty}(\gamma) \equiv \{\alpha \in \text{CQ}(\mathcal{C}) \mid \exists \xi \in \text{Exp}(\gamma), \ \xi \xrightarrow{\text{hom}} \alpha\}. \qquad \diamond$$

Our ExpSpace upper bound relies on the following technical lemma.

**Lemma IV.4.6.** If $\mathcal{C}$ is finite and closed under taking minors, then $\text{App}_{\mathcal{C}}^{\infty}(\Gamma)$ is equivalent to a query $\Delta \in \text{UCRPQ}(\mathcal{C})$ with exponentially many CRPQs, each CRPQ of $\Delta$ being of size polynomial in $\|\Gamma\| + \max\{\|G\|_{\text{at}} \mid G \in \mathcal{C}\}$. Further, the membership $\delta \in^? \Delta$ can be tested in NP. In particular, this UCRPQ $\Delta$ can be computed in exponential time from $\Gamma$.

*Proof.* We start by observing that $\text{App}_{\mathcal{C}}^{\infty}(\Gamma)$ admits an equivalent and more flexible definition in terms of refinements. This definition will allow us to effectively compute our desired equivalent query $\Delta$, by considering refinements of bounded lengths.

For each $m \in \mathbb{N} \cup \{+\infty\}$, we define the UCRPQ$^{\infty}$($\mathcal{C}$)

$$\text{App}_{\mathcal{C}}^{\leq m}(\Gamma) \doteq \{\alpha \in \text{CRPQ}(\mathcal{C}) \mid \exists \rho \in \text{Ref}^{\leq m}(\Gamma), \exists \eta \in \text{CRPQ } s.t. \ \rho \xrightarrow{\text{hom}} \eta$$
$$\text{and } \alpha \text{ is a contraction of } \eta\}.$$

We have that $\text{App}_{\mathcal{C}}^{\infty}(\Gamma) \equiv \text{App}_{\mathcal{C}}^{\leq +\infty}(\Gamma)$. Indeed, $\text{App}_{\mathcal{C}}^{\infty}(\Gamma) \subseteq \text{App}_{\mathcal{C}}^{\leq +\infty}(\Gamma)$ as the former is a subset of the latter. On the other hand, $\text{App}_{\mathcal{C}}^{\leq +\infty}(\Gamma) \subseteq \text{App}_{\mathcal{C}}^{\infty}(\Gamma)$ follows from Proposition IV.4.4 and the fact that $\text{App}_{\mathcal{C}}^{\leq +\infty}(\Gamma) \subseteq \Gamma$

[12] Note that the contraction of a CQ is a CRPQ: this is why $\text{App}_{\mathcal{C}}^{\infty}(\Gamma)$ is an infinite union of CRPQs and not of CQs.

by construction.

Let $\|\mathcal{C}\|_{at} \hat{=} \max\{\|G\|_{at} \mid G \in \mathcal{C}\}$ and $r_\Gamma$ be the maximum number of states over the NFAs describing the languages appearing in $\Gamma$. We claim that $\mathrm{App}_{\mathcal{C}}^{\leq +\infty}(\Gamma) \equiv \mathrm{App}_{\mathcal{C}}^{\leq \mathcal{O}(\|\Gamma\|_{at} \cdot r_\Gamma \cdot \|\mathcal{C}\|_{at})}(\Gamma)$. The right-to-left containment is trivial, so it suffices to show $\mathrm{App}_{\mathcal{C}}^{\leq +\infty}(\Gamma) \subseteq \mathrm{App}_{\mathcal{C}}^{\leq \mathcal{O}(\|\Gamma\|_{at} \cdot r_\Gamma \cdot \|\mathcal{C}\|_{at})}(\Gamma)$.

We define an *explicit approximation* of $\Gamma$ over $\mathcal{C}$ as a tuple

$$\vec{\alpha} = \langle \rho, \eta, \alpha, h, orig, contr \rangle$$

consisting of:

- queries $\rho \in \mathrm{Ref}(\Gamma)$, $\eta \in \mathrm{CRPQ}$ and $\alpha \in \mathrm{CRPQ}(\mathcal{C})$,
- a homomorphism $h \colon \rho \xrightarrow{\text{hom}} \eta$,
- witnesses that $\alpha$ is a contraction of $\eta$, in the form of a function *orig* from $\mathrm{vars}(\alpha)$ to $\mathrm{vars}(\eta)$, saying from which variable of $\eta$ a variable of $\alpha$ originates, and a function $contr \colon \mathrm{Atoms}(\eta) \to \mathrm{Atoms}(\alpha)$ saying onto which atom of $\alpha$ an atom of $\eta$ is contracted (*i.e.*, the functions *orig*, *contr* must meet the expected properties).

We say that an explicit approximation $\vec{\alpha_1}$ is *contained* in an explicit approximation $\vec{\alpha_2}$ if $\alpha_1 \subseteq \alpha_2$. For any $m$, to prove that $\mathrm{App}_{\mathcal{C}}^{\leq +\infty}(\Gamma) \subseteq \mathrm{App}_{\mathcal{C}}^{\leq m}(\Gamma)$, it suffices to prove that any explicit approximation $\vec{\alpha_1} = \langle \rho_1, \eta_1, \alpha_1, h_1, orig_1, contr_1 \rangle$ is contained in an explicit approximation $\vec{\alpha_2} = \langle \rho_2, \eta_2, \alpha_2, h_2, orig_2, contr_2 \rangle$ such that $\rho_2 \in \mathrm{Ref}^{\leq m}(\Gamma)$.

☙ *$1^{st}$ step: Bounding the size of $\eta$.* We define the *contraction length* of an explicit approximation $\vec{\alpha}$ as the size of the longest path in $\eta$ whose atoms are all sent on the same atom of $\alpha$ via *contr*. In symbols, this is $\max\{|contr^{-1}(\mu)| : \mu \in \mathrm{Atoms}(\alpha)\}$. We show that any explicit approximation $\vec{\alpha_1}$ is contained in an explicit approximation $\vec{\alpha_2}$ of bounded contraction length.

Let $\vec{\alpha_1}$ be an explicit approximation. Consider a path $x_0 \to x_1 \to \cdots \to x_k$ of $\eta_1$ whose atoms are all sent on the same atom of $\alpha_1$ via $contr_1$. If this path is very long, in particular greater than $\|\Gamma\|_{var}$, it must contain an internal variable $x_i$ such that all of its $h_1$-preimages are internal variables of $\rho_1$. Then we will be able to contract $x_i$ as well as the internal variables of the preimage, obtaining an explicit approximation which contains $\vec{\alpha_1}$.

We now formalize the previous claim: assume that for some $x_i$, with $i \in [\![1, k-1]\!]$, all variables in $h_1^{-1}(x_i)$ are internal in $\rho_1$. If $x_{i-1} \xrightarrow{L} x_i$ and $x_i \xrightarrow{L'} x_{i+1}$ are the only atoms containing $x_i$ in $\eta_1$, then for a variable $z \in h_1^{-1}(x_i)$, the only atoms in containing $z$ in $\rho_1$ must have the form $w \xrightarrow{L} z$ and $z \xrightarrow{L'} w'$. Let $\eta_2$ be the query resulting from $\eta_1$ by contracting the internal variable $x_i$ and replacing $L \cdot L'$ by $K$, where $K$ is defined as follows. Since $L$ and $L'$ appear consecutively in internal paths of the refinement $\rho_1$, there must be an NFA $\mathcal{A}$ in $\gamma$ and three states $p, q, r$ such that $L = \mathcal{A}[p, q]$ or $L = \{a\}$ with $a \in \mathcal{A}[p, q]$, and $L' = \mathcal{A}[q, r]$ or $L' = \{a\}$ with $a \in \mathcal{A}[q, r]$. We define $K = \mathcal{A}[p, r]$. Note that in any case, $L \cdot L' \subseteq K$. Similarly, define $\rho_2$ to be the query resulting from $\rho_1$ by contracting each internal variable $z \in h_1^{-1}(x_i)$ and replacing $L \cdot L'$ by $K$. Note that $\rho_2$ is still a refinement of $\gamma$ and that the homomorphism $h_1\colon \rho_1 \to \eta_1$ induces a homomorphism $f_2\colon \rho_2 \to \eta_2$. Define $\alpha_2$ be the contraction of $\eta_2$ obtained by contracting all the remaining internal variables as the contraction $\alpha_1$ is obtained from $\rho_1$. Since $\alpha_1 \subseteq \alpha_2$ as $L \cdot L' \subseteq K$, this defines an explicit approximation $\vec{\alpha_2}$ that contains $\vec{\alpha_1}$. Note that in the case $h_1^{-1}(x_i) = \varnothing$, we can take $K = L \cdot L'$, and $\alpha_1 \equiv \alpha_2$.

If $k - 1 > \|\Gamma\|_{\mathrm{var}}$, then path $x_0 \to x_1 \to \cdots \to x_k$ contains a variable satisfying the condition above, and hence we can apply the simplification. Overall, this shows that any explicit approximation is contained in an explicit approximation of contraction length at most $\mathcal{O}(\|\Gamma\|_{\mathrm{var}}) \le \mathcal{O}(\|\Gamma\|_{\mathrm{at}})$.

❧ *2nd step: bounding the size of $\rho$.* We now show that we can bound the refinement length of an explicit approximation, namely the maximal length of an atom refinement in $\rho$. Let $\vec{\alpha_1}$ be an explicit approximation of contraction length at most $\mathcal{O}(\|\Gamma\|_{\mathrm{at}})$. Then $\eta_1$ has at most $\mathcal{O}(\|\Gamma\|_{\mathrm{at}} \cdot \|\mathcal{C}\|_{\mathrm{at}})$ atoms. It follows then, by the pigeon-hole principle, that we can bound the refinement length of $\rho_1$ by $\mathcal{O}(\|\Gamma\|_{\mathrm{at}} \cdot r_\Gamma \cdot \|\alpha\|_{\mathrm{at}})$. Indeed, if the length of an atom refinement of $\rho_1$ is greater than this bound, there are two atoms in the refinement $x \xrightarrow{L_i} y$ and $x' \xrightarrow{L_j} y'$, with $i < j$, mapped to the same atom via $h_1$ and whose corresponding NFA states $q_i$ and $q_j$ in the definition of refinements are the same. We can then remove the path between $y$ and $y'$. In conclusion, this shows that $\mathrm{App}_{\mathcal{C}}^{\le +\infty}(\Gamma) \subseteq \mathrm{App}_{\mathcal{C}}^{\le \mathcal{O}(\|\Gamma\|_{\mathrm{at}} \cdot r_\Gamma \cdot \|\mathcal{C}\|_{\mathrm{at}})}(\Gamma)$.

❧ *Conclusion: Expressing & computing* $\mathrm{App}_{\mathcal{C}}^{\le \mathcal{O}(\|\Gamma\|_{\mathrm{at}} \cdot r_\Gamma \cdot \|\mathcal{C}\|_{\mathrm{at}})}(\Gamma)$ *as a UCRPQ.* In order to compute $\mathrm{App}_{\mathcal{C}}^{\le \mathcal{O}(\|\Gamma\|_{\mathrm{at}} \cdot r_\Gamma \cdot \|\mathcal{C}\|_{\mathrm{at}})}(\Gamma)$ we can enumerate the finitely many $m$-refinements $\rho$ of $\Gamma$, where $m = \mathcal{O}(\|\Gamma\|_{\mathrm{at}} \cdot r_\Gamma \cdot \|\mathcal{C}\|_{\mathrm{at}})$, and the finitely many CRPQs $\eta$ with at most $\mathcal{O}(\|\Gamma\|_{\mathrm{at}} \cdot \|\mathcal{C}\|_{\mathrm{at}})$ atoms such that $\rho \xrightarrow{\mathrm{hom}} \eta$. The only issue here is that we have infinitely many possibilities to choose languages labelling the atoms that are not in the homomorphic image of $\rho \xrightarrow{\mathrm{hom}} \eta$. However, we can choose the most general language $\mathbb{A}^*$ obtaining a query equivalent to $\mathrm{App}_{\mathcal{C}}^{\le m}(\Gamma)$. Note that each CRPQ has at most $\mathcal{O}(\|\Gamma\|_{\mathrm{at}} \cdot \|\mathcal{C}\|_{\mathrm{at}})$ atoms and its languages are concatenations of $\mathcal{O}(\|\Gamma\|_{\mathrm{at}} \cdot \|\mathcal{C}\|_{\mathrm{at}})$ sublanguages of $\Gamma$ or $\mathbb{A}^*$, and so they can be described by NFAs of polynomial size on $\|\Gamma\|$ and $\|\mathcal{C}\|_{\mathrm{at}}$. □

**Corollary IV.4.7.** Testing whether a UCRPQ is equivalent to a UCRPQ of at most $k$ atoms is ExpSpace-complete.

*Proof.* It suffices to test if the UCRPQ $\Gamma$ is equivalent to $\mathrm{App}_{\mathcal{C}}^{\le m}(\Gamma)$ where $\mathcal{C}$ is the class of all graphs with at most $k$ edges and $m = \mathcal{O}(\|\Gamma\|_{\mathrm{at}} \cdot r_\Gamma \cdot \|\mathcal{C}\|_{\mathrm{at}})$ as in the proof of Lemma IV.4.6. The correctness follows from Proposition IV.4.4 since $\mathcal{C}$ is trivially closed under minors. Each $\alpha \in \mathrm{App}_{\mathcal{C}}^{\le m}(\Gamma)$ has at most $k$ edges, and $\mathrm{App}_{\mathcal{C}}^{\le m}(\Gamma)$ contains exponentially many queries, so by [FM25, Proposition 3.11] (see also proof of Theorem III.3.1), it can be solved in ExpSpace. Finally, ExpSpace-hardness will follow from Theorem IV.5.1. □

### IV.4.3 CRPQs over Simple Regular Expressions

Let *UCRPQ(SRE)* (*resp. CRPQ(SRE)*) be the set of all UCRPQs (*resp.* CRPQs) whose languages are expressed via positive simple regular expressions, as defined in Section III.2.3. We show that if we restrict the regular expressions we obtain a much better complexity for the MINIMIZATION PROBLEM for UCRPQs. Throughout this chapter, we will shorten "positive simple regular expression" to "simple regular expression".

**Theorem IV.4.8.** The MINIMIZATION PROBLEM, restricted to the class of UCRPQs over simple regular expressions, is $\Pi_2^p$-complete.

*Proof.* We first begin with an easy small counterexample property.

*Claim* IV.4.9. Let $\Gamma, \Delta \in \text{UCRPQ}$ containing only atoms with expressions of the form *(i)* $a^+$, or *(ii)* $a_1 + \cdots + a_k$. Additionally, $\Delta$ may also have expressions of the form *(iii)* $\mathbb{A}^*$. If $\Gamma \not\subseteq \Delta$, then there exists $\xi \in \text{Exp}(\Gamma)$ such that (a) $\xi \not\subseteq \Delta$ and (b) $\|\xi\|_{\text{at}} \leq \mathcal{O}(\max_{\gamma \in \Gamma} \|\gamma\|_{\text{at}} \cdot \max_{\delta \in \Delta} \|\delta\|_{\text{at}})$.

The intuition is that if a counterexample includes an atom expansion $x \xrightarrow{a^n} y$ of some atom $x \xrightarrow{a^+} y$, where $n$ is greater than the maximum number of atoms in $\Delta$ (plus one), then the expansion obtained by replacing $x \xrightarrow{a^n} y$ with $x \xrightarrow{a^{n-1}} y$ must also be a counterexample. Hence, a minimal counterexample must have all atom expansions bounded by the maximum number of atoms in $\Delta$.

*Proof.* This fact follows from a standard technique as used in, *e.g.*, [Fig+20]. Take any counterexample $\xi \in \text{Exp}(\Gamma)$ as in the statement, and suppose it is of minimal size. By means of contradiction, assume $\|\xi\|_{\text{at}} > \max_{\gamma \in \Gamma} \|\gamma\|_{\text{at}} \cdot (\max_{\delta \in \Delta} \|\delta\|_{\text{at}} + 1)$. Then, it contains an atom expansion $x \xrightarrow{a^m} y$ of size $m > \max_{\delta \in \Delta} \|\delta\|_{\text{at}} + 1$. Consider removing one atom from such expansion (*i.e.*, replacing $x \xrightarrow{a^m} y$ with $x \xrightarrow{a^{m-1}} y$), obtaining some expansion $\xi' \in \text{Exp}(\Gamma)$ of smaller size. By minimality $\xi'$ is not a counterexample: in other words there is $\xi'' \in \text{Exp}(\delta)$ such that $h : \xi'' \xrightarrow{\text{hom}} \xi'$ for some $\delta \in \Delta$ and $h$. Since $x \xrightarrow{a^{m-1}} y$ contains more than $\|\delta\|_{\text{at}}$ atoms, there must be some $a$-atom of $x \xrightarrow{a^{m-1}} y$ which either (1) has no $h$-preimage or (2) every $h$-preimage is in an atom expansion of a $\mathbb{A}^*$-atom or a $a^+$-atom of $\delta$. We can then replace the $a$-atom with two $a$-atoms in $\xi'$ and do similarly in the atom expansions of $\xi''$ in the $h$-preimage, obtaining that $\xi' \subseteq \delta$. But this is in contradiction with our hypothesis, hence any minimal counterexample is of size smaller or equal to $\max_{\gamma \in \Gamma} \|\gamma\|_{\text{at}} \cdot (\max_{\delta \in \Delta} \|\delta\|_{\text{at}} + 1)$. □

Given a UCRPQ(SRE) $\Gamma$, the construction of Lemma IV.4.6 yields its maximal under-approximation by UCRPQs of at most $k$ atoms as a UCRPQ $\Delta_{\text{App}}$ whose every regular expression is a concatenation of expressions of the form (i), (ii) and (iii) above. It suffices then to test $\gamma \subseteq \Delta_{\text{App}}$ for every CRPQ(SRE) $\gamma$ in $\Gamma$. Due to Claim IV.4.9 (and observing that equivalent queries without concatenations can be obtained in polynomial time) its negation $\gamma \not\subseteq \Delta_{\text{App}}$ can be tested by guessing a polynomial sized expansion $\xi$ of $\gamma$ and then testing $\xi \not\subseteq \Delta_{App}$. In turn, $\xi \subseteq \Delta_{App}$ can be tested in NP by [Fig+20, Theorem 4.2].[13]

This yields a $\Pi_2^p$ algorithm for testing $\gamma \subseteq \Delta_{App}$, and thus also for $\Gamma \subseteq \Delta_{App}$. $\Pi_2^p$-hardness follows from Corollary IV.5.12. □

## IV.5  Lower Bounds

In this section we give some underlying ideas for showing lower bounds for the MINIMIZATION PROBLEMS.

### IV.5.1  Equivalence with a Single Atom

We shall use the hard instances of Proposition IV.2.4 to show that the MINIMIZATION PROBLEM for CRPQs is hard.

**Theorem IV.5.1.** The MINIMIZATION PROBLEM for CRPQs is ExpSpace-hard. Further, there is a fixed alphabet *s.t.* the problem of, given a Boolean CRPQ on this alphabet with only four variables, whether it is equivalent to a Boolean CRPQ with a single atom is ExpSpace-hard.

A formal proof follows the proof sketch.

*Proof sketch.* We reduce an instance of the problem of Proposition IV.2.4 to the instance $\delta$, where

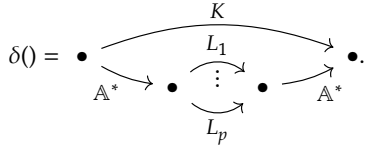

First, it is easy to see that if $\gamma_1 \subseteq \gamma_2$ then $\delta \equiv \gamma_1$. Conversely, if $\delta$ is equivalent to a Boolean CRPQ with at most one atom, then $\gamma_1 \subseteq \gamma_2$. The conditions imposed by Proposition IV.2.4 are necessary to discard one-atom queries which are self-loops and that whenever $\gamma_1 \not\subseteq \gamma_2$ there is an expansion of $\delta$ to which any $\delta$-equivalent single-atom query expansion cannot be mapped. □

*Proof details.* We use the reduction described in the proof sketch.

*Claim* IV.5.2. If $\gamma_1 \subseteq \gamma_2$ then $\delta \equiv \gamma_1$.

Note first that $\delta \subseteq \gamma_1$. Then, if $\gamma_1 \subseteq \gamma_2$, then every word of $K$ contains a factor which belongs to each $L_i$ for $i \in [\![1, p]\!]$, and hence $\gamma_1 \subseteq \delta$ *i.e.* $\delta \equiv \gamma_1$, and so $\delta$ is equivalent to a CRPQ with a single atom.

*Claim* IV.5.3. Conversely, if $\delta$ is equivalent to a Boolean CRPQ with at most one atom, then $\gamma_1 \subseteq \gamma_2$.

Let $\zeta$ be the Boolean CRPQ with at most one atom which is equivalent to $\delta$. Assume first, by contradiction, that it is a self-loop, *i.e.* $\zeta() = x \xrightarrow{M} x$ for some language $M$. Then by assumption on $K$, there exists a word $u \in K$ of size at least one. Since none of the $L_i$ are empty, there exists a canonical database $G_\delta^u$ where the atom $\bullet \xrightarrow{K} \bullet$ yielded a $u$-labelled path. Since $\delta \subseteq \zeta$, the database $G_\delta^u$ must satisfy $\zeta() = x \xrightarrow{M} x$. Since every strongly connected component of $G_\delta^u$ is trivial—we assumed that none of the languages of $\delta$ contained $\varepsilon$—, it must be that $\varepsilon \in M$, and hence $\zeta$ is the query which is always satisfied, which contradicts the equivalence $\delta \equiv \zeta$.

Similarly, it can be shown that $\zeta$ cannot have zero atoms since $\delta$ is non-trivial. Hence, $\zeta$ is exactly of the form $\zeta() = x \xrightarrow{M} y$ for some language $M$. First, note that from $\zeta \subseteq \delta$, it follows that

$$M \subseteq \mathbb{A}^*\Big(\bigcap_j L_j\Big)\mathbb{A}^*. \tag{IV.2}$$

Assume then, by contradiction, that there exists an $i$ *s.t.* every word of $L_i$ has a factor in $M$, *i.e.* $L_i \subseteq \mathbb{A}^* M \mathbb{A}^*$. Then (IV.2) implies $\mathbb{A}^* L_i \mathbb{A}^* = \mathbb{A}^*\big(\bigcap_j L_j\big)\mathbb{A}^*$ which contradicts the second assumption of Proposition IV.2.4. Therefore, for every $i$, there is a word $v_i \in L_i$ which contains no factor in $M$.

We are now ready to show that $\gamma_1 \subseteq \gamma_2$, by first observing that it boils down to showing $K \subseteq \mathbb{A}^*\big(\bigcap_j L_j\big)\mathbb{A}^*$. Let $u \in K$. Consider the following canonical database of $\delta$, where the $v_i$'s are words defined as in the paragraph above:

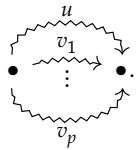

Since $\delta \subseteq \zeta$, it must contain a path labelled by a word of $M$. But no $v_i$ contains a factor in $M$, hence it has to be $u$ that does. Hence, $K \subseteq \mathbb{A}^* M \mathbb{A}^*$. Together with Equation (IV.2), we get $K \subseteq \mathbb{A}^*\big(\bigcap_j L_j\big)\mathbb{A}^*$, which concludes the proof of Claim IV.5.3.

Overall, we showed that $\gamma_1 \subseteq \gamma_2$ *iff* $\delta$ is equivalent to a CRPQ with at most one atom, which concludes the proof. $\square$

Note that the assumption of $\mathbb{A}^* L_i \mathbb{A}^* = \mathbb{A}^*\big(\bigcap_j L_j\big)\mathbb{A}^*$ in Proposition IV.2.4 in necessary for the reduction to be correct, otherwise $\delta()$ would be equivalent to

$$\delta'() \mathrel{\hat{=}} x \xrightarrow{K} y \wedge x \xrightarrow{\mathbb{A}^* L_i \mathbb{A}^*} y,$$

and so we could have $\mathbb{A}^* L_i \mathbb{A}^* \subsetneq K$, implying that (1) $K \not\subseteq \mathbb{A}^* L_i \mathbb{A}^*$ and hence $\gamma_1 \not\subseteq \gamma_2$ but (2) $\delta$ would be equivalent to a CRPQ with a single atom, namely $\delta''() \mathrel{\hat{=}} x \xrightarrow{\mathbb{A}^* L_i \mathbb{A}^*} y$.

### IV.5.2   Minimization is Harder than Containment

We show that, under some technical conditions, the containment problem can be reduced to the MINIMIZATION PROBLEM. This allows to transfer known lower bounds from the CONTAINMENT PROBLEM of CRPQ classes. Due to space constraints we only state the key definitions and lemmas. We first introduce an intermediary technical property called "canonization", which ensures the feasibility of the reduction.

*Canonization.* We say that an expansion $\xi$ of a CRPQ $\gamma$ is *non-degenerate* if no atom refinement in $\xi$ was obtained using the empty word.

**Fact IV.5.4.** *If $\xi$ is a non-degenerate expansion of $\gamma$ and $\gamma$ is fully contracted, then $\|\xi\|_{\mathrm{seg}} = \|\gamma\|_{\mathrm{at}}$.*

Given a *class of CRPQs $\mathcal{Q}$*, the *$\mathcal{Q}$-canonization problem* is the functional problem taking an alphabet $\mathbb{A}$ and two Boolean CRPQs $\langle \gamma_1, \gamma_2 \rangle$ in $\mathcal{Q}_{\mathbb{A}}$, and outputting an alphabet $\mathbb{M}$, two other Boolean CRPQs $\langle \delta_1, \delta_2 \rangle$, in $\mathcal{Q}_{\mathbb{A} \sqcup \mathbb{M}}$, such that:

(Cnz)$_{\mathrm{monotonic}}$: $\gamma_1 \subseteq \gamma_2$ iff $\delta_1 \subseteq \delta_2$,

(Cnz)$_{\mathrm{str\text{-}onto}}$: there exists a non-degenerate $D_1 \in \mathrm{Exp}(\delta_1)$ *s.t.*, for every $D_1' \in \mathrm{Exp}(\delta_1)$ and $f : D_1' \xrightarrow{\mathrm{hom}} D_1$ we have (i) $D_1'$ is non-degenerate, (ii) $f$ is strong onto, and (iii) $f(x) = x$ for every $x \in \mathrm{vars}(\delta_1)$,

(Cnz)$_{\mathrm{non\text{-}red}}$: for each $D_1 \in \mathrm{Exp}(\delta_1)$, for each $x, y \in \mathrm{vars}(\delta_1)$, there cannot be an atom refinement in $D_1$ from $x$ to $y$ and another path from $x$ to $y$ in $D_1$, disjoint from the atom refinement that share the same label,

(Cnz)$_{\mathrm{contracted}}$: $\delta_1$ is fully contracted,

(Cnz)$_{\mathrm{containment}}$: $\gamma_2 \subseteq \delta_2$, and

(Cnz)$_{\mathrm{marking}}$: each connected component of $\delta_1$ must contain at least one atom labelled by a language $L$ *s.t.* every word of $L$ must contain at least one letter from $\mathbb{M}$.

The *$\mathcal{Q}$-strong canonization problem* is defined similarly to the $\mathcal{Q}$-canonization problem, except that (Cnz)$_{\mathrm{str\text{-}onto}}$ is replaced by the axiom

(SCnz)$_{\mathrm{str\text{-}onto}}$: for every non-degenerate $D_1 \in \mathrm{Exp}(\delta_1)$, every $D_1' \in \mathrm{Exp}(\delta_1)$, and $f : D_1' \xrightarrow{\mathrm{hom}} D_1$ we have (i) $D_1'$ is non-degenerate, (ii) $f$ is strong onto, and (iii) $f(x) = x$ for every $x \in \mathrm{vars}(\delta_1)$,

We show that assuming we can solve the $\mathcal{Q}$-canonization problem (*resp.* $\mathcal{Q}$-strong canonization problem), then the CRPQ (resp. UCRPQ) MINIMIZATION PROBLEM restricted to CRPQs of $\mathcal{Q}$ (*resp.* to UCRPQs whose disjuncts are all in $\mathcal{Q}$) is harder than the CONTAINMENT PROBLEM over $\mathcal{Q}$.

In the following statement, a *$\mathcal{Q}$-canonization oracle* (resp. *$\mathcal{Q}$-strong canonization oracle*) is an oracle to any algorithm solving the $\mathcal{Q}$-canonization problem (resp. $\mathcal{Q}$-strong canonization problem).

**Lemma IV.5.5.** *For any class $\mathcal{Q}$ of CRPQs closed under disjoint conjunction, there is a polynomial-time reduction, using a $\mathcal{Q}$-canonization oracle (*resp.* $\mathcal{Q}$-strong canonization oracle), from the* CONTAINMENT PROBLEM *for Boolean queries of $\mathcal{Q}$ to the CRPQ (resp. UCRPQ)* MINIMIZATION PROBLEM *restricted to CRPQs in $\mathcal{Q}$ (resp. UCRPQs whose disjuncts are all in $\mathcal{Q}$). The reduction also applies under the restriction that the target query must also be in $\mathcal{Q}$.*

Once again, we start by giving a proof sketch before giving the full proof.

*Proof sketch.* The construction reduces some restriction of the CONTAINMENT PROBLEM to some variant of the MINIMIZATION PROBLEM. The main idea is, given an instance $\gamma_1 \subseteq^? \gamma_2$, to build CRPQs $\delta_1$ and $\delta_2$ with some desirable properties *s.t.* the following are equivalent: (i) $\gamma_1 \sqsubseteq \gamma_2$, (ii) $\delta_1 \sqsubseteq \delta_2$, (iii) $\delta_1 \oslash \delta_2 \equiv \delta_1$, where $\oslash$ denotes the *disjoint conjunction* (*i.e.*, the conjunction of atoms of both queries making sure that variables are disjoint), and (iv) $\delta_1 \oslash \delta_2$ is equivalent to a CRPQ whose size is at most the size of $\delta_1$. Of course, (ii) $\Leftrightarrow$ (iii) always holds, as well as (iii) $\Rightarrow$ (iv).

All the difficulty lies in guaranteeing the converse property: (iv) $\Rightarrow$ (iii). We will use the constructions $\gamma_i \mapsto \delta_i$, given by the canonization problem, to enforce it while respecting (i) $\Leftrightarrow$ (ii). The main idea of this approach is to add some 'marking' (with fresh alphabet symbols) of either the variables or the atoms of $\gamma_1$ in $\delta_1$, ensuring that $\delta_1$ has some strong structure implying that—loosely speaking—any query equivalent to $\delta_1 \oslash \delta_2$ must contain $\delta_1$ as a subquery. Using the assumption that $\delta_1 \oslash \delta_2$ is equivalent to a CRPQ whose size is at most the size of $\delta_1$, we conclude that in fact $\delta_1 \oslash \delta_2 \equiv \delta_1$, *i.e.* $\delta_1 \sqsubseteq \delta_2$ and so $\gamma_1 \sqsubseteq \gamma_2$. $\qquad\square$

*Proof.* ✎ *Minimization in the class of CRPQs.* Let $\gamma_1 \sqsubseteq^? \gamma_2$ be an instance of the CONTAINMENT PROBLEM for Boolean queries of $\mathcal{Q}$. We apply the $\mathcal{Q}$-canonization oracle to obtain a pair $\langle \delta_1, \delta_2 \rangle$ as in the axioms $(\text{Cnz})_*$. We then map the instance $\gamma_1 \sqsubseteq^? \gamma_2$ to $\langle \delta_1 \oslash \gamma_2, \|\delta_1\|_{\text{at}} \rangle$.

The reduction works in logarithmic space with a $\mathcal{Q}$-canonization oracle, and clearly $\delta_1 \oslash \delta_2 \in \mathcal{Q}$ since both $\delta_1$ and $\delta_2$ are in $\mathcal{Q}$ and $\mathcal{Q}$ is closed under disjoint conjunction. We need to show that $\gamma_1 \sqsubseteq \gamma_2$ *iff* $\delta_1 \oslash \gamma_2$ is equivalent to a CRPQ with at most $\|\delta_1\|_{\text{at}}$ atoms.

*Claim* IV.5.6. If $\gamma_1 \sqsubseteq \gamma_2$, then $\delta_1 \oslash \delta_2 \equiv \delta_1$ and so $\delta_1 \oslash \delta_2$ is equivalent to a CRPQ with at most $\|\delta_1\|_{\text{at}}$ atoms.

This follows from $(\text{Cnz})_{\text{monotonic}}$. The converse hold for the same reason, but we actually need a stronger property.

*Claim* IV.5.7. If $\delta_1 \oslash \delta_2$ is equivalent to a CRPQ with at most $\|\delta_1\|_{\text{at}}$ atoms, then $\gamma_1 \sqsubseteq \gamma_2$.

We write $\zeta$ as $\zeta_+ \oslash \zeta_-$ where $\zeta_+$ is the disjoint conjunction of all connected components of $\zeta$ containing an atom whose language contains a word containing a '$\mathbb{M}$'-letter, and $\zeta_-$ is the disjoint conjunction of all other components. We want to show that $\zeta_-$ is actually empty.

Let $D_1$ be a canonical database of $\delta_1$ as in $(\text{Cnz})_{\text{str-onto}}$. Then pick any canonical database $G_2$ of $\gamma_2$. By $(\text{Cnz})_{\text{containment}}$, there exists $D_2 \vDash^\star \delta_2$ *s.t.* $D_2 \xrightarrow{\text{hom}} G_2$. Then $D_1 \oplus D_2 \vDash^\star \delta_1 \oslash \delta_2$, so from $\delta_1 \oslash \delta_2 \equiv \zeta$ it follows that there exists $Z_+ \vDash^\star \zeta_+$, $Z_- \vDash^\star \zeta_-$, $D_1' \vDash^\star \delta_1$ and $D_2' \vDash^\star \delta_2$ such that we have homomorphisms

$$D_1' \oplus D_2' \xrightarrow{f} Z_+ \oplus Z_- \xrightarrow{g} D_1 \oplus D_2 \xrightarrow{h} D_1 \oplus G_2.$$

By $(\text{Cnz})_{\text{marking}}$, every connected component of $D_1'$ must contain at least one edge labelled by a letter of $\mathbb{M}$, and so the homomorphism $f_{\restriction D_1'} : D_1' \xrightarrow{\text{hom}} Z_+ \oplus Z_-$ is actually a homomorphism from $D_1'$ to $Z_+$. Note then that $(h \circ g)_{\restriction Z_+}$ maps $Z_+$ onto $D_1 \oplus G_2$ but since $G_2$ contains no letter $\mathbb{M}$, the image of this homomorphism is included in $D_1$. Overall, we have homomorphisms

$$D_1' \xrightarrow{i} Z_+ \xrightarrow{j} D_1.$$

By (Cnz)$_{\text{str-onto}}$, $j \circ i$ be strong onto and for every $x \in \text{vars}(\delta_1)$, $j(i(x)) = x$.

We claim that for each $x$ external in $D_1'$, then $i(x)$ is external. First, since $x \in \text{vars}(\delta_1)$ is external, then $x \in \text{vars}(\delta_1)$ and so $j(i(x)) = x$. Then $j \circ i$ is strong onto and so $j$ be also be strong onto. It follows that the in-degree (resp. out-degree) of $i(x)$ is lower bounded by the in-degree (resp. out-degree) $j(i(x)) = x$. So, if $x$ has in-degree or out-degree at least 2, so does $i(x)$. Moreover, if $x$ has in-degree (resp. out-degree 0), then so must $i(x)$ because otherwise, $j(i(x)) = x$ should also have an incoming edge. Overall, by letting $i[D_1']$ be the image of $D_1'$ by $i$, we get that the natural embedding $i[D_1'] \xrightarrow{\text{hom}} Z_+$ satisfies the assumption of Corollary IV.2.13 and so $\|i[D_1']\|_{\text{seg}} \leq \|Z_+\|_{\text{seg}}$.

Now observe that $i \colon D_1' \to i[D_1']$ is injective on $\text{vars}(\delta_1)$ because $j(i(x))$ for any $x \in \text{vars}(\delta_1)$. Moreover, by (Cnz)$_{\text{non-red}}$, $i$ cannot identity an atom with another path of atoms and so $D_1'$ is actually isomorphic to $i[D_1']$, from which we get $\|i[D_1']\|_{\text{seg}} = \|\delta_1\|_{\text{at}}$ and so $\|Z_+\|_{\text{seg}} \geq \|\delta_1\|_{\text{at}}$. By Proposition IV.2.8 $\|\zeta_+\|_{\text{at}} \geq \|Z_+\|_{\text{seg}}$, and so $\zeta_+$ has at least $\|\delta_1\|_{\text{at}}$ atoms, but since we assumed that $\zeta$ has at most $\|\delta_1\|_{\text{at}}$ atoms, it follows that $\zeta_-$ is trivial and $\zeta \equiv \zeta_+$.

We are now ready to prove that $\delta_1 \subseteq \delta_2$. Let $D_1 \vDash^\star \delta_1$. Pick any $G_2 \vDash^\star \gamma_2$. By (Cnz)$_{\text{containment}}$, there exists $D_2 \vDash^\star \delta_2$ s.t. $D_2 \xrightarrow{\text{hom}} G_2$. Then since $\delta_1 \oslash \delta_2 \equiv \zeta_+$, there exists $Z_+ \vDash^\star \zeta_+$, $D_1' \vDash^\star \delta_1$ and $D_2' \vDash^\star \delta_2$ s.t.

$$D_1' \oplus D_2' \xrightarrow{\text{hom}} Z_+ \xrightarrow{\text{hom}} D_1 \oplus D_2 \xrightarrow{\text{hom}} D_1 \oplus G_2.$$

Because of $\mathbb{M}$, the homomorphism $Z_+ \xrightarrow{\text{hom}} D_1 \oplus G_2$ must in fact be a homomorphism $Z_+ \xrightarrow{\text{hom}} D_1$, and so by composition we obtain a homomorphism $D_1' \oplus D_2' \xrightarrow{\text{hom}} D_1$, which can be restricted to $D_2'$, yielding $D_2' \xrightarrow{\text{hom}} D_1$. Hence, $\delta_1 \subseteq \delta_2$, and so by (Cnz)$_{\text{monotonic}}$, $\gamma_1 \subseteq \gamma_2$.

Putting Claims IV.5.6 and IV.5.7 together shows that the reduction is correct. We now prove that this reduction also works for the other variations of the problem.

❧ *Minimization in the class of UCRPQs.* If we allow $\zeta$ to be a UCRPQ, then Claim IV.5.6 still holds, and we need to adapt Claim IV.5.7. Assume that this "small" UCRPQ is of the form $\zeta_1 \lor \zeta_2 \lor \cdots \lor \zeta_k$ where each $\zeta_i$ has at most $\|\delta_1\|_{\text{at}}$ atoms. We say that a disjunct $\zeta_i$ is *relevant* when it has at least one canonical database $Z_i$ appearing in a pattern of the form

$$D_1' \oplus D_2' \xrightarrow{\text{hom}} Z_i \xrightarrow{\text{hom}} D_1 \oplus D_2 \xrightarrow{\text{hom}} D_1 \oplus G_2.$$

for some $D_1, D_1' \vDash^\star \delta_1$, $D_2, D_2' \vDash^\star \delta_2$ and $G_2 \vDash^\star \gamma_2$. Using (SCnz)$_{\text{str-onto}}$ on $D_1$, the same proof as in the case of CRPQs apply. We can then conclude that *w.l.o.g.* $\zeta_i \equiv (\zeta_i)_+$ for all relevant disjunct. The proof of $\delta_1 \subseteq \delta_2$—and hence $\gamma_1 \subseteq \gamma_2$—then goes through as before, which concludes the proof.

    ↬ *If $\zeta$ is restricted to be in $\mathcal{Q}$.* Then Claim IV.5.7 still holds. To adapt Claim IV.5.6, it suffices to remark that $\delta_1 \in \mathcal{Q}$. $\qquad\square$

Motivated by Lemma IV.5.5 we show that several reasonable classes admit a polynomial-time algorithm for the strong canonization problem—see Lemmas IV.5.8 and IV.5.11.

**Lemma IV.5.8.** The strong canonization problem can be solved in non-deterministic logarithmic space for the class of all CRPQs, or more generally for all classes of CRPQs defined by restricting the underlying multigraph class, provided that this class is closed under disjoint union.

*Proof.* Given a pair $\langle \gamma_1, \gamma_2 \rangle$ of Boolean queries we assume *w.l.o.g.* that $\gamma_1$ is fully contracted using Fact IV.2.2—which works in non-deterministic logarithmic space—, and we reduce it to the pair $\langle \delta_1, \delta_2 \rangle$, where $\delta_1$ is defined as

$$\delta_1() \mathrel{\hat=} \bigwedge_{\alpha = x \xrightarrow{L} y \in \gamma_1} x \xrightarrow{\triangleright_\alpha L \triangleleft_\alpha} y$$

where $\triangleright_\alpha$ and $\triangleleft_\alpha$ are fresh letters for each atom $\alpha$ of $\gamma_1$, and

$$\delta_2() \mathrel{\hat=} \bigwedge_{x \xrightarrow{L} y \in \gamma_2} x \xrightarrow{\phi_{-\mathbb{M}}^{-1}[L]} y,$$

where $\alpha_1, \dots, \alpha_k$ are all the atoms of $\gamma_1$, $\mathbb{M} \mathrel{\hat=} \{\triangleright_{\alpha_1}, \triangleleft_{\alpha_1}, \dots, \triangleright_{\alpha_k}, \triangleleft_{\alpha_k}\}$ and $\phi_{-\mathbb{M}} \colon (\mathbb{A} \cup \mathbb{M})^* \to \mathbb{A}^*$ is the monoid morphism that maps letters of $\mathbb{A}$ to themselves and letters of $\mathbb{M}$ to the empty word.

In other words, $\delta_1$ is similar to $\gamma_1$ except that it must read the special symbol $\triangleright_\alpha$ before satisfying atom $\alpha \in \gamma_1$, and read symbol $\triangleleft_\alpha$ after. On other hand, $\delta_2$ is obtained from $\gamma_2$ by relaxing the constraints: instead of having to read a path labelled by some language $L$, we now must read a path such that, when we ignore these new symbols $\triangleright_\alpha$ and $\triangleleft_\alpha$, then it belongs to $L$.

We now need to prove that properties (Cnz)∗ and (SCnz)str-onto hold.

*Claim* IV.5.9. If $\gamma_1 \subseteq \gamma_2$ then $\delta_1 \oslash \delta_2 \equiv \delta_1$.

Showing that $\delta_1 \oslash \delta_2 \equiv \delta_1$ amounts to showing $\delta_1 \subseteq \delta_2$. Let $D_1$ be a canonical database of $\delta_1$. Consider the canonical database of $\gamma_1$ obtained by removing every edge of the form $x \xrightarrow{\triangleright_\alpha} y$ or $x \xrightarrow{\triangleleft_\alpha} y$, and merging variables $x$ and $y$. Since $\gamma_1 \subseteq \gamma_2$, there exists a canonical database $G_2$ of $\gamma_2$ and a homomorphism $f \colon G_2 \to G_1$. We then define a canonical database $D_2$ of $\delta_2$ together with a homomorphism $g \colon D_2 \to D_1$ as follows: given an atom refinement

$$x_0 \xrightarrow{b_1} x_1 \xrightarrow{b_2} \cdots \xrightarrow{b_n} x_n \text{ in } G_2.$$

we look at its image

$$f(x_0) \xrightarrow{b_1} f(x_1) \xrightarrow{b_2} \cdots \xrightarrow{b_n} f(x_n) \text{ in } G_1.$$

Now some $f(x_i)$'s might be variables of $\gamma_1$ and hence this might not a path in $G_1$. We let $i_1 < \ldots < i_k$ denote the indices $i$ s.t. $x_i \in \gamma_1$, so that we can split the path in $G_1$ into multiples atom refinements of atoms of $\gamma_1$:

$$\underbrace{f(x_0) \xrightarrow{b_1} f(x_1) \xrightarrow{b_2} \cdots \xrightarrow{b_{i_1}} f(x_{i_1})}_{\text{end of an atom refinement of } \alpha_0} \underbrace{\xrightarrow{b_{i_1+1}} f(x_{i_1+1}) \xrightarrow{b_{i_1+2}} \cdots \xrightarrow{b_{i_2}} f(x_{i_2})}_{\text{atom refinement of } \alpha_1}$$

$$\xrightarrow{b_{i_2+1}} f(x_{i_2+1}) \xrightarrow{b_{i_2+2}} \cdots \xrightarrow{b_{i_k}} f(x_{i_k})$$

$$\underbrace{\xrightarrow{b_{i_k+1}} f(x_{i_k+1}) \xrightarrow{b_{i_k+2}} \cdots \xrightarrow{b_n} f(x_n)}_{\text{beginning of an atom refinement of } \alpha_k}$$

in $G_1$. For $i \in \llbracket 1, k \rrbracket$, we let $f(x_{i_j})^r$ (resp. $f(x_{i_j})^l$) denote the unique variable of $D_1$ s.t. there is an edge from $f(x_{i_j})$ to $f(x_{i_j})^r$ labelled by $\rhd_{\alpha_i}$ (resp. from $f(x_{i_j})^l$ to $f(x_{i_j})$ labelled by $\lhd_{\alpha_{i-1}}$), we obtain a path

$$\underbrace{f(x_0) \xrightarrow{b_1} f(x_1) \xrightarrow{b_2} \cdots \xrightarrow{b_{i_1}} f(x_{i_1})^l \xrightarrow{\lhd_{\alpha_0}} f(x_{i_1}) \xrightarrow{\rhd_{\alpha_1}} f(x_{i_1})^r}_{\text{end of an atom refinement of } \alpha_0}$$

$$\underbrace{\xrightarrow{b_{i_1+1}} f(x_{i_1+1}) \xrightarrow{b_{i_1+2}} \cdots \xrightarrow{b_{i_2}} f(x_{i_2})^l \xrightarrow{\lhd_{\alpha_1}} f(x_{i_2}) \xrightarrow{\rhd_{\alpha_1}} f(x_{i_2})^r}_{\text{atom refinement of } \alpha_1}$$

$$\xrightarrow{b_{i_2+1}} f(x_{i_2+1}) \xrightarrow{b_{i_2+2}} \cdots \xrightarrow{b_{i_k}} f(x_{i_k})^l \xrightarrow{\lhd_{\alpha_{k-1}}} f(x_{i_k}) \xrightarrow{\rhd_{\alpha_k}} f(x_{i_k})^r$$

$$\underbrace{\xrightarrow{b_{i_k+1}} f(x_{i_k+1}) \xrightarrow{b_{i_k+2}} \cdots \xrightarrow{b_n} f(x_n)}_{\text{beginning of an atom refinement of } \alpha_k}$$

in $D_1$. Hence, we build $D_2$ by replacing each atom refinement $x_0 \xrightarrow{b_1} x_1 \xrightarrow{b_2} \cdots \xrightarrow{b_n} x_n$ in $G_2$ by

$$x_0 \xrightarrow{b_1} x_1 \xrightarrow{b_2} \cdots \xrightarrow{b_{i_1}} x_{i_1}^l \xrightarrow{\lhd_{\alpha_0}} x_{i_1} \xrightarrow{\rhd_{\alpha_1}} x_{i_1}' \xrightarrow{b_{i_1+1}} x_{i_1+1} \xrightarrow{b_{i_1+2}} \cdots$$
$$\cdots \xrightarrow{b_{i_2}} x_{i_2}^l \xrightarrow{\lhd_{\alpha_1}} x_{i_2} \xrightarrow{\rhd_{\alpha_2}} x_{i_2}' \xrightarrow{b_{i_2+1}} x_{i_2+1} \xrightarrow{b_{i_2+2}} \cdots$$
$$\cdots \xrightarrow{b_{i_k}} x_{i_k}^l \xrightarrow{\lhd_{\alpha_{k-1}}} x_{i_k} \xrightarrow{\rhd_{\alpha_k}} x_{i_k}' \xrightarrow{b_{i_k+1}} x_{i_k+1} \xrightarrow{b_{i_k+2}} \cdots \xrightarrow{b_n} x_n,$$

where $x_{i_1}^l, x_{i_1}^r, \ldots, x_{i_k}^l, x_{i_k}^r$ are new fresh variables. By construction, $D_2$ comes equipped with a homomorphism $g \colon D_2 \to D_1$ which sends $x_i$ to $f(x_i)$, $x_{i_j}^l$ to $f(x_{i_j})^l$ and $x_{i_j}^r$ to $f(x_{i_j})^r$. Since $D_2$ is—by construction—a canonical database of $\delta_2$, this concludes the proof that $\delta_1 \subseteq \delta_2$.

*Claim* IV.5.10. If $\delta_1 \subseteq \delta_2$ then $\gamma_1 \subseteq \gamma_2$.

The construction is dual to Claim IV.5.9 and left to the reader. Both claims yield (Cnz)$_{\text{monotonic}}$.

We now show that (SCnz)$_{\text{str-onto}}$ holds: pick a canonical database $D_1 \vDash^\star \delta_1$. For any $D_1' \vDash^\star \delta_1$, if $D_1' \xrightarrow{\text{hom}} D_1$, then because of the letters in $\mathbb{M}$, it follows that for each atom $\alpha_i$ of $\gamma_1$, the atom refinement of $\alpha_i$ in $D_1'$ must be sent bijectively on the atom refinement of $\alpha_i$ in $D_1$, and so they are equal. It follows that the homomorphism $D_1' \to D_1$ must actually be the identity, and hence $D_1$ is hom-minimal. The same argument applied to $D_1' \triangleq D_1$ shows that the only homomorphism from $D_1$ itself is the identity, and so in particular $D_1'$ is a core. Lastly, because of the letters of $\mathbb{M}$, no atom of $\delta_1$ contains the empty word, and so in particular $D_1$ must be non-degenerate. Hence, $(\text{SCnz})_{\text{str-onto}}$ holds.

Since $\gamma_1$ is fully contracted, so is $\delta_1$, which proves $(\text{Cnz})_{\text{contracted}}$. For any language $L$, we have $L \subseteq \phi_{-\mathbb{M}}^{-1}[L]$, and so we have $(\text{Cnz})_{\text{containment}}$. Finally, $(\text{Cnz})_{\text{non-red}}$ and $(\text{Cnz})_{\text{marking}}$ trivially hold.

Together with that fact that $\langle \gamma_1, \gamma_2 \rangle \mapsto \langle \delta_1, \delta_2 \rangle$ preserves the underlying multigraphs, this shows that this is a solution to strong canonization problem for any class of CRPQs defined by restring the underlying class of multigraphs. Note also that an NFA for $\phi_{-\mathbb{M}}^{-1}[L]$ can be obtained from an NFA for $L$ by adding on every state a self-loop labelled by every possible letter of $\mathbb{M}$ and hence, this algorithm can be implemented in logarithmic space. □

Note however that if $L$ is a simple regular expression, then $\phi_{-\mathbb{M}}^{-1}[L]$ does not need to be. Hence, the construction above does not work for CRPQs over positive simple regular expressions.

**Lemma IV.5.11.** The strong canonization problem can be solved in polynomial time for the class of CRPQs over positive simple regular expressions.

Given a CRPQ $\gamma$, we say that an atom $x \xrightarrow{L} y$ is *locally redundant* if there exists a path of atoms $z_0 \xrightarrow{L_1} z_1 \xrightarrow{L_2} \cdots \xrightarrow{L_n} z_n$ in which $x \xrightarrow{L} y$ does not occur, and with $z_0 = x$ and $z_n = x$ where $L_1 L_2 \cdots L_n \subseteq L$.

*Proof of Lemma IV.5.11.* Fix a pair $\langle \gamma_1, \gamma_2 \rangle$ of CRPQs. From $\gamma_1$, we start by picking a locally redundant atom (if any), and remove it. We iterate this process, until we get a CRPQ with no locally redundant atom $\gamma_1'$. By construction, it is equivalent to $\gamma_1$.[14] Moreover, $\gamma_1'$ can be computed in polynomial time. We then refine in $\gamma_1'$ each atom so that each atom is either labelled by $a$ or $a^+$ for some $a \in \mathbb{A}$.

We then define $\langle \delta_1, \delta_2 \rangle$, where $\delta_2 \triangleq \gamma_2$ and

$$\delta_1 = \left( \bigwedge_{x \xrightarrow{L} y \in \gamma_1'} x \xrightarrow{L} y \right) \wedge \left( \bigwedge_{x \in \gamma_1'} x \xrightarrow{\triangledown_x} x \right),$$

and we let $\mathbb{M} \triangleq \{ \triangledown_x \mid x \in \gamma_1' \}$.

Next, we show that $(\text{Cnz})_{\text{monotonic}}$ holds: if $\gamma_1 \subseteq \gamma_2$ then $\delta_1 \subseteq \gamma_1' \equiv \gamma_1 \subseteq \gamma_2 = \delta_2$, and dually if $\delta_1 \subseteq \delta_2$ then let $G_1 \vDash^\star \gamma_1'$, and let $D_1$ be the associated canonical database. Since $\delta_1 \subseteq \delta_2$, there exists $D_2 \vDash^\star \delta_2$ s.t. $D_2 \xrightarrow{\text{hom}} D_1$ but since $D_2$ contains no letter from $\mathbb{M}$, we actually get a homomorphism $D_2 \xrightarrow{\text{hom}} G_1$, and so $\gamma_1' \subseteq \delta_2$ i.e. $\gamma_1 \subseteq \gamma_2$.

For $(\text{SCnz})_{\text{str-onto}}$, by definition of positive simple regular expressions, no language labelling an atom of $\delta_1$ contains the empty word, and hence every canonical database of $\delta_1$ is non-degenerate. Then, let $f \colon D_1' \to D_1$ be a homomorphism between canonical databases of $\delta_1$. Because of the letters of $\mathbb{M}$, $f$ must send $x \in \delta_1 \subseteq D_1$ onto $x \in D_1'$. We then claim that $f$ is strong onto. Let $\alpha \triangleq x \xrightarrow{L} y$ be an atom of $\delta_1$. We consider its atom refinement in $D_1$, and we want to show that it is *included in* the image of the atom refinement of $\alpha$ in $D_1'$:

- if $L = \{\triangledown_x\}$, this is trivial;
- if $L = \{a\}$ for some letter $a$, then since $\alpha$ is not locally redundant in $\gamma_1'$, there are no other $a$-edge from $x$ to $y$ in $D_1$ (or $D_1'$), and so the unique $a$-edge from $x$ to $y$ in $D_1'$ must be sent on the unique $a$-edge from $x$ to $y$ in $D_1$;
- if $L = a^+$ for some letter $a$, then the atom refinement of $\alpha$ in $D_1'$, say $x \xrightarrow{a^k} y$ ($k \geq 1$) is sent via $f$ on a path from $x$ to $y$ in $D_1$. If the atom refinement of $\alpha$ in $D_1$ is included in this path, we are done; otherwise, when lifting this path to $\delta_1$, we would obtain a path of atoms $x \xrightarrow{L_1} \cdots \xrightarrow{L_n} y$ s.t. $a^k \in L_1 \cdots L_n$. By definition of positive simple regular expressions, all $L_i$'s must be either $a$ or $a^+$, and hence in all cases $L_1 \cdots L_n \subseteq a^+$, contradicting that $\alpha$ is not locally redundant in $\gamma_1'$.

Thus, we have $(\text{SCnz})_{\text{str-onto}}$.

Similarly, $(\text{Cnz})_{\text{non-red}}$ holds because all atoms of $\gamma_1'$ are labelled by $a$ or $a^+$ and we removed locally redundant atoms. Thanks to the self-loops, $\delta_1$ is fully contracted and so $(\text{Cnz})_{\text{contracted}}$ holds. Moreover, $(\text{Cnz})_{\text{containment}}$ holds trivially since $\gamma_2 = \delta_2$, and so does $(\text{Cnz})_{\text{marking}}$ by definition of $\delta_1$. $\qquad \square$

**Corollary IV.5.12.** The CRPQ and UCRPQ MINIMIZATION PROBLEMS are:

1. ExpSpace-hard, even if restricted to queries of path-width at most 1,
2. PSpace-hard when restricted to forest-shaped CRPQs, or even a single atom,
3. $\Pi_2^p$-hard when restricted to CRPQs over positive simple regular expressions.

All hardness results are under polynomial-time reductions.

*Proof.* From Lemmas IV.5.5, IV.5.8 and IV.5.11 we can derive the stated hardness results when combined with known hardness results for the CONTAINMENT PROBLEM: Item 1 follows from the ExpSpace lower bound of [Fig20, Lemma 8] (or its strengthening Proposition IV.2.4). Item 2 follows from the trivial PSpace lower bound from regular language containment which is also the lower bound for one-atom CRPQs. Item 3 follows from the known $\Pi_2^p$-lower bound for CRPQ(SRE) queries implied by [Fig+20, Theorem 4.2]. $\qquad \square$

## IV.6  Discussion

Several open problems are left by our work, more prominently, the complexity gap for CRPQ MINIMIZATION. Below we discuss further avenues for future research.

### *IV.6.1   Variable minimization*

Another classical algorithm for query answering for (U)CRPQs consists in, given a query $\gamma$ and a graph database $G$, to first guess a variable assignment $f : \text{vars}(\gamma) \to G$ and then check, for each atom $x \xrightarrow{L} y$ of $\gamma$, that there is a path in $G$ from $f(x)$ to $f(y)$ with label in $L$.

This implementation approach privileges minimizing the number of *variables* as opposed to the number of *atoms* of a (U)CRPQ, and gives rise to the corresponding VARIABLE-MINIMIZATION PROBLEM(S). From a practical perspective, systems commonly evaluate CRPQs via join algorithms. Recent *worst-case optimal* joins algorithms work by ordering the variables and assigning potential values to these, and hence the number of variables may also be a relevant parameter in these cases [CRV23; Vrg+24].

---

| | |
|---|---|
| VARIABLE-MINIMIZATION PROBLEM for CRPQs (*resp.* for UCRPQ) | |
| *Input*: | A finite alphabet $\mathbb{A}$, a CRPQ (*resp.* UCRPQ) $\gamma$ over $\mathbb{A}$ and $k \in \mathbb{N}$. |
| *Question*: | Does there exist a CRPQ (*resp.* UCRPQ) $\delta$ over $\mathbb{A}$ with at most $k$ variables (*resp.* whose every CRPQ has at most $k$ variables) *s.t.* $\gamma \equiv \delta$? |

---

As before, a (U)CRPQ is *variable minimal* if there is no equivalent (U)CRPQ smaller in the number of variables.

It is worth observing that for conjunctive queries (and for tree patterns) minimizing the number of variables or minimizing the number of atoms is equivalent: a query is minimal in the number of variables *iff* it is minimal in the number of atoms. However, for CRPQs and UCRPQs it is not: $\gamma(x, y) = x \xrightarrow{a} y \wedge x \xrightarrow{a+b} y$ is variable minimal, but it is not (atom) minimal since it is equivalent to $\gamma'(x, y) = x \xrightarrow{a} y$. We further conjecture that there are (atom) minimal CRPQs which are not variable minimal.

**Conjecture IV.6.1.** There exist (atom) minimal CRPQs which are not variable minimal.

By adaptations of the algorithms of Sections IV.3 and IV.4 we can derive some upper bounds, which are likely to be sub-optimal.

**Theorem IV.6.2.** The VARIABLE-MINIMIZATION PROBLEM for CRPQs is in 4ExpSpace and for UCRPQs in 2ExpSpace. Both problems are ExpSpace-hard.

*Proof.* For the case of the CRPQ VARIABLE-MINIMIZATION PROBLEM, it suffices to observe, in the proof of Lemma IV.3.2, that since each NFA of the proof has size double-exponential, there there cannot be more than a triply-exponential number of distinct NFAs, and hence that the underlying multigraph of queries to be considered has a triply-exponential number of edges. Thus, there are 'only' a quadruply-exponential number of such triply-exponential queries. For each such query we test if it is equivalent to the original query in 4ExpSpace.

For the case of the UCRPQ VARIABLE-MINIMIZATION PROBLEM, let $\mathcal{C}_k$ be the (infinite) set of multigraphs having at most $k$ vertices. Via the same argument as in the proof of Lemma IV.4.6, we obtain that each $\text{CRPQ}(\mathcal{C}_k)$ in the union of CRPQs expressing $\text{App}_{\mathcal{C}_k}^{\infty}(\Gamma) = \text{App}_{\mathcal{C}_k}^{\leq \mathcal{O}(\|\Gamma\|_{\text{at}} \cdot r_\Gamma \|\mathcal{C}\|_{\text{at}})}(\Gamma)$ has atoms consisting in a concatenation of at most $\mathcal{O}(\|\Gamma\|_{\text{at}} \cdot \|\mathcal{C}\|_{\text{at}})$ sublanguages of $\Gamma$. Hence, there cannot be more than $2^{\mathcal{O}(\|\Gamma\|_{\text{at}} \cdot \|\mathcal{C}\|_{\text{at}})}$ distinct atoms between two variables, and $\text{App}_{\mathcal{C}_k}^{\leq \mathcal{O}(\|\gamma\|_{\text{at}} \cdot r_\Gamma \cdot \|\mathcal{C}\|_{\text{at}})}(\Gamma) \equiv \text{App}_{\mathcal{C}'_k}^{\leq \mathcal{O}(\|\Gamma\|_{\text{at}} \cdot r_\Gamma \cdot \|\mathcal{C}\|_{\text{at}})}(\Gamma)$ for $\mathcal{C}'_k$ being the finite subclass of $\mathcal{C}_k$ having graphs with no more than $2^{\mathcal{O}(\|\Gamma\|_{\text{at}} \cdot \|\mathcal{C}\|_{\text{at}})}$ parallel edges. Hence, there is a double-exponential number of exponential queries to test for equivalence with $\Gamma$, which yields a 2ExpSpace upper bound.

The lower bounds follow by a similar idea to Theorem IV.5.1 and can be found in Section IV.A. □

### IV.6.2 Tree patterns

We believe that the techniques of Section IV.4 should also yield a method to compute maximal under-approximations for unions of tree patterns, as well as a $\Pi_2^p$ upper bound for the minimization problem of unions of tree patterns, contrasting with the $\Sigma_2^p$-completeness of minimization of tree patterns proven by Czerwiński, Martens, Niewerth & Parys [CMNP18, Theorem 3.1].

A *tree pattern*—see *e.g.* [CMNP18, § 2.2]—over node variables $\mathbb{A}$ is a directed tree, whose nodes have a label from $\mathbb{A} \sqcup \{*\}$, and whose edges are partition into simple edges and transitive edges.

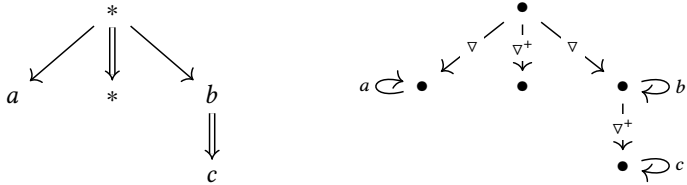

Figure IV.4: Encoding a tree pattern into a CRPQ.

(a) A tree pattern $\tau$. Double arrows represent transitive edges.

(b) Its encoding $\text{Enc}(\tau)$ as a CRPQ.

We *encode* a tree pattern $\tau$ into a CRPQ over $\mathbb{A} \sqcup \{\triangledown\}$, denoted by $\text{Enc}(\tau)$, obtained as follows:

- we start from the underlying tree of the tree pattern, and replace simple edges by an atom $\bullet \xrightarrow{\triangledown} \bullet$, and transitive edges by an atom $\bullet \xrightarrow{\triangledown^+} \bullet$;
- for any node $x$ with a node label $a \in \mathbb{A}$, we add an atom $x \xrightarrow{a} x$;
- for any node with a wildcard label $*$, we do not add any atom.

See Figure IV.4 for an example. Note that this encoding is injective.

**Proposition IV.6.3.** Given two tree patterns $\tau_1, \tau_2$ over $\mathbb{A}$, the following are equivalent:

- $\tau_1 \subseteq \tau_2$ as tree patterns—in the sense of [CMNP18, Definition 2.2], and
- $\text{Enc}(\tau_1) \subseteq \text{Enc}(\tau_2)$ as CRPQs.

*Proof sketch.* This follows from the characterization of containment for tree patterns using "canonical tree models"—see [CMNP18, §4.2][15]—and the characterization of containment for CRPQs via canonical databases (*a.k.a.* expansions)—see Proposition III.2.12. □

We do not fully understand the relation between tree pattern minimization and CRPQ MINIMIZATION, and conjecture that this encoding actually preserves minimality, but we have failed so far to prove this.

**Conjecture IV.6.4.** If a tree pattern is minimal among tree patterns, then its encoding as a CRPQ should also be minimal among CRPQs, up to contracting internal variables.

[15] Note however that the authors assume the set of labels to be infinite, and label '∗'-nodes by $z$-nodes where $z$ is a new label: this assumption can be removed by allowed unlabelled nodes in the model.

## *Appendices*

## *IV.A   Lower Bounds for Variable Minimization*

### *IV.A.1   Equivalence with a Single Variable*

**Theorem IV.A.1.** There is a fixed alphabet *s.t.* the problem of, given a Boolean CRPQ on this alphabet with only five variables, whether it is equivalent to a Boolean CRPQ with a single variable, is ExpSpace-hard.

*Proof.* We use same idea as in Theorem IV.5.1. We reduce the problem of Proposition IV.2.4 to the instance $\delta$, where

$$\delta() = \quad$$

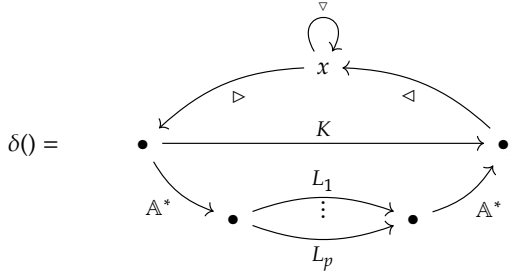

where $\triangleright$, $\triangledown$ and $\triangleleft$ are new symbols. Note that despite being named, variable $x$ is also existentially quantified.

*Claim* IV.A.2. If $\gamma_1 \subseteq \gamma_2$ then $\delta \equiv \gamma_1'$ where $\gamma_1'() \mathrel{\hat{=}} x \xrightarrow{\;\triangleright K \triangleleft\;} x \wedge x \xrightarrow{\;\triangledown\;} x$.

If $\gamma_1 \subseteq \gamma_2$ then any word of $K$ contains a factor which belongs to $\bigcap_j L_j$ and so $\gamma_1' \subseteq \delta$. The converse $\delta \subseteq \gamma_1'$ always holds.

*Claim* IV.A.3. Conversely, if $\delta$ is equivalent to a Boolean CRPQ with a single variable, then $\gamma_1 \subseteq \gamma_2$.

Let $\zeta() = \bigwedge_{i=0}^{n} x \xrightarrow{\;M_i\;} x$ be a single-variable Boolean CRPQ that is equivalent to $\delta$.

We first claim that there is some $i \in [\![0, n]\!]$ *s.t.* $M_i = \{\triangledown\}$. Every canonical database of $\delta$ contains a $\triangledown$-self loop and so from $\zeta \subseteq \delta$ it follows that any canonical database of $\zeta$ contains a $\triangledown$-self loop, which in turns implies that $M_i = \{\triangledown\}$ for some $i$. *W.l.o.g.*, assume that $M_0 = \{\triangledown\}$.

Observe that any evaluation map from $\zeta$ to a canonical database of $\delta$ must send $x \in \zeta$ to $x \in \delta$ because of the $\triangledown$-self loop, and conversely, any evaluation map from $\delta$ to a canonical database of $\zeta$ must send $x \in \delta$ to $x \in \zeta$.

We remove from $\zeta$ all atoms $x \xrightarrow{\;M_i\;} x$ *s.t.* $i \neq 0$ and $M_i \cap \triangledown^* \neq \varnothing$. Thanks to the $\triangledown$-self loop, this transformation preserve the semantics of $\zeta$. More generally, if $M_i$ contains a word in which the letter '$\triangledown$' occurs, we get remove the atom associated to $M_i$ altogether. The query obtained $\zeta'$ is clearly *s.t.* $\zeta \subseteq \zeta'$, but dually for any canonical database $G_{\zeta'}$ of $\zeta'$, extend it to a canonical database $G_\zeta$ of $\zeta$ by picking, for any atom that was removed, any word containing the letter '$\triangledown$'. Since $\zeta \subseteq \delta$, there is an evaluation map from $\delta$ to $G_\zeta$. Now the atoms of $\delta$ except the $\triangledown$-self loop do not use the letter $\triangledown$, and so it follows that the evaluation map from $\delta$ to $G_\zeta$ actually yields an evaluation map from $\delta$ to $G_{\zeta'}$. Hence, $\zeta' \sqsubseteq \delta$ and thus $\zeta' \equiv \delta$.

The same argument works for atoms containing a word that does not start with $\lhd$, or that does not end $\rhd$, or that contain strictly more than one occurrence of these symbols. Overall, it implies that *w.l.o.g.* $\zeta$ is equivalent to

$$x \xrightarrow{\triangledown} x \land \bigwedge_{j=1}^{m} x \xrightarrow{\rhd N_j \lhd} x$$

where $m \geq 0$ and the $N_j$'s are languages over $\mathbb{A}$.

Assume now, by contradiction, that for all $j \in [\![1, m]\!]$ *s.t.* $N_j \nsubseteq K \cap \mathbb{A}^*\big(\bigwedge_i L_i\big)\mathbb{A}^*$. Pick for each $j$ a word $u_j$ witnessing this. The canonical database of $\zeta$ induced by these words $\langle n_1, \dots, n_m \rangle$, namely

$$x \xrightarrow{\triangledown} x \land \bigwedge_{j=1}^{m} x \xrightarrow{\rhd n_j \lhd}$$

must satisfy $\delta$. But this implies that at least one $n_j$ must belong to $K \cap \mathbb{A}^*\big(\bigcap_i L_i\big)\mathbb{A}^*$. Contradiction.

In fact, a argument similar to what we claimed before shows that we can remove all atoms *s.t.* $N_j \nsubseteq K \cap \mathbb{A}^*\big(\bigcap_i L_i\big)\mathbb{A}^*$ without changing the semantics. Hence, *w.l.o.g.*, for each $j$, we have $N_j \subseteq K \cap \mathbb{A}^*\big(\bigcap_i L_i\big)\mathbb{A}^*$.

We then claim that each word of $K$ must belong to all $N_j$. Indeed, let $u$ be a word of $K$. Let $v_i$ be a word in $L_i \smallsetminus \mathbb{A}^*\big(\bigcap_k L_k\big)\mathbb{A}^*$—recall that such words exist by an assumption of Proposition IV.2.4—and consider the canonical database of $\delta$ obtained by expanding $K$ into $u$ and $L_i$ into $v_i$. Now $\delta \sqsubseteq \zeta$ so this database must satisfy $\zeta$. Hence, for each $j$, $N_j$ must contain one word among $u, v_1, \dots, v_m$. It cannot be any $v_i$ since otherwise we would have $v_i \in N_j \subseteq K \cap \mathbb{A}^*\big(\bigcap_k L_k\big)\mathbb{A}^* \subseteq \mathbb{A}^*\big(\bigcap_k L_k\big)\mathbb{A}^*$, which is a contradiction. And so $u \in N_j$. Therefore, we have

$$K \subseteq \bigcap_j N_j \subseteq K \cap \mathbb{A}^*\big(\bigcap_i L_i\big)\mathbb{A}^*$$

from which it follows that $K \subseteq \big(\bigcap_k L_k\big)$ and hence $\gamma_1 \sqsubseteq \gamma_2$.

Overall, Claims IV.A.2 and IV.A.3 imply that $\gamma_1 \sqsubseteq \gamma_2$ *iff* it is equivalent to a CRPQ with a single variable, in which case it is actually equivalent to

$$\gamma_1'() \triangleq x \xrightarrow{\rhd K \lhd} x \land x \xrightarrow{\triangledown} x,$$

which concludes the correctness of the reduction. $\qquad\square$

## IV.A.2   Variable Minimization is Harder than Containment

We say that a class $\mathcal{Q}$ is *closed under disjoint conjunction* if $\gamma \in \mathcal{Q}_\mathbb{A}$ and $\delta \in \mathcal{Q}_\mathbb{B}$ imply $\gamma \, \widehat{\land} \, \delta \in \mathcal{Q}_{\mathbb{A} \cup \mathbb{B}}$. Moreover, we say that the class is *closed under variable marking* if *one of* the three following properties holds:

$(\mathsf{VM})_{\mathrm{loop}}$ for any $\gamma \in \mathcal{Q}_{\mathbb{A}}$, if $y$ is a variable of $\gamma$, if $a \notin \mathbb{A}$, then $\gamma' \mathrel{\hat{=}} \gamma \wedge y \xrightarrow{a} y$
is in $\mathcal{Q}_{\mathbb{A} \sqcup \{a\}}$, or

$(\mathsf{VM})_{\mathrm{out}}$ for any $\gamma \in \mathcal{Q}_{\mathbb{A}}$, if $y$ is a variable of $\gamma$, if $a \notin \mathbb{A}$, then $\gamma' \mathrel{\hat{=}} \gamma \wedge y \xrightarrow{a} y'$
is in $\mathcal{Q}_{\mathbb{A} \sqcup \{a\}}$, where $y'$ is a new variable not occurring in $\gamma$, or

$(\mathsf{VM})_{\mathrm{in}}$ for any $\gamma \in \mathcal{Q}_{\mathbb{A}}$, if $y$ is a variable of $\gamma$, if $a \notin \mathbb{A}$, then $\gamma' \mathrel{\hat{=}} \gamma \wedge y' \xrightarrow{a} y$
is in $\mathcal{Q}_{\mathbb{A} \sqcup \{a\}}$, where $y'$ is a new variable not occurring in $\gamma$.

We will sometimes write $\gamma \in \mathcal{Q}$ to mean that $\gamma \in \mathcal{Q}_{\mathbb{A}}$ for some alphabet $\mathbb{A}$.

**Fact IV.A.4.** Any class defined by restricting the class of languages allowed to label the atoms is both closed under disjoint conjunction and closed under variable marking, assuming that languages of the form $\{a\}$ are allowed, where $a$ is a single letter.

**Theorem IV.A.5.** For any class of CRPQs closed under disjoint conjunction and closed under variable marking $\mathcal{Q}$, there is a polynomial-time reduction from the CONTAINMENT PROBLEM for Boolean queries of $\mathcal{Q}$ to the CRPQ MINIMIZATION PROBLEM restricted to queries of $\mathcal{Q}$. The same bound applies if we add the constraint that the target CRPQ must also belong to $\mathcal{Q}$.

Say that a CRPQ is *degenerate* if it contains an atom labelled the language $\{\varepsilon\}$. Equivalently, it is non-degenerate if it has at least one canonical database which is non-degenerate.

**Fact IV.A.6.** One can turn a degenerate CRPQ into a non-degenerate one by iteratively identifying variables adjacent to an atom $\xrightarrow{\{\varepsilon\}}$. This can be implemented in polynomial time.

*Proof of Theorem IV.A.5.* We assume for now that $\mathcal{Q}$ satisfies the axiom $(\mathsf{VM})_{\mathrm{loop}}$. Given an instance $\gamma_1() \subseteq^? \gamma_2()$ of the CONTAINMENT PROBLEM for Boolean queries of $\mathcal{Q}$, we assume *w.l.o.g.* that $\gamma_1$ is non-degenerate using Fact IV.A.6, and we reduce it to the instance $\langle \delta_1 \otimes \gamma_2, \|\delta_1\|_{\mathrm{var}} \rangle$, where $\delta_1$ is defined as:

$$\delta_1() \mathrel{\hat{=}} \gamma_1 \wedge \bigwedge_{x \in \gamma_1} x \xrightarrow{\triangledown_x} x$$

where $\triangledown_x$ is a fresh letter for each $x \in \gamma_1$. The reduction works clearly in logarithmic-space, and clearly $\delta_1 \otimes \gamma_2 \in \mathcal{Q}$ since $\mathcal{Q}$ is closed under disjoint conjunction and $(\mathsf{VM})_{\mathrm{loop}}$. Moreover, for it to be correct we need to show that $\gamma_1 \subseteq \gamma_2$ *iff* $\delta_1 \otimes \gamma_2$ is equivalent to a CRPQ with at most $\|\delta_1\|_{\mathrm{var}}$ variables.

*Claim IV.A.7.* If $\gamma_1 \subseteq \gamma_2$ then $\delta_1 \otimes \gamma_2 \equiv \delta_1$.

Indeed, $\gamma_1 \subseteq \gamma_2$ implies $\delta_1 \subseteq \gamma_1 \subseteq \gamma_2$ and so $\delta_1 \otimes \gamma_2 \equiv \delta_1$.

Actually this property is an "if and only if". For the converse, we will prove a stronger statement.

*Claim IV.A.8.* If $\delta_1 \otimes \gamma_2$ is equivalent to a CRPQ with at most $\|\delta_1\|_{\mathrm{var}}$ variables, then $\gamma_1 \subseteq \gamma_2$.

Let $\zeta$ be a CRPQ with at most $\|\delta_1\|_{\mathrm{var}}$ variables that is equivalent to $\delta_1 \otimes \gamma_2$. We claim first that for each $x \in \zeta$ there is a unique variable in $\zeta$ with a $\triangledown_x$- self-loop. Indeed, consider any canonical database $Z$ of $\zeta$: since $\zeta \subseteqq \delta_1 \oslash \gamma_2$, there exists a canonical database $D_1$ of $\delta_1$ and $G_2$ of $\gamma_2$ *s.t.* $D_1 \oplus G_2 \xrightarrow{\text{hom}} Z$ where $\oplus$ denotes the disjoint union. Since $D_1$ contains a $\triangledown_x$-self loop for each $x \in \gamma_1$, so does $Z$. Since this property holds for every $Z$, it follows that $\zeta$ must have a self-loop atom labelled by the singleton language $\{\triangledown_x\}$ for each $x \in \gamma_1$.

Now observe that no variable of $\zeta$ can be labelled by two $\triangledown_x$-self-loops with $x \in \gamma_1$. Indeed, $\gamma_1$ is non-degenerate, and so $\delta_1$ is also non-degenerate, and so there exists a canonical database $D_1$ of $\delta_1$ which is non-degenerate. Then, pick any canonical database $G_2$ of $\gamma_2$. $D_1 \oplus G_2$ is a canonical database of $\delta_1 \oslash \gamma_2$, which is equivalent to $\zeta$, so there is an evaluation map from $\zeta$ to $D_1 \oplus G_2$. If a variable of $\zeta$ had both a $\triangledown_x$- and a $\triangledown_y$-self-loop for $x \neq y \in \gamma_1$, then so would either $D_1$ or $G_2$. $G_2$ contains no such letters, and so it would have to be $D_1$. This contradicts the definition of $D_1$. Hence, no variable of $\zeta$ can be labelled by two $\triangledown_x$-self-loops. Together with the previous paragraph and the fact that $\zeta$ has at most $\|\delta_1\|_{\text{var}} = \|\gamma_1\|_{\text{var}}$ variables, it follows that we can assume *w.l.o.g.*—up to renaming the variables of $\zeta$—that $\zeta = \gamma_1$ and for each $x \in \gamma_1$, $x \xrightarrow{\triangledown_x} x$ is an atom of $\zeta$. Moreover, this is the only self-loop in $\zeta$ labelled by $\{\triangledown_x\}$, and for any self-loop atom $x \xrightarrow{L} x$ we cannot have $\triangledown_y \in L$ for any $y \neq x \in \gamma_1$.

We are now ready to prove that $\gamma_1 \subseteqq \gamma_2$. Let $G_1$ be a canonical database of $\gamma_1$, and let $D_1$ be the associated canonical database of $\delta_1$—it is obtained by adding an $\triangledown_x$-self-loop on every $x \in \gamma_1$. Pick any canonical database $G_2$ of $\gamma_2$. Since $\delta_1 \oslash \gamma_2 \subseteqq \zeta$, there exists a canonical database $Z$ of $\zeta$ *s.t.* $Z \xrightarrow{\text{hom}} D_1 \oplus G_2$. But then, since $\zeta \subseteqq \delta_1 \oslash \gamma_2$, there exists $D_1'$ and $G_2'$, which are canonical databases of $\delta_1$ and $\gamma_2$, respectively, *s.t.*

$$D_1' \oplus G_2' \xrightarrow{\text{hom}} Z \xrightarrow{\text{hom}} D_1 \oplus G_2.$$

Restrict this homomorphism to $G_2'$: we obtain

$$G_2' \xrightarrow{\text{hom}} Z \xrightarrow{\text{hom}} D_1 \oplus G_2.$$

Now note that, because of the previous paragraph, the homomorphism $Z \xrightarrow{\text{hom}} D_1 \oplus G_2$ must map $x \in Z$ to $x \in D_1$—because of the $\triangledown_x$-self-loop. Since $D_1 \oplus G_2$ is a disjoint union, it follows that image of this homomorphism is actually included in $D_1$, and so obtain a homomorphism

$$G_2' \xrightarrow{\text{hom}} Z \xrightarrow{\text{hom}} D_1.$$

Now of course $\triangledown_x$-self-loop will occur in the image of any homomorphism $Z \xrightarrow{\text{hom}} D_1$. However, in the composition $G_2' \xrightarrow{\text{hom}} Z \xrightarrow{\text{hom}} D_1$, since $G_2'$ does not use any letter of the form $\triangledown_x$, $x \in \gamma_1$, we conclude we actually get a homomorphism

$$G_2' \xrightarrow{\text{hom}} G_1,$$

which concludes the proof that $\gamma_1 \subseteqq \gamma_2$ and hence of Claim IV.A.8.

Now Claims IV.A.7 and IV.A.8 imply that $\gamma_1 \subseteq \gamma_2$ *iff* $\delta_1 \oslash \gamma_2$ is equivalent to a CRPQ with at most $\|\delta_1\|_{\mathrm{var}}$ variables, which concludes the reduction under the assumption that $\mathcal{Q}$ satisfies $(\mathsf{VM})_{\mathrm{loop}}$.

To conclude, note that if $\mathcal{Q}$ satisfies either $(\mathsf{VM})_{\mathrm{out}}$ or $(\mathsf{VM})_{\mathrm{in}}$ then exactly the same proof works, except that the definition of $\delta_1$ should be changed: variables will be marked using outgoing and incoming edges, respectively. Lastly, since $\delta_1 \in \mathcal{Q}$, then we have as a by-product of our proof that $\delta_1 \oslash \gamma_2$ is equivalent to a CRPQ with at most $k$ atoms *iff* it is equivalent to a CRPQ of $\mathcal{Q}$ with at most $k$ atoms. It follows that this reduction also works it we add the constraint that $\delta$ must be in $\mathcal{Q}$. $\qquad\square$

# Semantic Tree-Width and Path-Width of Conjunctive Regular Path Queries

### Abstract

We show that the problem of whether a query is equivalent to a query of tree-width $k$ is decidable, for the class of Unions of Conjunctive Regular Path Queries with two-way navigation (UC2RPQs). A previous result by Barceló, Romero, and Vardi [BRV16] has shown decidability for the case $k = 1$, and here we extend this result showing that decidability in fact holds for any arbitrary $k \geq 1$. The algorithm is in 2ExpSpace, but for the restricted but practically relevant case of simple regular expressions, we show that the complexity of the problem drops to $\Pi_2^p$.

We also investigate the related problem of approximating a UC2RPQ by queries of small tree-width. We exhibit an algorithm which, for any fixed number $k$, builds the maximal under-approximation of tree-width $k$ of a UC2RPQ. The maximal under-approximation of tree-width $k$ of a query $\gamma$ is a query $\gamma'$ of tree-width $k$ which is contained in $\gamma$ in a maximal and unique way, that is, such that for every query $\gamma''$ of tree-width $k$, if $\gamma''$ is contained in $\gamma$ then $\gamma''$ is also contained in $\gamma'$.

Our approach is shown to be robust, in the sense that it allows also to test equivalence with queries of a given path-width, it also covers the previously known result for $k = 1$, and it allows testing whether a (one-way) UCRPQ is equivalent to a UCRPQ of a given tree-width (or path-width).

### Acknowledgements

This chapter is mostly a reproduction of the eponymous paper that was published in LMCS [FM25], as a selected paper from ICDT '23 [FM23]. Part of the preliminaries and of the introduction have been moved to Chapter III. It is a joint work with Diego Figueira.

# Contents

## V.1 Introduction

### V.1.1 Conjunctive Regular Path Queries of Small Tree Width

In this chapter, we consider the problem of, given a C2RPQ, or a union of C2RPQs, deciding if it is equivalent to a query of small tree-width. We depict the relationship of the different notions of tree-width and path-width of C2RPQs in Figure V.1, where we denote by $\mathcal{T}w_k$ (resp. $1\mathcal{T}w_k$) the set of all C2RPQs (resp. CRPQs) of tree-width at most $k$. Similarly, $\mathcal{P}w_k$ (resp. $1\mathcal{P}w_k$) is the set of all C2RPQs (resp. CRPQs) of path-width at most $k$. Note that $1\mathcal{T}w_k$ and $1\mathcal{P}w_k$ are not explicitly drawn, but correspond to the intersection of $\mathcal{T}w_k$ (resp. $\mathcal{P}w_k$) with the class of CRPQs. Lastly, the *tree-width* (resp. *path-width*) of a UC2RPQ is simply the maximum of the tree-width (resp. path-width) of its disjuncts.

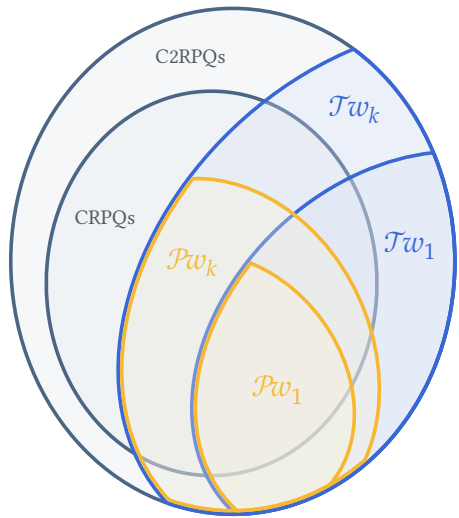

Figure V.1: Clickable taxonomy of syntactic classes studied in this chapter.

Similar statements of the following proposition can be considered folklore (see *e.g.* [RBV17, Theorem IV.3]). It can be proven in the same fashion as Proposition III.1.22.

**Proposition V.1.1.** For each $k \geq 1$, the EVALUATION PROBLEM for UC2RPQs of tree-width at most $k$ can be solved in time $\mathcal{O}(\|\Gamma\| \cdot |G|^{k+1} \cdot \log |G|)$ on a Turing machine, or $\mathcal{O}(\|\Gamma\| \cdot |G|^{k+1})$ using a RAM model, where $\Gamma$ and $G$ are the input UC2RPQ and graph database, respectively.

In practice, graph databases tend to be huge and often changing, while queries are in comparison very small and fixed. This motivates the following question, given some natural $k \geq 1$:

Given a UC2RPQ $\Gamma$, is it equivalent to a UC2RPQ $\Gamma'$ of tree-width at most $k$?
That is, does it have *semantic tree-width* at most $k$?

This problem is called the *semantic tree-width $k$ problem*. Should it be decidable in a constructive way—that is, decidable, and if the answer is positive, we can compute a witnessing $\Gamma'$ from $\Gamma$—, then one could, once and for all, compute $\Gamma'$ from $\Gamma$ and, whenever one wants to evaluate $\Gamma$ on a database, evaluate $\Gamma'$ instead.

We will also study the restriction of these notions to one-way queries: a UCRPQ has *one-way semantic tree-width* at most $k$ if it is equivalent to a UCRPQ of tree-width at most $k$. The *one-way semantic tree-width $k$ problem* is the problem of, given a UCRPQ $\Gamma$, whether it has one-way semantic tree-width at most $k$.

**Example V.1.2.** [1] Consider the following CRPQs, where $\bar{x} = (x_0, x_1, y, z)$:

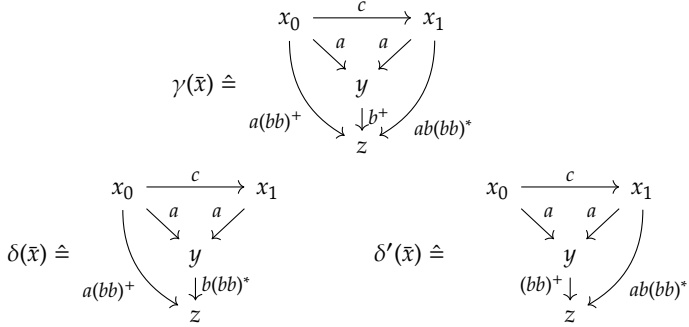

The underlying graph of $\gamma(\bar{x})$ being the directed 4-clique, $\gamma(\bar{x})$ has tree-width 3. We claim that $\gamma(\bar{x})$ is equivalent to the UCRPQ $\delta(\bar{x}) \vee \delta'(\bar{x})$, and hence has one-way semantic tree-width at most 2.

Indeed, given a graph database satisfying $\gamma(\bar{x})$ via some mapping $\mu$, it suffices to make a case disjunction on whether the number of $b$-labelled atoms in the path from $\mu(y)$ to $\mu(z)$ is even or odd. In the first case, the atom $x_0 \xrightarrow{a(bb)^*} z$ becomes redundant since we can deduce the existence of such a path from the conjunction $x \xrightarrow{a} y \xrightarrow{(bb)^*} z$, and hence the database satisfies $\delta(\bar{x})$ via $\mu$. Symmetrically, in the second case, the atom $x_1 \xrightarrow{b(bb)^*} z$ becomes redundant, and the database satisfies $\delta'(\bar{x})$ via $\mu$. Thus, $\gamma(\bar{x})$ is contained, and hence equivalent (the other containment being trivial), to the UCRPQ $\delta(\bar{x}) \vee \delta'(\bar{x})$ of tree-width 2. ◇

### V.1.2 Related Work

On the class of conjunctive queries, the semantic tree-width $k$ problem becomes the coNP-complete problem of finding out whether a query has a retraction of tree-width at most $k$. In fact, CQs enjoy the effective existence of unique minimal queries [CM77, Theorem 12] (the core), which happen to also minimize the tree-width. For CRPQs and UC2RPQs, the question is far more challenging, and it has only been solved for the case $k = 1$ by Barceló, Romero, and Vardi [BRV16, Theorem 6.1]; the case $k > 1$ was left widely open [BRV16, § 7].

Furthermore, having bounded semantic tree-width precisely describes the classes of CQs that have a tractable (both in the sense of polynomial time and FPT) evaluation problem [Gro07, Theorem 1.1]. This result is on bounded-arity schemas, which was later generalized [CGLP20, Theorem 1] for characterizing FPT evaluation on arbitrary schemas—by replacing semantic tree-width with semantic "submodular width" [Mar13].

The problem of computing maximal under-approximations of CQs of a given tree-width has been explored in [BLR14]. A maximal under-approximation of tree-width at most $k$ of a CQ $\gamma$ consists of a CQ $\delta_k$ of tree-width at most $k$, which under-approximates it, *i.e.* $\delta_k$ is contained in $\gamma$, and which is maximal, in the sense that for every CQ $\delta'$, if $\delta'$ has tree-width at most $k$ and is contained in $\gamma$, then $\delta'$ is contained in $\delta_k$. Maximal under-approximations of a given tree-width for CQs always exist [BLR14] and thus, a CQ is semantically equivalent to a CQ of tree-width at most $k$ if, and only if, it is equivalent to its maximal under-approximation of tree-width at most $k$. Our solution to decide the semantic tree-width $k$ problem for UC2RPQs is based on this idea.

While maximal under-approximations always exist for CQs, this is not the case for the dual notion of "minimal over-approximations". The problem of when these exist is still unknown to be decidable, aside for some the special cases of acyclic CQs and Boolean CQs over binary schemas [BRZ20].

### V.1.3 Contributions

Here we solve both the semantic tree-width $k$ problem and one-way semantic tree-width $k$ problem for every $k$ with one unifying approach.

**Theorem V.1.3.** For each $k \geq 1$, the semantic tree-width $k$ problem and the one-way semantic tree-width $k$ problem are decidable. Moreover, these problems are in 2ExpSpace and are ExpSpace-hard. When $k = 1$, the problems are in fact ExpSpace-complete.

In Section V.3 (Lemma V.3.10), we prove the upper bound for $k \geq 2$, by relying on the so-called "Key Lemma", which is our main technical result, and is proven in Sections V.4 and V.5. The upper bound for the case $k = 1$—which was already proven in [BRV16] for the (two-way) semantic tree-width 1 problem—is shown in Section V.7 (Corollary V.7.8). The lower bound is shown in Section V.9 (Lemma V.9.1).

The Key Lemma (Lemma V.3.8) essentially states that every UC2RPQ has a computable "maximal under-approximation" by a UC2RPQ of tree-width $k$ and that this approximation is well-behaved with respect to the class of languages used to label the queries under some mild assumptions on it (being "closed under sublanguages"). Let us first explain this assumption before formalizing the statement above (stated as Corollary V.3.9).

For a class $\mathcal{L}$ of languages, let UC2RPQ($\mathcal{L}$) denote the class of all UC2RPQs whose atoms are all labelled by languages from $\mathcal{L}$. We say that $\mathcal{L}$ is *closed under sublanguages* if (i) it contains every language of the form $\{a\}$, where $a \in \mathbb{A}$ is any (positive) letter such that either $a$ or $a^-$ occur in a word of a language of $\mathcal{L}$, and (ii) for every language $L \in \mathcal{L}$ there exists an NFA $\mathcal{A}_L$ accepting $L$ such that every sublanguage $\mathcal{A}_L[q, q']$ distinct from $\varnothing$ and $\{\varepsilon\}$ belongs to $\mathcal{L}$.

To the best of our knowledge, all classes of regular expressions that have been considered in the realm of regular path queries (see, *e.g.*, [Fig+20, §1]) are closed under sublanguages. In particular, this is the case for the class of simple regular expressions, which will be our focus of study in Section V.6. Moreover, even if some class $\mathcal{L}$ is not closed under sublanguages, such as $\{(aa)^*\}$, then it is contained in a minimal class closed under sublanguages—$\{a, a(aa)^*, (aa)^*\}$ in this example.

We can now state the main implication of the Key Lemma (whose formal statement requires some extra definitions).

**Corollary V.3.9** (Existence of the maximal under-approximation). For each $k \geq 2$, for each class $\mathcal{L}$ closed under sublanguages, and for each query $\Gamma \in \text{UC2RPQ}(\mathcal{L})$, there exists $\Gamma' \in \text{UC2RPQ}(\mathcal{L})$ of tree-width at most $k$ such that $\Gamma' \subseteq \Gamma$, and for every $\Delta \in \text{UC2RPQ}$, if $\Delta$ has tree-width at most $k$ and $\Delta \subseteq \Gamma$, then $\Delta \subseteq \Gamma'$. Moreover, $\Gamma'$ is computable from $\Gamma$ in ExpSpace.

As a consequence of Corollary V.3.9 and Proposition V.1.1, we have that queries of bounded semantic tree-width have tractable evaluation.

**Corollary V.3.16** (FPT evaluation for bounded semantic tree-width). For each $k \geq 1$, the EVALUATION PROBLEM for C2RPQs of semantic tree-width at most $k$ is fixed-parameter tractable—FPT—when parameterized in the size of the query. More precisely on input $\langle \Gamma, G \rangle$, the algorithm runs in time $\mathcal{O}(f(\|\Gamma\|) \cdot |G|^{k+1} \cdot \log |G|)$ on a Turing machine, where $f$ is a doubly-exponential function—or $\mathcal{O}(f(\|\Gamma\|) \cdot |G|^{k+1})$ under a RAM model.

Note that [FGM24, Theorem 22] shows that the statement above can be improved to have a single-exponential function $f$.

Moreover, we also show that for any class $\mathcal{L}$ of regular languages closed under sublanguages, if $\Gamma \in \text{UC2RPQ}(\mathcal{L})$ has semantic tree-width $k > 1$, then $\Gamma$ is equivalent to a $\text{UC2RPQ}(\mathcal{L})$ of tree-width at most $k$. Analogous characterizations hold for $k = 1$ and/or path-width, see Corollaries V.8.7 and V.7.9.

**Theorem V.3.13.** Assume that $\mathcal{L}$ is closed under sublanguages. For any $k > 1$ and any query $\Gamma \in \text{UC2RPQ}(\mathcal{L})$, the following are equivalent:

(1) $\Gamma$ is equivalent to an infinitary union of conjunctive queries of tree-width at most $k$;

(2) $\Gamma$ has semantic tree-width at most $k$;

(3) $\Gamma$ is equivalent to a $\text{UC2RPQ}(\mathcal{L})$ of tree-width at most $k$.

The implications $(3) \Rightarrow (2) \Rightarrow (1)$ immediately follow from the definition of the semantic tree-width. On the other hand, the implications $(1) \Rightarrow (2)$ and $(2) \Rightarrow (3)$ are surprising, since they are both trivially false when $k = 1$. We defer the proof of this last claim to Remark V.3.14 as we first need a few tools to manipulate CRPQs.

The previous theorem, together with the high complexity of semantic tree-width $k$ problem, motivates us to focus on the case of CRPQs using some simple regular expressions (SRE) in Section V.6, where we show that the complexity of this problem is much lower.

**Theorem V.6.1.** For $k \geq 2$, the semantic tree-width $k$ problem for UCRPQ(SRE) is in $\Pi_2^p$.

We then study the problem of $k = 1$: at first glance, our proof for $k \geq 2$ of Theorem V.1.3 does not capture this case, for a technical—yet crucial—reason. In Section V.7, we explain how to adapt our proof to capture it: and show the decidability the semantic tree-width 1 problem—which was already studied by Barceló, Romero and Vardi [BRV16]—and of the one-way semantic tree-width 1 problem.

Building on the same idea, we show in Section V.8 that our results extend to path-width.

**Theorem V.8.6.** For each $k \geq 1$, the semantic path-width $k$ problems are decidable. Moreover, they lie in 2ExpSpace and are ExpSpace-hard. Moreover, if $k = 1$, these problems are in fact ExpSpace-complete.

In turn, this leads to an evaluation algorithm with a remarkably low complexity.

**Theorem V.8.8.** For each $k \geq 1$, the EVALUATION PROBLEM, restricted to UC2RPQs of semantic path-width at most $k$ is in paraNL when parameterized in the size of the query. More precisely, the problem, on input $\langle \Gamma, G \rangle$, can be solved in non-deterministic space $f(|\Gamma|) + \log(|G|)$, where $f$ is a single exponential function.

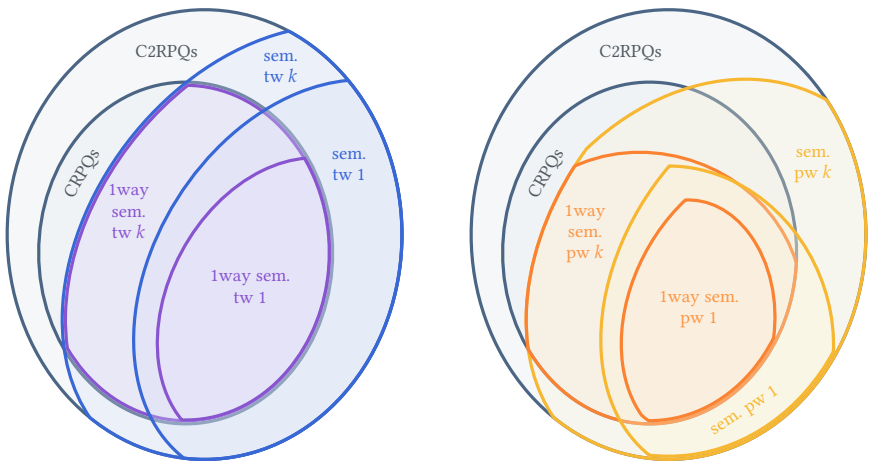

(a) Semantic classes of C2RPQs related to tree-width.

(b) Semantic classes of C2RPQs related to path-width.

Figure V.2: Clickable taxonomy of semantic classes studied in this paper, where $k \geq 2$.

Interestingly, the proof for tree-width 1 and path-width $k$ ($k \geq 1$) can be derived from the proof from tree-width $k \geq 2$ but necessitates an additional technical trick which yields different closure properties (or lack thereof). We show that a UCRPQ has semantic tree-width at most $k$ if, and only if, it has one-way semantic tree-width at most $k$ whenever $k \geq 2$ (Corollary V.3.15). In other words, if the original query does not use two-way navigation, then considering

UC2RPQs does not help to further minimize the tree-width. Interestingly, this is false for $k = 1$ (*cf.* Remark V.3.14, also [BRV16, Proposition 6.4]) and for path-width, no matter the value of $k \geq 1$ (see Section V.8.2). Overall, this leads to the landscape depicted in Figure V.2.

Finally, we conclude in Section V.10. We provide a *partial* characterization *à la* Grohe of classes of UC2RPQs which admit a tractable evaluation in Section V.10.2.

**Theorem V.10.5.** Assuming $W[1] \neq \text{FPT}$, for any recursively enumerable class $\mathcal{C}$ of finitely redundant Boolean UC2RPQs, the EVALUATION PROBLEM for $\mathcal{C}$ is FPT if, and only if, $\mathcal{C}$ has bounded semantic tree-width.

We also discuss open questions, ranging from complexity questions (Section V.10.1) to extensions of our results to bigger classes or larger settings (Sections V.10.3 and V.10.4).

## *V.2  Preliminaries*

*Contractions.*  We define *two-way contractions* (or *contractions* for short in this chapter) analogously to one-way contractions by redefining the notion of internal variable. In the context of one-wayness, it designated any variable with both in-degree and out-degree one. Now, for a variable to be internal, we only ask that its total degree is two. Concretely, this means that two-way contractions can also contract paths like $x \xrightarrow{K} y \xleftarrow{L} z$ into a single atom $x \xrightarrow{KL^{-1}} z$.[2]

*Some Intuitions on Maximal Under-Approximations.*  Given a conjunctive query $\gamma$, the union of all conjunctive queries that are contained in $\gamma$ is semantically equivalent to the union $\bigvee \{ \gamma' \mid \gamma \twoheadrightarrow \gamma' \}$. Naturally, this statement borders on the trivial since $\gamma'$ belongs to this union. It becomes interesting when we add a restriction: given a class $\mathcal{C}$ of CQs (to which $\gamma$ may not belong) closed under subqueries, then $\Gamma' \triangleq \bigvee \{ \gamma' \in \mathcal{C} \mid \gamma \twoheadrightarrow \gamma' \}$ is the maximal under-approximations of $\gamma$ by finite unions of conjunctive queries of $\mathcal{C}$, in the following sense:

i. (finite) $\Gamma'$ is a finite union of CQs of $\mathcal{C}$,

ii. (under-approximation) $\Gamma' \subseteq \gamma$, and

iii. (maximality) for any finite union $\Delta$ of CQs of $\mathcal{C}$, if $\Delta \subseteq \gamma$, then $\Delta \subseteq \Gamma'$.

*Proof.*  Only the last point is non-trivial, and follows from the fact that if $\Delta \subseteq \gamma$, then for each $\delta \in \Delta$, $\delta \subseteq \gamma$, so there is a homomorphism $f \colon \gamma \to \delta$. The image $\delta'$ of $f$ is a subquery of $\delta$, and $\mathcal{C}$ is closed under subqueries, so it belongs to $\mathcal{C}$, and hence to $\Gamma'$. Since there is a trivial homomorphism from $\delta'$ to $\delta$, we moreover have that $\delta \subseteq \delta'$. Hence, for each CQ $\delta \in \Delta$, there is a CQ $\delta' \in \Gamma'$ such that $\delta \subseteq \delta'$, and hence $\Delta \subseteq \Gamma'$. □

As a consequence, we deduce that for each $k \geq 1$, the maximal under-approximation of a CQ by a finite union of CQs of tree-width at most $k$ is computable, and hence we can effectively decide if some CQ is equivalent to a query of tree-width at most $k$ by testing the equivalence with this maximal under-approximation. For more details on approximations of CQs, see [BLR14]. Note that interestingly, changing $\Gamma'$ from $\bigvee\{\gamma' \in \mathcal{C} \mid \gamma \twoheadrightarrow \gamma'\}$ to $\bigvee\{\gamma' \in \mathcal{C} \mid \gamma' \sqsubseteq \gamma\}$ preserves both under-approximation and maximality, but $\Gamma'$ is now an infinite union of CQs of $\mathcal{C}$.

[2] Naturally, we can equally contract $x \xleftarrow{K} y \xrightarrow{L} z$ into the atom $x \xrightarrow{K^{-1}L} z$.

Unfortunately, these results cannot be straightforwardly extended to conjunctive regular path queries since the previous proof implicitly relied on two points:

1. the equivalence between the containment $\gamma' \sqsubseteq \gamma$ and the existence of a homomorphism $\gamma \xrightarrow{\text{hom}} \gamma'$, and

2. the possibility to restrict $\gamma'$ to its image $\gamma \xrightarrow{\text{hom}} \gamma'$ while obtaining a semantically bigger query.

These two crucial ingredients is what allows us to build a finite set $\Gamma'$ from $\gamma$. For CRPQs, the second point still holds, but not the first one. For instance, the CQ $\gamma(x, y) = x \xrightarrow{a} z \xrightarrow{b} y$ is contained in (in fact equivalent to) the CRPQ $\gamma'(x, y) = x \xrightarrow{ab} y$, but there is no homomorphism from $\gamma'(x, y)$ to $\gamma(x, y)$. Our main result shows that to find maximal under-approximations of C2RPQs, it suffices to take homomorphic images of so-called "refinements" of $\gamma$, instead of homomorphic images of $\gamma$ itself. The next paragraphs are devoted to introducing refinements and tools related to them.

*Refinements and Tree-Width.* Our approach to proving both Theorems V.1.3 and V.3.13 and the Key Lemma heavily rely on refinements. One crucial property that these objects satisfy is that they preserve tree-width $k$, unless $k = 1$, as illustrated in Figure V.3.

**Fact V.2.1.** Let $k \geq 2$ and let $\gamma$ be a C2RPQ of tree-width at most $k$. Then any refinement of $\gamma$ has tree-width at most $k$.

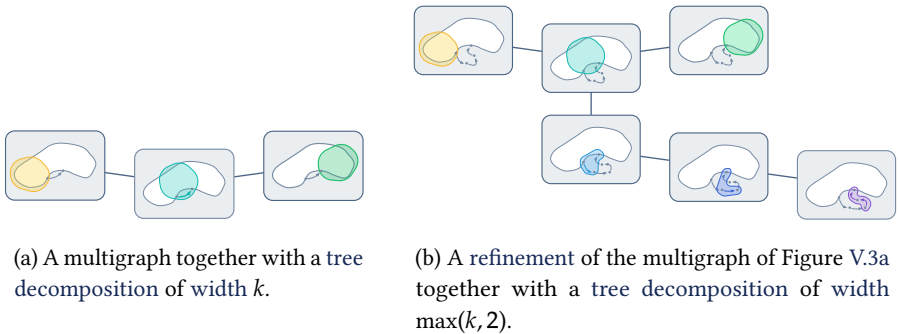

(a) A multigraph together with a tree decomposition of width $k$.

(b) A refinement of the multigraph of Figure V.3a together with a tree decomposition of width $\max(k, 2)$.

Figure V.3: Refinements and expansions preserve tree-width at most $k \geq 2$.

*Proof.* The underlying graph of a refinement of $\gamma$ is obtained from the underlying graph of $\gamma$ by either contracting some edges (when dealing with equality atoms), or by replacing a single edge by a path of edges (where the non-extremal nodes are new nodes).

This first operation preserves tree-width at most $k$ (even if $k = 1$), see *e.g.* [Bod98, Lemma 16]. The second operation preserves tree-width at most $k$, assuming $k > 1$: if a graph $G'$ is obtained from a graph $G$ by replacing an edge $x_0 \to x_n$ by a path $x_0 \to x_1 \to \cdots \to x_n$, then from a tree decomposition of $G$ it suffices to pick a bag containing both $x_0$ and $x_n$, and add a branch to the tree, rooted at this bag, and containing bags with nodes

$$\{x_0, x_1, x_n\}, \{x_1, x_2, x_n\}, \dots, \{x_i, x_{i+1}, x_n\}, \dots, \{x_{n-2}, x_{n-1}, x_n\},$$

as depicted in Figure V.3. All bags contain exactly three nodes, so we obtain tree decomposition of $G'$ whose width is the maximum between 2 and the width of the original tree decomposition of $G$. □

For $k = 1$, the property fails: for instance the CRPQ $\gamma(x) = x \xrightarrow{a^*} x$ has tree-width at most 1 (in fact it has tree-width 0), but its refinement $\rho(x) = x \xrightarrow{a^*} t_1 \xrightarrow{a^*} t_2 \xrightarrow{a^*} x$ has tree-width 2.

*Fine tree decompositions*  For technical reasons—the proof of Lemma V.5.4—, we will use a restrictive class of tree decompositions which we call "fine"[3]. A *fine tree decomposition* is a tree decomposition $\langle \mathbf{T}, \mathbf{v} \rangle$ in which:

> every non-root bag can be obtained from its parent
> bag by either adding or removing a non-empty set      (V.1)
> of vertices.

In the context of a fine tree decomposition of width $k$, a *full bag* is any bag of size $k + 1$.

A C2RPQ has tree-width $k$ if and only if it has a fine tree decomposition of width at most $k$. Indeed, from a tree decomposition, it suffices to:

1. first merge every consecutive pair of bags that contain exactly the same variables;

2. between every pair of bags that does not satisfy (V.1), add a bag whose set of vertices correspond to the intersection of the two adjacent bags.

## V.3   Maximal Under-Approximations

In this section, we state our key technical result, Lemma V.3.8, which we will refer to as the "Key Lemma". Essentially, we follow the same structure as Theorem V.3.13: given a C2RPQ $\gamma$ and a natural number $k > 1$, we start by considering its maximal under-approximation by infinitary unions of conjunctive queries of tree-width $k$ (Definition V.3.1), and then show that this query can in fact be expressed as a UC2RPQ of tree-width $k$ whose atoms contain sublanguages of those in $\gamma$ (Key Lemma V.3.8).

For the first definitions of this section, let us fix any class $\mathcal{C}$ of C2RPQs—we will later apply these results to the class $\mathcal{Tw}_k$ of C2RPQs of tree-width at most $k$.

**Definition V.3.1** (Maximal under-approximation). Let $\gamma$ be a C2RPQ. The

[3] This is similar—but orthogonal—to the classical notion of "nice tree decomposition", see *e.g.* [Klo94, Definition 13.1.4, page 149].

*maximal under-approximation* of $\gamma$ by infinitary unions of $\mathcal{C}$-queries is $\mathrm{App}_{\mathcal{C}}(\gamma) \doteq \{\alpha \in \mathcal{C} \mid \alpha \sqsubseteq \gamma\}$.

For intuition, we refer the reader back to paragraph "Some intuitions on maximal under-approximations" at the beginning of Section V.2.

**Remark V.3.2.** Observe that $\mathrm{App}_{\mathcal{C}}(\gamma)$ is an infinitary union of $\mathcal{C}$-queries, that $\mathrm{App}_{\mathcal{C}}(\gamma) \sqsubseteq \gamma$, and that for every infinitary union of $\mathcal{C}$-queries $\Delta$, if $\Delta \sqsubseteq \gamma$, then $\Delta \sqsubseteq \mathrm{App}_{\mathcal{C}}(\gamma)$ (*i.e.*, it is the unique maximal under-approximation up to semantical equivalence). Similarly, the maximal under-approximation of a UC2RPQ is simply the union of the maximal under-approximations of the C2RPQs thereof. $\diamond$

Unfortunately, the fact that a query $\alpha$ is part of this union, namely $\alpha \in \mathrm{App}_{\mathcal{C}}(\gamma)$, does not yield any useful information on the *shape* of $\alpha$—we merely know that $\alpha \sqsubseteq \gamma$. We thus introduce another infinitary union of $\mathcal{C}$-queries of a restricted shape, namely $\mathrm{App}_{\mathcal{C}}^{\star}(\gamma) \subseteq \mathrm{App}_{\mathcal{C}}(\gamma)$, in which queries $\alpha \in \mathrm{App}_{\mathcal{C}}^{\star}(\gamma)$ come together with a witness of their containment in $\gamma$.

**Definition V.3.3.** The maximal under-approximation of $\gamma$ by infinitary unions of homomorphically-smaller $\mathcal{C}$-queries is

$$\mathrm{App}_{\mathcal{C}}^{\star}(\gamma) \doteq \{\alpha \in \mathcal{C} \mid \exists \rho \in \mathrm{Ref}(\gamma), \exists f \colon \rho \twoheadrightarrow \alpha\}. \tag{V.2}$$

For a basic example of approximation (with no constraint on $\mathcal{C}$), we refer the reader to **??**. The resulting query $\alpha(x, y)$ is the homomorphic image of a refinement of $\gamma(x, y)$. Hence, $\alpha(x, y) \in \mathrm{App}_{\mathcal{C}}^{\star}(\gamma)$ if $\mathcal{C}$ is, for instance, the class of all C2RPQs—or more generally, if $\mathcal{C}$ contains $\alpha(x, y)$.

**Example V.3.4** (Example V.1.2, cont'd). Both $\delta(\bar{x})$ and $\delta'(\bar{x})$ are semantically equivalent to queries in $\mathrm{App}_{\mathcal{TW}_2}^{\star}(\gamma(\bar{x}))$. Indeed, starting from $\gamma(\bar{x})$, we can refine

$$x_0 \xrightarrow{a(bb)^+} z \quad \text{into} \quad x_1 \xrightarrow{a} t \xrightarrow{(bb)^+} z.$$

Denote by $\rho(\bar{x})$ the query obtained:

Then merge variables $t$ and $y$: this new query $\delta'_{\mathrm{app}}(\bar{x})$ is equivalent to $\delta'(\bar{x})$. Moreover, since $\delta'_{\mathrm{app}}(\bar{x})$ has tree-width at most 2 and was obtained as a homomorphic image of a refinement of $\gamma(\bar{x})$, we have that $\delta'_{\mathrm{app}}(\bar{x}) \in \mathrm{App}_{\mathcal{TW}_2}^{\star}(\gamma(\bar{x}))$. A similar argument applies to $\delta$, by refining the atom between $x_1$ and $z$ instead. $\diamond$

Clearly, $\mathrm{App}^\star_{\mathcal{C}}(\gamma)$—whose queries are informally called *approximations*—is included, and thus semantically contained, in $\mathrm{App}_{\mathcal{C}}(\gamma)$, since $\rho \subseteq \gamma$ and $\alpha \subseteq \rho$ in (V.2). In fact, under some assumptions on $\mathcal{C}$, the converse containment also holds.

**Fact V.3.5.** If $\mathcal{C}$ is closed under expansions and subqueries, then for any C2RPQ $\gamma$, we have $\mathrm{App}_{\mathcal{C}}(\gamma) \equiv \mathrm{App}^\star_{\mathcal{C}}(\gamma)$.

*Proof.* Since $\mathrm{App}_{\mathcal{C}}(\gamma) \supseteq \mathrm{App}^\star_{\mathcal{C}}(\gamma)$, it suffices to show that $\mathrm{App}_{\mathcal{C}}(\gamma) \subseteq \mathrm{App}^\star_{\mathcal{C}}(\gamma)$. Pick $\alpha \in \mathrm{App}_{\mathcal{C}}(\gamma)$. Let $\xi$ be an expansion of $\alpha$. Since $\alpha \subseteq \gamma$, there exists by Proposition III.2.12 an expansion $\xi_\gamma$ of $\gamma$ such that $\xi_\gamma \xrightarrow{\mathrm{hom}} \xi$. Consider the restriction $\xi'$ of $\xi$ to its homomorphic image. Since $\alpha \in \mathcal{C}$ and $\mathcal{C}$ is closed both under expansions and subqueries, $\xi' \in \mathcal{C}$. Since moreover, by construction, $\xi'$ is the (strong onto) homomorphic image of an expansion (hence refinement) of $\gamma$, then $\xi' \in \mathrm{App}^\star_{\mathcal{C}}(\gamma)$. Hence, we have shown that for every expansion of $\mathrm{App}_{\mathcal{C}}(\gamma)$, there is an expansion of $\mathrm{App}^\star_{\mathcal{C}}(\gamma)$ with a strong onto homomorphism from the latter to the former, which concludes the proof by Proposition III.2.12. □

Note that in the definition of $\mathrm{App}^\star_{\mathcal{C}}(\gamma)$ we work with strong onto homomorphisms: changing the definition to have any homomorphism would yield a slightly bigger but semantically equivalent class of queries—though having untamed shapes.

Observe then, by Fact V.2.1, that the class $\mathcal{Tw}_k$ of all C2RPQs of tree-width at most $k$ is closed under refinements and hence under expansions, provided that $k$ is greater or equal to 2. Moreover, $\mathcal{Tw}_k$ is always closed under subqueries for each $k$.

**Corollary V.3.6.** For $k \geq 2$, for all C2RPQ $\gamma$, $\mathrm{App}_{\mathcal{Tw}_k}(\gamma) \equiv \mathrm{App}^\star_{\mathcal{Tw}_k}(\gamma)$.

**Example V.3.7** (counterexample for $k = 1$). Consider the following query:

$$\gamma(x) \;\hat{=}\; \begin{array}{c} z \\ c \nearrow \; \nwarrow b \\ x \xrightarrow{\;a\;} y. \end{array}$$

We claim that $\mathrm{App}_{\mathcal{Tw}_1}(\gamma) \subsetneq \mathrm{App}^\star_{\mathcal{Tw}_1}(\gamma)$. First, we claim that

$$\gamma \in \mathrm{Exp}(\mathrm{App}_{\mathcal{Tw}_1}(\gamma))$$

since $\gamma$ is an expansion of $\delta(x) = x \xrightarrow{abc} x$, which clearly belongs to $\mathrm{App}_{\mathcal{Tw}_1}(\gamma)$. Then, observe that $\gamma(x)$ has a single refinement: itself! It follows that $\mathrm{App}^\star_{\mathcal{Tw}_1}(\gamma)$ is finite, and consists precisely of all homomorphic images of $\gamma(x)$ of tree-width at most 1, which are:

$$\alpha_1(w) \;\hat{=}\; a \circlearrowright w \overset{c}{\underset{b}{\rightleftarrows}} z \;, \qquad \alpha_2(w) \;\hat{=}\; c \circlearrowright w \overset{b}{\underset{a}{\rightleftarrows}} y$$

$$\alpha_3(x) \triangleq x \underset{a}{\overset{c}{\rightleftarrows}} w \supset b, \qquad \alpha_4(w) \triangleq a \circlearrowright w \supset c$$
$$\underset{b}{\circlearrowleft}$$

which correspond to the case when the following variable are merged: $\{x, y\}$, $\{x, z\}$, $\{y, z\}$ and $\{x, y, z\}$, respectively. Note that all of these queries are CQs, from which it follows that every expansion of a query in $\mathrm{App}^\star_{\mathcal{T}_{w_1}}(\gamma)$ is one of the $\alpha_i$, and has a self-loop. In particular, such an expansion cannot have a homomorphism to $\gamma$. Hence, we showed that there is an expansion of $\mathrm{App}_{\mathcal{T}_{w_1}}(\gamma)$ s.t. no expansion of $\mathrm{App}^\star_{\mathcal{T}_{w_1}}(\gamma)$ can be homomorphically mapped to it. Hence, by Proposition III.2.12, $\mathrm{App}_{\mathcal{T}_{w_1}}(\gamma) \not\sqsubseteq \mathrm{App}^\star_{\mathcal{T}_{w_1}}(\gamma)$. $\diamondsuit$

In general, by definition, $\mathrm{App}^\star_{\mathcal{T}_{w_k}}(\gamma)$ is an infinitary union of C2RPQs. Our main technical result shows that, in fact, $\mathrm{App}^\star_{\mathcal{T}_{w_k}}(\gamma)$ is always equivalent to a *finite* union of C2RPQs. This is done by bounding the length of the refinements occurring in the definition of $\mathrm{App}^\star_{\mathcal{T}_{w_k}}(\gamma)$. For any $m \geq 1$, we define:

$$\mathrm{App}^{\star, \leq m}_{\mathcal{C}}(\gamma) \triangleq \{\alpha \in \mathcal{C} \mid \exists \rho \in \mathrm{Ref}^{\leq m}(\gamma), \exists f\colon \rho \twoheadrightarrow \alpha\}.$$

**Lemma V.3.8** (*Key Lemma*)**.** For $k \geq 2$ and C2RPQ $\gamma$, we have $\mathrm{App}^\star_{\mathcal{T}_{w_k}}(\gamma) \equiv \mathrm{App}^{\star, \leq \ell}_{\mathcal{T}_{w_k}}(\gamma)$, where $\ell = \Theta(\|\gamma\|^2_{\mathrm{at}} \cdot (k+1)^{\|\gamma\|_{\mathrm{at}}})$.

By construction, $\mathrm{App}_{\mathcal{T}_{w_k}}(\gamma)$ is the maximal under-approximation of $\gamma$ by infinitary unions of C2RPQs of tree-width at most $k$. Using the equivalence above and Corollary V.3.6, it follows that it is also the maximal under-approximation of $\gamma$ by a UC2RPQ of tree-width at most $k$.

**Corollary V.3.9** (Existence of the maximal under-approximation)**.** For each $k \geq 2$, for each class $\mathcal{L}$ closed under sublanguages, and for each query $\Gamma \in \mathrm{UC2RPQ}(\mathcal{L})$, there exists $\Gamma' \in \mathrm{UC2RPQ}(\mathcal{L})$ of tree-width at most $k$ such that $\Gamma' \subseteq \Gamma$, and for every $\Delta \in \mathrm{UC2RPQ}$, if $\Delta$ has tree-width at most $k$ and $\Delta \subseteq \Gamma$, then $\Delta \subseteq \Gamma'$. Moreover, $\Gamma'$ is computable from $\Gamma$ in ExpSpace.

*Proof.* The algorithm to compute $\Gamma'$ is straightforward: it enumerates $\ell$-refinements, enumerates its homomorphic images, and keeps the result only if it has tree-width at most $k$—which can be done in linear time using Bodlaender's algorithm (Proposition III.1.23). $\square$

Using the Key Lemma as a black box—which will be proven in Section V.5—, we can now give a proof of the upper bound of Theorem V.1.3 for all cases $k \geq 2$—the case $k = 1$ will be the object of Section V.7.

**Lemma V.3.10** (Upper bound for Theorem V.1.3 for $k \geq 2$)**.** For $k \geq 2$, the semantic tree-width $k$ problem for UC2RPQ is in 2ExpSpace.

Note that $\mathrm{App}^{\star, \leq \ell}_{\mathcal{T}_{w_k}}(\gamma)$ has double-exponential size in $\|\gamma\|$, so testing equivalence of $\gamma$ with this UC2RPQ yields an algorithm in triple-exponential space in $\|\gamma\|$ since (U)C2RPQ equivalence is ExpSpace [CDLV00, Theorem 5] —see also [FLS98, § after Theorem 4.8] for a similar result on CRPQs without inverses but with an infinite alphabet. To get a better upper bound, we first need the following proposition:

**Proposition V.3.11.** The CONTAINMENT PROBLEM $\Gamma \subseteq \Delta$ between two UC2RPQs can be solved in non-deterministic space $\mathcal{O}(\|\Gamma\| + \|\Delta\|^{c \cdot n_\Delta})$, for some constant $c$, and where $n_\Delta$ is the maximal number of atoms of a disjunct of $\Delta$, namely $n_\Delta = \max \{\|\delta\|_{\mathrm{at}} \mid \delta \in \Delta\}$.

*Proof.* The proposition follows from the following claim.

*Claim* V.3.12 (implicit in [Fig20]). The CONTAINMENT PROBLEM $\Gamma \subseteq \Delta$ between two UC2RPQs can be solved in non-deterministic space $\mathcal{O}(\|\Gamma\| + \|\Delta\|^{c \cdot \mathrm{bw}(\Delta)})$, where $\mathrm{bw}(\Delta)$ is the bride-width of $\Delta$ and $c$ is a constant.

In the statement above, a *bridge* of a C2RPQ is a minimal set of atoms whose removal increases the number of connected components of the query, and the *bridge-width* of a C2RPQ is the maximum size of a bridge therein. The bridge-width of a union of C2RPQs is the maximum bridge-width among the C2RPQs it contains. In particular, the maximal number of atoms of a disjunct is an upper bound for bridge-width. $\qquad\square$

We provide an alternative upper bound in Proposition V.A.1 (Section V.A), which also yields a 2ExpSpace upper bound for Lemma V.3.10.

*Proof of Lemma V.3.10.* To test whether a query $\Gamma$ is of semantic tree-width $k$, it suffices to test the containment $\Gamma \subseteq \Gamma'$, where $\Gamma'$ is the maximal under-approximation $\bigcup_{\gamma \in \Gamma} \mathrm{App}_{\mathcal{T}w_k}^{\star, \leq \ell}(\gamma)$ given by Corollary V.3.9: a double-exponential union of single-exponential sized C2RPQs. Thus, by the bound of Proposition V.3.11 (and Savitch's Theorem), we obtain a double-exponential space upper bound. $\qquad\square$

Moreover, from the equivalences $\mathrm{App}_{\mathcal{T}w_k}(\gamma) \equiv \mathrm{App}_{\mathcal{T}w_k}^{\star}(\gamma)$ and $\mathrm{App}_{\mathcal{T}w_k}^{\star}(\gamma) \equiv \mathrm{App}_{\mathcal{T}w_k}^{\star, \leq \ell}(\gamma)$ of Corollary V.3.6 and Lemma V.3.8, we can derive new characterizations for queries of bounded semantic tree-width.

**Theorem V.3.13.** Assume that $\mathcal{L}$ is closed under sublanguages. For any $k > 1$ and any query $\Gamma \in \mathrm{UC2RPQ}(\mathcal{L})$, the following are equivalent:

(1) $\Gamma$ is equivalent to an infinitary union of conjunctive queries of tree-width at most $k$;

(2) $\Gamma$ has semantic tree-width at most $k$;

(3) $\Gamma$ is equivalent to a $\mathrm{UC2RPQ}(\mathcal{L})$ of tree-width at most $k$.

*Proof of Theorem V.3.13.* The implications (3) $\Rightarrow$ (2) $\Rightarrow$ (1) are straightforward: they follow directly from Fact V.2.1. For (1) $\Rightarrow$ (3), note that (1) implies that $\Gamma \equiv \mathrm{App}_{\mathcal{T}w_k}(\Gamma)$, and by Lemma V.3.8, $\mathrm{App}_{\mathcal{T}w_k}(\Gamma) \equiv \Delta \triangleq \bigvee_{\gamma \in \Gamma} \mathrm{App}_{\mathcal{T}w_k}^{\star, \leq \ell}(\gamma)$, so $\Gamma$ is equivalent to the latter. Since queries of $\Delta$ are obtained as homomorphic images of refinements of $\Gamma$, all of which are labelled by sublanguages of $\mathcal{L}$, and since $\mathcal{L}$ is closed under sublanguages, it follows that $\Gamma$ is equivalent to a UC2RPQ($\mathcal{L}$) of tree-width $k$. $\qquad\square$

**Remark V.3.14.** The statement of Theorem V.3.13 does not hold for $k = 1$.

(2) $\Rightarrow$ (1) when $k = 1$: consider the CRPQ $\gamma(x, y) = x \xrightarrow{a^*} y \wedge y \xrightarrow{b} x$ of tree-width 1, and hence of semantic tree-width 1, and observe that it is not equivalent to any infinitary union of conjunctive queries of tree-width 1—this can be proven by considering, for example, the expansion $x \xrightarrow{a} z \xrightarrow{a} y \wedge y \xrightarrow{b} x$ of $\gamma(x, y)$ and applying Proposition III.2.12.

(3) $\Rightarrow$ (2) when $k = 1$: by [BRV16, Proposition 6.4] the CRPQ of semantic tree-width 1 $\gamma(x) \triangleq x \xleftarrow{a} z \xrightarrow{a} y \wedge x \xrightarrow{b} y \equiv x \xrightarrow{ba^-a} x$ is not equivalent to any UCRPQ of tree-width 1. Hence, the implication is false when $\mathcal{L}$ is the class of regular languages over $\mathbb{A}^{\pm}$ that do not use any letter of the form $a^-$. $\qquad\diamond$

See Corollary V.7.9 for a similar (but different) characterization of queries of semantic tree-width at most 1. As an immediate corollary of Theorem V.3.13, by taking $\mathcal{L}$ to be the class of all regular languages over $\mathbb{A}$, we obtain the following result.

**Corollary V.3.15.** Let $k \geq 2$. A UCRPQ has semantic tree-width at most $k$ if and only if it has one-way semantic tree-width at most $k$.

Lastly, using Corollary V.3.9 as a black box, we can obtain an FPT algorithm for the EVALUATION PROBLEM.

**Corollary V.3.16** (FPT evaluation for bounded semantic tree-width). For each $k \geq 1$, the EVALUATION PROBLEM for C2RPQs of semantic tree-width at most $k$ is fixed-parameter tractable—FPT—when parameterized in the size of the query. More precisely on input $\langle \Gamma, G \rangle$, the algorithm runs in time $\mathcal{O}(f(\|\Gamma\|) \cdot |G|^{k+1} \cdot \log |G|)$ on a Turing machine, where $f$ is a doubly-exponential function—or $\mathcal{O}(f(\|\Gamma\|) \cdot |G|^{k+1})$ under a RAM model.

*Proof.* First, compute from $\Gamma$ its maximal under-approximation $\Gamma'$ using Corollary V.3.9 in single-exponential space, and hence double-exponential time. Then, evaluate $G$ on $\Gamma'$ using Proposition V.1.1. $\qquad\square$

This improves the database-dependency from the previously best (and first) known upper bound, which was $\mathcal{O}(f'(\|\Gamma\|) \cdot |G|^{2k+1})$ for a single-exponential $f'$ [RBV17, Theorem IV.11 & Lemma IV.13]. We discuss open questions related to this in Section V.10.2.

We are left with the proof of the Key Lemma. But before doing so, we will need to introduce in the next Section V.4 some basic notions that we will need in the proof, which is deferred to Section V.5.

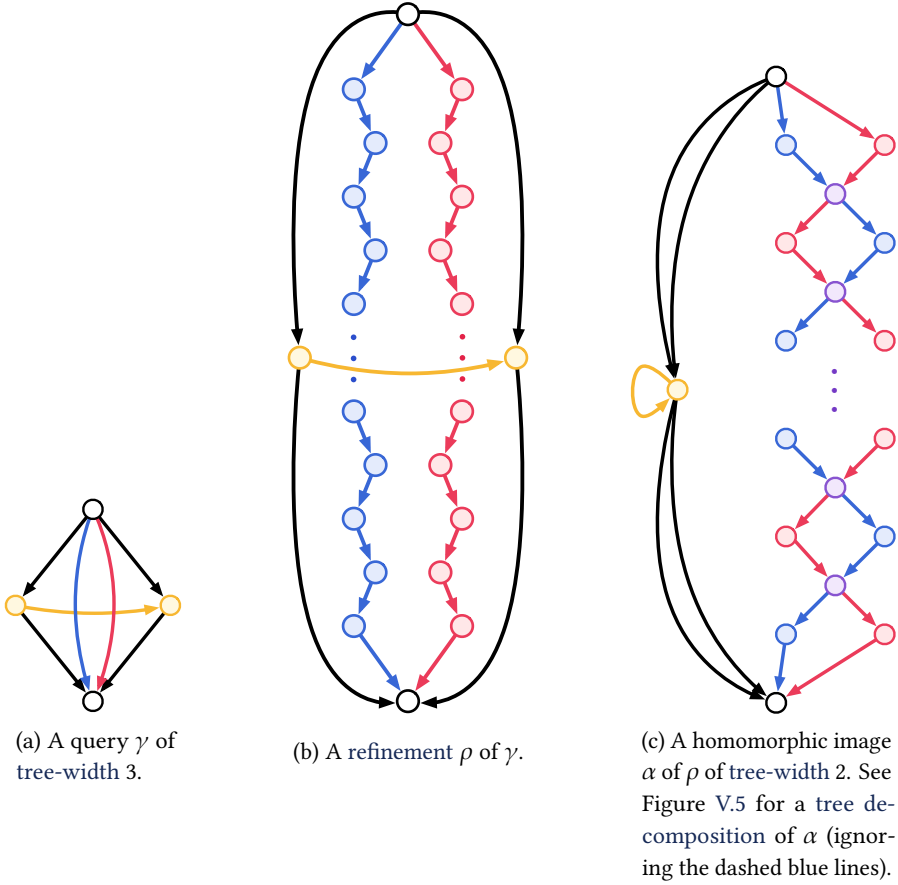

Figure V.4: An example of a homomorphism $f\colon \rho \twoheadrightarrow \alpha$. The strong onto homomorphism $f$ is implicitly defined: it sends the two yellow vertices of $\rho$ on the unique yellow vertex of $\alpha$, and identifies some blue and red vertices of $\rho$—thus creating purple vertices in $\alpha$.

(a) A query $\gamma$ of tree-width 3.

(b) A refinement $\rho$ of $\gamma$.

(c) A homomorphic image $\alpha$ of $\rho$ of tree-width 2. See Figure V.5 for a tree decomposition of $\alpha$ (ignoring the dashed blue lines).

## V.4  Intermezzo: Tagged Tree Decompositions

In this section we introduce some technical tools necessary for the proof of the Key Lemma. Remember that its statement deals with

$$\mathrm{App}^{\star,\leq m}_{\mathcal{T}w_k}(\gamma) \mathrel{\hat{=}} \{\alpha \in \mathcal{T}w_k \mid \exists \rho \in \mathrm{Ref}^{\leq m}(\gamma),\ \exists f\colon \rho \twoheadrightarrow \alpha\},$$

and consequently its proof needs to manipulate homomorphisms from refinements onto C2RPQs of tree-width $\leq k$. The proof will "massage" the homomorphism $f$ and queries $\alpha, \rho$ in order to reduce the size of $m$, while preserving (a) the existence of a homomorphism between the two queries, (b) the tree-width of the right-hand side, (c) the fact that the left-hand side is a refinement, and (d) some semantic properties of the queries. Our construction will be guided by the tree decomposition of $\alpha$, and more importantly by how $\rho$ is mapped onto such decomposition.

**Definition V.4.1.** Let $f\colon \rho \xrightarrow{\mathrm{hom}} \alpha$ be a homomorphism between two C2RPQs. A *tagged tree decomposition* of $f$ is a triple $(T, \mathbf{v}, \mathbf{t})$ where $\langle \mathbf{T}, \mathbf{v} \rangle$ is a tree decomposition of $\alpha$, and $\mathbf{t}$ is a mapping $\mathbf{t}\colon \mathrm{Atoms}(\rho) \to T$, called *tagging*, such that for each atom $e = x \xrightarrow{\lambda} y \in \mathrm{Atoms}(\rho)$, we have that $\mathbf{v}(\mathbf{t}(e))$ contains both $f(x)$ and $f(y)$.

In other words, $\mathbf{t}$ gives, for each atom of $\rho$, a witnessing bag that contains it, in the sense that it contains the image by $f$ of the atom's source and target.

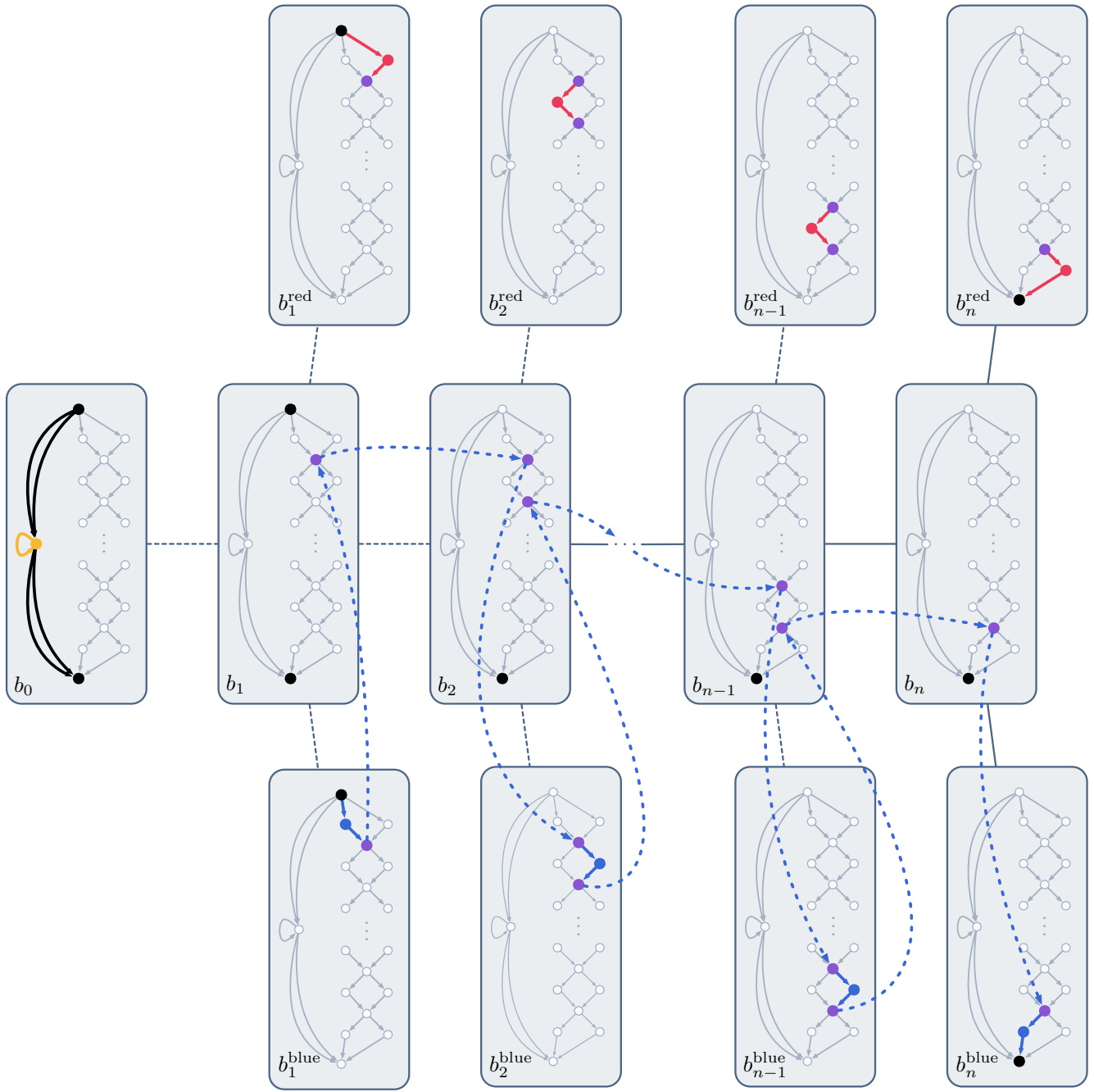

Figure V.5: A fine tagged tree decomposition of $\alpha$ (see Figure V.4) of width 2. (Recall that some bags are omitted for the sake of readability. These bags are there to make the decomposition fine.)

By definition, given a tree decomposition $\langle \mathbf{T}, \mathbf{v} \rangle$ of $\alpha$ and a homomorphism $f \colon \rho \twoheadrightarrow \alpha$, there is always one way (usually many) of extending $\langle \mathbf{T}, \mathbf{v} \rangle$ into a tagged tree decomposition of $f$.

We provide an example of homomorphism $f \colon \rho \twoheadrightarrow \alpha$ in Figure V.4. Note that in this example, $\rho$ is defined as the refinement of a query, and $f$ is strong onto—for now this is innocuous, but we will always work under these assumptions in Section V.5. In Figure V.5, we give a tagged tree decomposition of this homomorphism. Each bag is given a name, written in the bottom left corner. The tagging is represented as follows: if an atom is tagged in a bag, then it is drawn as a solid bold arrow in this bag. Note that by definition, a given atom is tagged in exactly one bag. For now, blue dashed arrow between bags can be ignored—they will illustrate Definition V.4.3.

**Fact V.4.2.** Let $(T, \mathbf{v}, \mathbf{t})$ be a tagged tree decomposition of some strong onto homomorphism $f \colon \rho \twoheadrightarrow \alpha$. Let $T'$ be the smallest connected subset of $T$ containing the image of $\mathbf{t}$. Then $(T', \mathbf{v}|_{T'}, \mathbf{t})$ is still a tagged tree decomposition of $f$, whose width is at most the width of $(T, \mathbf{v}, \mathbf{t})$.

In the following paragraphs, we extend the notion of tagging to paths. We illustrate this notion in Figure V.5, where we describe the path induced by the blue path of Figure V.4b—which starts at the top-most vertex, follows the blue atoms, and reaches the bottom-most vertex. Informally, in the context of a tagged tree decomposition $(T, \mathbf{v}, \mathbf{t})$ of $f \colon \rho \xrightarrow{\text{hom}} \alpha$, given a path $\pi$ of $\rho$, say $x_0 \xrightarrow{\lambda_1} x_1 \xrightarrow{\lambda_2} \cdots \xrightarrow{\lambda_n} x_n$, the path induced by $\pi$, denoted by $\mathbf{t}[\pi]$, is informally defined as the following "path" in $T \times \alpha$, seen as a sequence of pairs of bags and variables from $V(T) \times \text{vars}(\alpha)$:

- it starts with the bag $\mathbf{t}(x_0 \xrightarrow{\lambda_1} x_1)$ of $T$ and the variable $f(x_0)$ of $\alpha$; in Figure V.5, this corresponds to bag $b_1^{\text{blue}}$;
- it then goes to $\langle \mathbf{t}(x_0 \xrightarrow{\lambda_1} x_1), f(x_1) \rangle$;
- it then follows the shortest path in $T$ (unique, since it is a tree) that goes to the bag $\mathbf{t}(x_1 \xrightarrow{\lambda_2} x_2)$, while staying in $f(x_1)$ in $\alpha$—in Figure V.5, this bag is the same as before, namely $b_1^{\text{blue}}$, so we do nothing;
- then, it goes to $\langle \mathbf{t}(x_1 \xrightarrow{\lambda_2} x_2), f(x_2) \rangle$ in a single step;
- it then follows the shortest path in $T$ (unique, since it is a tree) that goes to the bag $\mathbf{t}(x_2 \xrightarrow{\lambda_3} x_3)$, while staying in $f(x_2)$ in $\alpha$—in our running example, we go from $b_i^{\text{blue}}$ to $b_i$, and then to $b_{i+1}$ before reaching $b_{i+1}^{\text{blue}}$;
- it continues in the same way for all other atoms of the path, ending up with the bag $\mathbf{t}(x_{n-1} \xrightarrow{\lambda_n} x_n)$ and the variable $f(x_n)$ of $\alpha$.

By construction, note that the constructed sequence $(b_i, z_i)_i$, also denoted by $(\binom{b_i}{z_i})_i$, is such that $z_i \in \mathbf{v}(b_i)$. Moreover, the values taken by the sequence $(z_i)_i$ are $(f(x_j))_{0 \le j \le n}$, in the same order but potentially with repetitions. Graphically, this sequence corresponds to a path in the tagged tree decomposition, where one can not only move along the bags, but also along the variables they contain. In our example, the path induced by the blue path of Figure V.4b corresponds in Figure V.5 to the blue path consisting of both solid and dashed edges. Moreover, note that a single atom $x_0 \xrightarrow{\lambda} x_1$ of $\rho$ induces the path:

$$\left\langle \left( t(x_0 \xrightarrow{\lambda} x_1) \atop x_0 \right), \left( t(x_0 \xrightarrow{\lambda} x_1) \atop x_1 \right) \right\rangle. \tag{V.3}$$

**Definition V.4.3** (Path induced in a tagged tree decomposition—formal definition)**.** Given a homomorphism $f \colon \rho \xrightarrow{\text{hom}} \alpha$ and a tagged tree decomposition $(T, \mathbf{v}, \mathbf{t})$ of $f$, the *link* from an atom $A = x \xrightarrow{\lambda} y$ to an atom $B = y \xrightarrow{\lambda'} z$ of $\rho$ is the unique (possibly empty) sequence $\left( b_1 \atop f(y) \right), \dots, \left( b_n \atop f(y) \right)$, where $\mathbf{t}(A), b_1, \dots, b_n, \mathbf{t}(B)$ is the unique simple path from $\mathbf{t}(A)$ to $\mathbf{t}(B)$ in $T$.

The *path induced* by a path $\pi = x_0 \xrightarrow{\lambda_1} x_1 \xrightarrow{\lambda_2} \cdots \xrightarrow{\lambda_n} x_n$ of $\rho$ is the unique sequence

$$\mathbf{t}[\pi] \triangleq \left( b_0 \atop f(x_0) \right) \left( b_0 \atop f(x_1) \right) L_1 \left( b_1 \atop f(x_1) \right) \left( b_1 \atop f(x_2) \right) L_2 \cdots \left( b_{n-2} \atop f(x_{n-2}) \right) L_{n-1} \left( b_{n-1} \atop f(x_{n-1}) \right) \left( b_{n-1} \atop f(x_n) \right)$$

where $b_i = \mathbf{t}(x_i \xrightarrow{\lambda_{i+1}} x_{i+1})$ and $L_i$ is the link from $x_{i-1} \xrightarrow{\lambda_i} x_i$ to $x_i \xrightarrow{\lambda_{i+1}} x_{i+1}$, for every $i$.

Moreover, given a bag $b$ of $T$ and a variable $z$ of $\alpha$, we say that $\mathbf{t}[\pi]$ *leaves* $b$ at $z$ when $\left( b \atop z \right)$ belongs to $\mathbf{t}[\pi]$, and this is either the last element of the sequence $\mathbf{t}[\pi]$, or the next element of the sequence has a bag distinct from $b$.

For example, in Figure V.5, $\mathbf{t}[\pi]$ leaves $b_1^{\text{blue}}$ at the first purple vertex. Similarly, it leaves $b_1$ and $b_2$ at this same vertex. Moreover, it also leaves $b_2$ at the second purple vertex.

We say that an induced path is *cyclic* if it contains two positions $i, j$ such that $i + 2 \leq j$ and $b_i = b_j$. We say that it is *acyclic* otherwise, meaning that if we visit a bag for the first time, we can visit it again at most once, in which case it must be precisely at the next time step. For instance, the path induced by the blue atom refinement in Figure V.5 is cyclic. However, the path induced by a single atom—see (V.3)— is always acyclic.

**Fact V.4.4.** If an induced path $\mathbf{t}[\pi]$ is acyclic, for any bag $b$, there is at most one variable $z$ of $\alpha$ such that $\mathbf{t}[\pi]$ leaves $b$ at $z$.

Lastly, we define a *fine tagged tree decomposition* of $f \colon \rho \xrightarrow{\text{hom}} \alpha$ to be a tagged tree decomposition of $f$ that is also a fine tree decomposition of $\alpha$. We abuse the notation and talk about the fine tagged tree decomposition of a C2RPQ $\gamma$ to talk about the fine tagged tree decomposition of the identity homomorphism $\text{id} \colon \gamma \twoheadrightarrow \gamma$.

One of the key properties of fine tagged tree decompositions is that in any of its *non-branching paths*—i.e. paths in $T$ whose non-extremal bags have degree exactly 2—, at least half of the bags are non-full, i.e. they contain at most $k$ variables[4]. Such bags will prove useful in the next section because of the following property.

**Proposition V.4.5.** Let $\gamma, \gamma'$ be C2RPQs, and $(T, \mathbf{v}, \mathbf{t})$—resp. $(T', \mathbf{v}', \mathbf{t}')$—be a fine tagged tree decomposition of width $k$ of $\gamma$—resp. of $\gamma'$. Let $b, b'$ be leaves of $T$ and $T'$ respectively, such that $b$ and $b'$ are non-full bags of the same

[4] Recall that in a decomposition of width $k$, bags are allowed to contain at most $k + 1$ variables.

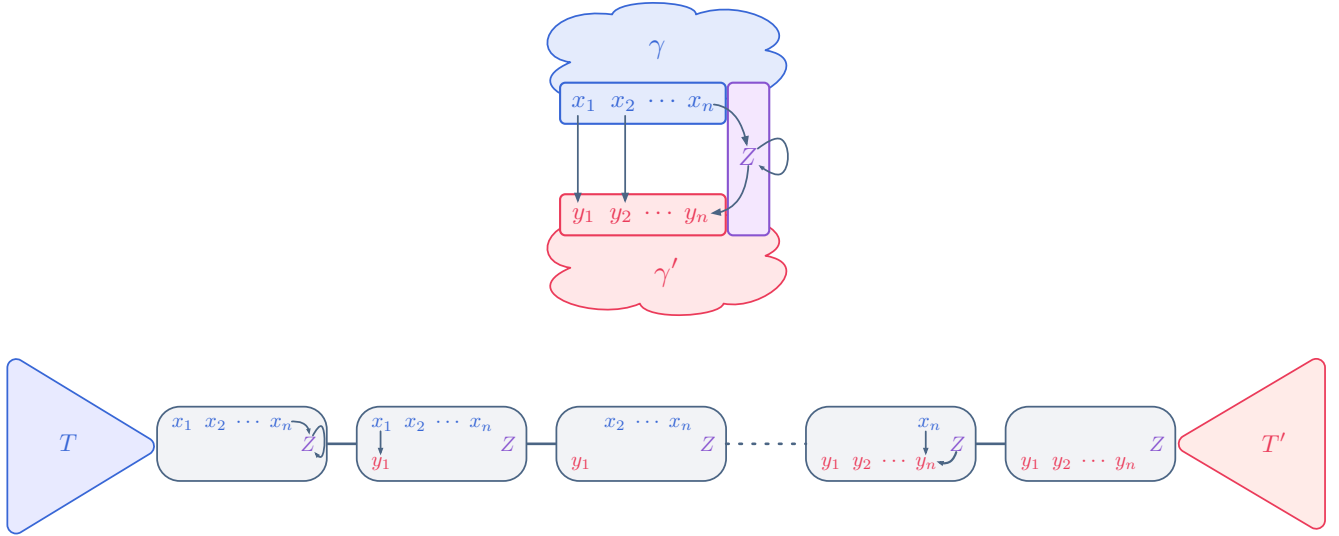

Figure V.6: The query $\gamma \wedge \gamma' \wedge \delta$ (top) and one of its fine tagged tree decomposition of width at most $k$ (bottom).

cardinality, and let $Z = \mathbf{v}(b) \cap \mathbf{v}'(b')$. In particular, we have

$$\mathbf{v}(b) = \{x_1, \dots, x_n\} \cup Z \quad \text{and} \quad \mathbf{v}'(b') = \{y_1, \dots, y_n\} \cup Z,$$

from some variables *s.t.* the $x_i$'s are disjoint from the $y_i$'s. Assume moreover that $\text{vars}(\gamma) \cap \text{vars}(\gamma') \subseteq Z$. Then, for any conjunction $\delta$ of atoms the form:

- $x_i \xrightarrow{L} y_i$ for some $i \in [\![1, n]\!]$,
- $x_i \xrightarrow{L} z$ for some $i \in [\![1, n]\!]$ and $z \in Z$,
- $z \xrightarrow{L} y_i$ for some $i \in [\![1, n]\!]$ and $z \in Z$,
- $z \xrightarrow{L} z'$ for some $z, z' \in Z$,

the query $\gamma \wedge \gamma' \wedge \delta$ has a fine tagged tree decomposition of width $k$ in which the length of the longest non-branching path is smaller than the sum of the longest non-branching paths of $T$ and of $T'$, plus $2n$.

The proof of Proposition V.4.5 is elementary and illustrated in Figure V.6.

*Proof.* We connect $T$ with $T'$ with $2n \leq 2k$ bags: start from $b_0 \triangleq b$, which contains $\{x_1, \dots, x_n\} \cup Z$. Then create the following bags:

- $\mathbf{v}(b_1) \triangleq \{x_1, x_2, \dots, x_n\} \cup \{y_1\} \cup Z = \mathbf{v}(b_0) \cup \{y_1\}$,
- $\mathbf{v}(b_2) \triangleq \{x_2, \dots, x_n\} \cup \{y_1\} \cup Z = \mathbf{v}(b_1) \smallsetminus \{x_1\}$,
- $\mathbf{v}(b_{2i-1}) \triangleq \{x_i, \dots, x_n\} \cup \{y_1, \dots, y_i\} \cup Z = \mathbf{v}(b_{2i-2}) \cup \{y_i\}$,
- $\mathbf{v}(b_{2i}) \triangleq \{x_{i+1}, \dots, x_n\} \cup \{y_1, \dots, y_i\} \cup Z = \mathbf{v}(b_{2i-1}) \smallsetminus \{x_i\}$

for $1 \leq i \leq n$, and observe that $\mathbf{v}(b_{2n}) = \mathbf{v}'(b')$. Then, tag every atom of $\delta$ in the first bag of $\langle b_1, \dots, b_{2n-1} \rangle$ containing both variables of the atom. Such a bag always exists:

- an atom of the form $x_i \xrightarrow{L} z$ is tagged in $b_0$;
- an atom of the form $z \xrightarrow{L} z'$ is tagged in $b_0$;
- an atom of the form $x_i \xrightarrow{L} y_i$ is tagged in $b_{2i-1}$;
- an atom of the form $z \xrightarrow{L} y_i$ is tagged in $b_{2i-1}$.

Observe that the decomposition obtained is indeed a fine tagged tree de-composition: in particular, it satisfies that for each variable $t$, the set of all $b \in T$ containing $t$ is a connected subtree of $T$, thanks to the assumption that $\mathrm{vars}(\gamma) \cap \mathrm{vars}(\gamma') \subseteq Z$. $\qquad\square$

## V.5 Key Lemma: Maximal Under Approximations are Semantically Finite

We can now start to describe the constructions used to prove the Key Lemma. Given a fixed C2RPQ $\gamma$ and a fixed $k \geq 1$, we call a *trio* any triple $(\alpha, \rho, f)$ such that $\alpha \in \mathcal{T}w_k$, $\rho \in \mathrm{Ref}(\gamma)$ and $f$ is a strong onto homomorphism from $\rho$ to $\alpha$. For clarity, we will denote such a trio by simply "$f \colon \rho \twoheadrightarrow \alpha$". Using this terminology, in order to prove Lemma V.3.8, it is sufficient (and necessary) to show that:

for every trio $f \colon \rho \twoheadrightarrow \alpha$, there exists another trio $f' \colon \rho' \twoheadrightarrow \alpha'$
s.t. $\alpha \subseteq \alpha'$ and $\rho' \in \mathrm{Ref}^{\leq \ell}(\gamma)$.

**Remark V.5.1.** Note that this section does not use the fact that $k \geq 2$. In particular, Lemma V.3.8 holds for $k = 1$. However, Corollary V.3.6 does not apply, and $\mathrm{App}_{\mathcal{T}w_1}(\gamma)$ (which we are interested in) is not equivalent to $\mathrm{App}^{\star}_{\mathcal{T}w_1}(\gamma)$ (which is shown to be computable by Lemma V.3.8). We discuss this case in further details in Section V.7. $\qquad\diamondsuit$

### V.5.1 Local Acyclicity

Our first construction, which will ultimately allow us to bound the size of atom refinements, shows that we can assume *w.l.o.g.* that they induce acyclic paths in a fine tagged tree decomposition of $f$.

**Lemma V.5.2.** For any trio $f \colon \rho \twoheadrightarrow \alpha$, there exists a trio $f' \colon \rho' \twoheadrightarrow \alpha'$ and a fine tagged tree decomposition $(T', \mathbf{v}', \mathbf{t}')$ of width at most $k$ of $f'$ such that $\alpha \subseteq \alpha'$, $\|\rho'\|_{\mathrm{at}} \leq \|\rho\|_{\mathrm{at}}$ and every atom refinement of $\rho'$ induces an acyclic path in the tree $T'$, in which case we say that $(T', \mathbf{v}', \mathbf{t}')$ is *locally acyclic w.r.t.* $f'$.

Note that the fact that $f'$ is a trio implies in particular that $\rho'$ is a refinement of $\gamma$. The construction behind Lemma V.5.2 is illustrated in Figure V.7.

**Notation V.5.3.** When two bags are linked by a dashed edge (as in Figures V.5 and V.7), it means that there is another bag in between them, which is there to ensure the fact that the decomposition is fine. The vertices contained in this extra bag are exactly the intersection of the vertices contained by its two neighbours, and no atom is tagged inside.

*Informal proof of Lemma V.5.2.* Start with a trio $f \colon \rho \twoheadrightarrow \alpha$, and let $(T, \mathbf{v}, \mathbf{t})$ be a fine tagged tree decomposition of $f$. Consider an atom refinement $\pi \triangleq z_0 \xrightarrow{L_1} z_1 \xrightarrow{L_2} \cdots \xrightarrow{L_n} z_n$ in $\rho$ of some atom $x \xrightarrow{L} y$ (with $z_0 \triangleq x$ and $z_n \triangleq y$), and assume that it induces a cyclic path in $T$— see *e.g.* Figure V.5.

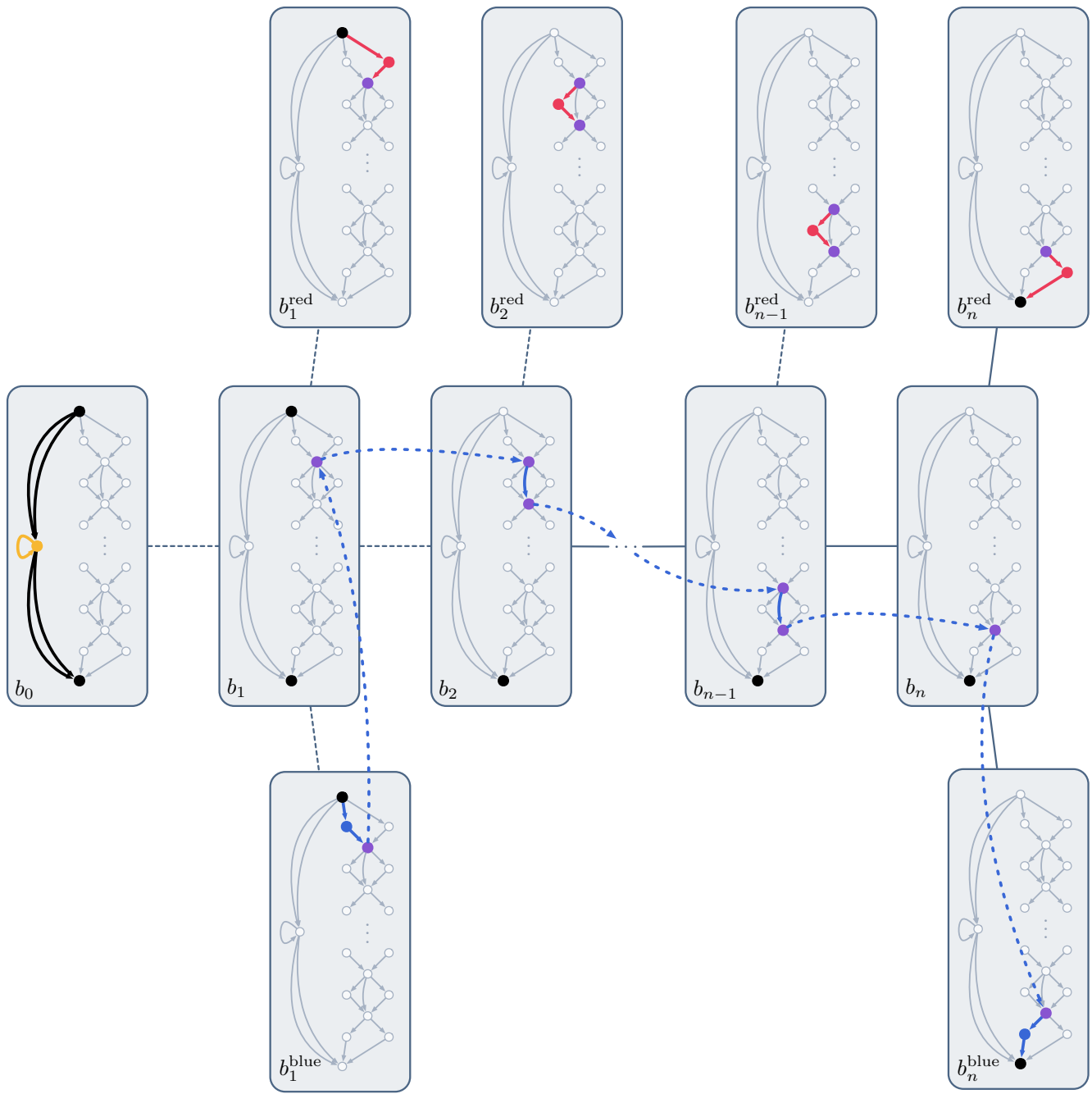

Figure V.7: Changing the blue atom refinement of $\gamma'$ (see Figure V.5) so that it induces an acyclic path.

It means that some variables $z_i$ and $z_j$ are mapped by $f$ to the same bag of $T$, somewhere along the path induced by $\pi$. It suffices then to condense $\rho$ by replacing the atoms $z_i \xrightarrow{L_{i+1}} \cdots \xrightarrow{L_j} z_j$ by a single atom $z_i \xrightarrow{[L_{i+1}\cdots L_j]} z_j$. We thus obtain a new refinement $\rho'$ of $\gamma$. Then define $\alpha'$ be simply adding an atom $f(z_i) \xrightarrow{L_{i+1}\cdots L_j} f(z_j)$. The definitions of $f'$ and $(T', \mathbf{v}', \mathbf{t}')$ are then straightforward—potentially, $\alpha'$ should be restricted to the image of $f': \rho' \xrightarrow{\text{hom}} \alpha'$ so that $f'$ is still strong onto by using Fact V.4.2. Crucially, $\alpha \subsetneq \alpha'$, and $\alpha'$ still has tree-width at most $k$ since we picked $f(z_i)$ and $f(z_j)$ so that they belonged to the same bag of $T$: therefore, adding an atom between them is innocuous. We then iterate this construction for every atom refinement. □

Figure V.7 shows the fine tagged tree decomposition $(T', \mathbf{v}', \mathbf{t}')$ obtained by applying the previous construction to the decomposition $(T, \mathbf{v}, \mathbf{t})$ of Figure V.5 for the blue atom refinement, followed by applying Fact V.4.2. In Figure V.5, the induced path was leaving the bag $b_2$ both at the first and at the second purple vertex. This leads in Figure V.7 to a new atom between these vertices. The same phenomenon happens to bags $b_3, \dots, b_{n-1}$. Lastly, note that because the atoms tagged in bags $b_2^{\text{blue}}, \dots, b_{n-1}^{\text{blue}}$ are not in the image of $f'$, these bags were removed by Fact V.4.2.

*Formal proof of Lemma V.5.2.* Let $\pi$ be an atom refinement in $\rho$ that induce a cyclic path in $T$, say

$$\pi = z_0 \xrightarrow{L_1} z_1 \xrightarrow{L_2} \cdots \xrightarrow{L_{n-1}} z_{n-1} \xrightarrow{L_n} z_n.$$

In order to build the trio $f': \rho' \twoheadrightarrow \alpha'$ and a fine tagged tree decomposition $T'$ of $f'$ of width at most $k$, we will mainly use the fact that if two vertices $(u, v)$ of some graph $G$ belong to the same bag of a tree decomposition $\langle \mathbf{T}, \mathbf{v} \rangle$ of $G$, then $\langle \mathbf{T}, \mathbf{v} \rangle$ is still also a tree decomposition of the graph obtained by adding an edge from $u$ to $v$.

By definition, the induced path $\mathbf{t}[\pi] = \left(\binom{b_i}{x_i}\right)_i$ is of the form

$$\mathbf{t}[\pi] = \left\langle \binom{b_{i_0}}{f(z_0)}, \binom{b_{i_0+1}}{f(z_1)}, \dots, \binom{b_{i_1}}{f(z_1)}, \binom{b_{i_1+1}}{f(z_2)}, \dots, \binom{b_{i_{n-1}}}{f(z_{n-1})}, \binom{b_{i_{n-1}+1}}{f(z_n)} \right\rangle,$$

where $i_0 \triangleq 0$, and for each $l$, $b_{i_l} = b_{i_l+1}$. Since it is not acyclic, there exists $(j, j')$ such that $j + 2 \leq j'$ and $b_j = b_{j'}$. Let $n(j)$ (resp. $n(j')$) denote the unique index such that $i_{n(j)-1} < j \leq i_{n(j)}$ (resp. $i_{n(j')-1} < j' \leq i_{n(j')}$). In particular, we have $f(z_{n(j)}) \in \mathbf{v}(b_j)$ and $f(z_{n(j')}) \in \mathbf{v}(b_{j'})$. We claim that $n(j) < n(j')$—otherwise, we would have twice the same bag in a link, which would contradict the fact that it is a simple path in $T$.

We can then define

$$\pi' \triangleq t_0 \xrightarrow{L_1} \cdots \xrightarrow{L_{n(j)}} t_{n(j)} \xrightarrow{K} t_{n(j')} \xrightarrow{L_{n(j')+1}} \cdots \xrightarrow{L_n} t_n,$$

where $K \triangleq [L_{n(j)+1} \cdots L_{n(j')}]$ (see Definition III.2.9) and let $\rho'$ be the query obtained from $\rho$ by replacing $\pi$ with $\pi'$. Then, define $\alpha'$ to be the query obtained from $\alpha$ by adding an atom $f(z_{n(j)}) \xrightarrow{K} f(z_{n(j')})$, so that by construction, we have $\alpha \subseteq \alpha'$, that $\rho' \in \mathrm{Ref}(\gamma)$ with $\|\rho'\|_{\mathrm{at}} \leq \|\rho\|_{\mathrm{at}}$ and $f$ induces a homomorphism $f' : \rho' \xrightarrow{\mathrm{hom}} \alpha'$.

We must then build a tagged tree decomposition $(T', \mathbf{v}', \mathbf{t}')$ of $f'$. First, we restrict $\alpha'$ to be the image of $f' : \rho' \xrightarrow{\mathrm{hom}} \alpha'$, in order to obtain a strong onto homomorphism. Then, starting from the tagged tree decomposition $(T, \mathbf{v}, \mathbf{t})$ of $f$, restrict $\mathbf{t}$ to the atoms $\mathrm{Atoms}(\rho') \setminus \{z_{n(j)} \xrightarrow{K} z_{n(j')}\} \subseteq \mathrm{Atoms}(\rho)$, and tag the atom $z_{n(j)} \xrightarrow{K} z_{n(j')}$ to the bag $b_j = b_{j'}$. This tree decomposition has the same width as $T$. Then, apply Fact V.4.2 to get rid of potentially useless bags.

Observe then that the path induced by $\pi'$ in $(T', \mathbf{v}', \mathbf{t}')$ is simply

$$\mathbf{t}'[\pi'] = \left\langle \binom{b_{i_0}}{f(z_0)}, \binom{b_{i_0+1}}{f(z_1)}, \dots, \binom{b_j}{f(z_{n(j)})}, \binom{b_{j'}}{f(z_{n(j')})}, \dots, \binom{b_{i_{n-1}}}{f(z_{n-1})}, \binom{b_{i_{n-1}+1}}{f(z_n)} \right\rangle$$

and thus $\mathbf{t}'[\pi']$ is strictly shorter than $\mathbf{t}[\pi]$ since $j + 2 \geq j'$, by definition of these indices. Finally, observe that if $(T, \mathbf{v}, \mathbf{t})$ is fine then so is $(T', \mathbf{v}', \mathbf{t}')$.

Overall, we built $f' : \rho' \twoheadrightarrow \alpha'$ together with a fine tagged tree decomposition $(T', \mathbf{v}', \mathbf{t}')$ of width at most $k$ where $\alpha \subseteq \alpha'$ (by Fact III.2.10), and $\rho' \in \mathrm{Ref}(\gamma)$ is such that $\|\rho'\|_{\mathrm{at}} \leq \|\rho\|_{\mathrm{at}}$, and for each atom of $\gamma$, the refinement of this atom in $\rho$ is exactly the same as the refinement of this atom in $\rho'$ except possibly for one atom, for which the path induced in $T'$ by its refinement in $\rho'$ is strictly shorter than the path induced in $T$ by its refinement in $\rho$. After iterating this construction as many times as needed, we obtain a trio as in the conclusion of Lemma V.5.2, which concludes our proof. □

### V.5.2 Short Paths

Ultimately, Lemma V.5.2 will allow us to give a bound on the number of leaves of a fine tagged tree decomposition of a trio. The following claim—which is significantly more technical than the foregoing—will give us a bound on the height of a decomposition.

**Lemma V.5.4.** Let $f : \rho \twoheadrightarrow \alpha$ be a trio and $(T, \mathbf{v}, \mathbf{t})$ be a locally acyclic fine tagged tree decomposition of width at most $k$ of $f$. Then there is a trio $f' : \rho' \twoheadrightarrow \alpha'$ and a fine tagged tree decomposition $(T', \mathbf{v}', \mathbf{t}')$ of width at most $k$ of $f'$ such that:

- $\alpha \subseteq \alpha'$,
- $(T', \mathbf{v}', \mathbf{t}')$ is locally acyclic w.r.t. $f'$, and
- the length of the longest non-branching path in $T$ is at most $\mathcal{O}(\|\gamma\|_{\mathrm{at}} \cdot (k + 1)^{\|\gamma\|_{\mathrm{at}}})$.

To prove Lemma V.5.4, we will try to find, in a long non-branching path, some kind of shortcut. The piece of information that is relevant to finding this shortcut is what we call the profile of a bag.

**Definition V.5.5** (Types and Profiles). Given a trio $f : \rho \twoheadrightarrow \alpha$ and a fine tagged tree decomposition $(T, \mathbf{v}, \mathbf{t})$ of $f$, for each bag $b$ of $T$, we say that:

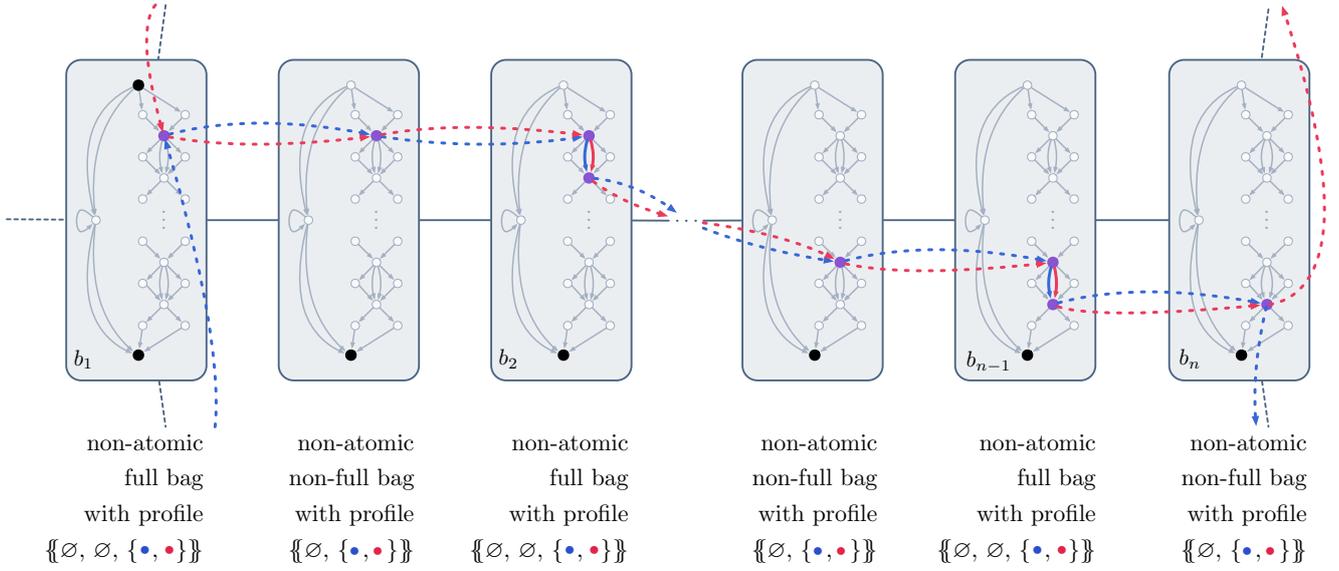

| non-atomic full bag with profile $\{\!\{\varnothing, \varnothing, \{\bullet, \bullet\}\}\!\}$ | non-atomic non-full bag with profile $\{\!\{\varnothing, \{\bullet, \bullet\}\}\!\}$ | non-atomic full bag with profile $\{\!\{\varnothing, \varnothing, \{\bullet, \bullet\}\}\!\}$ | non-atomic non-full bag with profile $\{\!\{\varnothing, \{\bullet, \bullet\}\}\!\}$ | non-atomic full bag with profile $\{\!\{\varnothing, \varnothing, \{\bullet, \bullet\}\}\!\}$ | non-atomic non-full bag with profile $\{\!\{\varnothing, \{\bullet, \bullet\}\}\!\}$ |

Figure V.8: Profiles of the bags in the non-branching path between $b_1$ and $b_n$ in the fine tagged tree decomposition obtained from Figure V.5 after applying Lemma V.5.2 and Fact V.4.2 to both the red and blue atom refinements.

- $b$ is "*atomic*" if there is at least one atom $e \in \mathbf{t}^{-1}[b]$ and at least one variable $x$ of $e$ such that $x \in \text{vars}(\gamma)$, *i.e.*, the atom $e$ is not in the 'middle' part of an atom refinement;

- otherwise, when $b$ is non-atomic, we assign to each variable $z \in \mathbf{v}(b) \subseteq \alpha$ a *type*

$$\text{type}_z^b \triangleq \left\{ x \xrightarrow{L} y \text{ atom of } \gamma \ \middle| \ \begin{array}{l} \text{the path induced by the atom refinement} \\ \text{of } x \xrightarrow{L} y \text{ in } \rho \text{ leaves } b \text{ at } z \end{array} \right\},$$

where each type is potentially the empty set. Then the *profile* of $b$ is the multiset of the types of $z$ when $z$ ranges over $\mathbf{v}(b)$.

Note that $\rho$ and $\alpha$ can have arbitrarily more atoms than the original query $\gamma$, and so the numbers of bags in $T$ can be arbitrarily high. However, only few of them can be atomic: an atom refinement of atom of $\gamma$ contains at most two atoms with a variable from $\gamma$—namely the first and the last atom in the refinement.

**Fact V.5.6.** There is at most $2\|\gamma\|_{\text{at}}$ atomic bags in $T$.

Consider the fine tree decomposition of Figure V.7, and now apply the construction of Lemma V.5.2 to the red atom refinement, followed by Fact V.4.2. We now obtain a non-branching path between bags $b_1$ and $b_n$. We depict it in Figure V.8: the implicit bags, hidden behind the dashed edges in Figure V.7 (see Notation V.5.3), are made explicit in this new figure, and, moreover, the rest of the fine tree decomposition is not drawn. Lastly, for each bag, we indicate if it is full and if it is atomic; when it is not atomic, we provide the profile of the bag.

The rest of the proof consists in two parts. First, we show that if two non-atomic bags $b$ and $b'$ occurring in some non-branching path of $T$ have the same profile, then we can essentially replace the path between $b$ and $b'$ by a path of constant length (Section V.5.2). And second, we show that in every sufficiently long non-branching path we can find $b$ and $b'$ satisfying the aforementioned property: this part simply relies on an enhanced "pigeon-hole principle" (Fact V.5.7).

Let $f\colon \rho \twoheadrightarrow \alpha$ be a trio, and consider a fine tagged tree decomposition of $f$ which is locally acyclic. Suppose there are two bags $b$ and $b'$ such that:

1. they contain at most $k$ nodes (*i.e.*, not full bags),
2. they have the same profile,
3. there is a non-branching path in $T$ between these bags, and
4. no bags of the path between $b$ and $b'$ (both included) are atomic.

Then, there exists a trio $f'\colon \rho' \twoheadrightarrow \alpha'$ and a fine tagged tree decomposition of $f'$ of width at most $k$ that can be obtained by replacing the non-branching path between $b$ and $b'$ in the fine tagged tree decomposition of $f$ by another non-branching path with at most $2k+1$ bags, such that $\alpha \subseteq \alpha'$. The proof of Section V.5.2 relies on the definition of profile, which was specifically designed so that we can condense every refinement between $b$ and $b'$, while preserving every needed property of the trio. We give first an informal and then a formal proof of Section V.5.2, which are illustrated in Figure V.9.

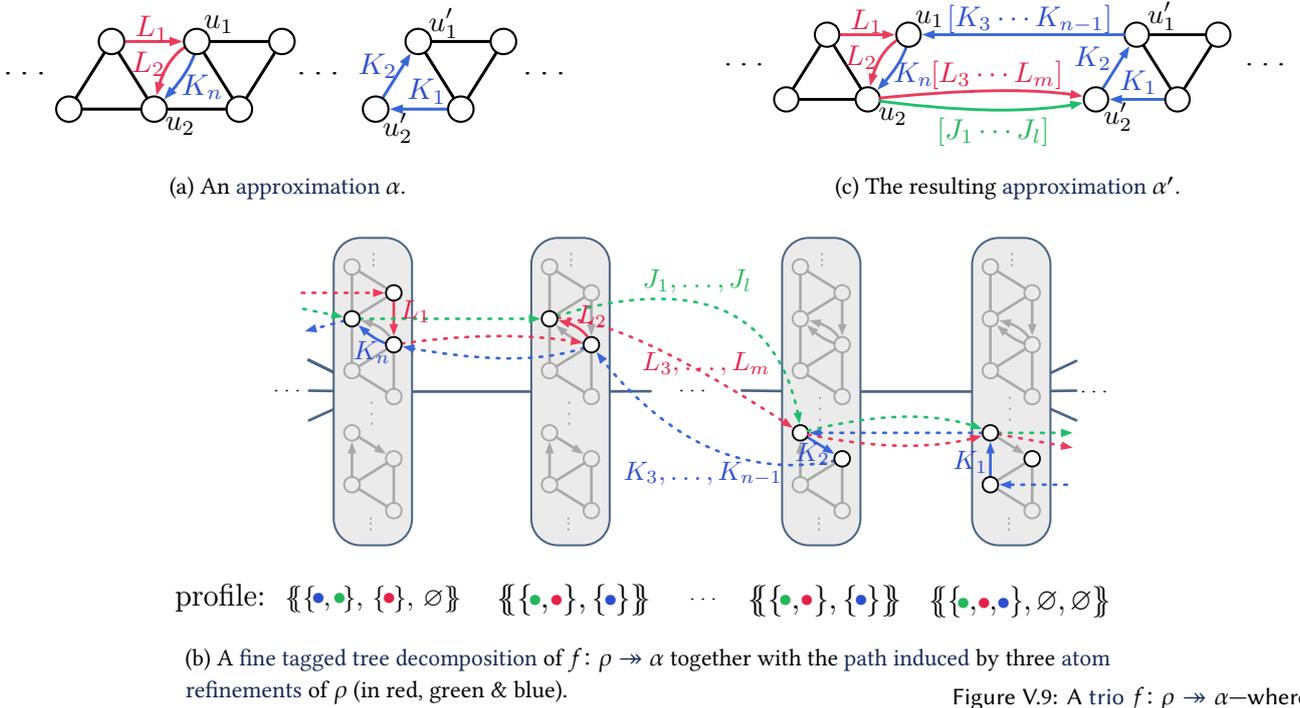

(a) An approximation $\alpha$.

(c) The resulting approximation $\alpha'$.

(b) A fine tagged tree decomposition of $f\colon \rho \twoheadrightarrow \alpha$ together with the path induced by three atom refinements of $\rho$ (in red, green & blue).

profile:   $\{\!\{\{\bullet,\bullet\}, \{\bullet\}, \varnothing\}\!\}$    $\{\!\{\{\bullet,\bullet\}, \{\bullet\}\}\!\}$   $\cdots$   $\{\!\{\{\bullet,\bullet\}, \{\bullet\}\}\!\}$   $\{\!\{\{\bullet,\bullet,\bullet\}, \varnothing, \varnothing\}\!\}$

Figure V.9: A trio $f\colon \rho \twoheadrightarrow \alpha$—where $\rho$ and $f$ are implicit— together with one of its fine tagged tree decomposition (Figures V.9a and V.9b). There are two non-full bags in the path with the same profile, and thus the query $\alpha$ can be simplified to $\alpha'$ (see Figure V.9c) by applying condensations to the atom refinements involved.

*Informal proof of Section V.5.2.* If $b$ and $b'$ have the same profile, then in particular they have the same cardinality $m$, which is smaller or equal to $k$ by assumption. Let $\mathbf{v}(b) = \{x_1, \dots, x_m\}$ and $\mathbf{v}(b') = \{y_1, \dots, y_m\}$ be such that: $\mathrm{type}_{x_i}^{b} = \mathrm{type}_{y_i}^{b'}$ for all $1 \leq i \leq m$. Note that the $x_i$'s don't need to be distinct from the $y_i$'s. Essentially, we can then condense every atom refinement in $\rho$ of some atom occurring in a set of the form $\mathrm{type}_{x_i}^{b} = \mathrm{type}_{y_i}^{b'}$ for some $i$. At this point, bags strictly comprised between $b$ and $b'$ are discarded, and so are variables of $\alpha$ that do not occur anywhere else. We are left with two halves of a fine tagged tree decomposition that we need to merge, which can easily be done by using Proposition V.4.5. The construction makes use of some crucial ingredients to guarantee its correctness.

- First, an atom $y \xrightarrow{L} y'$ of $\gamma$ cannot occur in two different types, allowing us to do the condensation of each atom refinement independently—this property is guaranteed by the fact that we started with a locally acyclic fine tagged tree decomposition, so an atom of $\gamma$ cannot leave a given bag at two different variables, by Fact V.4.4.

- Second, this condensation forces us to add new atoms in $\alpha$ (to preserve the existence of a homomorphism from the refinement to the approximation) from some variables of $\mathbf{v}(b)$ to some variables of $\mathbf{v}(b')$, but we only add edges from $x_i$ to $y_i$, and never from $x_i$ to $y_j$ with $i \neq j$. This allows us to preserve the tree-width of the approximation by using Proposition V.4.5. □

*Formal proof of Section V.5.2.* Let

$$\mathbf{v}(b) = \{x_1, \dots, x_m\} \text{ and } \mathbf{v}(b') = \{y_1, \dots, y_m\}$$

be as in the informal proof. Note that given an atom $x \xrightarrow{L} y$ of $\gamma$ and a bag, there is at most one variable of $\alpha$ s.t. $x \xrightarrow{L} y$ is in the type of this variable at this bag, by Fact V.4.4.

For every atom $x \xrightarrow{L} y$ of $\gamma$, let $\pi(x \xrightarrow{L} y) \mathrel{\hat=} t_0 \xrightarrow{L_1} \dots \xrightarrow{L_n} t_n$ be its refinement in $\rho$. If $x \xrightarrow{L} y$ is not in some type of the profile of $b$ (or equivalently, of $b'$), leave it as is. Otherwise, let $i$ (resp. $j$) be the unique index (by acyclicity) such that $\mathbf{t}[t_0 \xrightarrow{L_1} \dots \xrightarrow{L_n} t_n]$ leaves $b$ at $f(t_i)$ (resp. leaves $b'$ at $f(t_j)$). Define

$$\pi'(x \xrightarrow{L} y) \mathrel{\hat=} t_0 \xrightarrow{L_1} \dots \xrightarrow{L_i} t_i \xrightarrow{[L_{i+1} \cdots L_j]} t_j \xrightarrow{L_{j+1}} \dots \xrightarrow{L_n} t_n$$

when $i \leq j$ and otherwise the definition is symmetric. Then, let $\rho'$ be the refinement of $\gamma$ obtained by simultaneously substituting $\pi(x \xrightarrow{L} y)$ with $\pi'(x \xrightarrow{L} y)$ in $\rho$, for every atom $x \xrightarrow{L} y$ of $\gamma$.

Then, let $\alpha'$ be the query obtained by first adding the atoms

$$f(t_i) \xrightarrow{[L_{i+1} \cdots L_j]} f(t_j),$$

and observe that $f \colon \rho \twoheadrightarrow \alpha$ induces a homomorphism $f' \colon \rho' \xrightarrow{\text{hom}} \alpha'$—in particular, note that because of assumption (4) of our claim, we could not have removed images of free variables of $\gamma$. Moreover, by construction, $\alpha \subseteq \alpha'$ (by Fact III.2.10). As usual, we restrict $\alpha'$ to the image of $f'$ so that it becomes strong onto, while preserving that fact that $\alpha \subseteq \alpha'$. Finally, we build a tagged tree decomposition $(T', \mathbf{v}', \mathbf{t}')$ of $f'$ by applying Proposition V.4.5; it can be applied because:

- by assumption (1) and (2) of the claim, both bags have the same cardinality $m \leq k$;

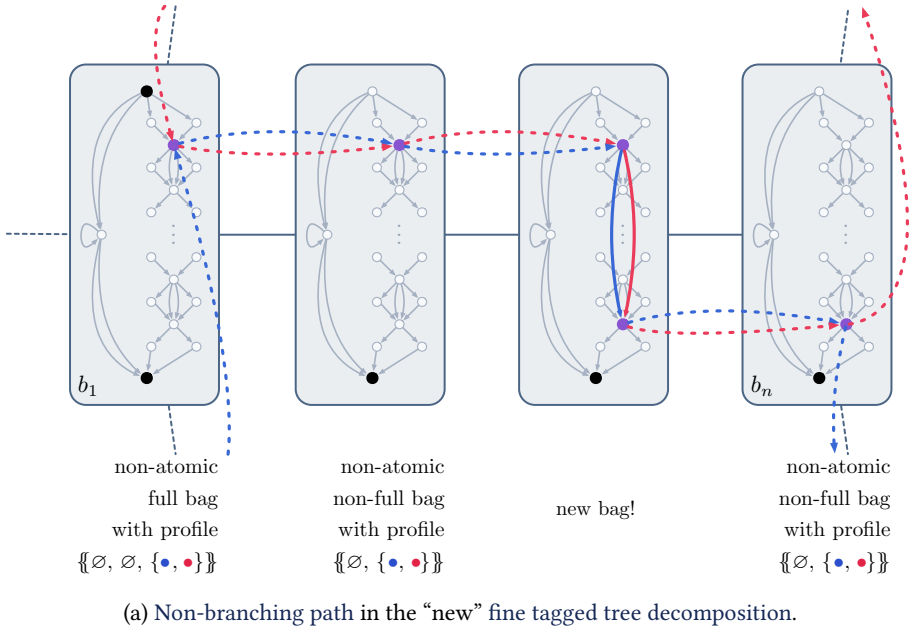

Figure V.10: Trio resulting from applying Section V.5.2 between the second and last bags of Figure V.8.

non-atomic
full bag
with profile
$\{\!\{\varnothing,\ \varnothing,\ \{\bullet,\bullet\}\}\!\}$

non-atomic
non-full bag
with profile
$\{\!\{\varnothing,\ \{\bullet,\bullet\}\}\!\}$

new bag!

non-atomic
non-full bag
with profile
$\{\!\{\varnothing,\ \{\bullet,\bullet\}\}\!\}$

(a) Non-branching path in the "new" fine tagged tree decomposition.

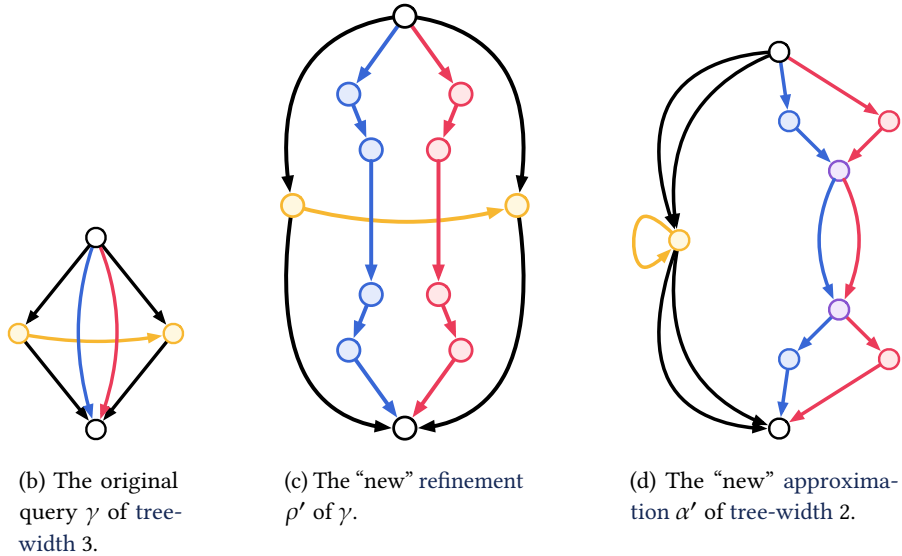

(b) The original query $\gamma$ of tree-width 3.

(c) The "new" refinement $\rho'$ of $\gamma$.

(d) The "new" approximation $\alpha'$ of tree-width 2.

- the variables in common between the first and second half of the decomposition are necessarily included in $Z \triangleq \mathbf{v}(b) \cap \mathbf{v}(b')$ since we started from a tree decomposition;
- we only add atoms from $x_i$ to $y_i$: depending on whether $x_i \in^? Z$, and whether $y_i \in^? Z$, we fall in one of the four types of atoms allowed by Proposition V.4.5.

This concludes the proof of Section V.5.2. □

In Figure V.10a, we depict the non-branching path (the rest of the fine tree decomposition is not depicted as it is left unchanged) obtained by applying the construction used to prove Section V.5.2 between the second and last bag of Figure V.8. Observe that a non-branching path of size $\mathcal{O}(n)$ is replaced, by this procedure, by a path with three bags. Then, after applying Fact V.4.2, we obtain a trio depicted in Figures V.10b to V.10d.

Before moving to the proof of Lemma V.5.4, we establish one last result.

**Fact V.5.7.** Let $n, d, t \in \mathbb{N}$. Let $P$ be a set with at most $n$ elements, and $\bar{P}$ be the disjoint union of $P$ and {trap, avoid}. For every natural number $m \geq 2(t+1)d(n+1) + 2t$, for every sequence $(p_i)_{0 \leq i < m} \in \bar{P}^m$ containing at most $t$ elements equal to trap, if at most half of the elements of the sequence are equal to avoid, then there exists $i < i'$ such that $p_i = p_{i'} \neq$ avoid, $i' - i \geq d$ and $p_j \neq$ trap for every $i \leq j \leq i'$.

*Proof.* First extract from $(x_i)_{0 \leq i < m}$ the subsequence of elements distinct from avoid, of length at least $\lceil \frac{m}{2} \rceil \geq (t+1)d(n+1) + t$. Then extract from it contiguous subsequences that avoid the trap element. Since there is at most $t+1$ subsequences like this, one of them must have size at least $d(n+1)$. Denote by $(y_i)_{0 \leq i < d(n+1)}$ the prefix of such a subsequence. Applying the pigeon-hole principle to $(y_{i\cdot d})_{0 \leq i < n+1}$ yields the desired result. $\square$

*Proof of Lemma V.5.4.* Let $f \colon \rho \twoheadrightarrow \alpha$ be a trio, and $(T, \mathbf{v}, \mathbf{t})$ be a locally acyclic fine tagged tree decomposition of $f$. If there is a non-branching path $(b_i)_{0 \leq i < m}$ in $T$ of length at least $m$, let $(p_i)_{0 \leq i < m}$ be the sequence defined by letting:

$$p_i \triangleq \begin{cases} \text{trap} & \text{if } b_i \text{ is atomic,} \\ \text{avoid} & \text{if } b_i \text{ contains } k+1 \text{ variables,} \\ \text{profile of } b_i & \text{otherwise.} \end{cases}$$

Observe that, by Fact V.4.4, profiles can be seen encoded as partial functions from the set of atoms of $\gamma$ to $[\![1, k]\!]$—of course this encoding is not surjective—, so there are at most $(k+1)^{\|\gamma\|_{\mathrm{at}}}$ different profiles on bags with at most $k$ variables. Applying Fact V.5.7 for $n = (k+1)^{\|\gamma\|_{\mathrm{at}}}$, $d = 2k+1$, $t = 2\|\gamma\|_{\mathrm{at}}$ yields, under the assumption that

$$m \geq m_0 \triangleq 2(2\|\gamma\|_{\mathrm{at}} + 1)(2k+1)((k+1)^{\|\gamma\|_{\mathrm{at}}} + 1) + 4\|\gamma\|_{\mathrm{at}},$$

the existence of indices $i < i'$ such that $i' - i \geq 2k+1$, and $b_i$ and $b_{i'}$ have the same profile, contain at most $k$ variables, and every bag $b_j$ for $i \leq j \leq i'$ is non-atomic—note that the hypothesis of Fact V.5.7 are satisfied since at most $t = 2\|\gamma\|_{\mathrm{at}}$ bags of $(b_i)_{0 \leq i < m}$ are atomic (*cf.* Fact V.5.6), and assuming *w.l.o.g.* that no two consecutive bags of $(b_i)_{0 \leq i < m}$ are identical, since the tagged tree decomposition $(T, \mathbf{v}, \mathbf{t})$ of width $k$ is fine, at most half of the bags contain $k+1$ variables. The assumption $i' - i \geq 2k+1$ means that the path from $b_i$ to $b_{i'}$ has length at least $2k+2$, and thus applying Section V.5.2 will strictly shorten this path. Note that Section V.5.2 preserves the fineness of the tagged tree decomposition, its local acyclicity, and that the size of this tree decomposition is strictly smaller (in number of nodes) than the original tree decomposition. By iteratively applying this construction, we obtain a trio $f' \colon \rho' \twoheadrightarrow \alpha'$ together with a locally acyclic fine tagged tree decomposition $T'$ of width at most $k$, such that $\alpha \subseteq \alpha'$ (by a variation of Fact III.2.10) and every non-branching path of $T'$ has length at most[5] $m_0 - 1 \in \mathcal{O}(\|\gamma\|_{\text{at}} \cdot (k+1)^{\|\gamma\|_{\text{at}}})$. □

[5] Recall that $k$ is fixed.

### V.5.3 Proof of Lemma V.3.8

Finally, our main lemma follows from Lemmas V.5.2 and V.5.4.

*Proof of Lemma V.3.8.* In order to show $\mathrm{App}_{\mathcal{T}w_k}(\gamma) \subseteq \mathrm{App}^{\star,\leq\ell}_{\mathcal{T}w_k}(\gamma)$—the other containment being trivial—, pick a trio $f\colon \rho \twoheadrightarrow \alpha$. Applying Lemma V.5.2 and then Lemma V.5.4 yields the existence of a trio $f'\colon \rho' \twoheadrightarrow \alpha'$ together with a fine tagged tree decomposition $(T', \mathbf{v}', \mathbf{t}')$ of $f'$ such that $\alpha \subseteq \alpha'$ and $(T', \mathbf{v}', \mathbf{t}')$ is locally acyclic, and any non-branching path in $T'$ has length at most $\mathcal{O}(\|\gamma\|_{\text{at}} \cdot (k+1)^{\|\gamma\|_{\text{at}}})$.

Moreover, we can assume *w.l.o.g.*, by applying Fact V.4.2, that every leaf of $T'$ is tagged by at least one atom of $\rho'$. The local acyclicity of $T'$ implies that if $b$ is a leaf of $T'$, and $\pi \triangleq x \xrightarrow{L_1} t_1 \xrightarrow{L_2} \cdots \xrightarrow{L_{n-1}} t_{n-1} \xrightarrow{L_n} y$ is an atom refinement in $\rho'$ of some atom $x \xrightarrow{L} y$ of $\gamma$, then if $b$ is tagged by one atom of $\pi$ this atom must either be $z_0 \xrightarrow{L_1} z_1$ or $z_{n-1} \xrightarrow{L_n} z_n$ by local acyclicity. The number of such atoms in $\rho'$ being bounded by $2\|\gamma\|_{\text{at}}$, we conclude that $T'$ has at most $2\|\gamma\|_{\text{at}}$ leaves.

Then, observe that a tree with at most $p$ leaves and whose non-branching paths have length at most $q$ is of height at most[6] $p \cdot q - 1$. We conclude that the height of $T'$ is $\mathcal{O}(\|\gamma\|^2_{\text{at}} \cdot (k+1)^{\|\gamma\|_{\text{at}}})$. Using again the local acyclicity of $T'$, observe that the refinement length of $\rho'$ is at most twice the height of $T'$, and hence $\rho' \in \mathrm{Ref}^{\leq\ell}(\gamma)$ where $\ell = \Theta(\|\gamma\|^2_{\text{at}} \cdot (k+1)^{\|\gamma\|_{\text{at}}})$. In other words, $\alpha' \in \mathrm{App}^{\star,\leq\ell}_{\mathcal{T}w_k}(\gamma)$. Hence, we have shown that for all $\alpha \in \mathrm{App}^{\star}_{\mathcal{T}w_k}(\gamma)$, there exists $\alpha' \in \mathrm{App}^{\star,\leq\ell}_{\mathcal{T}w_k}(\gamma)$ such that $\alpha \subseteq \alpha'$. □

[6] The length of a path being its number of nodes, and with the convention that the height of a single node is zero.

This concludes Section V.5 and the proof of the Key Lemma. The next four sections are independent of one another:

- in Section V.6, we show that the 2ExpSpace complexity of the semantic tree-width $k$ problem can be dropped down to $\Pi^p_2$ under assumptions on the regular languages;

- in Sections V.7 and V.8, we adapt the proofs of this section to deal with semantic tree-width 1 and semantic path-width $k$, respectively;

- in Section V.9, we prove an ExpSpace lower bound for the semantic tree-width $k$ problem and semantic path-width $k$ problems;

## V.6 Semantic Tree-Width for Simple Queries

We devote this section to showing the following result.

**Theorem V.6.1.** *For $k \geq 2$, the semantic tree-width $k$ problem for UCRPQ(SRE) is in $\Pi^p_2$.*

Observe that simple regular expressions are closed under sublanguages. Hence, in the light of Theorem V.3.13, the maximal under-approximation of a UCRPQ(SRE) query by infinitary unions of CQs of tree-width $k$ is always equivalent to a UCRPQ(SRE) query of tree-width $k$. We will see how the construction of the maximal under-approximation of the previous section can be exploited to improve the complexity from 2ExpSpace down to $\Pi_2^p$.

### V.6.1 Summary Queries

We will first show that the maximal under-approximation of tree-width $k$ of a UC2RPQ can be expressed as a union of polynomial sized "summary" queries. Each summary query represents a union of exponentially-bounded C2RPQs sharing some *common structure*. Summary queries are normal UC2RPQ queries extended with some special kind of atoms, called "path-$l$ approximations". Intuitively, they represent a maximal under-approximation of tree-width $l$ of queries of the form $\bigwedge_i x_i \xrightarrow{L_i} y_i$ such that $x_i \neq y_j$ for all $i, j$. Path-$l$ approximations may require an exponential size when represented as UC2RPQs. Formally, a *path-$l$ approximation* is a query of the form "$\mathbf{P}_l(X, Y, \delta)$" where:

1. $X, Y$, are two disjoint sets of variables of size at most $l$,

2. $\delta(\bar{z})$ is a conjunction of atoms $\bigwedge_{1 \leq i \leq n} A_i$ where $\bar{z}$ contains all variables of $X \cup Y$,

3. each $A_i$ is a C2RPQ atom of the form $x \xrightarrow{L} y$ or $y \xrightarrow{L} x$ such that $x \in X$, $y \in Y$, and $L_i$ is a regular language over $\mathbb{A}$.

We give the semantics of $\mathbf{P}_l(X, Y, \delta)$ in terms of infinitary unions of C2RPQs. A query like the one before is defined to be equivalent to the (infinitary) union of all queries $\alpha(\bar{z}) \in \mathrm{App}_{\mathcal{P}w_l}(\delta)$ such that

$$\alpha \text{ has a path decomposition of width } l \text{ where } X \text{ is the root and } Y \text{ is the leaf,} \tag{V.4}$$

that is, the root and leaf bags contain precisely the vertices of $X$ and $Y$, respectively. See Figure V.11 for an example.

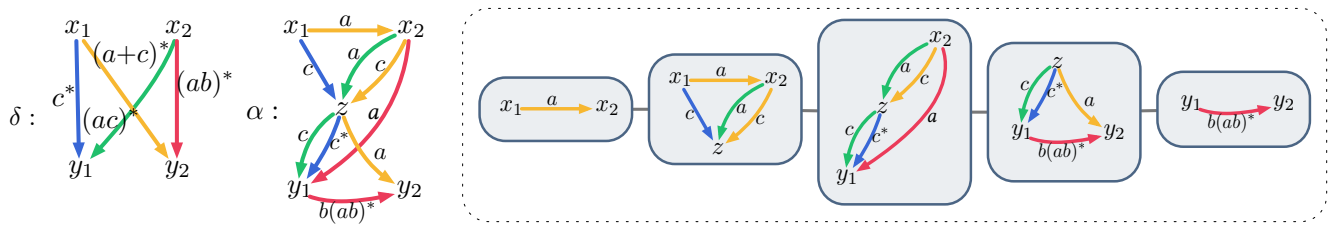

We now simply define a *k-summary query* as a C2RPQ extended with path-$l$ approximation atoms for any $l \leq k$, with the expected semantics. A *refinement* of a k-summary query is any C2RPQ obtained by replacing atoms with atom refinements, and each path-$l$ approximation $\mathbf{P}_l(X, Y, \delta)$ with any $\alpha(\bar{z}) \in \mathrm{App}_{\mathcal{P}w_k}(\delta)$ verifying (V.4). By definition, a database satisfies a k-summary query if and only if it satisfies one of its refinements.

A *tree decomposition* of a k-summary query $\gamma$ consists of a pair $\langle \mathbf{T}, \mathbf{v} \rangle$ with

Figure V.11: Consider the path-$l$ approximation $\mathbf{P}_l(\{x_1, x_2\}, \{y_1, y_2\}, \delta)$ where $l = 2$ and $\delta$ is depicted on the left. Its semantics contains the approximation $\alpha(x_1, x_2, y_1, y_2) \in \mathrm{App}_{\mathcal{P}w_l}(\delta)$ depicted in the middle because it has path decomposition of width $l$ verifying (V.4), as shown on the right.

$\mathbf{v}: T \to \mathfrak{P}(\text{vars}(\gamma))$ such that:

- for every classical atom $x \xrightarrow{L} y$ in $\gamma$, there is a bag $b \in T$ such that $\{x, y\} \subseteq \mathbf{v}(b)$;
- for every path-$l$ approximation $\mathbf{P}_l(X, Y, \delta)$ in $\gamma$, there are two adjacent bags $b, b' \in T$ such that $X \subseteq \mathbf{v}(b)$ and $Y \subseteq \mathbf{v}(b')$.

The width is defined as usual. Then, by Fact V.2.1, we obtain the following upper bound.

**Fact V.6.2.** For any $k \geq 2$, any refinement of a $k$-summary query with a tree decomposition of width at most $k$ is a C2RPQ of tree-width at most $k$.

Lastly, a *homomorphism* from a C2RPQ $\gamma(\bar{z}) = \bigwedge_i x_i \xrightarrow{L} y_i$ to a summary query $\delta(\bar{z}') = \left( \bigwedge_j x'_j \xrightarrow{L'_j} y'_j \right) \wedge \left( \bigwedge_{j'} \mathbf{P}_l(X_{j'}, Y_{j'}, \delta_{j'}) \right)$ consists of a mapping $f$ from variables of $\gamma$ to variables of $\delta$ such that $f(\bar{z}) = \bar{z}'$, and for each $i$, there is an atom $f(x_i) \xrightarrow{L} f(y_i)$ in $\delta$. Note that if there is a homomorphism from $\gamma(\bar{z})$ to $\delta(\bar{z}')$, then $\delta(\bar{z}') \subseteq \gamma(\bar{z})$.

Let us fix $\mathcal{L}$ to be any class closed under sublanguages. For every $\gamma \in$ C2RPQ($\mathcal{L}$), we define $\text{App}^{\text{zip}}_{\mathcal{T}w_k}(\gamma)$ as the set of all $k$-summary queries $\alpha$ such that:

i. $\alpha$ has a fine tagged tree decomposition $\langle \mathbf{T}, \mathbf{v} \rangle$ of width at most $k$,

ii. there exists a strong onto homomorphism from a refinement $\rho$ of $\gamma$ to $\alpha$,

iii. $T$ has at most $2\|\gamma\|_{\text{at}}$ leaves, and every non-branching path of $T$ consisting only of non-atomic bags must contain at most two non-full bags.

Note that since $\alpha$ is a homomorphism image a refinement of $\gamma$, and since $\mathcal{L}$ is closed under sublanguages, then $\alpha$ has only $\mathcal{L}$-labelled atoms.

**Lemma V.6.3.** Let $k \geq 2$. For every **finite** class $\mathcal{L}$ closed under sublanguages, and for every $\gamma \in$ C2RPQ($\mathcal{L}$), we have:

1. $\text{App}^{\text{zip}}_{\mathcal{T}w_k}(\gamma) \equiv \text{App}_{\mathcal{T}w_k}(\gamma)$,

2. $\text{App}^{\text{zip}}_{\mathcal{T}w_k}(\gamma)$ is a union of polynomial-sized $k$-summary queries having only $\mathcal{L}$-labelled atoms, and

3. one can test in NP if a summary query is part of this union.

*Proof.* Point (2) follows directly from the definition: there are few branches in the decomposition, branches are short, and each bag cannot contain more than $(k+1)^2 \cdot |\mathcal{L}|$ atoms labelled with $\mathcal{L}$-languages.

For point (3), recall that one can check if a query has tree-width at most $k$ in linear time, *e.g.* using Bodlaender's algorithm (Proposition III.1.23).

To prove (1), notice first that $\text{App}^{\text{zip}}_{\mathcal{T}w_k}(\gamma) \subseteq \text{App}_{\mathcal{T}w_k}(\gamma)$ as a consequence of Fact V.6.2.

For the converse containment, we use Corollary V.3.6 and prove instead $\text{App}^{\star}_{\mathcal{T}w_k}(\gamma) \subseteq \text{App}^{\text{zip}}_{\mathcal{T}w_k}(\gamma)$. Observe that, as corollary of the proof of Lemma V.3.8, we can assume to have $\text{App}^{\star}_{\mathcal{T}w_k}(\gamma)$ expressed as a union of C2RPQ($\mathcal{L}$) with a fine tagged tree decomposition of width $k$ with at most $2\|\gamma\|_{\text{at}}$ leaves, and hence it suffices to replace each non-branching paths having non-atomic bags with path-$l$ approximations.

Indeed, fix a fine tagged tree decomposition and a trio $f \colon \rho \twoheadrightarrow \alpha$. Given a long non-branching path from bag $b_X$ with variables $X$ to a bag $b_Y$ with variables $Y$, such that $b_X$ and $b_Y$ are non-full, and no bag in between is atomic, define $X' \triangleq X \setminus Y$ and $Y' \triangleq Y \setminus X$. Consider the set $\mathcal{S}$ of atoms $u \xrightarrow{L} v$ of $\gamma$, such that the path induced $\left( \binom{b_i}{z_i} \right)_i$ by the refinement, say

$$u = w_0 \xrightarrow{L_0} w_1 \xrightarrow{L_1} \cdots \xrightarrow{L_n} w_n = v$$

of $u \xrightarrow{L} v$ in $\rho$ goes through $b_X$ at some variable of $X'$ and through $b_Y$ at some variable of $Y'$, in the sense that $b_i = b_X$ and $z_i \in X'$ for some $i$, and $b_j = b_Y$ and $z_j \in Y'$ for some $j$. There exist $i', j'$ such that $f(w_{i'}) = z_i$ and $f(w_{j'}) = z_j$, and *w.l.o.g.* $i' < j'$. Now let $\alpha'$ be the query obtained from $\alpha$ by removing all atoms tagged in a bag between $b_X$ and $b_Y$, and add a path-$l$ approximation query

$$\mathbf{P}_l(X', Y', \delta)$$

where $\delta$ is the conjunction over $u \xrightarrow{L} v \in \mathcal{S}$ of $w_{i'} \xrightarrow{[L_{i'} \cdots L_{j'}]} w_{j'}$. Repeat this operation for every non-trivial non-branching path with non-atomic bags. We obtain $\alpha'' \in \mathrm{App}^{\mathrm{zip}}_{\mathcal{T}w_k}(\gamma)$ *s.t.* $\alpha \sqsubseteq \alpha''$, which concludes the proof that $\mathrm{App}^{\star}_{\mathcal{T}w_k}(\gamma) \subseteq \mathrm{App}^{\mathrm{zip}}_{\mathcal{T}w_k}(\gamma)$. □

## V.6.2 Semantic Tree-Width Problem

With the previous results in place, we now show that the semantic tree-width $k$ problem is in $\Pi_2^p$ for UCRPQ(SRE), for every $k > 1$.

**Theorem V.6.1.** *For $k \geq 2$, the semantic tree-width $k$ problem for UCRPQ(SRE) is in $\Pi_2^p$.*

*Proof.* It suffices to show the statement for any CRPQ(SRE) $\gamma$. Remember that $\gamma$ is of semantic tree-width $k$ if, and only if, $\gamma \sqsubseteq \mathrm{App}^{\mathrm{zip}}_{\mathcal{T}w_k}(\gamma)$. The first ingredient to this proof is the fact that this containment has a polynomial counterexample property.

*Claim* V.6.4. *If $\gamma \not\sqsubseteq \mathrm{App}^{\star}_{\mathcal{T}w_k}(\gamma)$ then there is a polynomial-sized expansion $\xi$ of $\gamma$ such that $\xi \not\sqsubseteq \mathrm{App}^{\star}_{\mathcal{T}w_k}(\gamma)$.*

*Proof.* Let us call any atom with a language of the form $a^*$ a *recursive atom*, and any other atom a *non-recursive atom*. Let $n$ be the number of non-recursive atoms of $\gamma$. Hence, any refinement $\rho \in \mathrm{Ref}(\gamma)$ has $n$ atoms deriving from non-recursive atom refinements, all the remaining ones derive from recursive atom refinements.

We will work with the infinary union of conjunctive queries

$$\mathcal{U} \triangleq \mathrm{App}^{\star}_{\mathcal{T}w_k}(\gamma) \cap \mathrm{CQ}.$$

Note that $\mathcal{U} = \{\alpha \in \mathrm{CQ} \mid \alpha \in \mathcal{T}w_k \text{ and there is } \xi \in \mathrm{Exp}(\gamma) \text{ s.t. } \xi \twoheadrightarrow \alpha\}$. It is easy to see that $\mathcal{U} \equiv \mathrm{App}^{\star}_{\mathcal{T}w_k}(\gamma)$ as a consequence of Fact V.2.1. By Proposition III.2.12, we have $\gamma \not\sqsubseteq \mathcal{U}$ if, and only if, there is some expansion $\xi$ of $\gamma$ such that $\xi \not\sqsubseteq \mathcal{U}$. In turn, this happens if, and only if, there is no $\delta \in \mathcal{U}$ such that $\delta \xrightarrow{\text{hom}} \xi$.

Take any such counterexample $\xi$ of minimal size (in number of atoms). We show that for any (two-way) internal path of $\xi$ of the shape

$$\pi = x_0 \xrightarrow{a} x_1 \xrightarrow{a} x_1 \cdots x_{m-1} \xrightarrow{a} x_m,$$

we have $m \leq n + 1$. Hence, since $\xi$ is an expansion of a CRPQ(SRE), this means that the size of each *atom expansion*—namely an expansion obtained from a query by only expanding one atom—of $\xi$ is linearly bounded in the size of $\gamma$, and thus that $\xi$ is quadratically bounded.

By means of contradiction, if $m > n + 1$ consider the expansion $\xi'$ resulting from "shrinking" the path $\pi$ to a path $\pi'$ of length $n + 1$. Hence, $\xi'$ is smaller than $\xi$, and since $\xi$ was assumed to be minimal, $\xi'$ cannot be a counterexample. Thus, there is some $\delta \in \mathcal{U}$ such that $f_1 : \delta \xrightarrow{\text{hom}} \xi'$ for some homomorphism $f_1$. Further, by definition of $\mathcal{U}$, we have $f_2 : \xi'' \twoheadrightarrow \delta$ for some $\xi'' \in \text{Exp}(\gamma)$. Consider then the composition $\xi'' \twoheadrightarrow \delta \xrightarrow{\text{hom}} \xi'$ of $f_2$ with $f_1$ and let us call it $g : \xi'' \xrightarrow{\text{hom}} \xi'$. By definition of $n$ there must be at least one atom $x_i \xrightarrow{a} x_{i+1}$ of the shrunken path $\pi'$ of $\xi'$ which either (i) is not in the image of $f_1$, or (ii) all its $g$-preimages proceed from atoms $z \xrightarrow{a} z'$ of $\xi''$ which are in the expansions of recursive atoms of $\gamma$. We show that, in both cases, we can replace $x_i \xrightarrow{a} x_{i+1}$ with a path of $a$'s of any arbitrary length $l > 0$, obtaining a conjunctive query $\xi'_{+l}$ which is—still—not a counterexample. In the first case (i), we actually obtain that $\delta \xrightarrow{\text{hom}} \xi'_{+l}$. In the second case (ii), we have to replace each atom $z \xrightarrow{a} z'$ in the $g$-preimage of $x_i \xrightarrow{a} x_{i+1}$ in $\xi''$ by an $a$-path of length $l$, obtaining some expansion $\xi''_{+l}$ of $\gamma$. We also replace each atom in the $f_1$-preimage of $x_i \xrightarrow{a} x_{i+1}$ by an $a$-path of length $l$ obtaining some $\delta_{+l}$ such that $\xi''_{+l} \twoheadrightarrow \delta_{+l} \xrightarrow{\text{hom}} \xi'_{+l}$. Further, $\delta_{+l} \in \mathcal{T}w_k$ since $\mathcal{T}w_k$ is closed under refinements by Fact V.2.1. In both cases this shows that $\xi'_{+l}$ is *not* a counterexample. In particular, for $l = m - n - 1$, we have $\xi'_{+l} = \xi$, and this would contradict the fact that $\xi$ is a counterexample. Therefore, there exists a counterexample of polynomial (quadratic) size whenever $\gamma \not\sqsubseteq \text{App}^{\star}_{\mathcal{T}w_k}(\gamma)$. □

The second ingredient is that testing whether a CQ is a counterexample is in coNP.

*Claim* V.6.5. The problem of testing, given a C2RPQ $\gamma$ and a CQ $\xi$, whether $\xi \sqsubseteq \text{App}^{\text{zip}}_{\mathcal{T}w_k}(\gamma)$, is in NP.

*Proof.* We first guess a polynomial-sized $k$-summary query $\delta_{\text{zip}}$ and test in NP that it is part of $\text{App}^{\text{zip}}_{\mathcal{T}w_k}(\gamma)$ by Lemma V.6.3. Let us call $\Delta$ be the equivalent UCRPQ(SRE) query, given by Lemma V.6.3 *cum* Lemma V.3.8. We have to check that there is some expansion $\delta$ of $\Delta$ such that there is a homomorphism $\delta \xrightarrow{\text{hom}} \xi$. We first guess a valuation $\mu : \text{vars}(\delta_{\text{zip}}) \to \text{vars}(\xi)$. Now it remains to check that:

1. For every CRPQ atom $x \xrightarrow{a} y$ of $\delta_{\text{zip}}$ there is an $a$-path in $\xi$ from $\mu(x)$ to $\mu(y)$.

2. Every path-$l$ approximation $\mathbf{P}_l(X, Y, \bigwedge_{1 \leq i \leq n} A_i(x_i, y_i))$ of $\delta_{\text{zip}}$ contains a CQ $\delta_{\text{path}}$ admitting a path decomposition of width $l$ which starts with the bag $X$ and ends with $Y$. And further, there is a homomorphism $h : \delta_{\text{path}} \xrightarrow{\text{hom}} \xi$ which coincides with $\mu$ on variables $X \cup Y$.

Observe that these two properties hold true if, and only if, there is some expansion $\delta$ of $\Delta$ such that $\delta \xrightarrow{\text{hom}} \gamma$. It is clear the first point can be achieved in polynomial time (actually, in NL) since it is a simple reachability query. The second point can also be achieved in polynomial time (or in NL), since the fine path decomposition of width $l$ can be guessed on-the-fly using $l + 1$ pointers to the variables of $\gamma$ (*cf.* Lemma V.8.9). An NL algorithm can advance down the path decomposition while simultaneously

1. guessing the conjunctive query $\delta_{\text{path}}$ via its fine path decomposition of width $k$,

2. checking that there is a partial homomorphism to $\gamma$ (*i.e.*, a homomorphism from the subquery of $\gamma$ restricted to current bag's variables to $\gamma$),

3. ensuring that the CQ $\delta_{\text{path}}$ being built is an element of

$$\mathbf{P}_l(X, Y, \bigwedge_{1 \leq i \leq n} A_i(x_i, y_i)),$$

which requires to also guess a homomorphism $\rho \xrightarrow{\text{hom}} \delta_{\text{path}}$ from a refinement $\rho$ of $\bigwedge_{1 \leq i \leq n} A_i(x_i, y_i)$.

Further, a simple test can ensure that the first and last bags of the decomposition coincide with the guessed assignment $\mu$. Since the number of pointers (bounded by $k + 1$) is fixed, this subroutine is in NL, and hence in polynomial time. This yields an NP algorithm for testing $\xi \subseteq \text{App}^{\text{zip}}_{\mathcal{T}w_k}(\gamma)$. □

As a consequence of the two claims, we obtain a $\Sigma_2^p$ algorithm for non-containment of $\gamma \subseteq \text{App}^{\text{zip}}_{\mathcal{T}w_k}(\gamma)$: We first guess an expansion $\xi$ of $\gamma$ of polynomial size, and we then test $\xi \not\subseteq \text{App}^{\text{zip}}_{\mathcal{T}w_k}(\gamma)$ in coNP. This gives a $\Pi_2^p$ algorithm for the semantic tree-width $k$ problem, which is correct by Lemma V.6.3 and Claim V.6.4. □

## V.7  Acyclic Queries: the Case $k = 1$

Observe that in the previous sections we have treated the cases of semantic tree-width $k$ for every $k \geq 2$. However, the case $k = 1$ remains rather elusive so far. While the Key Lemma holds for $k = 1$, it proves the computability of an object that is irrelevant to study semantic tree-width 1, see Remark V.5.1. The problem comes from Example V.3.7, namely that $\text{App}_{\mathcal{T}w_1}(\gamma) \not\equiv \text{App}^{\star}_{\mathcal{T}w_1}(\gamma)$. This is the main obstacle why our approach does not directly yield an algorithm for the case $k = 1$, which had been previously solved by Barceló, Romero and Vardi [BRV16]. However, as we argue in this section, a rather elegant modification on the notion of tree-width allows to use our approach as a unifying framework for both the case $k = 1$ and the cases $k \geq 2$. Concretely, we introduce a family of classes $\{\mathcal{C}tw_k\}_k$ such that $\mathrm{App}_{\mathcal{T}w_k}(\gamma) \equiv \mathrm{App}^{\star}_{\mathcal{C}tw_k}(\gamma)$ for every $\gamma$ and $k$, and where $\mathrm{App}^{\star}_{\mathcal{C}tw_1}(\gamma) \equiv \mathrm{App}^{\star,\leq\mathrm{poly}(\|\gamma\|)}_{\mathcal{C}tw_1}(\gamma)$. As a corollary, we reprove [BRV16, Theorem 6.1], namely that the semantic tree-width 1 problem is ExpSpace-complete. Further, we can also solve the one-way semantic tree-width 1 problem, which is outside the scope of [BRV16]. Remember that for $k = 1$, the semantic tree-width and one-way semantic tree-width 1 problems are two independent problems, since there are queries of semantic tree-width 1 but not of one-way semantic tree-width 1 (*cf.* Remark V.3.14).

### V.7.1 Contracted Tree-Width

We next formally define the notion of "contracted tree-width", meaning the tree-width of the graph obtained by contracting paths (or directed paths) into edges. This altered notion of tree-width will allow us to seamlessly prove the case of $k = 1$ for Theorem V.1.3. Recall that a one-way contraction of a CRPQ is defined analogously to a (two-way) contraction, except that we only contract one-way internal paths.

**Definition V.7.1.** Define the *contracted tree-width* (resp. *one-way contracted tree-width*) of a C2RPQ as the minimum of the tree-width among its contractions (resp. of its one-way contractions). Let $\mathcal{C}tw_k$ and $1\mathcal{C}tw_k$ be, respectively, the set of all C2RPQs of contracted tree-width at most $k$ and of CRPQs of one-way contracted tree-width at most $k$.

For instance, the query

$$\gamma(x_0, x_1) \hateq \quad \begin{array}{c} x_0 \xrightarrow{\;K\;} x_1 \\ {\scriptstyle L}\searrow \quad \swarrow{\scriptstyle M} \\ y \end{array}$$

has contracted tree-width one since the internal path $x_0 \xrightarrow{L} y \xleftarrow{M} x_1$ can be contracted into $x_0 \xrightarrow{LM^-} x_1$. On the other hand, its one-way contracted tree-width is two, since there is no non-trivial one-way internal path as $x_1$ is an output variable.

Note that, by definition:

- the contracted tree-width is at most the one-way contracted tree-width, which is in turn at most the tree-width;
- for $k \geq 2$, these notions collapse (by Fact V.2.1);
- for $k = 1$, both inequalities can be strict.

Moreover, for any $k \geq 1$, contracted tree-width at most $k$ and one-way contracted tree-width at most $k$ are both closed under refinements: if a query has tree-width at most $k$, so does any refinement thereof. In fact, the CQs of contracted tree-width 1 precisely correspond to what in [BRV16, § 5.2.1, p1358] is known as "pseudoacyclic graph databases".

**Fact V.7.2.** Let $k \geq 1$. For any CRPQ $\gamma$, we have $\mathrm{App}_{1\mathcal{T}w_k}(\gamma) \equiv \mathrm{App}^{\star}_{1\mathcal{C}tw_k}(\gamma)$.

Moreover, for any C2RPQ $\gamma$, $\mathrm{App}_{\mathcal{T}w_k}(\gamma) \equiv \mathrm{App}^{\star}_{\mathcal{C}tw_k}(\gamma)$.

*Proof.*

$$\mathrm{App}_{\mathcal{T}w_k}(\gamma) \equiv \mathrm{App}_{\mathcal{C}tw_k}(\gamma) \quad \text{since contractions preserve semantics,}$$

$$\equiv \mathrm{App}^{\star}_{\mathcal{C}tw_k}(\gamma) \quad \text{by Fact V.3.5.}$$

The same arguments work with one-wayness. □

### V.7.2 The Key Lemma for Contracted Tree-Width One

We show next that contracted tree-width 1 has all the needed properties for the analogue of Key Lemma for $k = 1$ to hold.

**Lemma V.7.3.** For any CRPQ $\gamma$, we have $\mathrm{App}^{\star}_{1\mathcal{C}tw_1}(\gamma) \equiv \mathrm{App}^{\star, \leq \ell_1}_{1\mathcal{C}tw_1}(\gamma)$, where $\ell_1 = \Theta(\|\gamma\|_{\mathrm{at}}^3)$. Similarly, for a C2RPQ $\gamma$, $\mathrm{App}^{\star}_{\mathcal{C}tw_1}(\gamma) \equiv \mathrm{App}^{\star, \leq \ell_1}_{\mathcal{C}tw_1}(\gamma)$.

*Proof.* Consider the proof of the Key Lemma (Lemma V.3.8). We claim that:

1. the constructions of Lemmas V.5.2 and V.5.4 both preserve contracted tree-width at most 1 and one-way contracted tree-width at most 1;

2. the upper bound $\ell \in \mathcal{O}(\|\gamma\|_{\mathrm{at}}^2 \cdot 2^{\|\gamma\|_{\mathrm{at}}})$ can be easily boiled down to $\ell_1 \in \mathcal{O}(\|\gamma\|_{\mathrm{at}}^3)$ in the special case of $k = 1$.

↬ (1) *Preservation of contracted tree-width.* We claim that all constructions of Section V.5 preserve contracted tree-width at most 1. The setting is similar, except that now, a trio consists of a triple $f\colon \rho \to \alpha$ where $\rho$ is a refinement of a fixed C2RPQ $\gamma$, and $\alpha$ is a C2RPQ of contracted tree-width 1. We now apply the constructions not to a decomposition of $\alpha$ but to a fine tagged tree decomposition of a contraction of $\alpha$ of tree-width 1.

**Fact V.7.4.** Let $\gamma$ be a C2RPQ, $\chi$ be a contraction of $\gamma$, and $(T, \mathbf{v}, \mathbf{t})$ be a fine tagged tree decomposition of $\chi$ of width at most 1. Let $z, z' \in \mathbf{v}(b)$ for some bag $b \in T$. Then $\gamma \wedge z \xrightarrow{\lambda} z'$ still has contracted tree-width at most 1.

As a consequence, the construction of Lemma V.5.2—which takes us from Figure V.5 to Figure V.7—preserves contracted tree-width 1. Then, Proposition V.4.5—illustrated in Figure V.6—can be trivially adapted to our setting as follows:

**Fact V.7.5.** Let $\gamma, \gamma'$ be two C2RPQ with a disjoint set of variables. Let $z$ (resp. $z'$) be a variable of $\gamma$ (resp. $\gamma'$). If both $\gamma$ and $\gamma'$ have contracted tree-width at most 1, then so does $\gamma \wedge \gamma' \wedge z \xrightarrow{\lambda} z'$.

As a consequence, the construction of Lemma V.5.4—which takes us from Figure V.7 to Figure V.10—preserves contracted tree-width 1.

↬ (2) *Improved upper bound.* In the proof claiming that in a sufficiently long non-branching path, we can always find two non-full, non-atomic bags with the same profile (see the proof of Lemma V.5.4), we obtain a bound of $\mathcal{O}(2^{\|\gamma\|_{\mathrm{at}}})$. We actually claim that it can be improved to obtain a polynomial bound. This is because, for width 1, a non-full bag $b$ contains exactly 1 variable $z_b$. So, its profile consists simply on a set of atoms of $\gamma$—namely the set of atoms whose refinement induces a path which leaves the bag $b$ at $z_b$. But we claim that in a non-branching path, not all of these $2^{\|\gamma\|_{at}}$ profiles can occur at the same time. Indeed, in tree decompositions, the set of bags containing a given variable must be connected. This property can be lifted to paths in tagged tree decompositions in the following way.

**Fact V.7.6.** Let $(T, \mathbf{v}, \mathbf{t})$ be a tagged tree decomposition of some homomorphism $f \colon \rho \xrightarrow{\text{hom}} \alpha$. Let $\pi$ be a path in $\rho$. Assume that:

- the simple path in $T$ from $b$ to $b''$ goes through $b'$,
- there exists some variable $z$ of $\alpha$ such that $\mathbf{t}[\pi]$ leaves $b$ at $z$, and
- there is no variable like that for the bag $b'$.

Then, there is no variable $z$ of $\alpha$ such that $\mathbf{t}[\pi]$ leaves $b''$ at $z$.

*Proof of Fact V.7.6.* Fix a tagged tree decomposition $(T, \mathbf{v}, \mathbf{t})$ of some homomorphism $f \colon \rho \xrightarrow{\text{hom}} \alpha$ and $\pi$ be a path in $\rho$. Let $b, b', b''$ be bags such that the simple path in $T$ from $b$ to $b''$ goes through $b'$. Say that an induced path $\mathbf{t}[\pi] = (\binom{b_i}{z_i})_i$ *visits* a bag $b$ if $b_i = b$ for some $i$. Note that this is equivalent to saying that there exists a variable $z$ of $\alpha$ s.t. $\mathbf{t}[\pi]$ leaves $b$ at $z$. Hence, Fact V.7.6 boils down to the following property: if $\mathbf{t}[\pi]$ visits both $b$ and $b''$, then it must also visit $b'$. This property holds because by construction, the sequence $(b_i)_i$—namely the projection of $\mathbf{t}[\pi]$ onto $T$—is a path in $T$, with some node repetition. □

As a consequence, if an atom occurs in a bag, but not in a latter one, then it can never occur again. Hence, the number of bags of size 1 in a non-branching path where each bag has a different profile must be at most $n = \|\gamma\|_{at}$. Hence, Fact V.5.7 yields a bound of $\mathcal{O}(\|\gamma\|_{at}^2)$. Finally, we can conclude like in Section V.5.3: we obtain a tree with at most $\mathcal{O}(\|\gamma\|_{at})$ leaves, and with non-branching paths of length at most $\mathcal{O}(\|\gamma\|_{at}^2)$, so the tree has size at most $\ell_1 \in \mathcal{O}(\|\gamma\|_{at}^3)$. By local acyclicity, this concludes the proof of Lemma V.7.3. The case of one-way contracted tree-width is completely similar. □

**Lemma V.7.7.** Let $k \geq 1$.

1. Given a UCRPQ $\Gamma$, it has one-way semantic tree-width at most 1 *iff* $\Gamma \equiv \text{App}_{1\mathcal{C}tw_1}^{\star, \leq \ell_1}(\Gamma)$;
2. Given a UC2RPQ $\Gamma$, it has semantic tree-width at most 1 *iff* $\Gamma \equiv \text{App}_{\mathcal{C}tw_1}^{\star, \leq \ell_1}(\Gamma)$;

where $\ell_1 \in \mathcal{O}(\|\gamma\|_{at}^3)$.

*Proof.* To prove the first point:

- if $\Gamma$ is equivalent to a UCRPQ $\Delta$ of tree-width at most 1, then $\Delta \subseteq \text{App}_{\mathcal{T}w_1}(\Gamma)$ and by Fact V.7.2 and Lemma V.7.3, $\Delta \subseteq \text{App}_{1\mathcal{C}tw_1}^{\star, \leq \ell_1}(\Gamma)$, and hence:

$$\Gamma \equiv \Delta \subseteq \text{App}_{1\mathcal{C}tw_1}^{\star, \leq \ell_1}(\Gamma) \subseteq \Gamma.$$

- If $\Gamma \equiv \mathrm{App}^{\star, \leq \ell_1}_{1\mathcal{C}tw_1}(\Gamma)$, then $\Gamma$ is equivalent to a UCRPQ of contracted tree-width at most 1, and hence (by contraction) it is equivalent a UCRPQ of tree-width at most 1.

The second point can be proven similarly. □

**Corollary V.7.8** (Upper bound of Theorem V.1.3 for $k = 1$)**.** The semantic tree-width 1 problem and one-way semantic tree-width 1 problem are in ExpSpace.

*Proof.* The fact that the semantic tree-width 1 problem is in ExpSpace is actually the main result of [BRV16, Theorem 6.1], but we show how the upper bound follows as a direct corollary of Lemma V.7.7 above. Since $\ell_1 \in \mathcal{O}(\|\gamma\|^3_{\mathrm{at}})$, $\mathrm{App}^{\star, \leq \ell_1}_{\mathcal{C}tw_1}(\gamma)$ is an exponential union of polynomial sized C2RPQs, and thus by Proposition V.3.11 the containment problem $\Gamma \subseteq \mathrm{App}^{\star, \leq \ell_1}_{\mathcal{C}tw_1}(\Gamma)$ is in ExpSpace, and so is the semantic tree-width 1 problem (since the converse containment $\mathrm{App}^{\star, \leq \ell_1}_{\mathcal{C}tw_1}(\Gamma) \subseteq \Gamma$ always holds, *cf.* Remark V.3.2). The proof for one-way semantic tree-width 1 problem is analogous. □

Lastly, note that we can derive from Lemma V.7.3 a characterization of semantic tree-width 1 somewhat similar to Theorem V.3.13.

**Corollary V.7.9.** Assume that $\mathcal{L}$ is closed under sublanguages.

❧ *Two-way queries:* For any query $\Gamma \in$ UC2RPQ($\mathcal{L}$), the following are equivalent:

1. $\Gamma$ is equivalent to an infinitary union of conjunctive queries of contracted tree-width at most 1;

2. $\Gamma$ has semantic tree-width at most 1;

3. $\Gamma$ is equivalent to a UC2RPQ($\mathcal{L}$) of contracted tree-width at most 1;

4. $\Gamma$ is equivalent to a UC2RPQ($\mathcal{L}'$) of tree-width at most 1, where $\mathcal{L}'$ is the closure of $\mathcal{L}$ under concatenation and inverses, *i.e.* $\mathcal{L}'$ is the smallest class containing $\mathcal{L}$ and such that if $K, L \in \mathcal{L}'$ then $K \cdot L \in \mathcal{L}'$ and $K^- \in \mathcal{L}'$.

❧ *One-way queries:* Similarly, if $\Gamma \in$ UCRPQ($\mathcal{L}$), then the following are equivalent:

1. $\Gamma$ is equivalent to an infinitary union of conjunctive queries of one-way contracted tree-width at most 1;

2. $\Gamma$ has one-way semantic tree-width at most 1;

3. $\Gamma$ is equivalent to a UCRPQ($\mathcal{L}$) of one-way contracted tree-width at most 1;

4. $\Gamma$ is equivalent to a UCRPQ($\mathcal{L}'$) of tree-width at most 1, where $\mathcal{L}'$ is the closure of $\mathcal{L}$ under concatenation, *i.e.* $\mathcal{L}'$ is the smallest class containing $\mathcal{L}$ and such that if $K, L \in \mathcal{L}'$ then $K \cdot L \in \mathcal{L}'$.

Note in particular how point (4) of each characterization reflects that a UCRPQ of semantic tree-width 1 can have one-way semantic tree-width at least 2—as we showed in Remark V.3.14. More generally, the differences between this last corollary and Theorem V.3.13 highlight the different combinatorial behaviour that semantic tree-width $k$ has, depending on whether

$k = 1$ or $k > 1$.

**Remark V.7.10.** Finally, note that results of Sections V.6 and V.7 can be joined in order to show that the semantic tree-width 1 problems are decidable in $\Pi_2^p$ for UC2RPQs over the closure under concatenation and inverses of SREs (resp. for UCRPQs over the closure under concatenation of SREs).[7]  ◇

## V.8   Semantic Path-Width

In this section, we extend our results to path-width. Our motivation lies in the fact that UC2RPQs of bounded semantic path-width admit a paraNL[8] algorithm for the EVALUATION PROBLEM—see Theorem V.8.8—to be compared with FPT for bounded semantic tree-width.

### V.8.1   Path-Width of Queries

Recall that for tree-width, for any $k \geq 2$, we proved that a CRPQ is equivalent to a finite union of C2RPQs of tree-width at most $k$ *iff* it is equivalent to finite union of CRPQs of tree-width at most $k$ (Theorem V.3.13). In other words, two-way navigation does not help to minimize further the semantic tree-width of a query that does not use two-way navigation. This property does not hold for $k = 1$ (Remark V.3.14). We show in Section V.8.2 that it also does not hold for path-width, no matter the value of $k \geq 1$.

This motivates the following two definitions:

- the *semantic path-width* of a UC2RPQ is the minimal path-width of a UC2RPQ equivalent to it
- the *one-way semantic path-width* of a UCRPQ is the minimal path-width of a UCRPQ equivalent to it.

For a given UCRPQ, the two natural numbers are well-defined, and the former is always less or equal to the letter. The *semantic path-width $k$ problems* ask, given a UCRPQ (resp. UC2RPQ), if it has semantic path-width (resp. one-way semantic path-width) at most $k$.

In this section, we first show that the semantic path-width $k$ problems are decidable (Theorem V.8.6), and then after showing that evaluation of UC2RPQs of bounded path-width is NL (Lemma V.8.9) we deduce that for the EVALUATION PROBLEM for UC2RPQs of bounded semantic path-width (in particular, this captures the case of UCRPQs of bounded one-way semantic path-width) is in paraNL when parameterized in the size of the query (Theorem V.8.8).

### V.8.2   Deciding Bounded Semantic Path-Width

The key (implicit) ingredient in the proof of Theorem V.3.13 and Corollary V.3.9 is that tree-width at most $k$ is closed under expansions (Fact V.2.1). Unfortunately, this property fails for path-width.

**Fact V.8.1.** For each $k \geq 1$, the class of graphs of path-width at most $k$ is not

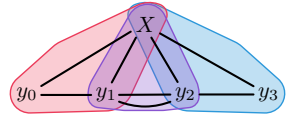

(a) Graph $\mathcal{G}_k$ with a path decomposition of width $k$, with three bags.

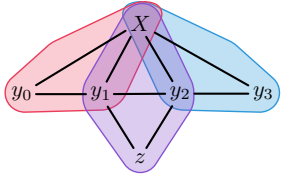

(b) Expansion $\mathcal{G}'_k$ of $\mathcal{G}_k$ with a path decomposition of width $k+1$, with three bags.

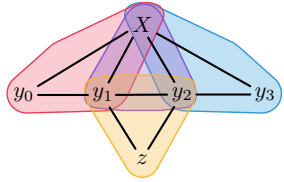

(c) Tree decomposition of $\mathcal{G}'_k$ of width $k$, with four bags.

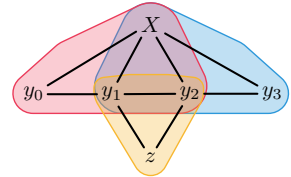

(d) Path decomposition of $\mathcal{G}'_k$ of width $k+1$, with three bags.

Figure V.12: The class of multigraphs with path-width at most $k \geq 1$ is not closed under expansions: illustration of a multigraph $\mathcal{G}_k$ of path-width $k$ (Figure V.12a) whose expansion $\mathcal{G}'_k$ has path-width $k+1$ (Figures V.12b and V.12d)—but tree-width $k$ (Figure V.12c). Set $X$ represents a $(k-1)$-clique.

closed under expansions.

The counterexample is illustrated in Figure V.12. A formal proof can be found in Section V.B. Contrary to the case of semantic tree-width, for every $k$ there are CRPQs which are of semantic path-width $k$ but not of one-way semantic path-width $k$.

*Proof.* Indeed, let

$$\gamma_k(\bar{x}, \bar{y}) \triangleq \Big( \bigwedge_{1 \leq i < j \leq k-1} x_i \xrightarrow{a} x_j \Big) \wedge \Big( \bigwedge_{1 \leq i \leq k-1} \bigwedge_{0 \leq j \leq 3} x_i \xrightarrow{b} y_i \Big)$$
$$\wedge \Big( \bigwedge_{0 \leq j < 3} y_i \xrightarrow{c} y_{i+1} \Big) \wedge y_1 \xrightarrow{d} z \wedge y_2 \xrightarrow{e} z,$$

whose underlying graph corresponds to Figure V.12b. Observe that it is a core and that only $z$ is existentially quantified. Then in $\gamma_k(\bar{x}, \bar{y})$, one can replace the two atoms $y_1 \xrightarrow{d} z \wedge y_2 \xrightarrow{e} z$ by $y_1 \xrightarrow{de^-} y_2$, while preserving the semantics. The underlying graph of this new query being Figure V.12a, it shows that $\gamma_k$ has semantic path-width $k$.

Finally, we claim that $\gamma_k$ has one-way semantic path-width $k+1$. The upper bound follows from Figure V.12d. For the lower bound, consider a UCRPQ $\Delta_k(\bar{x}, \bar{y})$ such that $\gamma_k \equiv \Delta_k$. Since $\gamma_k$ is a CQ, the equivalence implies that there exists an expansion $\xi$ of a CRPQ of $\Delta_k$ such $\gamma_k$ and $\xi$ are homomorphically equivalent. Since $\gamma_k$ is a core, it follows that $\xi$ contains it as a subgraph. Hence, the underlying directed multigraph of the CRPQ in $\Delta_k$ from which $\xi$ originated must contain a one-way contraction of $\xi$ as a subgraph. But the only one-way contraction of $\xi$ is itself, and so it follows that at least one CRPQ in $\Delta_k$ contains the underlying graph of $\xi$ as a subgraph. Therefore, $\Delta_k$ has path-width at least $k + 1$, which concludes the proof that the one-way semantic path-width of $\gamma_k$ is at least (and hence exactly) $k + 1$. □

As done for contracted tree-width, we define contracted path-width.

**Definition V.8.2.** Define the *contracted path-width* (resp. *one-way contracted path-width*) of a C2RPQ as the minimum of the path-width among its contractions (resp. of its one-way contractions). Let $\mathcal{C}pw_k$ and $1\mathcal{C}pw_k$ be, respectively, the set of all C2RPQs of contracted path-width at most $k$ and of CRPQs of one-way contracted path-width at most $k$.

The statements and proofs of this section are analogous to the ones of Section V.7 in the context of contracted tree-width 1. We keep the order and structure to make this correspondence evident.

Again, by definition, contracted path-width at most $k$ and one-way contracted path-width at most $k$ are both closed under refinements: if a query has width at most $k$, so does any refinement thereof.

**Fact V.8.3.** Let $k \geq 1$. For any CRPQ $\gamma$, we have $\mathrm{App}_{1\mathcal{P}w_k}(\gamma) \equiv \mathrm{App}_{1\mathcal{C}pw_k}^{\star}(\gamma)$. Moreover, for any C2RPQ $\gamma$, $\mathrm{App}_{\mathcal{P}w_k}(\gamma) \equiv \mathrm{App}_{\mathcal{C}pw_k}^{\star}(\gamma)$.

*Proof.*

$$\mathrm{App}_{1\mathcal{P}w_k}(\gamma) \equiv \mathrm{App}_{1\mathcal{C}pw_k}(\gamma) \quad \text{since contractions preserves the semantics,}$$
$$\equiv \mathrm{App}_{1\mathcal{C}pw_k}^{\star}(\gamma) \quad \text{by Fact V.3.5.}$$

The same arguments work for C2RPQs. □

**Lemma V.8.4.** For $k \geq 1$ and CRPQ $\gamma$, we have $\mathrm{App}_{1\mathcal{C}pw_k}^{\star}(\gamma) \equiv \mathrm{App}_{1\mathcal{C}pw_k}^{\star, \leq \ell}(\gamma)$, where $\ell = \Theta(\|\gamma\|_{\mathrm{at}}^2 \cdot (k+1)^{\|\gamma\|_{\mathrm{at}}})$. Similarly, for a C2RPQ $\gamma$, $\mathrm{App}_{\mathcal{C}pw_k}^{\star}(\gamma) \equiv \mathrm{App}_{\mathcal{C}pw_k}^{\star, \leq \ell}(\gamma)$.

*Proof.* Consider the proof of the Key Lemma (Lemma V.3.8): both constructions (Lemmas V.5.2 and V.5.4) preserve the contracted path-width of $\alpha$ if the operations are applied to a suitable path decomposition of a contraction of $\alpha$ of width $k$. □

**Lemma V.8.5.** Let $k \geq 1$.

1. Given UCRPQ $\Gamma$, it has one-way semantic path-width at most $k$ *iff* $\Gamma \equiv \mathrm{App}_{1\mathcal{C}pw_k}^{\star, \leq \ell}(\Gamma)$;

2. Given a UC2RPQ $\Gamma$, it has semantic path-width at most $k$ *iff* $\Gamma \equiv \mathrm{App}_{\mathcal{C}pw_k}^{\star, \leq \ell}(\Gamma)$.

*Proof.* To prove the first point:

- if $\gamma$ is equivalent to a UCRPQ $\Delta$ of path-width at most $k$, then $\Delta \subseteq \mathrm{App}_{\mathcal{P}w_k}(\gamma)$ and by Fact V.8.3 and Lemma V.8.4, $\Delta \subseteq \mathrm{App}_{1\mathcal{C}pw_k}^{\star, \leq \ell}(\gamma)$, and

hence:

$$\gamma \equiv \Delta \subseteq \mathrm{App}^{\star,\leq\ell}_{1\mathcal{C}pw_k}(\gamma) \subseteq \gamma.$$

- If $\gamma \equiv \mathrm{App}^{\star,\leq\ell}_{1\mathcal{C}pw_k}(\gamma)$, then $\gamma$ is equivalent to a UCRPQ of contracted path-width at most $k$, and hence it is equivalent a UCRPQ of path-width at most $k$.

The second point can be proven similarly. □

We can now prove the main theorem.

**Theorem V.8.6.** For each $k \geq 1$, the semantic path-width $k$ problems are decidable. Moreover, they lie in 2ExpSpace and are ExpSpace-hard. Moreover, if $k = 1$, these problems are in fact ExpSpace-complete.

*Proof.* The upper bounds follow from Lemma V.8.5. The lower bounds will be shown in Lemma V.9.1. Lastly, to prove the ExpSpace upper bound for $k = 1$, we can apply the same trick as in Corollary V.7.8. □

Similarly to Corollary V.7.9, we can derive from Lemma V.8.4 a characterization of semantic path-width at most $k$.

**Corollary V.8.7.** Assume that $\mathcal{L}$ is closed under sublanguages, and let $k \geq 1$.
☙ *Two-way queries:* For any query $\Gamma \in$ UC2RPQ($\mathcal{L}$), the following are equivalent:

1. $\Gamma$ is equivalent to an infinitary union of conjunctive queries of contracted path-width at most $k$;
2. $\Gamma$ has semantic path-width at most $k$;
3. $\Gamma$ is equivalent to a UC2RPQ($\mathcal{L}$) of contracted path-width at most $k$;
4. $\Gamma$ is equivalent to a UC2RPQ($\mathcal{L}'$) of path-width at most $k$, where $\mathcal{L}'$ is the closure of $\mathcal{L}$ under concatenation and inverses, *i.e.* $\mathcal{L}'$ is the smallest class containing $\mathcal{L}$ and such that if $K, L \in \mathcal{L}'$ then $K \cdot L \in \mathcal{L}'$ and $K^- \in \mathcal{L}'$.

☙ *One-way queries:* Similarly, if $\Gamma \in$ UCRPQ($\mathcal{L}$), then the following are equivalent:

1. $\Gamma$ is equivalent to an infinitary union of conjunctive queries of one-way contracted path-width at most $k$;
2. $\Gamma$ has one-way semantic path-width at most $k$;
3. $\Gamma$ is equivalent to a UCRPQ($\mathcal{L}$) of one-way contracted path-width at most $k$;
4. $\Gamma$ is equivalent to a UCRPQ($\mathcal{L}'$) of path-width at most $k$, where $\mathcal{L}'$ is the closure of $\mathcal{L}$ under concatenation, *i.e.* $\mathcal{L}'$ is the smallest class containing $\mathcal{L}$ and such that if $K, L \in \mathcal{L}'$ then $K \cdot L \in \mathcal{L}'$.

### V.8.3 Evaluation of Queries of Bounded Semantic Path-Width

In this section, we show that, as a consequence of Theorem V.8.6, we can obtain an efficient algorithm for the evaluation problem.

**Theorem V.8.8.** For each $k \geq 1$, the evaluation problem, restricted to UC2RPQs of semantic path-width at most $k$ is in paraNL when parameterized in the size of the query. More precisely, the problem, on input $\langle \Gamma, G \rangle$, can be solved in non-deterministic space $f(\|\Gamma\|) + \log(|G|)$, where $f$ is a single exponential function.

The class paraNL was introduced in [CCDF97, Definition, p. 123] under the name "uniform NL + advice". It was renamed paraNL in [FG03, Definition 1, p. 294]. For the sake of simplicity, instead of either of those definitions, we use the characterization of paraNL proven in [FG03, Theorem 4, p. 296].

We define *paraNL* as the class of parameterized languages $L \subseteq \Sigma^* \times \mathbb{N}$ for which there is a Turing machine $\mathcal{M}$ *s.t.*

$$\mathcal{M} \text{ accepts } \langle x, k \rangle \quad \textit{iff} \quad \langle x, k \rangle \in L,$$

$$\underbrace{\phantom{\langle x,k \rangle}}_{\text{input}} \qquad \underbrace{\phantom{\langle x, k \rangle}}_{\text{parameter}}$$

and, moreover, $\mathcal{M}$ runs in non-deterministic space $f(k) + \mathcal{O}(\log(|x|))$, where $f \colon \mathbb{N} \to \mathbb{N}$ is a computable function. A typical example of paraNL problem is the model-checking problem for first-order logic on finite structures, when parameterized by the maximum degree of the structure [FG03, Example 6].

To show Theorem V.8.8, we first focus on the evaluation of queries of bounded path-width.

**Lemma V.8.9.** For each $k \geq 1$, the EVALUATION PROBLEM, restricted to UC2RPQs of path-width at most $k$, is NL-complete.

*Proof.* The exact same proof as for CQs works, see Lemma V.8.9.[9] To get the NL complexity, we require an extra argument: namely that each atomic check—checking if there is an $L$-labelled path from $f(x)$ to $f(y)$—can be done in non-deterministic space. In fact, using a straightforward adaptation of the NL algorithm for graph reachability, we obtain an algorithm in $\mathcal{O}(\log(|D|) + \log(|\mathcal{A}_L|))$ where $\mathcal{A}_L$ is an NFA for $L$, ; note that these atomic checks are independent of one another, so we can reuse this space. $\square$

[9] For the full proof, see [FM25, Lemma 8.10].

We can now conclude and prove that the EVALUATION PROBLEM for UC2RPQs of semantic path-width $k$ is in paraNL.

*Proof of Theorem V.8.8.* Given a UC2RPQ $\Gamma(\bar{x})$ of semantic path-width at most $k$ and a database $G(\bar{u})$, we first compute $\mathrm{App}^{\star, \leq \ell}_{1\mathcal{C}pw_k}(\Gamma)$—where $\ell = \Theta(\|\Gamma\|^2_{\mathrm{at}} \cdot (k + 1)^{\|\Gamma\|_{\mathrm{at}}})$—, which is equivalent to $\Gamma$ by Lemma V.8.5. Then, we use Lemma V.8.9 to evaluate each $\delta(\bar{x}) \in \mathrm{App}^{\star, \leq \ell}_{1\mathcal{C}pw_k}(\Gamma)$ on $G(\bar{u})$. If one of the queries accepts, we accept. Otherwise, we reject.

The non-deterministic space needed by the algorithm is:

- $\mathcal{O}(\ell)$ bits to enumerate and store $\delta(\bar{x})$, where $\ell = \Theta(\|\Gamma\|^2_{\mathrm{at}} \cdot (k + 1)^{\|\Gamma\|_{\mathrm{at}}})$
- $\mathcal{O}(\log|G| + \log\|\delta\|) \subseteq \mathcal{O}(\log|G| + \log|\ell| + \log\|\Gamma\|)$ to evaluate $\delta(\bar{x})$ on $G(\bar{u})$, by Lemma V.8.9 and since $\|\delta\| \leq \|\Gamma\| \cdot \ell$.

Overall, we use non-deterministic space $f(\|\Gamma\|) + \mathcal{O}(\log(|G|))$ where $f$ is a single exponential, which concludes the proof. $\square$

## V.9  Lower Bounds for Deciding Semantic Tree-Width and Path-Width

An ExpSpace lower bound follows by a straightforward adaptation from the ExpSpace lower bound for the case $k = 1$ [BRV16, Proposition 6.2].

**Lemma V.9.1** (Lower bound of Theorem V.1.3). For every $k \geq 1$, the following problems are ExpSpace-hard, even if restricted to Boolean CRPQs:

- the semantic tree-width $k$ problem;
- the one-way semantic tree-width $k$ problem;
- the semantic path-width $k$ problem;
- the one-way semantic path-width $k$ problem.

We say that a C2RPQ is *connected* when its underlying undirected graph is connected. We first give a small useful fact.

**Fact V.9.2** (Implicit in [BRV16, Proof of Proposition 6.2]).

1. Let $G, G'$ be two databases and $\delta$ be a connected Boolean C2RPQ. If the disjoint union $G \uplus G'$ satisfies $\delta$, then either $G$ satisfies $\delta$ or $G'$ satisfies $\delta$.
2. Let $\gamma, \delta$ be Boolean C2RPQs. If $\delta$ is connected and $\gamma \subsetneqq \delta$, then there exists a subquery $\gamma'$ of $\gamma$, obtained as connected component of $\gamma$, such that $\gamma' \subsetneqq \delta$.

Notice first that if $\gamma$ and $\delta$ are CQs then the proof of Fact V.9.2 follows directly from the equivalence of $\gamma \subsetneqq \delta$ (resp. $G$ satisfies $\gamma$) and the existence of a homomorphism from $\delta$ to $\gamma$ (resp. $\gamma$ to $G$).

*Proof.* We first prove the second point. Write $\gamma \triangleq \gamma_1 \wedge ... \wedge \gamma_n$ where $\gamma_1, ..., \gamma_n$ are connected components of $\gamma$, and assume by contradiction that for all $i$, $\gamma_i \nsubseteq \delta$. Then there exists a database $G_i$ such that $G_i$ satisfies $\gamma_i$ but not $\delta$. Consider the disjoint union $G = G_1 \uplus ... \uplus G_n$.

On the one hand, since the $\gamma_i$'s have disjoint variables and $G_i$ satisfies $\gamma_i$ for each $i$, then $G$ satisfies $\gamma$. On the other hand, $G$ cannot satisfy $\delta$: if there was a homomorphism from $\delta$ to $G$, since $\delta$ is connected, there would exist an index $i$ such that $\delta$ is mapped on $G_i$, which would contradict the fact that $G_i$ does not satisfy $\delta$. Hence, $G$ does not satisfy $\delta$, which contradicts the containment $\gamma \subsetneqq \delta$.

To prove the first point, we simply apply the second one, by letting $\gamma$ be the conjunction of the canonical CQ associated with $G$ and $G'$—which is in fact the canonical CQ associated with $G \uplus G'$. From the assumption that $G \uplus G'$ satisfies $\delta$ it follows that $\gamma \subsetneqq \delta$ and so, by the first point, there is either a homomorphism from $\delta$ to $G$ or from $\delta$ to $G'$. $\qquad\square$

We can then prove Lemma V.9.1.

*Proof of Lemma V.9.1.* Fix $k \geq 1$. We focus on semantic tree-width, but the exact same reduction works for the other three problems. We introduce an intermediate problem, called the *asymmetric containment problem for tree-width $k$*: given two Boolean CRPQs $\gamma$ and $\gamma'$, where $\gamma$ has tree-width $k$, $\gamma'$ is connected and does *not* have semantic tree-width $k$, it asks whether $\gamma \subseteq \gamma'$. The proof of the lemma then contains two parts:

1. first, we reduce the asymmetric containment problem for tree-width $k$ to the semantic tree-width $k$ problem,

2. then, we prove that the asymmetric containment problem for tree-width $k$ is ExpSpace-hard.

☙ *(1)*: We reduce the instance $(\gamma, \gamma')$ of the asymmetric containment problem for tree-width $k$ to the instance $\gamma \wedge \gamma'$ of the semantic tree-width $k$ problem. We simply have to check that $\gamma \subseteq \gamma'$ if and only if $\gamma \wedge \gamma'$ has semantic tree-width $k$. The left-to-right implication is straightforward since $\gamma \subseteq \gamma'$ implies that $\gamma \wedge \gamma' \equiv \gamma$ and $\gamma$ was assumed to have tree-width $k$. For the converse implication, if $\gamma \wedge \gamma' \equiv \delta$ where $\delta$ is a UC2RPQ of tree-width $k$ then write $\delta = \bigvee_{i=1}^{n} \delta_i$ where the $\delta_i$'s are C2RPQs and let $\delta_{i,1}, \dots, \delta_{i,k_i}$ be the connected components of $\delta_i$.

Since for each $i$ we have $\delta_i \subseteq \delta \equiv \gamma \wedge \gamma' \subseteq \gamma'$, by Fact V.9.2, there exists $j_i$ such that $\delta_{i,j_i} \subseteq \gamma'$. Let $\delta' \triangleq \bigvee_{i=1}^{n} \delta_{i,j_i}$ so that, by construction $\delta' \subseteq \gamma'$. However, note that $\delta'$ has tree-width at most $k$ but $\gamma'$ was assumed not to have semantic tree-width $k$, hence $\delta' \subsetneq \gamma'$, so there exists $G'$ such that:

$$G' \text{ satisfies } \gamma' \quad \text{and} \quad G' \text{ does not satisfy } \delta'. \tag{V.5}$$

We now prove that $\gamma \subseteq \gamma'$. Let $G$ be a database satisfying $\gamma$. Then the disjoint union $G \uplus G'$ satisfies $\gamma \wedge \gamma'$ since $G$ satisfies $\gamma$, $G'$ satisfies $\gamma'$ and $\gamma$ and $\gamma'$ are Boolean so we can assume *w.l.o.g.* that they have disjoint variables. As a consequence, $G \uplus G'$ satisfies $\delta$ and hence $\delta'$, so there exists $i$ such that $G \uplus G'$ satisfies $\delta_{i,j_i}$. Since $\delta_{i,j_i}$ is connected, either $G$ satisfies $\delta_{i,j_i}$ or $G'$ satisfies $\delta_{i,j_i}$. By Equation (V.5), the latter cannot hold, so $G$ satisfies $\delta_{i,j_i}$ and hence $\gamma'$.

Therefore, we have shown that for each database $G$ that satisfies $\gamma$, then $G$ satisfies $\gamma'$, *i.e.*, $\gamma \subseteq \gamma'$. Overall, $\gamma \wedge \gamma'$ has semantic tree-width $k$ if and only if $\gamma \subseteq \gamma'$.

☙ *(2)*: We now show that the asymmetric containment problem for tree-width $k$ is ExpSpace-hard. It was shown in [Fig20, Lemma 8] that the containment of CRPQs was still ExpSpace-hard when restricted to inputs of the form:

where $K, L_1, \dots, L_p$ are regular languages over $\mathbb{A}$. We reduce it to the following problem:

where the right-hand side of $\delta'$ is a directed $(k + 2)$-clique and where # is a new symbol, i.e. # $\notin \mathbb{A}$.

We claim that $\gamma \subsetneqq \delta$ if and only if $\gamma' \subsetneqq \delta'$. The forward implication is direct and the converse implication simply relies on the fact that # # $\notin \mathbb{A}$.[10] Then, observe that $\gamma'$ has tree-width $1 \le k$, and that $\delta'$ is connected but do not have semantic tree-width at most $k$.

To prove the last point, consider a UC2RPQ $\Delta''$ that is equivalent to $\delta'$. Pick any expansion $\xi_1'$ of $\delta'$. Since $\Delta'' \subsetneqq \delta'$, there exists an expansion $\xi''$ of $\Delta''$ such that there is a homomorphism from $\xi_1'$ to $\xi''$. Dually, since $\delta' \subsetneqq \Delta''$, there exists an expansion $\xi_2'$ of $\delta'$ such that there is a homomorphism from $\xi'' \to \xi_2'$. Overall, we have homomorphisms $\xi_1' \to \xi'' \to \xi_2'$. Since $\xi_1'$ and $\xi_2'$ are both expansions of $\delta'$, they contain a #-labelled directed $(k + 2)$-clique, and the #-letter appears nowhere else. Should the homomorphism $\xi_1' \to \xi''$ not be injective, $\xi''$ would contain a #-labelled self-loop, and hence, the homomorphism $\xi'' \to \xi_2'$ would yield a #-self loop in $\xi_2'$, which does not exist! Hence, the homomorphism from $\xi_1'$ to $\xi''$ is injective on the $(k + 2)$-clique. As a result, $\xi''$ contains a $(k + 2)$-clique and has tree-width at least $k + 1$. We conclude that $\Delta''$ has tree-width at least $k + 1$ by Fact V.2.1, provided that $k \ge 2$.

Hence, we have shown that $\gamma \subsetneqq \delta$ if and only if $\gamma' \subsetneqq \delta'$ where $\gamma'$ has tree-width at most $k$, where $\delta'$ is connected and has semantic tree-width at least $k + 1$. Since our reduction can be implemented in polynomial time, we conclude that the problems of Lemma V.9.1 are ExpSpace-hard. □

[10] Indeed, the only possible homomorphisms from expansions of $\delta'$ to expansions of $\gamma'$ are the ones sending the expansions of atoms containing $L_1, \ldots, L_p$ inside the expansion of the atom on $K$.

## V.10   Discussion

### V.10.1   Complexity

We have studied the definability and approximation of UC2RPQ queries by queries of bounded tree-width and shown that the maximal under-approximation in terms of an infinitary union of conjunctive queries of tree-width $k$ can be always effectively expressed as a UC2RPQ of tree-width $k$ (Corollary V.3.9). However, while the semantic tree-width 1 problem is shown to be ExpSpace-complete (which was also established in [BRV16, Theorem 6.1, Proposition 6.2]), we have left a gap between our lower and upper bounds in Theorem V.1.3 for every $k > 1$.

**Question V.10.1.** For $k > 1$, is the semantic tree-width $k$ problem ExpSpace-complete?

A related question is whether the CONTAINMENT PROBLEM between a C2RPQ and a summary query is in ExpSpace. Should this be the case, then the semantic tree-width $k$ problem would be in ExpSpace. We also point out that since every path-$l$ approximation can be expressed by a polynomial UC2RPQ of tree-width $2k$—this is the same idea as in [RBV17, Lemma IV.13]—, one can produce, for every UC2RPQ $\Delta$ a union $\Gamma$ of poly-sized C2RPQ of tree-width $2k$ such that $\mathrm{App}_{\mathcal{T}w_k}(\Delta) \subseteq \Gamma \subseteq \Delta$. This implies that the following "promise" problem[11] is decidable in ExpSpace: given a UC2RPQ $\Gamma$, answer 'yes' if $\Gamma$ is of semantic tree-width $2k$, and answer 'no' if $\Gamma$ is not of semantic tree-width $k$. The fact that $\mathrm{App}_{\mathcal{T}w_k}(\Delta)$ can be approximated by an exponential query of tree-width $2k+1$ can also be seen as a corollary of the proof of [RBV17, Theorem V.1].

We also do not know whether the $\Pi_2^p$ bound on the semantic tree-width $k$ problem for UCRPQ(SRE) has a matching lower bound. The known lower bound for the UCRPQ(SRE) CONTAINMENT PROBLEM [Fig+20, Theorem 5.1] does not seem to be useful to be employed in a reduction in this context, since it necessitates queries of arbitrary high tree-width.

### V.10.2 Characterization of Tractability

Our result implies that for each $k$ the EVALUATION PROBLEM for UC2RPQs $\Gamma$ of semantic tree-width $k$ is fixed-parameter tractable when parameterized by the size of the query, *i.e.* it works in time $\mathcal{O}(|G|^c \cdot f(|\Gamma|))$ for a computable function $f$ and constant $c$, where $G$ is the database given as input. While this was a known fact [RBV17, Corollary IV.12], the dependence on the database was $c = 2k+1$. Our results show that the dependence can be improved to $c = k+1$, similarly to [BRV16, Theorem 6.3] for the case $k = 1$. It has been further shown by Feier, Gogacz and Murlak that the evaluation can be done with a single-exponential $f$ [FGM24, Theorem 22].

In a similar vein, our results show that the EVALUATION PROBLEM for UC2RPQs of semantic path-width $k$ is in paraNL. It is unknown whether the semantic bounded width properties characterize all FPT and paraNL classes.

**Question V.10.2.** Does every recursively enumerable class of CRPQs with paraNL evaluation have bounded semantic path-width?

**Question V.10.3** (Also mentioned in [RBV17, § IV-(4)])**.** Does every recursively enumerable class of CRPQs with FPT evaluation have bounded semantic tree-width?

Note that the classes of bounded contracted path-width or contracted tree-width are not counterexamples to Questions V.10.2 and V.10.3, since the path-width is upper bounded by one plus the contracted path-width, and lower bounded by the contracted path-width—and similarly for tree-width—and so a width is bounded *iff* its contracted variant is bounded.

In the case of CQs, the answer is 'yes' to Question V.10.3 [Gro07, Theorem 1] under standard complexity-theoretic hypotheses ($W[1] \neq$ FPT). For Question V.10.2, the answer is still 'yes' [CM13, Theorem 3.1] conditional to a less standard assumption[12] (no Tree-hard problem is in paraNL).

However, attempting at answering these questions for CRPQs is considerably more challenging. In particular, one important technical difficulty is that a class of CRPQs with unbounded tree-width may contain queries with no expansions which are hom-minimal in the sense of containment. That is, for every $k$, for every query $\gamma$ of semantic tree-width $> k$ and expansion $\xi$ of semantic tree-width $> k$ there may be another expansion $\xi'$ such that $\xi' \xrightarrow{\text{hom}} \xi$ (i.e., such that $\xi \subsetneq \xi'$). In fact, for classes of CRPQs avoiding such problematic behavior, Question V.10.3 can be positively answered. We next show why.

Let us call a UC2RPQ *finitely redundant* if there is no infinite chain $\xi_1(\bar{x}) \subsetneq \xi_2(\bar{x}) \subsetneq \cdots$ among its expansions. See Figure V.13 for a non-example. Observe that the classes of CQs and UCQs are finitely redundant, and also the class of *loopless CRPQs*, meaning no directed cycle in its underlying directed graph and no empty word $\varepsilon$ in the atom languages.

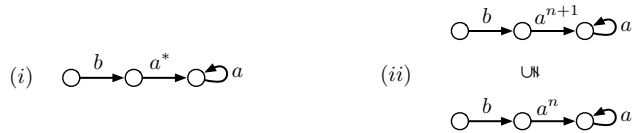

Figure V.13: (*i*) A simple non-finitely redundant Boolean CRPQ. (*ii*) For every $n$, there is a homomorphism from the $(n + 1)$-expansion to the $n$-expansion, but no homomorphism in the converse direction.

**Lemma V.10.4.** The class of loopless CRPQs is finitely redundant.

*Proof.* By means of contradiction, let $\gamma$ be loopless and suppose there is an infinite chain $\xi_1(\bar{x}) \subsetneq \xi_2(\bar{x}) \subsetneq \cdots$ of expansions of $\gamma$. Hence, there must be an atom expansion which grows arbitrarily in the chain. Take $\xi_i$ such that it contains an atom expansion of size bigger than $\xi_1$. Since such atom expansion is a directed path (as we are dealing with one-way CRPQs), the fact that $\xi_i \xrightarrow{\text{hom}} \xi_1$ implies that there is some cycle in $\xi_1$. Since $\gamma$ cannot contain the empty word in the atom languages, this is in contradiction with the hypothesis that there are no directed cycles in $\gamma$. $\square$

We next show that, restricted to classes of finitely redundant UC2RPQ, we can obtain a characterization of evaluation in FPT.

**Theorem V.10.5.** Assuming $W[1] \neq \text{FPT}$, for any recursively enumerable class $\mathcal{C}$ of finitely redundant Boolean UC2RPQs, the EVALUATION PROBLEM for $\mathcal{C}$ is FPT if, and only if, $\mathcal{C}$ has bounded semantic tree-width.

*Proof.* ⮑ *Left-to-right* By contraposition, we show that if $\mathcal{C}$ has unbounded tree-width, then its EVALUATION PROBLEM is $W[1]$-hard via an FPT-reduction from the *parameterized clique problem*. This is the problem of, given a parameter $k$ and a simple graph $G$, whether $G$ contains a $k$-clique. We do this by a simple adaptation of the proof of Grohe [Gro07, Theorem 4.1] for the case of CQs.

Given an instance $\langle G, k \rangle$ of the parameterized clique problem, the idea is to first search for a query $\gamma \in \mathcal{C}$ of "sufficiently large" semantic tree-width.

**Proposition V.10.6.** For any $k \geq 3$, if a finitely redundant C2RPQ has semantic tree-width $\geq k$, then there is a hom-minimal expansion thereof of semantic tree-width $\geq k$.

*Proof.* Let $\gamma$ be a finitely redundant C2RPQ. Consider the infinitary UCQ

$$\Xi \triangleq \{\check{\xi} \mid \xi \text{ is a hom-minimal expansion of } \gamma\}.$$

Since $\gamma$ is finitely redundant, we have $\gamma \equiv \Xi$. We prove the fact by contraposition. If all hom-minimal expansions of $\gamma$ have semantic tree-width $\leq k-1$, then all CQs of $\Xi$ have tree-width $\leq k-1$, and so by the implication (1) $\Rightarrow$ (3) of Theorem V.3.13, query $\gamma$ has semantic tree-width at most $k-1$. Note that for Theorem V.3.13 to apply, we need $k-1 > 1$ *i.e.* $k \geq 3$. □

**Proposition V.10.7.** The set of all hom-minimal expansions of queries from $\mathcal{C}$ is recursively enumerable.

*Proof.* We first show that given an expansion $\xi$ of some C2RPQ $\gamma$, it is decidable whether $\xi$ is hom-minimal. This follows from the following claim: there exists an expansion $\xi'$ of $\gamma$ s.t. $\xi' \xrightarrow{\text{hom}} \xi$ and $\xi \xcancel{\xrightarrow{\text{hom}}} \xi'$ iff there exists such an expansion whose atom expansions have length at most $2^m \cdot |\xi| \cdot |\mathbb{A}|^{2|\xi|}$ where $|\xi|$ is the number of variables of $\xi$ and $m$ is the greatest number of states of an NFA labelling an atom of $\gamma$. Decidability of hom-minimality clearly follows from this claim: it suffices to check if $\xi' \xrightarrow{\text{hom}} \xi$ implies $\xi \xrightarrow{\text{hom}} \xi'$ for all "small" $\xi'$.

To prove the claim, let $\xi'$ be an expansion of $\gamma$, and assume that there is a homomorphism $f \colon \xi' \to \xi$ and that $\xi \xcancel{\xrightarrow{\text{hom}}} \xi'$. Consider an atom expansion

$$\pi' = x_0 \xrightarrow{a_1} x_1 \xrightarrow{a_2} \cdots \xrightarrow{a_{n-1}} x_{n-1} \xrightarrow{a_n} x_n$$

of $\xi'$, and let $\mathcal{A}$ denote the NFA associated with the atom. For any index $i \in [\![0, n]\!]$ which is neither among the $|\xi|$ first positions nor the $|\xi|$ last positions, define its type $\tau_i$ as the word $a_{i-|\xi|} - 1 \cdots a_i a_{i+1} \cdots a_{i+|\xi|}$ of length $2|\xi|$—note that $\tau_i$ uniquely describes the ball of radius $|\xi|$ centred at $x_i$ in $\xi$. Consider the function which maps index $i \in [\![|\xi|, n - |\xi| + 1]\!]$ to the pair $\langle f(x_i), Q_i, \tau_i \rangle$, where $Q_i$ is the set of states $q$ of $\mathcal{A}$ which admit a path from an initial state to $q$ labelled by $a_1 \cdots a_i$. If $n \geq |\xi| \cdot 2^{|\mathcal{A}|} \cdot |\mathbb{A}|^{2|\xi|} + 2|\xi|$ then by the pigeon-hole principle, there exists $i, j \in [\![|\xi|, n - |\xi| + 1]\!]$ s.t. $i < j$, $f(x_i) = f(x_j)$, $Q_i = Q_j$ and $\tau_i = \tau_j$. Letting

$$\pi'' = x_0 \xrightarrow{a_1} x_1 \xrightarrow{a_2} \cdots \xrightarrow{a_{i-1}} x_{i-1} \xrightarrow{a_i} x_i = x_j \xrightarrow{a_{j+1}} x_{j+1} \xrightarrow{a_{j+2}} \cdots \xrightarrow{a_{n-1}} x_{n-1} \xrightarrow{a_n} x_n,$$

consider the query $\xi''$ obtained from $\xi'$ by replacing $\pi'$ with $\pi''$. Since $Q_i = Q_j$, $\xi''$ is still an expansion of $\gamma$. Moreover, $f(x_i) = f(x_j)$ implies that there is a homomorphism from $\xi''$ to $\xi$. Lastly, it there was a homomorphism from $\xi$ to $\xi''$, then this homomorphism should contain $x_i$ in its image—otherwise there would clearly be a homomorphism from $\xi$ to $\xi'$. Note that the image of this homomorphism is included in the ball of $\xi''$ centered at $x_i = x_j$ of radius $|\xi|$. But since $\tau_i = \tau_j$ this ball is equal to the ball of $\xi'$ centered at $x_i$ (or equivalently at $x_j$) of radius $|\xi|$, and so we found a homomorphism from $\xi$ to $\xi'$, which is not possible. Hence, there cannot be any homomorphism from $\xi$ to $\xi''$, which concludes the proof.

Finally, to enumerate all hom-minimal expansions of queries from $\mathcal{C}$, it suffices to enumerate all expansions of queries from $\mathcal{C}$—which is doable since $\mathcal{C}$ is recursively enumerable—and only keep those which are hom-minimal, using the previous algorithm. □

We proceed with the reduction. For any value of $k$ which is big enough, we enumerate all hom-minimal expansions of $\mathcal{C}$ until we find one such expansion $\xi$ whose core contains a $K \times K$ grid as a minor, for $K = \binom{k}{2}$. We know that this must happen by Proposition V.10.6 and the Excluded Minor Theorem [RS86], stating that there exists a function $f : \mathbb{N} \to \mathbb{N}$ such that for every $n \in \mathbb{N}$ every graph of tree-width at least $f(n)$ contains a $(n \times n)$-grid as a minor. Once we get hold of such a hom-minimal expansion $\xi$, we proceed as in [Gro07, proof of Theorem 4.1] to produce, in polynomial time, a graph database $G_\xi$ such that:

1. there is a homomorphism $G_\xi \xrightarrow{\text{hom}} \xi$, and
2. $G_\xi$ satisfies $\xi$ if, and only if, $G$ has a clique of size $k$.

Now consider the UC2RPQ $\Gamma \in \mathcal{C}$ of which $\xi$ is an expansion, and observe that if $G_\xi$ satisfies $\Gamma$, then we must have that $G_\xi$ also satisfies $\xi$, by the fact that $G_\xi \xrightarrow{\text{hom}} \xi$ and $\xi$ is hom-minimal. Hence, the following are equivalent:

- $G_\xi$ satisfies $\Gamma$,
- $G_\xi$ satisfies $\xi$,
- $G$ contains a $k$-clique.

This finishes the FPT-reduction.

☙ *Right-to-left*  This direction does not need any of the hypotheses (neither finite redundancy, $\mathsf{W}[1] \neq \mathsf{FPT}$, nor r.e.), by Corollary V.3.16. □

### V.10.3  Larger Classes

A natural and simple approach to extend the expressive power of CRPQs is to close the queries by transitive closure. That is, given a binary CRPQ $\gamma(x, y)$ we can consider CRPQ over the extended alphabet $\mathbb{A} \cup \{\gamma\}$, where the label $\gamma$ is interpreted as the binary relation defined by $\gamma(x, y)$. This is the principle behind *Regular Queries* [RRV17]. The notion of tree-width can be easily lifted to this class, and classes of bounded tree-width still have a polynomial-time EVALUATION PROBLEM. However, this class has not yet been studied in the context of the semantic tree-width. It is not known if the semantic tree-width $k$ problem is decidable, nor whether classes of bounded semantic tree-width have an FPT EVALUATION PROBLEM.

**Question V.10.8.** Is the semantic tree-width $k$ problem for Regular Queries decidable?

| Query class | Membership problem | EVALUATION PROBLEM |
|---|---|---|
| path-width $\leq k$<br>sem. path-w. $\leq k$<br><br>(Theorem V.8.6) | L-c [KM10, Theorem 1.3, p. 2]<br>2ExpSpace & ExpSpace-h | NL-c (Lemma V.8.9)<br>paraNL (Theorem V.8.8) |
| tree-width $\leq k$<br>sem. tree-w. $\leq k$<br><br>(Theorem V.1.3) | L-c [EJT10, Lemma 1.4]<br>2ExpSpace & ExpSpace-h[14] | P (Folklore)[13]<br>FPT [RBV17, Corollary V.2][15]<br>NP-c [RBV17, Theorem V.3] |

Table 14: Complexity of the membership and EVALUATION PROBLEM for some classes of UC2RPQs studied in this chapter, where $k \geq 1$ is fixed. The same results hold for the contracted variants. The abbreviation "-c" (resp. "-h") stands for "-complete" (resp. "-hard").

### V.10.4 Different Notions

CRPQs of small tree-width or path-width enjoy a tractable EVALUATION PROBLEM, see Table 14. However, it must be noticed that containment between tree-width $k$ or path-width $k$ queries is still very hard: ExpSpace-complete (even for $k = 1$) [CDLV00]. The more restrictive measure of "bridge-width" [Fig20] has been proposed as a more robust measure, which results in classes of queries which are well-behaved both for evaluation (since bridge-width $k$ implies tree-width $\leq k$) and for containment (since containment of bounded-bridge-width classes is in PSpace). It is not hard to see that bridge-width is closed under refinements, and thus that this notion is amenable to our approach (*cf.* Fact V.3.5).

**Question V.10.9.** Is the problem of whether a UC2RPQ is equivalent to a UC2RPQ of bridge-width at most $k$ decidable?

[13] Originally proven by Chekuri & Rajaraman [CR00, Theorem 3] for CQs. The generalization to UC2RPQs is trivial, see *e.g.* Proposition V.1.1 or [RBV17, Theorem IV.3].

[14] See also [BRV16, Theorem 6.1] for $k = 1$.

[15] See also Corollary V.3.16 and [FGM24, Theorem 22].

## *Appendices*

## *V.A    Alternative Upper Bound for Containment of UC2RPQs*

In Section V.3, in order to prove the 2ExpSpace upper bound to the semantic tree-width $k$ problem (Lemma V.3.10), we proved an upper bound on containment of UC2RPQs (Proposition V.3.11) by relying on the notion of bridge-width. In this section, we give a slightly different bound, which is more elementary (in the sense that it does not rely on bridge-width) and still yields a 2ExpSpace upper bound to the semantic tree-width $k$ problem.

**Proposition V.A.1.** The CONTAINMENT PROBLEM $\Gamma \subseteq \Delta$ for UC2RPQs can be solved in non-deterministic space $2^{c \cdot \|\Gamma\|} + p_\Delta \cdot 2^{c \cdot m_\Delta}$, where $m_\Delta$ is the size of the greatest disjunct of $\Delta$, namely $m_\Delta = \max\{\|\delta_\Delta\| \mid \delta \in \Delta\}$, $p_\Delta$ is the number of disjuncts of $\Delta$, and $c$ is a constant.

*Proof sketch.* The proposition can be shown by close inspection of the standard CONTAINMENT PROBLEM for UC2RPQs [CDLV00, Theorem 5]: the CONTAINMENT PROBLEM is reduced, in this instance, to checking the inclusion between NFAs of the form[16]

$$\mathcal{A}_\Gamma \subseteq^? \bigcup_{\delta \in \Delta} \mathcal{A}_\delta, \qquad\qquad \text{(V.6)}$$

where $A_\Gamma$ is a regular expression which is exponential in $\|\Gamma\|$, and $\mathcal{A}_\delta$ has size exponential in $\|\delta\| \leq m_\Delta$. Should (V.6) not hold, there must exist a counterexample of size at most

$$2^{|\mathcal{A}_\Gamma|} \times \prod_{\delta \in \Delta} 2^{|\mathcal{A}_\delta|}$$

Letting $p_\Delta$ be the number of queries in $\Delta$, we get that the logarithm of the expression above—representing the size of the non-deterministic space needed by the algorithm—is upper bounded by

$$c_0 \Big( |\mathcal{A}_\Gamma| + \sum_{\delta \in \Delta} |\mathcal{A}_\delta| \Big) \underset{\text{eventually}}{\leq} 2^{c \cdot \|\Gamma\|} + p_\Delta \cdot 2^{c \cdot n_\Delta},$$

for some constants $c_0$ and $c$.  □

## *V.B    Path-Width is not Closed under Refinements*

**Fact V.8.1.** For each $k \geq 1$, the class of graphs of path-width at most $k$ is not closed under expansions.

*Proof.* Let $X$ be a set of $k-1$ variables. Consider the undirected multigraph $\mathcal{G}_k$ whose set of nodes is $X \cup \{y_0, y_1, y_2, y_3\}$ with the following edge set:

- each $X \cup \{y_i\}$ ($i \in \{0, 1, 2, 3\}$) is a clique,
- there is an edge from $y_i$ to $y_{i+1}$ for $i \in \{0, 1, 2\}$, and

---

[16] $\mathcal{A}_\Gamma$ and $\mathcal{A}_\delta$ are denoted $A_1$ and $A_3$, respectively, in [CDLV00].

- there is a second edge from $y_1$ to $y_2$.

By definition, this graph has path-width exactly $k$: it is as least $k$ since it contains a $(k+1)$-clique—namely $X \cup \{y_i, y_{i+1}\}$—and, moreover the following sequence of bags—*cf.* Figure V.12a—defines a path decomposition of $\mathcal{G}_k$ of width $k$:

$$\langle X \cup \{y_0, y_1\}, \; X \cup \{y_1, y_2\}, \; X \cup \{y_2, y_3\} \rangle.$$

Let $\mathcal{G}'_k$ be the graph obtained by refining the second edge from $y_1$ to $y_2$, into two edges $\langle y_1, z \rangle$ and $\langle z, y_2 \rangle$, where $z$ is a new variable—see Figures V.12b and V.12d. We claim that $\mathcal{G}'_k$ has path-width at least $k+1$. Indeed, let $\langle T, \mathbf{v}, \mathbf{t} \rangle$ be a path decomposition of $\mathcal{G}'_k$.

Note that $X \cup \{y_0, y_1\}$, $X \cup \{y_1, y_2\}$, $X \cup \{y_2, y_3\}$ and $\{z, y_1, y_2\}$ are cliques, so there must be bags of $\langle T, \mathbf{v}, \mathbf{t} \rangle$ containing each of them. Let $b_{0,1}$, $b_{1,2}$, $b_{2,3}$ and $b_Z$ denote these bags—note that they do not have to be distinct.

1. If $b_Z$ appears in $T$ between at least two bags among $b_{0,1}$, $b_{1,2}$ and $b_{2,3}$ (as in Figure V.12b), since $X \subseteq \mathbf{v}(b_{i,j})$ for all $(i, j)$, then $X \subseteq \mathbf{v}(b_z)$. Hence $X \cup \{z, y_1, y_2\} \subseteq \mathbf{v}(b_z)$ and so $b_z$ has $k+2$ elements.

2. Otherwise, *w.l.o.g.* $b_z$ appears strictly before all three bags $b_{0,1}$, $b_{1,2}$ and $b_{2,3}$, as in Figure V.12d. We consider the way $b_{0,1}$, $b_{1,2}$ and $b_{2,3}$ are ordered in the path decomposition. If $b_{0,1}$ or $b_{2,3}$ appears first, then they are located between $b_z$ and $b_{1,2}$, which both contain $\{y_1, y_2\}$, and so this bag must also contain $\{y_1, y_2\}$, and so it has size at least $k+2$. Otherwise, if $b_{1,2}$ appears first, depending on the relative ordering of $b_{0,1}$ and $b_{2,3}$, we either get that $y_2 \in \mathbf{v}(b_{0,1})$ or that $y_1 \in \mathbf{v}(b_{2,3})$. In both cases, we have a bag with at least $k+2$ elements.

In all cases, the path decomposition has width at least $k+1$, showing that $\mathcal{G}'_k$ has path-width at least[17] $k+1$. □

[17] In fact it has path-width exactly $k+1$.

# Conclusion & Open Problems

## ABSTRACT

We conclude the first part of the thesis on graph databases. We recall the most interesting open problems and conjectures on minimization that we discussed in the previous chapters. We conclude by presenting the notion of profinite databases: we briefly explain that while the notion does seem promising to help understand the structure of CRPQs, we unfortunately failed to use it in an interesting manner.

## CONTENTS

## VI.1   Minimization Problems

Rather than closing the complexity gaps between our $k$-ExpSpace upper bounds and the ExpSpace lower bounds in the decision problems studied in Chapters IV and V, we believe that the most interesting questions *w.r.t.* to minimization are actually those related to *structure*, and more precisely those that try to connect the different notions of minimality together.

**Conjecture IV.6.1.** There exist (atom) minimal CRPQs which are not variable minimal.

*Minimization & Trees.*   Whether the seminal results of [CMNP18] could be lifted from tree patterns to their encoding as CRPQs remains open.

**Conjecture IV.6.4.** If a tree pattern is minimal among tree patterns, then its encoding as a CRPQ should also be minimal among CRPQs, up to contracting internal variables.

On a similar note, an interesting question is whether two goals (*e.g.* the number of variables and the number of atoms) can be simultaneously minimized. For CQs, this is always the case by Proposition III.1.19. However, this question seems quite hard for CRPQs, even on concrete examples. We say that a CRPQ is *forest-shaped* if its underlying directed graph is a disjoint union of directed trees.

**Question VI.1.1.** Is it the case that every Boolean CRPQ that is (1) equivalent to a forest-shaped CRPQ and (2) equivalent to a CRPQ with at most $k$ atoms is necessarily equivalent to a forest-shaped CRPQ with at most $k$ atoms?

We conjecture the answer to this question to be yes, but we were unable to prove it: we only managed to prove that, under these assumptions, the query should be equivalent to a forest-shaped CRPQ with at most $2^k$ atoms (Theorem VI.A.2).

*Maximal Under-Approximations.*   Both the result on UCRPQ minimization of Chapter IV and of semantic tree-width minimization of Chapter V rely on the existence—and computability—of maximal under-approximations. In the first case, the target class consists of finitely many graphs (Lemma IV.4.6), but in the second case, it is infinite (Lemma V.3.8): as such the proof is significantly harder. Having the remarkable genericity of Proposition III.1.19 in mind, we could only hope to be able to capture both Lemmas IV.4.6 and V.3.8.

**Question VI.1.2.** Given a class $\mathcal{C}$ of graphs that is closed under minors, do maximal under-approximations by UC(2)RPQs over $\mathcal{C}$ always exist? If so, are they computable?

## VI.2    Profinite Databases

As we have seen in Section III.1, duality—namely the existence of a dual isomorphism between queries and models—provides a remarkable framework to study conjunctive queries. For the more complex language of conjunctive regular path queries, however, this isomorphism fails, making static analysis much harder. We thought that part of the enjoyable properties of conjunctive queries could be recovered for CRPQs by considering the notion of profinite databases. In short, a *profinite database* consists of, a "projective system" of finite graph databases, *i.e.* an $\omega$-sequence of homomorphisms

$$\mathbf{G}_0 \xleftarrow{\text{hom}} \mathbf{G}_1 \xleftarrow{\text{hom}} ... \xleftarrow{\text{hom}} \mathbf{G}_n \xleftarrow{\text{hom}} \mathbf{G}_{n+1} \xleftarrow{\text{hom}} ...$$

that we denote by $\varprojlim_{n \in \mathbb{N}} \mathbf{G}_n$.

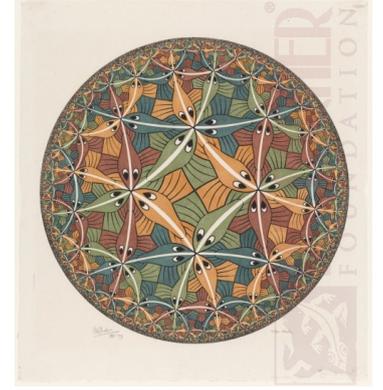

Figure VI.1: *Circle Limit III*, M. C. Escher, © The M.C. Escher Company.

A typical example of such a system can be obtained as follows: just like we did in Section III.1, we let $\mathbf{C}_n$ ($n \in \mathbb{N}_{>0}$) denote the directed cycle with domain $\mathbb{Z}/n\mathbb{Z}$ and with an edge from $i$ to $j$ *iff* $i+1 = j$, see Figure VI.2. Recall that $\mathbf{C}_n \xrightarrow{\text{hom}} \mathbf{C}_m$ *iff* $n$ is a multiple of $m$. In particular, we have

$$\mathbf{C}_1 \xleftarrow{\text{hom}} \mathbf{C}_2 \xleftarrow{\text{hom}} ... \xleftarrow{\text{hom}} \mathbf{C}_{2^n} \xleftarrow{\text{hom}} \mathbf{C}_{2^{n+1}} \xleftarrow{\text{hom}} ... .$$

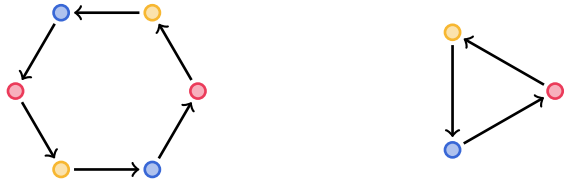

Figure VI.2:  The graphs $\mathbf{C}_6$ (left) and $\mathbf{C}_3$ (right) and a homomorphism from the former to the latter, described by colour coding. (Replica of Figure III.5.)

The crucial point is that projective system have a natural semantics: we define the semantics of

$$\varprojlim_{n \in \mathbb{N}} \mathbf{G}_n$$

as the set of points that above some element of this sequence, *i.e.* as

$$\mathbf{H} \vDash [\![\, \varprojlim_{n \in \mathbb{N}} \mathbf{G}_n \,]\!] \quad \text{when} \quad \exists n.\ \mathbf{G}_n \xrightarrow{\text{hom}} \mathbf{H}.$$

For instance

$$\varprojlim_{n \in \mathbb{N}} \mathbf{C}_{2^n}$$

has a simple semantical interpretation: letting $\mathbf{G}$ be a graph database, for any $n \in \mathbb{N}$, we get that $\mathbf{C}_{2^n} \xrightarrow{\text{hom}} \mathbf{G}$ if, and only if, $\mathbf{G}$ contains a directed cycle of length $2^n$. And so, there exists $n \in \mathbb{N}$ *s.t.* $\mathbf{C}_{2^n} \xrightarrow{\text{hom}} \mathbf{G}$ if, and only if, $\mathbf{G}$ contains a directed cycle whose length is a power of 2. In fact the true definition of projective system allows the sequence to be indexed by a directed set rather than $\mathbb{N}$: in the case we would obtain an object

$$\varprojlim_{n \in \langle \mathbb{N}_{>0}, | \rangle} \mathbf{C}_{n'}$$

where we order $\mathbb{N}_{>0}$ by divisibility, and whose semantics is "the database contains a directed cycle"!

*How Profinite Databases Arise.* Profinite models naturally arise by considering Stone duality. Roughly, Stone duality is a theory that, starting from a logic and its models, adds new idealized/abstract models which are necessary to make the theory compact, well-behaved and nicely describable. For instance, applied to first-order logic over all structures, this construction does nothing! This is precisely because this logic is already compact. On the other hand, when applied to regular languages of finite words, we obtained profinite words, which are exactly the models needed to describe pseudovarieties of monoids! We refer the reader to [GG24] for a not so short survey of the topic.

Second, profinite databases naturally appear when dealing with the semantics of CRPQs.[1] Indeed, let $\gamma()$ and $\delta()$ be CRPQs, and assume that $\gamma() \equiv \delta()$. Over CQs, it would mean that their core are isomorphic (by Proposition II.2.3 and duality), but over CRPQs, we do not have such a nice characterization. However, pick a canonical database $\mathbf{G}_0 \vDash^\star \gamma$. From the fact that $\gamma \subsetneqq \delta$, we get the existence of $\mathbf{D}_0 \vDash^\star \delta$ *s.t.* $\mathbf{G}_0 \xleftarrow{\text{hom}} \mathbf{D}_0$. In turn, using the converse containment $\delta \subsetneqq \gamma$, we get that there exists $\mathbf{G}_1 \vDash^\star \gamma_1$ *s.t.* $\mathbf{D}_0 \xleftarrow{\text{hom}} \mathbf{G}_1$. By induction, we obtain an infinite sequence

$$\mathbf{G}_0 \xleftarrow{\text{hom}} \mathbf{D}_0 \xleftarrow{\text{hom}} \dots \xleftarrow{\text{hom}} \mathbf{G}_n \xleftarrow{\text{hom}} \mathbf{D}_n \xleftarrow{\text{hom}} \mathbf{G}_{n+1} \xleftarrow{\text{hom}} \mathbf{D}_{n+1} \xleftarrow{\text{hom}} \dots .$$

In other words, we get that the profinite databases

$$\varprojlim_{n \in \mathbb{N}} \mathbf{G}_n \quad \text{and} \quad \varprojlim_{n \in \mathbb{N}} \mathbf{D}_n$$

are semantically equivalent!

*What Profinite Databases Could Help Us Achieve.* In the previous chapters, we presented two results that where conditional to the existence of hom-minimal expansions:

- the semantical structure theorem (Theorem IV.2.9), which provides lower bounds on the complexity required to express a CRPQ—and which is, to out knowledge, actually the only result of this form for CRPQs;
- a very partial generalization of Grohe's theorem to finitely redundant Boolean UC2RPQs (Theorem V.10.5).

However, a very simple CRPQ such as $\gamma() \triangleq x \xrightarrow{\mathbb{A}^*} x$, expressing that the database contains directed cycle no do not have hom-minimal expansions,

---

[1] They actually are at the heart of the proof of Theorem VI.A.2!

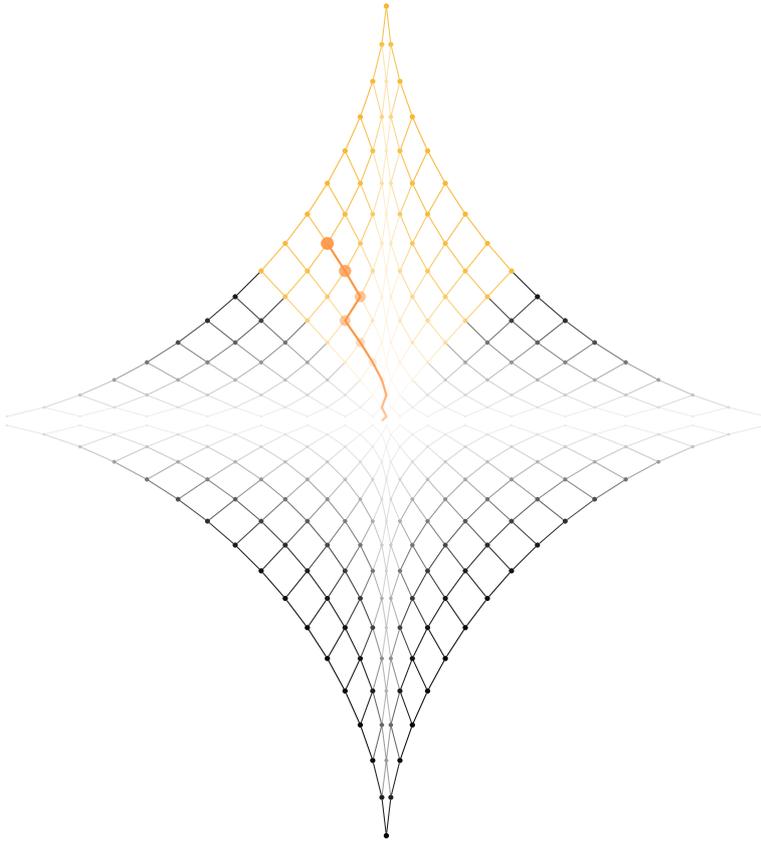

Figure VI.3: Semantics of a query that does admit a hom-minimal model (in yellow), together with a profinite database built out of its models in an effort to lift the opaque veil of the infinite structure of the distributive lattice of graph databases (in orange).

but it does have a hom-minimal profinite database, which is

$$\varprojlim_{n \in \langle \mathbb{N}_{>0}, | \rangle} \mathbf{C}_n.$$

More generally, by Zorn's lemma, every CRPQ admits at least one hom-minimal profinite database, see Figure VI.3!

**Question VI.2.1.** Can we generalize Theorems IV.2.9 and V.10.5 to handle hom-minimal profinite databases rather than hom-minimal finite databases?

A positive answer to this question would provide a necessary and sufficient condition on all CRPQs to be expressibly by simple CRPQs, and might lead to a characterization of tractable classes of CRPQs—which are results we can only dream of given our current knowledge.

**Question V.10.3** (Also mentioned in [RBV17, § IV-(4)]). Does every recursively enumerable class of CRPQs with FPT evaluation have bounded semantic tree-width?

One the positive side, we managed to prove—in an unpublished work with Sam van Gool—that the Stone dual of the Heyting algebra of conjunctive queries is isomorphic to the space of profinite databases, which seems to point towards the fact that profinite databases are natural objects.

On the negative side, we do not know what to do with this result... One of the main difficulties however is to find a reasonable definition for the notion of "$\mathcal{C}$-profinite databases". We would like this notion to be defined at least whenever $\mathcal{C}$ is a minor-closed class of CRPQs, and intuitively $\mathcal{C}$-profinite databases should generalize the $\mathcal{C}$-finite databases. For instance, for tree-width, we might have that homomorphisms

$$\mathbf{G} \xleftarrow{\text{hom}} \mathbf{D} \xleftarrow{\text{hom}} \mathbf{G}'$$

where $\mathbf{G}$ and $\mathbf{G}'$ have tree-width at most $k$, but where $\mathbf{D}$ is nowhere having semantic tree-width at most $k$... More abstractly: homomorphisms do not interact that well with the notion of minor.[2]

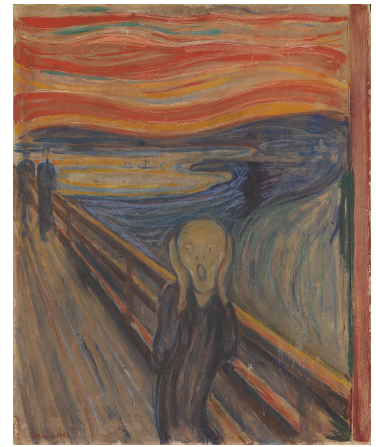

Figure VI.4: When your Ph.D. student talks about the Stone dual space of the Heyting algebra of conjunctive queries for the forth time this month. *Skrik*, by Edvard Munch.

[2] This contrasts with the fact that minor-closed classes are closed under taking cores.

*Appendices*

## VI.A    Tree-Like Queries

> HYPOTHESIS.  In this section, all CRPQs are assumed to be positive, meaning that no language can contain the empty word.

### VI.A.1    Forest-Shaped and DAG-Shaped Queries

We say that a CRPQ is *semantically forest-shaped* if it is semantically equivalent to a CRPQ which is forest-shaped.

Say that a CRPQ is *DAG-shaped* if its underlying directed multigraph is a DAG—parallel edges are allowed, not self loops. If $\delta$ is DAG-shaped, define its *unfolding*, denoted by $\mathcal{U}(\delta)$, as the following CRPQ:

- its variables are exactly labelled path of $\delta$ of the form $x_0 \xrightarrow{L_1} \cdots \xrightarrow{L_n} x_n$ with $n \in \mathbb{N}$ and $x_0$ is a vertex of $\delta$ with no predecessor;
- the atoms exactly go from $x_0 \xrightarrow{L_1} \cdots \xrightarrow{L_n} x_n$ to $x_0 \xrightarrow{L_1} \cdots \xrightarrow{L_n} x_n \xrightarrow{L_{n+1}} x_{n+1}$, with label $L_{n+1}$.

**Fact VI.A.1.**  If $\delta$ is a DAG-shaped CRPQ, then $\mathcal{U}(\delta)$ is a forest-shaped CRPQ, and moreover $\delta \subseteq \mathcal{U}(\delta)$.

The rest of this section is devoted to proving the following result.

**Theorem VI.A.2.**  Let $\delta$ be a CRPQ. The following are equivalent:

1. $\delta$ is semantically forest-shaped,
2. $\delta$ is DAG-shaped and for every hom-minimal canonical database $\mathbf{D}$ of $\delta$, the core of $\mathbf{D}$ is a forest,
3. $\delta$ is DAG-shaped and $\delta \equiv \mathcal{U}(\delta)$.

Note that since semantical equivalence is decidable in ExpSpace and since $\mathcal{U}(\delta)$ has exponential size, it follows that one can test if a CRPQ is semantically forest-shaped in 2ExpSpace.

### VI.A.2    Semantically DAG-Shaped Queries

We say that a *semantically DAG-shaped* when it is semantically equivalent to a DAG-shaped CRPQ.

**Fact VI.A.3.**  A CRPQ is semantically DAG-shaped *iff* it is DAG-shaped.

**Corollary VI.A.4.**  If a CRPQ is semantically forest-shaped, then it is DAG-shaped.

### VI.A.3    Semantically Forest-Shaped

From the fact that a CRPQ $\delta$ is equivalent to a forest-shaped query $\phi$ we know that for all canonical database $\mathbf{D}_0$ of $\delta$, since $\delta \subseteq \phi$, there exists a canonical database $\mathbf{F}_0$ of $\phi$ s.t. $\mathbf{D}_0 \xleftarrow{\text{hom}} \mathbf{F}_0$. But dually since $\phi \subseteq \delta$, there exists $\mathbf{D}_1 \vDash^\star \delta$ s.t. $\mathbf{F}_0 \xleftarrow{\text{hom}} \mathbf{D}_1$. By induction—and the axiom of choice—we obtain an infinite co-chain of homomorphisms

$$\mathbf{D}_0 \xleftarrow{\text{hom}} \mathbf{F}_0 \xleftarrow{\text{hom}} \mathbf{D}_1 \xleftarrow{\text{hom}} \mathbf{F}_1 \xleftarrow{\text{hom}} \cdots \xleftarrow{\text{hom}} \mathbf{D}_n \xleftarrow{\text{hom}} \mathbf{F}_n \xleftarrow{\text{hom}} \cdots.$$

We show that co-chains of forests are actually quite simple.

**Fact VI.A.5.** If $\mathbf{F}_0 \xleftarrow{\text{hom}} \mathbf{F}_1 \xleftarrow{\text{hom}} \cdots \xleftarrow{\text{hom}} \mathbf{F}_n \xleftarrow{\text{hom}} \cdots$ is an infinite co-chain of homomorphisms between forests, then there exists $n \in \mathbb{N}$ s.t. all $F_m$ with $m \geq n$ are hom-equivalent to one another.

*Proof.* For all $n \in \mathbb{N}$, from $F_n \xleftarrow{\text{hom}} F_{n+1}$ it follows that the maximal depth of a tree in $F_{n+1}$ is smaller or equal to the maximal depth of a tree in $F_n$. So, at some point this parameter must become stationary, say $d$. Then observe that there are finitely many forests with depth at most $d$, up to hom-equivalence, and hence, one of these must occur infinitely often in the co-chain. □

**Corollary VI.A.6.** If $\delta$ is semantically forest-shaped, then for any $\mathbf{D} \vDash^\star \delta$, there exists $\mathbf{D}' \vDash^\star \delta$ such that $\mathbf{D} \xleftarrow{\text{hom}} \mathbf{D}'$, $\mathbf{D}'$ is hom-minimal and the core of $\mathbf{D}'$ is a forest.

We can now proceed with the proof of Theorem VI.A.2, after giving a proposition that will prove useful.

**Proposition VI.A.7.** Let $\mathbf{F}$ be a forest and $\mathbf{D}$ a graph. If $\mathbf{F} \xrightarrow{\text{hom}} \mathbf{D}$ then $F \xrightarrow{\text{hom}} \mathcal{U}(D)$.

*Proof.* The homomorphism $\mathbf{F} \xrightarrow{\text{hom}} \mathcal{U}(\mathbf{D})$ can be defined by induction on $F$, from roots to leaves. □

*Proof of Theorem VI.A.2.* ❧ (1) $\Rightarrow$ (2). This follows from Corollaries VI.A.4 and VI.A.6.

❧ (2) $\Rightarrow$ (3). By Fact VI.A.1 we have $\delta \subseteq \mathcal{U}(\delta)$ so it suffices to prove the converse containment. Let $U$ be a canonical database of $\mathcal{U}(\delta)$. Then there exists $\mathbf{D} \vDash^\star \delta$ s.t. $\mathbf{U} = \mathcal{U}(\mathbf{D})$. By Corollary VI.A.6 there exists $\mathbf{D}' \vDash^\star \delta$ s.t. $\mathbf{D} \xleftarrow{\text{hom}} \mathbf{D}'$ and the core of $\mathbf{D}'$ is a forest. So $\mathbf{D} \xleftarrow{\text{hom}} \mathbf{\check{D}}'$, and so by Proposition VI.A.7, since $\mathbf{\check{D}}'$ is a forest, then $\mathcal{U}(\mathbf{D}) \xleftarrow{\text{hom}} \mathbf{\check{D}}'$ *i.e.* $\mathcal{U}(\mathbf{D}) \xleftarrow{\text{hom}} \mathbf{D}'$, which proves that $\mathcal{U}(\mathbf{D}) \vDash \delta$. Therefore, $\mathcal{U}(\delta) \subseteq \delta$.

❧ (3) $\Rightarrow$ (1). This is because $\mathcal{U}(\delta)$ is forest-shaped by Fact VI.A.1. □

# Entracte: What the Hare Said to Patroclus

Waiting for Achilles to return from his race, Patroclus stumbled onto a Hare that appeared to be lost in deep thought.

"Greetings, my friend. What seems to be bothering you?" inquired the Greek.

"Well, I have been thinking about Automedon—an automaton I mean" answered the Hare, "for I want to study rationality, but the notion of computation eludes me. After all, Babbage's machine hasn't been invented yet!"

"By far the most natural way of understanding rational languages is through the lens of algebra," Achilles' lover explained. "They simply consist of languages recognized by finite monoids."

"A monoid? What in Zeus's name is this?"

"Are you familiar with category theory? It is a branch of mathematics—"

"A branch of mathematics? What need do you have of having multiple branches? Have you found inconsistencies in Euclid's axioms?" interrupted the Hare.

"Not exactly. It provides a useful framework to abstractly study different structures. In geometry, points are considered atomic objects, and they are connected through shapes. In category theory, points are abstracted, and shapes are treated as first-class citizens, connected by shape transformations."

"This seems rather pointless... But if such abstract nonsense is what you have to endure to have provably consistent foundations of mathematics, I suppose it might be worth it," conceded the Hare.

"Anyway," added Patroclus after a brief pause, "a monoid can simply be defined as an Eilenberg-Moore algebra for the monad of finite words in the category of sets."

"A monad? Don't tell me that after disavowing points you abjured all gods but one."

"It has nothing to do with religion, put some faith in category theory: a monad is quite simple to define" said Patroclus. "Endofunctors of a category form themselves a category, whose morphisms are natural transformations. A monad consists of nothing else but a monoid in this category."

"A monoid? What in Zeus's name is this?" echoed the Hare.

At this point, the narrator, having to write his thesis, had to leave our two protagonists to their discussion. Several days passed after he submitted his manuscript when he went back to the spot, only to find the two companions arguing.

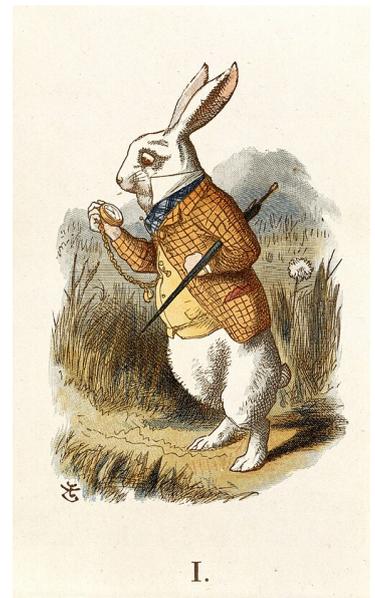

Figure E.1: Overconfident, the Hare preferred to think about algebraic language theory rather than how to win the upcoming race, see [Car95]. *The White Rabbit*, by John Tenniel.

"As I said, a monad is a monoid in the category of endofunctors!" Patroclus exclaimed, visibly irritated.

"But that doesn't help me understand what is a monoid..." the exhausted Hare replied, "at least we can hope for logicians not to be foolish enough to define rationality using monads."

# The Frontier of Decidability
# in Automatic Structures

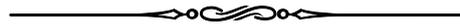

# Finite-Word Relations and Automatic Structures

### ABSTRACT

This preliminary chapter surveys the literature on the notion of *rationality* for finite word relations and automatic structures.

We start by reviewing competing definitions of rationality for $k$-ary relations of finite words ($k \geq 2$). We survey the corresponding hierarchy, which is mainly shaped by the expressive power of multitape automata. We also briefly mention other models such as transducers to emphasize their relationship with our hierarchy. Automatic relations naturally emerge as one of the most expressive class of relations having desirable closure properties and for which most "basic" problems are decidable.

After a brief logical interlude mostly dedicated to logical characterizations of automatic relations, we move on to automatic structures, which are infinite relational structures that can be finitely described using automatic relations. As a result, the first-order theory of such structures is decidable, explaining why automatic structures play a central role when looking for algorithms dealing with infinite structures. Unfortunately, other decision problems, such as the ISOMORPHISM PROBLEM, are undecidable, but become decidable when restricted to a specific subclass of structures. We hence conclude this chapter by discussing the computability status of natural decision problems on automatic structures.

### ACKNOWLEDGEMENTS

Parts of Section VII.1 come from [Mor25, § 1] and [Mor25, § B].

# Contents

## VII.1 The Landscape of Rationality for Relations over Finite Words

### VII.1.1 Regularity is Key

The study of classes of relations on words have always been a central topic in language theory [EM65; Niv68; Ber79; FS93; Cho06]. More recently, their study has also been motivated by database theory and verification, where they are used to build expressive languages. For instance, suitable classes of word relations where shown to be relevant for querying strings over relational databases [BLSS03], comparing paths in graph databases [BLLW12]—see also [Fig21b, § 8, p. 17] for more context & results on *extended* conjunctive regular path queries—, or defining string constraints for model checking [LB16]. We start by showing that, while the notion of regular language is canonical and admits numerous characterizations, these definitions are no longer equivalent when dealing with finite word relations, leading to a *hierarchy* of notions of *rationality*.

The class of regular languages is remarkably stable, and can be characterized as the languages recognized by either:

- deterministic or non-deterministic finite state automata,
  see *e.g.* [Pin21a, Proposition 1.2.3, p. 7];
- two-way finite state automata by Shepherdson-Rabin-Scott theorem
  [She59, Theorem 2, p. 198] [RS59, Theorem 15, p. 123];
- rational expressions[1] by Kleene's theorem,
  see *e.g.* [Pin21a, Theorem 1.5.11, p. 34];
- monadic second-order logic by Trakhtenbrot-Büchi-Elgot theorem,
  see *e.g.* [Boj20, Theorem 2.2, p. 32]; or
- finite monoids, see *e.g.* [Pin21a, § 1.4.2, p. 19].

Moreover, all transformations between these representations are effective—although some models are strictly more succinct than others.

These equivalences explain why the terms *recognizable language*—meaning implicitly "recognizable by a finite-state automaton" or "recognizable by a finite monoid"—and *rational language*—meaning "described by a rational expression"—are used interchangeably. In fact, in this thesis as well as in most of the literature, we will use the generic term of "regular language". However, in more complex settings, for instance subsets of non-free monoids,[2] the equivalence between these classes no longer holds [Pin21b].

The landscape of rationality for *k*-ary relations of finite words ($k \geq 2$) is far more complex than for languages,[3] as depicted in Figure VII.1. We will briefly present these classes, although this thesis will mostly deal with the two most restrictive ones, namely recognizable and automatic relations.[4]

We fix two alphabets $\Gamma$ and $\Sigma$. In the rest of this section, we focus on relations $\mathcal{R} \subseteq \Gamma^* \times \Sigma^*$. We will sometimes provide definitions for relations of higher arity when the generalization is not trivial.

[1] Usually called "regular expressions" by non-French speakers, however we use the terminology "rational" for its unambiguity.

[2] Recall that a language is nothing else but a subset of a free monoid.

[3] Which can be seen as unary relations of finite words.
[4] It should be noted that the names of these classes were often coined independently of one another and the terminology should be handled with care: for instance, "regular relations" do *not* correspond to the intersection of regular functions with functional relations.

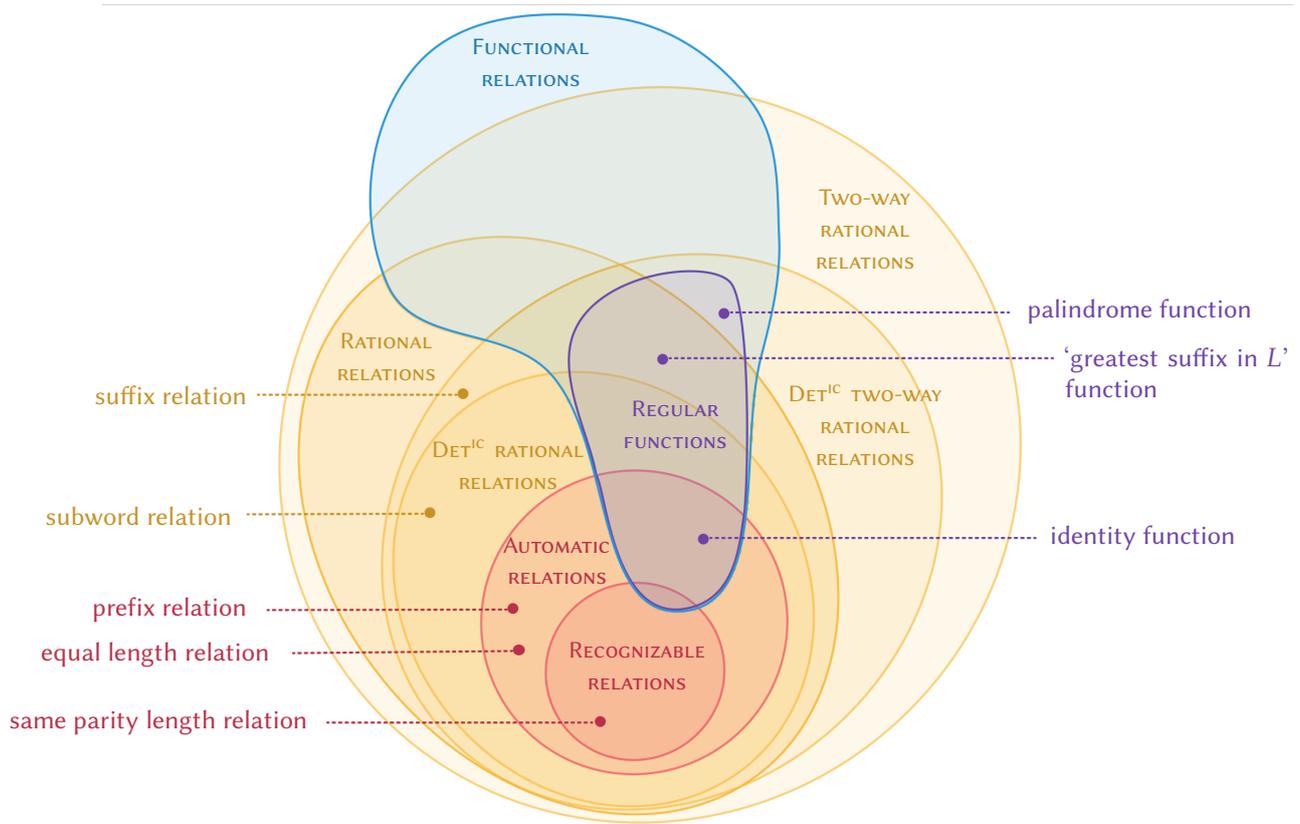

Figure VII.1: Clickable landscape of rationality for binary relations. See also Figures VII.8 and VII.9.

*Goal of this section.* Beyond introducing the classes of recognizable and automatic relations, the goal of this section is also to clarify the literature. When dealing with relations, there are essentially two families of machines for recognizing a relation $\mathcal{R} \subseteq \Gamma^* \times \Sigma^*$:

- *multitape automata:* this model reads a pair of words $\langle u, v \rangle \in \Gamma^* \times \Sigma^*$ and decides if it belongs to $\mathcal{R}$;
- *transducers:* this model reads a word $u \in \Gamma^*$ and produces a word (or no word, or multiple words) $v \in \Sigma^*$, in which case we say that $\langle u, v \rangle \in \mathcal{R}$.

The latter model is inherently passive while the former is more active: the output is produced, or *transduced*, from the input. However, these two families are far from being incomparable: when non-determinism is allowed and the model is expressive enough, a transducer can always guess the output and simulate a multitape automata. Despite this fact, the works on multitape automata and transducers rarely reference each other, and the relationship between classes of relations defined by these two families of models is often unclear.[5] For instance, we will see in Section VII.1.8 that deterministic two-way rational relations and deterministic two-way transductions actually differ: the distinction between *multitape automata* and *transducers* is hence of capital importance.

Thus, the goal of this section is to provide an overview of these classes of relations/transductions, explain how they relate to one another, and give pointers to suitable references. The aim of this section is not to be exhaustive, which would be a daunting task,[6] but rather to present a quick overview of the most common classes of relations studied in the literature.

*Relations vs. transductions.* Often, in the literature, when the terminology "transduction" is used, these relations are thought as functions from $\Gamma^*$ to $\mathfrak{P}(\Sigma^*)$. These functions are equivalent[7] to relations, *i.e.* subsets of $\Gamma^* \times \Sigma^*$. For deterministic transducers, the output is a singleton, and hence its semantics can be seen as a *partial function*—or equivalently as a functional relation—from $\Gamma^*$ to $\Sigma^*$.

When the terminology "relation" is used, we often take $\Gamma = \Sigma$.[8] Note also that most notions defined by multitape automata can be trivially extended to $k$-ary relations ($k \geq 3$): we simply work here with binary relations for the sake of simplicity. On the other hand, transducers intrinsically recognize functions $\Gamma^* \to \mathfrak{P}(\Sigma^*)$, *i.e.* binary relations.

Given two classes of finite-word relations $\mathcal{C}$ and $\mathcal{D}$ *s.t.* $\mathcal{C} \supseteq \mathcal{D}$,[9] we will often want to know if this inclusion is *effective*:

---

$\mathcal{D}$-MEMBERSHIP PROBLEM FOR $\mathcal{C}$-RELATIONS

*Input*: A $\mathcal{C}$-relation $\mathcal{R}$.
*Question*: Does $\mathcal{R} \in \mathcal{D}$?

---

This problem is also called the $\mathcal{C}/\mathcal{D}$-MEMBERSHIP PROBLEM. Since Part 2 will mostly focus on automatic relations, we use the terminology $\mathcal{D}$-MEMBERSHIP

[5] To witness this claim, I want to emphasize that I unwillingly included an erroneous version of Figure VII.1 in [Mor25]: the suffix relation was misplaced and, more importantly, the regular functions were claimed to correspond to the intersection of rational relations with deterministic two-way rational relations.

[6] Sakarovitch's book [Sak09] dedicates its second part to "Rationality in relations", which spans over more than two hundred pages—one can only imagine how extensive this part could have been if the book was not dedicated to presenting *elements* of automata theory but the full theory itself! Note in fact that the book does not even mention two-way models. What we call here multitape automata are called there "transducers" and our transducers are called "sequential transducers" there. We made this choice of terminology because it seems more consistent with the literature, and we find both more natural and more intuitive.

[7] Meaning formally that these categories are isomorphic.

[8] For instance this is somewhat important when dealing with logic or defining automatic structures in Sections VII.2 and VII.3.

[9] Formally, for the decision problem to make sense $\mathcal{C}$ and $\mathcal{D}$ should rather be classes of machines (recognizing relations), but we abuse the terminology for the sake of readability.

PROBLEM to refer to the problem above when $\mathcal{C}$ is the class of automatic relations.

A natural extension of this problem is the so-called separability problem, which has received a lot of attention.

---

$\mathcal{D}$-SEPARABILITY PROBLEM FOR $\mathcal{C}$-RELATIONS

*Input*: Two $\mathcal{C}$-relations $\mathcal{R}$ and $\mathcal{R}'$.

*Question*: Is there a $\mathcal{D}$-relation $\mathcal{S}$ *s.t.* $\mathcal{R} \subseteq \mathcal{S}$ and $\mathcal{R}' \cap \mathcal{S} = \varnothing$?

---

In this case, we say that $\mathcal{S}$ *separates* $\mathcal{R}$ from $\mathcal{R}'$. Note that when $\mathcal{D}$ is closed under complement, $\mathcal{R}$ and $\mathcal{R}'$ play a symmetric role in this problem. Moreover, if $\mathcal{D}$ is *effectively* closed under complement, then MEMBERSHIP reduces to SEPARABILITY since $\mathcal{R} \in \mathcal{D}$ if, and only if, $\mathcal{R}$ and its complement are separable by a $\mathcal{D}$-relation. Since this thesis is focused on automatic relations, when $\mathcal{C}$ is the class of automatic relations we simply call this problem the "$\mathcal{D}$-SEPARABILITY PROBLEM".

For a given class $\mathcal{C}$, the *inclusion problem* takes two (machines recognizing two) relations of $\mathcal{C}$ and asks whether one is included in the other, and the *equivalence problem* asks if these relations are equal.[10]

### VII.1.2 Recognizable Relations

A relation $\mathcal{R} \subseteq \Gamma^* \times \Sigma^*$ is *recognizable* if there exists a finite monoid **M** together with a monoid morphism

$$f \colon \Gamma^* \times \Sigma^* \to \mathbf{M},$$

as well as a subset $\mathrm{Acc} \subseteq M$ *s.t.* $\mathcal{R} = f^{-1}[\mathrm{Acc}]$. We denote by REC the class of recognizable relations.[11]

For instance the *same parity relation*

$$\approx_{\mathrm{mod}\,2} \triangleq \{\langle u, v \rangle \in \Gamma^* \times \Sigma^* \mid |u| = |v| \mod 2\}$$

is recognizable. Indeed, letting $f \colon \Gamma^* \times \Sigma^* \to \mathbb{Z}/2\mathbb{Z}$ be defined by $f(u, v) \triangleq |u| - |v| \mod 2$, then $\approx_{\mathrm{mod}\,2}$ can be written as $f^{-1}[0]$. These relations admit a remarkably simple characterization.

**Proposition VII.1.1** (*Mezei theorem.*).[12] A relation $\mathcal{R}$ is recognizable *iff* there exist $n \in \mathbb{N}$, regular languages $\langle K_i \rangle_{i \in [\![ 1, n ]\!]}$ over $\Gamma$ and regular languages $\langle L_i \rangle_{i \in [\![ 1, n ]\!]}$ over $\Sigma$ *s.t.*

$$\mathcal{R} = \bigcup_{i=1}^{n} K_i \times L_i.$$

In other words, recognizable relations are exactly the finite unions of Cartesian products of regular languages. For instance,

$$\approx_{\mathrm{mod}\,2} = (\Gamma\Gamma)^* \times (\Sigma\Sigma)^* \cup \Gamma(\Gamma\Gamma)^* \times \Sigma(\Sigma\Sigma)^*.$$

We provide a slightly more general statement of Mezei theorem.

[10] Note that functional relations represent *partial functions*, and hence one can be included in the other without the two relations being equal.

[11] To avoid preposterous set-theoretic paradoxes, we use the classical shenanigan of defining this as a function which maps any pair of alphabets $\Gamma$ and $\Sigma$ to the set of recognizable relations over $\Gamma$ and $\Sigma$.

[12] See *e.g.* [Sak09, Corollary II.2.20, p. 254]. [Sak09, § 2, "Notes & references"] mentions that this proposition is "unanimously ascribed to G. Mezei (unpublished)".

**Proposition VII.1.2.** Let $\mathbb{V}$ be a pseudovariety of monoids and $\mathcal{V}$ be the corresponding pseudovariety of regular languages. Let $\mathcal{R} \subseteq \Gamma^* \times \Sigma^*$ be a relation. The following are equivalent:

1. there exists a finite monoid $\mathbf{M} \in \mathbb{V}$, a monoid morphism $f \colon \Gamma^* \times \Sigma^* \to \mathbf{M}$ and $\mathrm{Acc} \subseteq \mathbf{M}$ *s.t.* $\mathcal{R} = f^{-1}[\mathrm{Acc}]$;

2. there exists $n \in \mathbb{N}$ and $K_1, \dots, K_n \in \mathcal{V}_\Gamma$ and $L_1, \dots, L_n \in \mathcal{V}_\Sigma$ *s.t.* $\mathcal{R} = \bigcup_{i=1}^n K_i \times L_i$,[13]

in which case we say that $\mathcal{R}$ is *$\mathcal{V}$-recognizable*.

When $\mathbb{V}$ is the pseudovariety of all regular languages, we get back Proposition VII.1.1.

*Proof.* ☙ *From monoids to products.* Assume that $\mathcal{R}$ is recognizable. Then by definition

$$\mathcal{R} = \bigcup_{z \in \mathrm{Acc}} f^{-1}[z].$$

Observe then that $f(u, v) = f(\langle u, \varepsilon \rangle \cdot \langle \varepsilon, v \rangle) = f(u, \varepsilon) \cdot f(\varepsilon, v)$ for all $u, v \in \Gamma^* \times \Sigma^*$, and hence:

$$\mathcal{R} = \bigcup_{\substack{x, y \in M \\ \text{s.t. } x \cdot y \in \mathrm{Acc}}} \underbrace{\{u \in \Gamma^* \mid f(u, \varepsilon) = x\}}_{\triangleq K_x} \times \underbrace{\{v \in \Sigma^* \mid f(\varepsilon, v) = y\}}_{\triangleq L_y}.$$

Since $M$ is finite, the union is finite, and moreover, each $K_x$ and $L_y$ is recognized by $M \in \mathbb{V}$, and hence belong to $\mathcal{V}$.

☙ *From products to monoids.* If $\mathcal{R} = \bigcup_{i=1}^n K_i \times L_i$ where all languages belong to $\mathcal{V}$, then let $M_i, N_i \in \mathbb{V}$ be their syntactic monoids, $g_i, h_i$ be their syntactic morphism, and $\mathrm{Acc}_i, \mathrm{Bcc}_i$ be their accepting sets. Consider the monoid morphism

$$
\begin{aligned}
\Gamma^* \times \Sigma^* &\to \textstyle\prod_i (M_i \times N_i) \\
\langle u, v \rangle &\mapsto \langle g_i(u), h_i(v) \rangle_i.
\end{aligned}
$$

Then $\mathcal{R}$ is the preimage by this morphism of

$$\bigcup_{i=1}^n \Big( \cdots \times (M_{i-1} \times N_{i-1}) \times (\mathrm{Acc}_i \times \mathrm{Bcc}_i) \times (M_{i+1} \times N_{i+1}) \times \cdots \Big).$$

The conclusion follows from the fact that $\mathbb{V}$ is closed under finite products. □

Both the algebraic definition of recognizable relations and Mezei theorem imply, informally, that all reasonable problems on recognizable relations are decidable. For instance, from Proposition VII.1.2, we get that $\mathcal{V}$ has decidable MEMBERSHIP *iff* the class of $\mathcal{V}$-recognizable relations has decidable MEMBERSHIP.

On the other hand, Proposition VII.1.1 proves that recognizable relations are not very expressive.

**Corollary VII.1.3.**[14] Let $\mathcal{R} \subseteq \Sigma^* \times \Sigma^*$ be a reflexive recognizable relation.

[13] $\mathcal{V}_\Gamma$ refers to all languages of $\mathcal{V}$ over the alphabet $\Gamma$.

[14] This property is far from being sufficient at characterizing reflexive relations: note that the proof does not even use the regularity of the languages at hand...

Then $\mathcal{R}$ contains an infinite clique, *i.e.* there exists an infinite language $L \subseteq \Sigma^*$ *s.t.* $\langle u, v \rangle \in \mathcal{R}$ for all $u, v \in L$.

*Proof.* Indeed, by Proposition VII.1.1, write $\mathcal{R}$ as $\bigcup_{i=1}^{n} K_i \times L_i$. Given a word $u \in \Sigma^*$, define $f(u) \in \mathbf{2}^{2n}$ where the $2i$-th (resp. $(2i+1)$-th) bit of $f(u)$ indicates if $u \in K_i$ (resp. $u \in L_i$) for all $i$. By the pigeon-hole principle, there exists a bit-sequence in $\mathbf{2}^{2n}$ whose preimage $L$ by $f$ is infinite. Then pick $u, v \in L$. Since $\mathcal{R}$ is reflexive, then $\langle u, u \rangle \in \mathcal{R}$ and so, since $f(u) = f(v)$, we have $u \in K_i$ *iff* $v \in K_i$ and $u \in L_i$ *iff* $v \in L_i$ for all $i$, and so $\langle u, v \rangle \in \mathcal{R}$.[15] □

In particular, this corollary implies that neither the *prefix relation* $\leqslant_{\text{pref}} \triangleq \{\langle u, v \rangle \in \Sigma^* \times \Sigma^* \mid u \text{ is a prefix of } v\}$, the *suffix relation* $\leqslant_{\text{suff}} \triangleq \{\langle u, v \rangle \in \Sigma^* \times \Sigma^* \mid u \text{ is a suffix of } v\}$, the equality relation, nor the *equal-length relation* $\approx_{\text{len}} \triangleq \{\langle u, v \rangle \in \Gamma^* \times \Sigma^* \mid |u| = |v|\}$ are recognizable.[16]

### VII.1.3 Automatic Relations

Automatic relations are a strictly larger class of relations, and trade some decidability properties to gain in expressiveness. This is precisely what makes this class interesting to us, and why we will focus on both automatic relations and automatic structures: while being relatively expressive, some problems remain decidable.[17]

Given a word $u \in \Gamma^*$ and $v \in \Sigma^*$, we define its *convolution* $u \otimes v$ to be the word $i \mapsto \langle u_i, v_i \rangle$ of length $\max(|u|, |v|)$, with the convention that $u_i = {}_-$ (resp. $v_i = {}_-$) if $u_i$ (resp. $v_i$) is undefined, where ${}_-$ is a new letter called *blank symbol* or *padding symbol*. In other words, $u \otimes v$ is obtained by writing $u$ and $v$ on two left-aligned horizontal tapes, adding padding symbols at the end of the shorter word if their length differ, and then reading pairs of letters from left to right. For this reason, the pair $\langle a, b \rangle$ is instead written $\binom{a}{b}$. We let $\Gamma \otimes \Sigma$ denote the alphabet

$$(\Gamma \times \Sigma) \cup (\Gamma \times \{{}_-\}) \cup (\{{}_-\} \times \Sigma),$$

and we let $\Sigma^2_\otimes \triangleq \Sigma \otimes \Sigma$.

By construction, if $u \in \Gamma^*$ and $v \in \Sigma^*$, then $u \otimes v \in (\Gamma \otimes \Sigma)^*$.[18] For instance

$$aba \otimes baa = \binom{a}{b}\binom{b}{a}\binom{a}{a} \quad \text{and} \quad abab \otimes cdd = \binom{a}{c}\binom{b}{d}\binom{a}{d}\binom{b}{{}_-}.$$

We denote by $\mathcal{R}^\otimes$ the language $\{u \otimes v \mid \langle u, v \rangle \in \mathcal{R}\} \subseteq (\Gamma \otimes \Sigma)^*$.

A (finite-state) *synchronous automaton* $\mathcal{A}$ over input alphabet $\Gamma$ and output alphabet $\Sigma$ is a finite-state automaton over $\Gamma \otimes \Sigma$.[19] We say that $\mathcal{A}$ accepts the pair of words $\langle u, v \rangle \in \Gamma^* \times \Sigma^*$ if it accepts $u \otimes v$ as a "classical automaton". Similarly, $\mathcal{A}$ *recognizes* $\mathcal{R}$ if $\mathcal{R}^\otimes$ is exactly the set of words of the form $u \otimes v$ that are accepted by $\mathcal{A}$ as a classical automaton. Figure VII.2 depicts a synchronous automaton for the prefix relation. Note that $\binom{{}_-}{a}\binom{a}{a}$ corresponds to a run of the automaton from an initial state to an accepting one. However, since $\binom{a}{a}\binom{a}{a}$ cannot be written as $u \otimes v$, then this run plays no role whatsoever in the semantics of this synchronous automaton.

---

[15] Another way of proving this result would be to apply Ramsey's infinite theorem to $f \colon \Sigma^* \times \Sigma^* \to \mathbf{M}$. Again, we do not use the fact that $f$ is a monoid morphism, but simply that it is a finite-domain function.

[16] Note however that Corollary VII.1.3 does not apply since we assumed there that the input and output alphabets are equal.

[17] They are known in the literature under many names: "automatic relations" *e.g.* in [CCG06, Definition 2.3], "regular relations" *e.g.* in [KN95, Definition 2.2], "automatic relations" *e.g.* in [LS19, § 2.1]. Frougny and Sakarovitch's work, which is often referred as the first one that extensively studied this class, refer to them as "synchronized rational relations" [FS93, § 4]. Of course, this class already appears in prior work, *e.g.* in Hodgson's work on automatic structures [Hod83], but no terminology was coined on these relations there.

[18] When dealing with relations of higher arity, note that $\otimes$ is associative up to a trivial alphabet relabelling.

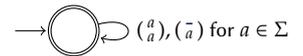

Figure VII.2: A deterministic one-state synchronous automaton recognizing the prefix relation.

[19] In particular, just like for classical automata, we allow for non-determinism unless otherwise specified.

We say that a relation is *automatic* if it is recognized by a finite-state synchronous automaton, and we denote by AUT the class of automatic relations.

**Remark VII.1.4.** Note that some words of $(\Gamma \otimes \Sigma)^*$ do *not* correspond to encodings of pairs of words, in the sense that they are not of the form $u \otimes v$ for some $u \in \Gamma^*$ and $v \in \Sigma^*$. This is for instance the case of $\binom{a}{\bot}\binom{\bot}{b}$. In fact, $(\Gamma^* \times \Sigma^*)^\otimes = \{u \otimes v \mid \in u \in \Gamma^* \wedge v \in \Sigma^*\}$ precisely corresponds to the words of $(\Gamma \otimes \Sigma)^*$ *s.t.* if some padding symbol is seen on some tape, then all subsequent symbols on this tape must also be padding symbols. These words are called *well-formed*, and the set of all well-formed words over $\Gamma$ and $\Sigma$ is denoted by WellFormed$_{\Gamma,\Sigma}$.[20]                                    $\diamond$

Because of this last remark, some automata that are not *classically* equivalent become equivalent when seen as synchronous automaton. For instance, both synchronous automata of Figures VII.2 and VII.3 recognize the prefix relation. However, they do not recognize the same language when seen as classical automaton. Generalizing this example, it is trivial to check that two synchronous automata have the same semantics if, and only if, they have the same intersection of their semantics, when seen as classical automata, with an automaton for WellFormed$_{\Gamma,\Sigma}$.

Note that, by definition of synchronous automata, everything that can be done on classical automata can be done with synchronous automata, including determinisation, removal of $\varepsilon$-transitions, completion, etc. Moreover, observe by the previous paragraph that the universality of a synchronous automaton $\mathcal{A}$ amounts to the universality of the disjoint union of $\mathcal{A}$ with an automaton for the complement of WellFormed$_{\Gamma,\Sigma}$. A similar construction works for the inclusion problem.

**Property VII.1.5.** Inclusion and universality of synchronous automata is PSpace-complete.

Similarly, the emptiness of a synchronous automaton $\mathcal{A}$ amounts to the emptiness of the classical automaton obtained by intersecting $\mathcal{A}$ with the set of well-formed words WellFormed$_{\Gamma,\Sigma}$.

**Property VII.1.6.** Emptiness and non-emptiness of synchronous automata are NL-complete.

It is straightforward to prove that the equal-length relation and the equality relation, *a.k.a.* the *identity relation*, denoted by $\mathcal{Id}$, are automatic, but, as already mentioned, not recognizable.

**Proposition VII.1.7.** Every recognizable relation is automatic.

*Proof.* Clearly, automatic relations are closed under union: it suffices to do the disjoint union of their automata. So, to prove this result, it suffices to show that if $K$ and $L$ are regular languages, then $K \times L$ is automatic. Let $\mathcal{A}$ (resp. $\mathcal{B}$) be an automaton recognizing $K$ (resp. $L$). We build a synchronous automaton $\mathcal{A} \otimes \mathcal{B}$ as follows:

- its states are the pairs of states of $\mathcal{A}$ and of states of $\mathcal{B}$,
- $\langle p, q \rangle$ is initial if both $p$ and $q$ are initial,
- $\langle p, q \rangle$ is accepting if both $p$ and $q$ are accepting, and
- transitions follow these rules:

$$\frac{p \xrightarrow{a} p' \in \mathcal{A} \qquad q \xrightarrow{b} q' \in \mathcal{B}}{\langle p, q \rangle \xrightarrow{\binom{a}{b}} \langle p', q' \rangle \in \mathcal{A} \otimes \mathcal{B}},$$

$$\frac{p \xrightarrow{a} p' \in \mathcal{A} \qquad q \in \mathcal{B}}{\langle p, q \rangle \xrightarrow{\binom{a}{\diamond}} \langle p', q \rangle \in \mathcal{A} \otimes \mathcal{B}} \quad \text{and} \quad \frac{p \in \mathcal{A} \qquad q \xrightarrow{b} q' \in \mathcal{B}}{\langle p, q \rangle \xrightarrow{\binom{\diamond}{b}} \langle p, q' \rangle \in \mathcal{A} \otimes \mathcal{B}}.$$

[20] When the alphabets are equal, we will write WellFormed$_\Sigma$ instead of WellFormed$_{\Sigma,\Sigma}$.

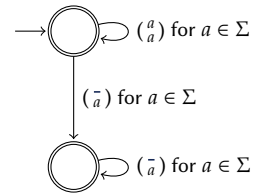

Figure VII.3: A deterministic two-state synchronous automaton recognizing the prefix relation.

By construction, for any $u \in \Gamma^*$ and $v \in \Sigma^*$, there is an accepting path in $\mathcal{A} \otimes \mathcal{B}$ labelled by $u \otimes v$ iff $u \in K$ and $v \in L$. $\qquad \square$

As often in automata theory, the pumping lemma provides a useful tool to prove non-regularity, or rather here non-automaticity. For instance, the suffix relation is not automatic: otherwise, using the pumping lemma on $a^n \otimes b^n a^n$ for some sufficiently big $n \in \mathbb{N}$ would imply that $a^{n+k} \otimes b^{n+k} a^n \in (\leqslant_{\text{suff}})^{\otimes}$ for some $k \in \mathbb{N}_{>0}$. Similarly, the *subword relation* $\leqslant_{\text{subw}}$ is also not automatic.[21]

Another useful tool to prove non-automaticity is the following one. Given $I \subseteq [\![1, k]\!]$, we say that a $k$-ary relation $\mathcal{R} \subseteq \Sigma_1^* \times \cdots \times \Sigma_k^*$ is *I-locally finite* if for every $\langle w_i \rangle_{i \in I} \in \prod_{i \in I} \Sigma_i^*$, there are only finitely many $\langle w_j \rangle_{j \notin I} \in \prod_{j \notin I} \Sigma_j^*$ *s.t.* $\langle w_1, \ldots, w_k \rangle \in \mathcal{R}$.[22] Note that every $k$-ary relation is trivially $[\![1, k]\!]$-locally finite.

**Proposition VII.1.8** (See *e.g.* [KNRS07, Proposition 3.1]).[23],[24] Let $\mathcal{R} \subseteq \Sigma_1^* \times \cdots \times \Sigma_k^*$ be *I-locally finite*. If $\mathcal{R}$ is automatic, then there exists a constant $n \in \mathbb{N}$ *s.t.* for every $\langle w_1, \ldots, w_k \rangle \in \mathcal{R}$,

$$\max_{j \notin I} |w_j| - \max_{i \in I} |w_i| \leq n.$$

For instance, consider the ternary relation of concatenation, consisting of all triples $\langle u, v, w \rangle$ *s.t.* $uv = w$, where $u, v, w \in \Sigma^*$. Then this relation is $\{3\}$-locally finite, and so by Proposition VII.1.8, there exists $n \in \mathbb{N}$ *s.t.* for all $u, v \in \Sigma^*$, $|uv| \leq \max(|u|, |v|) + n$. This is trivially false, and hence, concatenation is not automatic.

Despite being somewhat expressive, automatic relations retain some decidability properties.

**Proposition VII.1.9** ([BHLLN19, Theorem 1], see also [BGLZ22, Corollary 2.9]).[25] The Rec-membership problem for automatic relations is PSpace-complete.

On the other hand, the associated SEPARATION PROBLEM remains open. This part of this thesis is mostly motivated by this problem.

**Open Problem VII.1.10.** Is the Rec-separability problem for automatic

---

[21] Recall that this relation is defined by, for all $u, v \in \Sigma^*$, $u \leqslant_{\text{subw}} v$ iff there exists $j_1 < j_2 < \ldots < j_{|u|}$ *s.t.* $u_i = v_{j_i}$ for all $i \in [\![1, |u|]\!]$: in other words, $u$ can be obtained from $v$ by removing letters.

[22] Note that this is one of the very few statements that we spell out for $k$-ary relations since its generalization from binary to $k$-ary relations is not entirely trivial.

[23] [KNRS07] attributes this proposition to older work by Khoussainov, Nerode and Blumensath, but we failed to verify this claim.

[24] By convention $\max \varnothing = -\infty$, and so in the case of $I = [\![1, k]\!]$, this statement is trivially valid.

[25] Decidability, first in 3ExpTime and then in 2ExpTime was already known in the 1970s as a consequence of Proposition VII.1.13. An ExpTime upper bound was incorrectly claimed in [CCG06, Table 1], and proved wrong in [LS19, § 4.1]. Löding and Spinrath noticed that the previous algorithm was actually working in 2ExpTime, and improved the bound to ExpTime for binary relations in [LS19, Corollary 22].

RELATIONS decidable?

This problem goes back at least to 2006, when Choffrut and Grigorieff solved a special subcase of commutative relations. While the general problem is not explicitly mentioned in their paper, the question is rather obvious given their contribution the and formulation of the result: they show that given two rational relations of $\Sigma^* \times \mathbb{N}^m$, whether they are separable by a recognizable relation is decidable [CG06, Theorem 1]. Note that relations $\mathcal{R} \subseteq \Sigma^* \times \mathbb{N}^m$ are in natural bijection with the relations $\mathcal{R}' \subseteq \Sigma^* \times \Gamma^*$ where $\Gamma$ is an alphabet with $m$ letters, satisfying the property that for each $u \in \Sigma^*$ and $v \in \Gamma^*$, for any permutation $v'$ of the letters of $v$, we have $\langle u, v \rangle \in \mathcal{R}'$ iff $\langle u, v' \rangle \in \mathcal{R}'$.

*Closure under morphisms.* We say that a monoid morphism $f \colon \Gamma^* \to \Sigma^*$ is *length-multiplying* when $|f(a)| = |f(b)|$ for all $a, b \in \Gamma$.

**Proposition VII.1.11.** Automatic relations are closed under direct images and preimages of length-multiplying morphisms.[26]

*Proof.* Any length-multiplying morphism $f \colon \Gamma^* \to \Sigma^*$ induces a monoid morphism $\widehat{f} \colon (\Gamma_{\otimes}^2)^* \to (\Sigma_{\otimes}^2)^*$ by letting $\widehat{f}\binom{a}{b} \triangleq \binom{f(a)}{f(b)}$. Then for any relation, $(f[\mathcal{R}])^\otimes = \widehat{f}[\mathcal{R}^\otimes]$ and $(f^{-1}[\mathcal{R}])^\otimes = \widehat{f}^{-1}[\mathcal{R}^\otimes]$. The conclusion follows from the fact that regular languages are closed under direct images and preimages of monoid morphisms. □

This is false for arbitrary monoid morphisms. For instance, $\{\langle 0^n, 1^n \rangle \mid n \in \mathbb{N}\}$ is automatic, but its image by the monoid morphism $f \colon 2^* \to 2^*$ defined by $f(0) \triangleq 0$ and $f(1) \triangleq 11$ is $\{\langle 0^n, 1^{2n} \rangle \mid n \in \mathbb{N}\}$ which is clearly not automatic. Similarly, its preimage by the same morphism is $\{\langle 0^{2n}, 1^n \rangle \mid n \in \mathbb{N}\}$ which is not automatic either, by symmetry.

We will see more properties of automatic relations—or rather of automatic structures—in Sections VII.2 and VII.3.

### VII.1.4 Rational Relations

Rational relations are perhaps the most natural class of finite-word relations: they are defined similarly to automatic relations, except that the automaton used to defined them has one head per tape, and can move each head independently of one another—however each head can only move from left to right. For the sake of convenience, we assume that the automaton can move multiple heads at every step. This can be formalized as follows.

A *multitape automaton* over $\Gamma, \Sigma$ can be syntactically defined as a classical automaton over $\Gamma \otimes \Sigma$.[27] We then let $\pi_1 \colon (\Gamma \otimes \Sigma)^* \to \Gamma^*$ and $\pi_2 \colon (\Gamma \otimes \Sigma) \to \Sigma^*$ be the monoid morphisms defined by

$$\pi_1\binom{x}{y} \triangleq \begin{cases} a & \text{if } x = a \in \Gamma, \\ \varepsilon & \text{if } x = \_, \end{cases} \quad \text{and} \quad \pi_2\binom{x}{y} \triangleq \begin{cases} b & \text{if } y = b \in \Sigma, \\ \varepsilon & \text{if } y = \_, \end{cases}$$

[26] For the notion of "direct image" and "preimage" to make sense, we need the input and output alphabets to be equal.

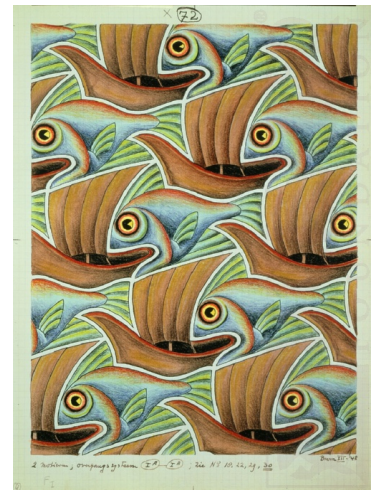

Figure VII.4: 2 motifs, transitional system I(a) - I(a), M. C. Escher, © The M.C. Escher Company.

[27] They share the syntax as synchronous automata, however be reassured: their semantics will differ!

for all $\left(\begin{smallmatrix} x \\ y \end{smallmatrix}\right) \in \Gamma \otimes \Sigma$. Then $\pi_{\text{word}} \triangleq \pi_1 \times \pi_2$ defines a monoid morphism from $(\Gamma \otimes \Sigma)^*$ to $\Gamma^* \times \Sigma^*$, that reads both tapes one after the other, while ignoring padding symbols. For instance:

$$\pi_{\text{word}}\left(\begin{smallmatrix} a\_b\_a\_c\_ \\ aabbaacc \end{smallmatrix}\right) = \langle abac, aabbaacc \rangle.$$

The semantics of the multitape automaton can be defined as the image by $\pi_{\text{word}}$ of its semantics when seen as a classical automaton.[28]

We say that a relation $\mathcal{R} \subseteq \Gamma^* \times \Sigma^*$ is *rational* if it is *recognized* by a multitape automaton, or equivalently if it is the image by $\pi_{\text{word}}$ of a regular language over $\Gamma \otimes \Sigma$. The class of all rational relations is denoted by RAT.

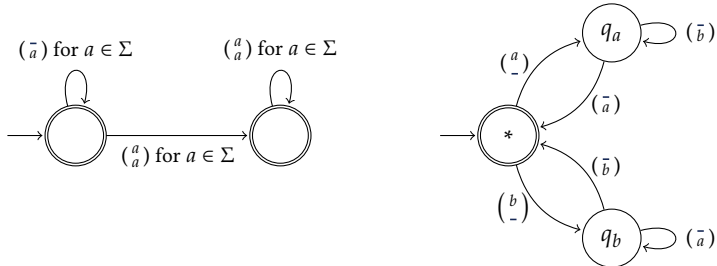

For instance, the suffix relation is rational: see Figure VII.5. Similarly, the subword relation $\leqslant_{\text{subw}}$ is rational: the idea behind the automaton recognizing it is that, when reading an $a$ on the first tape, it will only move its second head right, until it reads an $a$; it then reads the next letter on the first tape and iterates this process, see Figure VII.5.

Alternative definitions exist, but yield an equivalent definition of rational relations. For instance, in [CCG06, Definition 2.1], the multitape automaton are defined analogously (under the name "$k$-tape automaton") except that the alphabet is not $\Gamma \otimes \Sigma$ but $(\Gamma \times \{\_\}) \cup (\{\_\} \times \Sigma)$. In other words, their model can only move one head at a time. Of course, this makes little difference: we can simulate an $\left(\begin{smallmatrix} a \\ b \end{smallmatrix}\right)$-transition in our model by two transitions labelled by $\left(\begin{smallmatrix} a \\ \_ \end{smallmatrix}\right)$ and $\left(\begin{smallmatrix} \_ \\ b \end{smallmatrix}\right)$ in their model, at the cost of adding a new state. Moreover, rational relations can be equally characterized as the rational subsets of $\Gamma^* \times \Sigma^*$.[29]

*Undecidability.* While emptiness and finiteness of rational relations are decidable—see *e.g.* [Ber79, § III, Proposition 8.2]—unfortunately rational relations are maybe too expressive: most other decision problems on them are undecidable, including intersection non-emptiness, inclusion, equivalence, universality, co-finiteness and the RAT/REC-MEMBERSHIP PROBLEM, see [Ber79, § III, Theorem 8.4].[30]

*Closure properties.* More or less by construction, the class is closed under union. However, it is not closed under intersection (see *e.g.* [Ber79, § III, Example 2.5]), and hence, it is also not closed complementation.[31] We will see later that this has some interesting consequences, such as the undecidability of the intersection non-emptiness problem (Example VII.1.14). The class is also closed under concatenation, Kleene closure and reversal [FR68, § 3, Table I].

[28] These automata are often simply called "$k$-tape automata". They are sometimes referred to as "asynchronous automata", *e.g.* in [CL11, § 3] or [Pel97, § 3.1.2.2, p. 88]. We do *not* use this terminology following a suggestion of Anca Muscholl to avoid the confusion with Zielonka's "asynchronous automata" used in concurrency [Zie87, § 4].

Figure VII.5: Multitape automata for the suffix relation $\leqslant_{\text{suff}}$ (left-hand side) and for the subword relation $\leqslant_{\text{subw}}$ when $\Sigma = \{a, b\}$ (right-hand side).

[29] Meaning the smallest collection of subsets of $\Gamma^* \times \Sigma^*$ containing the empty set, all singletons, closed under union, concatenation and Kleene star, see *e.g.* [Ber79, § III.2, Definition].

[30] All these results essentially follow from an encoding of Post Correspondence Problem into rational relations.

[31] Indeed, closure under union and complementation implies closure under intersection, using De Morgan's law.

### VII.1.5 Deterministic Rational Relations

Because of the undecidability results above, a subclass of rational relations was introduced: deterministic rational relations. We will see that, while still strictly extending the class of automatic relations, its equivalence problem is decidable.

Note that the automaton of Figure VII.5 is deterministic as a classical automaton. However, we claim that it should not be considered to be "deterministic" as a multitape automaton. For instance, assume that we are running this automaton over the pair $\langle aab, ababaab \rangle$, and that we started reading the sequence $\binom{a}{a}, \binom{b}{b}$, while staying in the initial state. Should we take the $\binom{a}{a}$ and stay in the initial state, or take the transition $\binom{a}{a}$ and go to final state? In the first case, we can obtain an accepting run, but in the second case we cannot. In other word, assuming that $u \preccurlyeq_{\text{subw}} v$, we need to guess when we've reached the suffix of $v$ corresponding to $u$, and this is a source of non-determinism. Observe that the automaton for $\preccurlyeq_{\text{suff}}$ of Figure VII.5 is deterministic as a classical automaton because $\binom{a}{a}$ and $\binom{a}{a}$ are distinct letters of $\Gamma \otimes \Sigma$, although in the multitape setting, both pairs can be simultaneously admissible. In other words, the lack of injectivity of the function $\pi_{\text{word}}$ creates another source of non-determinism. This motivates an alternative definition for determinism.

Hence, we say that a $k$-ary multitape automaton over $\Sigma_1, \dots, \Sigma_k$ is *deterministic* when:

1. it has a unique initial state,
2. for any state $q$, for any $\bar{a} \in \Sigma_1 \otimes \cdots \otimes \Sigma_k$, there is at most one outgoing transition from $q$ labelled by $\bar{a}$, and
3. there exists a partition $\langle Q_H \rangle_{H \in \mathfrak{P}_+(\llbracket 1,k \rrbracket)}$ of its states *s.t.* for every $H \in \mathfrak{P}_+(\llbracket 1, k \rrbracket)$, for every $q \in Q_H$, every outgoing transition from $q$ can only be labelled by elements of the form $\bar{a} \in \Sigma_1 \otimes \cdots \otimes \Sigma_k$, with $a_h \in \Sigma_h$ for every $h \in H$ and $a_h =$ _ for every $h \notin H$.

In other words, we ask for classical determinism to hold, but moreover we need to know *a priori*, at each state, which set of heads we will be moving at the next step. For instance, in Figure VII.5, the automaton for $\preccurlyeq_{\text{subw}}$ is deterministic, but the automaton for $\preccurlyeq_{\text{suff}}$ is not: its initial state has outgoing transitions from both $\Sigma \times \Sigma$ and from $\{\_\} \times \Sigma$.

*Claim* VII.1.12. Let $\mathcal{A}$ be a $k$-ary deterministic multitape automaton over $\Sigma_1, \dots, \Sigma_k$. For any $\langle w_1, \dots, w_k \rangle \in \Sigma_1^* \times \cdots \times \Sigma_k^*$, for any state $q$ of $\mathcal{A}$, there exists at most one run of $\mathcal{A}$ starting at $q$ labelled by a word $w \in (\Sigma_1 \otimes \cdots \otimes \Sigma_k)^*$ *s.t.* $\pi_{\text{word}}(w) = \langle w_1, \dots, w_k \rangle$.[32]

We say that a relation $\mathcal{R} \subseteq \Gamma^* \times \Sigma^*$ is *deterministic rational* if the relation

$$\{\langle u \triangleleft, v \triangleleft \rangle \mid \langle u, v \rangle \in \mathcal{R}\}$$

is recognized by a deterministic multitape automaton, where $\triangleleft$ is a new

[32] On the other hand, observe that classical determinism only ensures at most one run for each $w \in (\Sigma_1 \otimes \cdots \otimes \Sigma_k)^*$, and not for each $\langle w_1, \dots, w_k \rangle \in \Sigma_1^* \times \cdots \Sigma_k^*$.

symbol. We denote by DRAT the class of all deterministic rational relations.

For instance, starting from Figure VII.5, it is easy to build a deterministic multitape automaton for $\{\langle u\triangleleft, v\triangleleft\rangle \mid u \leqslant_{\mathrm{subw}} v\}$, proving that the subword relation is deterministic rational. The motivation behind this somewhat strange definition is that, this way, deterministic rational relations generalize automatic relations.

**Proposition VII.1.13.** Every automatic relation is deterministic rational.

*Proof sketch.* Start from a synchronous automaton $\mathcal{A}$ recognizing $\mathcal{R}$: w.l.o.g. it can be assumed to be deterministic, when seen as a classical automaton. However, when seen as a multitape automaton, some non-determinism remains: for instance some state can have outgoing transitions labelled both by $\binom{a}{a}$ and $\binom{-}{a}$. To fix this, we build a multitape automaton $\mathcal{A}'$ by triplicating the states: each original state $q$ yields a state $q_H$ for $H \in \mathfrak{P}_+(\llbracket 1, 2\rrbracket)$. Transitions are built using the following rules:

$$\frac{p \xrightarrow{\binom{a}{a}} q \in \mathcal{A}}{p_{\{1,2\}} \xrightarrow{\binom{a}{a}} q_{\{1,2\}} \in \mathcal{A}'},$$

$$\frac{p \xrightarrow{\binom{a}{-}} q \in \mathcal{A}}{p_{\{1,2\}} \xrightarrow{\binom{a}{-}} q_{\{1\}} \in \mathcal{A}'}, \qquad \frac{p \xrightarrow{\binom{a}{-}} q \in \mathcal{A}}{p_{\{1\}} \xrightarrow{\binom{a}{-}} q_{\{1\}} \in \mathcal{A}'},$$

$$\frac{p \xrightarrow{\binom{-}{a}} q \in \mathcal{A}}{p_{\{1,2\}} \xrightarrow{\binom{-}{a}} q_{\{2\}} \in \mathcal{A}'}, \quad \text{and} \quad \frac{p \xrightarrow{\binom{-}{a}} q \in \mathcal{A}}{p_{\{2\}} \xrightarrow{\binom{-}{a}} q_{\{2\}} \in \mathcal{A}'}.$$

We also add to $\mathcal{A}'$ a single new state $q_\varnothing$, and for every accepting state $p$ of $\mathcal{A}$, we add a transition from $p_H$ ($H \in \mathfrak{P}_+(\llbracket 1, k\rrbracket)$) to $q_\varnothing$ labelled by the unique tuple having $\triangleleft$ at every index $h \in H$, and $\_$ at every other index. $q_\varnothing$ is the unique accepting state of $\mathcal{A}'$, and moreover $p_{\llbracket 1,k\rrbracket}$ is the unique initial state of $\mathcal{A}'$, where $p$ is the initial state of $\mathcal{A}$. See Figure VII.6 for an example.

By construction, $\mathcal{A}'$ is a deterministic multitape automaton. Using the fact that $\mathcal{A}$ is synchronous, it can be shown that if $\mathcal{R}$ is the relation recognized by $\mathcal{A}$, then $\mathcal{A}'$ precisely recognizes $\{\langle u\triangleleft, v\triangleleft\rangle \mid \langle u, v\rangle \in \mathcal{R}\}$. And hence, $\mathcal{R}$ is deterministic rational. $\square$

Observe how the addition of the end-of-tape symbols is crucial for the multitape automaton of the previous proof to be deterministic. For instance, while the prefix relation is deterministic rational, it can be shown that $\leqslant_{\mathrm{pref}}$ cannot be recognized by a deterministic multitape automaton.

Clearly, the automaton for $\leqslant_{\mathrm{suff}}$ of Figure VII.5 is not deterministic. But in fact, we can prove a stronger result: $\leqslant_{\mathrm{suff}}$ is not even deterministic rational.[33] However, the next example shows that deterministic rational relations are still quite expressive, leading to some undecidability results.

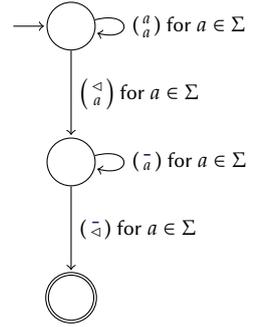

Figure VII.6: Construction of the proof of Proposition VII.1.13 applied to the deterministic multitape automaton built from the synchronous automaton of Figure VII.2 recognizing the prefix relation.

[33] This can be shown by using *e.g.* [Sak09, Lemma IV.5.15] on the pair of words $\langle a^n b\triangleleft, a^{n-1}ba^n b\triangleleft\rangle$ ($n \in \mathbb{N}$). We remark that there is a typo in the statement of [Sak09, Lemma IV.5.15] and that "$|(f, g)| \geq 0$" should be replaced by "$|(f, g)| > 0$".

**Example VII.1.14** (Post Correspondence Problem (PCP) [RS59, Theorem 18]). Deterministic multitape automata are actually quite expressive: given an instance $\langle u_1, \dots, u_k, v_1, \dots, v_k \rangle \in (\Sigma^*)^{2k}$ of PCP[34], we can define a relation $\mathcal{R}_u$ that essentially encodes the function $i \mapsto u_i$, and a relation $\mathcal{R}_v$ that encodes $i \mapsto v_i$, and show that their Kleene stars $(\mathcal{R}_u)^*$ and $(\mathcal{R}_v)^*$ are deterministic rational. By construction, any element in the intersection of $(\mathcal{R}_u)^*$ and $(\mathcal{R}_v)^*$ exactly corresponds to a solution of PCP. It follows that the *intersection non-emptiness problem* for deterministic rational relations, taking two such relations and asking if their intersection is non-empty, is undecidable. $\diamond$

Surprisingly, while the inclusion problem for deterministic multitape automata is undecidable [FR68], Harju and Karhumäki proved in 1991 that their equivalence problem is decidable [HK91, Theorem 3.11].[35]

While the class is closed under complementation, it is in fact neither closed under union, intersection [RS59, Theorem 17], concatenation, Kleene closure nor reversal [FR68, Table 1].

Lastly, the DRAT/REC-MEMBERSHIP PROBLEM is decidable. There is a double-exponential time algorithm for binary relations given by Valiant [Val75], improving Stearns's triple-exponential bound [Ste67]. The decidability result was later extended to relations of arbitrary arity by Carton, Choffrut and Grigorieff [CCG06, Theorem 3.7].

### VII.1.6    Restricted Head Movements

Observe that synchronous automata can be seen as multitape automata with a restriction on the head movements: at each step, all heads must move. This motivates the following generalization, introduced by Figueira and Libkin in [FL15].

Given a word $w \in (\Sigma_1 \otimes \dots \otimes \Sigma_k)^*$, we define $\pi_{\text{tape}}(w) \in \mathfrak{P}_+(\llbracket 1, k \rrbracket)^*$ to be the word whose $i$-th letter is the set of tape indices on which $w_i$ is distinct from $\lrcorner$. For instance,

$$\pi_{\text{tape}}\left( \begin{smallmatrix} a\_b\_a\_c\_ \\ aabbaacc \end{smallmatrix} \right) = \{1,2\} \cdot \{2\} \cdot \{1,2\} \cdot \{2\} \cdot \{1,2\} \cdot \{2\} \cdot \{1,2\} \cdot \{2\}.$$

Note that neither $\pi_{\text{tape}}$ nor $\pi_{\text{word}}$ are injective, but putting both information together

$$
\begin{array}{ccc}
(\Sigma_1 \otimes \dots \otimes \Sigma_k)^* & \to & (\Sigma_1^* \times \dots \times \Sigma_k^*) \times \mathfrak{P}_+(\llbracket 1, k \rrbracket)^* \\
w & \mapsto & \langle \pi_{\text{word}}(w), \pi_{\text{tape}}(w) \rangle
\end{array}
$$

yield an injective map. The word $\pi_{\text{tape}}(w)$ describes, starting from $\pi_{\text{word}}(w)$, which head movements have to be done to obtain a word of $(\Sigma_1 \otimes \dots \otimes \Sigma_k)^*$, which represent possible runs of a multitape automaton.[36]

Given a regular language $T$ over the alphabet $\mathfrak{P}_+(\llbracket 1, k \rrbracket)$, a *T-multitape automaton* is a multitape automaton over $\Sigma_1, \dots, \Sigma_k$ whose semantics is restricted in the following way: a tuple $\langle u_1, \dots, u_k \rangle \in \Sigma_1^* \times \dots \times \Sigma_k^*$ is accepted if there exists $w \in (\Sigma_1 \otimes \dots \otimes \Sigma_k)^*$ s.t. $\pi_{\text{word}}(w) = \langle u_1, \dots, u_k \rangle$ and $\pi_{\text{tape}}(w) \in T$.

[34] Recall that this undecidable problem asks if there exists $n \in \mathbb{N}$ and a sequence of (possibly repeating) indices $i_1, \dots, i_n \in \llbracket 1, k \rrbracket$ s.t.

$$u_{i_1} \cdots u_{i_n} = v_{i_1} \cdots v_{i_n}.$$

[35] Back then, this was one of the most prominent open questions in the field: recall that this model was introduced more than thirty years before by Rabin and Scott [RS59].

[36] The information $\pi_{\text{tape}}$ is not unlike what is called "origin information" for transducers [Boj14].

Binary synchronous automata exactly correspond to $T_{\text{sync}}$-multitape automaton when $T_{\text{sync}} \triangleq \{1,2\}^* \cdot (\{1\}^* + \{2\}^*)$:[37] we start by moving both heads synchronously, and then only of them—and we stick to the choice we made. Similarly, it can be shown that for $T_{\text{rec}} \triangleq \{1\}^* \cdot \{2\}^*$, then $T_{\text{rec}}$-multitape automaton exactly recognize recognizable relation [FL15, Proposition 1]. Both $\mathfrak{P}_+(\llbracket 1, k \rrbracket)^*$- and $\{\{i\} \mid i \in \llbracket 1, k \rrbracket\}^*$-multitape automata exactly recognize all rational relations—corresponding to the models where we allow any set of heads to move at a given time, or where we only want a single head to move at each time, respectively.

A *T-controlled relation* is any relation that can be recognized by a $T$-multitape automata.[38] The central question around this notion is to characterize the expressiveness of $T$-multitape automata. [FL15, Theorem 1] provides characterizations of $T$ s.t. all $T$-controlled relations are recognizable (resp. automatic), and this characterization is actually effective [FL15, Corollary 3]. Descotte, Figueira and Puppis showed that given regular languages $T_1$ and $T_2$, it is decidable if $T_1$-controlled relations exactly correspond to $T_2$-controlled relations for binary relations (*i.e.* assuming $k = 2$) [DFP18, Main Theorem]. Their algorithm fails to capture the general case, and the decidability of this problem is still open for relations of arbitrary arity [DFP18, § 9].

**Open Problem VII.1.15.** Given $k \in \mathbb{N}_{>0}$, given regular languages $T_1$ and $T_2$ over $\llbracket 1, k \rrbracket$, can we decide if $T_1$-controlled relations exactly correspond to $T_2$-controlled relations?

Similarly, Descotte, Figueira and Figueira showed that given a regular language $T$, it is decidable whether the class of $T$-controlled relations is closed under intersection, complement, concatenation, Kleene star and projection [DFF19, Theorem, p. 2] but the problem remains open for relation of higher arity [DFF19, § 6].

### VII.1.7   And All Other Multitape Automata

"Regular prefix relations", also known as "special relations", are a subclass of automatic relations introduced in 1984 that admits a nice logical characterization [Cho06, § "1984"].

*Right-automatic relations* can be defined analogously to automatic relations by putting padding symbol at the beginning of the words instead of at the end. The prefix relation is automatic but not right-automatic, and dually the suffix relation is right-automatic but not automatic. Hence, this class is orthogonal to the class of automatic relations, and shares all its closure properties and decidability results.

Multitape automaton have been extended to "two-way multitape automata" by allowing heads to move both ways by Rabin and Scott [RS59, § 12]. Similarly to what we did in the definition of deterministic rational relations, not only do we add an end marker ◁ at the end of words, but also a beginning marker ▷ at their beginning. We do not provide a formal definition but only an example in Figure VII.7.

[37] In the expression above $\{1,2\}$ denotes a single letter!

[38] This notion was introduced in [FL15, § 3]. In fact, the authors only allow one head to move at each step, *i.e.* $T$ is a regular language over $\llbracket 1, k \rrbracket$ and not over $\mathfrak{P}_+(\llbracket 1, k \rrbracket)$ as defined here. However both models are clearly equivalent: up to duplicating the states of the automaton, we can simulate the movement of multiple heads by moving one head at a time.

The resulting model, namely *two-way rational relations* contains the class of rational relations. In fact, this model is closed under union and intersection: since rational relations are not closed under intersection, it follows that two-way rational relations are strictly more expressive than them [RS59, § 12].

Of course, since equivalence is already undecidable on rational relations, it is also undecidable for this class.

It also makes sense to look at their deterministic models: informally, in a given state, we should not only know on which tapes we will read non-blank symbols, but also what head movements we will operate. This gives rise to the notion of *deterministic two-way multitape automata*. This model is in fact remarkably expressive: given two deterministic rational relations $\mathcal{R}_1$ and $\mathcal{R}_2$, it is possible to describe their intersection $\mathcal{R}_1 \cap \mathcal{R}_2$ by a deterministic two-way multitape automaton: the idea is to first simulate the first automaton from left to right, then, when reaching the end markers, to go back to the beginning of the tapes, and then simulate the second automaton from left to right. Since the intersection non-emptiness problem of deterministic rational relations is undecidable (see Example VII.1.14), it follows that non-emptiness of *deterministic two-way rational relations—i.e.* relations recognized by deterministic two-way multitape automata—is undecidable [RS59, Theorem 19]. It follows that both the inclusion problem and the equivalence problem are undecidable: this class is simply too expressive.

Generalizing the construction above for the intersection, shows that both deterministic two-way rational relations and two-way rational relations are closed under union and intersection.[39]

Deterministic two-way rational relations are closed under complement[40] We believe two-way rational relations are not be closed under complement, and that the proof is similar to rational relations; nonetheless, we could not find any formal reference to back this claim. Neither model are closed under composition [CES17, Theorem 5].

Two-way multitape automata where the two-wayness is restricted to moving all heads simultaneously back to the beginning of the words are known as "rewind automaton", and were introduced by Rosenberg in 1965, see *e.g.* [Cha80, § 4]. Alternating two-way two-tape automata have been introduced and studied by Carton, Exibard and Serre who showed that this class is not closed under complementation [CES17, Theorem 1].

### VII.1.8 The Surprisingly Strange World of Transducers

Recall that, by construction, multitape automata have access to a pair of words $\langle u, v \rangle$ and then can decide whether this pair should be accepted or not. Transducers differ in that they have access to $u$ and can potentially produce, from $u$, the word $v$.

Formally, *non-deterministic one-way transducers*, or *transducers* for short,

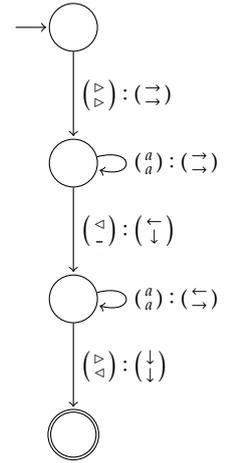

Figure VII.7: A "two-way multitape automaton" for the relation

$$\{\langle u, u \cdot u^r \rangle \mid u \in \Sigma^*\}$$

where $u^r$ denotes the reverse of $u$. Arrows in the transitions of the machine represent how the heads should move: for instance, $\left(\begin{smallmatrix} \leftarrow \\ \downarrow \end{smallmatrix}\right)$ means that the first head should move left and the second one should stay in place.

[39] The closure under union of deterministic two-way rational relations is slightly less immediate, see [CES17, Lemma 4] for more details.

[40] This can be proven like for classical deterministic automata, by simply complementing the set of accepting states. This construction also works for deterministic one-way rational relations [RS59, Theorem 17]. However, in the case of two-way automaton, we need to first assure that the automaton does not reject by looping forever: as mentioned in [CES17, Lemma 4] this can be dealt with using Sipser's trick [Sip80, Theorem 1].

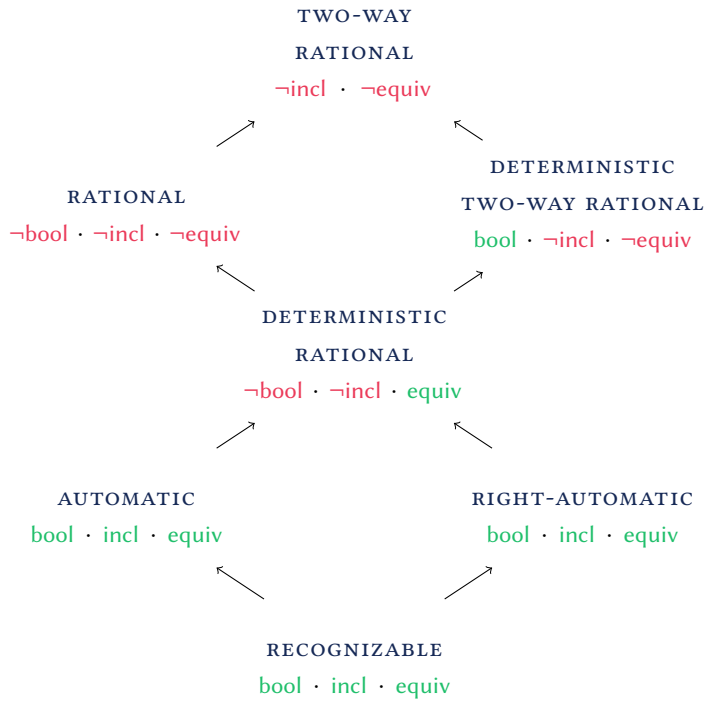

Figure VII.8: The hierarchy of rational relations. Arrows denote *strict* inclusions between classes of relations. Predicates bool, incl and equiv indicate whether the class is closed under Boolean operators, has decidable inclusion problem and equivalence problem, respectively.

are defined as multitape automata, but their transition relation is a subset of

$$\underbrace{Q}_{\text{source state}} \times \underbrace{\Gamma \cup \{\varepsilon\}}_{\text{input alphabet}} \times \underbrace{\Sigma^*}_{\text{output word}} \times \underbrace{Q}_{\text{target state}}.$$

We will also assume that the input word has markers at its extremities. This model is quite clearly equivalent to the one of multitape automata, and so *transductions* actually correspond to rational relations.[41]

However, because we think of transducers and multitape automata differently, the notion of deterministic transducer differs from the one of deterministic multitape automata: recall that the latter model has access to both the input and output, and so we say that it is deterministic when its behaviour it uniquely guided by both the input and the output. However, we say that a transducer is *deterministic* when its transition relation can actually be described by a partial function

$$\underbrace{Q}_{\text{source state}} \times \underbrace{\Gamma}_{\text{input alphabet}} \rightarrow \underbrace{Q}_{\text{target state}} \times \underbrace{\Sigma^*}_{\text{output word}}.$$

Hence, by construction, a deterministic transducer recognizes a partial function from $\Gamma^*$ to $\Sigma^*$. Note however that the deterministic multitape automaton of Figure VII.5 recognizes the subword relation, which is not functional. And hence, deterministic transducers are strictly less expressive than deterministic multitape automaton. But in fact we can prove a strictly stronger result: there are functional relations that can be recognized by a deterministic rational but that are not a *deterministic transduction*—*i.e.* recognizable by a deterministic one-way transducer. For instance, consider the function, where $\Sigma \triangleq \{a, b, c\}$,

[41] This is to ensure this equivalence that we allow a transducer to produce an output on $\varepsilon$. This feature is usually not allowed is most definitions of transducers.

$$f \colon \quad \Sigma^* \quad \rightarrow \quad \Sigma^*$$
$$a \cdot b^n \cdot c \quad \mapsto \quad c \cdot b^n \cdot a \quad (n \in \mathbb{N})$$
$$a \cdot b^n \cdot a \quad \mapsto \quad a \cdot b^n \cdot a \quad \text{``} \quad .$$

In fact, this relation is even automatic! A synchronous automaton can simply use two states to remember if the first pair it reads is $\binom{a}{c}$ or $\binom{a}{a}$, in which case it will know if it should expect to read $\binom{c}{a}$ or $\binom{a}{a}$ at the end of the word. On the other hand, a deterministic one-way transducer only has access to the input word, and while reading the first $a$, it has no choice but to non-deterministically guess if it should produce a '$c$' or an '$a$'.[42]

The same complication happens for two-wayness: a two-way transducer can move both left and right on its input tape, but not on the output, since the output is *produced* and not read. We call *two-way transductions* and *deterministic two-way transductions* the relations recognized by two-way transducers and deterministic two-way tranducers, respecitvely.

Deterministic two-way transductions are actually widely known as *regular functions*, and have been extensively studied in the past decade. They admit several characterizations, in the frameworks of multitape automata, transducers, logic, etc.[43]

Remarkably, the intersection of the classes of functional relations and of two-way transductions collapses exactly to the class of all regular functions [EH01, Theorem 22, p. 243]. In other words non-determinism does not increase the expressiveness of two-way multitape automata, when restricted to functional relations. As a consequence, the hierarchy of functional relations is surprisingly more linearly ordered than expected, as depicted in Figure VII.9.

Note however that the equivalence problem remains undecidable for (non-functional) two-way transductions because the class includes all rational relations. However, the problem becomes decidable—and even PSpace—when one wants to check "origin equivalence", which is a restricted form of equivalence: informally not only we want the transducers to describe the same relation, but we also require for the outputs to be produced in the same way [BMPP18, Theorem 1].[44]

The equivalence problem of regular functions was shown to be decidable by Gurari [Gur82, Theorem 1]. Note that for the subclass of functional rational relations, this result was proven one decade earlier, by Schützenberger in 1975—see *e.g.* [Ber79, Corollary IV.1.3].

*Polyregular functions.* Regular functions all have a linear growth, in the sense that if $\langle u, v \rangle$ belongs to the relation, then $|v| \le k \cdot |u|$ for some constant $k$.[45] This model has been generalized to *polyregular functions*—which have polynomial growth—see [Boj22] for a recent survey: one way of defining this class of functions is by allowing transducers to put and move a bounded number of *pebbles* on the input word. A typical example of polyregular function that is not regular is the squaring function, which takes a word $u$ and maps it to the concatenation $u^{|u|}$ of $u$ as many times as there are letters in the word. Since this function has quadratic growth, it cannot be regular, but it can easily be recognized by a deterministic two-way transducer with a single pebble, that puts a pebble on the first position, outputs the input word, moves the pebble one position to the right, and iterate this process until the pebble reaches the end of the word. However, the decidability of the equivalence problem for this class remains widely open [Boj22, § 8].

[42] One can prove formally that this function is not recognizable by a deterministic one-way transducer by using [Sak09, Theorem V.4.2] since $f$ is clearly not Lipschitz for the prefix distance.

[43] See [Boj22, 4th Paragraph] for pointers.

[44] This "origin information" is not unlike the function $\pi_{\text{tape}}$ introduced earlier for multitape automata.

[45] In the case of one-way deterministic transductions, this follows from [Sak09, Theorem V.4.2].

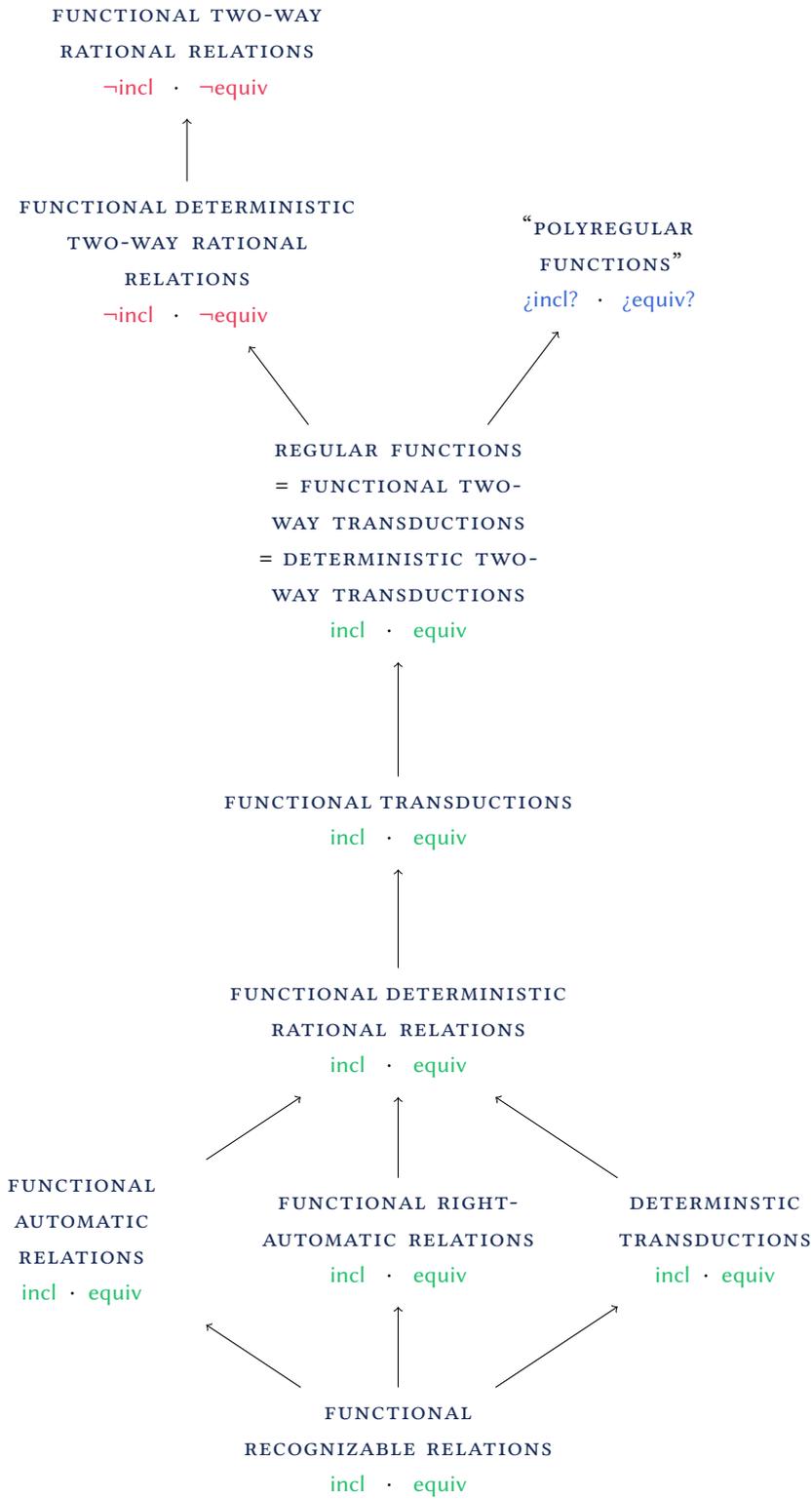

Figure VII.9: The hierarchy of functional relations. Arrows denote *strict* inclusions between classes of relations. "Functional *foo*" denotes the intersection of the class *foo* with the class of functional relations. Predicates incl and equiv indicate whether the class has decidable inclusion problem and equivalence problem, respectively.

We represent the hierarchy in Figure VII.9. Note that some classes are defined semantically and admit no machine characterization: for instance, the class of "two-way rational relations" is the intersection of functional relations with two-way rational relations. Given a two-way rational relation, it is unde­cidable whether the relation it recognizes is functional.[46] We refer the reader to Gauwin's habilitation [Gau20], and in particular to the splendid [Gau20, Figure 2.1, p. 16][47] for a clickable taxonomy of word-to-word transducers, and references towards decidability of their membership problem. See also Douéneau's Ph.D. thesis [Dou23, § 1] for variations on transducers models. Lastly, let us note that most classes of relations usually become quite "simple" when restricted to a unary alphabet, see *e.g.* [CG14, Theorem 1] for a simple characterization of deterministic two-way rational relations over a unary alphabet.

## VII.2  A Logical Excursion

### VII.2.1  First-Order Interpretation

A major construction in logic is to restrict the scope of the class of models—thereafter called "universe". Formally, given a universe $\mathcal{U}$ and a class of queries $\mathcal{Q}$,[48] given a subclass $\mathcal{V}$ of $\mathcal{U}$, we can consider the restriction

$$\mathcal{Q}|_{\mathcal{V}} \triangleq \{\phi \cap \mathcal{V} \mid \phi \in \mathcal{Q}\}.$$

For instance, letting $\mathcal{U}$ be the class of all $\sigma$-structures and $\mathcal{Q}$ be the set of all first-order sentences over $\sigma$, by taking $\mathcal{V}$ to be the set of all *finite* $\sigma$-structures, then $\mathcal{Q}|_{\mathcal{V}}$ precisely corresponds to the first-order sentences over finite structures. For instance, the class of all *finite* $\sigma$-structures belongs to $\mathcal{Q}|_{\mathcal{V}}$ but not to $\mathcal{Q}$.

However, this construction does not preserve any reasonable property on $\mathcal{Q}$, since $\mathcal{V}$ is an arbitrary subclass of $\mathcal{U}$. The idea behind first-order interpretation is precisely to circumvent this problem, by *interpreting* a class of structures inside another one by using first-order logic, allowing this way to preserve some properties.

Let $\sigma$ and $\tau$ be relational signatures. A *d-dimensional first-order interpreta­tion* $\mathcal{I}$ consists of the following tuple of first-order formulas over $\sigma$:
- $\mathrm{dom}_{\mathcal{I}}(x^1, \dots, x^d)$ (specifying the domain)
- $=_{\mathcal{I}}(\langle x^1, \dots, x^d\rangle, \langle y^1, \dots, y^d\rangle)$ (specifying equality)
- $\mathcal{R}_{\mathcal{I}}(\langle x_1^1, \dots, x_1^d\rangle, \dots, \langle x_k^1, \dots, x_k^d\rangle)$ for any predicate $\mathcal{R} \in \tau$ of arity $k$,

such that $=_{\mathcal{I}}$ defines an equivalence relation on the domain of $\mathcal{I}$.

[46] This can be proven by reduction from the inclusion of functional two-way rational relation since a function $f$ is included in another function $g$ iff $f \cup g$ is functional and the domain of $f$ is included in the domain of $g$. This reduction works for any class that is effectively closed under union.

[47] Note however that arrows are re­versed compared to Figure VII.8.

[48] Recall that we see queries here in a purely semantical way: it is nothing but a subclass of $\mathcal{U}$.

Given a $\sigma$-structure $\mathbf{A}$, the *$\mathcal{I}$-interpretation of* $\mathbf{A}$ is the $\tau$-structure denoted by $\mathcal{I}(\mathbf{A})$ defined as follows:[49]

- its domain is the quotient of $\{\bar{a} \in \mathbf{A}^d \mid \mathbf{A}, \bar{a} \vDash \mathrm{dom}_{\mathcal{I}}(\bar{x})\}$ by the equivalence relation $\{\langle \bar{a}, \bar{b} \rangle \mid \mathbf{A}, \bar{a}, \bar{b} \vDash =_{\mathcal{I}}(\bar{x}, \bar{y})\}$,
- for any predicate $\mathcal{R}_{(k)} \in \tau$, we have

$$\mathcal{R}_{(k)}(\mathcal{I}(\mathbf{A})) = \left\{ \langle \bar{x}_1, \ldots, \bar{x}_k \rangle \mid \exists \bar{y}_1. \ldots. \exists \bar{y}_k. \bigwedge_{i=1}^{k} \bar{x}_i =_{\mathcal{I}} \bar{y}_i \wedge \mathcal{R}_{\mathcal{I}}(\bar{y}_1, \ldots, \bar{y}_k) \right\}.$$

For instance, letting $\sigma = \tau$ be the graph signature, consider the one-dimensional interpretation $\mathcal{I}$ where:

- the domain formula keeps all vertices,
- the equality formula is proper equality,
- the formula for the edge predicate puts two variables $x$ and $y$ in relation if either there is an edge from $x$ to $y$, or if $x$ has no successor and $y$ has no predecessor.

Then, the $\mathcal{I}$-interpretation of a directed path is a cycle, see Figure VII.10.

The idea behind first-order interpretation is that first-order formulas about the interpretation $\mathcal{I}(\mathbf{A})$ can be translated into first-order formulas over $\mathbf{A}$. Formally, given a first-order formula $\phi(x_1, \ldots, x_k)$ over $\tau$, we can define a first-order formula $\phi^{\mathcal{I}}(\bar{x}_1, \ldots, \bar{x}_k)$ over $\sigma$ defined inductively by:

- replacing each variable $x$ by a $d$-tuple $\bar{x}$,
- replacing every occurrence of $\mathcal{R}(x_1, \ldots, x_k)$ (with $\mathcal{R}_{(k)} \in \tau$) by the formula

$$\exists \bar{x}_1'. \ldots. \exists \bar{x}_k'. \left( \bigwedge_{i=1}^{k} \bar{x}_i =_{\mathcal{I}} \bar{x}_i' \right) \wedge \mathcal{R}_{\mathcal{I}}(\bar{x}_1', \ldots, \bar{x}_k'),$$

- relativizing quantification with respect to $\mathrm{dom}_{\mathcal{I}}$, meaning that

$$(\exists x. \phi(x))^{\mathcal{I}} \;\hat{=}\; \exists \bar{x}. \mathrm{dom}_{\mathcal{I}}(\bar{x}) \wedge \phi^{\mathcal{I}}(\bar{x}), \text{ and}$$
$$(\forall x. \phi(x))^{\mathcal{I}} \;\hat{=}\; \forall \bar{x}. \mathrm{dom}_{\mathcal{I}}(\bar{x}) \Rightarrow \phi^{\mathcal{I}}(\bar{x}).$$

**Proposition VII.2.1.** For any first-order formula $\phi(x_1, \ldots, x_k)$ over $\tau$, for any pointed $\sigma$-structure $\langle \mathbf{A}, \bar{a}_1, \ldots, \bar{a}_k \rangle$, we have:

$$\langle \mathcal{I}(\mathbf{A}), [\bar{a}_1], \ldots, [\bar{a}_k] \rangle \vDash \phi(x_1, \ldots, x_k) \quad \textit{iff} \quad \langle \mathbf{A}, \bar{a}_1, \ldots, \bar{a}_k \rangle \vDash \phi^{\mathcal{I}}(\bar{x}_1, \ldots, \bar{x}_k).$$

And so, in particular, if $\mathcal{C}$ is a class of structures, then model checking (resp. satisfiability, resp. validity) over the class $\mathcal{I}(\mathcal{C})$ can be reduced to the same problem over $\mathcal{C}$.

Typical examples of one-dimensional interpretations include restricting a structure to a first-order definable subset, or taking the complement—that is, swapping hyperedges and non-hyperedges.

**Example VII.2.2.** We consider a 2-dimensional interpretation that we call "box product". Let $\sigma$ and $\tau$ be the graph signature. We define a two-dimensional

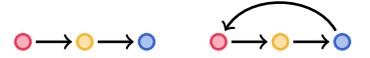

Figure VII.10: A directed path (left) and its interpretation by $\mathcal{I}$ (right), that adds an edge from any vertex with no successor to any vertex with no predecessor.

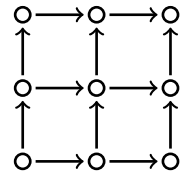

Figure VII.11: Interpretation of the directed path of Figure VII.10 by the "box product interpretation".

interpretation $\mathcal{J}$ with trivial domain and equality[50] and we put an edge from $\langle x^1, x^2 \rangle$ to $\langle y^1, y^2 \rangle$ if either there is an edge from $x^1$ to $y^1$ and $x^2 = y^2$, or if $x^1 = y^1$ and there is an edge from $x^2$ to $y^2$. Then the interpretation of a directed path is a directed grid, see Figure VII.11. ◇

An interpretation is said to be *injective* if the equality formula is the proper tuple-equality. Note that each interpretation $\mathcal{J}$ can be transformed into an injective interpretation $\mathcal{J}$ *s.t.* for any *totally ordered structure* **A**—meaning that there is a binary predicate in the signature that is interpreted as a total order in **A**, then $\mathcal{J}(\mathbf{A})$ and $\mathcal{J}(\mathbf{A})$ are isomorphic: for this, it suffices to encode any equivalence class in **A** by its minimal element—for more details, see Proposition VII.3.2.

We say that **A** is *first-order interpretable* in **B** if there exists a first-order interpretation $\mathcal{J}$ *s.t.* $\mathcal{J}(\mathbf{A})$ is isomorphic to **B**. We say that two structures are *first-order equivalent* if they are first-order interpretable in one another. Two structures that are isomorphic are clearly first-order equivalent. Moreover, if a $k$-ary relation $\mathcal{R}$ is first-order definable in a $\sigma$-structure **A**, then **A** is first-order equivalent to the structure **A** to which we add a new $k$-ary predicate interpreted as $\mathcal{R}$. For instance, $\langle \mathbb{N}, + \rangle$ and $\langle \mathbb{N}, +, <, 0, 1 \rangle$ are first-order equivalent.

### VII.2.2  First-Order Reduction and First-Order Model Checking

As we have seen, first-order interpretations preserve logical properties of classes of structures. In this part, we show that most complexity classes are closed under this construction.

We define a *first-order reductions* to be an injective first-order interpretation.[51] Complexity classes are defined in terms of decision problems, namely languages $L \subseteq 2^*$, rather than as classes of structures. However, it is known that:

- any language $L \subseteq 2^*$ can be seen as a class of structures over the signature of binary strings;
- for any relational signature $\sigma$, there is an encoding of finite $\sigma$-structures as a language $L \subseteq 2^*$, *i.e.* as a class of structure over the signature of binary strings—see *e.g.* [Imm98, § 2.2].

Importantly, these encodings between $\sigma$-structures and structures over the signature of binary strings are both first-order reductions. It follows that there is a first-order reduction between two classes of structures *iff* there is a first-order reduction between their encodings as languages. Overall, it implies that first-order reductions are proper reductions in the complexity-theoretic sense. We will see that first-order logic actually lives in the lowest levels of the hierarchy of complexity classes.

We denote by *FO$^{fin}$* the class of all problems over finite structures that are first-order definable.[52]

**Proposition VII.2.3** (Folklore). $\mathrm{FO}^{fin} \subseteq \mathsf{L}$.

[50] Meaning formally that

$$\mathrm{dom}_{\mathcal{J}}(x^1, x^2) \mathrel{\hat{=}} \top, \text{ and}$$

and $=_{\mathcal{J}}(\langle x^1, x^2 \rangle, \langle y^1, y^2 \rangle)$ is the formula $x^1 = y^1 \wedge x^2 = y^2$.

[51] The terminology "FO-interpretations" is usually employed in automata theory and graph theory, while "FO-reduction" is used in complexity theory—see *e.g.* [Imm98, Definition 2.11 & Definition 1.26]—however there is no good reason to distinguish these two notions.

[52] We can either assume the signature to be fixed, or to be part of the input.

*Proof sketch.* The naive recursive algorithm, recursing over the first-order formula, works in logarithmic space: it suffices to keep one pointer to the structure for every variable of the formula, but since this formula is fixed, we only require a constant number of pointers. □

Moreover, L-hardness is usually defined using first-order reductions. When the signature is sufficiently expressive, namely when it is able to express basic arithmetic [Imm98, Proviso 1.14], Immerman showed that $FO^{fin}$ corresponds to the circuit-class $AC^0$, and also to the whole logarithmic-time hierarchy [Imm98, Corollary 5.32].

### VII.2.3 A Model-Theoretic Perspective on Automatic Relations

Given an alphabet $\Sigma$, we define on $\Sigma^*$:
- a unary relation $l_a$ indicating that the last letter of a word is $a$,
- a binary relation $\approx_{len}$ indicating that two words have the same length,
- a binary relation $\leqslant_{pref}$ indicating that a word is a prefix of another.

We denote by $\sigma^{sync}_\Sigma$ the signature $\langle\langle l_a \rangle_{a \in \Sigma}, \approx_{len}, \leqslant_{pref}\rangle$[53] and by $\Sigma^*$ the $\sigma^{sync}_\Sigma$-structure over $\Sigma^*$ where the predicates are interpreted as above.[54]

**Proposition VII.2.4** ([EES69], Theorems 1 & 2]). *If $\Sigma$ has at least two letters, then a relation over $\Sigma^*$ is automatic iff it is first-order definable in $\Sigma^*$.*[55]

*Proof.* ↝ *From logic to automata.* This implication is easy: it suffices to observe first that each relation $l_a$, $\approx_{len}$ and $\leqslant_{pref}$ is automatic, and then to use the fact that automatic relations are closed under Boolean operations and quantification. This last point can be proven using a powerset construction on the automaton.

↝ *From automata to logic.* The converse implication is less straightforward. It was originally proven by Eilenberg, Elgot and Shepherdson starting from a rational expression, making the proof somewhat tedious. In [Cho06, § "1969"], Choffrut notes that "mimicking the automaton yields a much more intuitive proof and can be reconstructed by a good PhD student": we present here this proof.

We start with an example: we want to build a first-order formula $\phi$ over $\sigma^{sync}_2$ for the language $(00)^*$, in the sense that for all $u \in 2^*$,

$$\langle 2^*, u \rangle \vDash \phi(x) \quad \text{iff} \quad u \in (00)^*.$$

For this, it suffices to guess a word $v$ of the same length as $u$, which alternates zeroes and ones, and then to check that the first letter $v$ is distinct from its last letter.

Given a synchronous automaton $\mathcal{A}$ recognizing $\mathcal{R} \subseteq (\Sigma^*)^k$, we want to build a first-order formula $\phi_\mathcal{A}$ over $\sigma^{sync}_\Sigma$ *s.t.* for all $\langle u_1, \ldots, u_k \rangle \in (\Sigma^*)^k$:

$$\langle u_1, \ldots, u_k \rangle \in \mathcal{R} \quad \text{iff} \quad \langle \Sigma^*, u_1, \ldots, u_k \rangle \vDash \phi_\mathcal{A}(x_1, \ldots, x_k).$$

The formula $\phi_{\mathscr{A}}$ is built as follows:

- for each state $q$ of $\mathscr{A}$, we guess a word $v_q \in 2^*$ *s.t.* the length of $v_q$ is the maximal length of a word $u_i$ ($i \in [\![1, k]\!]$), *i.e.* $|u_1 \otimes \cdots \otimes u_k|$,

- we check that for each position $i$, there is exactly one state $q$ *s.t.* the $i$-th letter of $(v_q)$ is 1,

- this way, the tuple of words $\langle v_q \rangle_q$ represents a potential run of $\mathscr{A}$ over $\mathcal{A}$: we then check that this indeed defines an accepting run of the automaton.

It is routine to check that all these properties can be written in first-order logic using the predicates $l_a$, $\approx_{\text{len}}$ and $\preccurlyeq_{\text{pref}}$. Note in particular that this proof crucially relies on the fact that we can guess words in $2^*$, which is allowed by the assumption that $\Sigma$ has size at least 2.[56] Also, the formula $\phi_{\mathscr{A}}$ is of polynomial size in the size of $\mathscr{A}$. □

We want to highlight that Proposition VII.2.4 can be extended for $k = 0$, since the 0-ary relations are the subsets of $\varnothing^* = \{\varepsilon\}$, and so there is exactly two relations: $\{\bullet\}$ (identified with "true") and $\varnothing$ (identified with "false"). Both are first-order definable, and automatic: observe that synchronous automata over $\varnothing$ either return "true" if they have an initial state that is accepting, and otherwise they return "false". 0-ary relations naturally arise because they correspond to Boolean queries. For instance, assume that $\phi(x_1, x_2, x_3)$ is a first-order formula over $\sigma_\Sigma^{\text{sync}}$: then $\forall x_1. \exists x_2. \forall x_3. \phi(x_1, x_2, x_3)$ is a first-order sentence over $\sigma_\Sigma^{\text{sync}}$. The proof of Proposition VII.2.4 builds from a synchronous (3-ary) automaton for $\phi(x_1, x_2, x_3)$ a triple-exponential-sized synchronous (0-ary) automaton for $\forall x_1. \exists x_2. \forall x_3. \phi(x_1, x_2, x_3)$.

Let us point out that the formula built from the automaton is of the form $\exists^* \forall^* \exists^*$ followed by a quantifier-free formula. It shows that first-order logic over $\Sigma^*$ collapses to the $\Sigma^3$ level. Figueira, Ramanathan and Weil proved that this hierarchy does not collapse to a lower level [FRW19, Theorem 3], and provide effective characterization of these lower levels [FRW19, Theorem 9].

**Example VII.2.5.** Given an ordered alphabet $\Sigma = \{a_1, \ldots, a_n\}$ with $a_1 < \ldots < a_n$, we define the *lexicographic order* by $u \preccurlyeq_{\text{lex}} v$ *iff* when there is a prefix $w$ both of $u$ and of $v$, *s.t.* either $w$ has the same length as $u$, or the letter following $w$ in $u$ is strictly smaller than the letter following $w$ in $v$.

Note that this can be defined by a first-order formula over $\sigma_\Sigma^{\text{sync}}$ since saying that the letter following $w$ in $u$ is $a$ amounts to guessing the smallest prefix $w'$ of $u$ that must strictly contain $w$ as a prefix, and checking that $w'$ ends with letter $a$. From Proposition VII.2.4, it follows that the lexicographic order is an automatic relation. ◇

**Remark VII.2.6.** We want to highlight that, in the case of unary relations, Proposition VII.2.4 shows that regular languages are exactly the first-order definable sets of $\Sigma^*$ (if $\Sigma$ has at least two letters). This result contradicts in no way the fact that regular languages exactly those definable in monadic second-order logic, nor does it imply that monadic second-order logic collapses to first-order logic. Indeed, in this last characterization, the models are finite words, and its first-class citizens[57] are the positions in these words. In the case of Proposition VII.2.4, the model is fixed (namely $\Sigma^*$), and its first-class citizens are the finite words. In some sense, this logic is *external* to finite words—the logic does not have access to the "inner workings" of the words—, while monadic second-logic is *internal* as traditionally used. Overall, Proposition VII.2.4 can be rephrased as the equivalence between *external* first-order logic and *internal* monadic second-order logic.[58] Said otherwise, when going from FO over the structure $\Sigma^*$ to MSO over the structure of a finite word, we gain one level of quantification *via* the logic, but we lose one level of quantification *via* the change of model. ◇

[56] Indeed, we can then assume *w.l.o.g.* that $2 \subseteq \Sigma$, and then check in first-order logic that a word $v_q \in \Sigma^*$ actually belongs to $2^*$.

### VII.2.4 Logical Characterization of Other Classes of Relations

Benedikt, Libkin, Schwentick and Segoufin studied submodels of

$$\Sigma^* = \langle \Sigma^*, \langle l_a \rangle_{a \in \Sigma}, \approx_{\text{len}}, \leqslant_{\text{pref}} \rangle$$

in [BLSS03]. For instance, they show that by removing $\approx_{\text{len}}$, the definable sets are exactly the star-free regular languages [BLSS03, Corollary 3.7].

Adding a binary predicate for each regular language $L$, interpreted as $\{\langle u, uv \rangle \mid u, v \in \Sigma^*, v \in L\}$, yields a logic called $\mathbf{S}_{\text{reg}}$, whose definable relations are exactly the "regular prefix relations", also called "special relations" [BLSS03, Corollary 3.22]. This class lives between recognizable relations and automatic relations, see [Cho06, § "1984"].

## VII.3 Automatic Structures

### VII.3.1 Definitions

Automatic structures are a subclass of all relational structures. While some of them can be infinite, they are all finitely describable, by automata. We will see that, this way, some problems on finite structures remain decidable on this larger class of automatic structures.

This notion has a remarkably eventful history: it was introduced by Hodgson in his Ph.D. thesis in 1976 [Hod76][59], but the notion went largely unnoticed. The notion was rediscovered for groups in the late 1980s [Eps92]. Independently, Shapiro in 1992 [Sha92], Khoussainov and Nerode in 1995 [KN95] and Pélecq in his Ph.D. thesis from 1997 [Pel97, § 3.1.3, p. 91] reintroduced the notion of automatic structures.[60]

We fix a finite relational signature $\sigma$. An *automatic presentation $\mathcal{A}$ of a $\sigma$-structure* consists of:

- an alphabet $\Sigma$,
- a regular language $\text{dom}_{\mathcal{A}} \subseteq \Sigma^*$,
- for every relation $\mathcal{R}_{(k)} \in \sigma$, an automatic relation $\mathcal{R}_{\mathcal{A}} \subseteq (\Sigma^*)^k$, and
- an automatic relation $=_{\mathcal{A}} \subseteq \Sigma^* \times \Sigma^*$ that must be an equivalence relation.[61]

[57] By "first-class citizens" we mean the objects over which we can do first-order quantification.

[58] We thank Géraud Sénizergues for suggesting this terminology to us.

[59] Unfortunately the manuscript is not available online. See [Hod83] for the related journal article.

[60] What we call automatic structures correspond to "synchronous automatic structures" and "structures automatiques synchrones" in Shapiro's paper and Pélecq's thesis, respectively. Pélecq gives credit for the definition to his advisor Géraud Sénizergues. All three work by Shapiro, Khoussainov & Nerode and Pélecq claim to generalize the notion of automatic groups from [Eps92].

[61] Some readers already familiar with the notion of automatic structures might be somewhat surprised by the inclusion of this equality relation: we will soon see (Proposition VII.3.2) that in this specific setting, we do not need it. This is not true in general, see Footnote 65.

The *structure represented* **A** by an automatic presentation $\mathcal{A}$ has $\mathrm{dom}_{\mathcal{A}}/\!=_{\mathcal{A}}$ as its domain, and the predicate $\mathcal{R}_{(k)} \in \sigma$ is interpreted in such a way that a tuple $\langle X_1, \ldots, X_k \rangle$ of equivalence classes belongs to $\mathcal{R}(\mathbf{A})$ if, and only if, there exists $\langle u_i \rangle_{1 \leq i \leq k} \in \langle X_i \rangle_{1 \leq i \leq k}$ s.t. $\langle u_1, \ldots, u_k \rangle \in \mathcal{R}_{\mathcal{A}}$. Given $u \in \mathrm{dom}_{\mathcal{A}}$, we denote by $\mathcal{A}(s)$ the element of **A** it represents, namely the equivalence class of $u$ under $=_{\mathcal{A}}$.

We say that a $\sigma$-structure is *automatic* if it is represented by an automatic presentation.[62] For instance, $\langle \Sigma^*, \langle l_a \rangle_{a \in \Sigma}, \approx_{\mathrm{len}}, \preccurlyeq_{\mathrm{pref}} \rangle$ is an automatic structure.

Furthermore, the infinite binary tree can be represented by the automatic presentation $\mathcal{B}$ with $\mathrm{dom}_{\mathcal{B}} = 2^*$,

$$\mathcal{E}_{\mathcal{B}} = \{\langle u, u0 \rangle \mid u \in 2^* \} \cup \{\langle u, u1 \rangle \mid u \in 2^* \},$$

and $=_{\mathcal{B}}$ is equality, see Figure VII.12.

Presentations such as this one, where $=_{\mathcal{B}}$ is the equality relation, are called *injective presentations*.

**Example VII.3.1.** The structure $\langle \mathbb{N}, + \rangle$ is automatic.[63] We build an automatic presentation $\mathcal{N}$ by using a binary encoding $\mathrm{dom}_{\mathcal{N}} = 2^*$. A word $w \in 2^*$ will represent the number $\sum_{i=0}^{|w|-1} w_i \cdot 2^{i-1}$—notice that we write numbers with their least significant bit first. Naturally, it follows that $=_{\mathcal{N}}$ puts two words $u$ and $v$ in relation if they are equal after removing the trailing zeroes. We then need to describe $+_{\mathcal{N}}$: the idea is to simulate addition, reading words from left to right, by using two states (0 and 1) to remember the carry.

For instance, the transition $0 \xrightarrow{\langle 0, 1, 1 \rangle} 0$ can be understood as "when adding $0$ and $1$, with current carry $0$, the results equals $1$, with a carry of $0$ for the next bit". In general, we have a transition

$$p \xrightarrow{\langle x, y, z \rangle} q \quad \textit{iff} \quad x + y + p = z + 2q$$

for all $p, q \in 2$ and $\langle x, y, z \rangle \in 2 \otimes 2 \otimes 2$.[64] This gives the automaton of figure Figure VII.13. By construction, $\mathcal{N}$ represents $\langle \mathbb{N}, + \rangle$.    $\Diamond$

Recall that, using Proposition VII.2.4, a relation is automatic if and only if it is first-order definable over $\sigma_\Sigma^{\mathrm{sync}}$. This means that we can alternatively see automatic presentations as a collection of first-order formulas—one for the domain, one for equality, and one for each predicate. In turn, this view helps us prove the following result.

**Proposition VII.3.2.**[65] Every automatic structure admits an injective presentation.

*Proof.* We prove this property using logic. Let $\mathcal{A}$ be an arbitrary automatic presentation that represents **A**. The idea to build an injective automatic presentation $\mathcal{A}'$ of **A** is to represent an equivalence class $[u]^{=_{\mathcal{A}}} \subseteq \mathrm{dom}_{\mathcal{A}}$ by its length-lexicographic-minimal element, where the *length-lexicographic*

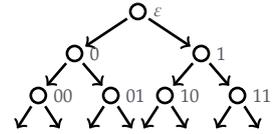

Figure VII.12: An automatic presentation of the infinite binary tree.

[63] We see $+$ as a ternary relation, given by $\{\langle x, y, x + y \rangle \mid x, y \in \mathbb{N}\}$.

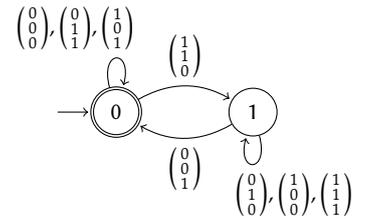

Figure VII.13: Synchronous automaton describing the addition of natural numbers. For the sake of readability, transitions involving the padding symbol are not represented: ␣ is treated as a zero.

[64] Of course, we identify ␣ with 0.

[65] This proposition is not true on the more general class of $\omega$-automatic structures [HKMN08, Theorem 6.6].

*order* is defined by first comparing the length of the word, and in case they have the same length, using the lexicographic order to compare them: this order is a well-founded total order.

Concretely, $\mathcal{A}'$ has the same alphabet as $\mathcal{A}$. Then, we define $\mathrm{dom}_{\mathcal{A}'}$ as the set of minimal elements under the length-lexicographic order of $[u]^{=_{\mathcal{A}}}$ ($u \in \mathrm{dom}_{\mathcal{A}}$). This can be described by a first-order formula since $\mathcal{A}$ is assumed to be automatic, and since $\leqslant_{\mathrm{lex}}$ (and hence the length-lexicographic order) is automatic by Example VII.2.5. Equality is interpreted as proper equality, and then

$$\mathcal{R}_{\mathcal{A}'}(x_1, \dots, x_k) \triangleq \exists y_1. \ \dots \ \exists y_k. \Big( \bigwedge_{i=1}^{k} x_i =_{\mathcal{A}} y_i \Big) \wedge \mathcal{R}_{\mathcal{A}}(y_1, \dots, y_k),$$

for any predicate $\mathcal{R}_{(k)} \in \sigma$. Overall, $\mathcal{A}'$ is an injective presentation that represents exactly the same structure as $\mathcal{A}$. □

**Proposition VII.3.3.** Any automatic structures admits a presentation with an alphabet of size 2.

*Proof.* The idea is to encode each letter of $\Sigma = \{a_1, \dots, a_n\}$ over $\mathbb{2}$ in such a way that each letter $a_i$ is encoded over a word $\widehat{a_i} \in \mathbb{2}^*$ s.t. all $\widehat{a_i}$'s have the same length—this is necessary to preserve automaticity. For instance, take $\widehat{a_i}$ to be the binary encoding of $i$ over $k$ bits, for some fixed value of $k \geqslant \log_2(n)$. Then, by Proposition VII.1.11, the relations we obtain are still automatic. □

Putting Propositions VII.3.2 and VII.3.3 together, we get the following result, that says that $\mathbb{2}^*$ is "universal", in the sense that it is not only automatic, but it "contains" all automatic structures.

**Proposition VII.3.4.** Let **A** be a $\sigma$-structure. The following are equivalent:

1. **A** is automatic,
2. **A** is an injective one-dimensional first-order interpretation of $\mathbb{2}^*$,
3. **A** is a first-order interpretation of $\mathbb{2}^*$.

Recall that $\mathbb{2}^*$ is the structure $\mathbb{2}^*$ equipped with $l_0, l_1, \approx_{\mathrm{len}}$ and $\leqslant_{\mathrm{pref}}$: it can be seen as the infinite binary tree, with unary relation saying if a node is a left or right child, and two binary relations saying if two nodes are at the same depth, and if one is an ancestor of the other.

*Proof.* (1) $\Rightarrow$ (2) follows from Propositions VII.3.2 and VII.3.3. (2) $\Rightarrow$ (3) is trivial. To prove (3) $\Rightarrow$ (1), consider a $d$-dimensional interpretation. An element of the domain of this interpretation is a set of $d$-tuples of words of $\mathbb{2}^*$. The idea is to encode a $d$-tuple of words of $\mathbb{2}^*$ as a word over the alphabet

$$\underbrace{\mathbb{2} \otimes \cdots \otimes \mathbb{2}}_{d \text{ times}}$$

by transforming the tuple $\langle u^1, \dots, u^d \rangle$ into $u^1 \otimes \dots \otimes u^d$. Formulas for equality and relations can be derived easily. We obtain an automatic presentation over the alphabet $2 \otimes \cdots \otimes 2$. $\qquad\square$

In the statement of Proposition VII.3.4, $2^*$ can be replaced by any automatic structure $\mathbf{U}$ *s.t.* there is an injective one-dimensional first-order interpretation of $2^*$ in $\mathbf{U}$. Another example of "universal" structure consists of the finite subsets of $\mathbb{N}$ equipped with inclusion and the preorder $\preceq$ defined by $X \preceq Y$ *iff* $X$ and $Y$ are singletons, say $\{x\}$ and $\{y\}$, respectively, and $x \leq y$, see [Blu24, Theorem XII.2.3].

**Remark VII.3.5.** In light of Figure VII.8, automatic relations and right-automatic relations emerge as the two maximal classes of relations that admit both closure under Boolean operations and the decidability of its "basic" decision problems. While these two classes are orthogonal, it should be noted that "right-automatic structures" and automatic structures are equally expressive: this can be shown by renaming any word $u_1 \ldots u_n$ to its reversal $u_n \ldots u_1$. Hence, automatic structures are maximal in this sense: substituting "automatic relations" for a notion of rationality of Section VII.1 either gives a less expressive class (recognizable relations), an equally expressive class (right-automatic relations), or a class that is too expressive leading, both to a lack of closure properties and undecidability (all other cases). $\qquad\diamond$

However, it should be noted that the notion of automatic structures can be generalized by substituting finite words for more complex models. It leads to the notions of

- *ω-automatic structures* by taking $\omega$-words,
- *tree-automatic structures* by taking finite trees,
- *ω-tree-automatic structures* by taking $\omega$-trees,

together with suitable notions of automata for these objects. We refer the reader to *e.g.* [Blu24, § XII] for more details.

> HYPOTHESIS. In light of Proposition VII.3.2, we will always assume the automatic presentations to be injective, unless specified otherwise.

## VII.3.2 Model-Checking

One of the key interest of automatic structures is that, while they are infinite, their model checking problem remains decidable.

> FIRST-ORDER MODEL CHECKING OF AUTOMATIC $\sigma$-STRUCTURES
>
> *Input*: A first-order sentence $\phi$ over $\sigma$ and an automatic presentation $\mathcal{A}$ of an automatic $\sigma$-structure $\mathbf{A}$.
> *Question*: Does $\mathbf{A} \vDash \phi$?

The *data complexity* of this problem refers to the complexity of the problem for a fixed $\phi$. More precisely, we say that it is $\mathcal{C}$-complete for a class $\mathcal{C}$ when:

- for every $\phi$, the problem over fixed $\phi$ is in $\mathcal{C}$, and

- there is at least one $\phi$ for which it is $\mathcal{C}$-hard.

**Proposition VII.3.6** (*Hodgson's theorem*).[66] FIRST-ORDER MODEL CHECKING OF AUTOMATIC STRUCTURES is decidable, and in fact is Tower-complete under polynomial-time reductions.

Propositions VII.2.1 and VII.2.4 prove this problem to be polynomial-time equivalent to its restriction to the structure $\mathbf{2}^*$.[67]

*Proof sketch.* ❧ *Tower upper bound.* The construction is similar to the easy implication of Proposition VII.2.4: for any automatic presentation $\mathcal{A}$, we build by induction on $\phi(x_1, \dots, x_k)$ a synchronous ($k$-ary) automaton $\mathcal{B}^\phi$ s.t. for all $u_1, \dots, u_k \in \Sigma^*$

$$\mathcal{B}^\phi \text{ accepts } u_1 \otimes \dots \otimes u_k \quad iff \quad \langle \mathbf{A}, u_1, \dots, u_k \rangle \vDash \phi(x_1, \dots, x_k).$$

In the end, we get a synchronous (0-ary) automaton $\mathcal{B}^\phi$ that accepts *iff* $\mathbf{A} \vDash \phi$. Notice that each quantifier alternation and negation implies to complement an automaton, *i.e.* to do a powerset construction, and so the size of $\mathcal{B}^\phi$ is a tower of exponentials in the number of quantifier alternations and nested negations in $\phi$. Hence, we get a Tower algorithm.

❧ *Tower lower bound.*[68] In 1990, Compton and Henson proved that there exists a constant $c > 0$ *s.t.* FIRST-ORDER MODEL-CHECKING restricted to the structure $\mathbf{2}^*$ admits $n \mapsto \text{tower}(cn)$ as a lower bound on the running time of a non-deterministic Turing machine solving the problem [CH90, Example 8.3]. In fact, applying [CH90, Theorem 6.1.(iv)] shows the problem to be hard under polynomial-time reductions for the class of problems which can be solved in non-deterministic time at most $\text{tower}(n^c)$ for some $c > 0$. This corresponds to the class Tower, and hence the problem is Tower-hard. □

**Proposition VII.3.7** (Folklore). The data complexity of the restriction of FIRST-ORDER MODEL CHECKING OF AUTOMATIC STRUCTURES to primitive-positive sentences is NL-complete.

*Proof.* Let again $\mathcal{A}$ be an automatic presentation and $\phi$ be a first-order formula, but we now assume that $\phi$ is of the form

$$\exists x_1. \ \dots \ \exists x_k. \ \psi(x_1, \dots, x_k),$$

where $\psi$ is a positive quantifier-free formula, the size of the automaton for $\psi$ is of the order $|\mathcal{A}|^{\|\psi\|}$: conjunctions and disjunctions only require products but no powerset construction. Now instead of explicitly building an automaton for $\exists x_1. \ \dots \ \exists x_k. \ \psi(x_1, \dots, x_k)$, we test if the automaton for $\psi$ accepts any word. The answer is "yes" *iff* $\mathbf{A} \vDash \phi$. Testing non-emptiness of the automaton amounts to checking if an accepting state can be reached from an initial state, which is NL. Since $\phi$ is fixed, the automaton in question is of polynomial size. To argue that we can effectively obtain an NL algorithm, it suffices to notice that we do not have to explicitly build the automaton for $\psi$, but it suffices to

[66] Decidability was originally proven in [Hod83, Théorème 3.5]. A slightly weaker form was independently re-proved by Pélecq in his thesis [Pel97, Théorème 61, p. 107].

[67] However, this does not work for data complexity since, when reducing $\mathbf{A} \vDash^? \phi$ to $\mathbf{2}^* \vDash^? \phi^{\mathcal{A}}$, the formula $\phi^{\mathcal{A}}$ depends on $\mathcal{A}$.

[68] We often found this result to be incorrectly credited to various papers in the literature.

work with pointers to the automata for $\mathcal{A}$. The number of pointers required only depends on $\psi$.

Lastly, the NL lower bound can be proven by a reduction from NFA non-emptiness, which is itself NL-hard by reduction from the FINITE GRAPH REACHABILITY PROBLEM. □

Recall that since NL is closed under complementation—see *e.g.* [Imm98, Corollary 9.23]—, we obtain the same complexity for negations of primitive-positive sentences.

We refer the reader to [BG00, § 3] for the detailed complexity of other variations of Hodgson's theorem. For instance, the data complexity of its restriction to existential sentences (here negation is allowed) is NP-complete [BG00, Theorem 3.7].

**Proposition VII.3.8** (Folklore). The image of an automatic structure by a first-order reduction is still an automatic structure.

*Proof.* This follows *e.g.* from Proposition VII.2.4. □

We define $FO^{aut}$ to be the class of all problems over automatic structures which are first-order definable. By Proposition VII.3.8, this class is closed under first-order reductions. Moreover, by Proposition VII.3.6, we obtain an upper bound.

**Corollary VII.3.9.**[69] $FO^{aut} \subseteq Tower$.

The goal of the corollary above is to highlight the difference with Proposition VII.2.3: first-order definable problems on automatic structures, while decidable, are not necessarily in L.

**Remark VII.3.10** (Presburger arithmetic). *Presburger arithmetic* is the first-order theory (over the signature $\langle +, 0, 1 \rangle$) derived from the following axioms:

- $\forall x. 0 \neq x + 1$;
- $\forall x. \forall y. x + 1 = y + 1 \Rightarrow x = y$;
- $\forall x. x + 0 = x$;
- $\forall x. \forall y. \forall z. x + (y + z) = (x + y) + z$;
- $(\phi(0) \wedge (\forall x. \phi(x) \Rightarrow \phi(x + 1))) \Rightarrow (\forall x. \phi(x))$, where $\phi(x)$ ranges over all first-order formulas.

Note that $\langle \mathbb{N}, +, 0, 1 \rangle$ is a model of this theory, which is moreover complete (see [Sta84]) and so, for a first-order sentence $\phi$, the following are equivalent:

- $\phi$ belongs to the theory, *i.e.* it is a logical consequence of the axioms above,
- $\phi$ holds in all models satisfying the axioms above, and
- $\langle \mathbb{N}, +, 0, 1 \rangle \vDash \phi$.

Since $\langle \mathbb{N}, +, 0, 1 \rangle$ is first-order equivalent to $\langle \mathbb{N}, + \rangle$, which is automatic by Example VII.3.1, it follows that we can decide Presburger arithmetic. ◇

The decidability result of Proposition VII.3.6 can be extended to $\omega$-tree-automatic structures. In fact it can even be extended to so-called *higher-order automatic structures*, see [Blu24, last remark of § XII.2]. Note that while automatic structures are always countable, higher-order automatic structures have at most the cardinality of the continuum. From this it follows that there exist structures with a decidable FIRST-ORDER MODEL-CHECKING that are not automatic, and not even higher-order automatic. It suffices for instance to take any non-standard model of Presburger arithmetic of cardinality strictly bigger than the continuum, which must exist by upward Löwenheim–Skolem theorem.[70] The same argument works to show that "elementary equivalence" does not preserve the notion of automaticity.

[69] Recall on the other hand that $FO^{fin} \subseteq L$ (Proposition VII.2.3).

*Order-Invariant First-Order Logic.* An *order-invariant first-order formula*[71] over $\sigma$ is any first-order formula[72] $\phi$ over the signature $\sigma \sqcup \{\leq\}$ *s.t.* for any $\sigma$-structure $\mathbf{A}$, for any *total orders* $\leq_1, \leq_2$ over $A$, we have $\langle \mathbf{A}, \leq_1 \rangle \vDash \phi$ *iff* $\langle \mathbf{A}, \leq_2 \rangle \vDash \phi$. In this case, we say that $\mathbf{A}$ models $\phi$. Similarly, we define $\omega$-*order-invariant first-order formulas* [Rub08, § 3.2] by restricting the total orders to have order-type $\omega$—*i.e.* to be isomorphic to $\langle \mathbb{N}, \leq \rangle$—or $|A|$ when $|A|$ is finite.

For instance, the formula

$$\forall x. \exists y. \, x \leq y \wedge x \neq y$$

is $\omega$-order-invariant: it expresses the fact that the model is infinite. However, it is not order-invariant.[73] *A priori*, it is non-trivial to check if an order-invariant first-order formula (resp. $\omega$-order-invariant first-order formula) holds in an automatic $\sigma$-structure: how should this order be interpreted? By order-invariance, any total order will do, and moreover by Example VII.2.5, the length-lexicographic order is always automatic, and has the required order-type!

**Proposition VII.3.11** ([Rub08, § 3.2])**.** Model-checking of order-invariant first-order formulas (resp. $\omega$-order-invariant first-order formulas) over automatic structures is decidable.

A restricted form of this result was originally proved by Blumensath and Grädel [BG04, Corollary 5.4] for the extension of first-order logic with the quantifier "there are infinitely many". Similarly, one can add counting quantifiers of the form "the number of $x$'s *s.t.* $\phi(x)$ holds is congruent to $k$ mod $n$" [Rub08, § 3.2] while preserving decidability.

Note however that it is undecidable whether a first-order formula is order-invariant or $\omega$-order-invariant, be it on finite structures or on all structures. This follows from the undecidability of first-order logic, see *e.g.* [Grä07, Exercise 3.1.12].[74]

*Interpretations.* One way of rephrasing—and proving—Hodgson's theorem would be to first observe that $\mathbf{2}^*$ has decidability first-order theory, and then than this property is preserved under first-order interpretations. The result would follow since all automatic structures are first-order interpretations of $\mathbf{2}^*$ by Proposition VII.3.4. Colcombet and Löding proposed in [CL07] an alternative vision by showing that:

- $\langle \mathbb{N}, \text{succ} \rangle$ has a decidable weak monadic second-order logic, where succ denotes the successor relation,[75]

- if a structure $\mathbf{U}$ has a decidable weak monadic second-order theory, then any structure $\mathbf{A}$ that admits a "weak monadic second-order interpretation"[76] from $\mathbf{U}$ has a decidable first-order theory [CL07, Corollary 2.5];

- automatic structures are exactly the structures that can be obtained from $\langle \mathbb{N}, \text{succ} \rangle$ using WMSO-interpretations [CL07, Proposition 3.1].

[70] These models form a supclass of the non-standard models of Peano's arithmetic, see [Hod93, Ex. 2, p. 36 & § 11.4].

[71] This is a standard notion in model theory, see *e.g.* [Grä07, Exercise 3.1.12].

[72] For the sake of simplicity we give all definitions for sentences, but they easily be generalized to handle free variables.

[73] Indeed, $\langle \mathbb{N}, \leq \rangle$ satisfies the formula but not $\langle \mathbb{N}, \geq \rangle$.

[74] Hence, Proposition VII.3.11 should be understood as a promise problem: we can decide the problem if we are given the promise that the input is order-invariant (resp. $\omega$-order-invariant).

### VII.3.3  Problems on Automatic Structures

From the notion of automaticity two questions naturally arise:

1. What are the *structural properties* of automatic structures? For instance, we have seen that all automatic structures are countable. This question has been somewhat extensively studied for algebraic structures.

2. Given a *decidable decision problem* on finite structures, is its generalization to automatic structures still decidable?

> **ISOMORPHISM PROBLEM FOR AUTOMATIC STRUCTURES**
> *Input*:  Two automatic presentations $\mathcal{A}$ and $\mathcal{B}$.
> *Question*:  Is $\mathbf{A}$ isomorphic to $\mathbf{B}$?

Blumensath and Grädel proved this problem to be undecidable [BG04, Theorem 5.15]. The problem was later shown to be complete for the first level of the analytical hierarchy [KNRS07, Theorem 5.9].

*Automatic Ordinals.*  Automatic ordinals are quite simple: using *e.g.* Cantor's normal form, it is straightforward to prove that any ordinal strictly smaller than $\omega^\omega$ is automatic.[77] Delhommé proved the converse property to be true: an ordinal is automatic *iff* it is strictly smaller than $\omega^\omega$ [Del04, Corollaire 2.2]. Using again Cantor's normal form, Khoussainov, Rubin & Stephan proved that the isomorphism problem is decidable for automatic ordinals [KRS05, Theorem 5.3]. These results have been generalized first to $\omega$-tree-automatic ordinals by Finkel and Todorčević [FT13] and then to tree-automatic ordinals which are moreover equipped with addition, by Jain, Khoussainov, Schlicht and Stephan [JKSS19].

*Automatic Boolean Algebras.*  Recall that a Boolean algebra can be either seen as a partially ordered set following a set of axioms ensuring among other the existence of a join, a meet and a negation, or a set equipped these three operations, together with a minimal and maximal element. Both definitions are actually first-order equivalent and so considering one or the other does not change the notion of automaticity.

The ISOMORPHISM PROBLEM for automatic Boolean algebras is decidable [KNRS07, Corollary 3.5] essentially because there are very few automatic

[75] This is *the* classical result of automata theory and the motivation for studying $\omega$-automata. Recall that weak monadic second-order logic consists of monadic second-order logic where monadic quantifiers are restricted to finite sets.

[76] This is the equivalent of FO-interpretation for weak monadic second-order logic. In particular, finite subsets elements of the domain of $\mathbf{U}$ are interpreted as elements of the new structure. See [CL07, § 2.3] for a formal definition.

[77] We see an ordinal as a structure with a binary relation describing its order.

Boolean algebras [KNRS07, Theorem 3.4]. However, when going to $\omega$-tree-automaticity, not only is the ISOMORPHISM PROBLEM undecidable, but Finkel and Todorčević exhibited two somewhat simple-looking $\omega$-tree-automatic Boolean algebras for which whether they were isomorphic is independent from ZFC [FT10, Theorem 6.1].

*Automatic Groups.*  The literature on automatic groups is remarkably extensive. Their study was introduced in the late 1980s, by showing that they have a solvable word problem [Eps92, Theorem 2.1.9].[78] Typical examples of infinite automatic group are the braid group [Eps92, Theorem 9.3.1] and the finitely-generated free groups. We refer the reader to [Ree22] for a recent and detailed account on the history of the development of the theory of automatic groups. Remarkably, despite being a very active research area, the decidability of the ISOMORPHISM PROBLEM remains open.

**Open Problem VII.3.12.** Is the ISOMORPHISM PROBLEM decidable for automatic groups?

*Automatic Semigroups.*  Following the success of automatic groups, automatic semigroups have been studied by Campbell, Robertson, Ruškuc and Thomas, who showed that any finitely generated subsemigroup of a free semigroup is automatic [CRRT01, Theorem 8.1].

*Automatic Rings and Fields.*  Richer algebraic structures, like rings or fields actually often fail to be captured by the notion of automaticity. For instance, $\langle \mathbb{N}, +, \cdot, 0, 1 \rangle$ (Peano's arithmetic) is not automatic. This essentially follows from Proposition VII.1.8, applied to multiplication, together with a counting argument, see [Blu24, Corollary XII.8.11]. Khoussainov, Nies, Rubin & Stephan moreover proved that no infinite field[79] can be automatic [JKSS19, Theorem 3.10 & Corollary 3.11].[80]

### VII.3.4   Automatic Graphs

Chapter VIII will mostly focus on automatic graphs, in which we will study the question of colourability and the homomorphism problem. One interesting source of undecidability for automatic graphs comes from the following construction.

**Example VII.3.13.** Consider a Turing machine $\mathcal{T} = \langle Q, \Gamma, \delta, q_0, \mathrm{Acc} \rangle$, where $Q$ is the set of states, $\Gamma$ is tape alphabet,

$$\delta \colon (Q \setminus \mathrm{Acc}) \times \Gamma_{\_} \to \mathfrak{P}(Q \times \Gamma \times \{\leftarrow, \downarrow, \rightarrow\})$$

is the transition relation, $\Gamma_{\_} \mathrel{\hat{=}} \Gamma \sqcup \{\_\}$, and $q_0$ and Acc are the initial and set of final states, respectively. We represent a configuration $\langle u, q, v \rangle$ by the word $uqv \in \Gamma^* Q \Gamma^*$: in light of this, we will henceforth denote by "configuration"

[78] Actually this last result is proved on the slightly bigger class of "regularly generated groups".

[79] In fact their result also holds for integral domains.

[80] Note that this contrasts with Tarski's result that the first-order theory of $\langle \mathbb{R}, +, \cdot, 0, 1 \rangle$, and more generally of any real closed fields is decidable since it admits quantifier elimination [Hod93, Theorem 8.4.4].

any string from the set $\text{Conf}_{\mathcal{T}} \triangleq \Gamma^* Q \Gamma^*$.[81] The *configuration graph* of $\mathcal{T}$ is the infinite graph $\mathscr{C}onf_{\mathcal{T}}$ having $\text{Conf}_{\mathcal{T}}$ as set of vertices and an edge from $\gamma$ to $\gamma'$, denoted $\gamma \to \gamma'$ if there is a one-step transition from $\gamma$ to $\gamma'$ in $\mathcal{T}$. The configuration graph $\mathscr{C}onf_{\mathcal{T}}$ of any Turing machine $\mathcal{T}$ is an effective automatic graph. ◇

As a consequence, by reduction from the halting problem of a universal Turing machine, we obtain the following result.

**Proposition VII.3.14.** There exists a fixed automatic graph $\mathbf{G}$ over the alphabet $\mathbb{2}$ *s.t.* the problem of whether, given two words $u$ and $v \in \mathbb{2}^*$, there is a path from $u$ to $v$ in $\mathbf{G}$ is undecidable.

Problems on automatic graphs have been mainly studied by Kuske and Lohrey, who studied the following problems over automatic graphs.

- *Highly undecidable problems:* the existence of a Hamiltonian path is undecidable, in fact it is complete for the first level of the *analytical* hierarchy [KL10, Theorem 3.2]; the existence of an infinite path in directed trees shares the same complexity [KL10, Theorem 3.6]; and so does the ISOMORPHISM PROBLEM [KNRS07, Theorem 5.9].
- *Moderately undecidable problems:* the existence of an Eulerian path is undecidable, but is only complete for the second level of the *arithmetical* hierarchy $\Pi_2^0$ [KL10, Theorem 4.13].

However, as a consequence of Proposition VII.3.11, the fact that the graph contains infinitely many edges—or isolated vertices, or any first-order definable property—, is decidable over automatic graphs as it can be expressed by an $\omega$-order-invariant property.

Some slightly more involved arguments, involving "Ramsey quantifiers", can show that whether the graph contains an infinite clique (or transitive tournament) is decidable [KL10, Corollary 5.5], see also [Rub08, Theorem 3.20].

**Proposition VII.3.15** ([Köc14, Proposition 6.5]). Whether an automatic graph is 2-colourable, or equivalently bipartite, is undecidable. More precisely it is coRE-complete.

For another survey on automatic structures, see [Grä20]. Lastly, we want to note that *recursive structures* have been extensively studied since the late XXth century: they are defined analogously to automatic structures, but synchronous automata are replaced by Turing machines. Unsurprisingly, all non-trivial problems are undecidable: hence, from a computability perspective, the main question is to characterize the Turing degree of these problems. The extent of the literature on this topic is too vast to be summarized here: we refer the reader to the two-volume handbook [EGNRM98a; EGNRM98b].

# A Dichotomy Theorem for Automatic Structures

## ABSTRACT

We study the separation problem of automatic relations, *a.k.a.* automatic relations, by recognizable relations—namely finite unions of Cartesian products of regular languages. We prove it to be computationally equivalent to the FINITE REGULAR COLOURABILITY OF AUTOMATIC GRAPHS, that takes an automatic presentation of a graph as input, and asks whether it admits a regular colouring—meaning that for each colour, the set of words representing the elements having this colour is a regular language—with finitely many colours.

While the decidability of this problem remains open, we first show that, if the number of colours is fixed to be any natural number $k \geq 2$, then this problem is undecidable. This implies the undecidability of the separation problem of automatic relations by recognizable relations where the number of unions allowed is bounded.

We then generalize this result, and prove a dichotomy theorem for automatic structures: for any finite relational structure **B**, the problem of whether an automatic structure admits a homomorphism to **B** is either decidable in NL, or is undecidable. We extend these results to regular homomorphisms, for which we require the homomorphisms to be regular, in the sense that every preimage of any element of the target structure **B** must be a regular language. In both cases, structures for which the problem is decidable exactly correspond to those with "finite duality".

## ACKNOWLEDGEMENTS

Section VIII.3 corresponds to [BFM23, §§ 3–6], which has been published at MFCS '23, and is a joint work with Pablo Barceló and Diego Figueira. Parts of Section VIII.1 also come from [BFM23].

Part of the dichotomy theorem, corresponding to Sections VIII.4.1, VIII.4.2 and VIII.5.2 and parts of Section VIII.5.3, was proven by Antoine Cuvelier during an internship I supervised in summer 2024.

We thank Joanna Fijalkow for helpful discussions on constraint satisfaction problems, and Edgar Baucher for pointing us to Mycielski's construction.

# Contents

## VIII.1   Introduction

### VIII.1.1   Classes of Relations

The complex landscape of rationality for finite-word relations described in Section VII.1 yields a natural question of *effectivity*: given a relation in some class, can we decide if it belongs to a given subclass?

Prior work has focused on the REC-MEMBERSHIP PROBLEM, which takes as input an $n$-ary rational relation $\mathcal{R}$ and asks whether it is equivalent to a recognizable relation $\bigcup_i L_{i,1} \times \cdots \times L_{i,n}$, where each $L_{i,j}$ is a regular language. Intuitively, the problem asks whether the different components of the rational relation $\mathcal{R}$ are almost independent of one another. The study of REC-MEMBERSHIP is relevant since relations enjoying this property are often amenable to some analysis including, *e.g.*, abstract interpretations in program verification, variable elimination in constraint logic programming, and query processing over constraint databases—see [BHLLN19, Introduction] for a more thorough discussion on this topic.

As mentioned in Section VII.1.5, in general, REC-MEMBERSHIP of rational relations is undecidable [Ber79, § III, Theorem 8.4], but it becomes decidable for the subclass of deterministic rational relations, which extends the class of automatic relations. For automatic relations, the decidability of REC-MEMBERSHIP can be obtained by a simple reduction to the problem of checking whether a finite automaton recognizes an infinite language [LS19]—which is decidable via a standard reachability argument. The precise complexity of the problem, however, was only recently pinned down. By applying techniques based on Ramsey Theorem over infinite graphs, it was shown that REC-MEMBERSHIP of automatic relations is PSpace-complete when relations are specified by non-deterministic automata [BHLLN19, Theorem 1] [BGLZ22, Corollary 2.9].

A natural generalization of this question is the RAT/REC-SEPARABILITY PROBLEM, which takes two $n$-ary rational relations $\mathcal{R}, \mathcal{R}' \subseteq \Sigma^* \times \Sigma^*$ and checks whether there is a recognizable relation $\mathcal{S} = \bigcup_i L_{i,1} \times \cdots \times L_{i,n}$ that separates $\mathcal{R}$ from $\mathcal{R}'$, meaning that $\mathcal{R} \subseteq \mathcal{S}$ and $\mathcal{R}' \cap \mathcal{S} = \varnothing$, in which case we say that $\mathcal{R}$ and $\mathcal{R}'$ are *separable* by a recognizable relation. In other words, this problem asks whether we can *overapproximate* $\mathcal{R}$ with a recognizable relation $\mathcal{S}$ that is constrained not to intersect with $\mathcal{R}'$. Separability problems of this kind abound in theoretical computer science, in particular in formal language theory where they have gained a lot of attention over the last few years, in a large variety of settings—see *e.g.* [PZ16] for first-order definable languages over finite and $\omega$-words, [CGM22] for their extension to countable ordinals, [CMRZZ17] for piecewise languages, [Kop16] for separability question over pushdown languages, or [CCLP17] for Parikh automata.

In this chapter, we focus on the following problem:

---

**AUT/REC-SEPARABILITY PROBLEM**

*Input:*   Two automatic relations $\mathcal{R}$ and $\mathcal{R}'$.

*Question:*   Does there exist a recognizable relation $\mathcal{S}$ s.t. $\mathcal{R} \subseteq \mathcal{S}$ and $\mathcal{R}' \cap \mathcal{S} = \varnothing$?

---

Notice that when $\mathcal{R}'$ is the complement of $\mathcal{R}$ this problem boils down to REC-MEMBERSHIP on automatic relations—which is decidable. However, AUT/REC-SEPARABILITY is more general than this problem, and to this day it is unknown whether it is decidable.

## VIII.1.2   Constraint Satisfaction Problems

Our work on the AUT/REC-SEPARABILITY PROBLEM will eventually lead up to the study of HOMOMORPHISM PROBLEMS where the target structure is fixed. These problems are the central topic in the domain of constraint satisfaction problems (CSPs).[1] When the target structure is finite, the HOMOMORPHISM PROBLEM is known to be decidable in NP. The precise complexity of the problem, however, depends on the structure of the target, and is tightly connected to the algebraic properties of the algebra associated to the structure.

A central result in the domain, originally conjectured by Feder and Vardi, and known as the "dichotomy theorem" (formerly "dichotomy conjecture"), states that any CSP whose target structure is finite is either in P or NP-complete [FV98, § 2, "Dichotomy question"],[2] and the characterization of structures for which the problem was in P is precisely formulated in algebraic terms.[3] This conjecture extended a result of Schaefer, who proved two decades earlier that CSPs over the Boolean domain[4] was either P or NP-complete [Sch78, Theorem 2.1]. Twenty-four years later, this conjecture was proven independently by Bulatov [Bul17, Theorem 1] and Zhuk [Zhu20, Theorem 1.4]. This algebraic approach is particularly fruitful but lies beyond the scope of this thesis: we refer the reader to [BKW17; Lar17] for surveys on the topic.

Beyond P and NP, the complexity of the homomorphism problem can reach some surprisingly low complexities. Even for Boolean CSPs[5], Allender, Bauland, Immerman, Schnoor and Vollmer extended Schaefer's theorem to prove that every not-so-easy problem—meaning that it is not solvable in coNLogTime—is complete for one class among NP, P, $\oplus$L, NL or L under $\text{AC}^0$-reductions [ABISV09, Theorem 3.1].

For non-Boolean CSPs, the landscape becomes even more complex. For instance, given $k \in \mathbb{N}$, we define the *$k$-transitive tournament*, denoted by $\mathbf{T}_k$ and illustrated on Figure VIII.1, to be the directed graph with vertices $0, 1, \dots, k$, and with an edge from $i$ to $j$ iff $i < j$. Similarly, the *$k$-path* $\mathbf{P}_k$ has $[\![0, k]\!]$ as its set of vertices, with an edge from $i$ to $j$ iff $j = i + 1$—see Figure VIII.1.

Then, a graph $\mathbf{G}$ admits a homomorphism to $\mathbf{T}_k$ iff it has no directed path of size $k + 1$, in the sense that this is no homomorphism from $\mathbf{P}_{k+1}$ to $\mathbf{G}$. Symbolically:

$$\forall \mathbf{G}, \quad \mathbf{P}_{k+1} \xrightarrow{\text{hom}} \mathbf{G} \quad \textit{iff} \quad \mathbf{G} \xrightarrow{\text{hom}} \mathbf{T}_k.$$

[1] A typical example of such a problem is to work on the graph signature, and to fix the target structure to be the $k$-clique ($k \in \mathbb{N}_{>0}$): it exactly corresponds to the $k$-COLOURABILITY PROBLEM.

[2] In other words, there is no NP-intermediate problem amongst these problems.

[3] Structures that admit a CSP problem in P are exactly those whose algebra admits a "weak near-unanimity operation"—see *e.g.* [Zhu20, § 1, p. 3].

[4] Meaning that the target structure can have only two elements—but recall that the signature $\sigma$ can be arbitrary complex.

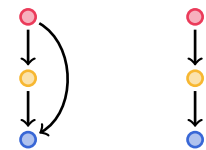

Figure VIII.1: The 2-transitive tournament $\mathbf{T}_2$ (left-hand side) and the 2-path $\mathbf{P}_2$ (right-hand side).

[5] Meaning again that the target structure has only two elements

As a result, to decide if, for a finite graph **G**, we have $\mathbf{G} \xrightarrow{\text{hom}} \mathbf{T}_k$, it suffices to test if $\mathbf{P}_{k+1} \xrightarrow{\text{hom}} \mathbf{G}$, which can be done in $\text{FO}^{\text{fin}}$!

Structures sharing the same property as $\mathbf{T}_k$, meaning that the existence of a homomorphism to them amounts to not containing any obstructions among a fixed finite set of obstructions are said to have finite duality. Like $\mathbf{T}_k$, their constraint satisfaction problem is in $\text{FO}^{\text{fin}}$.[6] Atserias proved the converse implication to this result [Ats08, Corollary 4].[7]

**Proposition VIII.1.1** (*Atserias' theorem*). Let **B** be a finite $\sigma$-structure. Then **B** has finite duality if, and only if, $\mathscr{H}om(\text{All}, \mathbf{B})$ is first-order definable.

Moreover, Larose and Tesson proved a dichotomy theorem for CSPs with small complexity.

**Proposition VIII.1.2** (*Larose-Tesson theorem*). Let **B** be a finite $\sigma$-structure. If **B** does not have finite duality, then $\mathscr{H}om(\text{Fin}, \mathbf{B})$ is L-hard under first-order reductions.

Observe that this last proposition only allows for finite structures on the source side of the problem. This chapter will generalize some results—including the latter—on the HOMOMORPHISM PROBLEM from finite structures to automatic structures. The study of CSPs beyond finite structures is far from a new topic, and has been the subject of many works in the past decade: see for instance [KKOT15] for CSPs over "nominal structures", *i.e.* structures with "atoms", or [KLOT16] for CSPs over $\sigma$-structures which are first-order definable over $\langle \mathbb{N}, = \rangle$.

### VIII.1.3  Contributions & Organization

In Section VIII.3, we start by rephrasing the AUT/REC-SEPARABILITY PROBLEM in graph-theoretic terms.

| FINITE REGULAR COLOURABILITY OF AUTOMATIC GRAPHS | |
|---|---|
| *Input*: | A presentation $\mathcal{G}$ of an automatic graph **G**. |
| *Question*: | Does $\mathcal{G}$ admit a regular colouring with finitely many colours? |

A regular colouring with finitely many colours (resp. a regular $k$-colouring) is a partition of the vertices of $\mathcal{G}$ into finitely many (resp. $k$) regular languages *s.t.* adjacent vertices cannot belong to the same set. Similarly, the REGULAR $k$-COLOURABILITY PROBLEM asks instead if $\mathcal{G}$ admits a regular $k$-colouring, in which case we say that $\mathcal{G}$ is *regularly $k$-colourable*.

**Theorem VIII.3.2.** There are polynomial-time reductions:

1. from AUT/REC-SEPARABILITY to FINITE REGULAR COLOURABILITY,
2. from FINITE REGULAR COLOURABILITY to AUT/REC-SEPARABILITY, and
3. from REGULAR $k$-COLOURABILITY to AUT/$k$-REC-SEPARABILITY, for every $k > 0$.

Further, the last two reductions are so that the second relation in the instance

[6] Recall that $\text{FO}^{\text{fin}} \subseteq \text{L}$, see Proposition VII.2.3.

[7] This result was followed in the same year by Rossman's theorem, that subsumes it.

of the separability problem is the identity $\mathcal{I}d$.

While we do not know how to solve the FINITE REGULAR COLOURABILITY, we prove the problem to be undecidable if the number of colours is fixed.

**Theorem VIII.3.7.** The REGULAR $k$-COLOURABILITY PROBLEM on automatic graphs is undecidable for every $k \geq 2$. More precisely, the problem is RE-complete. This holds also for connected automatic graphs.

Then, in Section VIII.3.3, we build on this result to prove the undecidability of some separability problems for automatic relations (Corollary VIII.3.10 and Proposition VIII.3.15).

Observing that regular $k$-colourability can be rephrased as the existence of a regular homomorphism—meaning, in this context, a homomorphism whose preimages are all regular languages—to the $k$-clique,[8] this motivates us to study the HOMOMORPHISM PROBLEM and REGULAR HOMOMORPHISM PROBLEM for automatic structures in Sections VIII.4 and VIII.5. Our main result is a dichotomy theorem for automatic structures.[9]

**Theorem VIII.4.1** (Dichotomy Theorem for Automatic Structures)**.** Let **B** be a finite $\sigma$-structure. The following are equivalent:

**(DT)$_{\mathsf{fin\text{-}dual}}$.** **B** has finite duality;

**(DT)$_{\mathsf{hom\text{-}dec}}$.** $\mathcal{H}om(\mathrm{Aut}, \mathbf{B})$ is decidable;

**(DT)$_{\mathsf{hom\text{-}reg\text{-}dec}}$.** $\mathcal{H}om^{\mathrm{reg}}(\mathrm{Aut}, \mathbf{B})$ is decidable;

**(DT)$_{\mathsf{equal}}$.** $\mathcal{H}om(\mathrm{Aut}, \mathbf{B}) = \mathcal{H}om^{\mathrm{reg}}(\mathrm{Aut}, \mathbf{B})$, *i.e.* for any automatic presentation $\mathcal{A}$ of a $\sigma$-structure **A**, there is a homomorphism from **A** to **B** *iff* there is a regular homomorphism from $\mathcal{A}$ to **B**;

**(DT)$_{\mathsf{first\text{-}order}}$.** $\mathcal{H}om(\mathrm{All}, \mathbf{B})$ has uniformly first-order definable homomorphisms.[10]

Moreover, when $\mathcal{H}om(\mathrm{Aut}, \mathbf{B})$ and $\mathcal{H}om^{\mathrm{reg}}(\mathrm{Aut}, \mathbf{B})$ are undecidable, they are coRE-complete and RE-complete, respectively. When they are decidable, they are NL.

The easy implications of this theorem are proven in Section VIII.4.1, and the undecidability results—one for homomorphisms, and another one for regular homomorphisms in the rest of Section VIII.4: both reductions for undecidability generalize Larose and Tesson's proof. In the case of regular homomorphisms, the full sequence of reductions to prove undecidability goes as follows:

- from the HALTING PROBLEM on deterministic reversible Turing machines to the REGULAR REACHABILITY PROBLEM for linear Turing machines;

- which is reduced in turn to REGULAR UNCONNECTIVITY IN AUTOMATIC GRAPHS;

- and lastly, this latter problem is reduced to $\mathcal{H}om^{\mathrm{reg}}(\mathrm{Aut}, \mathbf{B})$ when **B** does not have finite duality—this reduction is the adaptation of Larose and Tesson's proof.

To our knowledge, none of these problems—except the HALTING PROBLEM—have been studied before. Decidability results are given in Section VIII.5: in fact, the decidability of $\mathscr{H}om(\text{Aut}, \mathbf{B})$ when $\mathbf{B}$ has finite duality is trivial, so this section is dedicated to the problem $\mathscr{H}om^{\text{reg}}(\text{Aut}, \mathbf{B})$. In Section VIII.5.1 we give a succinct logic-based proof of the decidability of $\mathscr{H}om(\text{Aut}, \mathbf{B})$, however the proof is somewhat *abstract*. In Sections VIII.5.2 and VIII.5.3 we give a more visual solution, known as hyperedge consistency algorithm. It generalizes the eponymous algorithm for finite structures, that captures the HOMOMORPHISM PROBLEM when the target structure has tree duality—which is a supclass of finite duality. However, the proof of correctness of our algorithm is non-trivial and relies on providing a fine understanding of the behaviour of the algorithm for finite structures in the special case of target structures that have finite duality. We conclude in Section VIII.6, by discussing conjectures and related problems.

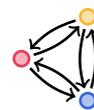

Figure VIII.2: The 3-clique $\mathbf{K}_3$. (Replica of Figure I.19.)

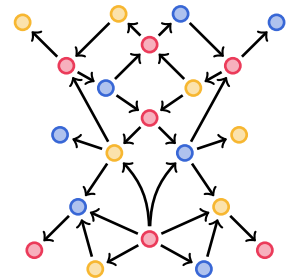

Figure VIII.3: A 3-colouring of some beetle-shaped graph. (Replica of Figure I.20.)

[8] See Figures VIII.2 and VIII.3.

[9] Note that while for finite structures the dichotomy theorems either separate $\mathrm{FO}^{\mathrm{fin}}$ from L—in the case of Larose and Tesson—, or P from NP—for Bulatov and Zhuk—, in our case the dichotomy is between NL and... undecidability!

[10] The notion of uniformly first-order definable homomorphisms is defined in Section VIII.5.1.

## VIII.2  Preliminaries

### VIII.2.1  Regular Homomorphisms

Given two regular languages $K$ and $L$,[11] a *regular function* from $K$ to $L$ is a function $f\colon K \to L$ *s.t.* the relation

$$\{\langle u, f(u)\rangle \mid u \in K\}$$

is automatic. A *regular homomorphism* between two presentations of automatic $\sigma$-structures $\mathcal{A}$ and $\mathcal{B}$ is a regular function from $\text{dom}_{\mathcal{A}}$ to $\text{dom}_{\mathcal{B}}$ that defines a homomorphism from $\mathbf{A}$ to $\mathbf{B}$.[12,13,14] We denote by $\mathcal{A} \xrightarrow{\text{reg hom}} \mathcal{B}$ the existence of a regular homomorphism from $\mathcal{A}$ to $\mathcal{B}$.

**Property VIII.2.1.** Let $f\colon K \to L$ be a function, where $L$ is finite. Then $f$ is a regular function *iff* for every $v \in L$, $f^{-1}[v]$ is a regular language.

*Proof.* For the left-to-right implication, if $f$ is a regular function, then there exists a first-order formula $\phi(x,y)$ *s.t.* for all $\langle u, v \rangle \in K \times L$, then

$$\Sigma^*, u, v \vDash \phi(x,y) \text{ iff } v = f(u).$$

Then given $v \in L$, the formula[15]

$$\exists y,\, \phi(x,y) \wedge y = v$$

describes $\{u \in K \mid f(u) = v\}$, which is hence regular.

Conversely, we consider first-order formulas $\phi_v$ describing each set $\{u \in K \mid f(u) = v\}$, with $v \in L$. Then

$$\phi(x) \triangleq \bigwedge_{v \in L} \phi_b(x) \wedge y = v$$

is a first-order formula—since $L$ is finite—describing $f$.  □

As a consequence, when $\mathbf{B}$ is finite, the existence of a regular homo-morphism to **B** is independent of its presentation, and we will hence write $\mathcal{A} \xrightarrow{\text{reg hom}} \mathbf{B}$ to mean that $\mathcal{A} \xrightarrow{\text{reg hom}} \mathcal{B}$. In particular, a *regular $k$-colouring* of a presentation of an automatic graph $\mathcal{G}$ is equivalently a regular homomorphism from $\mathcal{G}$ to $\mathbf{K}_k$, or a $k$-colouring of **G** *s.t.* for any colour, the set of words of $\text{dom}_{\mathcal{G}}$ sharing this colour is a regular language. Unsurprisingly, this property cannot be extended to all automatic structures: for instance letting $\mathcal{A}$ be the automatic presentation $\langle 0^*, \{\langle 0^n, 0^{n+1} \mid n \in \mathbb{N}\rangle\}\rangle$ of $\langle \mathbb{N}, \text{succ}\rangle$, and letting $\mathcal{A}'$ be the automatic presentation $\langle (00)^*, \{\langle 0^{2n}, 0^{2n+2} \mid n \in \mathbb{N}\rangle\}\rangle$ of the same structure, it can be shown that $\mathcal{A} \xrightarrow{\text{reg hom}} \mathcal{A}$ but $\mathcal{A} \xrightarrow{\text{reg hom}} \mathcal{A}'$.

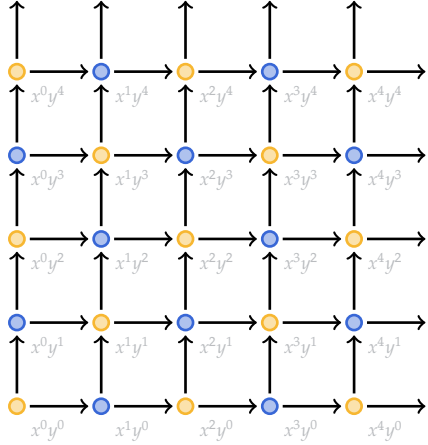

Figure VIII.4: A regular 2-colouring of an automatic presentation of the infinite quarter-grid.

In Figure VIII.4, we provide an example of a regular 2-colouring: we let $\mathcal{G} = \langle V, \mathcal{E}\rangle$ be the automatic presentation of the infinite quarter-grid, defined over the alphabet $\{x, y\}$ by $V \mathrel{\hat{=}} x^* y^*$ and

$$\mathcal{E} = \{\langle x^p y^q, x^{p+1} y^q\rangle \mid p, q \in \mathbb{N}\} \cup \{\langle x^p y^q, x^p y^{q+1}\rangle \mid p, q \in \mathbb{N}\}.$$

Then the unique 2-colouring of $\mathcal{G}$ assigns one colour[16] to the vertices of the form $x^p y^q$ *s.t.* $p - q$ is even. This colouring is of course regular.

[16] In yellow in Figure VIII.4.

Given a fixed signature $\sigma$ and a $\sigma$-structure **B**, we denote by:

- $\mathcal{H}om(\text{Fin}, \mathbf{B})$ (resp. $\mathcal{H}om(\text{Aut}, \mathbf{B})$, resp. $\mathcal{H}om(\text{All}, \mathbf{B})$) the class of all finite $\sigma$-structure (resp. automatic $\sigma$-structure, resp. arbitrary $\sigma$-structures) that admit a homomorphism to **B**,
- $\mathcal{H}om^{\text{reg}}(\text{Aut}, \mathbf{B})$ is the class of all automatic presentations of $\sigma$-structures that admit a regular homomorphism to **B**.

Somewhat abusively, we identify these classes with the associated decision problems—except for $\mathcal{H}om(\text{All}, \mathbf{B})$ since arbitrary $\sigma$-structures cannot be encoded as finite strings. For finite structures, we assume the input to be given using adjacency lists, and for automatic structures, we assume the input to be described by an automatic presentation. We call $\mathcal{H}om^{\text{reg}}(\text{Aut}, \mathbf{B})$ the REGULAR HOMOMORPHISM PROBLEM over **B**.

### VIII.2.2 Constructions on Structures

Given two structures **A** and **B**, we define the structure $\mathbf{B^A}$ as follows:

- its domain are homomorphisms $\mathbf{A} \to \mathbf{B}$,
- for every predicate $\mathcal{R}$ of arity $k$, for any homomorphism $f_1, \ldots, f_k$, we have $\langle f_1, \ldots, f_k \rangle \in \mathcal{R}(\mathbf{B^A})$ when

$$\text{for every } \langle a_1, \ldots, a_k \rangle \in \mathcal{R}(\mathbf{A}), \text{ we have } \langle f_1(a_1), \ldots, f_k(a_k) \rangle \in \mathcal{R}(\mathbf{B}).$$

**Proposition VIII.2.2** (Folklore: Currying Homomorphisms)**.** Given structures **A**, **B** and **C**, if $f \colon \mathbf{A} \times \mathbf{B} \to \mathbf{C}$ is a homomorphism, then $F \colon \mathbf{A} \to \mathbf{C^B}$, defined by $a \mapsto (b \mapsto f(a, b))$, is a homomorphism. In fact, this mapping $f \mapsto F$ is a bijection between homomorphisms $\mathbf{A} \times \mathbf{B} \to \mathbf{C}$ and homomorphisms $\mathbf{A} \to \mathbf{C^B}$.

*Proof.* Let $\mathcal{R}$ be a predicate of arity $k$, and let $\langle a_1, \ldots, a_k \rangle \in \mathcal{R}(\mathbf{A})$. We want to show that $\langle F(a_1), \ldots, F(a_k) \rangle \in \mathcal{R}(\mathbf{C^B})$: for any $\langle b_1, \ldots, b_k \rangle \in \mathcal{R}(\mathbf{B})$, we have

$$\langle F(a_1)(b_1), \ldots, F(a_k)(b_k) \rangle = \langle f(a_1, b_1), \ldots, f(a_k, b_k) \rangle \in \mathcal{R}(\mathbf{C})$$

since $f$ is a homomorphism from $\mathbf{A} \times \mathbf{B}$ to **C**. Hence, $F$ is indeed a homomorphism from **A** to $\mathbf{C^B}$.

Dually, if $F$ is a homomorphism from **A** to $\mathbf{C^B}$, we define $f \colon \mathbf{A} \times \mathbf{B} \to \mathbf{C}$ by $\langle a, b \rangle \mapsto F(a)(b)$, and claim that $f$ is a homomorphism. Indeed, if $\mathcal{R}$ be a predicate of arity $k$, for any $\langle a_1, \ldots, a_k \rangle \in \mathcal{R}(\mathbf{A})$ and $\langle b_1, \ldots, b_k \rangle \in \mathcal{R}(\mathbf{B})$, we have $\langle f(a_1, b_1), \ldots, f(a_k, b_k) \rangle = \langle F(a_1)(b_1), \ldots, F(a_k)(b_k) \rangle$. Since $\langle F(a_1), \ldots, F(a_k) \rangle \in \mathcal{R}(\mathbf{C^B})$ and $\langle b_1, \ldots, b_k \rangle \in \mathcal{R}(\mathbf{B})$ it follows that $\langle F(a_1)(b_1), \ldots, F(a_k)(b_k) \rangle \in \mathcal{R}(\mathbf{C})$. Therefore, $f$ is a homomorphism from $\mathbf{A} \times \mathbf{B}$ to **C**.

It is then routine to check that the maps $f \mapsto F$ and $F \mapsto f$ defined in the two previous paragraphs are mutually inverse bijections. □

### VIII.2.3 Constructions on Automatic Presentations

Let $\mathcal{A}$ and $\mathcal{B}$ be automatic presentations of some $\sigma$-structures **A** and **B**, over alphabets $\Sigma$ and $\Gamma$, respectively. We define $\mathcal{A} \times \mathcal{B}$ to be the presentation over the alphabet $(\Sigma \times \Gamma) \sqcup (\Sigma \times \{\_\}) \sqcup (\{\_\} \times \Gamma)$ such that:

$$\mathrm{dom}_{\mathcal{A} \times \mathcal{B}} \hat{=} \{u \otimes v \mid u \in \mathrm{dom}_{\mathcal{A}} \wedge v \in \mathrm{dom}_{\mathcal{B}}\}$$

$$\mathcal{R}_{\mathcal{A} \times \mathcal{B}} \hat{=} \{\langle u_1 \otimes v_1, \ldots, u_k \otimes v_k \rangle \mid \langle u_1, \ldots, u_k \rangle \in \mathcal{R}_{\mathcal{A}} \wedge \langle v_1, \ldots, v_k \rangle \in \mathcal{R}_{\mathcal{B}}\}$$

for each predicate $\mathcal{R}$ of arity $k$ in $\sigma$. It is an automatic presentation of $\mathbf{A} \times \mathbf{B}$. Indeed, given a first-order formula $\phi(x_1, \ldots, x_k)$ over $\sigma_{\Sigma}^{\mathrm{sync}}$, describing $\mathcal{R}_{\mathcal{A}}$, and a first-order formula $\psi(x_1, \ldots, x_k)$ over $\sigma_{\Sigma}^{\mathrm{sync}}$, describing $\mathcal{R}_{\mathcal{B}}$, we let $\phi^*$ (resp. $\psi^*$) be the formula obtained from $\phi$ (resp. $\psi$) by substituting $l_a(x)$ for $\bigvee_{b \in \Gamma \sqcup \{\_\}} l_{\langle a, b \rangle}(x)$ (resp. $\bigvee_{b \in \Sigma \sqcup \{\_\}} l_{\langle b, a \rangle}(x)$). Then $\phi^* \wedge \psi^*$ is a first-order formula, over the product alphabet, that describes $\mathcal{R}_{\mathcal{A} \times \mathcal{B}}$. The same construction works for $\mathrm{dom}_{\mathcal{A} \times \mathcal{B}}$. This shows that if **A** and **B** are automatic

$\sigma$-structures, then so is $\mathbf{A} \times \mathbf{B}$.[17]

**Proposition VIII.2.3.** Let $\mathbf{A}$, $\mathbf{B}$ and $\mathbf{C}$ be automatic $\sigma$-structures, such that $\mathbf{B}$ and $\mathbf{C}$ are finite. Let $\mathcal{A}$ (resp. $\mathcal{B}$ and $\mathcal{B}'$, resp. $\mathcal{C}$ and $\mathcal{C}'$) be an automatic presentation of $\mathbf{A}$ (resp. $\mathbf{B}$, resp. $\mathbf{C}$). Then $\mathcal{A} \times \mathcal{B} \xrightarrow{\text{reg hom}} \mathcal{C}$ iff $\mathcal{A} \times \mathcal{B}' \xrightarrow{\text{reg hom}} \mathcal{C}'$.

*Proof.* The proof follows from the following claim, which can be proven exactly in the same fashion as Property VIII.2.1.

*Claim* VIII.2.4. Assuming again that $\mathbf{B}$ and $\mathbf{C}$ are finite, a function $f\colon \mathcal{A} \times \mathcal{B} \to \mathcal{C}$ is a regular homomorphism *iff* for every $b \in \text{dom}_{\mathcal{B}}$, for every $c \in \text{dom}_{\mathcal{C}}$, $\{a \in \text{dom}_{\mathcal{A}} \mid f(a,b) = c\}$ is a regular language. □

In other words, the existence of a regular homomorphism does not depend on the automatic presentation of the *finite* structures that are involved, but only on the structure they represent. As a consequence of Proposition VIII.2.3, we write $\mathcal{A} \times \mathbf{B} \xrightarrow{\text{reg hom}} \mathbf{C}$ as a synonym for $\mathcal{A} \times \mathcal{B} \xrightarrow{\text{reg hom}} \mathcal{C}$.

**Corollary VIII.2.5** (Currying)**.** Let $\mathbf{A}$, $\mathbf{B}$ and $\mathbf{C}$ be automatic $\sigma$-structures, and let $\mathcal{A}$ be an automatic presentation of $\mathbf{A}$. Then $\mathcal{A} \times \mathbf{B} \xrightarrow{\text{reg hom}} \mathbf{C}$ iff $\mathcal{A} \xrightarrow{\text{reg hom}} \mathbf{C}^{\mathbf{B}}$.

*Proof.* This also follows from Claim VIII.2.4. □

## VIII.2.4 Idempotent Core

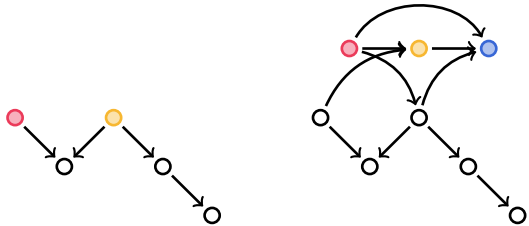

Figure VIII.5: Reduction from $\mathcal{H}om(\text{All}, \mathbf{B}^\dagger)$ to $\mathcal{H}om(\text{All}, \mathbf{B})$ when $\mathbf{B}$ is the 2-transitive tournament: we depict on the left-hand side, the $\sigma_{\mathbf{B}}$-structure $\mathbf{A} \notin \mathcal{H}om(\text{All}, \mathbf{T_2}^\dagger)$, and on the right-hand side, the $\sigma$-structure $\Phi(\mathbf{A}) \notin \mathcal{H}om(\text{All}, \mathbf{T_2})$ to which it is reduced. The interpretation of unary predicates in $\mathbf{A}$ are described using colours.

We fix a purely relational signature $\sigma$. Given a $\sigma$-structure $\mathbf{B}$, we denote by $\sigma_{\mathbf{B}}$ the signature obtained from $\sigma$ by adding a unary predicate $P_b$ for each $b \in B$. The *marked structure* $\mathbf{B}^\dagger$ of $\mathbf{B}$ is the $\sigma_{\mathbf{B}}$-structure obtained from $\mathbf{B}$ by interpreting each predicate $P_b$ as the singleton $\{b\}$.

**Proposition VIII.2.6** (Folklore)**.**[18,19] If $\mathbf{B}$ is a finite core, then the problems $\mathcal{H}om(\text{All}, \mathbf{B}^\dagger)$ and $\mathcal{H}om(\text{All}, \mathbf{B})$ are first-order equivalent. Moreover, this equivalence preserves finiteness, in the sense that finite structures are mapped to finite structures. Hence, by restricting this equivalence, we also obtain that $\mathcal{H}om(\text{Fin}, \mathbf{B}^\dagger)$ and $\mathcal{H}om(\text{Fin}, \mathbf{B})$ are first-order equivalent.

The non-easy part is to reduce $\mathcal{H}om(\text{Fin}, \mathbf{B}^\dagger)$ to $\mathcal{H}om(\text{Fin}, \mathbf{B})$: only this reduction requires the assumption that $\mathbf{B}$ is a core.

*Proof of Proposition VIII.2.6.* ↰ *Reduction from $\mathcal{H}om(\text{All}, \mathbf{B})$ to $\mathcal{H}om(\text{All}, \mathbf{B}^\dagger)$.* We reduce a $\sigma$-structure $\mathbf{A}$ to the $\sigma_{\mathbf{B}}$-structure $\mathbf{A}'$ obtained from $\mathbf{A}$ by inter-preting each predicate $P_b$ as the empty set. Clearly, a function from $A$ to $B$ is a homomorphism from $\mathbf{A}$ to $\mathbf{B}$ *iff* it is a homomorphism from $\mathbf{A}'$ to $\mathbf{B}^\dagger$, proving the correctness of the reduction. It is, by definition, first-order.

❧ *Reduction from* $\mathscr{H}om(\mathrm{All},\, \mathbf{B}^\dagger)$ *to* $\mathscr{H}om(\mathrm{All},\, \mathbf{B})$. We first define the reduction $\Phi$ and show its correctness; the fact that it is a first-order reduction is straightforward. We reduce a $\sigma_{\mathbf{B}}$-structure $\mathbf{A}$ to the $\sigma$-structure $\Phi(\mathbf{A})$ illustrated on Figure VIII.5 and defined as follows:

- its underlying universe is the disjoint union $A \sqcup B$,
- given a predicate $\mathscr{R}$ of arity $k$, its hyperedges are:
  - all $\mathscr{R}$-tuple of $\mathbf{A}$,
  - all $\mathscr{R}$-tuple of $\mathbf{B}$, and
  - all $\mathscr{R}$-tuple $\langle b_1, \ldots, b_{i-1}, a_i, b_{i+1}, \ldots, b_k\rangle$ *s.t.* there exists $b_i$ for which the $\mathscr{R}$-hyperedge $\langle b_1, \ldots, b_{i-1}, b_i, b_{i+1}, \ldots, b_k\rangle$ is in $\mathscr{R}(\mathbf{B})$, and $a_i$ belongs to the interpretation of $P_{b_i}$ in $\mathbf{A}$.

Note that by construction, the adjacency of $a \in A$ in $\Phi(\mathbf{A})$ is the union of its adjacency in $\mathbf{A}$, and the union of the adjacencies of $b$ in $\mathbf{B}$ for all $b$ *s.t.* $a \in P_b(\mathbf{A})$.

We show that $\mathbf{A} \in \mathscr{H}om(\mathrm{All},\, \mathbf{B}^\dagger)$ *iff* $\Phi(\mathbf{A}) \in \mathscr{H}om(\mathrm{All},\, \mathbf{B})$. So, assume that there exists a homomorphism $f \colon \mathbf{A} \to \mathbf{B}^\dagger$. Then we let $f' \colon A \sqcup B \to B$ be defined by $f'(a) = f(a)$ for all $a \in A$ and $f'(b) = b$ for all $b \in B$. We claim that $f'$ is a homomorphism from $\Phi(\mathbf{A})$ to $\mathbf{B}$. Indeed, consider a hyperedge of $\Phi(\mathbf{A})$:

- if it is a hyperedge of $\mathbf{A}$, its image by $f'$ is still a hyperedge of $\mathbf{B}$ since $f$ is a homomorphism from $\mathbf{A}$ to $\mathbf{B}^\dagger$;
- if it is a hyperedge of $\mathbf{B}$, then its image by $f'$ is itself, and is hence a hyperedge of $\mathbf{B}$;
- otherwise, it must be of the form

$$\langle b_1, \ldots, b_{i-1}, a_i, b_{i+1}, \ldots, b_k\rangle$$

*s.t.* there exists $b_i$ for which $\langle b_1, \ldots, b_{i-1}, b_i, b_{i+1}, \ldots, b_k\rangle \in \mathscr{R}(\mathbf{B})$ and $a_i \in P_{b_i}(\mathbf{A})$: in this case, its image by $f'$ is

$$f'(\langle b_1, \ldots, b_{i-1}, a_i, b_{i+1}, \ldots, b_k\rangle) = \langle b_1, \ldots, b_{i-1}, b_i, b_{i+1}, \ldots, b_k\rangle \in \mathscr{R}(\mathbf{B})$$

since $f'(b) = b$ for all $b \in B$ and $f'(a_i) = f(a_i) = b_i$ since $a_i \in P_{b_i}(\mathbf{A})$ and $f$ is a homomorphism from $\mathbf{A}$ to $\mathbf{B}^\dagger$.

And hence, $\Phi(\mathbf{A}) \xrightarrow{\text{hom}} \mathbf{B}$.

Conversely, now let $g \colon \Phi(\mathbf{A}) \to \mathbf{B}$ be a homomorphism. Its restriction to $\mathbf{B}$, namely $g|_B$ is a homomorphism from $\mathbf{B}$ to itself, and since $\mathbf{B}$ is a core, it must be an automorphism over $\mathbf{B}$ by Proposition II.2.2. We then define a map $g' \colon A \to B$ by sending $a$ to $(g|_B)^{-1} \circ g(a)$, and claim that it is a homomorphism from $\mathbf{A}$ to $\mathbf{B}^\dagger$. As a matter of fact, it clearly preserves $\mathscr{R}$-tuple for any $\mathscr{R}$ in $\sigma$, since $g$ and $(g|_B)^{-1}$ are homomorphisms. We must then show that it preserves all unary predicates $P_b$, with $b \in B$: let $a \in A$ *s.t.* $P_b$ holds, *i.e.* $a \in P_b(\mathbf{A})$.

Now, by construction of $\Phi(\mathbf{A})$, the adjacency of $g(a)$ in $\mathbf{B}$ and the adjacency of $g(b)$ in $\mathbf{B}$ are equal. Since $\mathbf{B}$ is a core, it follows by Proposition II.2.4 that $g(a) = g(b)$. By definition of $g'$, this rewrites as $g'(a) = b$, i.e. $g'(a) = P_b(\mathbf{B}^\dagger)$. Therefore, we have built a homomorphism from $\mathbf{A}$ to $\mathbf{B}^\dagger$.

Overall, this proves that $\Phi$ is correct. It is trivially a first-order reduction and moreover, it preserves finiteness since $\mathbf{B}$ is finite. □

### VIII.2.5 De Bruijn–Erdős Theorem

**Proposition VIII.2.7** (*De Bruijn–Erdős Theorem*).[20] Let $\mathbf{A}$ be an arbitrary $\sigma$-structure and $\mathbf{B}$ a finite $\sigma$-structure. There is a homomorphism from $\mathbf{A}$ to $\mathbf{B}$ *iff* for every finite substructure $\mathbf{A}'$ of $\mathbf{A}$, there is a homomorphism from $\mathbf{A}$ to $\mathbf{B}$.

*Proof.* The left-to-right implication is direct. We prove the converse by using the Tychonoff's compactness theorem.[21] So, assume that for every finite substructure $\mathbf{A}'$ of $\mathbf{A}$, there is a homomorphism from $\mathbf{A}$ to $\mathbf{B}$. Consider the topological space $B^A$, consisting of all functions from $A$ to $B$, together with the product topology.[22] By Tychonoff's compactness theorem, $B^A$ is compact. For each finite subset $X$ of $A$, let $H_X$ denote the set of all $f \in B^A$ s.t. $f|_X$ is a homomorphism from the substructure of $\mathbf{A}$ induced by $X$ to $\mathbf{B}$. Then, each $H_X$ is closed—indeed, whether $f \in B^A$ belongs to $H_X$ only depends on finitely many $f(x)$'s—, and moreover the intersection of finitely many $H_X$'s, say $H_{X_1} \cap \cdots \cap H_{X_n}$, is non-empty since $H_{X_1} \cap \cdots \cap H_{X_n} \supseteq H_{X_1 \cup \ldots \cup X_n}$ and by assumption $H_{X_1 \cup \ldots \cup X_n}$ is non-empty since $X_1 \cup \cdots \cup X_n$ is finite. Hence, by compactness of $B^A$ and the finite intersection property, it follows that $\bigcap_X H_X$ is non-empty, which means that there is a homomorphism from $\mathbf{A}$ to $\mathbf{B}$. □

**Corollary VIII.2.8.**[23] Given arbitrary $\sigma$-structures $\mathbf{B}_1$ and $\mathbf{B}_2$, the following are equivalent:

1. for every finite $\sigma$-structure $\mathbf{A}$, we have $\mathbf{A} \xrightarrow{\text{hom}} \mathbf{B}_1$ *iff* $\mathbf{A} \xrightarrow{\text{hom}} \mathbf{B}_2$;
2. for every arbitrary $\sigma$-structure $\mathbf{A}$, we have $\mathbf{A} \xrightarrow{\text{hom}} \mathbf{B}_1$ *iff* $\mathbf{A} \xrightarrow{\text{hom}} \mathbf{B}_2$;
3. $\mathbf{B}_1$ and $\mathbf{B}_2$ are homomorphically equivalent.

*Proof.* (2) $\Rightarrow$ (3) and (3) $\Rightarrow$ (1) are trivial. For (1) $\Rightarrow$ (2), we assume *w.l.o.g.* by contradiction that there is an arbitrary $\sigma$-structure $\mathbf{A}$ s.t. $\mathbf{A} \xrightarrow{\text{hom}} \mathbf{B}_1$ but $\mathbf{A} \xrightarrow{\text{hom}} \mathbf{B}_2$. Then by Proposition VIII.2.7, there exists a finite substructure $\mathbf{A}_0$ of $\mathbf{A}$ s.t. $\mathbf{A}_0 \xrightarrow{\text{hom}} \mathbf{B}_2$. But then $\mathbf{A}_0 \xrightarrow{\text{hom}} \mathbf{A} \xrightarrow{\text{hom}} \mathbf{B}_1$, which contradicts (1). □

### VIII.2.6 Obstructions and Finite Duality

Let $\mathbf{B}$ and $\mathbf{D}$ be finite $\sigma$-structures. We say that $\mathbf{D}$ is an *obstruction* of $\mathbf{B}$ when $\mathbf{D} \xrightarrow{\text{hom}} \mathbf{B}$. In this case, note that finding $\mathbf{D}$ inside $\mathbf{A}$—in the sense that $\mathbf{D} \xrightarrow{\text{hom}} \mathbf{A}$—implies that $\mathbf{A}$ cannot have a homomorphism to $\mathbf{B}$: in this sense, the presence of $\mathbf{D}$ in an *obstruction* to the existence of a homomorphism to $\mathbf{B}$.

[20] It is straightforward to note that one can replace "every finite substructure" by "every finite induced substructure" in the statement of the theorem. The original theorem is about graph colouring, but the generalization is straightforward.

[21] This is a direct adaptation from [Wik24, § "Proof"].

[22] We equip $B$ with the discrete topology, making it compact since $B$ is finite.

[23] Another important consequence of the De Bruijn–Erdős Theorem is that, for instance, the notion of dual does not depend on whether we are considering finite or arbitrary $\sigma$-structures.

A *dual* of **B** is any arbitrary set $\mathcal{D}$ of finite $\sigma$-structures *s.t.* for any finite $\sigma$-structure **A**:

$$\mathbf{A} \xrightarrow{\text{hom}} \mathbf{B} \; \textit{iff} \; \forall \mathbf{D} \in \mathcal{D}, \; \mathbf{D} \xrightarrow{\text{hom}} \mathbf{A}.$$

Note that any dual must only contain obstructions of **B**.[24] The set of all obstructions of **B** is a dual of **B**.

**Example VIII.2.9.** Let $k \in \mathbb{N}$. The set $\{\mathbf{P}_{k+1}\}$ is a dual of the $k$-transitive tournament $\mathbf{T}_k$—see Figure VIII.1.

Indeed, $\mathbf{P}_{k+1} \xrightarrow{\text{hom}} \mathbf{T}_k$. Dually, if **G** if a finite graph *s.t.* $\mathbf{G} \xrightarrow{\text{hom}} \mathbf{T}_k$ then letting $f$ be a homomorphism from **G** to $\mathbf{T}_k$, we have that any edge from $u \in G$ to $v \in G$, we must have $f(u) < f(v)$, and so $\mathbf{P}_k \xrightarrow{\text{hom}} \mathbf{G}$. $\diamond$

Moreover, if $\mathcal{D}$ and $\mathcal{D}'$ are sets of obstructions of **B** and $\mathcal{D} \subseteq \mathcal{D}'$, then if $\mathcal{D}$ is a dual, so is $\mathcal{D}'$: hence, the goal is to find *small* duals. For this reason, duals are also called *complete sets of obstructions*. We say that **B** has *finite duality* if is admits a finite dual, *i.e.* consisting of finitely many structures. For instance, $\mathbf{T}_k$ has finite duality.

An obstruction **D** of **B** is *critical*[25] when for every proper substructure **D'** of **D**, we have $\mathbf{D}' \xrightarrow{\text{hom}} \mathbf{B}$. Clearly, every critical obstruction must be a core.

Note first that the set of all critical obstructions of **B** is a dual of **B**. Indeed, if $\mathbf{A} \xrightarrow{\text{hom}} \mathbf{B}$, then **A** is an obstruction and so it must contain—by well-foundedness of $\mathbb{N}$—a critical obstruction **D** as a substructure.

**Proposition VIII.2.10.** Let **B** be a finite $\sigma$-structure. **B** has finite duality *iff* it has finitely many critical obstructions.

*Proof.* The right-to-left implication is trivial. For the converse one, let $\mathcal{D} = \{D_1, \dots, D_m\}$ be a finite dual of **B**. If **C** is a critical obstruction of **B** then in particular $\mathbf{C} \xrightarrow{\text{hom}} \mathbf{B}$ and so there exists $i \in [\![1, m]\!]$ *s.t.* there is a homomorphism $f$ from $\mathbf{D}_i$ to **C**. Now the image of $f$ is again an obstruction of **B**, and since **C** is critical, it follows that it must be **C** itself. In other words, $f$ is strong onto. In particular, we obtain that **C** is a quotient of $\mathbf{D}_i$. Hence, there are only finitely many critical obstructions of **B**. $\square$

One of the key interest of Proposition VIII.2.10 is to prove that some structures don't have finite duality.

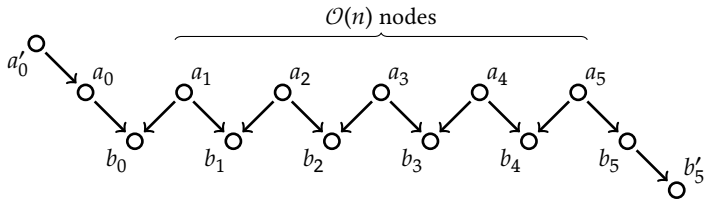

**Example VIII.2.11** (Example VIII.2.9, continued.)**.** Let $n \in \mathbf{N}$. We define the *zigzag graph* $\mathbf{Z}_2^{(n)}$ of width $n$ and length 2 to be graph whose vertices are $a_0, \dots, a_n, b_0, \dots, b_n, a_0'$ and $b_n'$, with edges from $a_i$ to $b_{i-1}$ and to $b_i$ (for $i \in [\![0, n]\!]$, whenever the nodes exist), and with an edge from $a_0'$ to $a_0$ and from $b_n$ to $b_n'$. See Figure VIII.6 for an illustration.

---

Figure VIII.6: The zigzag graph $\mathbf{Z}_2^{(5)}$.

Note that $\mathbf{Z}_2^{(n)}$ does not admit a homomorphism to the 2-path—indeed, such a homomorphism should send $a_0'$, $a_0$ and $b_0$ onto 0, 1, and 2, respectively, and so all $a_i$'s (resp. $b_i$'s) must be sent onto 1 (resp. 2), but then $b_n'$ cannot be mapped anywhere.

We claim that each $\mathbf{Z}_2^{(n)}$ ($n \in \mathbb{N}$) is a critical obstruction of $\mathbf{P}_2$. We have already seen that $\mathbf{Z}_2^{(n)}$ is an obstruction of $\mathbf{P}_2$. But then notice that to obtain a proper substructure of $\mathbf{Z}_2^{(n)}$, we must either:

- remove the edge from $a_0'$ to $a_0$ or the edge from $b_n$ to $b_n'$, in which case it admits a homomorphism to $\mathbf{P}_2$, or
- remove any other edge, in which case the resulting substructure is not connected, and both parts admit a homomorphism to $\mathbf{P}_2$.

And hence, by Proposition VIII.2.10, it follows that $\mathbf{P}_2$ does not have finite duality.

On the other hand, each $\mathbf{Z}_2^{(n)}$ with $n \in \mathbb{N}_{>0}$ admits a homomorphism to the 2-transitive tournament, as witnessed by Figure VIII.7.[26] In fact, this homomorphism is far from being unique: each vertex $a_1, a_2, \ldots, a_{n-1}$ can be sent on either 0 or 1 (the red and yellow vertices), and similarly, each vertex $b_1, b_2, \ldots, b_{n-1}$ can be sent on either 1 or 2 (the yellow and blue vertices).[27]

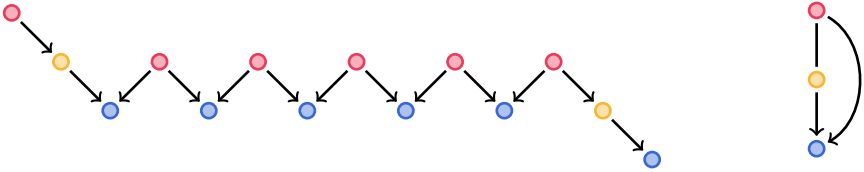

$\diamond$

While $\mathbf{P}_k$ and $\mathbf{T}_k$ are similar structures, one has finite duality and the other does not.

Recall that by Atserias' theorem, this implies that $\mathcal{H}om(\mathrm{Fin}, \mathbf{T}_k)$ is first-order definable[28] but $\mathcal{H}om(\mathrm{Fin}, \mathbf{P}_k)$ is not.[29]

**Remark VIII.2.12** (From finite to infinite structures). In the property

$$\mathbf{A} \xrightarrow{\mathrm{hom}} \mathbf{B} \quad iff \quad \forall \mathbf{D} \in \mathcal{D}, \ \mathbf{D} \xrightarrow{\mathrm{hom}} \mathbf{A}$$

defining a dual, we quantified $\mathbf{A}$ over finite $\sigma$-structures. We could equally quantify over all $\sigma$-structures without changing the notion of dual by De Bruijn-Erdős theorem. $\diamond$

We say that a $\sigma$-structure is *rigid* if its only automorphism is the identity.

**Proposition VIII.2.13** ([LLT07, Lemma 4.1]).[30] If a finite core has finite duality, then it is rigid.

### VIII.2.7 Trees and Tree Duality

A $\sigma$-structure is *strongly acyclic* if its incidence graph is acyclic. A *$\sigma$-tree* is a $\sigma$-structure that is both connected and strongly acyclic. Do not confuse

---

[26] However, observe that $\mathbf{Z}_2^{(0)} = \mathbf{P}_3$ is an obstruction of $\mathbf{T}_2$.

[27] Note that it is straightforward to extend the fact that

$$\mathbf{Z}_2^{(n)} \xrightarrow{\mathrm{hom}} \mathbf{T}_2 \text{ and } \mathbf{Z}_2^{(n)} \xrightarrow{\mathrm{hom}} \mathbf{P}_2$$

to arbitrary values of $k \in \mathbb{N}$, in the sense that

$$\mathbf{Z}_k^{(n)} \xrightarrow{\mathrm{hom}} \mathbf{T}_k \text{ and } \mathbf{Z}_k^{(n)} \xrightarrow{\mathrm{hom}} \mathbf{P}_k,$$

by letting $\mathbf{Z}_k^{(n)}$ be the graph obtained from $\mathbf{Z}_2^{(n)}$ by replacing the path leading to $a_0$ by a path of length $k-1$.

Figure VIII.7: A homomorphism from the zigzag graph (left-hand side) to the 2-transitive tournament (right-hand side).

[28] By the formula saying that there are no variables $x_0, \ldots, x_{k+1}$ *s.t.* for all $i$ there is an edge from $x_i$ to $x_{i+1}$.

[29] This can also be proven by hand: for instance we let $\bar{\mathbf{Z}}_2^{(n)}$ be defined analogously to $\mathbf{Z}_2^{(n)}$ except that we remove the edge from $a_k$ to $b_k$ for $k = \lceil \frac{n}{2} \rceil$. Then for a well-chosen single-exponential function $f: \mathbb{N} \to \mathbb{N}$, for $n \in \mathbb{N}$, we have that Duplicator wins the Ehrenfeucht-Fraïssé game on $\mathbf{Z}_2^{(f(n))}$ and $\bar{\mathbf{Z}}_2^{(f(n))}$. Since $\mathbf{Z}_2^{(f(n))} \xrightarrow{\mathrm{hom}} \mathbf{P}_2$ but $\bar{\mathbf{Z}}_2^{(n)} \xrightarrow{\mathrm{hom}} \mathbf{P}_2$, this implies that $\mathcal{H}om(\mathrm{Fin}, \mathbf{P}_2)$ is not first-order definable. See *e.g.* [Kol07, § 3] for details on Ehrenfeucht-Fraïssé games.

[30] In fact only assumes that $\mathbf{B}$ has tree duality: as we will see in Proposition VIII.2.14, this is a weaker condition than having finite duality.

this notion with the classical notion of directed trees: every directed tree is a $\sigma$-tree for the graph signature $\sigma$, but $\mathbf{Z}_2^{(n)}$—see Figure VIII.6—is a $\sigma$-tree while it is not a directed tree.

Given a $\sigma$-tree $\mathbf{T}$ and a vertex $t \in T$, the *height* of $\mathbf{T}$ when rooted at $t$ is the maximal distance between $t$ and any other vertex of $\mathbf{T}$.

We say that a finite $\sigma$-structure $\mathbf{B}$ has *tree duality* if it admits a (potentially infinite) dual consisting only of $\sigma$-trees. Somewhat surprisingly, Nešetřil and Tardif showed that this notion generalizes finite duality.

**Proposition VIII.2.14** ([NT00, Theorem 3.1]).[31] If a finite $\sigma$-structure $\mathbf{B}$ has finite duality, then it has tree duality.

The converse does not hold.

**Proposition VIII.2.15.** The 2-path $\mathbf{P}_2$ has tree duality.

*Proof sketch.* It can be shown that $\{\mathbf{Z}_|^{(n)} | n \in \mathbb{N}\}$ is a dual of $\mathbf{P}_2$. Moreover, each $\mathbf{Z}_2^{(n)}$ ($n \in \mathbb{N}$) is a $\sigma$-tree. □

Feder and Vardi introduced a construction to decide if a finite structure has tree duality: given a $\sigma$-structure $\mathbf{B}$, we let $\mathfrak{U}(\mathbf{B})$ be the $\sigma$-structure whose domain is $\mathcal{P}_+(B)$, and for every $\mathcal{R}_{(k)} \in \sigma$, we have $\langle Y_1, \ldots, Y_k \rangle \in \mathcal{R}(\mathfrak{U}(\mathbf{B}))$ (with $Y_1, \ldots, Y_k \in \mathcal{P}_+(B)$) precisely when for every $i \in [\![1, k]\!]$, for every $b_i \in Y_i$, there exists $b_j \in Y_j$ for every $j \neq i$ s.t. $\langle y_1, \ldots, y_k \rangle \in \mathcal{R}(\mathbf{B})$.[32] In the case of graphs, the nodes of $\mathfrak{U}(\mathbf{H})$ are non-empty subsets of vertices of $\mathbf{H}$, and there is an edge from $X$ to $Y$ when:

- for every $x \in X$, there exists $y \in Y$ s.t. $\langle x, y \rangle$ is an edge of $\mathbf{H}$, and
- for every $y \in Y$, there exists $x \in X$ s.t. $\langle x, y \rangle$ is an edge of $\mathbf{H}$.

By construction, note that $b \mapsto \{B\}$ defines a homomorphism from $\mathbf{B}$ to $\mathfrak{U}(\mathbf{B})$.

**Proposition VIII.2.16** ([FV98, Theorem 21]). A finite $\sigma$-structure $\mathbf{B}$ has tree duality if, and only if, $\mathfrak{U}(\mathbf{B}) \xrightarrow{\text{hom}} \mathbf{B}$, or, equivalently, if $\mathbf{B}$ and $\mathfrak{U}(\mathbf{B})$ are homomorphically equivalent.

Note that $\mathfrak{U}(\mathfrak{U}(\mathbf{B}))$ is always homomorphically equivalent to $\mathfrak{U}(\mathbf{B})$—see e.g. [NO12a, § 9.2.2, Proposition 9.1]— and so $\mathfrak{U}(\mathbf{B})$ *always* has tree duality, no matter if $\mathbf{B}$ has this property.

**Example VIII.2.17** (Example VIII.2.11, continued). Using Proposition VIII.2.16, we can prove for instance that if a graph $\mathbf{G}$ has tree duality, then either it is a DAG, or $\mathbf{G}$ contains a self-loop.

Indeed, let $v_0 \to v_1 \to \cdots \to v_m$ be a directed cycle in $\mathbf{G}$, with $v_0 = v_m$. Then in $\mathfrak{U}(\mathbf{G})$, there is an edge from $\{v_0, v_1, \ldots, v_m\}$ to itself. Since $\mathbf{G}$ has tree duality, it follows by Proposition VIII.2.16 that $\mathfrak{U}(\mathbf{G}) \xrightarrow{\text{hom}} \mathbf{G}$ and hence $\mathbf{G}$ also contains a self-loop.

Note that, if $\mathbf{G}$ contains a self-loop, then actually it is homomorphically equivalent to the graph consisting of a single self-loop—that admits $\varnothing$ as a dual and hence has tree duality. In other words, what we showed can be rephrased as: any non-trivial graph with tree duality is a DAG. For instance, this implies that $\mathbf{K}_2$ does not have tree duality.

[31] The statement of [NT00, Theorem 3.1] is somewhat cryptic: the relationship with duals is given by [NT00, Lemma 2.5]. We refer the reader to Foniok's Ph.D. for statements that use a terminology closer to ours: [Fon07, Theorem 2.1.12] shows that if {A} is a dual of $\mathbf{B}$, then $\mathbf{A}$ is homomorphically equivalent to a tree, *i.e.* if a structure has "singleton duality", then it has tree duality. The generalization to finite duality then follows from [Fon07, Theorem 2.4.4].

[32] In fact $\mathfrak{U}(-)$ can easily be extended to be a functor in the category of $\sigma$-structures, see *e.g.* [NO12a, § 9.2.2].

On the other hand, we saw in Example VIII.2.11 that $\mathbf{P}_2$ does not have finite duality. However, it has tree duality: indeed, see Figure VIII.8 and observe that $\mathfrak{u}(\mathbf{P}_2) \xrightarrow{\text{hom}} \mathbf{P}_2$. ◇

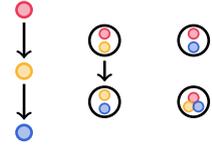

Figure VIII.8: The Feder-Vardi construction $\mathfrak{u}(\mathbf{P}_2)$ on the 2-path.

## VIII.3 From Separation to Colouring of Automatic Graphs

### VIII.3.1 Separability is Equivalent to Regular Colourability

We start by showing that the Rec-separability problem is equivalent, under polynomial time reductions, to the regular colourability problem. To make our statement precise, we need some terminology.

Let $k$-Rec be the class of languages expressed by unions of products of $k$ regular languages which form a partition: in the binary case, this corresponds to relations of the form $(L_{i_1} \times L_{j_1}) \cup \cdots \cup (L_{i_\ell} \times L_{j_\ell})$, with $i_1, j_1, \dots, i_\ell, j_\ell \in [\![1, k]\!]$, for some regular partition[33] $L_1, \dots, L_k$ of $\Sigma^*$ and $\ell \in \mathbb{N}$. Note that Rec $= \bigcup_k k$-Rec.

**Example VIII.3.1.** For instance, $(aa)^* \times (aaa)^*$ belongs to 4-Rec but not to 3-Rec since, for any $k \in \mathbb{N}$, we have that $(aa)^* \times (aaa)^* \in k$-Rec *iff* there exists a regular partition of $a^*$ into $k$ languages *s.t.* both $(aa)^*$ and $(aaa)^*$ can be expressed as the union of some of these languages. To get the upper bound $k = 4$, consider the partition $\langle L_6, L_3, L_2, L_\perp \rangle$ where $L_6 = (a^6)^*$, $L_3 = (a^3)^* \smallsetminus L_6$, $L_2 = (a^2)^* \smallsetminus L_6$ and $L_\perp = a^* \smallsetminus (L_2 \cup L_3 \cup L_6)$. ◇

**Theorem VIII.3.2.** There are polynomial-time reductions:

1. from Aut/Rec-separability to finite regular colourability,
2. from finite regular colourability to Aut/Rec-separability, and
3. from regular $k$-colourability to Aut/$k$-Rec-separability, for every $k > 0$.

Further, the last two reductions are so that the second relation in the instance of the separability problem is the identity $\mathcal{Id}$.

*Proof of* (2) *and* (3). We start with the last two reductions. Given an automatic graph $\langle V, \mathcal{E} \rangle$ over an alphabet $\Sigma$, consider the instance $\langle \mathcal{R}_1, \mathcal{R}_2 \rangle$ for the Rec-separability problem, where $\mathcal{R}_1 = \mathcal{E}$ and $\mathcal{R}_2 = \mathcal{Id}$. If $\langle V, \mathcal{E} \rangle$ is regularly $k$-colourable via the colouring $V_1, \dots, V_k$ then the $k$-Rec relation $\bigcup_{i \neq j} V_i \times V_j$ separates $\mathcal{R}_1 = \mathcal{E}$ and $\mathcal{R}_2 = \mathcal{Id}$. Conversely, if a $k$-Rec relation $\mathcal{R} \subseteq \Sigma^* \times \Sigma^*$ on the regular partition $V_1 \sqcup \cdots \sqcup V_k = \Sigma^*$ separates $\mathcal{R}_1$ and $\mathcal{R}_2$, then $\bigcup_{i \neq j} V_i \times V_j$ also separates $\mathcal{R}_1$ and $\mathcal{R}_2$, and this implies that $V_1, \dots, V_k$ is a $k$-colouring for $\langle \Sigma^*, \mathcal{E} \rangle$, and in particular for $\langle V, \mathcal{E} \rangle$. □

For the first reduction, let us introduce some terminology. Given two relations $\mathcal{R}_1, \mathcal{R}_2$ over $\Sigma^*$, say that $u \in \Sigma^*$ is *compatible* with $u' \in \Sigma^*$ when for all words $v \in \Sigma^*$:

[33] By "regular partition", we mean that $L_1, \dots, L_k$ is a partition of $\Sigma^*$, and that moreover each $L_i$ is a regular language.

(COMP$_\ell$): $\langle u, v \rangle \in \mathcal{R}_1 \Rightarrow \langle u', v \rangle \notin \mathcal{R}_2$,     (COMP$_r$): $\langle v, u \rangle \in \mathcal{R}_1 \Rightarrow \langle v, u' \rangle \notin \mathcal{R}_2$,

(COMP$'_\ell$): $\langle u', v \rangle \in \mathcal{R}_1 \Rightarrow \langle u, v \rangle \notin \mathcal{R}_2$  &  (COMP$'_r$): $\langle v, u' \rangle \in \mathcal{R}_1 \Rightarrow \langle v, u \rangle \notin \mathcal{R}_2$.

Define the *incompatibility graph* $\mathcal{I}nc_{\mathcal{R}_1, \mathcal{R}_2}$ as the graph whose vertices are all words of $\Sigma^*$, and with an edge from $u$ to $v$ whenever $u$ is not compatible with $v$. Note that $\mathcal{I}nc_{\mathcal{R}, \mathcal{I}d}$ is exactly the graph $\langle \Sigma^*, \mathcal{R} \rangle$.

**Example VIII.3.3.** Let $\Sigma = \{a, b\}$, $\mathcal{R}_1$ be the equal-length relation, and

$$\mathcal{R}_2 = \{\langle u, ua \rangle \mid u \in \Sigma^*\} \cup \{\langle u, ub \rangle \mid u \in \Sigma^*\},$$

that we depict in Figure VIII.9. Then, $u$ is incompatible with $u'$ if $|u| = |u'| + 1$ (this is given by (COMP$_\ell$) or (COMP$_r$)), or if $|u'| = |u| + 1$ (this is given by (COMP$'_\ell$) or (COMP$'_r$)). This gives rise to the incompatibility graph of Figure VIII.10.

Note that while neither $\mathcal{R}_1$ nor $\mathcal{R}_2$ are recognizable, they are separable by the recognizable relation $\mathcal{S}$ consisting of all pairs $\langle u, v \rangle$ such that $|u|$ and $|v|$ have the same parity. Moreover, $\mathcal{I}nc_{\mathcal{R}_1, \mathcal{R}_2}$ is regularly 2-colourable, the two colours being the words of even and odd length. ◇

**Proposition VIII.3.4.** If $\mathcal{R}_1$ and $\mathcal{R}_2$ are automatic, then so is $\mathcal{I}nc_{\mathcal{R}_1, \mathcal{R}_2}$. Moreover, we can build an automaton for $\mathcal{I}nc_{\mathcal{R}_1, \mathcal{R}_2}$ in polynomial time in the size of the automata for $\mathcal{R}_1$ and $\mathcal{R}_2$.

*Proof.* By definition, the incompatibility relation $\mathcal{I}nc_{\mathcal{R}_1, \mathcal{R}_2}$ can be written as $\mathcal{R}_{\neg(\text{COMP}_\ell)} \cup \mathcal{R}_{\neg(\text{COMP}'_\ell)} \cup \mathcal{R}_{\neg(\text{COMP}_r)} \cup \mathcal{R}_{\neg(\text{COMP}'_r)}$, where:

$$\mathcal{R}_{\neg(\text{COMP}_\ell)} \triangleq \left\{\langle u, u' \rangle \in \Sigma^* \times \Sigma^* \mid \exists v \in \Sigma^*, \ \langle u, v \rangle \in \mathcal{R}_1 \wedge \langle u', v \rangle \in \mathcal{R}_2\right\},$$

$$\mathcal{R}_{\neg(\text{COMP}'_\ell)} \triangleq \left\{\langle u, u' \rangle \in \Sigma^* \times \Sigma^* \mid \exists v \in \Sigma^*, \ \langle u', v \rangle \in \mathcal{R}_1 \wedge \langle u, v \rangle \in \mathcal{R}_2\right\},$$

$$\mathcal{R}_{\neg(\text{COMP}_r)} \triangleq \left\{\langle u, u' \rangle \in \Sigma^* \times \Sigma^* \mid \exists v \in \Sigma^*, \ \langle v, u \rangle \in \mathcal{R}_1 \wedge \langle v, u' \rangle \in \mathcal{R}_2\right\}, \text{ and}$$

$$\mathcal{R}_{\neg(\text{COMP}'_r)} \triangleq \left\{\langle u, u' \rangle \in \Sigma^* \times \Sigma^* \mid \exists v \in \Sigma^*, \ \langle v, u' \rangle \in \mathcal{R}_1 \wedge \langle v, u \rangle \in \mathcal{R}_2\right\}$$

Observe that starting from automata for $\mathcal{R}_1$ and $\mathcal{R}_2$, then for each of the relation $\mathcal{R}_{\neg(\text{COMP}_\ell)}$, $\mathcal{R}_{\neg(\text{COMP}'_\ell)}$, $\mathcal{R}_{\neg(\text{COMP}_r)}$ or $\mathcal{R}_{\neg(\text{COMP}'_r)}$, we can build an automaton recognizing them using a product construction, which can be implemented in polynomial time.[34] It then follows that we can build a polynomial-size automaton recognizing $\mathcal{I}nc_{\mathcal{R}_1, \mathcal{R}_2}$ in polynomial time. □

We can now finish the proof of Theorem VIII.3.2.

*Proof of* (1). Given an instance $\langle \mathcal{R}_1, \mathcal{R}_2 \rangle$ of the REC-SEPARABILITY PROBLEM, we reduce it to the REGULAR COLOURABILITY PROBLEM on its incompatibility graph $\mathcal{I}nc_{\mathcal{R}_1, \mathcal{R}_2}$.

☞ *Left-to-right implication.* Assume that there exists $\mathcal{S}$ in $k$-REC that separates $\mathcal{R}_1$ from $\mathcal{R}_2$. Then $\mathcal{S}$ can be written as $(L_{i_1} \times L_{j_1}) \cup \cdots \cup (L_{i_\ell} \times L_{j_\ell})$, where $(L_1, \ldots, L_k)$ is a partition of $\Sigma^*$ in $k$ regular languages. We define the

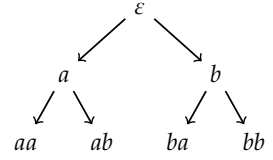

Figure VIII.9: The relation $\mathcal{R}_2$ of Example VIII.3.3, restricted to words of length at most 2.

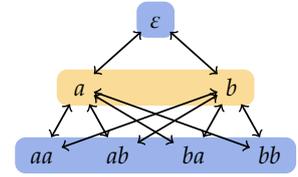

Figure VIII.10: Incompatibility graph $\mathcal{I}nc_{\mathcal{R}_1, \mathcal{R}_2}$ and its regular 2-colouring.

[34] More precisely, in this product construction the states are $Q_1 \times Q_2$ where $Q_i$ is the set of states of an automaton $\mathcal{A}_i$ for $\mathcal{R}_i$. Then, for each transition

$$q_i \xrightarrow{\langle a, b \rangle} q'_i$$

in $\mathcal{A}_i$ ($i \in \{1, 2\}$), we put a transition

$$\langle q_1, q_2 \rangle \xrightarrow{\langle a, c \rangle} \langle q'_1, q'_2 \rangle,$$

with $a, b, c \in \Sigma \sqcup \{\bot\}$. Potentially, this can produce transitions labelled by $\langle \bot, \bot \rangle$: we can get rid of those using the standard elimination of $\varepsilon$-transitions. Finally, a state $\langle q_1, q_2 \rangle$ is accepting if $q_1$ and $q_2$ are accepting in $\mathcal{A}_1$ and $\mathcal{A}_2$, respectively.

colour of a word $u \in \Sigma^*$ as the unique $i \in [\![1, k]\!]$ s.t. $u \in L_i$. In other words, the colouring is simply $\langle L_1, \dots, L_k \rangle$.

This is indeed a proper colouring: if $u$ and $u'$ have the same colour, we claim that $\langle u, u' \rangle \notin \mathcal{I}nc_{\mathcal{R}_1, \mathcal{R}_2}$, i.e. that $u$ compatible with $u'$. Indeed, take any $v \in \Sigma^*$: if $\langle u, v \rangle \in \mathcal{R}_1$, then $\langle u, v \rangle \in \mathcal{S}$, so $\langle u, v \rangle \in L_{i_m} \times L_{j_m}$ for some $m \in [\![1, \ell]\!]$. But since $u$ has the same colour as $u'$, the fact that $u \in L_{i_m}$ implies $u' \in L_{i_m}$, and hence $\langle u', v \rangle \in L_{i_m} \times L_{j_m} \subseteq \mathcal{S}$. But $\mathcal{S}$ separates $\mathcal{R}_1$ from $\mathcal{R}_2$, and therefore $\langle u', v \rangle \notin \mathcal{R}_2$. This tells us that (COMP$_\ell$) holds. The other conditions hold by symmetry. We conclude that $\langle L_1, \dots, L_k \rangle$ defines a proper $k$-colouring of $\mathcal{I}nc_{\mathcal{R}_1, \mathcal{R}_2}$, that is regular since the $L_i$'s are regular languages by definition.

↬ *Right-to-left implication.* Assume that $\mathcal{I}nc_{\mathcal{R}_1, \mathcal{R}_2}$ is finitely regularly colourable, say by $\langle L_1, \dots, L_k \rangle$. Then let $\mathcal{S}$ be the union of all $\mathcal{S}_i$'s ($i \in [\![1, k]\!]$) where

$$\mathcal{S}_i \triangleq \{ \langle u, v \rangle \mid u \in L_i \text{ and } \langle u', v \rangle \in \mathcal{R}_1 \text{ for some } u' \in L_i \}$$
$$\cup \{ \langle u, v \rangle \mid v \in L_i \text{ and } \langle u, v' \rangle \in \mathcal{R}_1 \text{ for some } v' \in L_i \}.$$

Since $\bigcup_i L_i = \Sigma^*$, we get $\mathcal{R}_1 \subseteq \mathcal{S}$. Moreover, we claim that $\mathcal{R}_2 \cap \mathcal{S} = \varnothing$. Indeed, if $\langle u, v \rangle \in \mathcal{S}$, then $\langle u, v \rangle \in \mathcal{S}_i$ for some $i \in [\![1, k]\!]$. It either means that (1) $u \in L_i$ and $\langle u', v \rangle \in \mathcal{R}_1$ for some $u' \in L_i$, or (2) $v \in L_i$ and $\langle u, v' \rangle \in \mathcal{R}_2$ for some $v' \in L_i$. In case (1), $u$ and $u'$ have the same colour, and since $\langle L_1, \dots, L_k \rangle$ is a colouring of $\mathcal{I}nc_{\mathcal{R}_1, \mathcal{R}_2}$, $u$ must be compatible with $u'$. The assumption $\langle u', v \rangle \in \mathcal{R}_1$ together with (COMP$'_\ell$) then yields that $\langle u, v \rangle \notin \mathcal{R}_2$. The other case is symmetric. Therefore, $\langle u, v \rangle \notin \mathcal{R}_2$, and thus $\mathcal{S}$ separates $\mathcal{R}_1$ from $\mathcal{R}_2$.

Finally, we show that $\mathcal{S}$ is recognizable. In fact,

$$\mathcal{S} = \bigcup_{i=1}^{k} \Big( L_i \times \mathcal{R}_1[L_i] \Big) \cup \big( \mathcal{R}_1^{-1}[L_i] \times L_i \big),$$

where for any set $X \subseteq \Sigma^*$ we define $\mathcal{R}_1[X]$ (resp. $\mathcal{R}_1^{-1}[X]$) to be the set of $v \in \Sigma^*$ (resp. $u \in \Sigma^*$) such that $\langle u, v \rangle \in \mathcal{R}_1$ for some $u \in X$ (resp. $v \in X$). Hence, $\mathcal{R}_1$ and $\mathcal{R}_2$ are REC-separable.[35]  □

Motivated by these reductions, we focus our attention to regular colourings of automatic graphs, eventually proving that both the REGULAR $k$-COLOURABILITY and REGULAR COLOURABILITY PROBLEMS are undecidable.

### VIII.3.2   Regular $k$-Colourability Problem

We show that the REGULAR $k$-COLOURABILITY PROBLEM is undecidable for $k \geq 2$.[36] This is proven by a reduction from a suitable problem on reversible Turing machines with certain restrictions, which we call "well-founded".

*Regularity of Reachability for Turing Machines.*   We say that a Turing machine $\mathcal{T}$ is *reversible*, if every node of $\mathcal{C}onf_{\mathcal{T}}$ has in-degree at most 1, in other words if the machine is co-deterministic. Note that ehe proof of undecidability of the isomorphism problem for automatic structures [KNRS07, § 5] also relies on the use of reversible Turing machines. For more details and pointers on reversible Turing machines, see [Mor17, Chapter 5].

---

[35] Note however that $\mathcal{S}$ does not necessarily belong to $k$-REC, but *a priori* to $8^k$-REC: indeed, from the $L_i$'s, $\mathcal{R}_1[L_i]$'s and $\mathcal{R}_1^{-1}[L_i]$'s, one can produce using a powerset construction a partition of $\Sigma^*$ in $2^{3k} = 8^k$ regular languages s.t. any $L_i$ (resp. $\mathcal{R}_1[L_i]$, resp. $\mathcal{R}_1^{-1}[L_i]$) can be written as the union of some of these languages.

[36] Using this, we obtain in the next subsection the undecidability for the separability problem on two natural classes of recognizable relations.

We say that a Turing machine $\mathcal{T}$ is *well-founded* if its configuration graph is such that:

1. the initial configuration has in-degree zero, and
2. there are no infinite backward paths $\gamma_0 \leftarrow \gamma_1 \leftarrow \cdots$ in $\mathcal{Conf}_{\mathcal{T}}$.

We say that a Turing machine is *linear* if it is well-founded, deterministic and reversible. By construction, a Turing machine is linear *iff* (1) its configuration graph consists of a possibly infinite disjoint union of directed paths, which are all finite, or isomorphic to $\langle \mathbb{N}, \mathrm{succ}\rangle$ and (2) the initial configuration has in-degree zero. Such a configuration graph is depicted on Figure VIII.13.[37]

The REGULAR REACHABILITY PROBLEM is the problem of, given a Turing machine $\mathcal{T}$, to decide whether its set of reachable configurations $\mathrm{Reach}_{\mathcal{T}}$ is a regular language. To show that is it undecidable, we exhibit a reduction from the halting problem on deterministic reversible Turing machines.

**Proposition VIII.3.5** ([Lec63, Theorem 1])**.** The HALTING PROBLEM on deterministic reversible Turing machines is RE-complete.

**Lemma VIII.3.6.** The REGULAR REACHABILITY PROBLEM is RE-complete, even if restricted to linear Turing machines.

*Proof sketch.* The upper bound is trivial. For the converse one, we reduce the HALTING PROBLEM on deterministic reversible Turing machines, in such a way that the reachable configurations whose state $q$ coincide with the state of the original machine are of the form $u \cdot q \cdot v \triangleright^n \triangleleft^n$ where $u \cdot q \cdot v$ is a configuration of the original machine, $\triangleright$ and $\triangleleft$ are new symbols, and $n \in \mathbb{N}$. Transitions are defined in such a way that the new machine is linear: this is implemented by having, for every transition $u \cdot q \cdot v \to u' \cdot q' \cdot v'$ of the original machine and every $n, m \in \mathbb{N}$, a multistep transition

$$u \cdot q \cdot v \triangleright^n \triangleleft^m \to^* u' \cdot q' \cdot v' \triangleright^{n+1} \triangleleft^{m+1}.$$

The construction is illustrated in Figure VIII.11.

Moreover:

- if the original machine was halting, then the reachable configurations of the new one are finite and hence regular;
- otherwise, the set of reachable configurations is not regular, which follows from the non-regularity of any infinite subset of $\{\triangleright^n \triangleleft^n \mid n \in \mathbb{N}\}$.   □

*Some proof details.* Letting $\mathcal{T}$ denote the instance of the HALTING PROBLEM, which runs on the empty word, we denote by $\mathcal{T}'$ the instance of the REGULAR REACHABILITY PROBLEM to which it is reduced.

The Turing machine $\mathcal{T}'$ is defined as follows: every time there is a transition $u \cdot p \cdot v \to u' \cdot q \cdot v'$ in $\mathcal{T}$, we simulate this transition in $T'$: to achieve this, '$\triangleright$' and '$\triangleleft$'s should be treated as blank symbols, and then we rewrite $\triangleright^n \triangleleft^m$ into $\triangleright^{n+1} \triangleleft^{m+1}$. When $\mathcal{T}$ writes on a blank symbol that was actually an '$\triangleright$' in $\mathcal{T}'$, we must also add an extra $\triangleright$ (to account for the one that was overwritten): this case is depicted in Figure VIII.11. Moreover, when $\mathcal{T}$ deletes a symbol at the end of the tape, we must shift the $\triangleright^n \triangleleft^m$ suffix. This can be done by replacing the blank with an '$\triangleright$', the last '$\triangleright$' with a '$\triangleleft$', and deleting the last '$\triangleleft$'.

[37] Note that it is decidable whether a Turing machine is linear. In fact, condition (1) can be expressed in first-order logic over $\Sigma^*$.

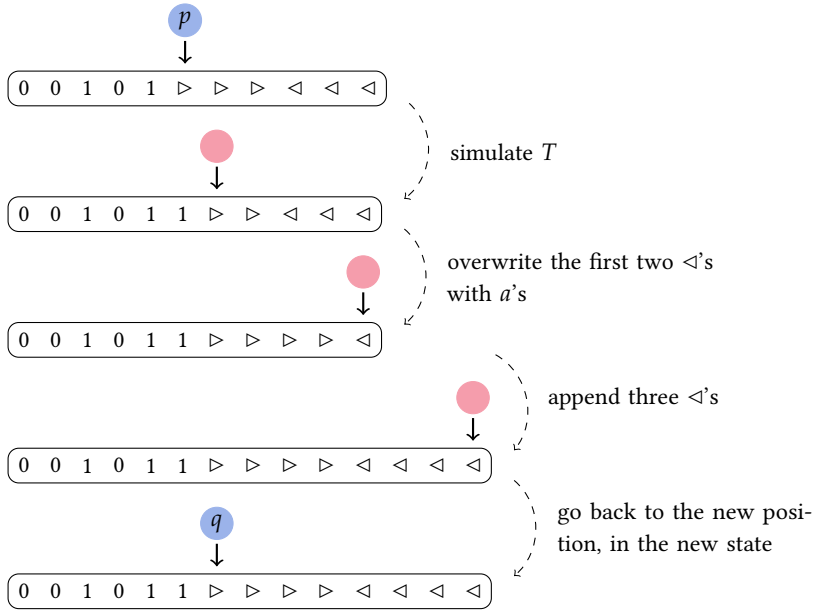

Figure VIII.11: Encoding of a single transition of the form "when reading a blank in state $p$, write a 1, go in state $q$ and move right" of the machine $\mathcal{T}$ in the machine $\mathcal{T}'$ in the proof of Lemma VIII.3.6. Red unlabelled states represent states of $\mathcal{T}'$ that are not originally present in $\mathcal{T}$.

We now prove that $\mathcal{T}'$ is linear:

1. it is deterministic and reversible:
   - every configuration inside a path

     $$u \cdot q \cdot v \triangleright^n \triangleleft^m \to^* u' \cdot q' \cdot v' \triangleright^{n+1} \triangleleft^{m+1}$$

     has, by definition, exactly in- and out-degree one;
   - every configuration of the form $u \cdot q \cdot va^n b^m$ has as many predecessors (resp. successors) in $\mathcal{T}'$ as $u \cdot q \cdot v$ in $\mathcal{T}$, namely one since $\mathcal{T}$ was assumed to be deterministic and reversible;

2. the initial configuration $\_ \cdot q_0 \cdot \_$ has no predecessor;

3. it has no infinite backward path since $\mathbb{N}$ is well-founded,

Moreover, $\mathcal{T}'$ has no cycle,[38] and so if $\mathcal{T}$ is halting on an empty input, then the set of reachable configurations of $\mathcal{T}'$ is finite, and thus regular. If $\mathcal{T}$ is not halting, the set of reachable configurations of $\mathcal{T}'$ is infinite and its projection onto $\{\triangleright, \triangleleft\}$ is an infinite set of words of the form $a^n b^m$ where $n - 2 \le m \le n + 2$. Hence, since regular languages are closed under homomorphic images, the reachable configurations of $\mathcal{T}'$ cannot be regular. $\square$

[38] Indeed, we encoded a strictly increasing counter inside the configurations of $\mathcal{T}'$.

*Undecidability of the regular $k$-colourability Problem.* We can now show undecidability for the REGULAR $k$-COLOURABILITY PROBLEM by reduction from the REGULAR REACHABILITY PROBLEM restricted to linear Turing machines.

A configuration of a Turing machine—or more generally the node of an

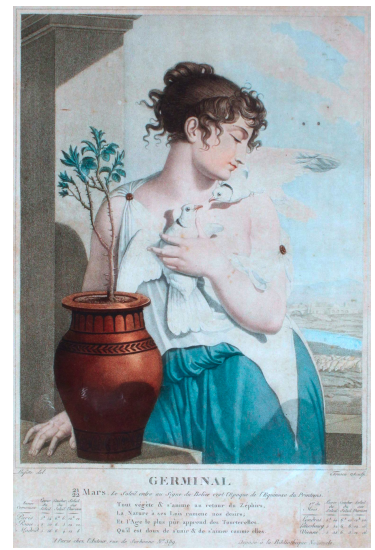

Figure VIII.12: *Allégorie pour le mois de Germinal*, Louis Lafitte.

automatic graph—is said to be *germinal* if it has in-degree 0.[39]

**Theorem VIII.3.7.** The REGULAR $k$-COLOURABILITY PROBLEM on automatic graphs is undecidable for every $k \geq 2$. More precisely, the problem is RE-complete. This holds also for connected automatic graphs.

*Proof.* ❧ *Lower bound.* By reduction from the REGULAR REACHABILITY PROBLEM for linear Turing machines (Lemma VIII.3.6). We first show it for $k = 2$.

Given a linear Turing machine $\mathcal{T}$, observe that the set $\mathrm{Germ}_{\mathcal{T}}$ of all germinal configurations of $\mathcal{C}onf_{\mathcal{T}}$.

*Claim* VIII.3.8. $\mathrm{Germ}_{\mathcal{T}}$ is effectively a regular language.

Observe moreover that, by definition of linear Turing machines, the initial configuration $\_ \cdot q_0 \cdot \_$ is germinal. Let $b$ and $r$ be fresh symbols. Consider the automatic graph $\langle V, \mathcal{E} \rangle$ for $V \triangleq \mathcal{C}onf_{\mathcal{T}} \cdot (b + r)$, having an edge from $\gamma \cdot c$ to $\gamma' \cdot c$ if either

1. $\langle c, c' \rangle = \langle b, r \rangle$ and $\gamma = \gamma'$;
2. $\langle c, c' \rangle = \langle r, b \rangle$ and there is an edge from $\gamma$ to $\gamma'$ in $\mathcal{C}onf_{\mathcal{T}}$; or
3. $\langle c, c' \rangle = \langle b, b \rangle$, $\gamma$ is the initial configuration, and $\gamma' \neq \gamma$ is germinal.

Symbols $b$ and $r$ are utilized to represent two versions of each configuration. This graph is depicted in Figure VIII.14. Note that $\langle V, \mathcal{E} \rangle$ is connected and 2-colourable: in fact, it is a directed tree whose root is $\_ \cdot q_0 \cdot \_ \cdot b$.

We claim that $\langle V, \mathcal{E} \rangle$ is regularly 2-colourable if, and only if, the set of reachable configurations of $T$ is a regular language. In fact, up to permuting the two-colours, $\langle V, \mathcal{E} \rangle$ admits a unique 2-colouring $\langle C_1, C_2 \rangle$, defined by:

$$C_1 \triangleq \mathrm{Reach}_{\mathcal{T}} \cdot b \cup (\mathrm{Conf}_{\mathcal{T}} \setminus \mathrm{Reach}_{\mathcal{T}}) \cdot r$$

and $C_2$ is the complement of $C_1$. If $\mathrm{Reach}_{\mathcal{T}}$ is regular, then so is $C_1$. Dually, if $C_1$ is regular, then $\mathrm{Reach}_{\mathcal{T}}$ is exactly the set of configurations $\gamma$ such that $\gamma \cdot b \in C_1$ and hence is regular. It follows that $\langle V, \mathcal{E} \rangle$ is regularly 2-colourable if and only if the reachable configurations of $\mathcal{T}$ are regular, which concludes the proof for $k = 2$.

To prove the statement for any $k > 2$, we define $\langle V, \mathcal{E}_k \rangle$ as the result of adding a $(k-2)$-clique to $\langle V, \mathcal{E} \rangle$ and adding an edge from every vertex of the clique to every vertex incident to an edge of $\mathcal{E}$. This forces the clique to use $k-2$ colours that cannot be used in the remaining part of the graph and the proof is then analogous.

❧ *Upper-bound.* We show that the problem is RE. Let us define a *$k$-coloured automaton* like a regular (complete) DFA, except that instead of having a set of final states, it has a partition $\langle C_1, \ldots, C_k \rangle$ of its states. Such an automaton recognizes a regular colouring $\Sigma^* \to [\![1, k]\!]$. Given an automatic graph $\mathcal{G} = \langle V, \mathcal{E} \rangle$—whose edge relations is given by a synchronous automaton $\mathcal{E}_{\mathcal{G}}$—and a $k$-coloured automaton $\mathcal{B}$, we can build, by a product construction, an automaton $\mathcal{A}'$ which accepts all $u \otimes v \in R^{\otimes}$ such that the colour of $u$ is distinct from the colour of $v$. Then, $\mathcal{A}'$ is equivalent to $\mathcal{E}_{\mathcal{G}}$ if, and only if, $\mathcal{B}$

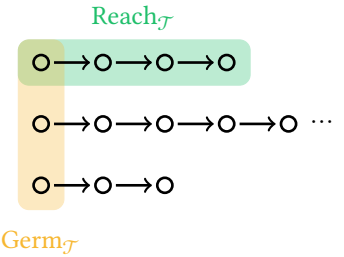

$\mathrm{Reach}_{\mathcal{T}}$

$\mathrm{Germ}_{\mathcal{T}}$

Figure VIII.13: Configuration graph of a linear Turing machine.

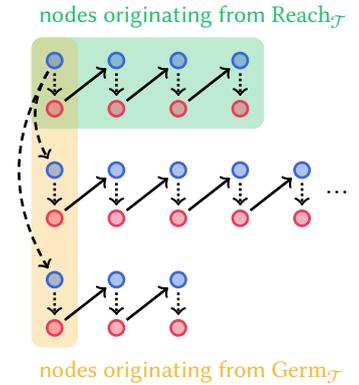

nodes originating from $\mathrm{Reach}_{\mathcal{T}}$

nodes originating from $\mathrm{Germ}_{\mathcal{T}}$

Figure VIII.14: The automatic graph to which the configuration graph of Figure VIII.13 is reduced.

describes a proper $k$-colouring of $\langle V, \mathcal{E} \rangle$. The RE upper bound of the REGULAR $k$-COLOURABILITY PROBLEM follows: it suffices to enumerate all $k$-coloured automata and check for equivalence. $\qquad\square$

Note that this reduction provides an easy way of building graphs in the shape of Figure VIII.14 that are 2-colourable (in fact, they are trees) but not regularly 2-colourable. In fact, we can provide a slightly more direct construction.

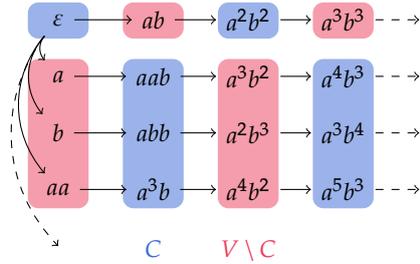

Figure VIII.15: The automatic tree $\mathcal{T}$ of Example VIII.3.9, and its unique 2-colouring $(C, V \setminus C)$, which is not regular.

**Example VIII.3.9.** On the alphabet $\Sigma = \{a, b\}$, the tree $\mathcal{T}$ depicted in Figure VIII.15 whose set of vertices is $V = a^* b^*$ and whose set of edges is $\mathcal{E} = \mathcal{E}_{\text{incr}} \cup \mathcal{E}_{\text{init}}$, with

$$\mathcal{E}_{\text{incr}} = \{(a^p b^q, a^{p+1} b^{q+1}) \mid p, q \in \mathbb{N}\}$$
$$\mathcal{E}_{\text{init}} = \{(\varepsilon, a^p) \mid p \in \mathbb{N}\} \cup \{(\varepsilon, b^q) \mid q \in \mathbb{N}\},$$

is automatic but not regularly 2-colourable. Indeed, its only 2-colouring consists in partitioning the vertices of $\mathcal{T}$ into

$$C = \{a^n b^n \mid n \in 2\mathbb{N}\}$$
$$\cup \{a^p b^q \mid p > q \text{ and } q \text{ is odd}\}$$
$$\cup \{a^p b^q \mid p < q \text{ and } p \text{ is odd}\}$$

and its complement $V \setminus C$. Let $P = \{a^p b^q \mid p, q \in 2\mathbb{N}\} = (aa)^*(bb)^*$: $P$ is regular, yet $C \cap P = \{a^n b^n \mid n \in 2\mathbb{N}\}$ is not. Hence, $C$ is not regular, and thus $\mathcal{T}$ is not regularly 2-colourable. $\qquad\diamondsuit$

### VIII.3.3 Bounded Recognizable Relations

*Separability for Bounded Recognizable Relations.* In this part, we capitalize on the undecidability result of Section VIII.3.2, showing how this implies the undecidability for the separability problem on two natural classes of bounded recognizable relations, namely $k$-Rec and $k$-Prod. For any $k$, $k$-Prod is the subclass of Rec consisting of unions of $k$ Cartesian products of regular languages (which is a subclass of $2^{2k}$-Rec).

First, observe that the 1-Rec-separability problem is trivially decidable, since the only possible separator is $\Sigma^* \times \Sigma^*$. However, for any other $k > 1$, the problem is undecidable.

**Corollary VIII.3.10.** The $k$-REC-SEPARABILITY PROBLEM is undecidable, for every $k > 1$.

*Proof.* This is a consequence of the reduction from the REGULAR $k$-COLOURABILITY PROBLEM of Theorem VIII.3.2, combined with the undecidability of the latter for every $k > 1$ (Theorem VIII.3.7). □

On the $k$-PROD hierarchy we will find the same phenomenon. In particular the case $k = 1$ is also trivially decidable.

**Proposition VIII.3.11.** The 1-PROD-SEPARABILITY PROBLEM is decidable.

*Proof.* Given two automatic relations $\mathcal{R}_1, \mathcal{R}_2$, there exists $S \in$ 1-PROD that separates $\mathcal{R}_1$ from $\mathcal{R}_2$ if and only if $\pi_1(\mathcal{R}_1) \times \pi_2(\mathcal{R}_1)$ separates $\mathcal{R}_1$ from $\mathcal{R}_2$.[40] □

As soon as $k > 1$, the $k$-PROD-SEPARABILITY PROBLEM becomes undecidable.

**Lemma VIII.3.12.** A symmetric automatic relation $\mathcal{R}$ and the identity $\mathcal{Id}$ are separable by a relation in 2-PROD iff they have a separator of the form $(A \times B) \cup (B \times A)$.

*Proof.* Assume that $S \in$ 2-PROD separates $\mathcal{R}$ from $\mathcal{Id}$. Then $\mathcal{R} \subseteq S$, but since $\mathcal{R}$ is symmetric, $\mathcal{R} = \mathcal{R}^{-1} \subseteq S^{-1}$, and so $\mathcal{R} \subseteq S \cap S^{-1}$. Moreover, since $S$ has empty intersection with $\mathcal{Id}$, so does $S \cap S^{-1}$. Hence, $S \cap S^{-1}$ separates $\mathcal{R}$ from $\mathcal{Id}$.

Since $S \in$ 2-PROD, there exists $A_1, A_2, B_1, B_2 \subseteq \Sigma^*$ such that $S = A_1 \times B_1 \cup B_2 \times A_2$. Note that $S \cap \mathcal{Id} = \varnothing$ yields $A_i \cap B_i = \varnothing$ for each $i \in \{1, 2\}$. Finally:

$$\begin{aligned}
S \cap S^{-1} &= \Big(A_1 \times B_1 \cup B_2 \times A_2\Big) \cap \Big(B_1 \times A_1 \cup A_2 \times B_2\Big) \\
&= \Big((A_1 \times B_1) \cap (B_1 \times A_1)\Big) \cup \Big((A_1 \times B_1) \cap (A_2 \times B_2)\Big) \\
&\quad \cup \Big((B_2 \times A_2) \cap (B_1 \times A_1)\Big) \cup \Big((B_2 \times A_2) \cap (A_2 \times B_2)\Big) \\
&= \underbrace{\Big((A_1 \cap B_1) \times (A_1 \cap B_1)\Big)}_{=\varnothing} \cup \Big((A_1 \cap A_2) \times (B_1 \cap B_2)\Big) \\
&\quad \cup \Big((B_1 \cap B_2) \times (A_1 \cap A_2)\Big) \cup \underbrace{\Big((A_2 \cap B_2) \times (A_2 \cap B_2)\Big)}_{=\varnothing} \\
&= \Big((A_1 \cap A_2) \times (B_1 \cap B_2)\Big) \cup \Big((B_1 \cap B_2) \times (A_1 \cap A_2)\Big).
\end{aligned}$$

Therefore, $S \cap S^{-1}$ is a separator of $\mathcal{R}$ and $\mathcal{Id}$ of the desired shape. □

**Corollary VIII.3.13.** A symmetric automatic relation $\mathcal{R}$ and $\mathcal{Id}$ are separable by a relation in 2-PROD iff $\langle \Sigma^*, \mathcal{R} \rangle$ is regularly 2-colourable.

*Proof.* By observing that for any symmetric relation $\mathcal{R} \subseteq \Sigma^* \times \Sigma^*$, we have that $A, B \subseteq \Sigma^*$ is a colouring of $\langle \Sigma^*, \mathcal{R} \rangle$ if, and only if, $(A \times B) \cup (B \times A)$ separates $\mathcal{R}$ from $\mathcal{Id}$. □

[40] Here, $\pi_1(\mathcal{R}_1) \stackrel{\wedge}{=} \{u \in \Sigma^* \mid \exists v \in \Sigma^*, \langle u, v \rangle \in \mathcal{R}_1\}$, and similarly, $\pi_2(\mathcal{R}_1) \stackrel{\wedge}{=} \{v \in \Sigma^* \mid \exists u \in \Sigma^*, \langle u, v \rangle \in \mathcal{R}_1\}$. Both languages can be effectively computed from $\mathcal{R}_1$.

We can now easily show undecidability for the 2-Prod-separability problem by reduction from the regular 2-colourability problem.

**Lemma VIII.3.14.** The 2-Prod-separability problem is undecidable.

*Proof.* By reduction from the regular 2-colourability problem on automatic graphs, which is undecidable by Theorem VIII.3.7. Let $\langle V, \mathcal{E} \rangle$ be an automatic graph and $\langle V, \mathcal{E}' \rangle$ the symmetric closure of $\langle V, \mathcal{E} \rangle$. It follows that $\langle V, \mathcal{E}' \rangle$ is still automatic and that there is a regular 2-colouring for $\langle V, \mathcal{E}' \rangle$ *iff* there is a regular 2-colouring for $\langle V, \mathcal{E} \rangle$—take the same colouring. Thus, by Corollary VIII.3.13, $\langle V, \mathcal{E} \rangle$ is regularly 2-colourable *iff* there is a 2-Prod relation that separates $\mathcal{E}'$ from $\mathcal{I}d$. □

Further, this implies undecidability for every larger $k$.

**Proposition VIII.3.15.** The $k$-Prod-separability problem is undecidable, for every $k \geq 2$.

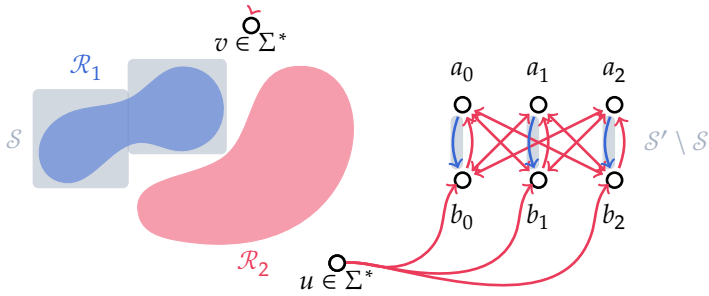

Figure VIII.16: Construction in the proof of Proposition VIII.3.15 for $k = 5$. $\mathcal{S}$ is depicted as the union of two grey rectangles since $\mathcal{S} \in$ 2-Prod. The relation $\mathcal{R}'_1$ is obtained from $\mathcal{R}_1$ (blue shape) by adding all blue edges, namely $a_i \to b_i$ for $i \in [\![1, k-2]\!]$. The relation $\mathcal{R}'_2$ is obtained from $\mathcal{R}_2$ (red shape) by adding all red edges, namely every other non-self-loop edge involving a vertex $a_i$ or $b_i$. Finally, $\mathcal{S}'$ ($\mathcal{S}$ plus three grey rectangles) is obtained from $S$ by adding each $\{a_i\} \times \{b_i\}$.

*Proof.* The case $k = 2$ is shown in Lemma VIII.3.14, so suppose $k > 2$. The proof goes by reduction from the 2-Prod-separability problem. Let $\mathcal{R}_1, \mathcal{R}_2$ be a pair of automatic relations over an alphabet $\Sigma$. Consider the alphabet extended with $2(k-2)$ fresh symbols $\Sigma' = \Sigma \sqcup \{a_1, \cdots, a_{k-2}, b_1, \cdots, b_{k-2}\}$. We build automatic relations $\mathcal{R}'_1, \mathcal{R}'_2$ over $\Sigma'$ such that $\langle \mathcal{R}_1, \mathcal{R}_2 \rangle$ are 2-Prod separable over $\Sigma$ *iff* $\langle \mathcal{R}'_1, \mathcal{R}'_2 \rangle$ are $k$-Prod separable over $\Sigma'$.

Let $\mathcal{R}'_1 \triangleq \mathcal{R}_1 \sqcup \{\langle a_i, b_i \rangle : 1 \leq i \leq k-2\}$ and

$$\mathcal{R}'_2 = \mathcal{R}_2 \sqcup \{\langle a_i, v \rangle \mid v \in \Sigma^*, \, i \in [\![1, k-2]\!]\}$$
$$\sqcup \{\langle u, b_i \rangle \mid u \in \Sigma^*, \, i \in [\![1, k-2]\!]\}$$
$$\sqcup \{\langle a_i, b_j \rangle \mid i, j \in [\![1, k-2]\!] \text{ and } i \neq j\}$$
$$\sqcup \{\langle b_i, a_j \rangle \mid i, j \in [\![1, k-2]\!]\}$$

If $\mathcal{R}_1$ and $\mathcal{R}_2$ have a 2-Prod separator $\mathcal{S}$, then $\mathcal{S} \sqcup \{\langle a_i, b_i \rangle \mid i \in [\![1, k-2]\!]\}$ is a $k$-Prod separator of $\mathcal{R}'_1$ and $\mathcal{R}'_2$.

Conversely, if $\mathcal{S}' = (A_1 \times B_1) \cup \cdots \cup (A_k \times B_k)$ is a $k$-Prod separator of $\mathcal{R}'_1$ and $\mathcal{R}'_2$, then for every $i$ there must be some $j_i$ such that $A_{j_i} \times B_{j_i}$ contains $(a_i, b_i)$. From $\mathcal{S}' \cap \mathcal{R}'_2 = \varnothing$, we get:

- $A_{j_i} \cup B_{j_i}$ cannot contain any $a_{i'}$ or $b_{i'}$ for $i' \neq i$, and

- $A_{j_i} \cup B_{j_i}$ cannot contain any $w \in \Sigma^*$;

since otherwise we would have $(A_{j_i} \times B_{j_i}) \cap \mathcal{R}'_2 \neq \emptyset$. Hence, $\{i \mapsto j_i\}_i$ is injective, and thus $\mathcal{S}'$ is of the form $\mathcal{S}' = (A_1 \times B_1) \cup (A_2 \times B_2) \cup (\{a_1\} \times \{b_1\}) \cup \cdots \cup (\{a_{k-2}\} \times \{b_{k-2}\})$. We can further assume that $A_1, B_1, A_2, B_2$ do not contain any $a_i$ or $b_i$ since otherwise we can remove them preserving the property of being a $k$-Prod separator of $\mathcal{R}'_1$ and $\mathcal{R}'_2$. Hence, $\mathcal{S} \hat{=} (A_1 \times B_1) \cup (A_2 \times B_2)$ must cover $\mathcal{R}_1$ and be disjoint from $\mathcal{R}_2$, obtaining that $\mathcal{S}$ is a 2-Prod separator of $\mathcal{R}_1$ and $\mathcal{R}_2$. □

*Membership for Bounded Recognizable Relations.* Up until now, we have examined two hierarchies of bounded recognizable relations, namely $k$-Prod and $k$-Rec. Our previous analysis demonstrated that, for any element in these hierarchies (where $k > 1$), their separability problem is undecidable. Nevertheless, we will now establish that their membership problem are decidable.

Given an automatic relation $\mathcal{R} \subseteq \Sigma^* \times \Sigma^*$, consider the automatic equivalence relation $\sim_{\mathcal{R}} \subseteq \Sigma^* \times \Sigma^*$, defined as $w \sim_{\mathcal{R}} w'$ if for every $v \in \Sigma^*$ we have

1. $(w, v) \in \mathcal{R}$ iff $(w', v) \in \mathcal{R}$, and
2. $(v, w) \in \mathcal{R}$ iff $(v, w') \in \mathcal{R}$.

It turns out that equivalence classes of $\sim_{\mathcal{R}}$ define the coarsest partition onto which $\mathcal{R}$ can be recognized in terms of $k$-Rec.

**Lemma VIII.3.16.** For every automatic $\mathcal{R} \subseteq \Sigma^* \times \Sigma^*$, $\sim_{\mathcal{R}}$ has index at most $k$ if, and only if, $\mathcal{R}$ is in $k$-Rec.[41]

[41] Recall that the index of an equivalence relation is its number of equivalence classes.

*Proof.* ↪ *Left-to-right.* Assume that $\sim_{\mathcal{R}}$ has the equivalence classes $C_1, \cdots, C_k$. Consider the set $P \subseteq [\![1, k]\!]^2$ of all pairs $\langle i, j \rangle$ such that there are $u_i \in C_i$ and $u_j \in C_j$ with $\langle u_i, u_j \rangle \in \mathcal{R}$. Define the $k$-Rec relation $\mathcal{R}' = \bigcup_{(i,j) \in P} C_i \times C_j$. We claim that $\mathcal{R} = \mathcal{R}'$. In fact, by definition of $\sim_{\mathcal{R}}$, note that if there are $u_i \in C_i$ and $u_j \in C_j$ with $\langle u_i, u_j \rangle \in \mathcal{R}$, then $C_i \times C_j \subseteq \mathcal{R}$. Hence, $\mathcal{R}' \subseteq \mathcal{R}$. On the other hand, for every pair $\langle u, v \rangle \in \mathcal{R}$ there exists $\langle i, j \rangle \in P$ such that $u \in C_i$, $v \in C_j$ implying $\langle u, v \rangle \in \mathcal{R}'$. Hence, $\mathcal{R} \subseteq \mathcal{R}'$.

↪ *Right-to-left.* If $\mathcal{R}$ is a union of products of sets from the partition $C_1 \sqcup ... \sqcup C_k = \Sigma^*$, then every two elements of each $C_i$ are $\sim_{\mathcal{R}}$-related, and thus $\sim_{\mathcal{R}}$ has index at most $k$. □

We can then conclude that the membership problem for $k$-Rec is decidable.

**Corollary VIII.3.17.** The $k$-Rec-membership problem is decidable, for every $k \in \mathbb{N}_{>0}$.

*Proof.* An automatic relation $\mathcal{R}$ is in $k$-Rec iff $\sim_{\mathcal{R}}$ has at most $k$ equivalence classes by Lemma VIII.3.16. In other words, an automatic relation $\mathcal{R}$ is not in $k$-Rec iff the complement of $\sim_{\mathcal{R}}$ contains a $(k+1)$-clique, which can be easily tested. □

The relation $\sim_{\mathcal{R}}$ can also be used to characterize which automatic relations are definable in the class $k$-PROD.

**Proposition VIII.3.18.** An automatic relation $\mathcal{R}$ is in $k$-PROD if, and only if, $\mathcal{R} = (A_1 \times B_1) \cup ... \cup (A_k \times B_k)$ where each $A_i$ and $B_i$ is a union of equivalence classes of $\sim_{\mathcal{R}}$.

*Proof.* The right-to-left implication is trivial. For the converse implication, assume that $\mathcal{R}$ is in $k$-PROD, say

$$\mathcal{R} = (A_1 \times B_1) \cup ... \cup (A_k \times B_k)$$

for some arbitrary regular languages $A_1, ... , A_k$ and $B_1, ... , B_k$. By definition of $\sim_{\mathcal{R}}$, we also have

$$\mathcal{R} = ([A_1]^{\sim_{\mathcal{R}}} \times [B_1]^{\sim_{\mathcal{R}}}) \cup ... \cup ([A_k]^{\sim_{\mathcal{R}}} \times [B_k]^{\sim_{\mathcal{R}}}). \qquad \square$$

Again, this characterization allows us to show that membership in the class $k$-PROD is decidable.

**Corollary VIII.3.19.** The $k$-PROD-MEMBERSHIP PROBLEM is decidable, for every $k > 0$.

*Proof.* By brute force testing whether the automatic relation $\mathcal{R}$ is equivalent to $(A_1 \times B_1) \cup ... \cup (A_k \times B_k)$ for every possible $A_i, B_i$ which is a union of equivalence classes of $\sim_{\mathcal{R}}$. $\qquad \square$

## VIII.4  Undecidability of the Homomorphism Problems

Recall that Theorem VIII.3.7 show that the REGULAR $k$-COLOURABILITY PROBLEM is undecidable for all automatic graphs. This problem can be rephrased in terms of homomorphisms, and naturally leads us to conjecture that for most target structure **B**, the problem of given an automatic presentation $\mathcal{A}$, to decide if there is a regular homomorphism from $\mathcal{A}$ to **B** is undecidable. This is formalized in Theorem VIII.4.1, and we devote the rest of this chapter to proving it.

### VIII.4.1  Overview & Easy Implications of the Dichotomy Theorem

**Theorem VIII.4.1** (*Dichotomy Theorem for Automatic Structures*). Let **B** be a finite $\sigma$-structure. The following are equivalent:

**(DT)_fin-dual**.  **B** has finite duality;

**(DT)_hom-dec**.  $\mathcal{H}om(\mathrm{Aut}, \mathbf{B})$ is decidable;

**(DT)_hom-reg-dec**.  $\mathcal{H}om^{\mathrm{reg}}(\mathrm{Aut}, \mathbf{B})$ is decidable;

**(DT)_equal**.  $\mathcal{H}om(\mathrm{Aut}, \mathbf{B}) = \mathcal{H}om^{\mathrm{reg}}(\mathrm{Aut}, \mathbf{B})$, *i.e.* for any automatic presentation $\mathcal{A}$ of a $\sigma$-structure **A**, there is a homomorphism from **A** to **B** *iff* there is a regular homomorphism from $\mathcal{A}$ to **B**;

**(DT)**$_{\text{first-order}}$. $\mathcal{H}om(\text{All}, \mathbf{B})$ has uniformly first-order definable homomorphisms.[42]

Moreover, when $\mathcal{H}om(\text{Aut}, \mathbf{B})$ and $\mathcal{H}om^{\text{reg}}(\text{Aut}, \mathbf{B})$ are undecidable, they are coRE-complete and RE-complete, respectively. When they are decidable, they are NL.

**Remark VIII.4.2** (Beyond automatic relations). In the statement of Theorem VIII.4.1, automatic structures can be replaced by $\omega$-tree-automatic structures, or even higher-order automatic structures and the statement of the theorem would remain true. The undecidability results obviously remain true when defined on a larger class. Moreover, to prove decidability, notice that the key lemma of Section VIII.5.1, namely Lemma VIII.5.2, actually deals with $\mathcal{H}om(\text{All}, \mathbf{B})$ and not $\mathcal{H}om(\text{Aut}, \mathbf{B})$. The decidability follows from the fact that the first-order theory of any higher-order automatic structure is decidable. ◇

We prove Theorem VIII.4.1 by showing the implications depicted in Figure VIII.17. The most difficult implications are **(DT)**$_{\text{fin-dual}}$ ⇒ **(DT)**$_{\text{first-order}}$, which we prove in Section VIII.5, and the implications **(DT)**$_{\text{hom-dec}}$ ⇒ **(DT)**$_{\text{fin-dual}}$ and **(DT)**$_{\text{hom-reg-dec}}$ ⇒ **(DT)**$_{\text{fin-dual}}$, which we prove by contraposition in Sections VIII.4.2 and VIII.4.3.

On the other hand, the implications **(DT)**$_{\text{first-order}}$ ⇒ **(DT)**$_{\text{equal}}$, **(DT)**$_{\text{equal}}$ ⇒ **(DT)**$_{\text{hom-reg-dec}}$ and **(DT)**$_{\text{equal}}$ ⇒ **(DT)**$_{\text{hom-dec}}$ are straightforward: we prove the first one in Section VIII.5, and the last two in this section. Before showing these implications, we start by proving **(DT)**$_{\text{fin-dual}}$ ⇒ **(DT)**$_{\text{hom-dec}}$.[43]

*Decidability of the Homomorphism Problem.*

**Proposition VIII.4.3.**[44] Let $\mathbf{B}$ be a finite $\sigma$-structure. If $\mathbf{B}$ has finite duality, then $\mathcal{H}om(\text{Aut}, \mathbf{B})$ is decidable in NL.

*Proof.* Given a finite $\sigma$-structure $\mathbf{D}$ with domain $\{d_1, \dots, d_n\}$, we build the first-order sentence

$$\phi_{\mathbf{D}} \triangleq \exists x_1. \; \cdots \; \exists x_n. \bigwedge_{\mathcal{R}_{(k)} \in \sigma} \bigwedge_{\substack{\langle i_1, \dots, i_k \rangle \in [\![1,n]\!]^k \\ \text{s.t. } \langle d_{i_1}, \dots, d_{i_k} \rangle \in \mathcal{R}(\mathbf{D})}} \mathcal{R}(x_{i_1}, \dots, x_{i_k}).$$

By construction, for any arbitrary $\sigma$-structure $\mathbf{A}$, we have $\mathbf{A} \vDash \phi_{\mathbf{D}}$ *iff* $\mathbf{D} \xrightarrow{\text{hom}} \mathbf{A}$. Then, since $\mathbf{B}$ has finite duality, it admits a finite dual $\mathbf{D}_1, \dots, \mathbf{D}_m$. Then

$$\mathbf{A} \vDash \bigwedge_{i=1}^{m} \neg \phi_{\mathbf{D}_i} \quad \textit{iff} \quad \mathbf{A} \xrightarrow{\text{hom}} \mathbf{B}.$$

The conclusion follows from the fact that the data complexity of FIRST-ORDER MODEL CHECKING OF AUTOMATIC STRUCTURES is NL by Proposition VII.3.6. □

Note that Proposition VIII.4.3 still holds when the HOMOMORPHISM PROBLEM takes as input higher-order automatic structures since such structures have a decidable first-order theory.

[42] The notion of uniformly first-order definable homomorphisms is defined in Section VIII.5.1.

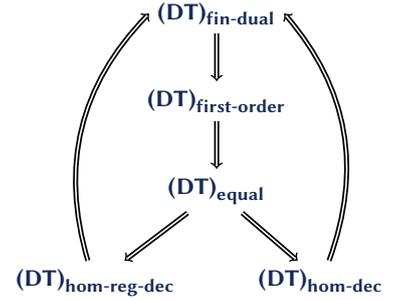

Figure VIII.17: Implications shown in the chapter to prove Theorem VIII.4.1.

[43] While it is redundant with the implications of Figure VIII.17, we prove this implication since not only is it straightforward, but it is also the implication which, together with the fact that both $\mathcal{H}om(\text{Aut}, \mathbf{K}_2)$ and $\mathcal{H}om^{\text{reg}}(\text{Aut}, \mathbf{K}_2)$ are undecidable by [Köc14, Proposition 6.5] and Theorem VIII.3.7, that lead us to conjecture Theorem VIII.4.1.

[44] This corresponds to the implication **(DT)**$_{\text{fin-dual}}$ ⇒ **(DT)**$_{\text{hom-dec}}$ of Theorem VIII.4.1.

*Equality of the Homomorphism Problems Imply their Decidability.*

**Proposition VIII.4.4.** [45] Let $\mathbf{B}$ be a finite $\sigma$-structure. If $\mathcal{H}om(\mathrm{Aut}, \mathbf{B}) = \mathcal{H}om^{\mathrm{reg}}(\mathrm{Aut}, \mathbf{B})$ then $\mathcal{H}om(\mathrm{Aut}, \mathbf{B})$ and $\mathcal{H}om^{\mathrm{reg}}(\mathrm{Aut}, \mathbf{B})$ are decidable.

> [45] This corresponds to the implication $(\mathbf{DT})_{\mathbf{equal}} \Rightarrow (\mathbf{DT})_{\mathbf{hom\text{-}dec}}$ of Theorem VIII.4.1.

To prove this, we first give an upper bound on the homomorphism problems independently of any assumption on $\mathbf{B}$.

**Proposition VIII.4.5.** Let $\mathbf{B}$ be a finite $\sigma$-structure. Then $\mathcal{H}om(\mathrm{Aut}, \mathbf{B})$ is coRE and $\mathcal{H}om^{\mathrm{reg}}(\mathrm{Aut}, \mathbf{B})$ is RE.

*Proof.* ➥ $\mathcal{H}om(\mathrm{Aut}, \mathbf{B})$ *is coRE.* By the De Bruijn–Erdős Theorem, for any arbitrary $\sigma$-structure $\mathbf{A}$, we have $\mathbf{A} \xrightarrow{\mathrm{hom}} \mathbf{B}$ *iff* there exists a finite substructure $\mathbf{A}'$ of $\mathbf{A}$ *s.t.* $\mathbf{A}' \xrightarrow{\mathrm{hom}} \mathbf{B}$. Given a finite $\sigma$-structure $\mathbf{A}'$ and an automatic $\sigma$-structure, it is decidable to test whether $\mathbf{A}'$ is a substructure of $\mathbf{A}$: indeed, it suffices to check, using Proposition VII.3.6 if

$$\mathbf{A} \vDash^? \left( \exists x_1. \; \cdots \; \exists x_n. \bigwedge_{\mathcal{R}_{(k)} \in \sigma} \bigwedge_{\substack{\langle i_1, \ldots, i_k \rangle \in [\![1,n]\!]^k \\ s.t. \; \langle a_{i_1}, \ldots, a_{i_k} \rangle \in \mathcal{R}(\mathbf{A}')}} \mathcal{R}(x_{i_1}, \ldots, x_{i_k}) \right),$$

by letting $\{a_1, \ldots, a_n\} = \mathbf{A}'$. Moreover, whether $\mathbf{A}' \xrightarrow{\mathrm{hom}^?} \mathbf{B}$ is also decidable in coNP. Overall, this provides a co-semi-algorithm for $\mathcal{H}om(\mathrm{Aut}, \mathbf{B})$: we enumerate finite $\sigma$-structure $\mathbf{A}'$, and test if (1) $\mathbf{A}'$ is a substructure of $\mathbf{A}$ and if (2) $\mathbf{A}' \xrightarrow{\mathrm{hom}} \mathbf{B}$. And hence $\mathcal{H}om(\mathrm{Aut}, \mathbf{B})$ is coRE.

➥ $\mathcal{H}om^{\mathrm{reg}}(\mathrm{Aut}, \mathbf{B})$ *is RE.* This is an easy generalization of Theorem VIII.3.7: instead of having $k$-coloured automaton, we define the notion of "$\mathbf{B}$-automata": the notion of accepting state is replaced by a partition $\langle C_b \rangle_{b \in B}$ of the states. The semantics of such an automaton is a partial function $f \colon \Sigma^* \to B$. Given an automatic structure $\mathcal{A}$, we can then effectively test if $f$ is defined on $\mathrm{dom}_\mathcal{A}$, and if $f$ defines a homomorphism from $\mathbf{A}$ to $\mathbf{B}$—see the proof of Theorem VIII.3.7. If so, $\mathcal{A} \xrightarrow{\mathrm{reg\,hom}} \mathbf{B}$. Dually, any regular homomorphism $\mathcal{A} \xrightarrow{\mathrm{reg\,hom}} \mathbf{B}$ can be described by such a $\mathbf{B}$-automaton. Therefore, $\mathcal{H}om^{\mathrm{reg}}(\mathrm{Aut}, \mathbf{B})$ is RE. □

*Proof of Proposition VIII.4.4.* If $\mathcal{H}om(\mathrm{Aut}, \mathbf{B}) = \mathcal{H}om^{\mathrm{reg}}(\mathrm{Aut}, \mathbf{B})$, then by Proposition VIII.4.5, these problems are both RE and coRE, and are hence decidable. □

### VIII.4.2 Undecidability of $\mathcal{H}om(\mathrm{Aut}, \mathbf{B})$

We now prove the undecidability of $\mathcal{H}om(\mathrm{Aut}, \mathbf{B})$ and $\mathcal{H}om^{\mathrm{reg}}(\mathrm{Aut}, \mathbf{B})$ when $\mathbf{B}$ does not have finite duality. Both reductions are direct adaptations of the proof that $\mathcal{H}om(\mathrm{Fin}, \mathbf{B})$ is L-hard when $\mathbf{B}$ does not have finite duality by Larose and Tesson [LT09, Theorem 3.2]. However, proving the undecidability of the problem that is reduced to $\mathcal{H}om^{\mathrm{reg}}(\mathrm{Aut}, \mathbf{B})$ is not entirely trivial and requires some work.

- For $\mathcal{H}om(\text{Fin}, \mathbf{B})$, we reduce the complement of CONNECTIVITY IN AUTOMATIC GRAPHS, providing a coRE-lower bound.
- For $\mathcal{H}om^{\text{reg}}(\text{Aut}, \mathbf{B})$, we reduce REGULAR UNCONNECTIVITY IN AUTOMATIC GRAPHS, which in turn is reduced from the REGULAR REACHABILITY PROBLEM.

For $n \in \mathbb{N}$, we define the *n-link* $\mathbf{L}_n$ be the $\sigma$-structure[46] whose domain is $[\![0, n]\!]$, and every *predicate* $\mathcal{R}$ of arity $k$, is interpreted as the set of tuples $\langle a_1, \ldots, a_k \rangle$ s.t. $|a_i - a_j| \leq 1$ for all $i, j \in [\![0, n]\!]$. See Figure VIII.18.

Given a $\sigma$-structure $\mathbf{B}$, say that $b \in \mathbf{B}$ and $b'$ are *n-linked* if there exists a homomorphism from $\mathbf{L}_n$ to $\mathbf{B}$ that sends 0 to $b$ and $n$ to $b'$. We say that $b$ and $b'$ are *linked* if they are *n*-linked for some $n \in \mathbb{N}$.

Note that the fact that $k \mapsto n - k$ defines an automorphism of $\mathbf{L}_n$ implies that the relation of being *n*-linked—and to a greater extent of being linked—is symmetric. Moreover, being linked is transitive, but not necessarily reflexive.

**Proposition VIII.4.6** ([LLT07, Theorem 4.7]).[47] An arbitrary $\sigma$-structure $\mathbf{B}$ has *finite duality iff* $\pi_1$ and $\pi_2$ are linked in $\mathbf{B}^{(\mathbf{B}^2)}$.

Equipped with the previous proposition, we can now show the undecidability of $\mathcal{H}om(\text{Aut}, \mathbf{B})$ by reduction from the following problem.

---

CONNECTIVITY IN AUTOMATIC GRAPHS
*Input*:    An automatic presentation $\mathcal{G}$ of a directed graph, and two elements $s, t \in \Sigma^*$.
*Question*:  Are $\mathcal{G}(s)$ and $\mathcal{G}(t)$ connected in $\mathbf{G}$?

---

**Property VIII.4.7.** CONNECTIVITY IN AUTOMATIC GRAPHS is RE-complete.

*Proof.* This follows from the fact that the configuration graph of a Turing machine is always automatic by Example VII.3.13. Indeed, a Turing machine halts on the empty word *iff* there is, in its configuration graph, a path from the initial configuration to $\bullet$, where $\bullet$ is a newly added node, *s.t.* we add an edge from any accepting configuration to $\bullet$. $\qquad \square$

**Lemma VIII.4.8.** Assume that $\sigma$ contains at least one predicate of arity at least 2, and let $\mathbf{B}$ be a finite $\sigma$-structure. If $\mathbf{B}$ does not have finite duality, then there is a first-order reduction from the complement of CONNECTIVITY IN AUTOMATIC GRAPHS to $\mathcal{H}om(\text{Aut}, \mathbf{B}^\dagger)$.

*Proof.* Given an instance $\langle \mathcal{G}, s, t \rangle$ of CONNECTIVITY IN AUTOMATIC GRAPHS, we first define the $\sigma$-structure $\mathbf{A}$ with automatic presentation $\mathcal{A}$ obtained by replacing every edge of $\mathcal{G}$ by a 1-link. Formally, $\mathbf{A}$ has the same domain as $\mathbf{G}$, and for any predicate $\mathcal{R} \in \sigma$ of arity $k$, $\langle g_1, \ldots, g_k \rangle \in \mathcal{R}(\mathbf{A})$ *iff* $\{g_1, \ldots, g_k\} = \{g, g'\}$ for some $g, g' \in G$ *s.t.* there is an edge from $g$ to $g'$ in $\mathbf{G}$. See Figures VIII.19 and VIII.20.

*Claim* VIII.4.9.  $\mathcal{G}(s)$ and $\mathcal{G}(t)$ are connected *iff* $\mathcal{A}(s)$ and $\mathcal{A}(t)$ are linked.

[46] From [LLT07, § 2].

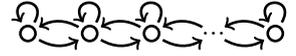

Figure VIII.18: The *n*-link $\mathbf{L}_n$ over the graph signature.

[47] Actually [LLT07, Theorem 4.7] assumes that $\mathcal{H}om(\text{Fin}, \mathbf{B})$ is first-order definable, but this condition is equivalent to $\mathbf{B}$ having finite duality by Atserias' result [Ats08, Corollary 4].

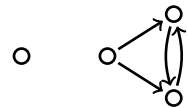

Figure VIII.19: A graph $\mathbf{G}$.

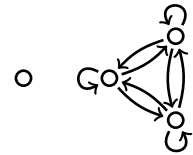

Figure VIII.20: The structure $\mathbf{A}$ defined from $\mathbf{G}$ (in Figure VIII.19), using the construction done in the proof of Lemma VIII.4.8, when $\sigma$ consists of a single binary relation.

For the left-to-right implication: if there is an edge between two elements in $\mathbf{G}$, then they are 1-linked in $\mathbf{A}$. Since being linked is reflexive and transitive, the conclusion follows. Conversely, if two elements $a$ and $a'$ of $\mathbf{A}$ are 1-linked, then pick a predicate $\mathcal{R} \in \sigma$ of arity at least 2. Then $\langle a, \ldots, a, a' \rangle \in \mathcal{R}(\mathbf{A})$, and so by definition of $\mathbf{A}$ there is either an edge from $a$ to $a'$ or from $a'$ to $a$ in $\mathbf{G}$.[48]

We then consider the automatic $\sigma$-structure $\mathbf{A} \times \mathbf{B}^2$, and extend it to a $\sigma_{\mathbf{B}}$-structure $(\mathbf{A} \times \mathbf{B}^2)^\star$ in which for each $b_0 \in B$, we interpret the unary predicate $P_{b_0}$ as

$$\left\{ \langle \mathcal{A}(s), b_0, b \rangle \mid b \in B \right\} \cup \left\{ \langle \mathcal{A}(t), b, b_0 \rangle \mid b \in B \right\}$$

To construct an automatic presentation for this structure, see Section VIII.2.3.

*Claim* VIII.4.10. If there is a homomorphism from $(\mathbf{A} \times \mathbf{B}^2)^\star$ to $\mathbf{B}^\dagger$, then $\mathcal{G}(s)$ and $\mathcal{G}(t)$ are not connected in $\mathbf{G}$.

Let $f \colon (\mathbf{A} \times \mathbf{B}^2)^\star \xrightarrow{\text{hom}} \mathbf{B}^\dagger$ be a homomorphism.[49] It induces a homomorphism

$$\bar{f} \colon \mathbf{A} \times \mathbf{B}^2 \to \mathbf{B}$$

between $\sigma$-structures, and by currying (Proposition VIII.2.2), $\bar{f}$ can be seen as a homomorphism

$$F \colon \mathbf{A} \to \mathbf{B}^{(\mathbf{B}^2)}.$$

Note moreover that because $\bar{f}$ comes from a homomorphism between $\sigma_{\mathbf{B}}$-structures then we must have $f(\mathcal{A}(s), b, b') = b$ and $f(\mathcal{A}(t), b, b') = b'$ for all $b, b' \in B$. This implies that $F(\mathcal{A}(s)) = \pi_1$ and $F(\mathcal{A}(t)) = \pi_2$.

We now assume by contradiction that $\mathcal{G}(s)$ and $\mathcal{G}(t)$ are connected, and hence by Claim VIII.4.9 there is some $n \in \mathbb{N}$ *s.t.* there is a homomorphism $g \colon \mathbf{L}_n \to \mathbf{A}$ with $g(0) = \mathcal{A}(s)$ and $g(n) = \mathcal{A}(t)$. Then by composition, we obtain a homomorphism

$$F \circ g \colon \mathbf{L}_n \to \mathbf{B}^{(\mathbf{B}^2)},$$

which sends 0 to $F(g(0)) = F(\mathcal{A}(s)) = \pi_1$ and sends $n$ to $F(g(n)) = F(\mathcal{A}(t)) = \pi_2$. So, by Proposition VIII.4.6, $\mathbf{B}$ would have finite duality, which is a contradiction. Hence, $\mathcal{A}(s)$ and $\mathcal{A}(t)$ are not linked, and so by Claim VIII.4.9, $\mathcal{G}(s)$ and $\mathcal{G}(t)$ are not connected.

*Claim* VIII.4.11. If $\mathcal{G}(s)$ and $\mathcal{G}(t)$ are not connected in $\mathbf{G}$, then there is a homomorphism from $(\mathbf{A} \times \mathbf{B}^2)^\star$ to $\mathbf{B}^\dagger$.

We define a homomorphism $f \colon (\mathbf{A} \times \mathbf{B}^2)^\star \to \mathbf{B}^\dagger$ by:

$$f(a, b, b') \mathrel{\hat{=}} \begin{cases} b & \text{if } \mathcal{A}(s) \text{ and } a \text{ are linked,} \\ b' & \text{otherwise.} \end{cases}$$

We show that this is indeed a homomorphism: for any predicate $\mathcal{R}$ of arity $k$ in $\sigma$, if

$$\langle a_1, b_1, b'_1 \rangle, \langle a_2, b_2, b'_2 \rangle, \dots, \langle a_k, b_k, b'_k \rangle$$

are all $\mathcal{R}$-tuples of $(\mathbf{A} \times \mathbf{B}^2)^\star$, then by definition of $\mathbf{A}$, we have that either (1) all $a_i$'s are equal, or (2) $\{a_1, \dots, a_k\} = a, a'$ for some $a \neq a' \in A$ and there is an edge from $a$ to $a'$ or from $a'$ to $a$ in $\mathbf{G}$. In both cases, it follows that $\mathcal{A}(s)$ and $a_i$ are linked iff $\mathcal{A}(s)$ and $a_j$ are linked, for all $i, j \in [\![1, k]\!]$. Hence, either:

- $f(a_i, b_i, b'_i) = b_i$ for all $i \in [\![1, k]\!]$ (if all $a_i$'s are connected to $\mathcal{A}(s)$), or
- $f(a_i, b_i, b'_i) = b'_i$ for all $i \in [\![1, k]\!]$ (otherwise).

In both cases, we get that

$$\left\langle f(a_1, b_1, b'_1), f(a_2, b_2, b'_2), \dots, f(a_k, b_k, b'_k) \right\rangle \in \mathcal{R}(\mathbf{B}).$$

We also need to show that this map preserves the new unary predicates of $\sigma_{\mathbf{B}}$: this follows from—and is in fact equivalent to—the fact that $\mathcal{A}(s)$ and $\mathcal{A}(t)$ are not linked by Claim VIII.4.9: indeed, the currying of $f$ behaves like $\pi_1$ on the connected component of $\mathcal{A}(s)$ and like $\pi_2$ on its complement. Overall, this proves that $(\mathbf{A} \times \mathbf{B}^2)^\star \xrightarrow{\text{hom}} \mathbf{B}^\dagger$.

Putting Claims VIII.4.10 and VIII.4.11 together, we get that the reduction is correct. Lastly, note that it is a first-order reduction: clearly, one can go from $\mathbf{G}$ to $\mathbf{A}$ via a (one-dimensional) first-order reduction, and then from $\mathbf{A}$ to $\mathbf{A} \times \mathbf{B}^2$ via a (multi-dimensional) first-order reduction since $\mathbf{B}$ is finite, and lastly $(\mathbf{A} \times \mathbf{B}^2)^\star$ can be obtained by a first-order reduction from the latter structure since first-order logic can test equality with a fixed element. □

By Property VIII.4.7, the complement of CONNECTIVITY IN AUTOMATIC GRAPHS is coRE-complete, and assuming that $\sigma$ contains at least one predicate of arity 2, it reduces by Lemma VIII.4.8 to any problem $\mathcal{H}om(\text{Aut}, \mathbf{B}^\dagger)$ when $\mathbf{B}$ has finite duality. In turn, by Proposition VIII.2.6, it reduces to $\mathcal{H}om(\text{Aut}, \mathbf{B})$, which is thus coRE-hard. It remains to deal with signatures consisting of only unary predicates.[50]

**Property VIII.4.12.** If $\sigma$ only consists of unary predicates, then all $\sigma$-structures have finite duality.

*Proof.* Fix a $\sigma$-structure $\mathbf{B}$. We define the *unary type* $\mu_{\mathbf{B}}(b)$ of $b \in \mathbf{B}$ to be the set of predicates $\mathcal{P}$ s.t. $b \in \mathcal{P}(\mathbf{B})$.

Given $\tau \subseteq \sigma$, define $\mathbf{1}_\tau$ to be the $\sigma$-structure consisting of a single element $*$, and s.t. $* \in \mathcal{P}(\mathbf{1}_\tau)$ iff $\mathcal{P} \in \tau$. We say that $\tau$ is *obstructing* if $\tau \not\subseteq \mu_{\mathbf{B}}(b)$ for all $b \in \mathbf{B}$.

*Claim* VIII.4.13. If $\tau$ is obstructing, then $\mathbf{1}_\tau \xcancel{\xrightarrow{\text{hom}}} \mathbf{B}$.

We prove the result by contraposition. Any homomorphism from $\mathbf{1}_\tau$ to $\mathbf{B}$ should send $*$ on some element $b$ of $\mathbf{B}$ s.t. $b \in \mathcal{P}(\mathbf{B})$ for all $\mathcal{P} \in \tau$, and hence $\tau \subseteq \mu_{\mathbf{B}}(b)$.

[50] It is not clear to us whether this case was properly handled in [LLT07].

*Claim* VIII.4.14. If $\mathbf{A} \xrightarrow{\text{hom}} \mathbf{B}$ then there exists an obstructing $\tau \subseteq \sigma$ *s.t.* $\mathbf{1}_\tau \xrightarrow{\text{hom}} \mathbf{A}$.

We define a partial homomorphism $f$ from $A$ to $B$, by sending $a \in A$ to any $b \in B$ *s.t.* the unary type of $a$ is included in the unary type of $b$. This is clearly a (partial) homomorphism, and so since $\mathbf{A} \xrightarrow{\text{hom}} \mathbf{B}$, it follows that it must be partial, *i.e.* that some element $a \in \mathbf{A}$ *s.t.* $\mu_{\mathbf{A}}(a) \nsubseteq \mu_{\mathbf{B}}(b)$ for every $b \in B$. It follows that $\mu_{\mathbf{A}}(a)$ is obstructing. Since $\mathbf{1}_{\mu_{\mathbf{A}}(a)} \xrightarrow{\text{hom}} \mathbf{A}$ *via* $* \mapsto a$, the conclusion follows.

Putting Claims VIII.4.13 and VIII.4.14 together, we get that

$$\left\{ \, \mathbf{1}_\tau \; \middle| \; \tau \subseteq \sigma \text{ is obstructing} \, \right\}$$

is a finite dual for $\mathbf{B}$. □

**Corollary VIII.4.15.** [51] If $\mathbf{B}$ does not have finite duality, then $\mathcal{H}om(\text{Aut}, \mathbf{B})$ is coRE-hard.

*Proof.* By Property VIII.4.12, since $\mathbf{B}$ does not have finite duality, then $\sigma$ has at least one predicate of arity at least 2. The conclusion follows from Proposition VIII.2.6, Property VIII.4.7, and Lemma VIII.4.8. □

### VIII.4.3 Undecidability of $\mathcal{H}om^{\text{reg}}(\text{Aut}, \mathbf{B})$

The reduction to show undecidability of is nearly identical to Lemma VIII.4.8, but the input problem differs quite a lot.

---

REGULAR UNCONNECTIVITY IN AUTOMATIC GRAPHS

*Input*: An automatic presentation $\mathcal{G}$ of a directed graph $\mathbf{G}$, and two elements $s, t \in \Sigma^*$.

*Question*: Is there a regular language $L \subseteq \Sigma^*$ such that $s \in L$, $t \notin L$ and $L$ is a union of connected components of $\mathcal{G}$?[52] In this case we say that $s$ and $t$ are *regularly unconnected*.

---

We will first reduce this problem to $\mathcal{H}om^{\text{reg}}(\text{Aut}, \mathbf{B})$, and will later settle its complexity.

**Lemma VIII.4.16.** Assume that $\sigma$ contains at least one predicate of arity at least 2. If $\mathbf{B}$ does not have finite duality, then there is a first-order reduction from REGULAR UNCONNECTIVITY IN AUTOMATIC GRAPHS to $\mathcal{H}om^{\text{reg}}(\text{Aut}, \mathbf{B}^\dagger)$.

*Proof.* Given an instance $\langle \mathcal{G}, s, t \rangle$ of REGULAR UNCONNECTIVITY IN AUTOMATIC GRAPHS, we first define the $\sigma$-structure $\mathbf{A}$ with automatic presentation $\mathcal{A}$ obtained by replacing every edge by a 1-link, as in Lemma VIII.4.8.

*Claim* VIII.4.17.[53] $\mathcal{G}(s)$ and $\mathcal{G}(t)$ are regularly unconnected *iff* there is a regular language $L \subseteq \Sigma^*$ *s.t.* $\mathcal{A}(s) \in L$ and $t \notin L$, and $L$ is a union of equivalences classes of $\text{dom}_{\mathcal{A}}$ under "being linked".

The proof is similar to Claim VIII.4.17. Then again, we reduce the instance $\langle \mathcal{G}, s, t \rangle$ to an automatic presentation of $(\mathbf{A} \times \mathbf{B}^2)^\star$, as in Lemma VIII.4.8.

---

[51] In the case of Larose and Tesson, they study the problem $\mathcal{H}om(\text{Fin}, -)$, and prove in [LT09, Theorem 3.2] that there is a first-order reduction from CONNECTIVITY IN FINITE GRAPHS to $\mathcal{H}om(\text{Fin}, \mathbf{B}^\dagger)$ for any $\mathbf{B}$ that does not have finite duality. Together with Proposition VIII.2.6, this shows that $\mathcal{H}om(\text{Fin}, \mathbf{B})$ is L-hard under first-order reductions.

[52] Formally, we mean that $L = \mathcal{G}^{-1}[U]$ for some union $U$ of connected components of $\mathbf{G}$.

[53] While "being linked" is not reflexive in general, it is over the structure $\mathbf{A}$, by reflexivity of "being connected" in $\mathcal{G}$.

*Claim* VIII.4.18. If $(\mathcal{A} \underline{\times} \mathbf{B}^2)^\star \xrightarrow{\text{reg hom}} \mathbf{B}^\dagger$, then $\mathcal{G}(s)$ and $\mathcal{G}(t)$ are regularly unconnected in **G**.

Let $f : (\mathcal{A} \underline{\times} \mathbf{B}^2)^\star \to \mathbf{B}^\dagger$ be a regular homomorphism. By currying—see Corollary VIII.2.5—of the underlying homomorphism between $\sigma$-structures, we obtain a regular homomorphism

$$F : \mathcal{A} \to \mathbf{B}^{(\mathbf{B}^2)}.$$

Moreover, using the predicates $P_b$, $b \in B$, we get that $F(\mathcal{A}(s)) = \pi_1$ and $F(\mathcal{A}(s)) = \pi_2$.

We then define

$$\mathcal{X} \triangleq \{ g \in \mathbf{B}^{(\mathbf{B}^2)} \mid g \text{ and } \pi_1 \text{ are linked or } g = \pi_1 \}.$$

We claim that $F^{-1}[\mathcal{X}]$ witnesses the fact that $\mathcal{G}(s)$ and $\mathcal{G}(t)$ are regularly unconnected. First, $\pi_1 \in \mathcal{X}$ so $\mathcal{A}(s) \in F^{-1}[\mathcal{X}]$. Since **B** has finite duality, by Proposition VIII.4.6, $\pi_2 \notin \mathcal{X}$ and so $\mathcal{A}(t) \notin F^{-1}[\mathcal{X}]$. Then, $F^{-1}[\mathcal{X}]$ is regular since $F$ is a regular homomorphism. Finally, $F^{-1}[\mathcal{X}]$ is a union of equivalences classes of $\text{dom}_\mathcal{A}$ under "being linked".[54] Hence, by Claim VIII.4.17, $\mathcal{G}(s)$ and $\mathcal{G}(t)$ are regularly unconnected.

*Claim* VIII.4.19. If $\mathcal{G}(s)$ and $\mathcal{G}(t)$ are regularly unconnected in **G**, then $\mathcal{A} \underline{\times} \mathbf{B}^2 \xrightarrow{\text{reg hom}} \mathbf{B}^\dagger$.

Since $\mathcal{G}(s)$ and $\mathcal{G}(t)$ are regularly unconnected in **G**, by Claim VIII.4.17 there is a regular language $L \subseteq \Sigma^*$ s.t. $\mathcal{A}(s) \in L$ and $\mathcal{A}(t) \notin L$, and $L$ is a union of equivalences classes of $\text{dom}_\mathcal{A}$ under "being linked". We define a function $f : \text{dom}_\mathcal{A} \times B^2 \to B$ by

$$f(a, b, b') \triangleq \begin{cases} b & \text{if } \mathcal{A}(s) \in L, \\ b' & \text{otherwise,} \end{cases}$$

and we claim that $f$ is a regular homomorphism from $\mathcal{A} \underline{\times} \mathbf{B}^2$ to $\mathbf{B}^\dagger$. The proof that it is a homomorphism is similar to Claim VIII.4.11: in particular, we use the fact that $\mathcal{G}(s)$ and $\mathcal{G}(t)$ are not connected in **G**, which is a consequence of the fact that they are regularly unconnected. Regularity follows from the regularity of $L$. Hence, $\mathcal{A} \underline{\times} \mathbf{B}^2 \xrightarrow{\text{reg hom}} \mathbf{B}^\dagger$.

Putting Claims VIII.4.18 and VIII.4.19 together, we get that the reduction is correct. Lastly, note that this is a first-order reduction for the same reason as Lemma VIII.4.8. □

We then prove a lower bound on the complexity of REGULAR UNCONNEC- TIVITY IN AUTOMATIC GRAPHS.

**Lemma VIII.4.20.** REGULAR UNCONNECTIVITY IN AUTOMATIC GRAPHS is RE-hard.

*Proof.* We reduce the REGULAR REACHABILITY PROBLEM on linear Turing ma-

[54] Indeed, if $c_1, c_2 \in \mathbf{C}$ are linked in some structure **C** and if $f : \mathbf{C} \to \mathbf{D}$ is a homomorphism, then $f(c_1)$ and $f(c_2)$ are linked in **D**.

chines, which is RE-hard by Lemma VIII.3.6, to REGULAR UNCONNECTIVITY IN
AUTOMATIC GRAPHS.

Given a linear Turing machine $\mathcal{T}$ with configuration graph $\mathcal{C}onf_\mathcal{T} = \langle V, \mathcal{E} \rangle$,
we reduce it to the automatic graph $\mathcal{G} = \langle V', \mathcal{E}' \rangle$ where: $V' \doteq V \sqcup \{\bullet\}$ and
$\mathcal{E}'$ is the union of $\mathcal{E}$ with the clique that puts in relation all vertices that are
either of the form $\bullet$ or that are germinal but not the initial configuration. We
then pick $s$ to be the initial configuration, and $t$ to be $\bullet$.

By construction, $\mathcal{G}$ is automatic and has exactly two connected components.
Reach$_\mathcal{T}$ (containing $s$) and its complement (containing $t$). Hence, the only
union of connected components of $\mathcal{G}$ that contains $s$ but not $t$ is Reach$_\mathcal{T}$. And
hence, $\mathcal{T}$ is a positive instance of the REGULAR REACHABILITY PROBLEM if,
and only if, Reach$_\mathcal{T}$ is regular, i.e. $\langle \mathcal{G}, s, t \rangle$ is a positive instance of REGULAR
UNCONNECTIVITY IN AUTOMATIC GRAPHS. □

**Corollary VIII.4.21.** *If* **B** *does not have finite duality, then* $\mathcal{H}om($Aut, **B**$)$ *is*
RE-*hard.*

*Proof.* Recall that $\mathcal{H}om($Aut, **B**$) = \mathcal{H}om($Aut, **B̌**$)$, so we assume *w.l.o.g.* that
**B** is a core. By Lemma VIII.4.20, REGULAR UNCONNECTIVITY IN AUTOMATIC
GRAPHS is RE-hard. Then by Property VIII.4.12, since **B** does not have fi-
nite duality, $\sigma$ does not consist only of unary predicates, and hence by
Lemma VIII.4.16, we get a reduction from REGULAR UNCONNECTIVITY IN AUTO-
MATIC GRAPHS to $\mathcal{H}om^{\text{reg}}($Aut, **B**$^\dagger$$)$, which in turns reduces to $\mathcal{H}om^{\text{reg}}($Aut, **B**$)$
by Proposition VIII.2.6 since **B** was assumed to be a core. Indeed, first-order
reductions preserves regularity, by Proposition VII.2.4. □

## VIII.5 Decidability of the Regular Homomorphism Problem

In this section, we show that if **B** has finite duality, then $\mathcal{H}om^{\text{reg}}($Aut, **B**$)$
is decidable. We provide two alternative proofs: a logic-based one, and a
graph-based one. In Section VIII.5.1 we provide a logic-based proof that if **B**
has finite duality, then $\mathcal{H}om^{\text{reg}}($Aut, **B**$) = \mathcal{H}om($Aut, **B**$)$. In turns, not only
does this implies by Proposition VIII.4.4 that $\mathcal{H}om^{\text{reg}}($Aut, **B**$)$ is decidable, but
actually that the problem is in NL, by Proposition VIII.4.3. Independently, in
Sections VIII.5.2 and VIII.5.3, we introduce the hyperedge consistency algo-
rithm for automatic structures, which is a variation of the classical hyperedge
consistency algorithm for finite structures. We start by explaining the later
algorithm, which solves $\mathcal{H}om($Fin, **B**$)$ for some **B**'s.[55] Then, we will use
the former algorithm to prove that assuming that **B** has finite duality, then
$\mathcal{H}om^{\text{reg}}($Aut, **B**$)$ is decidable.[56]

### VIII.5.1 Uniformly First-Order Definable Homomorphisms

We say that $\mathcal{H}om($Fin, **B**$)$ (resp. $\mathcal{H}om($All, **B**$)$) has *uniformly first-order defin-*
*able homomorphisms* if there exists first-order formulas $\langle \phi_b(x) \rangle_{b \in B}$ over $\sigma$ *s.t.*
for any finite (resp. arbitrary) $\sigma$-structure **A**, for any $a \in A$, there is at most

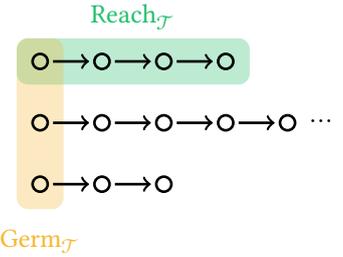

Figure VIII.21: Configuration graph
of a linear Turing machine. (Replica
of Figure VIII.13.)

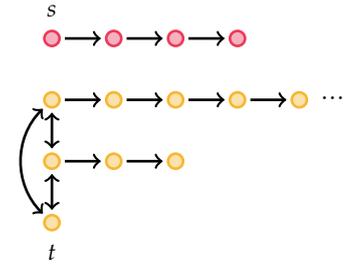

Figure VIII.22: The instance of REGU-
LAR UNCONNECTIVITY IN AUTOMATIC
GRAPHS to which the Turing machine
of Figure VIII.21 is reduced. Colours
indicate the different connected com-
ponents.

[55] We will see later that the algorithm
is correct for $\sigma$-structures with so-
called tree duality, which is a super-
class of the structures with finite du-
ality.

[56] Interestingly, this algorithm can-
not solve $\mathcal{H}om^{\text{reg}}($Aut, **B**$)$ when **B**
has tree duality but not finite duality.





one $b \in B$, denoted by $b(a)$ s.t. $\langle \mathbf{A}, a \rangle \vDash \phi_b(x)$, and moreover if $\mathbf{A} \xrightarrow{\text{hom}} \mathbf{B}$ then $a \mapsto b(a)$ is a homomorphism from $\mathbf{A}$ to $\mathbf{B}$.[57]

**Example VIII.5.1** (Example VIII.2.11, continued). For instance, $\mathcal{H}om(\text{All}, \mathbf{T}_2)$ has uniformly first-order definable homomorphisms, by letting $\phi_0(x)$ be the set of vertices with no predecessors, $\phi_2(x)$ be the set of vertices with no successor (but at least one predecessor), and $\phi_1(x)$ be the set of vertices that satisfy neither $\phi_0$ not $\phi_2$.

On the other hand, looking at the zigzag graph of Example VIII.2.11 and Figure VIII.6 for long enough will convince the reader that no such strategy can work for $\mathbf{P}_2$. ◇

**Lemma VIII.5.2.** Let $\mathbf{B}$ be a finite structure. Then $\mathcal{H}om(\text{All}, \mathbf{B})$ is first-order definable *iff* $\mathcal{H}om(\text{All}, \mathbf{B})$ has uniformly first-order definable homomorphisms.[58]

Before proving this lemma, we show an intermediate result.

**Fact VIII.5.3.** If $\mathbf{B}$ is a finite core, then $\mathcal{H}om(\text{All}, \mathbf{B})$ is first-order definable *iff* $\mathcal{H}om(\text{All}, \mathbf{B}^\dagger)$ is first-order definable.

*Proof.* By Proposition VIII.2.6 the problems are first-order equivalent and so one is first-order definable *iff* the other is. □

*Proof of Lemma VIII.5.2.* ☛ *Converse implication.* Assume that $\mathcal{H}om(\text{All}, \mathbf{B})$ has uniformly first-order definable homomorphisms, say by $\langle \phi_b(x) \rangle_{b \in B}$. Then the conjunctions of the properties "every $x$ must satisfy exactly one $\phi_b(x)$ ($b \in \mathbf{B}$)", and "for every predicate $\mathcal{R}$ of arity $k$, for any $\langle x_1, \dots, x_k \rangle$ in $\mathcal{R}$, there exists $\langle b_1, \dots, b_k \rangle \in \mathcal{R}(\mathbf{B})$ s.t. each $x_i$ satisfies $b_i$ ($i \in [\![1, k]\!]$)" is a first-order sentence describing all $\sigma$-structures of $\mathcal{H}om(\text{All}, \mathbf{B})$.

☛ *Direct implication.* Let $\mathbf{B}$ be such that $\mathcal{H}om(\text{All}, \mathbf{B})$ is first-order definable. Given an arbitrary $\sigma$-structure $\mathbf{A}$, we define a function $F \colon A \to \mathfrak{P}(B)$ by mapping each $a$ to the set of $b$'s ($b \in B$) s.t. there is a homomorphism from $\mathbf{A}$ to $\mathbf{B}$ that maps $a$ to $b$.

*Claim* VIII.5.4. If $\mathbf{A} \xrightarrow{\text{hom}} \mathbf{B}$ then $F$ is a homomorphism from $\mathbf{A}$ to $\mathfrak{U}(\mathbf{B})$.

Indeed, since $\mathbf{A} \xrightarrow{\text{hom}} \mathbf{B}$, for each $a \in A$ the set $F(a)$ is non-empty subset of $B$—and hence an element of the domain of $\mathfrak{U}(\mathbf{B})$. We then prove that it is a homomorphism: let $\mathcal{R}$ be a predicate of arity $l$, and let $\langle a_1, \dots, a_l \rangle \in \mathcal{R}(\mathbf{A})$. Then for each $i \in [\![1, l]\!]$, for every $b_i \in F(a_i)$, there exists a homomorphism $f$ from $\mathbf{A}$ to $\mathbf{B}$ that sends $a_i$ to $b_i$. Then $f(a_j) \in F(a_j)$ for every $j \in [\![1, l]\!]$ and moreover $\langle f(a_1), \dots, f(a_l) \rangle \in \mathcal{R}(\mathbf{B})$. Hence, $\langle F(a_1), \dots, F(a_l) \rangle \in \mathcal{R}(\mathfrak{U}(\mathbf{B}))$, which concludes the proof that $F$ is a homomorphism from $\mathbf{A}$ to $\mathfrak{U}(\mathbf{B})$.

By Atserias' theorem, since $\mathcal{H}om(\text{All}, \mathbf{B})$ is first-order definable, then $\mathbf{B}$ has finite duality, and in particular it has tree duality (by Proposition VIII.2.14) and so by Proposition VIII.2.16, there exists a homomorphism $g \colon \mathfrak{U}(\mathbf{B}) \to \mathbf{B}$. We will now produce first-order formulas to describe $g \circ F$.

If $\mathcal{H}om(\text{All}, \mathbf{B})$ is first-order definable, then so is $\mathcal{H}om(\text{All}, \mathbf{B}^\dagger)$ by Fact VIII.5.3.


[57] The adverb "uniformly" in "uniformly first-order definable homomorphisms" refers to the fact that the formulas do not depend on the source structure.

[58] The same equivalence holds if one replaces $\mathcal{H}om(\text{All}, \mathbf{B})$ with $\mathcal{H}om(\text{Fin}, \mathbf{B})$. In both cases, these conditions are equivalent, by Atserias' theorem, to asking whether $\mathbf{B}$ has finite duality. This corresponds to the implication $(\mathbf{DT})_{\text{fin-dual}} \Rightarrow (\mathbf{DT})_{\text{first-order}}$ of Theorem VIII.4.1 and its converse implication.






So, let $\phi$ be a first-order formula over $\sigma_{\mathbf{B}}$ that describes $\mathcal{H}om(\mathrm{All}, \mathbf{B}^{\dagger})$. We let $B = \{b_1, \ldots, b_k\}$. We define a first-order formula $\phi_i^*(x_i)$ over $\sigma$, by substituting each occurrence of $P_{b_i}(y)$ in $\phi$ for $y = x_i$, and $P_{b_j}(y)$ $(j \neq i)$ for $\bot$. Let $\mathbf{A}$ be a finite $\sigma$-structure, $a \in A$ and $i \in [\![1, k]\!]$ and $\mathbf{A}_{a,i}$ be the $\sigma_{\mathbf{B}}$-structure obtained by letting $P_{b_i}(\mathbf{A}_{a,i}) \mathrel{\hat{=}} \{a\}$ and $P_{b_j}(\mathbf{A}_{a,i}) \mathrel{\hat{=}} \varnothing$ for all $j \neq i$.

*Claim* VIII.5.5. $\mathbf{A}_{a,i} \vDash \phi$ iff $\langle \mathbf{A}, a \rangle \vDash \phi_i^*(x_i)$.

We prove it by induction on formulas $\psi(\bar{x})$ that $\langle \mathbf{A}_{a,i}, \bar{a} \rangle \vDash \psi$ iff $\langle \mathbf{A}, \bar{a}, a \rangle \vDash \psi_i^*(x_i)$. The base case $P_{b_i}(y)$ is trivial since $\langle \mathbf{A}_{a,i}, a' \rangle \vDash P_{b_i}(y)$ iff $a' = a$ i.e. $\langle \mathbf{A}, a' , a \rangle \vDash y = x_i$. Similarly, for $P_{b_j}(y)$ $(j \neq i)$, we have $\langle \mathbf{A}_{a,i}, a' \rangle \nvDash P_{b_j}(y)$ and so this is equivalent to $\langle \mathbf{A}, a', a \rangle \vDash \bot$. The other atomic cases, and inductive cases are trivial.

*Claim* VIII.5.6. There exist first-order formulas $\langle \chi_Y(x) \rangle_{Y \in \mathfrak{P}(B)}$, that do not depend on $\mathbf{A}$, s.t. for every arbitrary $\sigma$-structure $\mathbf{A}$ and for every $a \in A$, we have $\langle \mathbf{A}, a \rangle \vDash \chi_Y(x)$ iff $F(a) = Y$.

Indeed, given $a \in A$ and $i \in [\![1, k]\!]$, there is a homomorphism from $\mathbf{A}$ to $\mathbf{B}$ that sends $a$ to $b_i$ iff $\mathbf{A}_{a,i} \vDash \phi$, and so by Claim VIII.5.5, this is equivalent to $\langle \mathbf{A}, a \rangle \vDash \phi_i^*(x_i)$. Hence, each $\chi_Y(x)$ can be defined as a Boolean combination of the $\phi_i^*(x_i)$'s, after renaming $x_i$ to $x$.[59]

We can now prove that $\mathcal{H}om(\mathrm{All}, \mathbf{B})$ has uniformly first-order definable homomorphisms. For each $b \in B$, we let $\psi_b(x) \mathrel{\hat{=}} \bigvee_{Y \in g^{-1}[b]} \chi_Y(x)$. Now for any arbitrary $\sigma$-structure $\mathbf{A}$, for any $a \in A$, there is at most one $b \in B$ s.t. $\langle \mathbf{A}, a \rangle \vDash \psi_b(x)$—indeed, there is a unique $Y \in \mathfrak{P}(B)$ (and so at most one $Y \in \mathfrak{P}_+(B)$) s.t. $\langle \mathbf{A}, a \rangle \vDash \chi_Y(x)$ by Claim VIII.5.6. Furthermore, if $\mathbf{A} \xrightarrow{\mathrm{hom}} \mathbf{B}$, then for each $a$ there is a unique $b(a) \in B$ s.t. $\langle \mathbf{A}, a \rangle \vDash \psi_{b(a)}(x)$, and moreover $a \mapsto b(a)$ is a homomorphism by Claim VIII.5.4. In turns, using Claim VIII.5.4, we get that for each $a \in A$, there is exactly one $b(a) \in B$ s.t. $\langle \mathbf{A}, a \rangle \vDash \phi_{b(a)}(x)$, and that if $\mathbf{A} \xrightarrow{\mathrm{hom}} \mathbf{B}$, then $a \mapsto b(a)$ is a homomorphism—that is equal to $g \circ F$. And hence, $\mathcal{H}om(\mathrm{All}, \mathbf{B})$ has uniformly first-order definable homomorphisms. $\qquad\square$

Uniformly first-order definable homomorphisms are actually a very strong restriction: we show that such homomorphisms are always regular.

**Proposition VIII.5.7.**[60] Let $\mathbf{B}$ be a finite $\sigma$-structure. If $\mathcal{H}om(\mathrm{All}, \mathbf{B})$ has uniformly first-order definable homomorphisms, then $\mathcal{H}om^{\mathrm{reg}}(\mathrm{Aut}, \mathbf{B}) = \mathcal{H}om(\mathrm{Aut}, \mathbf{B})$.

*Proof.* Let $\mathcal{A}$ be an automatic presentation of a $\sigma$-structure $\mathbf{A}$, and assume that $\mathbf{A} \xrightarrow{\mathrm{hom}} \mathbf{B}$. We need to show that $\mathcal{A} \xrightarrow{\mathrm{reg\,hom}} \mathbf{B}$. Let $\langle \phi_b(x) \rangle_{b \in B}$ be first-order formulas over $\sigma$ as in the definition of uniformly first-order definable homomorphisms.

Since $\mathcal{A}$ is an automatic presentation over $\Sigma$, for each predicate $\mathcal{R}$ of arity $k$ of $\sigma$, there exists a first-order formula $\psi_{\mathcal{R}}(x_1, \ldots, x_k)$ over $\sigma_{\Sigma}^{\mathrm{sync}}$ describing each relation $\mathcal{R}$. We then define $\phi_b^*(x)$ as the formulas obtained from $\phi_b(x)$

---

[59] In particular, note that $\forall x.\neg \chi_{\varnothing}(x)$ is a first-order formula that defines $\mathcal{H}om(\mathrm{All}, \mathbf{B})$ since for any $\sigma$-structure, $\mathbf{A} \xrightarrow{\mathrm{hom}} \mathbf{B}$ iff $F(a) \neq \varnothing$ for all $a \in A$.

[60] This corresponds to the implication $(\mathbf{DT})_{\text{first-order}} \Rightarrow (\mathbf{DT})_{\text{equal}}$ of Theorem VIII.4.1.





by substituting $\mathcal{R}(x_1, \ldots, x_k)$ for $\psi_{\mathcal{R}}(x_1, \ldots, x_k)$.

Then, for each $b \in B$, $\phi_b^*(x)$ is a first-order formula over $\sigma_{\Sigma}^{\mathrm{sync}}$, and so

$$\{u \in \Sigma^* \mid \langle \Sigma^*, u \rangle \vDash \phi_b^*(x)\}$$

is regular by Proposition VII.2.4. Clearly, these sets are disjoint and cover $\mathrm{dom}_{\mathcal{A}}$, and the function that maps $u \in \mathrm{dom}_{\mathcal{A}}$ to the unique $b$ s.t. $\langle \Sigma^*, u \rangle \vDash \phi_b^*(x)$ is a homomorphism. Hence, we have built a regular homomorphism from $\mathcal{A}$ to $\mathbf{B}$, which concludes the proof. □

**Corollary VIII.5.8** (of Atserias' theorem, Lemma VIII.5.2 and Proposition VIII.5.7). If $\mathbf{B}$ has finite duality, then $\mathcal{Hom}^{\mathrm{reg}}(\mathrm{Aut}, \mathbf{B}) = \mathcal{Hom}(\mathrm{Aut}, \mathbf{B})$.

In turn, since $\mathcal{Hom}(\mathrm{Aut}, \mathbf{B})$ is coRE and $\mathcal{Hom}^{\mathrm{reg}}(\mathrm{Aut}, \mathbf{B})$ is RE (Proposition VIII.4.5), this implies that $\mathcal{Hom}^{\mathrm{reg}}(\mathrm{Aut}, \mathbf{B})$ is decidable. In fact, using the formulas $\phi_b^*(x)$, we can build a first-order formula saying "every $x$ satisfies exactly one $\phi_b^*(x)$, and moreover if $\langle x_1, \ldots, x_k \rangle$ is an $\mathcal{R}$-tuple then $\langle b_1, \ldots, b_k \rangle$ is an $\mathcal{R}$-tuple, where $b_i$ is the unique element of $B$ s.t. $\phi_{b_i}(x_i)$ holds". Each property "$\langle x_1, \ldots, x_k \rangle$ is an $\mathcal{R}$-tuple" can be expressed using a first-order formula expressing the relations of the automatic presentation given as input.

**Corollary VIII.5.9** (of Atserias' theorem and Lemma VIII.5.2). Let $\mathbf{B}$ be a finite $\sigma$-structure with finite duality. For each automatic presentation $\mathcal{A}$ over alphabet $\Sigma$, there exists a first-order formula $\phi$, whose size is linear in $\mathcal{A}$, s.t. $\Sigma^* \vDash \phi$ iff $\mathcal{A} \xrightarrow{\mathrm{reg\ hom}} \mathbf{B}$.

In particular, this implies the decidability of $\mathcal{Hom}^{\mathrm{reg}}(\mathrm{Aut}, \mathbf{B})$.

## VIII.5.2  Hyperedge Consistency for Finite Structures

Given a homomorphism $f \colon \mathbf{G} \to \mathbf{H}$ between graphs, note that if $g \in G$ has at least one successor in $\mathbf{G}$, then $f(g)$ must also have one successor in $\mathbf{H}$. As a consequence, such an $g$ cannot be mapped by any homomorphism to a vertex of $\mathbf{H}$ with no successor. The idea behind hyperedge consistency is precisely to identify for each $g \in G$ the set $\mathrm{Im}_g$ of all elements of $\mathbf{H}$ to which it can be mapped: initially this set is $H$, and we try to find some "obstructions". These obstructions take the following form: if $g \in G$ has a successor (resp. predecessor) $g' \in G$, then any vertex of $\mathrm{Im}_g$ must have a successor (resp. predecessor) in $\mathbf{H}$ that lives in $\mathrm{Im}_{g'}$—see Figure VIII.23.

**Example VIII.5.10** (Example VIII.5.1, continued). We depict in Figure VIII.24 the first steps of the hyperedge consistency algorithm—that we will define formally after this example—, when the target structure if $\mathbf{T}_2$ and the source structure is $\mathbf{Z}_2^{(n)}$. The second step is a fixpoint, and so the procedure stops there. Note also that each $\mathrm{Im}_g$ ($g \in \mathbf{Z}_2^{(n)}$) is non-empty. ◇

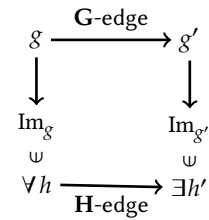

Figure VIII.23: Diagrammatic representation of the hyperedge consistency algorithm.





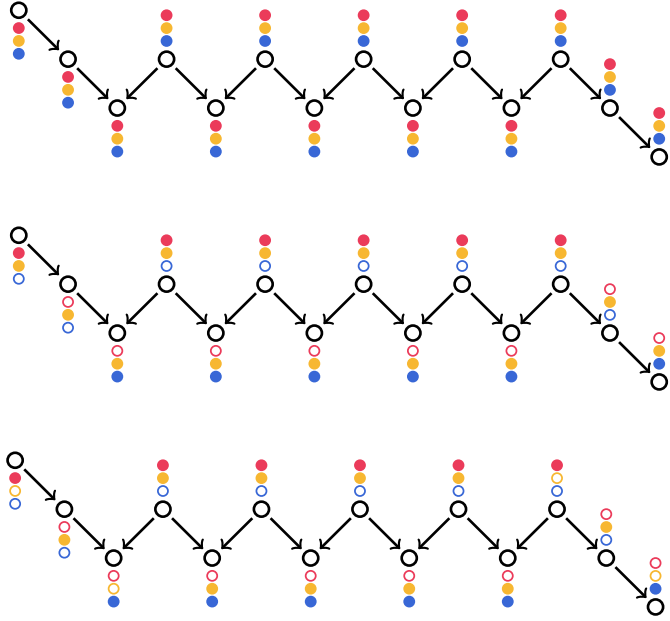

Figure VIII.24: Zeroth (top), first (middle) and second step (bottom) of the hyperedge consistency algorithm on $\mathbf{Z}_2^{(n)}$ when the target structure is $\mathbf{T}_2$, depicted in Figure VIII.25. Next to each vertex $g$ of $\mathbf{Z}_2^{(n)}$ we represent all vertices $h$ of $\mathbf{T}_2$: the vertex is filled when $h \in \mathrm{Im}_g$.

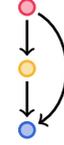

Figure VIII.25: The 2-transitive tournament $\mathbf{T}_2$.

We formalize this algorithm as the greatest fixpoint of some operator. Given a finite $\sigma$-structure $\mathbf{B}$, and an arbitrary[61] $\sigma$-structure $\mathbf{A}$, we say that a function $F \colon A \to \mathfrak{P}(B)$ is subsumed by $G \colon A \to \mathfrak{P}(B)$, denoted by $F \sqsubseteq G$, if $F(a) \subseteq G(a)$ for each $a \in A$. We denote by $\langle \mathfrak{P}(B)^A, \sqsubseteq \rangle$ the set of functions $A \to \mathfrak{P}(B)$ under this order.[62]

We then define an operator on this space, which corresponds to one step of the hyperedge consistency algorithm:

$$\mathcal{HC}_{\mathbf{A},\mathbf{B}} \colon \quad \langle \mathfrak{P}(B)^A, \sqsubseteq \rangle \quad \to \quad \langle \mathfrak{P}(B)^A, \sqsubseteq \rangle$$
$$F \quad \mapsto \quad \mathcal{HC}_{\mathbf{A},\mathbf{B}}(F),$$

where for each $a \in A$, $\mathcal{HC}_{\mathbf{A},\mathbf{B}}(F)(a)$ is the set of $b \in F(a)$ s.t. for every $\mathcal{R}_{(k)} \in \sigma$, for every $i \in [\![1, k]\!]$, if $\langle a_1, \dots, a_{k-1} \rangle \in \mathcal{A}dj_{\mathbf{A}}^{\mathcal{R},i}(a)$, then there exists $b_1 \in F(a_1)$, $\dots, b_{k-1} \in F(a_{k-1})$ s.t. $\langle b_1, \dots, b_{k-1} \rangle \in \mathcal{A}dj_{\mathbf{B}}^{\mathcal{R},i}(b)$.[63]

**Fact VIII.5.11.** The ordered set $\langle \mathfrak{P}(B)^A, \sqsubseteq \rangle$ is a complete lattice, and moreover $\mathcal{HC}$ is monotonic.

As a consequence of the Knaster-Tarski theorem, $\mathcal{HC}$ admits a greatest fixpoint, that we denote by $H_{\mathbf{A},\mathbf{B}}^\star \in \langle \mathfrak{P}(B)^A, \sqsubseteq \rangle$.[64]

**Proposition VIII.5.12.** If $f \colon \mathbf{A} \to \mathbf{B}$ is a homomorphism then $f(a) \in H_{\mathbf{A},\mathbf{B}}^\star(a)$ for each $a \in A$.

*Proof.* The property "$f(a) \in F(a)$ for each $a \in A$" holds for the greatest element of $\langle \mathfrak{P}(B)^A, \sqsubseteq \rangle$, is stable under application of $\mathcal{HC}$ and under arbitrary meets.[65] Hence, by ordinal induction, it holds for $H_{\mathbf{A},\mathbf{B}}^\star$. □

**Corollary VIII.5.13.** If $H_{\mathbf{A},\mathbf{B}}^\star(a) = \varnothing$ for some $a \in A$, then $\mathbf{A} \xnrightarrow{\mathrm{hom}} \mathbf{B}$.

In general, the converse property does not hold. For instance, if $\sigma$ is the graph signature and $\mathbf{B}$ is the 2-clique—or more generally any clique—, then

---

[61] Note that in this part, while some results—mostly complexity/decidability results—require the assumption that the source structure is finite, some results do not, and are stated for arbitrary structures.

[62] Equivalently, $\langle \mathfrak{P}(B)^A, \sqsubseteq \rangle$ is the set of binary relations between $A$ and $B$, ordered by inclusion.

[63] We write $\mathcal{HC}$ for $\mathcal{HC}_{\mathbf{A},\mathbf{B}}$ when there is no ambiguity on the structures involved.

[64] Recall that this greatest fixpoint can be obtained by ordinal induction by starting from the greatest element of $\langle \mathfrak{P}(B)^A, \sqsubseteq \rangle$, namely the map $a \mapsto B$, and iterating $\mathcal{HC}$.

[65] Meaning that if all $F_i$ ($i \in I$ for some arbitrary set $I$) satisfy the property, then so does $a \mapsto \bigcap_{i \in I} F(a)$





$H^*_{\mathbf{A},\mathbf{B}}$ is always the map $a \mapsto B$, no matter whether there is a homomorphism from $\mathbf{A}$ to $\mathbf{B}$.

Yet, Larose, Loten and Tardif managed to identify a necessary and sufficient condition on $\mathbf{B}$ for the hyperedge consistency algorithm to decide whether $\mathbf{A} \in \mathscr{H}om(\mathrm{Fin}, \mathbf{B})$.

**Proposition VIII.5.14** ([LLT07, Theorem 3.3]). *If $\mathbf{B}$ has tree duality then $\mathbf{A} \xrightarrow{\mathrm{hom}} \mathbf{B}$ iff $H^*_{\mathbf{A},\mathbf{B}}(a) \neq \varnothing$ for all $a \in A$.*

When $\mathbf{A}$ is moreover finite, this immediately gives an algorithm to decide $\mathbf{A} \xrightarrow{\mathrm{hom}} \mathbf{B}$ since $H^*_{\mathbf{A},\mathbf{B}}$ can be computed not only by an ordinal induction but with a finite induction.

As an example, Example VIII.5.10 witnesses that $\mathbf{Z}_2^{(n)} \xrightarrow{\mathrm{hom}} \mathbf{T}_2$ since $\mathbf{T}_2$ has tree duality by Proposition VIII.2.14.

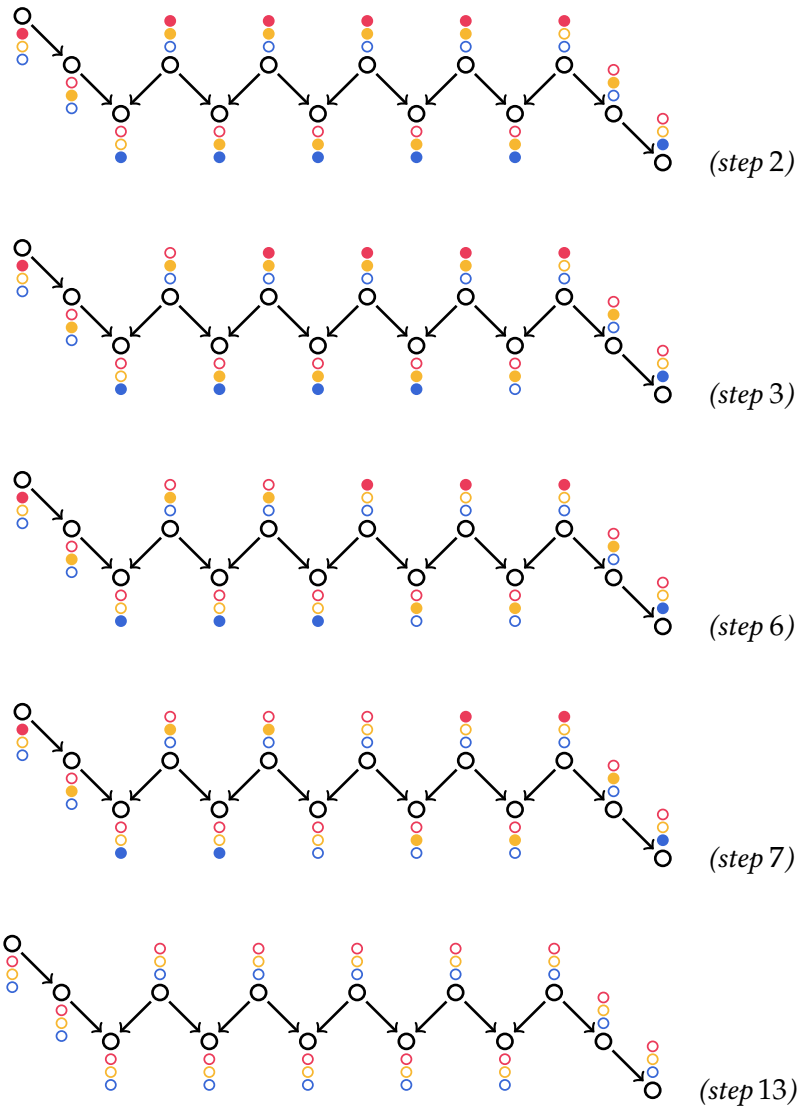

Figure VIII.26: Steps 2, 3, 6, 7 and 13 of the hyperedge consistency algorithm on $\mathbf{Z}_2^{(5)}$, when the target structure is $\mathbf{P}_2$, depicted in Figure VIII.27.

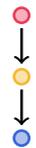

Figure VIII.27: The 2-path $\mathbf{P}_2$.

**Example VIII.5.15** (Example VIII.5.10, continued). While $\mathbf{P}_2$ does not have finite duality (Example VIII.2.11), it has tree duality (Proposition VIII.2.15), and so the hyperedge consistency algorithm decides whether a finite $\sigma$-structure has a homomorphism to $\mathbf{P}_2$. We represent some steps of the algorithm in





Figure VIII.26, on the source structure $\mathbf{Z}_2^{(5)}$. Steps 0, 1 and 2 of the hyper-edge consistency algorithm are identical to Example VIII.5.10. Yet, in step 2, we have not reached the fixpoint. In step 7, this is the first time we have $\mathcal{HC}^7(\Lambda_{\mathbf{P}_2})(g) = \varnothing$ for some $g \in \mathbf{Z}_2^{(5)}$. This propagates until step 13, when $\mathcal{HC}^{13}(\Lambda_{\mathbf{P}_2})(g) = \varnothing$ for all $g \in \mathbf{Z}_2^{(5)}$. This is of course the fixpoint of $\mathcal{HC}$, proving that $H_{\mathbf{Z}_2^{(5)}, \mathbf{P}_2}^*$ is the constant map equal to $\varnothing$, and by Corollary VIII.5.13 that $\mathbf{Z}_2^{(5)} \xrightarrow{\text{hom}} \mathbf{P}_2$.

In general, on source structure $\mathbf{Z}_2^{(n)}$ (with $n \in \mathbb{N}_{>0}$), the smallest $k$ s.t. $\mathcal{HC}^k(\Lambda_{\mathbf{P}_2})(g) = \varnothing$ for some $g \in \mathbf{Z}_2^{(n)}$ is of size $\frac{n}{2} + \mathcal{O}(1)$, and if we want to the property to hold for *all* $g$'s, then $k$ has size $n + \mathcal{O}(1)$. $\diamondsuit$

**Corollary VIII.5.16.** If $\mathbf{B}$ has tree duality, then $\mathcal{H}om(\text{Fin}, \mathbf{B})$ can be solved in polynomial time using the algorithm of Figure VIII.28.

However, note that the example of $\mathbf{K}_2$ shows that the property of having tree duality, is not necessary for $\mathcal{H}om(\text{Fin}, \mathbf{B})$ to be decidable in P.

**Input:** Two finite $\sigma$-structures $\mathbf{A}$ and $\mathbf{B}$.
$\text{Im}_a^0 \leftarrow B$ for $a \in A$;
$n \leftarrow 0$;
**do**
> **for** $a \in A$ **do**
>> $\text{Im}_a^{n+1} \leftarrow \{b \in \text{Im}_a^n \mid$
>>> $\forall \langle a_1, \ldots, a_{k-1} \rangle \in \mathcal{A}dj_{\mathbf{A}}^{\mathcal{R},i}(a)$,
>>> $\exists \langle b_1, \ldots, b_{k-1} \rangle \in \mathcal{A}dj_{\mathbf{B}}^{\mathcal{R},i}(b)$,
>>> $b_1 \in F(a_1) \wedge \ldots \wedge b_{k-1} \in F(a_{k-1})$
>> $\}$;
>> **if** $\text{Im}_a^{n+1} = \varnothing$ **then**
>>> **return** *false*;
> $n \leftarrow n+1$;
**while** *some $\text{Im}_a$ has been updated*;
**return** *true*

Figure VIII.28: The *hyperedge consistency algorithm for finite structures*.

### VIII.5.3 Hyperedge Consistency for Automatic Structures

When $\mathbf{A}$ is automatic, Proposition VIII.5.14 still applies, however it is not clear how to compute $H_{\mathbf{A},\mathbf{B}}^*$ since this element cannot necessarily be obtained by finite induction, namely as $\mathcal{HC}^n(\Lambda_B)$ for some $n \in \mathbb{N}$, where $\Lambda_B$ is the maximum element of $\langle \mathfrak{P}(B)^A, \sqsubseteq \rangle$, namely the constant map $a \mapsto B$. Another issue is to have a finite representation of the functions of $\langle \mathfrak{P}(B)^A, \sqsubseteq \rangle$ since $A$ can be infinite. This last point is easy to address.

Given an automatic presentation $\mathcal{A}$ of $\mathbf{A}$, we extend $\mathcal{HC}_{\mathcal{A},\mathbf{B}}$ and $H_{\mathcal{A},\mathbf{B}}^*$ to be defined on $\text{dom}_{\mathcal{A}}$ instead of $\mathbf{A}$.

**Lemma VIII.5.17** ($\mathcal{HC}$ preserves regularity). Let $\mathbf{A}$ be an arbitrary $\sigma$-structure and $\mathbf{B}$ a finite $\sigma$-structure. For any $F \in \langle \mathfrak{P}(B)^A, \sqsubseteq \rangle$, if $F$ is regular, then $\mathcal{HC}(F)$ is regular.





*Proof.* Let $F \in \langle \mathfrak{P}(B)^{\mathcal{A}}, \sqsubseteq \rangle$ be regular, so for each $Y \in \mathfrak{P}(B)$, $F^{-1}[Y]$ is regular, and so by Proposition VII.2.4, there exists a first-order formula $\phi_Y(x)$ over $\sigma_{\Sigma}^{\text{sync}}$ s.t. $F^{-1}[Y] = [\![\phi_Y(x)]\!]^{\Sigma^*}$. Also, since $\mathcal{A}$ is an automatic presentation, for any $\mathcal{R} \in \sigma$ of arity $k$, there exists by Proposition VII.2.4 a first-order formula $\psi_{\mathcal{R}}(x_1, \ldots, x_k)$ over $\sigma_{\Sigma}^{\text{sync}}$ s.t. $\mathcal{R}_{\mathcal{A}} = [\![\psi_{\mathcal{R}}(x_1, \ldots, x_k)]\!]^{\Sigma^*}$. Similarly, $\text{dom}_{\mathcal{A}} = [\![\psi_{\text{dom}}(x)]\!]^{\Sigma^*}$ for some formula $\psi_{\text{dom}}(x)$.

It is then easy to prove that $\mathcal{HC}(F)$ is regular by providing a first-order formula $\widehat{\phi}_Y(x)$ for each $Y \in \mathfrak{P}(B)$, describing $\mathcal{HC}(F)^{-1}[Y]$, using both the formulas above, and the definition of $\mathcal{HC}$. Indeed, recall that an element $u \in \text{dom}_{\mathcal{A}}$ should be sent via $\mathcal{HC}(F)$ onto $Y \in \mathfrak{P}(B)$ if $Y$ is exactly the set of elements $b \in B$ s.t. for every $\mathcal{R}_{(k)} \in \sigma$, for every $i \in [\![1, k]\!]$, if $\langle a_1, \ldots, a_{k-1} \rangle \in Adj_{\mathcal{A}}^{\mathcal{R}, i}(a)$, then there exists $b_1 \in F(a_1), \ldots, b_{k-1} \in F(a_{k-1})$ s.t. $\langle b_1, \ldots, b_{k-1} \rangle \in Adj_{\mathbf{B}}^{\mathcal{R}, i}(b)$. Symbolically the set of such $u$'s can be written as $[\![\widehat{\phi}_Y(x)]\!]^{\Sigma^*}$, where

$$\widehat{\phi}_Y(x) \triangleq \psi_{\text{dom}}(x) \wedge \Big( \bigwedge_{b \in Y} \chi_b(x) \wedge \bigwedge_{b \notin Y} \neg \chi_b(x) \Big)$$

where $\chi_b(x)$ is the formula[66]

$$\chi_b(x) \triangleq \bigwedge_{\mathcal{R}_{(k)} \in \sigma} \bigwedge_{i \in [\![1, k]\!]} \forall x_1. \ldots \forall x_{i-1}. \forall x_{i+1}. \ldots \forall x_k.$$

$$\psi_{\mathcal{R}}(x_1, \ldots, x_{i-1}, x, x_{i+1}, \ldots, x_k)$$

$$\Rightarrow \Big( \bigvee_{\langle b_1, \ldots, b_{i-1}, b_{i+1}, \ldots b_k \rangle \in Adj_{\mathbf{B}}^{\mathcal{R}, i}(b)} \bigwedge_{i \in [\![1, k]\!] \smallsetminus \{i\}} \underbrace{\bigvee_{\substack{Y' \in \mathfrak{P}(B) \\ b_i \in Y'}} \phi_{Y'}(x_i)}_{b_i \in F(x_i)} \Big). \quad \square$$



Notice that $\Lambda_B$ is trivially regular, and so by immediate induction, each $\mathcal{HC}^n(\Lambda_B)$ with $n \in \mathbb{N}$ is also regular. While this opens the door to solving $\mathcal{H}om^{\text{reg}}(\text{Aut}, \mathbf{B})$ when $\mathbf{B}$ has tree duality using the hyperedge consistency algorithm, the problem of finite convergence remains.

First, we show that finite iterations are enough to detect the absence of homomorphism.

**Proposition VIII.5.18.** Let $\mathbf{A}$ be an arbitrary $\sigma$-structure and $\mathbf{B}$ a finite $\sigma$-structure with tree duality. If $\mathbf{A} \xrightarrow{\text{hom}} \mathbf{B}$, then there exists $n \in \mathbb{N}$ and $a \in A$ s.t. $\mathcal{HC}^n(\Lambda_B)(a) = \varnothing$.

In order to prove this proposition, we rely on the following property.

**Property VIII.5.19** (Monotonicity of $\mathcal{HC}$).[67] Let $\mathbf{A}, \mathbf{A}'$ be arbitrary $\sigma$-structures s.t. there is a homomorphism $h \colon \mathbf{A}' \to \mathbf{A}$. Let $\mathbf{B}$ be a finite $\sigma$-structure. For any $F \in \langle \mathfrak{P}(B)^A, \sqsubseteq \rangle$ and $F' \in \langle \mathfrak{P}(B)^{A'}, \sqsubseteq \rangle$, if $F(h(a)) \subseteq F'(a)$ for all $a \in A'$, then $\mathcal{HC}(F)(h(a)) \subseteq \mathcal{HC}(F')(a)$ for all $a \in A'$.[68]



*Proof.* Assume that $F|_{A'} \sqsubseteq F'$, and let us show that $\mathcal{HC}(F)|_{A'} \sqsubseteq \mathcal{HC}(F')$. Let $a \in A'$, and let $b \in \mathcal{HC}(F)(h(a))$. By definition of $\mathcal{HC}$, for every $\mathcal{R}_{(k)} \in \sigma$, for every $i \in [\![1, k]\!]$, if $\langle a_1, \ldots, a_{k-1} \rangle \in Adj_{\mathbf{A}}^{\mathcal{R}, i}(h(a))$, then there exists $b_1 \in F(a_1)$,





..., $b_{k-1} \in F(a_{k-1})$ s.t. $\langle b_1, \ldots, b_{k-1} \rangle \in \mathcal{A}dj_{\mathbf{B}}^{\cdot;\mathcal{R},i}(b)$. Now, let $\mathcal{R}_{(k)} \in \sigma$ and $i \in [\![1,k]\!]$, and let $\langle a_1, \ldots, a_{k-1} \rangle \in \mathcal{A}dj_{\mathbf{A}'}^{\mathcal{R},i}(a)$. Since $h$ is a homomorphism, we have $\langle h(a_1), \ldots, h(a_{k-1}) \rangle \in \mathcal{A}dj_{\mathbf{A}}^{\mathcal{R},i}(h(a))$, and so there exists $b_1 \in F(h(a_1))$, ..., $b_{k-1} \in F(h(a_{k-1}))$ s.t. $\langle b_1, \ldots, b_{k-1} \rangle \in \mathcal{A}dj_{\mathbf{B}}^{\cdot;\mathcal{R},i}(b)$. Since by hypothesis $F(h(a_i)) \subseteq F'(a_i)$ (for $a_i \in A'$), it follows that for every $\mathcal{R}_{(k)} \in \sigma$, for every $i \in [\![1,k]\!]$, if $\langle a_1, \ldots, a_{k-1} \rangle \in \mathcal{A}dj_{\mathbf{A}'}^{\cdot;\mathcal{R},i}(a)$, then there exists $b_1 \in F'(a_1)$, ..., $b_{k-1} \in F'(a_{k-1})$ s.t. $\langle b_1, \ldots, b_{k-1} \rangle \in \mathcal{A}dj_{\mathbf{B}}^{\cdot;\mathcal{R},i}(b)$. And hence $b \in \mathcal{HC}(F')(a)$, which concludes the proof. □

*Proof of Proposition VIII.5.18.* Let $\mathbf{A}$ be an arbitrary $\sigma$-structure and $\mathbf{B}$ a finite $\sigma$-structure with tree duality. Assume that $\mathbf{A} \xrightarrow{\text{hom}} \mathbf{B}$. Then by Proposition VIII.2.7 there exists a finite substructure $\mathbf{A}'$ of $\mathbf{A}$ s.t. $\mathbf{A}' \xrightarrow{\text{hom}} \mathbf{B}$, and so by Proposition VIII.5.14, there exists some $a \in A$ s.t. $H^\star_{\mathbf{A}',\mathbf{B}}(a) = \varnothing$. But since $\mathbf{A}'$ is finite, $H^\star_{\mathbf{A}',\mathbf{B}} = \mathcal{HC}^n(\Lambda_{\mathbf{B}})$ for some $n \in \mathbb{N}$. Then using Property VIII.5.19,

$$\mathcal{HC}^n_{\mathbf{A},\mathbf{B}}(\Lambda_B)(a) \subseteq \mathcal{HC}^n_{\mathbf{A}',\mathbf{B}}(\Lambda_B)(a) = H^\star_{\mathbf{A}',\mathbf{B}}(a) = \varnothing. \qquad \square$$

So, if $\mathbf{A} \xrightarrow{\text{hom}} \mathbf{B}$, then the hyperedge consistency algorithm will detect it in finite time assuming that $\mathbf{B}$ has tree duality. We will then show the dual implication, under the stronger assumption that $\mathbf{B}$ has finite duality: the reason we need this stronger assumption is that while the hyperedge consistency algorithm converges for finite $\sigma$-structures when $\mathbf{B}$ has tree duality, the number of iterations needed to reach the fixpoint depends on $\mathbf{A}$: for instance, in Example VIII.5.15, we showed that when the target structure is $\mathbf{P}_2$, then hyperedge consistency algorithm converges on $\mathbf{Z}_2^{(n)}$ in $n + \mathcal{O}(1)$ steps. On the other hand, for structures with finite duality, we show that this is not the case—and hence, convergence generalizes to infinite $\sigma$-structures: for instance, we showed in Example VIII.5.15 that over the target structure $\mathbf{T}_2$, which has finite duality, then the hyperedge consistency algorithm converges on $\mathbf{Z}_5^{(n)}$ in only 2 steps.

**Lemma VIII.5.20** (Uniform Convergence of Hyperedge Consistency for Structures with Finite Duality)**.** Let $\mathbf{B}$ be a finite $\sigma$-structure. The following are equivalent:

1. $\mathbf{B}$ has finite duality;

2. there exists $n \in \mathbb{N}$ s.t. for every *finite* $\sigma$-structure $\mathbf{A}$:
   - $\mathcal{HC}^n_{\mathbf{A},\mathbf{B}}(\Lambda_B) = H^\star_{\mathbf{A},\mathbf{B}}$ when $\mathbf{A} \xrightarrow{\text{hom}} \mathbf{B}$, and
   - $\mathcal{HC}^n_{\mathbf{A},\mathbf{B}}(\Lambda_B)(a) = \varnothing$ for some $a \in A$ when $\mathbf{A} \not\xrightarrow{\text{hom}} \mathbf{B}$;

3. there exists $n \in \mathbb{N}$ s.t. for every *arbitrary* $\sigma$-structure $\mathbf{A}$:
   - $\mathcal{HC}^n_{\mathbf{A},\mathbf{B}}(\Lambda_B) = H^\star_{\mathbf{A},\mathbf{B}}$ when $\mathbf{A} \xrightarrow{\text{hom}} \mathbf{B}$, and
   - $\mathcal{HC}^n_{\mathbf{A},\mathbf{B}}(\Lambda_B)(a) = \varnothing$ for some $a \in A$ when $\mathbf{A} \not\xrightarrow{\text{hom}} \mathbf{B}$;

*Proof.* We prove the implications $(1) \Rightarrow (2) \Rightarrow (3) \Rightarrow (1)$.

✎ $(2) \Rightarrow (3)$. Let $n \in \mathbb{N}$ s.t. $\mathcal{HC}^n_{\mathbf{A}',\mathbf{B}}(\Lambda_B) = H^\star_{\mathbf{A}',\mathbf{B}}$ for every finite $\sigma$-structure $\mathbf{A}'$. Let $\mathbf{A}$ be an arbitrary $\sigma$-structure. Note that for any $F \in \langle \mathfrak{P}(B)^A, \sqsubseteq \rangle$, for any substructure $\mathbf{A}'$ of $\mathbf{A}$ containing $a$, then by Prop-





erty VIII.5.19 $\mathcal{HC}_{\mathbf{A},\mathbf{B}}(F)(a) \subseteq \mathcal{HC}_{\mathbf{A}',\mathbf{B}}(F)(a)$. We show that equality is reached by a particular finite substructure.

*Claim* VIII.5.21. For any $F \in \langle \mathfrak{P}(B)^A, \sqsubseteq \rangle$ and $m \in \mathbb{N}$, there exists a finite substructure[69] $\mathbf{A}_{a,m}$ of $\mathcal{B}_{\mathbf{A}}^m(a)$ *s.t.*

$$\mathcal{HC}_{\mathbf{A},\mathbf{B}}^m(F)(a) = \mathcal{HC}_{\mathbf{A}_{a,m},\mathbf{B}}^m(F)(a).$$

We give a proof sketch of this claim. Note that by definition, $\mathcal{HC}_{\mathbf{A},\mathbf{B}}(F)(a)$ only depends on the values of $F(a')$ where $a'$ is at distance 1. More precisely, it only depends on the values $\langle F(a_1), \dots, F(a_{k-1}) \rangle$, where $\mathcal{R}$ is any predicate of arity $k$, $i \in [\![1,k]\!]$ and $\langle a_1, \dots, a_{k-1} \rangle \in \mathcal{A}dj_{\mathbf{A}}^{\mathcal{R},i}(a)$. Since $B$ is finite, there are finitely many tuples of the form $\langle F(a_1), \dots, F(a_{k-1}) \rangle$, and so for each of them it suffices to keep (for distance $m = 1$) only one tuple $\langle a_1, \dots, a_{k-1} \rangle \in \mathcal{A}dj_{\mathbf{A}}^{\mathcal{R},i}(a)$. By induction on $m$, we obtain a finite substructure $\mathbf{A}_{a,m}$ of $\mathcal{B}_{\mathbf{A}}^m(a)$ as in Claim VIII.5.21.

We now show that if $\mathbf{A} \xrightarrow{\text{hom}} \mathbf{B}$, then $\mathcal{HC}_{\mathbf{A},\mathbf{B}}^n(\Lambda_B) = H_{\mathbf{A},\mathbf{B}}^\star$, and if $\mathbf{A} \xrightarrow{\text{hom}} \mathbf{B}$, then $\mathcal{HC}_{\mathbf{A},\mathbf{B}}^n(\Lambda_B)(a) = \varnothing$ for some $a \in A$. We assume that $\mathbf{A} \xrightarrow{\text{hom}} \mathbf{B}$: by Claim VIII.5.21 and Property VIII.5.19

$$\mathcal{HC}_{\mathbf{A},\mathbf{B}}^n(\Lambda_B)(a) = \mathcal{HC}_{\mathbf{A}_{a,n+1},\mathbf{B}}^n(\Lambda_B)(a).$$

Since $\mathbf{A}_{a,n+1}$ is finite, by (2), the right-hand side of the equality above equals $H_{\mathbf{A}_{a,n},\mathbf{B}}^\star(a)$. But then again by Claim VIII.5.21 and (2),

$$\mathcal{HC}_{\mathbf{A},\mathbf{B}}^{n+1}(\Lambda_B)(a) = H_{\mathbf{A}_{a,n+1},\mathbf{B}}^\star(a).$$

And hence $\mathcal{HC}_{\mathbf{A},\mathbf{B}}^n(\Lambda_B)(a) = \mathcal{HC}_{\mathbf{A},\mathbf{B}}^{n+1}(\Lambda_B)(a)$. Since this property holds for arbitrary values of $a \in A$, it follows that $\mathcal{HC}_{\mathbf{A},\mathbf{B}}^n(\Lambda_B) = H_{\mathbf{A},\mathbf{B}}^\star$. The case when $\mathbf{A} \xrightarrow{\text{hom}} \mathbf{B}$ is handled similarly.

🕭 $(3) \Rightarrow (1)$. One can show—exactly as in the proof of Lemma VIII.5.17—by induction on $m \in \mathbb{N}$ that $\mathcal{HC}_{\mathbf{A},\mathbf{B}}^m(\Lambda_B)$ is first-order definable, in the sense that for all $Y \in \mathfrak{P}(B)$, there exists a first-order formula $\phi_{m,Y}(x)$ over $\sigma$ *s.t.* $\mathcal{HC}_{\mathbf{A},\mathbf{B}}^m(\Lambda_B)^{-1}[Y] = [\![\phi_{m,Y}(x)]\!]^{\mathbf{A}}$, *i.e.* for any $a \in A$, we have

$$\langle \mathbf{A}, a \rangle \vDash \phi_{m,Y}(x) \text{ iff } \mathcal{HC}_{\mathbf{A},\mathbf{B}}^m(\Lambda_B)(a) = Y.$$

We then claim that

$$\mathbf{A} \vDash \forall x. \neg\phi_{n,\varnothing}(x) \text{ iff } \mathbf{A} \xrightarrow{\text{hom}} \mathbf{B}.$$

The left-to-right implication can be proven by contraposition, since if $\mathbf{A} \xrightarrow{\text{hom}} \mathbf{B}$ then by (3) we have $\mathcal{HC}_{\mathbf{A},\mathbf{B}}^n(\Lambda_B)(a) = \varnothing$ for some $a \in A$. For the right-to-left implication, we again use (3), which yields that $\mathcal{HC}_{\mathbf{A},\mathbf{B}}^n(\Lambda_B) = H_{\mathbf{A},\mathbf{B}}^\star$. Together with the contraposition of Corollary VIII.5.13, this implies that $\mathcal{HC}_{\mathbf{A},\mathbf{B}}^n(\Lambda_B)(a) \neq \varnothing$ for all $a \in A$. The conclusion that $\mathbf{B}$ has finite duality follows from Atserias' theorem.







❧ $(1) \Rightarrow (2)$. To prove this implication, we first need a characterization of what it means for an element $b \in B$ not to be in $H^*_{\mathbf{A},\mathbf{B}}(a)$, where $a \in A$. We denote by $n(\mathbf{B})$ the maximal diameter of a critical obstruction of $\mathbf{B}$—which must be finite since $\mathbf{B}$ has finite duality.

*Claim* VIII.5.22. Let $a \in A$ and $b \in B$. Assume that there exists a $\sigma$-tree $\mathbf{T}$ *s.t.* there is a homomorphism from $\mathbf{T}$ to $\mathbf{A}$ that maps some element $t \in T$ to $a$, but no homomorphism from $\mathbf{T}$ to $\mathbf{B}$ can map $t$ to $b$. Then, letting $m$ denote the height of $\mathbf{T}$ when rooted at $t$, we have $b \notin \mathcal{HC}^m_{\mathbf{A},\mathbf{B}}(\Lambda_B)(a)$.

We can prove by induction on $\mathbf{T}$ that if there are no homomorphism from $\mathbf{T}$ to $\mathbf{B}$ that can map $t$ to $b$, then $b \notin \mathcal{HC}^m_{\mathbf{T},\mathbf{B}}(\Lambda_B)(t)$ where $m$ is the height of $\mathbf{T}$ rooted at $t$. Then by Property VIII.5.19, $\mathcal{HC}^m_{\mathbf{A},\mathbf{B}}(\Lambda_B)(a) \subseteq \mathcal{HC}^m_{\mathbf{T},\mathbf{B}}(\Lambda_B)(t)$ and so $b \notin \mathcal{HC}^m_{\mathbf{A},\mathbf{B}}(\Lambda_B)(a)$.

*Claim* VIII.5.23. Let $a \in A$ and $b \in B$. If $\mathbf{A} \xrightarrow{\text{hom}} \mathbf{B}$, then $\mathcal{HC}^{n(\mathbf{B})}_{\mathbf{A},\mathbf{B}}(\Lambda_B)(a) = \varnothing$ for some $a \in A$.

Indeed, since $\mathbf{A} \xrightarrow{\text{hom}} \mathbf{B}$, there exists a critical obstruction $\mathbf{T}$ of $\mathbf{B}$ *s.t.* $\mathbf{T} \xrightarrow{\text{hom}} \mathbf{A}$. By Proposition VIII.2.14, *w.l.o.g.* $\mathbf{T}$ is a $\sigma$-tree. Since $\mathbf{T} \xrightarrow{\text{hom}} \mathbf{B}$, for any $t \in T$ and $a \in A$ *s.t.* $t$ is mapped on $a$, we have by Claim VIII.5.22 that for any $b \in B$, $b \notin \mathcal{HC}^m_{\mathbf{A},\mathbf{B}}(\Lambda_B)(a)$, where $m$ is the height of $\mathbf{T}$ when rooted at $t$. And hence $\mathcal{HC}^m_{\mathbf{A},\mathbf{B}}(\Lambda_B)(a) = \varnothing$. Since $m \leq n(\mathbf{B})$, this concludes the proof of the first part of (2). We will now handle the more tricky case of $\mathbf{A} \xrightarrow{\text{hom}} \mathbf{B}$.

*Claim* VIII.5.24.[70] Let $a \in A$ and $b \in B$. If $\mathbf{A} \xrightarrow{\text{hom}} \mathbf{B}$ and $b \notin H^*_{\mathbf{A},\mathbf{B}}(a)$, then $b \notin \mathcal{HC}^{n(\mathbf{B})}_{\mathbf{A},\mathbf{B}}(\Lambda_B)(a)$.

To prove this claim, we use a construction that is similar to Proposition VIII.2.6. Fix $a \in A$ and $b \in B$ *s.t.* $b \notin H^*_{\mathbf{A},\mathbf{B}}(a)$. Let $\mathbf{C}$ be defined by first taking the disjoint union of $\mathbf{A}$ and $\mathbf{B}$, and then identifying $a$ and $b$. Note that for any $a' \in A$, we have:

$$\mathcal{A}dj^{\mathcal{R},i}_{\mathbf{C}}(a') = \begin{cases} \mathcal{A}dj^{\mathcal{R},i}_{\mathbf{A}}(a') & \text{if } a' \neq a, \\ \mathcal{A}dj^{\mathcal{R},i}_{\mathbf{A}}(a) \cup \mathcal{A}dj^{\mathcal{R},i}_{\mathbf{B}}(b) & \text{if } a' = a = b. \end{cases}$$

The goal of this construction is that, when running the hyperedge consistency algorithm on $\mathbf{C}$, vertex $b$ will be removed as a potential image for $a$ since $b \notin H^*_{\mathbf{A},\mathbf{B}}(a)$, but because of the copy of $\mathbf{B}$ included in $\mathbf{C}$, any homomorphism from $\mathbf{C}$ to $\mathbf{B}$ must map $a$ to $b$. See Figure VIII.29 for an example.

*Claim* VIII.5.25. $\mathbf{C} \xrightarrow{\text{hom}} \mathbf{B}$.

Indeed, if there was a homomorphism from $\mathbf{C}$ to $\mathbf{B}$, say $f$, then $f|_{\mathbf{B}}$ would be a homomorphism from $\mathbf{B}$ to $\mathbf{B}$. Since $\mathbf{B}$ has finite duality, it is rigid by Proposition VIII.2.13, and so in particular $f(b) = b$. Hence, we would get that $f|_A$ is a homomorphism from $\mathbf{A}$ to $\mathbf{B}$ which sends $a$ to $b$, and so by Proposition VIII.5.12 we would have $b \in H^*_{\mathbf{A},\mathbf{B}}(a)$, which is a contradiction.

So, since $\mathbf{C} \xrightarrow{\text{hom}} \mathbf{B}$, there exists a critical obstruction $\mathbf{T}$ of $\mathbf{B}$ *s.t.* there is a homomorphism $f$ from $\mathbf{T}$ to $\mathbf{C}$. We claim that $a = b$ must be in the image of



[70] In fact, Larose, Loten & Tardif implicitly showed a weaker result, by adapting [FV98, Theorem 21], in [LLT07, Proof of Lemma 3.2]. It states that if $b \notin H^*_{\mathbf{A},\mathbf{B}}(a)$, then there exists a $\sigma$-tree $\mathbf{T}$ *s.t.* there is a homomorphism from $\mathbf{T}$ to $\mathbf{A}$ that maps some element $t \in T$ to $a$, but no homomorphism from $\mathbf{T}$ to $\mathbf{B}$ can map $t$ to $b$. Moreover, the height of their $\sigma$-tree $\mathbf{T}$ is linearly bounded by the least $n \in \mathbb{N}$ *s.t.* $b \notin \mathcal{HC}^n_{\mathbf{A},\mathbf{B}}(\Lambda_B)(a)$. This property is true without any duality assumption on $\mathbf{B}$, and only follows from the inner workings of the hyperedge consistency algorithm.

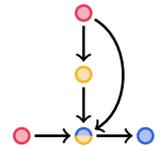

Figure VIII.29: Construction of $\mathbf{C}$ as in the proof of Claim VIII.5.24 when $\mathbf{A}$ is the 2-path and $\mathbf{B}$ is the 2-transitive tournament.



$f$. Indeed, since $a = b$ is the only vertex of $\mathbf{A}$ that is adjacent to $\mathbf{B}$ in $\mathbf{C}$, and since $\mathbf{T}$ is connected as a critical obstruction, we would otherwise get that either $\mathbf{T} \xrightarrow{\text{hom}} \mathbf{B}$—contradicting that $\mathbf{T}$ is a critical obstruction of $\mathbf{B}$—or that $\mathbf{T} \xrightarrow{\text{hom}} \mathbf{A}$—contradicting that $\mathbf{A} \xrightarrow{\text{hom}} \mathbf{B}$.

And so, there exists $t \in T$ s.t. $f(t) = a = b$. We let $\mathbf{U}$ be the quotient structure of $\mathbf{T}$ by the congruence induced by $f$, namely the smallest congruence containing $\{\langle t_1, t_2 \rangle \mid f(t_1) = f(t_2)\}$. Then, we let $\mathbf{U_A}$ be the substructure of $\mathbf{U}$ induced by the elements that are sent via $f$ on $\mathbf{A}$.

*Claim* VIII.5.26. $\mathbf{U_A}$ is a $\sigma$-tree.

*Claim* VIII.5.27. There is a homomorphism from $\mathbf{T_A}$ to $\mathbf{A}$ that maps $t$ to $a$, but no homomorphism from $\mathbf{T_A}$ to $\mathbf{B}$ that maps $t$ to $b$.

The first point is trivial: it suffices to consider the restriction of $f$ to $T_A$. For the second point, assume by contradiction that there is a homomorphism $g$ from $\mathbf{T_A}$ to $\mathbf{B}$ that maps $t$ to $b$. Then the function

$$t' \in T \mapsto \begin{cases} g(t') & \text{if } t' \in T_A, \\ f(t') & \text{if } t' \in T_B, \end{cases}$$

is well-defined—if $t' \in T_A \cap T_B$, then $f(t') = b = g(t)$—and is a homomorphism from $\mathbf{T}$ to $\mathbf{B}$. This contradicts that $\mathbf{T}$ is a critical obstruction of $\mathbf{B}$. And hence, no homomorphism from $\mathbf{T_A}$ to $\mathbf{B}$ can map $t$ to $b$. We then apply Claim VIII.5.22 to get that $b \notin \mathcal{HC}_{\mathbf{A},\mathbf{B}}^{n(\mathbf{B})}(\Lambda_B)(a)$, concluding the proof of Claim VIII.5.24.

Overall, Claims VIII.5.23 and VIII.5.24 prove (2), which concludes the proof of Lemma VIII.5.20. □

Putting Lemma VIII.5.17 (for computations) and Lemma VIII.5.20 (for correctness), we get (1) that the algorithm of Figure VIII.30 decides $\mathcal{H}om^{\text{reg}}(\text{Aut}, \mathbf{B})$ when $\mathbf{B}$ has finite duality and (2) that $\mathcal{H}om^{\text{reg}}(\text{Aut}, \mathbf{B}) = \mathcal{H}om(\text{Aut}, \mathbf{B})$ under the same assumption.

**Input:** An automatic presentation $\mathcal{A}$ of a $\sigma$-structure $\mathbf{A}$ and a finite $\sigma$-structure $\mathbf{B}$.

**Invariant:** First-order formulas $\phi_Y(x)$ ($Y \subseteq B$) defining the set of $a \in A$ s.t. $\text{Im}_a = Y$, where $\text{Im}_a$ is as in Figure VIII.28.

$\phi_B^0(x) \leftarrow \top$;
$\phi_Y^0(x) \leftarrow \bot$ for all $Y \subsetneq B$;
$n \leftarrow 0$;
**do**
   | compute $\langle \phi_Y^{n+1}(x) \rangle_{Y \subseteq B}$ from the $\langle \phi_Y^n(x) \rangle_{Y \subseteq B}$ using
   | Lemma VIII.5.17;
   | **if** $[\![\phi_\varnothing^{n+1}(x)]\!]^\mathbf{A} \neq \varnothing$ **then**
   |   | **return** *false*;
   | $n \leftarrow n + 1$;
**while** *some* $\phi_Y(x)$ *has been semantically updated*;
**return** *true*

Figure VIII.30: The *hyperedge consistency algorithm for automatic structures*.



## VIII.6  Discussion

### VIII.6.1  Undecidability of Finite Regular Colourability

First and foremost, our original problem, namely the AUT/REC-SEPARABILITY PROBLEM, or equivalently by Theorem VIII.3.2, the FINITE REGULAR COLOURABILITY PROBLEM, remains open.

**Conjecture VIII.6.1.**[71] FINITE REGULAR COLOURABILITY OF AUTOMATIC GRAPHS is RE-complete.

We briefly explain here why the techniques developed in this chapter, cannot immediately solve this problem.

We define $\mathbf{K}_{<\omega}$ to be the disjoint union $\bigoplus_{k\in\mathbb{N}} \mathbf{K}_{2^k}$ of all finite cliques. Note that it is homomorphically equivalent to $\bigoplus_{k\in\mathbb{N}} \mathbf{K}_k$. Moreover, it admits a simple automatic presentation $\mathcal{K}_{<\omega}$: take the binary alphabet $\Sigma \triangleq \mathbf{2}$ and let[72]

$$\mathrm{dom}_{\mathcal{K}_{<\omega}} \triangleq \Sigma^*,$$
$$\mathcal{E}_{\mathcal{K}_{<\omega}} \triangleq \{\langle u, v\rangle \in \Sigma^* \times \Sigma^* \mid |u| = |v| \text{ and } u \neq v\}.$$

Of course, if some graph $\mathcal{G}$ is finitely colourable, then it has a homomorphism to $\mathbf{K}_{<\omega}$, but the converse implication does not hold—e.g. $\mathbf{K}_{<\omega}$ itself is not finitely colourable. However, the equivalence holds for a large class of graphs.

**Property VIII.6.2.** Let $\mathbf{G}$ be an arbitrary connected graph. $\mathbf{G} \xrightarrow{\text{hom}} \mathbf{K}_{<\omega}$, iff $\mathbf{G}$ is finitely colourable. Similarly, let $\mathcal{G}$ be a connected automatic graph. $\mathcal{G} \xrightarrow{\text{reg hom}} \mathcal{K}_{<\omega}$, iff $\mathbf{G}$ is finitely regularly colourable.

*Proof.* A homomorphism must send pairs of connected vertices to pairs of connected vertices, and hence a homomorphism from $\mathbf{G}$ to $\mathbf{K}_{<\omega}$ actually yields a homomorphism from $\mathbf{G}$ to some $\mathbf{K}_{2^n}$ for some $n$, and hence a finite colouring of $\mathbf{G}$. □

**Proposition VIII.6.3.** The FINITE REGULAR COLOURABILITY PROBLEM and its restriction to connected automatic graphs are computationally equivalent.

*Proof.* The reduction from the restricted to the general problem is straightforward. For the converse one, we add a new element $*$ to the structure $\mathcal{A}$,[73] and put an edge from $*$ to any element $u$ of the original structure. Clearly, the original structure is finitely regularly colourable iff the new one is—one direction is trivial, for the other one it suffices to assign to $*$ a new colour. □

Putting Property VIII.6.2 and Proposition VIII.6.3 together, we obtain the following corollary.

**Corollary VIII.6.4.** The FINITE REGULAR COLOURABILITY PROBLEM and the restriction of $\mathcal{H}om^{\text{reg}}(\text{Aut}, \mathcal{K}_{<\omega})$ to connected graphs are computationally

[71] The upper bound is trivial: given an automatic graph $\mathcal{G}$, we guess some $k \in \mathbb{N}$ and check if $\mathcal{G} \in \mathcal{H}om^{\text{reg}}(\text{Aut}, \mathbf{K}_k)$, which is RE by Proposition VIII.4.5.

[72] In this presentation, words of length $k$ are used to encode $\mathbf{K}_{2^k}$.

[73] In particular, we extend the alphabet of the presentation.





equivalent.

On the other hand, we strongly believe that the proof of Lemma VIII.4.16 can be adapted to obtain a lower bound on $\mathcal{H}om^{\mathsf{reg}}(\mathrm{Aut},\ \mathcal{K}_{<\omega}^{\dagger})$.

**Conjecture VIII.6.5.** There is a reduction from REGULAR UNCONNECTIVITY IN AUTOMATIC GRAPHS to $\mathcal{H}om^{\mathsf{reg}}(\mathrm{Aut},\ \mathcal{K}_{<\omega}^{\dagger})$.

Note however that Lemma VIII.4.16 does not directly apply since the target structure is not finite—a fact that we use throughout the proof of Lemma VIII.4.16.

Putting these last two statements together, to prove the RE-hardness of FINITE REGULAR COLOURABILITY OF AUTOMATIC GRAPHS, it would "suffice" to build the following reduction.

**Conjecture VIII.6.6.** There is a reduction from $\mathcal{H}om^{\mathsf{reg}}(\mathrm{Aut},\ \mathcal{K}_{<\omega}^{\dagger})$ to the restriction of $\mathcal{H}om^{\mathsf{reg}}(\mathrm{Aut},\ \mathcal{K}_{<\omega})$ to connected graphs.

This question seems however quite challenging. Note first that the reduction from REGULAR UNCONNECTIVITY IN AUTOMATIC GRAPHS to

$$\mathcal{H}om^{\mathsf{reg}}(\mathrm{Aut},\ \mathcal{K}_{<\omega}^{\dagger})$$

heavily uses the fact that the source structure can be unconnected. Moreover, beyond this issue of connectivity, whether $\mathcal{H}om^{\mathsf{reg}}(\mathrm{Aut},\ \mathcal{K}_{<\omega}^{\dagger})$ and $\mathcal{H}om^{\mathsf{reg}}(\mathrm{Aut},\ \mathcal{K}_{<\omega})$ are equivalent is also not straightforward: note in particular that Proposition VIII.2.6 does not apply: not only $\mathbf{K}_{<\omega}$ is not finite, but more importantly it is not a core.[74]

[74] Indeed, $\bigoplus_{k\in\mathbb{N}}\mathbf{K}_k$ is homomorphically equivalent to $\bigoplus_{k\in I}\mathbf{K}_k$ for any infinite subset $I$ of $\mathbb{N}$.

### VIII.6.2   Invariance under Graph Isomorphisms

Note that given an automatic presentation $\mathcal{A}$ of some $\sigma$-structure $\mathbf{A}$, the property of whether $\mathcal{A}\xrightarrow{\mathsf{reg\ hom}}\mathbf{B}$, where $\mathbf{B}$ is a finite $\sigma$-structure, does not depend only on the structure $\mathbf{A}$, but on its presentation $\mathcal{A}$—see Example VIII.3.9 for an example; it is trivial to come up with a presentation of the same graph that admits a regular 2-colouring.

On the other hand, the implication $(\mathsf{DT})_{\mathsf{fin\text{-}dual}}\Rightarrow(\mathsf{DT})_{\mathsf{equal}}$ proves that if $\mathbf{B}$ has finite duality, then the property of whether $\mathcal{A}\xrightarrow{\mathsf{reg\ hom}}\mathbf{B}$ is invariant under graph isomorphisms, in the sense that for any presentations $\mathcal{A}_1$ and $\mathcal{A}_2$ of the structures $\mathbf{A}_1$ and $\mathbf{A}_2$, respectively, if $\mathbf{A}_1$ and $\mathbf{A}_2$ are isomorphic, then $\mathcal{A}_1\xrightarrow{\mathsf{reg\ hom}}\mathbf{B}$ iff $\mathcal{A}_2\xrightarrow{\mathsf{reg\ hom}}\mathbf{B}$. We do not know whether the converse implication holds.

**Conjecture VIII.6.7.** For any finite $\sigma$-structure $\mathbf{B}$, $\mathcal{H}om^{\mathsf{reg}}(\mathrm{Aut},\ \mathbf{B})$ is invariant under graph isomorphisms iff $\mathbf{B}$ has finite duality.

### VIII.6.3   Obstacles to Finite Colourability

We also do not know whether FINITE COLOURABILITY PROBLEM is undecidable.





---

FINITE COLOURABILITY OF AUTOMATIC GRAPHS

*Input*:      A presentation $\mathcal{G}$ of an automatic graph **G**.

*Question*:   Does **G** admit a colouring with finitely many colours?

---

**Conjecture VIII.6.8.**[75] FINITE COLOURABILITY OF AUTOMATIC GRAPHS is $\Sigma_2^0$-complete.

Given that we do not know whether FINITE COLOURABILITY OF AUTOMATIC GRAPHS is decidable, a natural question would be find algorithm to identify sufficient conditions for a graph not to be finitely colourable. The first natural idea is to ask the graph to contain an infinite transitive tournament, and actually this property is decidable by Proposition VII.3.11 since it can be expressed by the order-invariant first-order formula

$$\forall x. \, \exists y. \, x < y \wedge (\mathcal{E}(x,y) \vee \mathcal{E}(y,x)).$$

A typical example of such a condition is to *contain unbounded tournaments*—meaning that $\mathbf{K}_k \xrightarrow{\text{hom}} \mathbf{A}$ for all $k \in \mathbb{N}$.

**Conjecture VIII.6.9.**[76] The problem of whether an automatic graph has bounded tournaments is decidable.

In [BFM23, Conjecture 7.2], we conjectured that there was some automatic graphs that were not finitely colourable, but did not contain unbounded tournaments. We pointed out that for arbitrary graphs, the property was clearly true since there are *triangle-free graphs*[77] **G** that are not finitely colourable [UD54]. However, we believe(d) that the infinite graph built using Ungar-Descartes' technique is not automatic. Since then, we managed to prove this conjecture, by relying on another classical construction of triangle-free graphs with arbitrary large chromatic number.

**Proposition VIII.6.10.** There exists a triangle-free automatic graph that is not finitely colourable.

Mycielski's construction is an operator $\mathfrak{M}$ on undirected graphs, introduced in [Myc55], with the property that if **G** is triangle-free then so is $\mathfrak{M}(\mathbf{G})$, and moreover the chromatic number of $\mathfrak{M}(\mathbf{G})$ is exactly one more than the chromatic number of $G$. Iterating this operator on a triangle-free graph shows that there exists triangle-free graphs with arbitrarily high chromatic number. To prove Proposition VIII.6.10, we will build an automatic graph whose underlying undirected graph is $\bigsqcup_{n \in \mathbb{N}} \mathfrak{M}^n(\mathbf{0})$ where $\mathbf{0}$ is the graph on a single vertex with no edge. Note that usually Mycielski's construction is not iterated on $\mathbf{0}$ but on the undirected path of size 1: we made this choice to make the proof of automaticity of the graph easier.

**Definition VIII.6.11.** Given an undirected graph $\mathbf{G} = \langle V, \mathcal{E} \rangle$, its *Mycielskian*, denoted by $\mathfrak{M}(\mathbf{G})$, is the undirected graph whose set of vertices is $V \times 2 \sqcup \{\bullet\}$, with the following edges:

---

[75] Note again that the upper bound is trivial, since by De Bruijn-Erdős theorem, this problem is equivalent to asking if there exists $k \in \mathbb{N}$ *s.t.* every finite subgraph of the source is $k$-colourable.

[76] This conjecture corresponds to [BFM23, Conjecture 7.3].

[77] Meaning that its underlying undirected graph is *triangle-free*, which means that there are no three vertices that are pairwise adjacent.





- $\{\langle u, 0\rangle, \langle v, 0\rangle\}$ for every edge $\{u, v\} \in \mathcal{E}$,
- $\{\langle u, 0\rangle, \langle v, 1\rangle\}$ for every edge $\{u, v\} \in \mathcal{E}$, and
- $\{\bullet, \langle v, 1\rangle\}$ for every $v \in \mathcal{E}$.

Note that $u \mapsto \langle u, 0\rangle$ always defines an embedding of $\mathbf{G}$ into $\mathfrak{M}(\mathbf{G})$.

**Property VIII.6.12.** If $\mathbf{G}$ is triangle-free, then so is $\mathfrak{M}(\mathbf{G})$.

*Proof.* All adjacent elements of $\bullet$ are of the form $\langle -, 1\rangle$, and two vertices of the form $\langle -, 1\rangle$ are never adjacent. Hence, any potential triangle in $\mathfrak{M}(\mathbf{G})$ must be of the form

$$\{\langle u, 0\rangle, \langle v, 0\rangle, \langle w, 0\rangle\} \quad \text{or} \quad \{\langle u, 0\rangle, \langle v, 1\rangle, \langle w, 0\rangle\}.$$

In both cases, this would imply that $\{u, v, w\}$ is a triangle in $\mathbf{G}$: contradiction. □

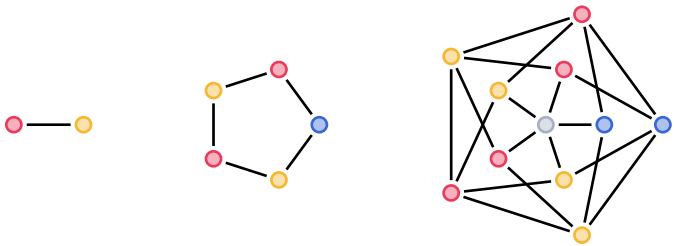

Figure VIII.31: The undirected path of length 1, and two iterations of Mycielski's construction $\mathfrak{M}$ on this graph. Each of these graphs is equipped with a 2-, 3-, and 4-colouring, respectively, which are built using the construction described in the proof of Property VIII.6.13.

**Property VIII.6.13.** The chromatic number of $\mathfrak{M}(\mathbf{G})$ is exactly one more than the chromatic number of $\mathbf{G}$.

*Proof.* ↝ *Upper bound.* Let $f\colon \mathbf{G} \to [\![1, k]\!]$ be a colouring of $\mathbf{G}$ for some $k \in \mathbb{N}$. Then mapping $\langle u, i\rangle$ to $f(u)$ for $u \in V$ and $i \in 2$ and mapping $\bullet$ to colour $k + 1$ defines a $(k + 1)$-colouring of $\mathfrak{M}(\mathbf{G})$.

↝ *Lower bound.* Let $k \in \mathbb{N}_{>0}$ and let $g\colon \mathfrak{M}(\mathbf{G}) \to [\![1, k]\!]$ be a $k$-colouring of $\mathfrak{M}(\mathbf{G})$. Let $g'\colon \mathfrak{M}(\mathbf{G}) \to [\![1, k]\!]$ be defined by $g'\langle u, i\rangle \triangleq g\langle u, 1\rangle$, and $g'(\bullet) \triangleq g(\bullet)$. Since the adjacency of $\langle u, 0\rangle$ is included in the one of $\langle u, 1\rangle$, it follows that $g'$ is still a $k$-colouring of $\mathfrak{M}(\mathbf{G})$. Since $\bullet$ is adjacent to all vertices of the form $\langle u, 1\rangle$ where $u \in \mathbf{G}$, then $\{g'\langle u, 1\rangle \mid u \in \mathbf{G}\}$ can only contain at most $k - 1$ colours. By construction of $g'$, $g'\langle u, 1\rangle = g'\langle u, 0\rangle$. And hence $u \mapsto g'\langle u, 0\rangle$ defines a $(k - 1)$-colouring of $\mathbf{G}$. □

We can now prove our result.

*Proof of Proposition VIII.6.10.* Let $\Sigma \triangleq \{0, 1, \bullet\}$, and let $\mathbf{0}$ be the directed graph on a single vertex with no edge. We are going to define an automatic graph whose underlying undirected graph is $\bigsqcup_{n \in \mathbb{N}} \mathfrak{M}^n(\mathbf{0})$, in such a way that the vertices of $\mathfrak{M}^n(\mathbf{0})$ will be encoded by words of length $n$.

We define disjoint subsets $\langle V_i\rangle_{i \in \mathbb{N}}$ of $\Sigma^*$ and $(\mathcal{E}_i)_{i \in \mathbb{N}}$ of $V_i \times V_i$, by induction on $i \in \mathbb{N}$, such that $(V_i, \mathcal{E}_i)$ is a directed version of $\mathfrak{M}^i(\mathbf{0})$.





Define $V_0 \triangleq \{\varepsilon\}$ and $\mathcal{E}_0 \triangleq \varnothing$. Then for $n \in \mathbb{N}$, let

$$V_{n+1} \triangleq \{u0 \mid u \in V_n\} \cup \{u1 \mid u \in V_n\} \cup \{\bullet^{n+1}\}$$
$$\mathcal{E}_{n+1} \triangleq \{\langle u0, v0\rangle \mid \langle u, v\rangle \in \mathcal{E}_n\} \cup \{\langle v0, u0\rangle \mid \langle u, v\rangle \in \mathcal{E}_n\}$$
$$\cup \{\langle u0, v1\rangle \mid \langle u, v\rangle \in \mathcal{E}_n\} \cup \{\langle v1, u0\rangle \mid \langle u, v\rangle \in \mathcal{E}_n\}$$
$$\cup \{\langle u1, \bullet^{n+1}\rangle \mid u \in V_n\} \cup \{\langle \bullet^{n+1}, u1\rangle \mid u \in V_n\}.$$

Then let $\mathbf{M} \triangleq \langle V, \mathcal{E}\rangle$ be the *infinite Mycielski graph* where $V \triangleq \bigcup_{n\in\mathbb{N}} V_n$ and $E \triangleq \bigcup_{n\in\mathbb{N}} E_n$. By construction, each $\mathcal{E}_n$ is symmetric,[78] and moreover $V_n \subseteq \Sigma^n$ by immediate induction on $n \in \mathbb{N}$.

Also, the underlying undirected graph of $\mathbf{M}$ is $\bigsqcup_{n\in\mathbb{N}} \mathfrak{M}^n(\mathbf{0})$, and so it is triangle-free—since $\mathbf{0}$ is triangle-free—by Property VIII.6.12 but has infinite chromatic number by Property VIII.6.13.

It remains to show that $\mathbf{M}$ is automatic. First, notice that $V = \bullet^*(0+1)^*$ since by trivial induction on $n \in \mathbb{N}$ we have $V_n = (\bullet^*(0+1)^*) \cap \Sigma^n$. And hence, $V$ is regular.

Now we claim that there is an edge from $u \in V$ to $v \in V$ if and only if

- $u$ and $v$ have the same length,
- $u$ contains at least one $\bullet$ and letting $i$ be the index of the last occurrence of $\bullet$ in $u$, we have $v_i = 1$, and
- for all $j \in [\![i+1, n]\!]$, $u_j = 0$ and $v_j \in 2$.

Call the property above $\mathcal{P}$. To prove it, observe first that $\mathcal{E}$ contains no edge between vertices of distinct length, and then we prove by induction on $n \in \mathbb{N}$ that $\mathcal{E}_n = \{\langle u, v\rangle \in V_n \mid (u, v) \vDash \mathcal{P}\}$. For $n = 0$ the result is trivial, and if it holds at rank $n \in \mathbb{N}$, then it is clear that $\mathcal{E}_{n+1} \subseteq \{\langle u, v\rangle \in V_{n+1} \mid (u, v) \vDash \mathcal{P}\}$. Conversely, if $u, v \in V_{n+1}$ satisfies $\mathcal{P}$, then either:

- the last letter of $u$ is $\bullet$ and so $v$ ends with a 1, and thus $\langle u, v\rangle \in \{\langle \bullet^{n+1}, v'1\rangle \mid v' \in V_n\} \subseteq \mathcal{E}_{n+1}$, or
- the last letter of $u$ is 0, and the last letter of $v$ is 0 or 1; letting $u = u'0$ and $v = v'a$ with $a \in 2$, we have that $\langle u', v'\rangle$ satisfy $\mathcal{P}$, so by induction hypothesis, $\langle u', v'\rangle \in \mathcal{E}_n$ and hence

$$\langle u, v\rangle \in \{\langle u'0, v'0\rangle \mid \langle u', v'\rangle \in \mathcal{E}_n\} \cup \{\langle u'0, v'1\rangle \mid \langle u', v'\rangle \in E_n\} \subseteq \mathcal{E}_{n+1}.$$

In all cases, $\langle u, v\rangle \in \mathcal{E}_{n+1}$ which concludes the induction. To conclude, it suffices to notice that the condition $\mathcal{P}$ is clearly automatic. $\square$

### VIII.6.4 Beyond Finite Duality

The notion of duality has been generalized in graph theory to *restricted dualities*, in which the quantification over structures is restricted to a fixed class of finite graphs. For instance, Naserasr showed [Nas07, Theorem 11] that for any finite planar undirected graph $\mathbf{G}$, then $\mathbf{G}$ is triangle-free *iff* $\mathbf{G}$ admits a homomorphism to the so-called Clebsch graph. Letting $\mathbf{H}_5$ denote the directed graph induced by it—meaning that if $\{u, v\}$ is an edge, we put an

[78] In fact some of the sets defining $\mathcal{E}_{n+1}$ are redundant because of these, but we keep them in the definition to emphasize this symmetry.





edge from $u$ to $v$ and from $v$ to $u$—it follows that $\mathcal{H}om(\text{Aut}, \mathbf{H}_5)$ is decidable when restricted to planar graphs, even though the full problem is undecidable.

This opens a wide class of problems that are undecidable by the dichotomy theorem for automatic structures, but that admit non-trivial restrictions which are decidable. We refer the reader to [NO12b] for more details on restricted dualities.





# *The Algebras for Automatic Relations*


### Abstract

We introduce "synchronous algebras", an algebraic structure tailored to recognize automatic relations (*a.k.a.* automatic relations, or regular relations). They are the equivalent of monoids for regular languages, however they conceptually differ in two points: first, they are typed and second, they are equipped with a dependency relation expressing constraints between elements of different types.

The interest of the proposed definition is that it allows to lift, in an effective way, pseudovarieties of regular languages to that of automatic relations, and we show how algebraic characterizations of pseudovarieties of regular languages can be lifted to the pseudovarieties of automatic relations that they induce. Since this construction is effective, this implies that the membership problem is decidable for (infinitely) many natural classes of automatic relations. A typical example of such a pseudovariety is the class of "group relations", defined as the relations recognized by finite-state synchronous permutation automata.

In order to prove this result, we adapt two pillars of algebraic language theory to synchronous algebras: (a) any relation admits a syntactic synchronous algebra recognizing it, and moreover, the relation is synchronous if, and only if, its syntactic algebra is finite and (b) classes of automatic relations with desirable closure properties (*i.e.* pseudovarieties) correspond to pseudovarieties of synchronous algebras.



### Acknowledgements

This article is mostly a reproduction of [Mor25], which was published at CSL '25, with minor modifications to the introduction and conclusion, and appendices have been incorporated to the main body. Section IX.5.3 is entirely new. We thank Pablo Barceló, Mikołaj Bojańczyk, and Diego Figueira for helpful discussions, and some anonymous reviewers for valuable feedback.




# Contents







## IX.1 Introduction

### IX.1.1 Motivation

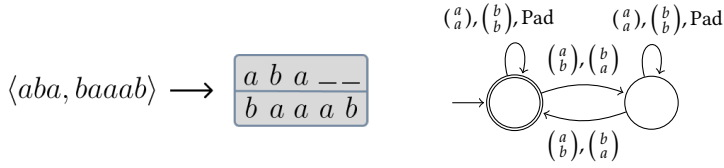

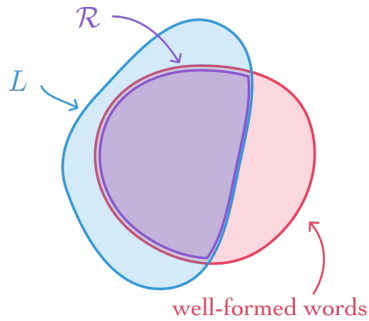

Figure IX.1: Encoding a pair of words of $\Sigma^* \times \Sigma^*$ into an element of $(\Sigma_\otimes^2)^*$ where $\Sigma_\otimes^2 \triangleq (\Sigma \times \Sigma) \cup (\Sigma \times \{\_\}) \cup (\{\_\} \times \Sigma)$ (left) and a deterministic complete synchronous automaton (right) over $\Sigma = \{a, b\}$ accepting the binary relation of pairs $(u, v)$ such that the number of $a$'s in $u_1 \dots u_k$ and in $v_1 \dots v_k$ are the same mod 2, where $k = \min(|u|, |v|)$. Pad denotes the set of transitions $\{(\begin{smallmatrix} a \\ \_ \end{smallmatrix}), (\begin{smallmatrix} b \\ \_ \end{smallmatrix}), (\begin{smallmatrix} \_ \\ a \end{smallmatrix}), (\begin{smallmatrix} \_ \\ b \end{smallmatrix})\}$.

As mentioned in Section VII.1, automatic relation can be seen as a regular language over the alphabet $\Sigma_\otimes^2 \triangleq (\Sigma \times \Sigma) \cup (\Sigma \times \{\_\}) \cup (\{\_\} \times \Sigma)$ of pairs. On the other hand any regular language $L$ over $\Sigma_\otimes^2$ produces an automatic relation when intersected with the language of all well-formed words—namely words where the padding symbols are consistently placed; see Section IX.2 for precise definitions. In fact, the semantics of synchronous automata such as the one in Figure IX.1 is precisely defined this way: it is the intersection of the "classical semantic" of the automaton, seen as an NFA, intersected with well-formed words.

Figure IX.2: Drawing in $(\Sigma_\otimes^2)^*$ of a $\mathcal{V}$-relation $\mathcal{R}$ and $\neg\mathcal{R} \triangleq \{(u, v) \in \Sigma^* \times \Sigma^* \mid (u, v) \notin \mathcal{R}\}$, where $\mathcal{R}$ is defined as $L \cap \mathrm{WellFormed}_\Sigma$ with $L \in \mathcal{V}$.

In particular, a class $\mathcal{V}$ of regular languages over $\Sigma_\otimes^2$ (e.g. first-order definable languages, group languages, etc.) induces a class of so-called $\mathcal{V}$-relations, defined as the relations over $\Sigma$ obtained as the intersection of some language of $\mathcal{V}$ with well-formed words, see Figure IX.2. For instance, the relation of Figure IX.1 is a $\mathcal{V}$-relation where $\mathcal{V}$ is the class of all group languages—these relations can be alternatively described as those recognized by a deterministic complete synchronous automaton whose transitions functions are permutations of states.

**Question IX.1.1.** Given a class $\mathcal{V}$ of languages, can we characterize and decide the class of $\mathcal{V}$-relations?

As we will see in Example IX.2.4, being in $\mathcal{V}_{\Sigma_\otimes^2}$, as a language, is a sufficient but not necessary condition for a relation to be a $\mathcal{V}$-relation.





## IX.1.2  Contributions

We answer this question positively. For this we first need to develop an algebraic theory of automatic relations, which enables us to prove the lifting theorem. In short, the lifting theorem states that algebraic characterizations of classes of word languages can be lifted in a canonical way to algebraic characterizations of classes of word relations.

The algebraic approach usually provides more than decidability: it attaches canonical algebras to languages/relations (*e.g.* monoids for languages of finite words), and often simple ways to characterize complex properties (*e.g.* first-order definability, see *e.g.* [Boj20], Theorem 2.6, p. 40]). Our synchronous algebras differ from monoids in two points:

- they are typed—a quite common feature in algebraic language theory, shared *e.g.* by $\omega$-semigroups [PP04, § 4.1, p. 91];
- they are equipped with a dependency relation, which expresses constraints between elements of different types—to our knowledge, this feature is entirely novel.[1]

To understand how this dependency relation appears, we find that looking at the syntactic congruence helps: recall that the syntactic congruence of a language $L \subseteq \Sigma^*$ is defined by

$$u \sim_L v \quad \textit{iff} \quad \forall x, y \in \Sigma^*, \; xuy \in L \Leftrightarrow xvy \in L.$$

This relation is, by definition, an equivalence relation. We claim that the same definition cannot capture the essence of automatic relations, for the simple reason that it might be that $u$, $v$, $x$ and $y$ are all well-formed, and that $xuy$ is well-formed but $xvy$ is not. In this case, asking for the property $xuy \in \mathcal{R} \Leftrightarrow xvy \in \mathcal{R}$ to hold does not make sense since the object $xvy$ simply does not exist in our universe. To tackle this problem, we need to relativize this property with the assumption that both $xuy$ and $xvy$ are well-formed. Doing so yields a notion of syntactic congruence that makes sense—see Section IX.3.4 for the formal definition—, however because of this relativization, it is no longer an equivalence relation: the non-trivial structure of this relation is precisely captured by the notion of dependency relation!

**Importantly, some variations are possible on the definition of synchronous algebras: in particular, one could get rid of the notion of dependency relation. However, we show in Section IX.4.2 that these simplified synchronous algebras cannot characterize the property of being a $\mathcal{V}$-relation. Therefore, the notion of dependency seems necessary to tackle Question IX.1.1.** Moreover, we show that these algebras arise from a monad, but to our knowledge none of the meta-theorems developing algebraic language theories over monads apply to it, see Section IX.A for more details.

We show that assuming that $\mathcal{V}$ is a *-pseudovariety of regular languages—in short, a class of regular languages with desirable closure properties—, then the

[1] Note that algebras equipped with binary relations have been studied before, *e.g.* Pin's ordered $\omega$-semigroups—see [Pin98, § 2.4, p. 7]—but the constraints (here the ordering) are always defined between elements of the *same type*.





algebraic characterization of $\mathcal{V}$ can be easily lifted to characterize $\mathcal{V}$-relations.

**Theorem IX.4.2** (*Lifting theorem: Elementary Formulation*). Given a relation $\mathcal{R}$ and a *-pseudovariety of regular languages $\mathcal{V}$ corresponding to a pseudovariety of monoids $\mathbb{V}$, the following are equivalent:

1. $\mathcal{R}$ is a $\mathcal{V}$-relation,
2. $\mathcal{R}$ is recognized by a finite synchronous algebra $\mathbf{A}$ whose underlying monoids are all in $\mathbb{V}$,
3. all underlying monoids of the syntactic synchronous algebras $\mathbf{A}_{\mathcal{R}}$ of $\mathcal{R}$ are in $\mathbb{V}$.

This theorem rests on a solid algebraic theory. First, we show the existence of syntactic algebras (Lemma IX.3.11): each relation $\mathcal{R}$ admits a unique canonical and minimal algebra $\mathbf{A}_{\mathcal{R}}$, which is finite *iff* the relation is automatic, and then, we exhibit a correspondence between classes of finite algebras and classes of automatic relations (Lemma IX.4.11)—we assume suitable closure properties; these classes are called "pseudovarieties". While the proof structures of Lemmas IX.3.11 and IX.4.11 follow the classic proofs, see *e.g.* [Pin22], the dependency relation has to be taken into account quite carefully, leading for instance to a surprising definition of residuals, see Definition IX.4.7.

**Remark IX.1.2.** All our results are described for binary relations, but can be extended to $k$-ary automatic relations. $\diamondsuit$

*Organization.* After giving preliminary results in Section IX.2, we introduce the synchronous algebras in Section IX.3 and show the existence of syntactic algebras. We then proceed to prove the lifting theorem for *-pseudovarieties in Section IX.4, and after introducing *-pseudovarieties of automatic relations, we provide a more algebraic reformulation of the lifting theorem (Theorem IX.4.13). We conclude this chapter with a short discussion in Section IX.5.

### IX.1.3 Related Work

The algebraic framework has been extended far beyond languages of finite words. Let us cite amongst others:

- Reutenauer's "algèbre associative syntactique"—pardon my French—for weighted languages [Reu80, Théorème I.2.1, p. 451] and their associated Eilenberg theorem [Reu80, Théorème III.1.1, p. 469];
- for languages of $\omega$-words, Wilke's algebras and $\omega$-semigroups, see [PP04, § II, pp. 75–131 & § VI, pp. 265–306];
- for languages over countable ordinals, Bedon defined "$\omega_1$-semigroupes syntaxiques" [Bed98, §3, pp. 49–109] and their Eilenberg theorem [BC98, Theorem 22, p. 62];
- for languages over countable scattered orderings, see Rispal's "$\diamondsuit$-semigroupe syntaxique" [Ris04, § 4.4, pp. 82–86] and their Eilenberg theorem [BR05, Theorem 6, p. 144];
- more generally, for languages over countable linear orderings, see Carton,





Colcombet & Puppis' "⊛-monoids" and "⊛-algebras" [CCP18, § 3, p. 7];

- Bojańczyk & Walukiewicz's forest algebras [BW08, § 1.3, p. 4] [Boj20, § 5, p. 159] dealing with tree languages;

- Engelfriet's hyperedge replacement algebras for graph languages [CE12, § 2.3, p. 100] [Boj15, § 6.2, p. 194].

A systemic approach has been recently developed using monads, see Section IX.A. For relations over words, recognizable relations are exactly the ones recognized by monoid morphisms $\Sigma^* \times \Sigma^* \to M$ where $M$ is finite. This can be trivially generalized to show that a relation $\mathcal{R}$ is a finite union of Cartesian products of languages in $\mathcal{V}$ if, and only if, it is recognized by a monoid from $\mathbb{V}$, the pseudovariety of monoids corresponding to $\mathcal{V}$, see Proposition VII.1.2. In 2023, Bojańczyk & Nguyễn managed to develop an algebraic structure called "transducer semigroups" for "regular functions" [BN23, Theorem 3.2, p. 6], an orthogonal class of relations to ours—see Figure VII.1.

The counterpart of $\mathcal{V}$-relations for deterministic transductions —that we call here *deterministic $\mathcal{V}$-transductions*—was studied by Filiot, Gauwin & Lhote [FGL19]: they show that if $\mathcal{V}$ has decidable membership, then "*deterministic $\mathcal{V}$-transductions*" also have decidable membership [FGL19, Theorem 2.1]—which is proven via the use of a minimal object (a deterministic transducer). They extend their result to functional transductions by showing the decidability of membership for *functional $\mathcal{V}$-transductions*, under the same assumption [FGL19, Theorem 4.10, p. 26]; however this proof does not rely on a canonical construction but rather on a finite set of minimal objects. These results are orthogonal to our lifting theorem: deterministic transductions and functional transductions are both orthogonal to the class of automatic relations, see Section VII.1. A different problem—focussing more on the semantics of the transduction—, called "$\mathcal{V}$-continuity" was studied by Cadilhac, Carton & Paperman [CCP20, Theorem 1.3, p. 3], although it has to be noted that their results only concern a finite number of pseudovarieties.

## IX.2  Preliminaries

### IX.2.1  Automata & Relations

We assume familiarity with basic algebraic language theory over finite words, see [Boj20, § 1, 2, 4, pp. 3–66 & pp. 107–156] for a succinct and monad-driven approach, or [Pin22, § I–XIV, pp. 3–247] for a more detailed presentation of the domain. We also refer to [SW21] for a presentation on pseudovarieties.[2] More precise pointers are given in ??.

Recall that relation is a subset of $\Sigma^* \times \Sigma^*$, where $\Sigma$ is an alphabet—*i.e.* a non-empty finite set. We define its *complement* $\neg\mathcal{R}$ as the relation $\{(u,v) \in \Sigma^* \times \Sigma^* \mid (u,v) \notin \mathcal{R}\}$. Letting $\Sigma_\otimes^2 \triangleq (\Sigma \times \Sigma) \cup (\Sigma \times \{\_\}) \cup (\{\_\} \times \Sigma)$, a synchronous automaton is a finite-state machine with initial states, final states, and non-deterministic transitions labelled by elements of $\Sigma_\otimes^2$. As

[2] "Pseudovarieties of *foo*" and "varieties of finite *foo*"—where *foo* is *e.g.* "groups" or "semigroups"—are used interchangeably in the literature.





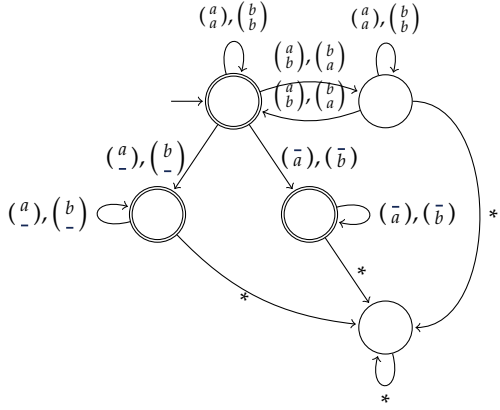



mentioned in Chapter VII, we denote by WellFormed$_\Sigma$ the set of well-formed words over $\Sigma_\otimes^2$ where the padding symbols are placed consistently, namely: if some padding symbol occurs on a tape/component, then the following symbols of this tape/component must all be padding symbols. From this constraint, and since $\left(\begin{smallmatrix} - \\ - \end{smallmatrix}\right) \notin \Sigma_\otimes^2$, there can never be padding symbols on both tapes.

Note that elements of WellFormed$_\Sigma$ are in natural bijection with $\Sigma^* \times \Sigma^*$—see Figure IX.1. The relation recognized by a synchronous automaton is the set of pairs $(u, v) \in \Sigma^* \times \Sigma^*$ such that their corresponding element in WellFormed$_\Sigma$ is the label of an accepting run of the automaton. Recall that we say that a relation is automatic if it is recognized by such a machine.

**Remark IX.2.1.** Crucially, in the semantics of synchronous automata we *never* try to feed them inputs where the padding symbols are not consistent: for instance, while

$$\left(\begin{smallmatrix} aab \\ b\_a \end{smallmatrix}\right), \text{ or } \left(\begin{smallmatrix} aba\_ \\ a\_\_b \end{smallmatrix}\right)$$

are sequences in $(\Sigma_\otimes^2)^*$, the behaviour of a synchronous automaton on such sequences is completely disregarded to define the relation it recognizes. $\Diamond$

We can then reformulate the definition of the semantics of a synchronous automaton, to make the connection with $\mathcal{V}$-relations—see the next subsection—explicit.

**Fact IX.2.2.** Given a synchronous automaton, its semantics as a synchronous automaton can be written as the intersection of its semantics as a classical automaton over $\Sigma_\otimes^2$ with WellFormed$_\Sigma$.

In particular a relation $\mathcal{R}$ is automatic if, and only if, it is a regular language when seen as a subset of $(\Sigma_\otimes^2)^*$.

### IX.2.2 Induced Relations

Given a class $\mathcal{V}$ of regular languages, the class of $\mathcal{V}$-relations over $\Sigma$ consists of all relations of the form $L \cap$ WellFormed$_\Sigma$ for some $L \in \mathcal{V}_{\Sigma_\otimes^2}$—see Figure IX.2.[3]

For instance, if $\mathcal{V}$ is the class of all regular languages, then by Fact IX.2.2, $\mathcal{V}$-relations are exactly the automatic relations! However, because of Re-

[3] The notation $L \in \mathcal{V}_{\Sigma_\otimes^2}$ means that $L$ is a language over the alphabet $\Sigma_\otimes^2$. See [Pin22, introduction of §XIII.1] for why classes of regular languages are defined in such a way.





mark IX.2.1, the minimal automaton for a relation, seen as a language over $\Sigma_\otimes^2$, can be significantly more complex than a deterministic complete synchronous automaton recognizing it, see Figure IX.3 in page 303—while the size blow-up is only polynomial, it breaks many of the structural properties of the automaton, such as the property of being a permutation automaton.

Note that if $\mathcal{R}$ belongs to $\mathcal{V}$ when $\mathcal{R}$ is seen as a language over $\Sigma_\otimes^2$, then $\mathcal{R}$ is a $\mathcal{V}$-relation. The converse implication holds under some strong assumption on $\mathcal{V}$ (Fact IX.2.3), but is not true in general (Example IX.2.4).

**Fact IX.2.3.** If $\mathcal{V}$ is a class of languages closed under intersection and that contains WellFormed$_\Sigma$, then a relation $\mathcal{R}$ is a $\mathcal{V}$-relation if, and only if, it belongs to $\mathcal{V}$ when seen as a language over $\Sigma_\otimes^2$.

Classes of languages $\mathcal{V}$ satisfying the previous assumption (*e.g.* first-order definable languages, piecewise-testable languages, etc.) are easy to capture when it comes to $\mathcal{V}$-relations since this class reduces to $\mathcal{V}$-languages. So, in the remaining of the chapter, we will focus on classes $\mathcal{V}$ which do not satisfy the assumptions of Fact IX.2.3, such as group languages.

**Example IX.2.4** (Group relations). If $\mathcal{V}$ is the class of group languages, namely languages recognized by permutation automata[4] or equivalently by a finite group, then we call $\mathcal{V}$-relations "*group relations*". They can be characterized as relations recognized by permutation synchronous automata. For instance, the relation of Figure IX.1 is a group relation as witnessed by the permutation synchronous automaton of Figure IX.1. Note however that it is not a group language, when seen as a language over $\Sigma_\otimes^2$, since its minimal automaton over $\Sigma_\otimes^2$ is not a permutation automaton, see Figure IX.3 on Page 303. $\diamond$

[4] A permutation automaton is a finite-state deterministic complete automaton whose transition functions are all permutations of states.

**Fact IX.2.5.** Given a relation $\mathcal{R}$ and a class $\mathcal{V}$ of languages, the following are equivalent:

1. $\mathcal{R}$ is a $\mathcal{V}$-relation;
2. $\mathcal{R}$ and $\neg\mathcal{R}$ are $\mathcal{V}$-separable as languages over $\Sigma_\otimes^2$, *i.e.* there is a language in $\mathcal{V}$ which contains $\mathcal{R}$ and does not intersect $\neg\mathcal{R}$.

*Proof.* By definition, see Figure IX.2, on page 299. □

And so, if the $\mathcal{V}$-separability problem is decidable, then the class of $\mathcal{V}$-relations is decidable. However, there are pseudovarieties $\mathcal{V}$ with decidable membership but undecidable separability problem [RS11, Corollary 1.6, p. 478].[5] Moreover, some of these classes do not contain WellFormed$_\Sigma$ [RS11, Corollary 1.7, p. 478]. But beyond this, even when a separation algorithm exists, it can be conceptually much harder than its membership counterpart: for instance, deciding membership for group languages is trivial—it boils down to checking if a monoid is a group—, yet the decidability of the separation problem for group languages is considered to be one of the major results in semigroup theory: it follows from Ash's famous type II theorem [Ash91, Theorem 2.1, p. 129], see [HMPR91, Theorem 1.1, p. 3] for a presentation of

[5] The paper cited only claims undecidability of pointlikes, but it was noted in [GS19, § 1, pp. 1–2] that undecidability of the 2-pointlikes also holds, which is a problem equivalent to separability by [Alm99, Proposition 3.4, p. 6].





the result in terms of pointlike sets, see also [PZ23, § III, Theorem 8, p. 5] for an elegant automata-theoretic reformulation.

## IX.3 Synchronous Algebras

In this section, we introduce and study the "elementary" properties of synchronous algebras.

### IX.3.1 Types & Dependent Sets

*Motivation.* The axiomatization of a semigroup reflects the algebraic structure of finite words: these objects can be concatenated, in an associative way—reflecting the linearity of words. Now observe that elements of $\mathrm{WellFormed}_\Sigma$ are still linear, but not all words can be concatenated together: for instance, $\binom{a}{\_}$ cannot be followed by $\binom{a}{b}$. Formally, given two words $u, v \in \mathrm{WellFormed}_\Sigma$, to decide if $uv \in \mathrm{WellFormed}_\Sigma$ it is necessary and sufficient to know if the last pair of $u$ and first pair of $v$ consists of a pair of proper letters (denoted by $\mathsf{L/L}$), a pair of a proper letter and a blank/padding symbol ($\mathsf{L/B}$) or a pair of a blank/padding symbol and a proper letter ($\mathsf{B/L}$). This information is called the *letter-type* of an element of $\Sigma_\otimes^2$.

We then define the *type* of a word of $(\Sigma_\otimes^2)^+$ as the pair $(\alpha, \beta)$, usually written $\alpha \to \beta$, of the letter-types of its first and last letters. It is then routine to check that the possible types of well-formed words are

$$\mathcal{T} \triangleq \left\{ \mathsf{L/L} \to \mathsf{L/L}, \ \mathsf{L/L} \to \mathsf{L/B}, \ \mathsf{L/B} \to \mathsf{L/B}, \ \mathsf{L/L} \to \mathsf{B/L}, \ \mathsf{B/L} \to \mathsf{B/L} \right\}.$$

For the sake of readability, we will write $\alpha$ instead of $\alpha \to \alpha$ for $\alpha \in \{\mathsf{L/L}, \mathsf{L/B}, \mathsf{B/L}\}$.

One non-trivial point lies in the following innocuous question: what is the type of the empty word? Any type of $\mathcal{T}$ sounds like an acceptable answer. But then it would be natural to say that the concatenation of $\binom{aaa}{aaa}$ of type $\mathsf{L/L}$ with the empty word of type $\mathsf{L/L} \to \mathsf{L/B}$ should be $\binom{aaa}{aaa}$ of type $\mathsf{L/L} \to \mathsf{L/B}$. Automata-wise, this would represent a sequence of transitions $\binom{a}{a}, \binom{a}{a}, \binom{a}{a}$ together with the promise that the next transition would have a padding symbol on its second tape. But then, one has to formalize the idea that the two elements $\binom{aaa}{aaa}$ of type $\mathsf{L/L}$ and $\mathsf{L/L} \to \mathsf{L/B}$ represent the same underlying pair of words of $\Sigma^* \times \Sigma^*$: this idea will be captured by what we call a dependency relation. A more natural solution would be to simply introduce a new type for the empty word (or to forbid it), but we show in Section IX.4.2 that the resulting notion of algebras cannot capture the property of being a $\mathcal{V}$-relation.

A *$\mathcal{T}$-typed set* (or *typed set* for short) consists of a tuple $\mathbf{X} = (X_\tau)_{\tau \in \mathcal{T}}$, where each $X_\tau$ is a set. Instead of $x \in X_\tau$, we will often write $x_\tau \in \mathbf{X}$. A *map between typed sets* $\mathbf{X}$ and $\mathbf{Y}$ is a collection of functions $X_\tau \to Y_\tau$ for each type $\tau$. Similarly, a subset of $\mathbf{X}$ is a tuple of subsets of $X_\tau$ for each type $\tau$. To make the notations less heavy, we will often think of typed sets as sets with





type annotations rather than tuples, and ask that all operators/constructions should preserve this type.

**Definition IX.3.1.** A *dependency relation* over a typed set **X** consists of a reflexive and symmetric relation $\asymp$ over $\biguplus \mathbf{X} \triangleq \bigcup_{\tau \in \mathcal{T}} X_\tau \times \{\tau\}$, such that for all $x_\sigma, y_\sigma \in \mathbf{X}$, if $x_\sigma \asymp y_\sigma$, then $x_\sigma = y_\tau$.

Crucially, we do not ask for this relation to be transitive—in some examples the dependency relation will be an equivalence relation, but not always (Example IX.3.15), and this non-transitivity is actually an important feature, motivated amongst other by the syntactic congruence and Corollary IX.3.14.

A *dependent set* is a $\mathcal{T}$-typed set together with a dependency relation over it. A *closed subset* of a dependent set $\langle \mathbf{X}, \asymp \rangle$ is a subset $C \subseteq \mathbf{X}$ such that for all $x, x' \in \mathbf{X}$, if $x \asymp x'$ then $x \in C \iff x' \in C$.[6]

> [6] In other words, $C$ is a union of equivalence classes of the transitive closure of $\asymp$.

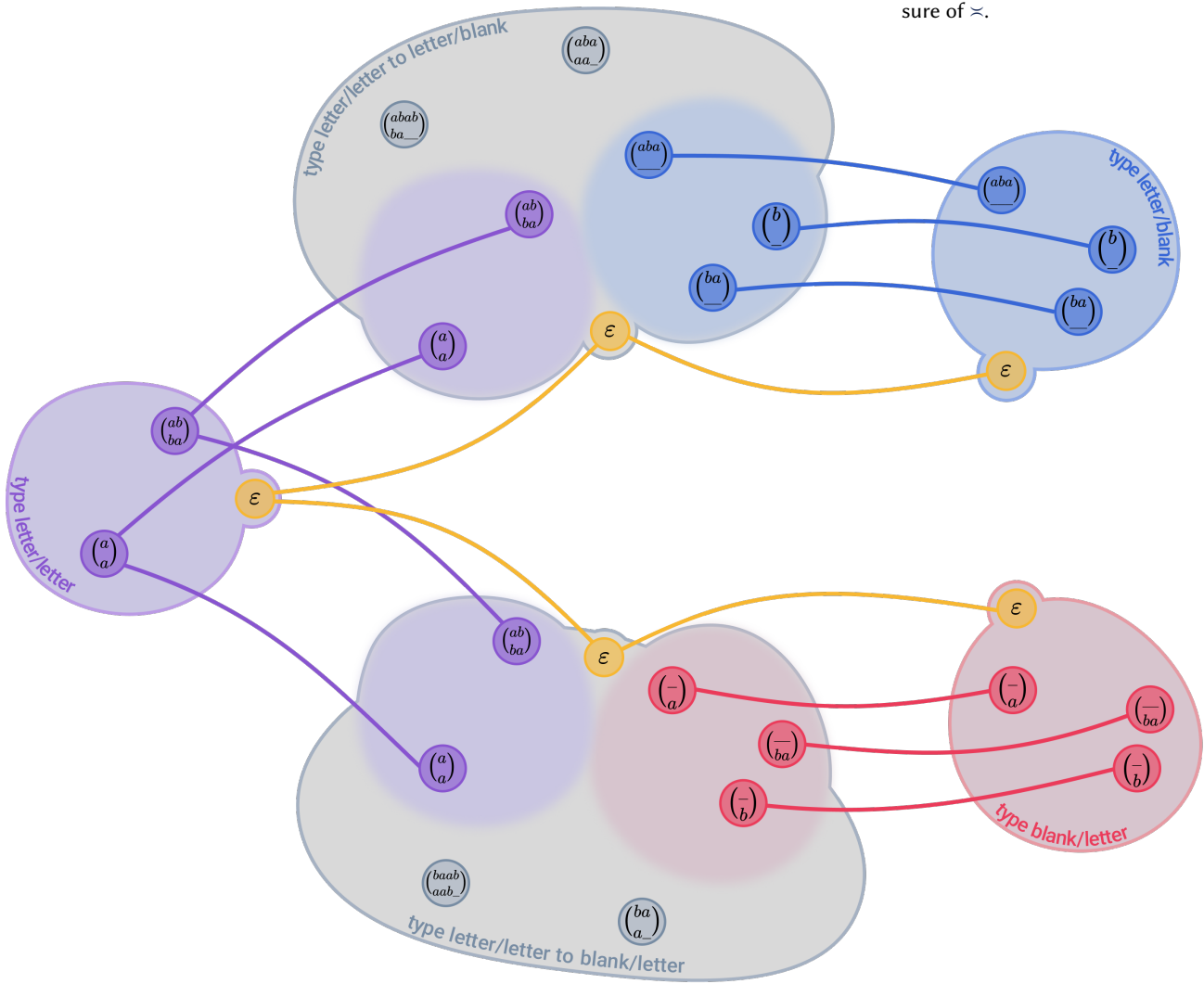

Figure IX.4: Representation of the dependent set $\mathbf{S}_2\Sigma$ of synchronous words. Coloured edges represent the dependency relation, and self-loops are not drawn.





**Example IX.3.2.** Given a finite alphabet $\Sigma$, let $\mathbf{S}_2\Sigma$ be the dependent set of *synchronous words* defined by:

- $(\mathbf{S}_2\Sigma)_{\llcorner/\llcorner} \triangleq (\Sigma \times \Sigma)^*$,
- $(\mathbf{S}_2\Sigma)_{\llcorner/\llcorner \to \llcorner/\text{в}} \triangleq (\Sigma \times \Sigma)^*(\Sigma \times \_)^*$,
- $(\mathbf{S}_2\Sigma)_{\llcorner/\text{в}} \triangleq (\Sigma \times \_)^*$,
- $(\mathbf{S}_2\Sigma)_{\llcorner/\llcorner \to \text{в}/\llcorner} \triangleq (\Sigma \times \Sigma)^*(\_ \times \Sigma)^*$,
- $(\mathbf{S}_2\Sigma)_{\text{в}/\llcorner} \triangleq (\_ \times \Sigma)^*$.

Moreover, $\asymp$ is the reflexive and symmetric closure of the relation that identifies $u_{\llcorner/\llcorner}$ with $u_{\llcorner/\llcorner \to \beta}$ for all $u \in (\Sigma \times \Sigma^*)$ and $\beta \in \{\llcorner/\text{в}, \text{в}/\llcorner\}$, and $u_{\llcorner/\llcorner \to \llcorner/\text{в}}$ with $u_{\llcorner/\text{в}}$ for $u \in (\Sigma \times \_)^*$, and $u_{\llcorner/\llcorner \to \text{в}/\llcorner}$ with $u_{\text{в}/\llcorner}$ for $u \in (\_ \times \Sigma)^*$. This structure is depicted in Figure IX.4.[7]  $\diamondsuit$

Given a relation $\mathcal{R} \subseteq \Sigma^* \times \Sigma^*$, we denote by $\underline{\mathcal{R}} \triangleq \{(u, v)_\tau \mid (u, v)_\tau \in \mathbf{S}_2\Sigma$ and $(u, v) \in \mathcal{R}\}$ the closed subset of $\mathbf{S}_2\Sigma$ induced by $\mathcal{R}$.

**Fact IX.3.3.** The map $\mathcal{R} \mapsto \underline{\mathcal{R}}$ is a bijection between binary relations over $\Sigma^*$ and closed subsets of $\mathbf{S}_2\Sigma$.

*Proof.* Let $f$ be the function which maps a closed subset $C$ of $\mathbf{S}_2\Sigma$ to $\{(u, v) \in \Sigma^* \times \Sigma^* \mid (u, v)_\tau \in C$ for some $\tau \in \mathcal{T}\}$. It then follows that $f \circ \underline{-}$ (resp. $\underline{f(-)}$) is the identity on subsets of $\Sigma^* \times \Sigma^*$ (resp. closed subsets of $\mathbf{S}_2\Sigma$).  $\square$

### IX.3.2   Synchronous Algebras

One key property of types is that some of them can be concatenated to produce other types. We say that two types $\sigma, \tau \in \mathcal{T}$ are *compatible* when there exists non-empty words $u, v \in \text{WellFormed}_\Sigma$ of type $\sigma$ and $\tau$, respectively, such that $uv$ is well-formed. Said otherwise, $\alpha \to \beta$ is compatible with $\beta' \to \gamma$ if either $\beta = \beta'$ or $\beta = \llcorner/\llcorner$—indeed, for this last case note that *e.g.* the concatenation of $\left(\begin{smallmatrix} aaa \\ aaa \end{smallmatrix}\right)$ of type $\llcorner/\llcorner$ with $\left(\begin{smallmatrix} \bar{a}\bar{a} \\ aa \end{smallmatrix}\right)$ of type $\text{в}/\llcorner$ is well-formed. Lastly, if $\alpha \to \beta$ is compatible with $\beta' \to \gamma$, we define their product as $(\alpha \to \beta) \cdot (\beta' \to \gamma) \triangleq \alpha \to \gamma$. Note that this partial operation is associative, in the following sense: for $\rho, \sigma, \tau \in \mathcal{T}$, $(\rho \cdot \sigma) \cdot \tau$ is well-defined if and only if $\rho \cdot (\sigma \cdot \tau)$ is well-defined, in which case both types are equal. This implies that the notion of compatibility of types can be unambiguously lifted to finite lists of types $\tau_1, \dots, \tau_n$.

**Definition IX.3.4.** A *synchronous algebra* $\langle \mathbf{A}, \cdot, \asymp \rangle$ consists of a dependent set $\langle \mathbf{A}, \asymp \rangle$ together with a partial binary operation $\cdot$ on $\mathbf{A}$, called *product* such that:

- for $x_\sigma, y_\tau \in \mathbf{A}$, $x_\sigma \cdot y_\tau$ is defined *iff* $\sigma$ and $\tau$ are compatible,
- *associativity:* for all $x_\rho, y_\sigma, z_\tau \in \mathbf{A}$, if $\rho, \sigma, \tau$ are compatible:

$$(x_\rho \cdot y_\sigma) \cdot z_\tau = x_\rho \cdot (y_\sigma \cdot z_\tau),$$

- *"monotonicity":* for all $x_\sigma, x'_{\sigma'}, y_\tau \in \mathbf{A}$, if $x_\sigma \asymp x'_{\sigma'}$ and both $\sigma, \tau$ and $\sigma', \tau$ are compatible, then $x_\sigma \cdot y_\tau \asymp x'_{\sigma'} \cdot y_\tau$, and dually if $\tau, \sigma$ and $\tau, \sigma'$ are compatible, then $y_\tau \cdot x_\sigma \asymp y_\tau \cdot x'_{\sigma'}$,
- *units:* for each type $\tau$ there is an element $1_\tau \in \mathbf{A}$ such that for any $x_\sigma \in \mathbf{A}$, then $1_\tau \cdot x_\sigma \asymp x_\sigma$ if $\tau$ and $\sigma$ are compatible, and $x_\sigma \cdot 1_\tau \asymp x_\sigma$ if $\sigma$ and $\tau$ are

[7] The index refers to the arity of the relations we are considering: here we focus on binary relations, but all constructions can be generalized to higher arities.





compatible, and moreover, $1_{\llcorner/\llcorner \to \beta} = 1_{\llcorner/\llcorner} \cdot 1_{\beta}$ for $\beta \in \{\llcorner/\mathtt{B}, \mathtt{B}/\llcorner\}$.

Note in particular that for any type $\tau \in \{\llcorner/\llcorner, \llcorner/\mathtt{B}, \mathtt{B}/\llcorner\}$, then $1_{\tau} \cdot x_{\tau} \asymp x_{\tau}$ but since $1_{\tau} \cdot x_{\tau}$ has type $\tau$ and $\asymp$ is a dependency relation, then $1_{\tau} \cdot x_{\tau} = x_{\tau}$. This implies in particular that restricting $\langle \mathbf{A}, \cdot \rangle$ to a type $\llcorner/\llcorner$, $\llcorner/\mathtt{B}$ or $\mathtt{B}/\llcorner$ yields a monoid. These are called the three *underlying monoids* of $\mathbf{A}$. The canonical example of synchronous algebras is synchronous words $\mathbf{S}_2\Sigma$ under concatenation. Its underlying monoids are $(\Sigma \times \Sigma)^*$, $(\Sigma \times \{\_\})^*$ and $(\{\_\} \times \Sigma)^*$.

**Fact IX.3.5.** Any closed subset of $\mathbf{A}$ either contains all units, or none of them.

*Proof.* From $1_{\llcorner/\llcorner \to \llcorner/\mathtt{B}} = 1_{\llcorner/\llcorner} \cdot 1_{\llcorner/\mathtt{B}}$ we have $1_{\llcorner/\llcorner} \asymp 1_{\llcorner/\llcorner \to \llcorner/\mathtt{B}}$ and $1_{\llcorner/\llcorner \to \llcorner/\mathtt{B}} \asymp 1_{\llcorner/\mathtt{B}}$. By symmetry between $\llcorner/\mathtt{B}$ and $\mathtt{B}/\llcorner$, we also have $1_{\llcorner/\llcorner} \asymp 1_{\llcorner/\llcorner \to \mathtt{B}/\llcorner}$ and $1_{\llcorner/\llcorner \to \mathtt{B}/\llcorner} \asymp 1_{\mathtt{B}/\llcorner}$. Hence, if a closed subset of $\mathbf{A}$ contains at least one unit, then it must contain them all. $\square$

Note that the product induces a monoid left (resp. right) action of the underlying monoid $\mathbf{A}_{\llcorner/\llcorner}$ (resp. $\mathbf{A}_{\llcorner/\mathtt{B}}$) on the set $\mathbf{A}_{\llcorner/\llcorner \to \llcorner/\mathtt{B}}$. Moreover, $x_{\llcorner/\llcorner} \mapsto x_{\llcorner/\llcorner} \cdot 1_{\llcorner/\mathtt{B}}$ identifies any element of type $\llcorner/\llcorner$ with an element of type $\llcorner/\llcorner \to \llcorner/\mathtt{B}$. Over $\mathbf{S}_2\Sigma$, these identifications are injective, but it need not be the case in general. Note also that in general, $x_{\llcorner/\llcorner} \cdot 1_{\llcorner/\llcorner \to \llcorner/\mathtt{B}} = x_{\llcorner/\llcorner} \cdot 1_{\llcorner/\llcorner} \cdot 1_{\llcorner/\mathtt{B}} = x_{\llcorner/\llcorner} \cdot 1_{\llcorner/\mathtt{B}}$.

**Remark IX.3.6.** There exists a monad over the category of dependent sets whose Eilenberg-Moore algebras exactly correspond to synchronous algebras, see Section IX.A. $\diamond$

*Morphisms of synchronous algebras* are defined naturally as maps that preserve the type, units, the product and the dependency relation.

*Free algebras* $\mathbf{S}_2\Sigma$ is free in the sense that for any synchronous algebra $\mathbf{A}$, there is a natural bijection between synchronous algebra morphisms $\mathbf{S}_2\Sigma \to \mathbf{A}$ and maps of typed sets $\Sigma_\otimes^2 \to \mathbf{A}$. Said otherwise, synchronous algebra morphisms are uniquely defined by their value on $\Sigma_\otimes^2$.

### IX.3.3 Recognizability

Given a synchronous algebra $\mathbf{A}$, a morphism $\phi \colon \mathbf{S}_2\Sigma \to \mathbf{A}$ and a closed subset Acc $\subseteq \mathbf{A}$ called "accepting set", we say that $\langle \phi, \mathbf{A}, \mathrm{Acc} \rangle$ *recognizes* a relation $\mathcal{R} \subseteq \Sigma^* \times \Sigma^*$ when $\underline{\mathcal{R}} = \phi^{-1}[\mathrm{Acc}]$. We extend the notion of recognizability to $\langle \phi, \mathbf{A} \rangle$ or to simply $\mathbf{A}$ by existential quantification over the missing elements in the tuple $\langle \phi, \mathbf{A}, \mathrm{Acc} \rangle$.

*Synchronous algebra induced by a monoid* A monoid morphism $\phi \colon (\Sigma_\otimes^2)^* \to M$ naturally *induces* a synchronous algebra morphism $\tilde{\phi} \colon \mathbf{S}_2\Sigma \to \mathbf{A}_M$, where:

- $\mathbf{A}_M$ has for every type $\tau$ a copy of $M$, and $\asymp$ is $\{(x_\sigma, x_\tau) \mid x \in M, \sigma, \tau \in \mathcal{T}\}$,
- for all $x_\sigma, y_\tau \in \mathbf{A}_M$ with compatible type, $x_\sigma \cdot y_\tau \triangleq (x \cdot y)_{\sigma \cdot \tau}$,
- $\tilde{\phi}\left(\begin{smallmatrix} a \\ b \end{smallmatrix}\right) \triangleq \left(\phi\left(\begin{smallmatrix} a \\ b \end{smallmatrix}\right)\right)_{\llcorner/\llcorner}$, $\tilde{\phi}\left(\begin{smallmatrix} a \\ \_ \end{smallmatrix}\right) \triangleq \left(\phi\left(\begin{smallmatrix} a \\ \_ \end{smallmatrix}\right)\right)_{\llcorner/\mathtt{B}}$, and $\tilde{\phi}\left(\begin{smallmatrix} \_ \\ a \end{smallmatrix}\right) \triangleq \left(\phi\left(\begin{smallmatrix} \_ \\ a \end{smallmatrix}\right)\right)_{\mathtt{B}/\llcorner}$.





The algebra simply duplicates $M$ as many times as needed and identifies two elements together when they originated from the same element of $M$.

**Fact IX.3.7.** If $\phi$ recognizes $\mathcal{R}$ for some relation $\mathcal{R} \subseteq \Sigma^* \times \Sigma^*$ seen as a language over $\Sigma^2_\otimes$, then $\tilde{\phi}$ recognizes $\mathcal{R}$.

*Consolidation of a synchronous algebra* Given a synchronous algebra morphism $\phi \colon \mathbf{S}_2\Sigma \to \mathbf{A}$, define its *consolidation*[8] as the semigroup morphism $\phi^0 \colon (\Sigma^2_\otimes)^* \to \mathbf{A}^0$, where $\mathbf{A}^0$ is the monoid obtained from $\biguplus \mathbf{A}$ by first merging units, by adding a zero (denoted by 0), and extending $\cdot$ to be a total function by letting all missing products equal 0, and $\phi^0$ sends a word $u \in (\Sigma^2_\otimes)^*$ to



- $0$ if $u$ is not well-formed,
- $\phi(u_{\llcorner\lrcorner})$ if $u \in (\Sigma \times \Sigma)^*$,
- $\phi(u_{\llcorner B})$ if $u \in (\Sigma \times \_)^+$,
- $\phi(u_{B\lrcorner})$ if $u \in (\_ \times \Sigma)^+$,
- $\phi(u_{\llcorner\lrcorner \to \llcorner B})$ if $u \in (\Sigma \times \Sigma)^+(\Sigma \times \_)^+$,
- $\phi(u_{\llcorner\lrcorner \to B\lrcorner})$ if $u \in (\Sigma \times \Sigma)^+(\_ \times \Sigma)^+$.

Note that this operation disregards the dependency relation of $\mathbf{A}$.

**Fact IX.3.8.** If $\phi$ recognizes some relation $\mathcal{R}$, then $\phi^0$ recognizes $\mathcal{R}$, when seen as a language over $\Sigma^2_\otimes$.

The following result follows from Facts IX.2.2, IX.3.7 and IX.3.8.

**Proposition IX.3.9.** A relation is automatic if and only if it is recognized by a finite synchronous algebra.

Let us continue with a slightly less trivial example of algebra.

**Example IX.3.10** (Group relations: Example IX.2.4, cont'd.). Fix $p, q \in \mathbb{N}_{>0}$. Let $\mathbf{Z}_{p,q}$ denote the algebra whose underlying monoids are:
- the trivial monoid $(0, +)$ for type $\llcorner/\lrcorner$,
- the cyclic monoid $(\mathbb{Z}/p\mathbb{Z}, +)$ for type $\llcorner/B$,
- the cyclic monoid $(\mathbb{Z}/q\mathbb{Z}, +)$ for type $B/\lrcorner$.

Moreover, the sets $Z_{\llcorner\lrcorner \to \llcorner B}$ and $Z_{\llcorner\lrcorner \to B\lrcorner}$ are defined as $\mathbb{Z}/p\mathbb{Z}$ and $\mathbb{Z}/q\mathbb{Z}$, respectively. The product is addition—we identify $0_{\llcorner\lrcorner}$ with the zero of $\mathbb{Z}/p\mathbb{Z}$ and of $\mathbb{Z}/q\mathbb{Z}$. We denote by $\bar{k}$ the equivalence class of $k \in \mathbb{Z}$ in $\mathbb{Z}/n\mathbb{Z}$ when $n$ is clear from context. The dependency relation identifies (1) all units together and (2) $x_\sigma$ with $1_\tau \cdot x_\sigma$ and $x_\sigma \cdot 1_\tau$ when the types are compatible.

Let $\phi \colon \mathbf{S}_2\Sigma \to \mathbf{Z}_{p,q}$ be the synchronous algebra morphism defined by

$$\phi\left(\begin{smallmatrix} a \\ b \end{smallmatrix}\right) \triangleq \bar{0}_{\llcorner\lrcorner}, \quad \phi\left(\begin{smallmatrix} a \\ \_ \end{smallmatrix}\right) \triangleq \bar{1}_{\llcorner B}, \quad \phi\left(\begin{smallmatrix} \_ \\ a \end{smallmatrix}\right) \triangleq \bar{1}_{B\lrcorner} \quad \text{and} \quad \phi(\varepsilon_\tau) \triangleq \bar{0}_\tau \text{ for } \tau \in \mathcal{T}.$$

This morphism recognizes any relation of the form

$$\mathcal{R}^{I,J} \triangleq \Big\{ (u, v) \,\Big|\, |u| > |v| \text{ and } (|u| - |v| \bmod p) \in I, \text{ or}$$
$$|u| < |v| \text{ and } (|v| - |u| \bmod q) \in J. \Big\},$$

where $I \subseteq \mathbb{Z}/p\mathbb{Z}$ and $J \subseteq \mathbb{Z}/q\mathbb{Z}$ are such that $\bar{0} \notin I$ and $\bar{0} \notin J$. This last





condition is necessary because the accepting set has to be a closed subset of $\mathbf{Z}_{p,q}$: if $\bar{0}$ was in $I$, then we would need $\bar{0} \in J$, but also to add $\bar{0}_{\text{\tiny L,L}}$ to the accepting set: this would recognize

$$\Big\{(u, v) \mid |u| > |v| \text{ and } (|u| - |v| \bmod p) \in I, \text{ or}$$
$$|u| < |v| \text{ and } (|v| - |u| \bmod q) \in J, \text{ or } |u| = |v|\Big\}.$$

Note also that all relations $\mathcal{R}^{I,J}$ with $\bar{0} \notin I$ and $\bar{0} \notin J$ are group relations: letting $G$ be the group $\mathbb{Z}/p\mathbb{Z} \times \mathbb{Z}/q\mathbb{Z}$, $\mathcal{R}$ can be written as $\text{WellFormed}_\Sigma \cap \psi^{-1}[I \times \{0\} \cup \{0\} \times J]$ where $\psi\colon (\Sigma_\otimes^2)^* \to G$ is the monoid morphism defined by $\psi\big(\begin{smallmatrix} a \\ b \end{smallmatrix}\big) \triangleq (\bar{0}, \bar{0})$, $\psi\big(\begin{smallmatrix} a \\ \_ \end{smallmatrix}\big) \triangleq (\bar{1}, \bar{0})$ and $\psi\big(\begin{smallmatrix} \_ \\ a \end{smallmatrix}\big) \triangleq (\bar{0}, \bar{1})$. $\diamond$

### IX.3.4  Syntactic Morphisms & Algebras

**Lemma IX.3.11** (*Syntactic morphism theorem*). For each relation $\mathcal{R}$, there exists a surjective synchronous algebra morphism

$$\eta_\mathcal{R}\colon \mathbf{S}_2\Sigma \twoheadrightarrow \mathbf{A}_\mathcal{R}$$

that recognizes $\mathcal{R}$ and is such that for any other surjective synchronous algebra morphism $\phi\colon \mathbf{S}_2\Sigma \twoheadrightarrow \mathbf{B}$ recognizing $\mathcal{R}$, there exists a synchronous algebra morphism $\psi\colon \mathbf{B} \twoheadrightarrow \mathbf{A}_\mathcal{R}$ such that the diagram

commutes. The objects $\eta_\mathcal{R}$ and $\mathbf{A}_\mathcal{R}$ are called the *syntactic synchronous algebra morphism* and *syntactic synchronous algebra* of $\mathcal{R}$, respectively. Moreover, these objects are unique up to isomorphisms of the algebra.

**Corollary IX.3.12** (of Proposition IX.3.9 and Lemma IX.3.11). A relation is automatic if and only if its syntactic synchronous algebra is finite.

The proof of Lemma IX.3.11 relies, as in the case of monoids, on the notion of congruence.

Given a synchronous algebra $\langle \mathbf{A}, \asymp, \cdot \rangle$, a *congruence* is any reflexive, symmetric relation $\equiv$ over $\mathbf{A}$ which is coarser than $\asymp$, and which is *locally transitive*, meaning that for all $x_\sigma, x'_\sigma, y_\tau, y'_\tau \in \mathbf{X}$, if $x'_\sigma \equiv x_\sigma$, $x_\sigma \equiv y_\tau$ and $y_\tau \equiv y'_\tau$, then $x'_\sigma \equiv y'_\tau$.[9]

The *quotient structure* $\mathbf{A}/\!\!\equiv$ of $\mathbf{A}$ by a congruence $\equiv$ is defined as follows:

- its underlying typed set consists of the equivalence classes of $\mathbf{A}$ under the equivalence relation $\{(x_\sigma, y_\sigma) \mid x_\sigma \equiv y_\sigma\}$, such a class being abusively denoted by $[x]_\equiv$,
- its product is the product induced by $\mathbf{A}$, in the sense that $[x]_\equiv \cdot [y]_\equiv \triangleq [xy]_\equiv$, and
- its dependency relation is the relation induced by $\asymp$, *i.e.* $[x]_\equiv \asymp [y]_\equiv$ whenever $x \equiv y$,
- its units are defined as the equivalence classes of the units of $\mathbf{A}$.

[9] In particular, it implies that $\equiv$ is transitive when restricted to elements of the same type.





Moreover, $x \mapsto [x]_{\equiv}$ defines a surjective morphism of synchronous algebras from $\mathbf{A}$ to $\mathbf{A}/_{\equiv}$.

Given a synchronous algebra $\langle \mathbf{A}, \asymp, \cdot \rangle$ and a closed subset $C \subseteq \mathbf{A}$, we define a congruence $\equiv_C$, called *syntactic congruence* of $C$ over $\mathbf{A}$ by letting $a_\sigma \equiv_C b_\tau$ when for all $x, y \in \mathbf{A}$

- if both $xa_\sigma y$ and $xb_\tau y$ are defined, then $xa_\sigma y \in C$ *iff* $xb_\tau y \in C$, and
- if both $xa_\sigma$ and $xb_\tau$ are defined, then $xa_\sigma \in C$ *iff* $xb_\tau \in C$, and
- if both $a_\sigma y$ and $b_\tau y$ are defined, then $a_\sigma y \in C$ *iff* $b_\tau y \in C$.

It is routine to check that the syntactic congruence is indeed a congruence. For instance, to prove that $\equiv_C$ is coarser than $\asymp$, observe that if $a_\sigma \asymp b_\tau$, then for all $x, y$ s.t. both $xa_\sigma y$ and $xb_\tau y$ are defined, then $xa_\sigma y \asymp xb_\tau y$, and since $C$ is a closed subset of $\mathbf{A}$, $xa_\sigma y \in C$ *iff* $xb_\tau y \in C$. The other two conditions are proven in the same fashion. Note however that while the relation is locally transitive, it is not transitive in general.

When $\mathcal{R} \subseteq \Sigma^* \times \Sigma^*$ is a relation, we abuse the notation and write $\equiv_{\mathcal{R}}$ to denote the syntactic congruence $\equiv_{\underline{\mathcal{R}}}$ of $\underline{\mathcal{R}}$ in $\mathbf{S}_2\Sigma$. The existence of the syntactic morphism then follows from the next proposition.

**Proposition IX.3.13.** Let $\phi \colon \mathbf{S}_2\Sigma \twoheadrightarrow \mathbf{A}$ be a surjective synchronous algebra morphism that recognizes $\mathcal{R}$, say $\underline{\mathcal{R}} = \phi^{-1}[\mathrm{Acc}]$ for some closed subset $\mathrm{Acc} \subseteq \mathbf{A}$, then

$$
\begin{aligned}
\phi/_{\equiv_{\mathrm{Acc}}} \colon \quad \mathbf{S}_2\Sigma \quad &\to \quad \mathbf{A}/_{\equiv_{\mathrm{Acc}}} \\
u \quad &\mapsto \quad [\phi(u)]_{\equiv_{\mathrm{Acc}}}
\end{aligned}
$$

is the syntactic morphism of $\mathcal{R}$.

*Proof.* We claim that $\phi/_{\equiv_{\mathrm{Acc}}}$ is *a*, and hence *the*, syntactic synchronous algebra morphism of $\mathcal{R}$. First, if $\phi(u) \equiv_{\mathrm{Acc}} \phi(v)$ then in particular $\phi(u) \in \mathrm{Acc}$ *iff* $\phi(v) \in \mathrm{Acc}$. It follows that the preimage of $\mathrm{Acc}' \triangleq [\mathrm{Acc}]_{\equiv_{\mathrm{Acc}}}$ by $\phi/_{\equiv_{\mathrm{Acc}}}$ is indeed $\underline{\mathcal{R}}$. Moreover $\mathrm{Acc}'$ is a closed subset of $\mathbf{A}/_{\equiv_{\mathrm{Acc}}}$ since $\mathrm{Acc}$ is a closed subset of $\mathbf{A}$. Hence, $\phi/_{\equiv_{\mathrm{Acc}}}$ is a surjective morphism which recognizes $\mathcal{R}$.

Then, given another surjective morphism $\psi \colon \mathbf{S}_2\Sigma \twoheadrightarrow \mathbf{B}$ which recognizes $\mathcal{R}$, say $\underline{\mathcal{R}} = \psi^{-1}[\mathrm{Bcc}]$, then for each $b_\sigma \in \mathbf{B}$, there exists $u_\sigma \in \mathbf{S}_2\Sigma$ such that $\psi(u_\sigma) = b_\sigma$. This defines a map $\chi \colon \mathbf{B} \to \mathbf{A}/_{\equiv_{\mathrm{Acc}}}$ which sends $b_\sigma$ to $\phi/_{\equiv_{\mathrm{Acc}}}(u_\sigma)$.

We claim that $\phi/_{\equiv_{\mathrm{Acc}}} = \chi \circ \psi$, meaning that the following diagram commutes

Indeed, for $u_\sigma \in \mathbf{S}_2\Sigma$, $\chi(\psi(u_\sigma))$ equals $\phi/_{\equiv_{\mathrm{Acc}}}(v_\sigma)$ for some $v_\sigma \in \mathbf{S}_2\Sigma$ such that $\psi(v_\sigma) = \psi(u_\sigma)$. This in turns implies that $u_\sigma \equiv_{\mathcal{R}} v_\sigma$ since for all $x, y \in \mathbf{S}_2\Sigma$, if $xuy$ and $xvy$ are well-defined, then so are $\psi(x)\psi(u)\psi(v)$ and $\psi(x)\psi(v)\psi(y)$,





and both elements are equal, so one belongs to Bcc *iff* the other does. It follows that $xuy \in \mathcal{R}$ *iff* $xvy \in \mathcal{R}$, and hence $u_\sigma \equiv_{\mathcal{R}} v_\sigma$. By surjectivity of $\phi$, it follows that $\phi(u_\sigma) \equiv_{\mathrm{Acc}} \phi(v_\sigma)$, *i.e.* $\phi/\equiv_{\mathrm{Acc}}(u_\sigma) = \phi/\equiv_{\mathrm{Acc}}(v_\sigma)$. And hence $\chi(\psi(u)) = \phi/\equiv_{\mathrm{Acc}}(u_\sigma)$.

We now show that $\chi$ is a morphism. From $\phi/\equiv_{\mathrm{Acc}} = \chi \circ \psi$ it follows that the map $\chi$ preserves the product[10] and is surjective. Lastly, we claim that it preserves the dependency relation. Indeed, by surjectivity of $\psi$, it boils down to proving that for all $u_\sigma, v_\tau \in \mathbf{S}_2\Sigma$, if $\psi(u) \asymp \psi(v)$ then $\chi(\psi(u)) \asymp \chi(\psi(v))$, namely $\phi/\equiv_{\mathrm{Acc}}(u_\sigma) \asymp \phi/\equiv_{\mathrm{Acc}}(v_\tau)$, which rewrites as $\phi(u_\sigma) \equiv_{\mathrm{Acc}} \phi(v_\tau)$. So pick $x, y$ such that both $xuy$ and $xvy$ are well-defined. Then $\psi(x)\psi(u)\psi(y)$ and $\psi(u)\psi(v)\psi(y)$ have the same type as $xuy$ and $xvy$, respectively, so they are well-defined, but since $\psi(u) \asymp \psi(v)$, then $\psi(x)\psi(u)\psi(y) \asymp \psi(x)\psi(u)\psi(y)$ and since Bcc is a closed subset, one of these elements belongs in it *iff* the other ones does too, from which it follows that $xuv \in \mathcal{R}$ *iff* $xvy \in \mathcal{R}$ *i.e.* $u_\sigma \equiv_{\mathcal{R}} v_\tau$ and hence $\phi(u_\sigma) \equiv_{\mathrm{Acc}} \phi(v_\tau)$. Overall, this proves that $\chi$ is a surjective synchronous algebra morphism, and concludes our proof. □

*Proof of Lemma IX.3.11.* ↪ *Uniqueness.* Consider two potential syntactic morphisms, say $\eta_1 \colon \mathbf{S}_2\Sigma \twoheadrightarrow \mathbf{A}_1$ and $\eta_2 \colon \mathbf{S}_2\Sigma \twoheadrightarrow \mathbf{A}_2$. Then by the universal property of $\eta_1$ (resp. $\eta_2$), there exists $\psi_1 \colon \mathbf{A}_2 \twoheadrightarrow \mathbf{A}_1$ and $\psi_2 \colon \mathbf{A}_1 \twoheadrightarrow \mathbf{A}_2$ such that $\eta_1 = \psi_1 \circ \eta_2$ and $\eta_2 = \psi_2 \circ \eta_1$. Overall, it implies that the following digram commutes

$$
\begin{array}{ccc}
 & & \mathbf{A}_1 \\
 & \overset{\eta_1}{\nearrow} & \uparrow \psi_1 \\
\mathbf{S}_2\Sigma & \overset{\eta_2}{\longrightarrow} & \mathbf{A}_2 \\
 & \overset{\eta_1}{\searrow} & \uparrow \psi_2 \\
 & & \mathbf{A}_1,
\end{array}
$$

and so $\psi_1 \circ \psi_2 \circ \eta_1 = \eta_1$. Since $\eta_1$ is surjective, and hence right-cancellative, $\psi_1 \circ \psi_2 = \mathrm{id}_{\mathbf{A}_1}$. Symmetrically, $\psi_2 \circ \psi_1 = \mathrm{id}_{\mathbf{A}_2}$, showing that $\psi_1$ and $\psi_2$ are mutually inverse isomorphisms of synchronous algebras.

↪ *Existence.* It follows from Proposition IX.3.13—which we will prove just afterwards—applied to the identity morphism $\mathrm{id} \colon \mathbf{S}_2\Sigma \twoheadrightarrow \mathbf{S}_2\Sigma$, which recognizes $\mathcal{R}$ since $\mathcal{R} = \mathrm{id}^{-1}[\mathcal{R}]$ and $\mathcal{R}$ is closed. □

**Corollary IX.3.14.** In the syntactic synchronous algebra $\mathbf{A}_{\mathcal{R}}$, the syntactic congruence $\equiv_{\mathrm{Acc}}$ and the dependency relation $\asymp$ coincide.

*Proof.* By Proposition IX.3.13 applied to the syntactic morphism, $x \mapsto [x]_{\equiv_{\mathrm{Acc}}}$ is an isomorphism from $\mathbf{A}_{\mathcal{R}}$ to $\mathbf{A}_{\mathcal{R}}/\equiv_{\mathrm{Acc}}$. Hence, $[x]_{\equiv_{\mathrm{Acc}}} \asymp [y]_{\equiv_{\mathrm{Acc}}}$ in $\mathbf{A}_{\mathcal{R}}/\equiv_{\mathrm{Acc}}$ *iff* $x \asymp y$ in $\mathbf{A}_{\mathcal{R}}$, for all $x, y \in \mathbf{A}_{\mathcal{R}}$. But then, the dependency relation $\asymp$ of $\mathbf{A}_{\mathcal{R}}/\equiv_{\mathrm{Acc}}$ is, by definition, such that $[x]_{\equiv_{\mathrm{Acc}}} \asymp [y]_{\equiv_{\mathrm{Acc}}}$ *iff* $x \equiv_{\mathrm{Acc}} y$. Putting both equivalences together, we get that $x \equiv_{\mathrm{Acc}} y$ *iff* $x \asymp y$ for all $x, y \in \mathbf{A}_{\mathcal{R}}$. □

We now provide a simple example of syntactic synchronous algebra whose dependency relation is not an equivalence relation.







**Example IX.3.15.** We now provide an example of syntactic synchronous algebra whose dependency relation is not an equivalence relation. Consider the relation

$$\mathcal{R} = \left(\tfrac{\Sigma}{\Sigma}\right)^* \left[\left(\tfrac{\Sigma}{\_}\right) + \left(\tfrac{\Sigma}{\_}\right)^+ \left(\tfrac{a}{\_}\right) + \left(\tfrac{\_}{\Sigma}\right) + \left(\tfrac{\_}{\Sigma}\right)^+ \left(\tfrac{\_}{a}\right)\right],$$

where $\Sigma = \{a, b\}$. In other words, a pair $(u, v)$ belongs to $\mathcal{R}$ if either: the length difference between $u$ and $v$ is one, or it is at least two and the longer words ends with an '$a$'. We are going to compute the syntactic congruence $\equiv_{\mathcal{R}}$ of $\underline{\mathcal{R}}$ in $\mathbf{S}_2\Sigma$.

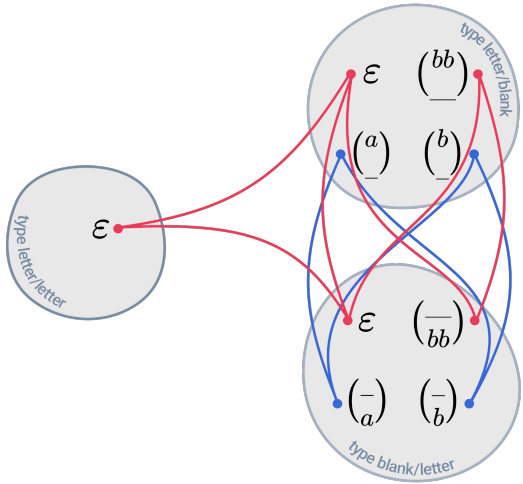

Figure IX.5: The three underlying monoids of the syntactic synchronous algebra of the relation $\mathcal{R}$ of Example IX.3.15, together with its dependency relation. The two equivalence classes of its transitive closure are drawn in red and blue.

Recall that the restriction of $\equiv_{\mathcal{R}}$ to a single type is an equivalence relation. For ᴸ/ᴸ, there is a single equivalence class, denoted by/identified with $\varepsilon_{\text{L/L}}$. For type ᴸ/ʙ, we claim that for any $u \in \Sigma^*$:

- $\left(\tfrac{u}{\_}\right)_{\text{L/B}} \equiv_{\mathcal{R}} \left(\tfrac{a}{\_}\right)_{\text{L/B}}$ *iff* $u \in \Sigma^* a$;
- $\left(\tfrac{u}{\_}\right)_{\text{L/B}} \equiv_{\mathcal{R}} \left(\tfrac{b}{\_}\right)_{\text{L/B}}$ *iff* $u = b$;
- $\varepsilon_{\text{L/B}}$ is alone in its equivalence class;

- all elements of the form $\left(\tfrac{u}{\_}\right)_{\text{L/B}}$ where $u$ is a word of length at least 2 whose last letter is a '$b$' are pairwise equivalent, and the class is identified with $\left(\tfrac{bb}{\_}\right)_{\text{L/B}}$.

Since $\mathcal{R}$ is invariant under $(u, v) \mapsto (v, u)$, the situation is symmetric for type ʙ/ᴸ. Moreover, $\mathcal{R}$ is invariant under adding/removing prefixes of type ᴸ/ᴸ, so types ᴸ/ᴸ → ᴸ/ʙ and ᴸ/ᴸ → ʙ/ᴸ also have four elements each.

In the end, we obtain the synchronous algebra drawn in Figure IX.5 (elements of type ᴸ/ᴸ → ᴸ/ʙ and ᴸ/ᴸ → ʙ/ᴸ are omitted for the sake of simplicity). Note that the dependency relation is not transitive: for instance $\left(\tfrac{a}{\_}\right)_{\text{L/B}} \equiv_{\mathcal{R}} \left(\tfrac{\_}{b}\right)_{\text{B/L}}$ and $\left(\tfrac{\_}{b}\right)_{\text{B/L}} \equiv_{\mathcal{R}} \left(\tfrac{b}{\_}\right)_{\text{L/B}}$ but $\left(\tfrac{a}{\_}\right)_{\text{L/B}} \not\equiv_{\mathcal{R}} \left(\tfrac{b}{\_}\right)_{\text{L/B}}$.

Given two elements of distinct type, we want to determine when they are dependent. For the sake of simplicity, we will focus on types ᴸ/ᴸ, ᴸ/ʙ and ʙ/ᴸ.

🐾 *Types ᴸ/ᴸ and ᴸ/ʙ.* Since dependent elements must either both belong





to $\mathcal{R}$ or both to $\neg\mathcal{R}$, we have $\varepsilon_{\text{L/L}} \not\approx_{\mathcal{R}} \left(\begin{smallmatrix} a \\ - \end{smallmatrix}\right)_{\text{L/B}}$ and $\varepsilon_{\text{L/L}} \not\approx_{\mathcal{R}} \left(\begin{smallmatrix} b \\ - \end{smallmatrix}\right)_{\text{L/B}}$. Because $\underline{\mathcal{R}}$ is a closed subset, then $\varepsilon_{\text{L/L}} \approx_{\mathcal{R}} \varepsilon_{\text{L/B}}$. Moreover, $\varepsilon_{\text{L/L}} \not\approx_{\mathcal{R}} \left(\begin{smallmatrix} bb \\ - \end{smallmatrix}\right)_{\text{L/B}}$ since $\varepsilon_{\text{L/L}} \left(\begin{smallmatrix} b \\ - \end{smallmatrix}\right)_{\text{L/B}} \in \mathcal{R}$ but $\left(\begin{smallmatrix} bb \\ - \end{smallmatrix}\right)_{\text{L/B}} \left(\begin{smallmatrix} b \\ - \end{smallmatrix}\right)_{\text{L/B}} \notin \mathcal{R}$.

☞ *Types L/B and B/L.* Again, using the fact that dependent elements must either both belong to $\mathcal{R}$ or both to $\neg\mathcal{R}$, we have

$$\left(\begin{smallmatrix} a \\ - \end{smallmatrix}\right)_{\text{L/B}} \not\approx_{\mathcal{R}} \varepsilon_{\text{B/L}}, \qquad\qquad \left(\begin{smallmatrix} a \\ - \end{smallmatrix}\right)_{\text{L/B}} \not\approx_{\mathcal{R}} \left(\bar{b}\bar{b}\right)_{\text{B/L}}$$

$$\left(\begin{smallmatrix} b \\ - \end{smallmatrix}\right)_{\text{L/B}} \not\approx_{\mathcal{R}} \varepsilon_{\text{B/L}}, \qquad\qquad \left(\begin{smallmatrix} b \\ - \end{smallmatrix}\right)_{\text{L/B}} \not\approx_{\mathcal{R}} \left(\bar{b}\bar{b}\right)_{\text{B/L}},$$

$$\varepsilon_{\text{L/B}} \not\approx_{\mathcal{R}} \left(\bar{a}\right)_{\text{B/L}}, \qquad\qquad \varepsilon_{\text{L/B}} \not\approx_{\mathcal{R}} \left(\bar{b}\right)_{\text{B/L}}$$

$$\left(\begin{smallmatrix} bb \\ -- \end{smallmatrix}\right)_{\text{L/B}} \not\approx_{\mathcal{R}} \left(\bar{a}\right)_{\text{B/L}}, \qquad\qquad \left(\begin{smallmatrix} bb \\ -- \end{smallmatrix}\right)_{\text{L/B}} \not\approx_{\mathcal{R}} \left(\bar{b}\right)_{\text{B/L}}.$$

We claim that $\left(\begin{smallmatrix} a \\ - \end{smallmatrix}\right)_{\text{L/B}} \approx_{\mathcal{R}} \left(\bar{a}\right)_{\text{B/L}}$. Note that there is no, $y_\tau$ such that $\left(\begin{smallmatrix} a \\ - \end{smallmatrix}\right)_{\text{L/B}} y_\tau$ and $\left(\bar{a}\right)_{\text{B/L}} y_\tau$ are well-formed. So, let $x_\sigma$ be such that $x_\sigma \left(\begin{smallmatrix} a \\ - \end{smallmatrix}\right)_{\text{L/B}}$ and $x_\sigma \left(\bar{a}\right)_{\text{B/L}}$ are well-formed: $\sigma$ must be of type L/L, but then $\mathcal{R}$ is invariant under removing prefixes of type L/L, and so $x_\sigma \left(\begin{smallmatrix} a \\ - \end{smallmatrix}\right)_{\text{L/B}} \in \mathcal{R}$ and $x_\sigma \left(\bar{a}\right)_{\text{B/L}} \in \mathcal{R}$. And hence, $\left(\begin{smallmatrix} a \\ - \end{smallmatrix}\right)_{\text{L/B}} \approx_{\mathcal{R}} \left(\bar{a}\right)_{\text{B/L}}$. Similarly,

$$\left(\begin{smallmatrix} a \\ - \end{smallmatrix}\right)_{\text{L/B}} \approx_{\mathcal{R}} \left(\bar{b}\right)_{\text{B/L}}, \quad \left(\begin{smallmatrix} b \\ - \end{smallmatrix}\right)_{\text{L/B}} \approx_{\mathcal{R}} \left(\bar{a}\right)_{\text{B/L}} \quad \text{and} \quad \left(\begin{smallmatrix} b \\ - \end{smallmatrix}\right)_{\text{L/B}} \approx_{\mathcal{R}} \left(\bar{b}\right)_{\text{B/L}}.$$

The dual argument (using now the fact that both sides do *not* belong to $\mathcal{R}$) shows that

$$\varepsilon_{\text{L/B}} \approx_{\mathcal{R}} \varepsilon_{\text{B/L}}, \quad \varepsilon_{\text{L/B}} \approx_{\mathcal{R}} \left(\bar{b}\bar{b}\right)_{\text{B/L}}, \quad \left(\begin{smallmatrix} bb \\ -- \end{smallmatrix}\right)_{\text{L/B}} \approx_{\mathcal{R}} \varepsilon_{\text{B/L}} \text{ and } \left(\begin{smallmatrix} bb \\ -- \end{smallmatrix}\right)_{\text{L/B}} \approx_{\mathcal{R}} \left(\bar{b}\bar{b}\right)_{\text{B/L}}.$$

The case of types L/L and L/B follows by symmetry. $\diamond$

*Boolean operations* Given two synchronous algebras **A** and **B**, define their *Cartesian product* $\mathbf{A} \times \mathbf{B}$ by taking, for each type $\tau$, the Cartesian product $A_\tau \times B_\tau$. Units, product are defined naturally, and the dependency relation is defined by taking the conjunction over each component. Then $\neg\mathcal{R}$ is recognized by **A**, and $\mathcal{R} \cup \mathcal{S}$ and $\mathcal{R} \cap \mathcal{S}$ are recognized by $\mathbf{A} \times \mathbf{B}$.

## IX.4 The Lifting Theorem & Pseudovarieties

### IX.4.1 Elementary Formulation

**Example IX.4.1** (Group relations: Example IX.3.10 cont'd). We want to decide when the relation

$$\mathcal{R}^{I,J} \triangleq \Big\{ (u,v) \ \Big| \ |u| > |v| \text{ and } (|u| - |v| \bmod p) \in I, \text{ or }$$
$$|u| < |v| \text{ and } (|v| - |u| \bmod q) \in J. \ \Big\}$$

from Example IX.3.10 is a group relation. By definition this happens if and only if there exists a finite group $G$, together with a monoid morphism $\phi\colon (\Sigma_\otimes^2)^* \to G$ and a subset $\text{Acc} \subseteq G$ s.t. $\forall u \in \text{WellFormed}_\Sigma$, $u \in \mathcal{R}^{I,J}$ iff $\phi(u) \in \text{Acc}$. We





claim:

$$\mathcal{R}^{I,J} \text{ is a group relation} \quad \textit{iff} \quad \left(\bar{0} \notin I \text{ and } \bar{0} \notin J\right). \tag{IX.1}$$

The right-to-left implication was shown in Example IX.3.10. We prove the implication from left to right: let $n$ be the order of $G$ so that $x^n = 1$ for all $x \in G$. In particular, we have: $\phi\left(\binom{a}{-}^{pqn}\right) = 1 = \phi\left(\binom{a}{a}^{pqn}\right)$. Since $\phi\left(\binom{a}{a}^{pqn}\right) \notin \mathcal{R}^{I,J}$, it follows that $\binom{a}{-}^{pqn} \notin \mathcal{R}^{I,J}$ *i.e.* $\bar{0} \notin I$. Also, $\bar{0} \notin J$ by symmetry, which concludes the proof. $\diamond$

Even more generally, we can decide if a relation $\mathcal{R}$ is a group relation by simply looking at the syntactic synchronous algebra of $\mathcal{R}$.

**Theorem IX.4.2** (*Lifting theorem: Elementary Formulation*). Given a relation $\mathcal{R}$ and a $*$-pseudovariety of regular languages $\mathcal{V}$ corresponding to a pseudovariety of monoids $\mathbb{V}$, the following are equivalent:

1. $\mathcal{R}$ is a $\mathcal{V}$-relation,

2. $\mathcal{R}$ is recognized by a finite synchronous algebra $\mathbf{A}$ whose underlying monoids are all in $\mathbb{V}$,

3. all underlying monoids of the syntactic synchronous algebras $\mathbf{A}_{\mathcal{R}}$ of $\mathcal{R}$ are in $\mathbb{V}$.

*Proof.* ☙ *(1) $\Rightarrow$ (2).* Since $\mathcal{R}$ is a $\mathcal{V}$-relation, there exists $\mathcal{L} \in \mathcal{V}_{\Sigma_{\otimes}^2}$ such that $\underline{\mathcal{R}} = \mathcal{L} \cap \text{WellFormed}_\Sigma$. Hence, there exists a morphism of monoids $\phi: (\Sigma_{\otimes}^2)^* \to M$ such that $M \in \mathbb{V}$ and $\mathcal{L} = \phi^{-1}[\text{Acc}]$ for some Acc $\subseteq M$. It follows that $\underline{\mathcal{R}} = \mathcal{L} \cap \text{WellFormed}_\Sigma$ rewrites as "for all $u \in \text{WellFormed}_\Sigma$, $\phi(u) \in \text{Acc}$ *iff* $u \in \underline{\mathcal{R}}$". Letting $\mathbf{A}_M$ be the synchronous algebra induced by the monoid $M$, define $\psi: \mathbf{S}_2\Sigma \to \mathbf{A}_M$ by $\psi(u_\sigma) \triangleq (\phi(u))_\sigma$ for $u_\sigma \in \mathbf{S}_2\Sigma$. Let Acc' $\triangleq \{x_\sigma \mid x \in \text{Acc} \wedge \sigma \in \mathcal{T}\}$. We claim that $\psi^{-1}[\text{Acc}'] = \underline{\mathcal{R}}$. Indeed, for $u_\sigma \in \mathbf{S}_2\Sigma, u_\sigma \in \underline{\mathcal{R}}$ *iff* $u \in \mathcal{L}$, *i.e.* $\phi(u) \in \text{Acc}$, that is $\psi(u_\sigma) = (\phi(u))_\sigma \in \text{Acc}'$. Notice then that all underlying monoids of $\mathbf{A}_M$ are $M$, and hence they belong to $\mathbb{V}$.

☙ *(2) $\Rightarrow$ (3).* By Lemma IX.3.11, the syntactic synchronous algebra of $\mathcal{R}$ divides any algebra $\mathbf{B}$ recognizing $\mathcal{R}$. In particular, its underlying monoids divide the underlying monoids of $\mathbf{B}$. The conclusion follows since $\mathbb{V}$ is closed under division.

☙ *(3) $\Rightarrow$ (1).* Denote by $M_{\text{L/L}}, M_{\text{L/B}}$ and $M_{\text{B/L}}$ the underlying monoids of $\mathbf{A}_{\mathcal{R}}$. Let Acc $\subseteq \mathbf{A}_{\mathcal{R}}$ be the accepting set such that $\underline{\mathcal{R}} = \eta_{\mathcal{R}}^{-1}[\text{Acc}]$. Define $M \triangleq M_{\text{L/L}} \times M_{\text{L/B}} \times M_{\text{B/L}}$, and

$$
\begin{aligned}
\phi: \quad (\Sigma_{\otimes}^2)^* &\to M \\
\binom{a}{b} &\mapsto \langle \eta_{\mathcal{R}}\binom{a}{b}_{\text{L/L}}, 1_{\text{L/B}}, 1_{\text{B/L}} \rangle \\
\binom{a}{-} &\mapsto \langle 1_{\text{L/L}}, \eta_{\mathcal{R}}\binom{a}{-}_{\text{L/B}}, 1_{\text{B/L}} \rangle \\
\binom{-}{a} &\mapsto \langle 1_{\text{L/L}}, 1_{\text{L/B}}, \eta_{\mathcal{R}}\binom{-}{a}_{\text{B/L}} \rangle
\end{aligned}
$$





and finally, let

$$\mathrm{Acc}' \cup \left\{ \langle 1_{\mathsf{L} / \mathsf{L}}, 1_{\mathsf{L} / \mathsf{B}}, z_{\mathsf{B} / \mathsf{L}} \rangle \mid z_{\mathsf{B} / \mathsf{L}} \in \mathrm{Acc} \right\}$$
$$\cup \left\{ \langle x_{\mathsf{L} / \mathsf{L}}, y_{\mathsf{L} / \mathsf{B}}, 1_{\mathsf{B} / \mathsf{L}} \rangle \mid x_{\mathsf{L} / \mathsf{L}} \cdot y_{\mathsf{L} / \mathsf{B}} \in \mathrm{Acc} \right\}$$
$$\cup \left\{ \langle x_{\mathsf{L} / \mathsf{L}}, 1_{\mathsf{L} / \mathsf{B}}, z_{\mathsf{B} / \mathsf{L}} \rangle \mid x_{\mathsf{L} / \mathsf{L}} \cdot z_{\mathsf{B} / \mathsf{L}} \in \mathrm{Acc} \right\}.$$

We first claim that

> For every $u_{\mathsf{L} / \mathsf{L} \to \mathsf{L} / \mathsf{B}} \in \mathbf{S}_2 \Sigma$,
> $$\phi(u) \text{ is of the form } \langle a, b, 1 \rangle \text{ and moreover,} \qquad \text{(IX.2)}$$
> $$\eta_{\mathcal{R}}(u_{\mathsf{L} / \mathsf{L} \to \mathsf{L} / \mathsf{B}}) = a \cdot b,$$

which can trivially be proven by induction on $u$. Analogous results hold for words of different type. We then prove that for each $u_\sigma \in \mathbf{S}_2 \Sigma$,

$$\eta_{\mathcal{R}}(u_\sigma) \in \mathrm{Acc} \quad \textit{iff} \quad \phi(u) \in \mathrm{Acc}'. \qquad \text{(IX.3)}$$

The direct implication is straightforward, using Equation (IX.2). The converse implication is more tricky: assume *e.g.* that $\sigma = \mathsf{L} / \mathsf{L} \to \mathsf{L} / \mathsf{B}$, say $t_\sigma = u_{\mathsf{L} / \mathsf{L}} v_{\mathsf{L} / \mathsf{B}}$. If $\phi(t) \in \mathrm{Acc}'$, using Equation (IX.2) then it implies either that

1. $\eta_{\mathcal{R}}(u_{\mathsf{L} / \mathsf{L}}) = 1_{\mathsf{L} / \mathsf{L}}$ and $\eta_{\mathcal{R}}(v_{\mathsf{L} / \mathsf{B}}) = 1_{\mathsf{L} / \mathsf{L}}$, and $1_{\mathsf{B} / \mathsf{L}} \in \mathrm{Acc}$, or
2. $\eta_{\mathcal{R}}(u_{\mathsf{L} / \mathsf{L}}) \cdot \eta_{\mathcal{R}}(v_{\mathsf{L} / \mathsf{B}}) \in \mathrm{Acc}$, or even
3. $\eta_{\mathcal{R}}(v_{\mathsf{L} / \mathsf{B}}) = 1_{\mathsf{L} / \mathsf{B}}$ and $\eta_{\mathcal{R}}(u_{\mathsf{L} / \mathsf{L}}) \cdot 1_{\mathsf{B} / \mathsf{L}} \in \mathrm{Acc}$.

Clearly, (2) implies the desired conclusion, namely that

$$\eta_{\mathcal{R}}(t_\sigma) = \eta_{\mathcal{R}}(u_{\mathsf{L} / \mathsf{L}}) \eta_{\mathcal{R}}(v_{\mathsf{L} / \mathsf{B}}) \in \mathrm{Acc}.$$

In all other cases, we will make heavy use of the dependency relation. For case (1), we have that $\eta_{\mathcal{R}}(t_\sigma) = 1_{\mathsf{L} / \mathsf{L} \to \mathsf{L} / \mathsf{B}}$. From $1_{\mathsf{B} / \mathsf{L}} \in \mathrm{Acc}$, Fact IX.3.5 yields $1_{\mathsf{L} / \mathsf{L}} \cdot 1_{\mathsf{L} / \mathsf{B}} = 1_{\mathsf{L} / \mathsf{L} \to \mathsf{L} / \mathsf{B}} \in \mathrm{Acc}$, since Acc is closed. Lastly, in case (3), $\eta_{\mathcal{R}}(u_{\mathsf{L} / \mathsf{L}}) \asymp \eta_{\mathcal{R}}(u_{\mathsf{L} / \mathsf{L}}) \cdot 1_{\mathsf{B} / \mathsf{L}} \in \mathrm{Acc}$ so $\eta_{\mathcal{R}}(u_{\mathsf{L} / \mathsf{L}}) \in \mathrm{Acc}$ and hence $\eta_{\mathcal{R}}(t_{\mathsf{L} / \mathsf{L} \to \mathsf{L} / \mathsf{B}}) = \eta_{\mathcal{R}}(u_{\mathsf{L} / \mathsf{L}}) \cdot 1_{\mathsf{L} / \mathsf{B}} \asymp \eta_{\mathcal{R}}(u_{\mathsf{L} / \mathsf{L}}) \in \mathrm{Acc}$ which yields $\eta_{\mathcal{R}}(t_{\mathsf{L} / \mathsf{L} \to \mathsf{L} / \mathsf{B}}) \in \mathrm{Acc}$. This concludes the proof of (IX.3) for type $\sigma = \mathsf{L} / \mathsf{L} \to \mathsf{L} / \mathsf{B}$. Other types are handled similarly, and hence $\underline{\mathcal{R}} = \phi^{-1}[\mathrm{Acc}'] \cap \mathrm{WellFormed}_\Sigma$. $\qquad \square$

**Remark IX.4.3.** In light of Theorem IX.4.2, one can wonder whether the notion of synchronous algebra is necessary to characterize $\mathcal{V}$-relations, or if it is enough to look at the languages corresponding to the underlying monoids. Said otherwise, is the membership of $\mathcal{R}$ in the class of $\mathcal{V}$-relations uniquely determined by the regular languages $\mathcal{R} \cap (\Sigma \times \Sigma)^*$, $\mathcal{R} \cap (\Sigma \times \{\_\})^*$ and $\mathcal{R} \cap (\{\_\} \times \Sigma)^*$? Unsurprisingly, synchronous algebras are indeed necessary, as there are relations $\mathcal{R}$ such that:

$$\underline{\mathcal{R}} \cap (\Sigma \times \Sigma)^* \in \mathcal{V}_{\Sigma \times \Sigma}, \quad \underline{\mathcal{R}} \cap (\Sigma \times \_)^* \in \mathcal{V}_{\Sigma \times \_} \quad \text{and} \quad \underline{\mathcal{R}} \cap (\_ \times \Sigma)^* \in \mathcal{V}_{\_ \times \Sigma}, \qquad \text{(IX.4)}$$

but $\mathcal{R}$ is *not* a $\mathcal{V}$-relation. This can happen even if $\mathcal{V}$ is the $*$-pseudovariety of





all regular languages: for instance for the relation

$$\mathcal{R} \triangleq \{(u, v) \mid |u| > |v| > 0 \text{ and } |u| - |v| \text{ is prime}\}.$$

Notice that there is a subtle but crucially important difference between (IX.4) and the second item of the Lifting Theorem: while the underlying monoids of a synchronous algebra **A** recognizing $\mathcal{R}$ only accept words of the form $(\Sigma \times \Sigma)^*$, $(\Sigma \times \_)^*$ or $(\_ \times \Sigma)^*$, elements of $(\Sigma \times \Sigma)^+ (\Sigma \times \_)^+$ or $(\Sigma \times \Sigma)^+ (\_ \times \Sigma)^+$ influence the underlying monoids of **A** via the axioms of synchronous algebras. $\diamondsuit$

Also, note that the existence the Lifting Theorem follows from the careful definition of synchronous algebras: more naive definitions of these algebras simply cannot characterize $\mathcal{V}$-relations, see Section IX.4.2.

From Theorem IX.4.2 and the implicit fact that all our constructions are effective, we obtain a decidability (meta-)result for $\mathcal{V}$-relations.

**Corollary IX.4.4.** The class of $\mathcal{V}$-relations has decidable membership if, and only if, $\mathcal{V}$ has decidable membership.

For instance, a relation is a group relation if, and only if, all underlying monoids of its syntactic synchronous algebra are groups.

### IX.4.2 Synchronous Algebras Require a Dependency Relation

In this part, we introduce the notion of synchronous algebra with no dependency relation, called naive synchronous algebra. This notion is more natural—or naive—than Definition IX.3.4, and share some of its enjoyable properties, such as the existence of syntactic algebras. Yet, we show that these algebras cannot characterize some natural classes of automatic relation. More precisely, we show that there is a *-pseudovariety of regular languages $\mathcal{V}$ and two automatic relations $\mathcal{R}_0$ and $\mathcal{R}_1$, such that:

- $\mathcal{R}_0$ is a $\mathcal{V}$-relation,
- $\mathcal{R}_1$ is not a $\mathcal{V}$-relation,
- $\mathcal{R}_0$ and $\mathcal{R}_1$ have the same syntactic naive synchronous algebra.

**Definition IX.4.5** (Naive synchronous algebra). Let $\mathcal{T}_1 \triangleq \mathcal{T} \cup \{1\}$. We extend the notion of compatibility so that every $\sigma \in \mathcal{T}_1$ is compatible with 1 and 1 is compatible with $\sigma$. A *naive synchronous algebra* **A** consists of a $\mathcal{T}_1$-typed sets, together with a partial binary operator $\cdot$ such that:

- $\cdot$ is defined exactly on compatible elements and is associative, and
- there is a unique element of type 1, denoted by 1, and it satisfies $x_\sigma \cdot 1 = x_\sigma = 1 \cdot x_\sigma$ for all $x_\sigma \in \mathbf{A}$.

The set of all synchronous words is naturally a naive synchronous algebra under concatenation. Moreover, any automatic relation admits a syntactic naive synchronous algebra—this can be shown in the same fashion as Lemma IX.3.11.

**Example IX.4.6** (Group relations: Example IX.4.1 cont'd)**.** Consider the rela-





tions

$$\mathcal{R}_0 \triangleq \Big\{(u, v) \;\Big|\; |u| > |v| \text{ and } (|u| - |v| \bmod p) = 0\Big\}$$
$$\mathcal{R}_1 \triangleq \Big\{(u, v) \;\Big|\; |u| > |v| \text{ and } (|u| - |v| \bmod p) = 1\Big\}.$$

Then by Example IX.4.1, $\mathcal{R}_0$ is not a group relation but $\mathcal{R}_1$ is. Yet, we claim that both relations have the same syntactic naive synchronous algebra **A**, described as follows:

- it has a unit, denoted by 0, of type 1,
- $\mathbf{A}_{L/L}$, $\mathbf{A}_{B/L}$ and $\mathbf{A}_{L/L \to B/L}$ are all reduced to a single element, denoted by $0_{L/L}$, $0_{B/L}$ and $0_{L/L \to B/L}$,
- $\mathbf{A}_{L/L \to L/B}$ and $\mathbf{A}_{L/B}$ contain the elements $\mathbb{Z}/p\mathbb{Z}$,
- $\cdot$ is defined as the addition over $\mathbb{Z}/p\mathbb{Z}$, by identifying $0_1$, $0_{L/L}$, $0_{L/L \to B/L}$ and $0_{B/L}$ with the zero of $\mathbb{Z}/p\mathbb{Z}$.

Then $\mathcal{R}_0$ and $\mathcal{R}_1$ are the preimages of $\{0_{L/L \to L/B}, 0_{L/B}\}$ and $\{1_{L/L \to L/B}, 1_{L/B}\}$, respectively, by the natural morphism onto **A**. And hence $\mathcal{R}_0$ and $\mathcal{R}_1$ are recognized by **A**. It is easy to show that it is in fact the syntactic naive synchronous algebra of these relations: by surjectivity of the morphism above, it suffices to show that no two elements of **A** can be identified and still recognize the same relation. $\diamond$

And so, from this example is follows that "being a $\mathcal{V}$-relations" cannot be characterized by the syntactic naive synchronous algebra of the relation, which shows how crucial the dependency relation of Definition IX.3.4 is in order to get Theorem IX.4.2.

The same result can be used to prove that "naive positive synchronous algebras"—defined analogously to naive synchronous algebra except that there is no type 1 and no unit, and hence no empty word in the free algebra—are also unable to capture the property of "being a $\mathcal{V}$-relations".

### IX.4.3  Pseudovarieties of Automatic Relations

We introduce the notion of pseudovariety of synchronous algebras and $*$-pseudovariety of automatic relations. We show an Eilenberg-Schützenberger correspondence between these two notions. We then reformulate the Lifting Theorem to show that any Eilenberg-Schützenberger correspondence between monoids and regular languages lifts to an Eilenberg-Schützenberger correspondence between synchronous algebras and automatic relations.

Recall that a synchronous algebra **A** is a quotient of **B** when there exists a surjective synchronous algebra morphism from **B** to **A**. A *subalgebra* of **B** is any closed subset of **B** closed under product and containing the units. We then say that synchronous algebra **A** *divides* **B** when **A** is a quotient of a subalgebra of **B**.

Observe that $\mathbf{S}_2\Sigma$ admits the following property: elements of type $L/L \to L/B$ and $L/L \to B/L$ are generated by the underlying monoids. Since syntactic





synchronous algebras are homomorphic images of $\mathbf{S}_2\Sigma$, they also satisfy this property. In general, we say that a synchronous algebra $\mathbf{A}$ is *locally generated* if every element of type $^L/_L \rightarrow {}^L/_B$ (resp. $^L/_L \rightarrow {}^B/_L$) can be written as the product of an element of type $^L/_L$ with an element of type $^L/_B$ (resp. $^B/_L$).

A *pseudovariety of synchronous algebras* is any class $\mathbb{V}$ of locally generated finite synchronous algebras closed under

- *finite product:* if $\mathbf{A}, \mathbf{B} \in \mathbb{V}$ then $\mathbf{A} \times \mathbf{B} \in \mathbb{V}$,

- *division:* if some finite locally generated algebra $\mathbf{A}$ divides $\mathbf{B}$ for some $\mathbf{B} \in \mathbb{V}$, then $\mathbf{A} \in \mathbb{V}$.

Because of Lemma IX.3.11, an automatic relation is recognized by a finite synchronous algebra of a pseudovariety $\mathbb{V}$ *iff* its syntactic synchronous algebra belongs to $\mathbb{V}$.

A *∗-pseudovariety of automatic relations* is a function $\mathcal{V} \colon \Sigma \mapsto \mathcal{V}_\Sigma$ such that for any finite alphabet $\Sigma$, $\mathcal{V}_\Sigma$ is a set of automatic relations over $\Sigma$ such that $\mathcal{V}$ is closed under

- *Boolean combinations:* if $\mathcal{R}, \mathcal{S} \in \mathcal{V}_\Sigma$, then $\neg\mathcal{R}$, $\mathcal{R} \cup \mathcal{S}$ and $\mathcal{R} \cap \mathcal{S}$ belong to $\mathcal{V}_\Sigma$ too,

- *Syntactic derivatives:* if $\mathcal{R} \in \mathcal{V}_\Sigma$, then any relation recognized by the syntactic synchronous algebra morphism of $\mathcal{R}$ also belongs to $\mathcal{V}_\Sigma$.

- *Inverse morphisms:* if $\phi \colon \mathbf{S}_2\Gamma \rightarrow \mathbf{S}_2\Sigma$ is a synchronous algebra morphism and $\mathcal{R} \in \mathcal{V}_\Sigma$ then $\phi^{-1}[\mathcal{R}] \in \mathcal{V}_\Gamma$.

To recover a more traditional definition (of the form "closure under Boolean operations, residuals[11] and inverse morphisms"), we need to properly define what are the residuals of a relation. It turns out that the answer is quite surprising and less trivial than what one would expect.

[11] Also called "quotient" *e.g.* in [Pin22, § III.1.3, p. 39], or "polynomial derivative" in [Boj15, § 4, p. 19].

**Definition IX.4.7** (Residuals)**.** Let $\mathbf{A}$ be a synchronous algebra, $x_\sigma \in \mathbf{A}$, and $C \subseteq \mathbf{A}$ be a closed subset. The *left residual* and *right residual* of $C$ by $x_\sigma$ are defined by

$$x_\sigma^{-1}C \triangleq \left\{ y_\tau \in \mathbf{A} \mid \exists y'_{\tau'} \equiv_C y_\tau, \; x_\sigma y'_{\tau'} \in C \right\}, \text{ and}$$
$$Cx_\sigma^{-1} \triangleq \left\{ y_\tau \in \mathbf{A} \mid \exists y'_{\tau'} \equiv_C y_\tau, \; y'_{\tau'}x_\sigma \in C \right\},$$

respectively. We refer indiscriminately to both these notions as *residuals*. We extend these notions to sets, by letting $X^{-1}C \triangleq \bigcup_{x \in X} x^{-1}C$ and $CX^{-1} \triangleq \bigcup_{x \in X} Cx^{-1}$.

For the sake of readability, we will sometimes drop the type of elements when dealing with residuals. It is routine to check that residuals are always closed subsets (since $\equiv_C$ is coarser than the dependency relation), or that $(x^{-1}C)y^{-1} = x^{-1}(Cy^{-1})$. Equivalently, $Cx_\sigma^{-1}$ can be defined as the smallest closed subset containing the "naive residual" $\left\{ y_\tau \in \mathbf{A} \mid y_\tau x_\sigma \in C \right\}$. This latter set is always contained in $Cx_\sigma^{-1}$ (by reflexivity of $\equiv_C$), and moreover, if it is empty, then so is $Cx_\sigma^{-1}$.

As an example, consider the relation $\mathcal{R}$ from Example IX.3.15. Then the





"naive right residual" of $\underline{\mathcal{R}}$ by $\left(\genfrac{}{}{0pt}{}{a}{\_}\right)_{\llcorner/\rrcorner}$ consists of $\varepsilon_{\llcorner/\llcorner}$ and all elements of type $\llcorner/\rrcorner$ and $\llcorner/\llcorner \to \llcorner/\rrcorner$. But it does not contain any element of type $\rrcorner/\llcorner$ or $\llcorner/\llcorner \to \rrcorner/\llcorner$ because such elements cannot be concatenated with $\left(\genfrac{}{}{0pt}{}{a}{\_}\right)_{\llcorner/\rrcorner}$ on the right. Yet, the residual $\underline{\mathcal{R}}\left(\genfrac{}{}{0pt}{}{a}{\_}\right)_{\llcorner/\rrcorner}^{-1}$ contains all elements of type $\rrcorner/\llcorner$ (and also $\llcorner/\llcorner \to \rrcorner/\llcorner$): for instance, $\left(\genfrac{}{}{0pt}{}{a}{\_}\right)_{\rrcorner/\llcorner} \in \underline{\mathcal{R}}\left(\genfrac{}{}{0pt}{}{a}{\_}\right)_{\llcorner/\rrcorner}^{-1}$ since $\left(\genfrac{}{}{0pt}{}{a}{\_}\right)_{\rrcorner/\llcorner} \equiv_{\mathcal{R}} \left(\genfrac{}{}{0pt}{}{a}{\_}\right)_{\llcorner/\rrcorner}$ and $\left(\genfrac{}{}{0pt}{}{a}{\_}\right)_{\llcorner/\rrcorner}\left(\genfrac{}{}{0pt}{}{a}{\_}\right)_{\llcorner/\rrcorner} \in \mathcal{R}$.

On the other hand, in the algebra $\mathbf{S}_2 a$ consider the relation $\mathcal{S} = (aa)^* \times a(aa)^*$. Then $\underline{\mathcal{S}}\left(\genfrac{}{}{0pt}{}{a}{a}\right)_{\llcorner/\llcorner}^{-1}$ is empty since its "naive residual" $\{y_\tau \in \mathbf{S}_2 a \mid y_\tau \cdot \left(\genfrac{}{}{0pt}{}{a}{a}\right) \in \mathcal{S}\}$ is empty. Indeed, for $y_\tau \cdot \left(\genfrac{}{}{0pt}{}{a}{a}\right)_{\llcorner/\llcorner}$ to be well-defined, one needs $\tau$ to be $\llcorner/\llcorner$, $i.e.$ $y$ encodes a pair of two words $(u, v)$ of the same length. But then $(ua, va) \notin \mathcal{S}$.

**Lemma IX.4.8.** A class $\mathcal{V} \colon \Sigma \mapsto \mathcal{V}_\Sigma$ is a $*$-pseudovariety of automatic relations if, and only if, it is closed under Boolean combinations, residuals and inverse morphisms.

*Proof.* We first need two propositions.

*Claim* IX.4.9. Let $\phi \colon \mathbf{A} \twoheadrightarrow \mathbf{B}$ be a surjective morphism, and Acc be a closed subset of $\mathbf{B}$. Let $a, a' \in \mathbf{A}$. Then

$$a \equiv_{\phi^{-1}[\text{Acc}]} a' \quad \textit{iff} \quad \phi(a) \equiv_{\text{Acc}} \phi(a').$$

�señ *Direct implication.* Pick any $b_\ell, b_r \in \mathbf{B}$ such that both $b_\ell \phi(a) b_r$ and $b_\ell \phi(a') b_r$ are well-defined. By surjectivity of $\phi$, there exists $a_\ell, a_r \in \mathbf{A}$ such that $\phi(a_\ell) = b_\ell$ and $\phi(a_r) = b_r$. Then both $a_\ell a a_r$ and $a_\ell a' a_r$ are well-defined since they have the same type as $b_\ell \phi(a) b_r$ and $b_\ell \phi(a') b_r$, respectively. From $a \equiv_{\phi^{-1}[\text{Acc}]} a'$, it follows that $a_\ell a a_r$ belongs to $\phi^{-1}[\text{Acc}]$ $iff$ $a_\ell a' a_r$ does. And hence

$$b_\ell \phi(a) b_r \in \text{Acc} \quad \textit{iff} \quad b_\ell \phi(a') b_r \in \text{Acc}.$$

🌿 *Converse implication.* Dually, pick any $a_\ell, a_r \in \mathbf{A}$ such that both $a_\ell a a_r$ and $a_\ell a' a_r$ are well-defined. Then $\phi(a_\ell)\phi(a)\phi(a_r)$ and $\phi(a_\ell)\phi(a')\phi(a_r)$ are also well-defined since they have the same type as their preimage, and $\phi(a) \equiv_{\text{Acc}} \phi(a')$ implies that the element $\phi(a_\ell)\phi(a)\phi(a_r)$ belongs to Acc $iff$ $\phi(a_\ell)\phi(a')\phi(a_r)$ does. It follows that $a_\ell a a_r \in \phi^{-1}[\text{Acc}]$ $iff$ $a_\ell a' a_r \in \phi^{-1}[\text{Acc}]$. This concludes the proof of Claim IX.4.9.

*Claim* IX.4.10 (Inverse images of surjective morphisms preserve residuals). Let $\phi \colon \mathbf{A} \twoheadrightarrow \mathbf{B}$ be a surjective morphism, and Acc $\subseteq \mathbf{B}$ be a closed subset. Let $u \in \mathbf{A}$. Then

$$u^{-1}\phi^{-1}[\text{Acc}] = \phi^{-1}[\phi(u)^{-1}\text{Acc}].$$

🌿 *Left-to-right inclusion.* Let $a \in u^{-1}\phi^{-1}[\text{Acc}]$. Then there exists $a' \in \mathbf{A}$ such that $a \equiv_{\phi^{-1}[\text{Acc}]} a'$ and $ua' \in \phi^{-1}[\text{Acc}]$. By Claim IX.4.9 $a \equiv_{\phi^{-1}[\text{Acc}]} a'$ implies $\phi(a) \equiv_{\text{Acc}} \phi(a')$, and $ua' \in \phi^{-1}[\text{Acc}]$ yields $\phi(u)\phi(a') \in \text{Acc}$. Overall, this shows that $a \in \phi^{-1}[\phi(u)^{-1}\text{Acc}]$.

🌿 *Right-to-left inclusion.* Let $a \in \phi^{-1}[\phi(u)^{-1}\text{Acc}]$. Then $\phi(a) \in \phi(u)^{-1}\text{Acc}$, so there exists $b' \in \mathbf{B}$ such that $\phi(a) \equiv_{\text{Acc}} b'$ and $\phi(u)b' \in \text{Acc}$. By surjectivity of $\phi$ and Claim IX.4.9, there exists $a' \in \mathbf{A}$ such that $\phi(a') = b'$ and $a \equiv_{\phi^{-1}[\text{Acc}]}$





$a'$.

Being done with the proof of Claim IX.4.10, we now proceed to prove Lemma IX.4.8. ☙ *Direct implication.* By Claim IX.4.10, the residual of any relation recognized by some morphism $\phi$ is also recognized by $\phi$. Hence, being closed under syntactic derivatives implies being closed under residuals.

☙ *Converse implication.* Consider some relation $\mathcal{R}$. We will show that any relation recognized by $\eta_{\mathcal{R}}$ can be expressed as a Boolean combination of residuals of $\mathcal{R}$.[12] Let Acc be the closed subset of $\mathbf{A}_{\mathcal{R}}$ such that $\underline{\mathcal{R}} = \eta_{\mathcal{R}}^{-1}[\text{Acc}]$. Pick $x \in \mathbf{A}_{\mathcal{R}}$. Let $\Lambda \triangleq \{s, t \in \mathbf{A}_{\mathcal{R}} \mid \exists x' \in \mathbf{A}_{\mathcal{R}}, \ x' \asymp x \text{ and } sx't \in \text{Acc}\}$. We claim that

$$[x]_{\asymp \mathbf{A}_{\mathcal{R}}} = \left( \bigcap_{(s,t) \in \Lambda} s^{-1} \text{Acc} \, t^{-1} \right) \smallsetminus \left( \bigcup_{(s,t) \notin \Lambda} s^{-1} \text{Acc} \, t^{-1} \right). \tag{IX.5}$$

To prove the inclusion from left-to-right, first notice that $x \in s^{-1} \text{Acc} \, t^{-1}$ for all $(s, t) \in \Lambda$. Then, assume by contradiction that there exists $(s, t) \notin \Lambda$ s.t. $x \in s^{-1} \text{Acc} \, t^{-1}$. Then there would exist $x' \equiv_{\text{Acc}} x$ such that such that $sx't \in \text{Acc}$. But since $\eta_{\mathcal{R}}$ is the syntactic synchronous algebra of $\mathcal{R}$, $\equiv_{\text{Acc}}$ is precisely the relation $\asymp$ by Corollary IX.3.14. Contradiction. Hence, $x$ belongs to the right-hand side (RHS). But then, this latter set is a Boolean combination of residuals of a closed subset, so it is also closed, and hence $[x]_{\asymp \mathbf{A}_{\mathcal{R}}}$ is included in the RHS.

Dually, any element $y$ of the RHS satisfies that for all $s, t \in \mathbf{A}_{\mathcal{R}}$, $x \in s^{-1} \text{Acc} \, t^{-1}$ iff $y \in s^{-1} \text{Acc} \, t^{-1}$. We claim that $x \equiv_{\text{Acc}} y$. Pick $s, t \in \mathbf{B}$ and assume that both $sxt$ and $syt$ are well-defined. If $sxt \in \text{Acc}$ then $x \in s^{-1} \text{Acc} \, t^{-1}$ so $y \in s^{-1} \text{Acc} \, t^{-1}$ and hence, there exists $y' \asymp_{\mathbf{A}_{\mathcal{R}}} y$ s.t. $sy't \in \text{Acc}$. But $syt$ is also well-defined so $y \asymp_{\mathbf{A}_{\mathcal{R}}} y'$ so $syt \in \text{Acc}$. By symmetry, we have shown that $sxt \in \text{Acc}$ *iff* $syt \in \text{Acc}$, and hence $x \equiv_{\text{Acc}} y$. Using again the fact that $\mathbf{A}_{\mathcal{R}}$ is the syntactic algebra of $\mathcal{R}$, it follows that $x \asymp_{\mathbf{A}_{\mathcal{R}}} y$. This concludes the proof of (IX.5). By taking the union, it follows that any closed subset of $\mathbf{A}_{\mathcal{R}}$ is a Boolean combination of residuals of Acc. Applying Claim IX.4.10 then yields that any relation recognized by $\phi$ is a Boolean combination of residuals of $\mathcal{R}$. Hence, any class closed under Boolean combinations and residuals is also closed under syntactic derivatives. □

Let $\mathbb{V} \to \mathcal{V}$ denote the map (called *correspondence*) that takes a pseudovariety of synchronous algebras and maps it to

$$\mathcal{V} : \Sigma \mapsto \{\mathcal{R} \subseteq \Sigma^* \times \Sigma^* \mid \mathbf{A}_{\mathcal{R}} \in \mathbb{V}\}.$$

Dually, let $\mathcal{V} \to \mathbb{V}$ denote the *correspondence* that takes a *-pseudovariety of automatic relations $\mathcal{V}$ and maps it to the pseudovariety of synchronous algebras generated by all $\mathbf{A}_{\mathcal{R}}$ for some $\mathcal{R} \in \mathcal{V}_{\Sigma}$. Here, the *pseudovariety generated* by a class $C$ of finite locally generated synchronous algebras is the smallest pseudovariety containing all finite locally generated algebras of $C$, or equivalently,[13] the class of all finite locally generated synchronous algebras

[12] This result can be put in perspective with [Pin22, Lemma XIII.4.11, p. 229] which proves a similar result in the context of monoids.

[13] The proof is straightforward, see *e.g.* [Pin22, Proposition XI.1.1, p. 190] for a proof in the context of semigroups.





that divide a finite product of algebras of $C$.[14]

**Lemma IX.4.11** (*An Eilenberg theorem for automatic relations*). The correspondences $\mathbb{V} \rightarrow \mathcal{V}$ and $\mathcal{V} \rightarrow \mathbb{V}$ define mutually inverse bijections between pseudovarieties of synchronous algebras and ∗-pseudovarieties of automatic relations.



*Proof.* We very roughly follow the proof schema of [Pin22, § XIII.4, pp. 226–229], which is a proof of Eilenberg's theorem in the context of monoids.

↳ *The correspondence $\mathbb{V} \rightarrow \mathcal{V}$ produces varieties.* First we have to show that if $\mathbb{V}$ is a pseudovariety of synchronous algebras and $\mathbb{V} \rightarrow \mathcal{V}$, then $\mathcal{V}$ is a ∗-pseudovarieties of automatic relations. Since $\mathbb{V}$ is closed under finite products, $\mathcal{V}$ is closed under Boolean operations.

*Syntactic derivatives:* Then let $\mathcal{R} \in \mathcal{V}_\Sigma$, and let $\mathcal{S}$ be any other relation recognized by $\mathbf{A}_\mathcal{R}$. This implies that $\mathbf{A}_\mathcal{S}$ divides $\mathbf{A}_\mathcal{R}$, and so $\mathbf{A}_\mathcal{S} \in \mathbb{V}$, from which we have $\mathbf{A}_\mathcal{S} \in \mathcal{V}_\Sigma$.

*Inverse morphisms:* Lastly, if $\mathcal{R} \in \mathcal{V}_\Sigma$, say $\underline{\mathcal{R}} = \eta_\mathcal{R}^{-1}[\text{Acc}]$, if $\psi \colon \mathbf{S}_2\Gamma \rightarrow \mathbf{S}_2\Sigma$ is a synchronous algebra morphism, then $\psi^{-1}[\mathcal{R}] = (\eta_\mathcal{R} \circ \psi)^{-1}[\text{Acc}]$, so $\psi^{-1}[\mathcal{R}]$ is recognized by $\mathbf{A}_\mathcal{R}$, that is $\mathbf{A}_{\psi^{-1}[\mathcal{R}]}$ divides $\mathbf{A}_\mathcal{R}$. Since $\mathbf{A}_\mathcal{R} \in \mathbb{V}$ and $\mathbb{V}$ is closed by division, it follows that $\mathbf{A}_{\psi^{-1}[\mathcal{R}]} \in \mathbb{V}$ and hence $\psi^{-1}[\mathcal{R}] \in \mathcal{V}_\Gamma$. This concludes the proof that $\mathcal{V}$ is a ∗-pseudovariety of automatic relations.

↳ *Inverse bijections: part 1.* Assume that $\mathbb{V} \rightarrow \mathcal{V}$ and $\mathcal{V} \rightarrow \mathbb{W}$. Then

$$\mathcal{V} \colon \Sigma \mapsto \{\mathcal{R} \subseteq \Sigma^* \times \Sigma^* \mid \mathbf{A}_\mathcal{R} \in \mathbb{V}\},$$

and so $\mathbb{W}$ is the pseudovariety generated by all syntactic synchronous algebras that belong to $\mathbb{V}$. It follows that $\mathbb{W} \subseteq \mathbb{V}$. To prove that $\mathbb{V} \subseteq \mathbb{W}$, let $\mathbf{A} \in \mathbb{V}$. Let $\Sigma_\mathbf{A}$ be an alphabet big enough so that there are injections from $\mathbf{A}_{\llcorner \llcorner}$ to $\Sigma_\mathbf{A} \times \Sigma_\mathbf{A}$, and from $\mathbf{A}_{\llcorner \mathcal{B}}$ and $\mathbf{A}_{\mathcal{B} \llcorner}$ to $\Sigma_\mathbf{A} \times \_$ and $\_ \times \Sigma_\mathbf{A}$, respectively. Since $\mathbf{A}$ is locally generated, this allows us to define a surjective synchronous algebra morphism $\phi \colon \mathbf{S}_2\Sigma_\mathbf{A} \twoheadrightarrow \mathbf{A}$. We then claim that $\mathbf{A}$ divides $\mathbf{B} \triangleq \prod_{x_\tau \in \mathbf{A}} \mathbf{B}_{x_\tau}$ where $\mathbf{B}_{x_\tau}$ is the syntactic synchronous algebra of $\phi^{-1}[x_\tau]$. Indeed, let $\psi_{x_\tau} \colon \mathbf{S}_2\Sigma_\mathbf{A} \twoheadrightarrow \mathbf{B}_{x_\tau}$ be the syntactic synchronous algebra morphism of $\phi^{-1}[x_\tau]$, say $\phi^{-1}[x_\tau] = \psi_{x_\tau}^{-1}[\text{Acc}_{x_\tau}]$. Then consider

$$\Psi \colon \quad \mathbf{S}_2\Sigma_\mathbf{A} \quad \rightarrow \quad \mathbf{B}$$
$$u_\sigma \quad \mapsto \quad \langle \psi_{x_\tau}(u_\sigma) \rangle_{x_\tau \in \mathbf{A}},$$

and let $\mathbf{B}_0$ be its image. Observe that for each $u_\sigma \in \mathbf{S}_2\Sigma_\mathbf{A}$, $\psi_{x_\tau}(u_\sigma) \in \text{Acc}_\tau$ *iff* $u_\sigma \in \phi^{-1}[x_\tau]$ *i.e.* $\phi(u_\sigma) = x_\tau$—note by the way that it implies $\sigma = \tau$. This implies that for any $(\langle y_{x_\tau} \rangle_{x_\tau \in \mathbf{A}})_\sigma \in \mathbf{B}_0$, there exists a unique $x_\tau$ *s.t.* $y_{x_\tau} \in \text{Acc}_{x_\tau}$. This defines a map $\chi \colon \mathbf{B}_0 \rightarrow \mathbf{A}$. Since moreover it makes the following diagram commute





$$\begin{array}{ccc} \mathbf{S_2\Sigma_A} & \xrightarrow{\;\Psi\;} & \mathbf{B_0} \\ & \phi\searrow & \downarrow\chi \\ & & \mathbf{A} \end{array}$$

it follows that $\chi$ is in fact a surjective synchronous algebra morphism.[15] Hence, $\mathbf{A}$ is a quotient of $\mathbf{B_0}$, which is a subalgebra of $\mathbf{B}$, which in turns in a product of algebras from $\mathbb{W}$, and so $\mathbf{A} \in \mathbb{W}$. It concludes the proof that $\mathbb{V} = \mathbb{W}$.



☙ *Inverse bijections: part 2.* Assume now that $\mathcal{V} \to \mathbb{V}$ and $\mathbb{V} \to \mathbb{W}$. Then for each $\Sigma$, for each $\mathcal{R} \in \mathcal{V}_\Sigma$, $\mathbf{A}_\mathcal{R} \in \mathbb{V}$ so $\mathcal{R} \in \mathcal{W}_\Sigma$, and hence $\mathcal{V} \subseteq \mathcal{W}$.

We then want to show the converse inclusion, namely $\mathcal{W} \subseteq \mathcal{V}$. Let $\mathcal{R} \in \mathcal{W}_\Sigma$ for some $\Sigma$, *i.e.* $\mathbf{A}_\mathcal{R} \in \mathbb{V}$. Hence there exists $\Gamma$ and relations $\mathcal{S}_1 \in \mathcal{V}_{\Gamma_1}, \dots \mathcal{S}_k \in \mathcal{V}_{\Gamma_k}$ such that $\mathbf{A}_\mathcal{R}$ divides $\mathbf{B} \triangleq \mathbf{A}_{\mathcal{S}_1} \times \cdots \times \mathbf{A}_{\mathcal{S}_k}$, *i.e.* there is a subalgebra $\mathbf{C} \subseteq \mathbf{B}$ which is a quotient of $\mathbf{B}$. Then $\mathbf{C}$ also recognizes $\mathcal{R}$, say $\mathcal{R} = \phi^{-1}[\mathrm{Acc}]$ for some morphism $\phi \colon \mathbf{S_2\Sigma} \twoheadrightarrow \mathbf{C}$ and $\mathrm{Acc} \subseteq \mathbf{C}$. Let $\iota \colon \mathbf{C} \to \mathbf{B}$ be the canonical embedding, $\pi_i \colon \mathbf{B} \twoheadrightarrow \mathbf{A}_{\mathcal{S}_i}$ be the canonical projection, and $\phi_i \triangleq \pi_i \circ \iota \circ \phi \colon \mathbf{S_2\Sigma} \to \mathbf{A}_{\mathcal{S}_i}$ for $i \in [\![1, k]\!]$. Then notice that since $\eta_{\mathcal{S}_i} \colon \mathbf{S_2\Gamma_i} \twoheadrightarrow \mathbf{A}_{\mathcal{S}_i}$ is surjective, then there exists $\psi_i \colon \mathbf{S_2\Sigma} \to \mathbf{S_2\Gamma_i}$ such that $\eta_{\mathcal{S}_i} \circ \psi_i = \phi_i$. Indeed, it suffices to send $\binom{a}{b}$ (resp. $\binom{a}{-}$, resp. $\binom{a}{}\!\vec{}$) on any element $u_{\mathsf{L}\mathsf{L}} \in \mathbf{S_2\Gamma_i}$ (resp. $u_{\mathsf{L}\mathsf{B}}$, resp. $u_{\mathsf{B}\mathsf{L}}$) such that $\eta_{\mathcal{S}_i}(u_{\mathsf{L}\mathsf{L}}) = \phi\binom{a}{b}$ (resp. $\eta_{\mathcal{S}_i}(u_{\mathsf{L}\mathsf{B}}) = \phi\binom{a}{-}$, resp. $\eta_{\mathcal{S}_i}(u_{\mathsf{B}\mathsf{L}}) = \phi(\vec{a})$). Overall, the following diagram commutes

$$\begin{array}{ccc} \mathbf{S_2\Sigma} & \xrightarrow{\;\psi_i\;} & \mathbf{S_2\Gamma_i} \\ \phi\downarrow & \phi_i\searrow & \downarrow\eta_{\mathcal{S}_i} \\ \mathbf{C} & & \\ & \iota\searrow\quad \mathbf{B} \xrightarrow{\;\pi_i\;} & \mathbf{A}_{\mathcal{S}_i}. \end{array}$$

Our goal is to show that $\mathcal{R} \in \mathcal{V}_\Sigma$. Observe that:

$$\underline{\mathcal{R}} = \phi^{-1}[\mathrm{Acc}] = \bigcup_{x \in \mathrm{Acc}} \phi^{-1}[x]$$

but then $\mathrm{Acc} \subseteq \mathbf{B}$, so $x$ is a tuple $\langle x_1, \dots, x_n \rangle$ (all elements having the same type), and by definition:

$$\phi^{-1}[x] = \bigcap_{i=1}^{n} \phi^{-1}[\iota^{-1}[\pi_i^{-1}[x_i]]] = \bigcap_{i=1}^{n} \phi_i^{-1}[x_i].$$

But then $\phi_i^{-1}[x_i] = \psi_i^{-1}[\eta_{\mathcal{S}_i}^{-1}[x_i]]$. Since $\mathcal{V}$ is closed under syntactic derivatives and $\mathcal{S}_i \in \mathcal{V}_{\Gamma_i}$, we have $\eta_{\mathcal{S}_i}^{-1}[x_i] \in \mathcal{V}_{\Gamma_i}$, and then since $\mathcal{V}$ is closed under Inverse morphisms and $\psi_i \colon \mathbf{S_2\Sigma} \to \mathbf{S_2\Gamma_i}$ is a morphism between free algebras, $\psi_i^{-1}[\eta_{\mathcal{S}_i}^{-1}[x_i]] \in \mathcal{V}_\Sigma$. Thus $\underline{\mathcal{R}}$ is a Boolean combination of elements of $\mathcal{V}_\Sigma$, and hence it also belongs to $\mathcal{V}_\Sigma$. This concludes the proof of $\mathcal{W} \subseteq \mathcal{V}$. □

As consequence of Lemma IX.4.11, if $\mathcal{V}$ is a ∗-pseudovariety of automatic relations and $\mathbb{V}$ is a pseudovariety of synchronous algebras, we write $\mathcal{V} \leftrightarrow \mathbb{V}$





to mean that either $\mathcal{V} \to \mathbb{V}$ or, equivalently, $\mathbb{V} \to \mathcal{V}$. This relation is called an *Eilenberg-Schützenberger correspondence*.

**Proposition IX.4.12.** If $\mathbb{V}$ is a pseudovariety of monoids, then

$$\mathbb{V}^{\mathrm{sync}} \triangleq \{\mathbf{A} \text{ locally generated finite synchronous algebra}$$
$$\text{s.t. all underlying monoids of } \mathbf{A} \text{ are in } \mathbb{V}\}$$

is a pseudovariety of synchronous algebras. Similarly, if $\mathcal{V}$ is an $*$-pseudovariety of regular languages, then the class of $\mathcal{V}$-relations, namely

$$\mathcal{V}^{\mathrm{sync}} \colon \Sigma \mapsto \{\mathcal{R} \subseteq \Sigma^* \times \Sigma^* \mid \exists L \in \mathcal{V}_{\Sigma_{\otimes}^2}, \; \underline{\mathcal{R}} = L \cap \mathrm{WellFormed}_{\Sigma}\},$$

is a $*$-pseudovariety of automatic relations.

*Proof.* The first point is straightforward. The second one follows from it and Lemma IX.4.11 and Theorem IX.4.2. □

Finally, Theorem IX.4.2 can be elegantly rephrased by saying that correspondences between pseudovarieties of monoids and $*$-pseudovarieties of regular languages lift to correspondences between pseudovarieties of synchronous algebras and $*$-pseudovarieties of automatic relations.

**Theorem IX.4.13** (*Lifting Theorem: Pseudovariety Formulation*)**.** If, in the Eilenberg-Schützenberger correspondence between pseudovarieties of monoids and $*$-pseudovarieties of regular languages we have $\mathcal{V} \leftrightarrow \mathbb{V}$, then in the Eilenberg-Schützenberger correspondence between the pseudovariety of synchronous algebras $\mathbb{V}^{\mathrm{sync}}$ and the $*$-pseudovariety of automatic relations $\mathcal{V}^{\mathrm{sync}}$, we have $\mathcal{V}^{\mathrm{sync}} \leftrightarrow \mathbb{V}^{\mathrm{sync}}$.

## IX.5   Discussion

### IX.5.1   Path Algebras and the Lifting Theorem

A natural next step is to generalize Question IX.1.1 by replacing $\mathrm{WellFormed}_{\Sigma}$ by a fixed regular language $\Omega$.

**Question IX.5.1.** Given a class of regular languages $\mathcal{V}$, can we characterize the class $\mathcal{V}^{\Omega}$ of all languages of the form $L \cap \Omega$ for some $L \in \mathcal{V}$, in a way that preserves decidability?

**Remark IX.5.2.** There is no (meta$^2$-)algorithm taking as input a regular language $\Omega$, and returning a (meta-)algorithm *s.t.*, given a membership algorithm for $\mathcal{V}$, returns a membership algorithm for $\mathcal{V}^{\Omega}$.

Indeed, let $\mathcal{V}$ be a pseudovariety of regular languages with decidable membership but undecidable separation—see [RS11, Corollary 1.6, p. 478] and Footnote 5. We reduce the $\mathcal{V}$-separability problem to our problem.

Given two regular languages $L_1$ and $L_2$, to decide if they are $\mathcal{V}$-separable, we first test if $L_1 \cap L_2 = \varnothing$: if not, we reject. Otherwise, we let $\Omega \triangleq L_1 \cap L_2$,





and ask whether $L_1$ is in $\mathcal{V}^\Omega$. By definition, this happens *iff* $L_1$ can be written as $S \cap \Omega = S \cap (L_1 \cup L_2)$ for some $S \in \mathcal{V}$. Since $L_1$ and $L_2$ are disjoint, the equality $L_1 = S \cap (L_1 \cup L_2)$ is precisely equivalent to having $L_1 \subseteq S$ and $S \cap L_2 = \varnothing$. Hence, $L_1$ is in $\mathcal{V}^\Omega$ if, and only if, $L_1$ and $L_2$ are $\mathcal{V}$-separable. $\diamondsuit$

What this remark shows is that actually the language WellFormed$_\Sigma$ is special, in the sense that we relied on some of its specific properties to obtain the Lifting theorem.

We claim that the construction of synchronous algebras can be generalized for any $\Omega$, giving rise to the notion of "path algebras".[16] The lifting theorem for monoids can be shown to hold for some $\Omega$, including well-formed words for $n$-ary relations with $n \geq 3$, and that it cannot effectively hold for all $\Omega$. We believe that a necessary condition for the Lifting theorem to hold would be that $\Omega$ is *fully-preordered*, in the sense that there exists a preorder $\leqslant$ on the alphabet $\Sigma$ *s.t.* $\Omega = \{u_1 \dots u_n \in \Sigma^* \mid u_1 \leqslant \dots \leqslant u_n\}$.[17]

### IX.5.2   Path Algebras and Restricted Head Movements

A natural next step would then be to study the relationship between "path algebras" and Figueira & Libkin's $T$-controlled relations defined in Section VII.1.6. For any regular language $T \subseteq [\![1, k]\!]^*$, we claim that we can define a regular language $\Omega_T$ such that finite $\Omega_T$-path algebras exactly recognize $T$-controlled relations.

**Conjecture IX.5.3.** There is a way of defining a map $T \mapsto \Omega_T$ *s.t.* for any regular languages $T_1, T_2 \subseteq [\![1, k]\!]^*$, $T_1$-controlled relations are included in $T_2$-controlled relations if, and only if, there is an adjunction from the category of $\Omega_{T_1}$-path algebras to the category of $\Omega_{T_2}$-path algebras.

### IX.5.3   Theorem Projection via Monad Adjunction

In fact, Conjecture IX.5.3 is part of a larger idea we have, about relating the expressiveness of universe—formalized by a monad—and its categorical properties. In algebraic language theory, monads play an essential role—see Section IX.A. Informally, a monad describes a universe. For instance, the monad

$$\Sigma \mapsto \Sigma^*$$

describes the universe of finite words, and gives rise to the notion of monoids. Similarly, weighted words over a field $\mathbb{K}$ can be described by

$$\Sigma \mapsto \mathbb{K}\Sigma^*,$$

where $\mathbb{K}X$ denotes the free vector space over set $X$. Again, this monad gives rise to Reutenauer's "algèbres associatives" [Reu80], *a.k.a.* "$\mathbb{K}$-weighted monoids", which is essentially an algebraic structure that is both a monoid and a $\mathbb{K}$-vector space. Our synchronous algebras are no exception—see Section IX.A—, and we refer the reader to [Boj20, § 4] for more examples.

[16] In short, they are the adaptation of the free category generated by a graph to dependent sets. See also Section IX.A.

[17] The important property about these languages is that the monad defining their path-algebras are strongly acyclic, like the monad of Section IX.A.





For well-behaved monads, "languages" recognized by their finite algebras coincide with monadic second-order-definable languages.[18] This gives rise to a natural set of questions, consisting in finding algebraic characterizations of fragments of monadic second-order logic. The perhaps most celebrated example is that of the Schützenberger-McNaughton-Papert theorem, which proves that first-order definable languages of finite words are exactly those recognized by aperiodic monoids [Sch65; MP71]. The statement was extended to $\omega$-words [Per84], to countable ordinal words [Bed01], and to countable scattered words [BC11]. In the latter case, asking only for aperiodicity is not enough: the algebra should also be "gap-insensitive".[19] This can be explained as follows: by going from the simpler setting of finite words (or countable ordinal words) to countable scattered words, there are new phenomena that appear and that first-order logic is not able to describe.

However, intuitively, going in the converse direction should be easy! Given an algebraic characterization of first-order logic over a "richer" monad (*e.g.* countable scatter words), it should be easy to reprove the Schützenberger-McNaughton-Papert theorem over finite words, using the "rich" algebraic characterization as a black box.

Surprisingly, while it is easy to "project" the algebraic characterization of first-order logic over $\omega$-words to get its counterpart for finite words, the task becomes less straightforward—but still possible—when projecting from countable ordinal words or countable scattered words to finite ones. In both cases, the proof relies on two ingredients:

- a function, encoding a language of finite words into a language in the richer setting;
- a function, projecting a language in the richer setting to a language of finite words,

see Figure IX.6. Moreover, both functions should preserve the properties that we are studying—either being first-order definable, or satisfying the algebraic characterization.

In the case of $\omega$-words, finding these functions is easy: for instance, the projection simply consists in restricting a language of words of length at most $\omega$ to the finite words. On the other hand, the projection from countable scattered words to finite ones cannot simply do this, since finite words are not first-order definable amongst countable scattered ones.[20] Hence, the projection in question needs to be tailor-made to preserve first-order logic.[21]

In fact, not only does the projection of $\omega$-words to finite words preserve the property of being first-order definable, it also preserve membership in any pseudovariety! In this sense, words of length at most $\omega$ form a *conservative extension* of finite words, while countable scattered words do not: this is why we can find a single construction to project algebraic characterizations[22] from $\omega$-words to finite ones, but not from countable scattered words to finite ones.

This begs the question: how can we formalize this idea of a monad being a *conservative extension* of another one? In light of Figure IX.6, we propose



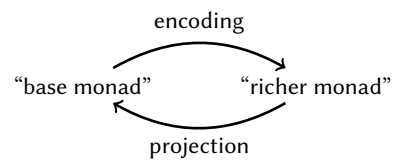

Figure IX.6: Tools to project algebraic characterizations from one universe to another.







to look at the notion of monad adjunction. Before introducing it, we need a preliminary definition.

Let $\mathbf{S} = (S, \mathrm{unit}, \mathrm{mult})$ be a C-monad and $\mathbf{T} = (T, \mathrm{unit}, \mathrm{mult})$ be a D-monad.[23] We assume C and D to be both categories of typed sets.

A *monad functor*,[24] from $\mathbf{S}$ to $\mathbf{T}$ is a pair $(\Phi, F)$, where:

- $F\colon \mathrm{C} \to \mathrm{D}$ is a functor,
- $\Phi\colon TF \Rightarrow FS$ is a natural transformation,

such that, for every object $X \in \mathrm{C}$, the following diagrams commute:

$$(MF)_{unit}\colon \quad \begin{array}{ccc} & FX & \\ {\scriptstyle \mathrm{unit}_{FX}}\swarrow & & \searrow{\scriptstyle F\mathrm{unit}_X} \\ TFX & \xrightarrow{\ \phi_X\ } & FSX, \end{array} \qquad (MF)_{mult}\colon \quad \begin{array}{ccc} TFSX & \xrightarrow{\ \phi_{SX}\ } & FSSX \\ {\scriptstyle T\phi_X}\nearrow & & \searrow{\scriptstyle F\mathrm{mult}_X} \\ TTFX & & FSX. \\ {\scriptstyle \mathrm{mult}_{FX}}\searrow & & \nearrow{\scriptstyle \phi_X} \\ & TFX & \end{array}$$

Such a pair is denoted by $\Phi_F\colon \mathbf{S} \Rightarrow \mathbf{T}$.[25]

**Example IX.5.4** (Finite words and $\omega$-words). Let $\mathbf{K}^+$ be *Kleene's monad* over the category Set of non-empty finite words, defined by $X \mapsto X^+$. Moreover, let $\mathbf{W}$ be *Wilke's monad*, defined over the category of 2-typed sets Set$^{+,\omega}$, by $\langle F, I \rangle \mapsto \langle F^+, F^+I \cup F^{\omega\text{-up}} \rangle$, where $F^{\omega\text{-up}}$ denotes the set of ultimately-periodic words of length $\omega$ over $F$.

We start by defining a monad functor from $\mathbf{W}$ to $\mathbf{K}^+$: it represents the natural surjection from words of length at most $\omega$ to non-empty finite words. Let proj$\colon$ Set$^{+,\omega} \to$ Set$^S$ be the functor that only keeps the first component, defined by $\langle F, I \rangle \mapsto F$, and $\Phi$ be the natural transformation $\mathbf{K}^+$ proj $\Rightarrow$ proj $\mathbf{W}$, defined for $\langle F, I \rangle \in$ Set$^{+,\omega}$ by letting

$$\phi_{F,I}\colon \underbrace{\mathbf{K}^+\,\mathrm{proj}\langle F, I\rangle}_{=F^+} \longrightarrow \underbrace{\mathrm{proj}\,\mathbf{W}\langle F, I\rangle}_{=F^+}$$

be the identity. Then $\Phi_{\mathrm{proj}}$ is a monad functor from Wilke's monad to Kleene's monad.

Dually, we then define a monad functor from $\mathbf{K}^+$ to $\mathbf{W}$ that corresponds to the natural embedding of non-empty finite words into words of length at most $\omega$. We let zero$\colon$ Set $\to$ Set$^{+,\omega}$ be the functor that transforms a set $X$ into a pair $\langle X, \{\bullet\} \rangle$, and $\Psi$ be the natural transformation $\mathbf{W}$ zero $\Rightarrow$ zero $\mathbf{K}^+$, defined for $X \in$ Set by letting

$$\psi_X\colon \underbrace{\mathbf{W}\,\mathrm{zero}\,X}_{=\langle X^+, X^+\{\bullet\}\cup X^{\omega\text{-up}}\rangle} \longrightarrow \underbrace{\mathrm{zero}\,\mathbf{K}^+ X}_{=\langle X^+,\{\bullet\}\rangle}$$

send $u \in X^+$ to itself, and $u \in X^+\{\bullet\} \cup X^{\omega\text{-up}}$ to $\bullet \in \{\bullet\}$. Then $\Psi_{\mathrm{zero}}$ is a monad functor from Kleene's monad to Wilke's monad, see Figure IX.7. ◇

Unsurprisingly, monad functors can be composed. Moreover, it can be


[23] Following Street, we use the same notation for the unit and multiplication in $\mathbf{S}$ and $\mathbf{T}$.

[24] The notion was introduced by Street [Str72, § 1] in the slightly more general context of 2-categories. We found this reference thanks to [Rez12]. Monad functors are also known as "lax maps of monads".

[25] Note that, somewhat surprisingly, one element of a monad functor from $\mathbf{S}$ to $\mathbf{T}$ is covariant—namely the functor $F\colon \mathrm{C} \to \mathrm{D}$—while the other one is contravariant—namely the natural transformation $\Phi\colon TF \Rightarrow FS$. The next example will illustrate why this definition makes sense.


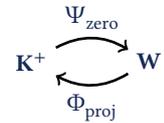

Figure IX.7: Monads functors between Kleene's monad and Wilke's monad.





shown that any monad functor from **S** to **T** induces a functor from the category of **S**-algebras to **T**-algebras. For instance, $\Phi_{\text{proj}} \colon \mathbf{W} \Rightarrow \mathbf{K}^+$ induces the functor from Wilke's algebras to semigroups that sends $\langle S^+, S^\omega \rangle$ to $S^+$. Dually, $\Psi_{\text{zero}} \colon \mathbf{K}^+ \Rightarrow \mathbf{W}$ induces the functor from semigroups to Wilke's algebras that sends $S$ to $\langle S, \{\bullet\} \rangle$.

We conjecture that we can naturally define a notion of "monad (functor) adjunction", which generalizes the notion of "functor adjunctions", in such a way that a *monad adjunction* between **S** and **T** induces an adjunction between **S**-algebras and **T**-algebras. A typical example of such a monad adjunction would be Figure IX.7.

"Monad (functor) adjunctions" are high-order objects, and so, for the sake of simplicity, we will work with the adjunction they induce between the categories algebras. Going back to Example IX.5.4 and Fig. IX.7, we represent in Figure IX.8 the two functors induced between their Eilenberg-Moore categories. It is indeed an adjunction, with $\langle L^+, L^\omega \rangle \mapsto L^+$ acting as the left adjoint and $R \mapsto \langle R, \{\bullet\} \rangle$ as the right adjoint, and there are *natural* bijections

$$\hom_{\text{Sgp}}(L^+, R) \cong \hom_{\text{Wilke}}(\langle L^+, L^\omega \rangle, \langle R, \bullet \rangle).$$

We believe that this adjunction is precisely what allows us to automatically project any algebraic characterization of $\omega$-regular languages to an algebraic characterization of regular languages, for the good reason that the two maps $\langle L^+, L^\omega \rangle \mapsto L^+$ and $R \mapsto \langle R, \{\bullet\} \rangle$ are the key ingredients of the proof!

Interestingly, sending a semigroup $S$ to the ordinal semigroup $S \sqcup \{\bullet\}$, defined by letting any infinite product be equal to the zero element $\bullet$ also defines a functor from the category of semigroups to the category of ordinal semigroups. Dually, the canonical surjection sending an ordinal semigroup to its underlying semigroup is a functor from the category of ordinal semigroups to the category of semigroups. However, they do not form an adjunction! We actually believe that there are no adjunction between these categories.

These observations lead us to the last conjecture of this thesis.

**Conjecture IX.5.5** (Projecting Algebraic Characterizations). **Let S and T be two monads. Any monad adjunction between them induces a "natural function" from** Eilenberg-Schützenberger correspondences **over T to** Eilenberg-Schützenberger correspondences **over S.**

In the case of finite words and $\omega$-words, this surjection sends for instance the theorem "an $\omega$-regular language is first-order definable *iff* its syntactic Wilke's algebra is aperiodic" onto the theorem "a regular language is first-order definable *iff* its syntactic semigroup is aperiodic"! We believe that the lifting theorem (Theorem IX.4.13) could be a consequence of Conjecture IX.5.5.

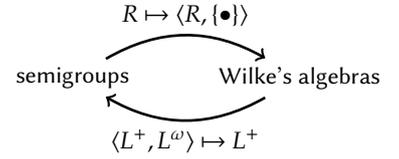

Figure IX.8: An adjunction between the category of **W**-algebras and the category of **K**⁺-algebras.





*Appendices*

## IX.A    *Monads Everywhere!*

We denote by Set$^{\mathcal{S}}$ and Pos$^{\mathcal{S}}$ the categories of $\mathcal{S}$-typed sets and $\mathcal{S}$-partially ordered sets—note that in this model, each type is equipped with its own order and that elements of different types cannot be compared. Similarly, let Dep$^{\mathcal{S}}$ be the category of $\mathcal{S}$-dependent sets.

We claim that synchronous algebras correspond to Eilenberg-Moore algebras of some monad over the category Dep$^{\mathcal{T}}$. For the sake of readability, we represent the underlying typed set of a $\mathcal{T}$-dependent set

$$\mathbf{X} = \langle X_{\text{L/L}},\, X_{\text{L/L}\to\text{L/B}},\, X_{\text{L/B}},\, X_{\text{L/L}\to\text{B/L}},\, X_{\text{B/L}} \rangle$$

as follows:

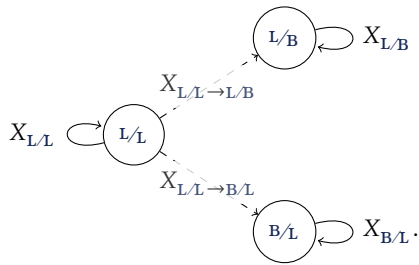

We define the *synchronous monad* $\mathbf{S}_2$ over Dep$^{\mathcal{T}}$ as the functor which maps equipped with a dependency relation $\asymp$ to the dependent set and two words are dependent if their domain are isomorphic and their letters are pairwise dependent. The unit and free multiplication are defined naturally.

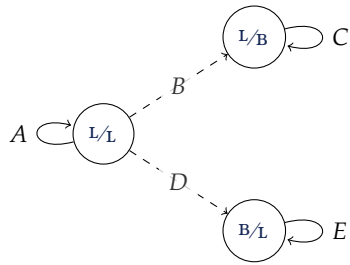

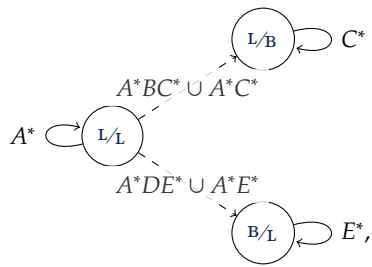

Note in particular that all five empty words are mutually dependent, and that synchronous words $\mathbf{S}_2\Sigma$ correspond to applying $\mathbf{S}_2$ to equipped with equality. Moreover, synchronous algebras exactly correspond to $\mathbf{S}_2$-algebras.

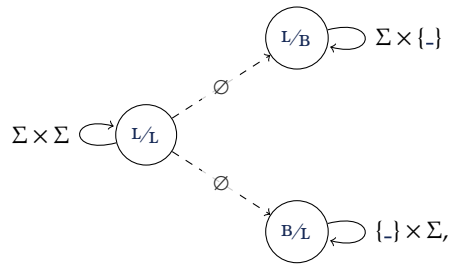

A systemic approach to algebraic language theory was proposed by Bojańczyk using the formalism of monads [Boj15], for monads over finitely typed sets $\mathrm{Set}^{\mathcal{S}}$. Urbat, Adámek, Chen & Milius then extended these results to capture monads over varieties of typed (ordered) algebras [UACM17]. Lastly, Blumensath extended those results to monads over the category of typed posets $\mathrm{Pos}^{\mathcal{S}}$ when the set $\mathcal{S}$ of types is infinite [Blu21].

Observe that the category of dependent sets is not captured by any of the results above since the dependency relation can compare elements of different types, contrary to typed posets & co.

# Chapter X

# Conclusion & Open Problems

## Abstract

This chapter concludes Part 2 of this thesis. We recall some open problems mentioned previously, and highlight a new research direction relating the structural properties of a language-theoretic framework with its expressiveness.

## Contents

## X.1   Separating Automatic Relations by Recognizable Ones

The problem introduced in Chapter VII remains open.

**Open Problem VII.1.10.** Is the REC-SEPARABILITY PROBLEM FOR AUTOMATIC RELATIONS decidable?

In Chapter VIII, we proved this problem to be equivalent to FINITE REGULAR COLOURABILITY OF AUTOMATIC GRAPHS (Theorem VIII.3.2), and showed that when the number of colours is fixed, the problem is undecidable (Theorem VIII.3.7). In fact, we showed that most problems of this form are undecidable (Theorem VIII.4.1). In turn, it implies that this separability problem becomes undecidable when the separator is restricted to be a union of $k$ products of regular languages for some fixed $k \geq 2$. However, as explained in Section VIII.6, some gaps remain to be able to use our techniques to prove the undecidability of the AUT/REC-SEPARABILITY PROBLEM.

On the other hand, we introduced in Chapter IX an algebraic approach for automatic relations, hoping to prove the decidability of this problem. Algebraic language theory is a powerful tool to prove the decidability of separation over finite words [PZ16], but also in more complex settings such as countable ordinal words [CGM22].

Alas, using the theory we developped to tackle Open Problem VII.1.10 seems non-trivial. The main obstacle being that, while recognizable relations have some desirable closure properties, they do not form a pseudovariety of automatic relations.

However, should Open Problem VII.1.10 problem be decidable, the question of the decidability of its generalization to larger class of relations would be a natural next step.

**Open Problem X.1.1.** Are the DRAT/REC-SEPARABILITY and RAT/REC-SEPARABILITY PROBLEMS decidable?

In [BFM23, § 1], we incorrectly stated that "As for definability[1], the REC-SEPARABILITY PROBLEM for rational relations is in general undecidable", which is unfounded as we do not currently know if it is true. Indeed, as mentioned in Chapter VII, the RAT/REC-MEMBERSHIP PROBLEM is undecidable, by [Ber79, § III, Theorem 8.4]. Moreover, in general, MEMBERSHIP PROBLEMS reduced to SEPARABILITY PROBLEMS: a relation $\mathcal{R}$ belongs to a class $\mathcal{V}$ if, and only if, $\mathcal{R}$ and $\neg\mathcal{R}$ are $\mathcal{V}$-separable. However, for this argument to work, the larger class of relation needs to be effectively closed under complement. This is not the case of rational relations, see Section VII.1.4.

For deterministic rational relations, while they are effectively closed under complement, the DRAT/REC-MEMBERSHIP PROBLEM is decidable—see Section VII.1.5!

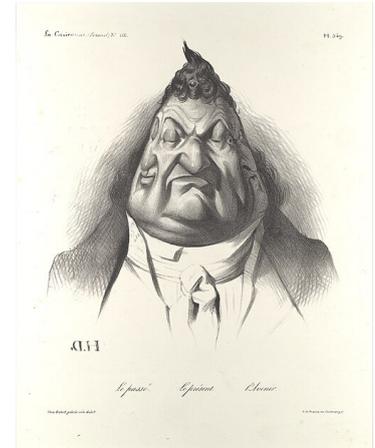

Figure X.1: *Le passé – Le présent – L'Avenir*, by Honoré Daumier.

[1] Definability is the same as the MEMBERSHIP PROBLEM.

## X.2  Colouring Problems on Automatic Graphs

Let us quickly recall some of the conjectures that have been discussed in Section VIII.6. The first one is about finding an equivalent characterization in the dichotomy theorem for automatic structures (Theorem VIII.4.1).

**Conjecture VIII.6.7.** For any finite $\sigma$-structure **B**, $\mathcal{H}om^{\mathrm{reg}}(\mathrm{Aut}, \mathbf{B})$ is invariant under graph isomorphisms *iff* **B** has finite duality.

In other words, this conjecture could be rephrased as follows: "a problem of the form $\mathcal{H}om^{\mathrm{reg}}(\mathrm{Aut}, \mathbf{B})$ is decidable if, and only if, its output only depends on the structure represented by the automatic presentation, and not on the presentation itself"! Going back to our original question, while we do not know if finite regular colourability of automatic graphs is decidable, it is natural to study variations of this problem, as well as sufficient or necessary conditions to ensure colourability or non-colourability. However, our understanding of these problems are also somewhat limited.

**Conjecture VIII.6.8.**[2] Finite colourability of automatic graphs is $\Sigma_2^0$-complete.

**Conjecture VIII.6.9.**[3] The problem of whether an automatic graph has bounded tournaments is decidable.

## X.3  An Algebraic Approach Beyond Automatic Relations

The algebraic theory developed in Chapter IX can be generalized, by replacing the constraint of "being well-formed" with an arbitrary regular language $\Omega$. This would result in the notion of $\Omega$-*path algebras*—with WellFormed$_\Sigma$-path algebras corresponding to our synchronous algebras.

While there is no hope to get a version of the lifting theorem (Theorem IX.4.2) for every $\Omega$—see Remark IX.5.2—, we believe it holds for fully-preordered languages. See Section IX.5 for more details.

Interestingly, we believe this question to be related to Figueira & Libkin's *T-controlled relations*, presented in Section VII.1.6. Each language $T$, specifying the head movements that are allowed, gives rise to a language $\Omega_T$, such that relations recognized by finite $\Omega_T$-path algebras exactly correspond to the $T$-controlled relations.

**Open Problem X.3.1.** Given $k \in \mathbb{N}_{>0}$, given regular languages $T_1$ and $T_2$ over $[\![1, k]\!]$, can we decide if $T_1$-controlled relations exactly correspond to $T_2$-controlled relations?

We conjecture that, one way of tackling this problem would be to study properties of the category of $\Omega_T$-path algebras.

**Conjecture IX.5.3.** There is a way of defining a map $T \mapsto \Omega_T$ *s.t.* for any regular languages $T_1, T_2 \subseteq [\![1, k]\!]^*$, $T_1$-controlled relations are included in $T_2$-controlled relations if, and only if, there is an adjunction from the category

[2] Note again that the upper bound is trivial, since by De Bruijn-Erdős theorem, this problem is equivalent to asking if there exists $k \in \mathbb{N}$ *s.t.* every finite subgraph of the source is $k$-colourable.

[3] This conjecture corresponds to [BFM23, Conjecture 7.3].

of $\Omega_{T_1}$-path algebras to the category of $\Omega_{T_2}$-path algebras.

Hopefully, the latter condition could maybe be massaged to get decidability. Finally, Section IX.5.3 explains how the notion of monad adjunction could provide a way of automatically projecting—or lifting, depending on the point of view—an algebraic characterization from a universe to another.

**Conjecture IX.5.5** (Projecting Algebraic Characterizations). Let **S** and **T** be two monads. Any monad adjunction between them induces a "natural function" from Eilenberg-Schützenberger correspondences over **T** to Eilenberg-Schützenberger correspondences over **S**.

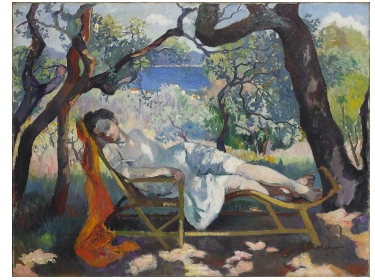

Figure X.2: *La Sieste*, by Henri Manguin.

# Bibliography

**[AB09]** Sanjeev Arora and Boaz Barak. *Computational Complexity: A Modern Approach*. 2009 (Cited on pp. 58–60, 87).
DOI: 10.1017/CBO9780511804090

**[ABISV09]** Eric Allender, Michael Bauland, Neil Immerman, Henning Schnoor, and Heribert Vollmer. "The complexity of satisfiability problems: Refining Schaefer's theorem." In *Journal of Computer and System Sciences (JCSS)* 75.4 (2009), pp. 245–254 (Cited on p. 248).
DOI: 10.1016/j.jcss.2008.11.001

**[ACP87]** Stefan Arnborg, Derek G Corneil, and Andrzej Proskurowski. "Complexity of finding embeddings in a k-tree." In 8.2 (1987), pp. 277–284 (Cited on p. 83).
DOI: 10.1137/0608024

**[AHV95]** Serge Abiteboul, Richard Hull, and Victor Vianu. *Foundations of Databases*. 1995. ISBN: 0-201-53771-0 (Cited on pp. 67–69).
URL: http://webdam.inria.fr/Alice/

**[Alm99]** Jorge Almeida. "Some algorithmic problems for pseudovarieties." In *Publ. Math. Debrecen* 52.1 (1999). Consulted version: https://www.researchgate.net/profile/Jorge-Almeida-14/publication/2510507_Some_Algorithmic_Problems_for_Pseudovarieties/links/02e7e531d968b4fe8f000000/Some-Algorithmic-Problems-for-Pseudovarieties.pdf, pp. 531–552 (Cited on p. 304)

**[Ang+17]** Renzo Angles, Marcelo Arenas, Pablo Barceló, Aidan Hogan, Juan Reutter, and Domagoj Vrgoč. "Foundations of Modern Query Languages for Graph Databases." In *ACM Computing Surveys* 50.5 (2017) (Cited on p. 90).
DOI: 10.1145/3104031

**[AP89]** Stefan Arnborg and Andrzej Proskurowski. "Linear time algorithms for NP-hard problems restricted to partial k-trees." In 23.1 (1989), pp. 11–24 (Cited on p. 82).
DOI: 10.1016/0166-218X(89)90031-0

**[Ash91]** C. J. Ash. "Inevitable graphs: a proof of the type II conjecture and some related decision procedures." In *International Journal of Algebra and Computation* 01.01 (1991), pp. 127–146 (Cited on p. 304).
DOI: 10.1142/S0218196791000079

**[Ats08]** Albert Atserias. "On digraph coloring problems and treewidth duality." In *European Journal of Combinatorics* 29.4 (2008), pp. 796–820 (Cited on pp. 45, 249, 273).
DOI: 10.1016/j.ejc.2007.11.004

**[AW12]** Isolde Adler and Mark Weyer. "Tree-width for first order formulae." In *Logical Methods in Computer Science (LMCS)* Volume 8, Issue 1 (2012) (Cited on p. 69).
DOI: 10.2168/LMCS-8(1:32)2012

**[Bar09]** Pablo Barceló. "Locality of Queries." In *Encyclopedia of Database Systems*. 2009, pp. 1637–1638 (Cited on p. 91).
DOI: 10.1007/978-0-387-39940-9_1270

**[Bar13]** Pablo Barceló Baeza. "Querying Graph Databases." In *ACM Symposium on Principles of Database Systems (PODS)*. 2013, pp. 175–188 (Cited on p. 90).
DOI: 10.1145/2463664.2465216

[BC11]  Alexis Bès and Olivier Carton. "Algebraic Characterization of FO for Scattered Linear Orderings." In *EACSL Annual Conference on Computer Science Logic (CSL)*. Vol. 12. 2011, pp. 67–81 (Cited on p. 326).
DOI: 10.4230/LIPIcs.CSL.2011.67

[BC98]  Nicolas Bedon and Olivier Carton. "An Eilenberg theorem for words on countable ordinals." In *Latin American Theoretical Informatics Symposium (LATIN)*. Lecture Notes in Computer Science. 1998, pp. 53–64 (Cited on p. 301).
DOI: 10.1007/BFb0054310

[Bed01]  Nicolas Bedon. "Logic over Words on Denumerable Ordinals." In *Journal of Computer and System Sciences* 63.3 (2001), pp. 394–431 (Cited on p. 326).
DOI: 10.1006/jcss.2001.1782

[Bed98]  Nicolas Bedon. "Langages reconnaissables de mots indexés par des ordinaux." PhD thesis. Université de Marne la Vallée, 1998 (Cited on p. 301).
URL: https://theses.hal.science/tel-00003586

[Ber79]  Jean Berstel. *Transductions and Context-Free Languages*. Consulted version: http://www-igm.univ-mlv.fr/~berstel/LivreTransductions/LivreTransductions.pdf. 1979 (Cited on pp. 211, 220, 227, 247, 332)

[BFM23]  Pablo Barceló, Diego Figueira, and Rémi Morvan. "Separating Automatic Relations." In *International Symposium on Mathematical Foundations of Computer Science (MFCS)*. Vol. 272. 2023, 17:1–17:15 (Cited on pp. 245, 292, 332, 333).
DOI: 10.4230/LIPIcs.MFCS.2023.17

[BFR19]  Pablo Barceló, Diego Figueira, and Miguel Romero. "Boundedness of Conjunctive Regular Path Queries." In *International Colloquium on Automata, Languages and Programming (ICALP)*. Vol. 132. 2019, 104:1–104:15 (Cited on p. 101).
DOI: 10.4230/LIPIcs.ICALP.2019.104

[BG00]  A. Blumensath and E. Gradel. "Automatic structures." In *Annual Symposium on Logic in Computer Science (LICS)*. 2000, pp. 51–62 (Cited on p. 239).
DOI: 10.1109/LICS.2000.855755

[BG04]  Achim Blumensath and Erich Grädel. "Finite Presentations of Infinite Structures: Automata and Interpretations." en. In *Theory of Computing Systems* 37.6 (2004), pp. 641–674 (Cited on pp. 240, 241).
DOI: 10.1007/s00224-004-1133-y

[BG21]  Joshua Brakensiek and Venkatesan Guruswami. "Promise Constraint Satisfaction: Algebraic Structure and a Symmetric Boolean Dichotomy." In *SIAM Journal on computing* 50.6 (2021), pp. 1663–1700 (Cited on p. 190).
DOI: 10.1137/19M128212X

[BGLZ22]  Pascal Bergsträßer, Moses Ganardi, Anthony W. Lin, and Georg Zetzsche. "Ramsey Quantifiers over Automatic Structures: Complexity and Applications to Verification." In *Annual Symposium on Logic in Computer Science (LICS)*. 2022, 28:1–28:14 (Cited on pp. 218, 247).
DOI: 10.1145/3531130.3533346

[BHLLN19]  Pablo Barceló, Chih-Duo Hong, Xuan Bach Le, Anthony W. Lin, and Reino Niskanen. "Monadic Decomposability of Regular Relations." In *International Colloquium on Automata, Languages and Programming (ICALP)*. 2019, 103:1–103:14 (Cited on pp. 218, 247).
DOI: 10.4230/LIPIcs.ICALP.2019.103

[BKW17]  Libor Barto, Andrei Krokhin, and Ross Willard. "Polymorphisms, and How to Use Them." In *The Constraint Satisfaction Problem: Complexity and Approximability*. Ed. by Andrei Krokhin and Stanislav Zivny. Vol. 7. Dagstuhl Follow-Ups. 2017, pp. 1–44 (Cited on p. 248).
DOI: 10.4230/DFU.Vol7.15301.1

[BLLW12]  Pablo Barceló, Leonid Libkin, Anthony Widjaja Lin, and Peter T. Wood. "Expressive Languages for Path Queries over Graph-Structured Data." In *ACM Transactions on Database Systems (TODS)* 37.4 (2012), p. 31 (Cited on p. 211).
DOI: 10.1145/2389241.2389250

**[BLR14]**   Pablo Barceló, Leonid Libkin, and Miguel Romero. "Efficient Approximations of Conjunctive Queries." In *SIAM Journal on computing* 43.3 (2014), pp. 1085–1130 (Cited on pp. 147, 151).
DOI: 10.1137/130911731

**[BLSS03]**   Michael Benedikt, Leonid Libkin, Thomas Schwentick, and Luc Segoufin. "Definable relations and first-order query languages over strings." In *Journal of the ACM* 50.5 (2003), pp. 694–751 (Cited on pp. 211, 232, 234).
DOI: 10.1145/876638.876642

**[Blu21]**   Achim Blumensath. "Algebraic Language Theory for Eilenberg–Moore Algebras." In *Logical Methods in Computer Science (LMCS)* Volume 17, Issue 2 (2021) (Cited on p. 330).
DOI: 10.23638/LMCS-17(2:6)2021

**[Blu24]**   Achim Blumensath. *Monadic Second-Order Model Theory*. Unpublished monograph. Version of 2024-07-30. 2024 (Cited on pp. 237, 239, 242).
URL: https://www.fi.muni.cz/~blumens/MSO.pdf

**[BMPP18]**   Sougata Bose, Anca Muscholl, Gabriele Puppis, and Vincent Penelle. "Origin-equivalence of two-way word transducers is in PSPACE." In *IARCS Annual Conference on Foundations of Software Technology and Theoretical Computer Science (FST&TCS)*. Consulted version: https://hal.science/hal-02415557. 2018 (Cited on p. 227).
DOI: 10.4230/LIPIcs.FSTTCS.2018.22

**[BMT20]**   Angela Bonifati, Wim Martens, and Thomas Timm. "An Analytical Study of Large SPARQL Query Logs." In *The VLDB Journal* 29.2 (2020), pp. 655–679 (Cited on p. 101).
DOI: 10.1007/s00778-019-00558-9

**[BN23]**   Mikołaj Bojańczyk and Lê Thành Dũng (Tito) Nguyễn. "Algebraic Recognition of Regular Functions." In *International Colloquium on Automata, Languages and Programming (ICALP)*. Vol. 261. Consulted version: https://hal.science/hal-03985883v2. 2023, 117:1–117:19 (Cited on p. 302).
DOI: 10.4230/LIPIcs.ICALP.2023.117

**[Bod96]**   Hans L. Bodlaender. "A Linear-Time Algorithm for Finding Tree-Decompositions of Small Treewidth." In 25.6 (1996), pp. 1305–1317 (Cited on p. 83).
DOI: 10.1137/S0097539793251219

**[Bod98]**   Hans L. Bodlaender. "A partial k-arboretum of graphs with bounded treewidth." In 209.1 (1998), pp. 1–45 (Cited on p. 151).
DOI: 10.1016/S0304-3975(97)00228-4

**[Boj14]**   Mikołaj Bojańczyk. "Transducers with Origin Information." In *International Colloquium on Automata, Languages and Programming (ICALP)*. Consulted version: https://arxiv.org/pdf/1309.6124. 2014, pp. 26–37 (Cited on p. 223)

**[Boj15]**   Mikołaj Bojańczyk. "Recognisable Languages over Monads." In *International Conference on Developments in Language Theory*. Lecture Notes in Computer Science. Consulted version: https://arxiv.org/abs/1502.04898v1. 2015, pp. 1–13 (Cited on pp. 302, 312, 319, 323, 330).
DOI: 10.1007/978-3-319-21500-6_1

**[Boj20]**   Mikołaj Bojańczyk. *Languages recognised by finite semigroups, and their generalisations to objects such as trees and graphs, with an emphasis on definability in monadic second-order logic*. Lecture notes. 2020 (Cited on pp. 58, 211, 300, 302, 325).
DOI: 10.48550/arXiv.2008.11635

**[Boj22]**   Mikołaj Bojańczyk. "Transducers of polynomial growth." In *Annual Symposium on Logic in Computer Science (LICS)*. 2022 (Cited on pp. 227, 229).
DOI: 10.1145/3531130.3533326

**[Bou06]**   Nicolas Bourbaki. "Description de la mathématique formelle." In *Théorie des ensembles*. 2006, pp. 8–45 (Cited on p. 43).
DOI: 10.1007/978-3-540-34035-5_2

**[BR05]**   Nicolas Bedon and Chloé Rispal. "Schützenberger and Eilenberg Theorems for Words on Linear Orderings." In *International Conference on Developments in Language Theory*. Lecture Notes in Computer Science. 2005, pp. 134–145 (Cited on p. 301).
DOI: 10.1007/11505877_12

[BRV16]   Pablo Barceló, Miguel Romero, and Moshe Y. Vardi. "Semantic Acyclicity on Graph Databases." In *SIAM Journal on computing* 45.4 (2016), pp. 1339–1376 (Cited on pp. 38, 101, 143, 146, 147, 149, 150, 157, 177, 178, 181, 187, 189, 190, 194).
DOI: 10.1137/15M1034714

[BRZ20]   Pablo Barceló, Miguel Romero, and Thomas Zeume. "A More General Theory of Static Approximations for Conjunctive Queries." In 64.5 (2020), pp. 916–964 (Cited on p. 147).
DOI: 10.1007/s00224-019-09924-0

[Bul17]   Andrei A. Bulatov. "A Dichotomy Theorem for Nonuniform CSPs." In *Annual Symposium on Foundations of Computer Science (FOCS)*. 2017, pp. 319–330 (Cited on pp. 42, 248).
DOI: 10.1109/FOCS.2017.37

[BW08]   Mikołaj Bojańczyk and Igor Walukiewicz. "Forest algebras." In *Logic and Automata: History and Perspectives [in Honor of Wolfgang Thomas]*. Vol. 2. Texts in Logic and Games. Consulted version: https://hal.science/hal-00105796v1. 2008, pp. 107–132 (Cited on p. 302)

[Car95]   Lewis Carroll. *What the Tortoise Said to Achilles*. 1895 (Cited on p. 205).
URL: https://www.ditext.com/carroll/tortoise.html

[CCDF97]   Liming Cai, Jianer Chen, Rodney G. Downey, and Michael R. Fellows. "Advice classes of parameterized tractability." In *Annals of Pure and Applied Logic (APAL)* 84.1 (1997), pp. 119–138 (Cited on p. 186).
DOI: 10.1016/S0168-0072(95)00020-8

[CCG06]   Olivier Carton, Christian Choffrut, and Serge Grigorieff. "Decision problems among the main subfamilies of rational relations." In *RAIRO Theoretical Informatics and Applications* 40.2 (2006), pp. 255–275 (Cited on pp. 216, 218, 220, 223).
DOI: 10.1051/ita:2006005

[CCLP17]   Lorenzo Clemente, Wojciech Czerwiński, Sławomir Lasota, and Charles Paperman. "Regular Separability of Parikh Automata." In *International Colloquium on Automata, Languages and Programming (ICALP)*. 2017, 117:1–117:13 (Cited on p. 247).
DOI: 10.4230/LIPIcs.ICALP.2017.117

[CCP18]   Olivier Carton, Thomas Colcombet, and Gabriele Puppis. "An algebraic approach to MSO-definability on countable linear orderings." en. In *Journal of Symbolic Logic* 83.3 (2018). Consulted version: https://arxiv.org/abs/1702.05342v2, pp. 1147–1189 (Cited on p. 302).
DOI: 10.1017/jsl.2018.7

[CCP20]   Michaël Cadilhac, Olivier Carton, and Charles Paperman. "Continuity of Functional Transducers: A Profinite Study of Rational Functions." In *Logical Methods in Computer Science (LMCS)* Volume 16, Issue 1 (2020) (Cited on p. 302).
DOI: 10.23638/LMCS-16(1:24)2020

[CD15]   Balder ten Cate and Victor Dalmau. "The Product Homomorphism Problem and Applications." In *International Conference on Database Theory (ICDT)*. Vol. 31. 2015, pp. 161–176 (Cited on p. 81).
DOI: 10.4230/LIPIcs.ICDT.2015.161

[CD21]   Balder ten Cate and Victor Dalmau. "Conjunctive Queries: Unique Characterizations and Exact Learnability." In *International Conference on Database Theory (ICDT)*. Vol. 186. 2021, 9:1–9:24 (Cited on p. 82).
DOI: 10.4230/LIPIcs.ICDT.2021.9

[CDK13]   Balder ten Cate, Víctor Dalmau, and Phokion G. Kolaitis. "Learning schema mappings." In *ACM Transactions on Database Systems (TODS)* 38.4 (2013) (Cited on p. 82).
DOI: 10.1145/2539032.2539035

[CDLV00]   Diego Calvanese, Giuseppe De Giacomo, Maurizio Lenzerini, and Moshe Y. Vardi. "Containment of Conjunctive Regular Path Queries with Inverse." In *Principles of Knowledge Representation and Reasoning (KR)*. 2000, pp. 176–185 (Cited on pp. 98, 99, 155, 194, 195)

[CE12]   Bruno Courcelle and Joost Engelfriet. *Graph Structure and Monadic Second-Order Logic: A Language-Theoretic Approach*. Encyclopedia of

Mathematics and its Applications. 2012 (Cited on p. 302).

DOI: 10.1017/CBO9780511977619

[CES17]   Olivier Carton, Léo Exibard, and Olivier Serre. "Two-Way Two-Tape Automata." In *International Conference on Developments in Language Theory*. 2017, pp. 147–159 (Cited on p. 225).

DOI: 10.1007/978-3-319-62809-7_10

[CG06]   Christian Choffrut and Serge Grigorieff. "Separability of rational relations in A∗∗Nm by recognizable relations is decidable." In *Information Processing Letters (IPL)* 99.1 (2006). Consulted version: https : / / www . irif . fr / ~seg / _2006IPLsepar/Choffrut_Grigorieff_2006_ RecognizableSeparability . pdf, pp. 27–32 (Cited on p. 219).

DOI: 10.1016/j.ipl.2005.09.018

[CG14]   Christian Choffrut and Bruno Guillon. "An Algebraic Characterization of Unary Two-Way Transducers." In *International Symposium on Mathematical Foundations of Computer Science (MFCS)*. 2014, pp. 196–207 (Cited on p. 229).

[CGLP20]   Hubie Chen, Georg Gottlob, Matthias Lanzinger, and Reinhard Pichler. "Semantic Width and the Fixed-Parameter Tractability of Constraint Satisfaction Problems." In *International Joint Conference on Artificial Intelligence (IJCAI)*. 2020, pp. 1726–1733 (Cited on pp. 86, 146).

DOI: 10.24963/ijcai.2020/239

[CGM22]   Thomas Colcombet, Sam van Gool, and Rémi Morvan. "First-order separation over countable ordinals." In *International Conference on Foundations of Software Science and Computational Structures (FOSSACS)*. Consulted version: https : / / arxiv.org/abs/2201.03089v1. 2022, pp. 264–284 (Cited on pp. 247, 332).

DOI: 10.1007/978-3-030-99253-8_14

[CH90]   Kevin J. Compton and C. Ward Henson. "A uniform method for proving lower bounds on the computational complexity of logical theories." In *Annals of Pure and Applied Logic* 48.1 (1990), pp. 1–79 (Cited on p. 238).

DOI: 10.1016/0168-0072(90)90080-L

[Cha80]   Tat-hung Chan. "Reversal-Bounded Computations." Consulted version: https://scispace. com / pdf / reversal - bounded - computations - 25s73w61fj.pdf. PhD thesis. Cornell University, USA, 1980 (Cited on p. 225)

[Cho06]   Christian Choffrut. "Relations over words and logic: A chronology." In *Bulletin of the EATCS* 89 (2006), pp. 159–163 (Cited on pp. 211, 224, 232, 234)

[CL07]   Thomas Colcombet and Christof Löding. "Transforming structures by set interpretations." In *Logical Methods in Computer Science (LMCS)* Volume 3, Issue 2, 4 (2007) (Cited on p. 241).

DOI: 10.2168/LMCS-3(2:4)2007

[CL11]   Arnaud Carayol and Christof Löding. "Uniformization in automata theory." In *Proceedings of the 14th Congress of Logic, Methodology and Philosophy of Science Nancy*. Vol. 2. Consulted version: https://igm.univ-mlv.fr/~carayol/Papers/ x23ada2cef7606c52.pdf. 2011 (Cited on p. 220)

[CM13]   Hubie Chen and Moritz Müller. "The fine classification of conjunctive queries and parameterized logarithmic space complexity." In *ACM Symposium on Principles of Database Systems (PODS)*. 2013, pp. 309–320 (Cited on p. 190).

DOI: 10.1145/2463664.2463669

[CM77]   Ashok K. Chandra and Philip M. Merlin. "Optimal Implementation of Conjunctive Queries in Relational Data Bases." In *Symposium on Theory of Computing (STOC)*. 1977, pp. 77–90 (Cited on pp. 71, 72, 146).

DOI: 10.1145/800105.803397

[CMNP18]   Wojciech Czerwiński, Wim Martens, Matthias Niewerth, and Paweł Parys. "Minimization of Tree Patterns." In *Journal of the ACM* 65.4 (2018) (Cited on pp. 106, 107, 136, 137, 198)

[CMRZZ17]   Wojciech Czerwiński, Wim Martens, Lorijn van Rooijen, Marc Zeitoun, and Georg Zetzsche. "A Characterization for Decidable Separability by Piecewise Testable Languages." In *Discrete*

*Mathematics & Theoretical Computer Science* 19.4 (2017) (Cited on p. 247).

DOI: 10.23638/DMTCS-19-4-1

[Cod72]   E. F. Codd. "Relational Completeness of Data Base Sublanguages." In *Research Report / RJ / IBM / San Jose, California* (1972) (Cited on p. 68)

[CR00]   Chandra Chekuri and Anand Rajaraman. "Conjunctive query containment revisited." In *Theoretical Computer Science* 239.2 (2000), pp. 211–229 (Cited on pp. 33, 83, 194).

DOI: 10.1016/S0304-3975(99)00220-0

[CRRT01]   Colin M. Campbell, Edmund F. Robertson, Nikola Ruškuc, and Richard M. Thomas. "Automatic semigroups." In *Theoretical Computer Science* 250.1 (2001), pp. 365–391 (Cited on p. 242).

DOI: 10.1016/S0304-3975(99)00151-6

[CRV23]   Tamara Cucumides, Juan Reutter, and Domagoj Vrgoč. "Size Bounds and Algorithms for Conjunctive Regular Path Queries." In *International Conference on Database Theory (ICDT)*. 2023, 13:1–13:17 (Cited on pp. 107, 135).

DOI: 10.4230/LIPIcs.ICDT.2023.13

[CS15]   Thomas Colcombet and A. V. Sreejith. "Limited Set Quantifiers over Countable Linear Orderings." In *International Colloquium on Automata, Languages and Programming (ICALP)*. 2015, pp. 146–158 (Cited on p. 326).

DOI: 10.1007/978-3-662-47666-6_12

[CS25]   Thomas Colcombet and A V Sreejith. *Regular expressions over countable words*. 2025 (Cited on p. 326).

DOI: 10.48550/arXiv.2505.01039

[Cyg+15]   Marek Cygan, Fedor V. Fomin, Łukasz Kowalik, Daniel Lokshtanov, Dániel Marx, Marcin Pilipczuk, Michał Pilipczuk, and Saket Saurabh. "Fixed-parameter intractability." In *Parameterized Algorithms*. 2015, pp. 421–465 (Cited on p. 60).

DOI: 10.1007/978-3-319-21275-3_13

[Del04]   Christian Delhommé. "Automaticité des ordinaux et des graphes homogènes." In *Comptes Rendus Mathematique* 339.1 (2004), pp. 5–10 (Cited on p. 241).

DOI: 10.1016/j.crma.2004.03.035

[DFF19]   María Emilia Descotte, Diego Figueira, and Santiago Figueira. "Closure Properties of Synchronized Relations." In *International Symposium on Theoretical Aspects of Computer Science (STACS)*. Vol. 126. 2019, 22:1–22:17 (Cited on p. 224).

DOI: 10.4230/LIPIcs.STACS.2019.22

[DFP18]   María Emilia Descotte, Diego Figueira, and Gabriele Puppis. "Resynchronizing Classes of Word Relations." In *International Colloquium on Automata, Languages and Programming (ICALP)*. Vol. 107. Consulted version: https://hal.science/hal-01721046v2. 2018, 123:1–123:13 (Cited on p. 224).

DOI: 10.4230/LIPIcs.ICALP.2018.123

[Die70]   Jean A. Dieudonné. "The Work of Nicholas Bourbaki." In *The American Mathematical Monthly* 77.2 (1970), pp. 134–145 (Cited on p. 44).

URL: http://www.jstor.org/stable/2317325

[DKV02]   Victor Dalmau, Phokion G. Kolaitis, and Moshe Y. Vardi. "Constraint Satisfaction, Bounded Treewidth, and Finite-Variable Logics." In *Principles and Practice of Constraint Programming (CP)*. Vol. 2470. Lecture Notes in Computer Science. 2002, pp. 310–326 (Cited on p. 85).

DOI: 10.1007/3-540-46135-3\_21

[Dou23]   Gaëtan Douéneau. "Optimization of string transducers." PhD thesis. Université Paris Cité, 2023 (Cited on p. 229).

URL: http://www.theses.fr/2023UNIP7217/document

[DPPDP09]   Apostolos K. Doxiadēs, Christos H. Papadimitriou, Alekos Papadatos, Annie Di Donna, and Alekos Papadatos, eds. *Logicomix: an epic search for truth*. 2009. ISBN: 978-0-7475-9720-9 (Cited on p. 43)

[EES69]   S. Eilenberg, C.C. Elgot, and J.C Shepherdson. "Sets recognized by *n*-tape automata." In *Journal of Algebra* 13.4 (1969), pp. 447–464 (Cited on p. 232).

DOI: 10.1016/0021-8693(69)90107-0

[EGNRM98a]   Yu. L. Ershov, S. S. Goncharov, A. Nerode, J. B. Remmel, and V. W. Marek. *Handbook of recursive mathematics. Volume 1: Recursive Model Theory*. Studies in logic and the foundations of mathematics v. 138. 1998 (Cited on p. 243).
URL: https : / / www . sciencedirect . com / bookseries / studies - in - logic - and - the - foundations-of-mathematics/vol/138/

[EGNRM98b]   Yu. L. Ershov, S. S. Goncharov, A. Nerode, J. B. Remmel, and V. W. Marek. *Handbook of recursive mathematics. Volume 2: Recursive Algebra, Analysis and Combinatorics*. Studies in logic and the foundations of mathematics v. 139. 1998 (Cited on p. 243).
URL: https : / / www . sciencedirect . com / bookseries / studies - in - logic - and - the - foundations-of-mathematics/vol/139/

[EH01]   Joost Engelfriet and Hendrik Jan Hoogeboom. "MSO definable string transductions and two-way finite-state transducers." In *ACM Transactions on Computational Logic* 2.2 (2001), pp. 216–254 (Cited on p. 227).
DOI: 10.1145/371316.371512

[EJT10]   Michael Elberfeld, Andreas Jakoby, and Till Tantau. "Logspace Versions of the Theorems of Bodlaender and Courcelle." In *Annual Symposium on Foundations of Computer Science (FOCS)*. Available on Scihub. 2010, pp. 143–152 (Cited on pp. 87, 194).
DOI: 10.1109/FOCS.2010.21

[EM65]   Calvin C. Elgot and Jorge E. Mezei. "On Relations Defined by Generalized Finite Automata." In *IBM Journal of Research and Development* 9.1 (1965), pp. 47–68 (Cited on p. 211).
DOI: 10.1147/rd.91.0047

[Eps92]   David B.A. Epstein. *Word Processing in Groups*. 1992 (Cited on pp. 234, 242).
DOI: 10.1201/9781439865699

[Ete97]   Kousha Etessami. "Counting Quantifiers, Successor Relations, and Logarithmic Space." In *Journal of Computer and System Sciences (JCSS)* 54.3 (1997), pp. 400–411 (Cited on p. 60).
DOI: 10.1006/jcss.1997.1485

[FFM08]   S. Flesca, F. Furfaro, and E. Masciari. "On the minimization of XPath queries." In *Journal of the ACM* 55.1 (2008) (Cited on p. 107).
DOI: 10.1145/1326554.1326556

[FG03]   Jörg Flum and Martin Grohe. "Describing parameterized complexity classes." In 187.2 (2003), pp. 291–319 (Cited on p. 186).
DOI: 10.1016/S0890-5401(03)00161-5

[FGL19]   Emmanuel Filiot, Olivier Gauwin, and Nathan Lhote. "Logical and Algebraic Characterizations of Rational Transductions." In *Logical Methods in Computer Science (LMCS)* Volume 15, Issue 4 (2019) (Cited on p. 302).
DOI: 10.23638/LMCS-15(4:16)2019

[FGM24]   Cristina Feier, Tomasz Gogacz, and Filip Murlak. "Evaluating Graph Queries Using Semantic Treewidth." In *International Conference on Database Theory (ICDT)*. 2024 (Cited on pp. 148, 190, 194).
DOI: 10.4230/LIPIcs.ICDT.2024.22

[Fig+20]   Diego Figueira, Adwait Godbole, S. Krishna, Wim Martens, Matthias Niewerth, and Tina Trautner. "Containment of Simple Conjunctive Regular Path Queries." In *Principles of Knowledge Representation and Reasoning (KR)*. 2020 (Cited on pp. 100, 101, 125, 126, 134, 148, 190).
DOI: 10.24963/kr.2020/38

[Fig20]   Diego Figueira. "Containment of UC2RPQ: the hard and easy cases." In *International Conference on Database Theory (ICDT)*. 2020 (Cited on pp. 100, 101, 109, 134, 156, 188, 194).
DOI: 10.4230/LIPIcs.ICDT.2020.9

[Fig21a]   Diego Figueira. "Foundations of Graph Path Query Languages." In *Reasoning Web. Declarative Artificial Intelligence - 17th International Summer School 2021*. Vol. 13100. Lecture Notes in Computer Science. 2021, pp. 1–21 (Cited on p. 90).
DOI: 10.1007/978-3-030-95481-9\_1

[Fig21b]   Diego Figueira. "Foundations of Graph Path Query Languages (Course Notes)." In *Reasoning Web Summer School 2021*. Vol. 13100. Reasoning Web. Declarative Artificial Intelligence - 17th International Summer School 2021, Leuven, Belgium,

September 8-15, 2021, Tutorial Lectures. Consulted version: https://hal.science/hal-03349901. 2021, pp. 1–21 (Cited on p. 211).

DOI: 10.1007/978-3-030-95481-9\_1

[FKMP24]  Diego Figueira, S. Krishna, Om Swostik Mishra, and Anantha Padmanabha. "Boundedness for Unions of Conjunctive Regular Path Queries over Simple Regular Expressions." In *Principles of Knowledge Representation and Reasoning (KR)*. 2024 (Cited on p. 101).

DOI: doi.org/10.24963/kr.2024/34

[FL15]  Diego Figueira and Leonid Libkin. "Synchronizing Relations on Words." en. In *Theory of Computing Systems* 57.2 (2015). Consulted version: https://hal.science/hal-01793633v1/, pp. 287–318 (Cited on pp. 223, 224).

DOI: 10.1007/s00224-014-9584-2

[FLS98]  Daniela Florescu, Alon Levy, and Dan Suciu. "Query Containment for Conjunctive Queries with Regular Expressions." In *ACM Symposium on Principles of Database Systems (PODS)*. 1998, pp. 139–148 (Cited on pp. 95, 96, 98, 99, 156).

DOI: 10.1145/275487.275503

[FM23]  Diego Figueira and Rémi Morvan. "Approximation and Semantic Tree-Width of Conjunctive Regular Path Queries." In *International Conference on Database Theory (ICDT)*. Vol. 255. 2023, 15:1–15:19 (Cited on p. 143).

DOI: 10.4230/LIPIcs.ICDT.2023.15

[FM25]  Diego Figueira and Rémi Morvan. "Semantic Tree-Width and Path-Width of Conjunctive Regular Path Queries." In *Logical Methods in Computer Science (LMCS)* Volume 21, Issue 1, 21 (2025) (Cited on pp. 65, 86, 107, 113, 120, 124, 143, 186).

DOI: 10.46298/lmcs-21(1:21)2025

[FMR25]  Diego Figueira, Rémi Morvan, and Miguel Romero. *Minimizing Conjunctive Regular Path Queries*. 2025 (Cited on pp. 65, 103).

arXiv: 2504.00612 [cs.DB]

[Fon07]  Jan Foniok. "Homomorphisms and Structural Properties of Relational Systems." PhD thesis. Charles University in Prague, 2007 (Cited on p. 259).

URL: https://arxiv.org/pdf/0710.4477

[FR23]  Diego Figueira and Miguel Romero. "Conjunctive Regular Path Queries under Injective Semantics." In *ACM Symposium on Principles of Database Systems (PODS)*. 2023, pp. 231–240 (Cited on p. 101).

DOI: 10.1145/3584372.3588664

[FR68]  Patrick C. Fischer and Arnold L. Rosenberg. "Multitape one-way nonwriting automata." In *Journal of Computer and System Sciences (JCSS)* 2.1 (1968), pp. 88–101 (Cited on pp. 220, 223).

DOI: 10.1016/S0022-0000(68)80006-6

[Fra+23a]  Nadime Francis, Amélie Gheerbrant, Paolo Guagliardo, Leonid Libkin, Victor Marsault, Wim Martens, Filip Murlak, Liat Peterfreund, Alexandra Rogova, and Domagoj Vrgoč. "A Researcher's Digest of GQL (Invited Talk)." In *International Conference on Database Theory (ICDT)*. Vol. 255. 2023, 1:1–1:22 (Cited on p. 95).

DOI: 10.4230/LIPICS.ICDT.2023.1

[Fra+23b]  Nadime Francis, Amélie Gheerbrant, Paolo Guagliardo, Leonid Libkin, Victor Marsault, Wim Martens, Liat Peterfreund, Alexandra Rogova, and Domagoj Vrgoc. "GPC: A Pattern Calculus for Property Graphs." In *ACM Symposium on Principles of Database Systems (PODS)*. 2023, pp. 241–250 (Cited on p. 95).

DOI: 10.1145/3584372.3588662

[Fre90]  Eugene C. Freuder. "Complexity of K-tree structured constraint satisfaction problems." In *AAAI Conference on Artificial Intelligence*. 1990, pp. 4–9 (Cited on p. 83)

[FRW19]  Diego Figueira, Varun Ramanathan, and Pascal Weil. "The Quantifier Alternation Hierarchy of Automatic Relations." In *International Symposium on Mathematical Foundations of Computer Science (MFCS)*. Vol. 138. 2019, 29:1–29:14 (Cited on p. 233).

DOI: 10.4230/LIPICS.MFCS.2019.29

[FS93]  Christiane Frougny and Jacques Sakarovitch. "Synchronized Rational Relations of Finite and Infi-

nite Words." In *Theoretical Computer Science* 108.1 (1993), pp. 45–82 (Cited on pp. 211, 216).

DOI: 10.1016/0304-3975(93)90230-Q

[FT10]   Olivier Finkel and Stevo Todorčević. "The isomorphism relation between tree-automatic Structures." In *Open Mathematics* 8.2 (2010), pp. 299–313 (Cited on p. 242).

DOI: 10.2478/s11533-010-0014-7

[FT13]   Olivier Finkel and Stevo Todorčević. "Automatic Ordinals." In *International Journal of Unconventional Computing* 9.1-2 (2013), pp. 61–70 (Cited on p. 241).

URL: http://www.oldcitypublishing.com/journals/ijuc-home/ijuc-issue-contents/ijuc-volume-9-number-1-2-2013/ijuc-9-1-2-p-61-70/

[FV98]   Tomás Feder and Moshe Y. Vardi. "The Computational Structure of Monotone Monadic SNP and Constraint Satisfaction: A Study through Datalog and Group Theory." In *SIAM Journal on computing* 28.1 (1998), pp. 57–104 (Cited on pp. 42, 248, 259, 288).

DOI: 10.1137/S0097539794266766

[Gau20]   Olivier Gauwin. "Transductions: resources and characterizations." PhD thesis. Université de Bordeaux, 2020 (Cited on p. 229).

URL: https://theses.hal.science/tel-03118919

[GG24]   Mai Gehrke and Sam van Gool. *Topological Duality for Distributive Lattices: Theory and Applications.* 2024 (Cited on p. 200).

DOI: 10.1017/9781009349680

[GGIM22]   Víctor Gutiérrez-Basulto, Albert Gutowski, Yazmín Ibáñez-García, and Filip Murlak. "Finite Entailment of UCRPQs over ALC Ontologies." In *Principles of Knowledge Representation and Reasoning (KR).* 2022, pp. 184–194 (Cited on p. 101).

DOI: 10.24963/kr.2022/19

[GGIM24]   Víctor Gutiérrez-Basulto, Albert Gutowski, Yazmín Ibáñez-García, and Filip Murlak. "Containment of Graph Queries Modulo Schema." In *ACM Symposium on Principles of Database Systems (PODS).* 2024, pp. 1–26 (Cited on p. 101).

DOI: 10.1145/3651140

[Grä07]   Erich Grädel. "Finite Model Theory and Descriptive Complexity." In *Finite Model Theory and Its Applications.* 2007, pp. 125–230 (Cited on p. 240).

DOI: 10.1007/3-540-68804-8_3

[Grä20]   Erich Grädel. "Automatic Structures: Twenty Years Later." In *Annual Symposium on Logic in Computer Science (LICS).* LICS '20. 2020, pp. 21–34 (Cited on p. 243).

DOI: 10.1145/3373718.3394734

[Gro07]   Martin Grohe. "The complexity of homomorphism and constraint satisfaction problems seen from the other side." In *Journal of the ACM* 54.1 (2007), 1:1–1:24 (Cited on pp. 33, 85, 146, 190, 191, 193).

DOI: 10.1145/1206035.1206036

[GS19]   S. J. v. Gool and B. Steinberg. "Pointlike sets for varieties determined by groups." In *Advances in Mathematics* 348 (2019). Consulted version: https://arxiv.org/abs/1801.04638v1, pp. 18–50 (Cited on p. 304).

DOI: 10.1016/j.aim.2019.03.020

[GSS01]   Martin Grohe, Thomas Schwentick, and Luc Segoufin. "When is the evaluation of conjunctive queries tractable?" In *Symposium on Theory of Computing (STOC).* 2001, pp. 657–666 (Cited on p. 85).

DOI: 10.1145/380752.380867

[Gur82]   Eitan M. Gurari. "The Equivalence Problem for Deterministic Two-Way Sequential Transducers is Decidable." In *SIAM Journal on computing* 11.3 (1982), pp. 448–452 (Cited on p. 227).

DOI: 10.1137/0211035

[HK91]   T. Harju and J. Karhumäki. "The equivalence problem of multitape finite automata." In *Theoretical Computer Science* 78.2 (1991), pp. 347–355 (Cited on p. 223).

DOI: 10.1016/0304-3975(91)90356-7

[HKMN08]   Greg Hjorth, Bakh Khoussainov, Antonio Montalbán, and André Nies. "From Automatic Structures to Borel Structures." In *Annual Sympo-*

*sium on Logic in Computer Science (LICS)*. 2008, pp. 431–441 (Cited on p. 235).

DOI: 10.1109/LICS.2008.28

[HMPR91]   Karsten Henckell, Stuart W. Margolis, Jean-Éric Pin, and John Rhodes. "Ash's type II theorem, profinite topology and Malcev products: part I." In *International Journal of Algebra and Computation* 01.04 (1991). Consulted version: https://www.irif.fr/~jep/PDF/HMPR.pdf (saved on http://web.archive.org/), pp. 411–436 (Cited on p. 304).

DOI: 10.1142/S0218196791000298

[HN90]   Pavol Hell and Jaroslav Nešetřil. "On the complexity of H-coloring." In *Journal of Combinatorial Theory, Series B* 48.1 (1990), pp. 92–110 (Cited on p. 42).

DOI: 10.1016/0095-8956(90)90132-J

[HN92]   Pavol Hell and Jaroslav Nešetřil. "The core of a graph." In *Discrete Mathematics* 109.1 (1992), pp. 117–126 (Cited on pp. 80, 85).

DOI: 10.1016/0012-365X(92)90282-K

[Hod76]   Bernard R. Hodgson. "Théories décidables par automate fini." Not available online. PhD thesis. Université de Montréal, 1976 (Cited on p. 234)

[Hod83]   Bernard R. Hodgson. "Décidabilité par Automate Fini." In *Annales des Sciences Mathématiques du Québec* 7.1 (1983). Consulted version: https://www.labmath.uqam.ca/~annales/volumes/07-1/PDF/039-057.pdf (saved on http://web.archive.org/), pp. 39–57 (Cited on pp. 216, 234, 238)

[Hod93]   Wilfrid Hodges. *Model Theory*. Encyclopedia of Mathematics and its Applications. 1993 (Cited on pp. 240, 242)

[Hog94]   Mark Hogarth. "Non-Turing Computers and Non-Turing Computability." In *PSA: Proceedings of the Biennial Meeting of the Philosophy of Science Association* 1 (1994), pp. 126–138 (Cited on p. 6).

DOI: 10.1086/psaprocbienmeetp.1994.1.193018

[Imm98]   Neil Immerman. *Descriptive complexity*. 1998 (Cited on pp. 231, 232, 239).

DOI: 10.1007/978-1-4612-0539-5

[ISO23]   ISO: International Organization for Standardization. *ISO/IEC 9075-16:2023 Part 16: Property Graph Queries (SQL/PGQ)*. https://www.iso.org/standard/79473.html. 2023 (Cited on p. 95)

[ISO24]   ISO: International Organization for Standardization. *ISO/IEC 39075:2024 GQL*. https://www.iso.org/standard/76120.html. 2024 (Cited on p. 95)

[JKSS19]   Sanjay Jain, Bakhadyr Khoussainov, Philipp Schlicht, and Frank Stephan. "The isomorphism problem for tree-automatic ordinals with addition." In *Information Processing Letters (IPL)* 149 (2019), pp. 19–24 (Cited on pp. 241, 242).

DOI: 10.1016/j.ipl.2019.05.004

[KBHSN24]   Nikolaos Karalis, Alexander Bigerl, Liss Heidrich, Mohamed Ahmed Sherif, and Axel-Cyrille Ngonga Ngomo. "Efficient Evaluation of Conjunctive Regular Path Queries Using Multi-way Joins." In 2024, pp. 218–235 (Cited on p. 107).

DOI: 10.1007/978-3-031-60626-7_12

[KKOT15]   Bartek Klin, Eryk Kopczynski, Joanna Ochremiak, and Szymon Torunczyk. "Locally Finite Constraint Satisfaction Problems." In *Annual Symposium on Logic in Computer Science (LICS)*. 2015, pp. 475–486 (Cited on p. 249).

DOI: 10.1109/LICS.2015.51

[KL10]   Dietrich Kuske and Markus Lohrey. "Some natural decision problems in rational graphs." In *Journal of Symbolic Logic* 75.2 (2010), pp. 678–710 (Cited on p. 243).

DOI: 10.2178/jsl/1268917499

[Klo94]   Ton Kloks. *Treewidth: computations and approximations*. Lecture Notes in Computer Science. 1994 (Cited on p. 152).

DOI: 10.1007/BFb0045375

[KLOT16]   Bartek Klin, Slawomir Lasota, Joanna Ochremiak, and Szymon Torunczyk. "Homomorphism Problems for First-Order Definable Structures." In *IARCS Annual Conference on Foundations*

*of Software Technology and Theoretical Computer Science (FST&TCS)*. Vol. 65. 2016, 14:1–14:15 (Cited on pp. 249, 251).

DOI: 10.4230/LIPIcs.FSTTCS.2016.14

[KM10]   Shiva Kintali and Sinziana Munteanu. "Computing Bounded Path Decompositions in Logspace." In vol. 19. 126. 2010 (Cited on pp. 87, 194).

URL: https://eccc.weizmann.ac.il/report/2012/126/

[KN95]   Bakhadyr Khoussainov and Anil Nerode. "Automatic presentations of structures." In *Logic and Computational Complexity*. 1995, pp. 367–392 (Cited on pp. 216, 234, 251).

DOI: 10.1007/3-540-60178-3_93

[KNRS07]   Bakhadyr Khoussainov, Andre Nies, Sasha Rubin, and Frank Stephan. "Automatic Structures: Richness and Limitations." In *Logical Methods in Computer Science (LMCS)* Volume 3, Issue 2, 2 (2007) (Cited on pp. 218, 241–243, 263).

DOI: 10.2168/LMCS-3(2:2)2007

[Köc14]   Chris Köcher. "Analyse der Entscheidbarkeit diverser Probleme in automatischen Graphen." de. Bachelor's Thesis. Ilmenau: Technische Universität Ilmenau, 2014 (Cited on pp. 44, 243, 271).

URL: https://people.mpi-sws.org/~ckoecher/files/theses/bsc-thesis.pdf

[Kol07]   Phokion G. Kolaitis. "On the Expressive Power of Logics on Finite Models." In *Finite Model Theory and Its Applications*. 2007, pp. 27–123 (Cited on pp. 91, 258).

DOI: 10.1007/3-540-68804-8_2

[Kop16]   Eryk Kopczyński. "Invisible Pushdown Languages." In *Annual Symposium on Logic in Computer Science (LICS)*. 2016, pp. 867–872 (Cited on p. 247).

DOI: 10.1145/2933575.2933579

[KRS05]   Bakhadyr Khoussainov, Sasha Rubin, and Frank Stephan. "Automatic linear orders and trees." In *ACM Transactions on Computational Logic* 6.4 (2005), pp. 675–700 (Cited on p. 241).

DOI: 10.1145/1094622.1094625

[KS08]   Benny Kimelfeld and Yehoshua Sagiv. "Revisiting redundancy and minimization in an XPath fragment." In *International Conference on Extending Database Technology (EDBT)*. 2008, pp. 61–72 (Cited on p. 107).

DOI: 10.1145/1353343.1353355

[Lar17]   Benoit Larose. "Algebra and the Complexity of Digraph CSPs: a Survey." In *The Constraint Satisfaction Problem: Complexity and Approximability*. Ed. by Andrei Krokhin and Stanislav Zivny. Vol. 7. Dagstuhl Follow-Ups. 2017, pp. 267–285 (Cited on p. 248).

DOI: 10.4230/DFU.Vol7.15301.267

[LB16]   Anthony W. Lin and Pablo Barceló. "String Solving with Word Equations and Transducers: Towards a Logic for Analysing Mutation XSS." In *Annual Symposium on Principles of Programming Languages (POPL)*. 2016, pp. 123–136 (Cited on p. 211).

DOI: 10.1145/2837614.2837641

[Lec63]   Yves Lecerf. "Machines de Turing réversibles. Récursive insolubilité en $n \in \mathbf{N}$ de l'équation $u = \theta^n u$, ou $\theta$ est un « isomorphisme de codes »." In *Comptes rendus hebdomadaires des séances de l'Académie des sciences* 257 (1963), pp. 2597–2600 (Cited on p. 263)

[LLT07]   Benoit Larose, Cynthia Loten, and Claude Tardif. "A Characterisation of First-Order Constraint Satisfaction Problems." In *Logical Methods in Computer Science (LMCS)* Volume 3, Issue 4, 6 (2007) (Cited on pp. 257, 258, 273, 275, 283, 288).

DOI: 10.2168/LMCS-3(4:6)2007

[LS19]   Christof Löding and Christopher Spinrath. "Decision Problems for Subclasses of Rational Relations over Finite and Infinite Words." In *Discrete Mathematics & Theoretical Computer Science* 21.3 (2019) (Cited on pp. 216, 218, 247).

DOI: 10.23638/DMTCS-21-3-4

[LT09]   Benoît Larose and Pascal Tesson. "Universal algebra and hardness results for constraint satisfaction problems." In *Theoretical Computer Science* 410.18 (2009), pp. 1629–1647 (Cited on pp. 42, 45, 254, 272, 276).

DOI: 10.1016/j.tcs.2008.12.048

[Man11]  Paolo Mancosu. "Book review: Logicomix by Apostolos Doxiadis, Christos H. Papadimitriou, Alecos Papadatos, and Annie di Donna." In *Journal of Humanistic Mathematics* 1.1 (2011), pp. 137–152 (Cited on p. 43)

[Mar13]  Dániel Marx. "Tractable Hypergraph Properties for Constraint Satisfaction and Conjunctive Queries." In *Journal of the ACM* 60.6 (2013), 42:1–42:51 (Cited on pp. 86, 147).
DOI: 10.1145/2535926

[Mor17]  Kenichi Morita. "Reversible Turing Machines." In *Theory of Reversible Computing.* 2017, pp. 103–156 (Cited on p. 263).
DOI: 10.1007/978-4-431-56606-9_5

[Mor25]  Rémi Morvan. "The Algebras for Automatic Relations." In *EACSL Annual Conference on Computer Science Logic (CSL).* Vol. 326. 2025, 21:1–21:21 (Cited on pp. 209, 213, 297).
DOI: 10.4230/LIPIcs.CSL.2025.21

[MP71]  Robert McNaughton and Seymour A. Papert. *Counter-Free Automata.* 1971. ISBN: 9780262130769 (Cited on p. 326)

[Myc55]  Jan Mycielski. "Sur le coloriage des graphs." fre. In *Colloquium Mathematicae* 3.2 (1955), pp. 161–162 (Cited on p. 292).
URL: http://eudml.org/doc/210000

[Nas07]  Reza Naserasr. "Homomorphisms and edge-colourings of planar graphs." In *Journal of Combinatorial Theory, Series B* 97.3 (2007), pp. 394–400 (Cited on p. 294).
DOI: 10.1016/j.jctb.2006.07.001

[Neo]  Neo4j. *ICIJ Empowers Investigative Journalists with Neo4j Graph Technology* (Cited on pp. 29, 91, 94).
URL: https://neo4j.com/customer-stories/icij/

[Niv68]  Maurice Nivat. "Transduction des langages de Chomsky." In *Annales de l'Institut Fourier* 18 (1968), pp. 339–455 (Cited on p. 211)

[NM12]  Jaroslav Nešetřil and Patrice Ossona de Mendez. "Prolegomena." In *Sparsity: Graphs, Structures, and Algorithms.* 2012, pp. 21–60 (Cited on p. 82).
DOI: 10.1007/978-3-642-27875-4_3

[NO12a]  Jaroslav Nešetřil and Patrice Ossona de Mendez. "First-Order Constraint Satisfaction Problems, Limits and Homomorphism Dualities." In *Sparsity: Graphs, Structures, and Algorithms.* 2012, pp. 195–226 (Cited on p. 259).
DOI: 10.1007/978-3-642-27875-4_9

[NO12b]  Jaroslav Nešetřil and Patrice Ossona de Mendez. "Restricted Homomorphism Dualities." In *Sparsity: Graphs, Structures, and Algorithms.* 2012, pp. 253–275 (Cited on p. 295).
DOI: 10.1007/978-3-642-27875-4_11

[NP78]  Jaroslav Nešetřil and Aleš Pultr. "On classes of relations and graphs determined by subobjects and factorobjects." In *Discrete Mathematics* 22.3 (1978), pp. 287–300 (Cited on p. 257).
DOI: 10.1016/0012-365X(78)90062-6

[NT00]  Jaroslav Nešetřil and Claude Tardif. "Duality Theorems for Finite Structures (Characterising Gaps and Good Characterisations)." In *Journal of Combinatorial Theory, Series B* 80.1 (2000), pp. 80–97 (Cited on p. 259).
DOI: 10.1006/jctb.2000.1970

[Pel97]  Laurent Pelecq. "Isomorphismes et automorphismes des graphes context-free, équationnels et automatiques." Available at Bibliothèque universitaire des sciences et techniques, Univ. Bordeaux. PhD thesis. Université Bordeaux 1, 1997 (Cited on pp. 220, 234, 238)

[Per84]  Dominique Perrin. "Recent results on automata and infinite words." In *International Symposium on Mathematical Foundations of Computer Science.* Springer. 1984, pp. 134–148 (Cited on p. 326).
DOI: 10.1007/BFb0030294

[Pin21a]  Jean-Éric Pin. "Finite automata." In *Handbook of Automata Theory. Volume I. Theoretical Foundations.* 2021, pp. 3–38 (Cited on p. 211).
DOI: 10.4171/automata-1/1

**[Pin21b]** Jean-Éric Pin. *Regular expression vs rational expression*. Computer Science Stack Exchange. Version: 2021-08-20. 2021 (Cited on p. 211).
URL: https://cs.stackexchange.com/q/143318

**[Pin22]** Jean-Éric Pin. *Mathematical Foundations of Automata Theory*. Version of February 18, 2022 (saved on http://web.archive.org/); MPRI lecture notes. 2022 (Cited on pp. 57, 58, 301–303, 319, 321, 322).
URL: https://www.irif.fr/~jep/PDF/MPRI/MPRI.pdf

**[Pin98]** Jean-Éric Pin. "Positive varieties and infinite words." In *Latin American Theoretical Informatics Symposium (LATIN)*. Lecture Notes in Computer Science. Consulted version: https://hal.science/hal-00113768v1. 1998, pp. 76–87 (Cited on p. 300).
DOI: 10.1007/BFb0054312

**[PP04]** Dominique Perrin and Jean-Éric Pin. *Infinite Words, Automata, Semigroups, Logic and Games*. Vol. 141. 2004 (Cited on pp. 300, 301).

**[PZ16]** Thomas Place and Marc Zeitoun. "Separating Regular Languages with First-Order Logic." In *Logical Methods in Computer Science (LMCS)* 12.1 (2016) (Cited on pp. 247, 332).
DOI: 10.2168/LMCS-12(1:5)2016

**[PZ23]** Thomas Place and Marc Zeitoun. "Group Separation Strikes Back." In *Annual Symposium on Logic in Computer Science (LICS)*. Consulted version: https://arxiv.org/abs/2205.01632v2. 2023, pp. 1–13 (Cited on p. 305).
DOI: 10.1109/LICS56636.2023.10175683

**[RBV17]** Miguel Romero, Pablo Barceló, and Moshe Y. Vardi. "The homomorphism problem for regular graph patterns." In *Annual Symposium on Logic in Computer Science (LICS)*. 2017, pp. 1–12 (Cited on pp. 39, 145, 157, 189, 190, 194, 201).
DOI: 10.1109/LICS.2017.8005106

**[Ree22]** Sarah Rees. *The development of the theory of automatic groups*. 2022 (Cited on p. 242).
arXiv: 2205.14911 [math.GR]

**[Reu80]** Christophe Reutenauer. "Séries formelles et algèbres syntactiques." In *Journal of Algebra* 66.2 (1980), pp. 448–483 (Cited on pp. 301, 325).
DOI: 10.1016/0021-8693(80)90097-6

**[Rez12]** Charles Rezk. *"Functors between monads": what are these really called?* MathOverflow. Version: 2012-03-24. 2012 (Cited on p. 327).
URL: https://mathoverflow.net/q/92093

**[Ris04]** Chloé Rispal. "Automates sur les ordres linéaires : Complémentation." PhD thesis. Université de Marne la Vallée, 2004 (Cited on p. 301).
URL: https://theses.hal.science/tel-00720658

**[Ros08]** Benjamin Rossman. "Homomorphism preservation theorems." In *Journal of the ACM* 55.3 (2008) (Cited on p. 89).
DOI: 10.1145/1379759.1379763

**[Ros25]** Benjamin Rossman. "Equi-Rank Homomorphism Preservation Theorem on Finite Structures." In *EACSL Annual Conference on Computer Science Logic (CSL)*. Vol. 326. 2025, 6:1–6:17 (Cited on p. 89).
DOI: 10.4230/LIPIcs.CSL.2025.6

**[RRV17]** Juan L. Reutter, Miguel Romero, and Moshe Y. Vardi. "Regular Queries on Graph Databases." In *Theory of Computing Systems* 61.1 (2017), pp. 31–83 (Cited on p. 193).
DOI: 10.1007/s00224-016-9676-2

**[RS11]** John Rhodes and Benjamin Steinberg. "Point-like sets, hyperdecidability and the identity problem for finite semigroups." In *International Journal of Algebra and Computation* (2011) (Cited on pp. 304, 324).
DOI: 10.1142/S021819679900028X

**[RS59]** M. O. Rabin and D. Scott. "Finite Automata and Their Decision Problems." In *IBM Journal of Research and Development* 3.2 (1959), pp. 114–125 (Cited on pp. 211, 223–225).
DOI: 10.1147/rd.32.0114

**[RS86]** Neil Robertson and Paul D. Seymour. "Graph minors. V. Excluding a planar graph." In 41.1 (1986), pp. 92–114 (Cited on pp. 86, 193).
DOI: 10.1016/0095-8956(86)90030-4

[Rub08]    Sasha Rubin. "Automata Presenting Structures: A Survey of the Finite String Case." In *Bulletin of Symbolic Logic* 14.2 (2008), pp. 169–209 (Cited on pp. 240, 243).
DOI: 10.2178/bsl/1208442827

[Sak09]    Jacques Sakarovitch. *Elements of Automata Theory*. Ed. by Reuben Thomas. 2009 (Cited on pp. 213, 214, 222, 227).
DOI: 10.1017/CBO9781139195218

[Sch16]    Sylvain Schmitz. "Complexity Hierarchies beyond Elementary." In *ACM Transactions on Computation Theory* 8.1 (2016) (Cited on p. 59).
DOI: 10.1145/2858784

[Sch65]    M.P. Schützenberger. "On finite monoids having only trivial subgroups." In *Information and Control* 8.2 (1965), pp. 190–194 (Cited on p. 326).
DOI: 10.1016/S0019-9958(65)90108-7

[Sch78]    Thomas J. Schaefer. "The complexity of satisfiability problems." In *Symposium on Theory of Computing (STOC)*. 1978, pp. 216–226 (Cited on pp. 42, 248).
DOI: 10.1145/800133.804350

[Sha92]    Michael Shapiro. "Deterministic and non-deterministic asynchronous automatic structures." In *International Journal of Algebra and Computation* 02.03 (1992), pp. 297–305 (Cited on p. 234).
DOI: 10.1142/S0218196792000189

[She59]    J. C. Shepherdson. "The Reduction of Two-Way Automata to One-Way Automata." In *IBM Journal of Research and Development* 3.2 (1959), pp. 198–200 (Cited on p. 211).
DOI: 10.1147/rd.32.0198

[Sip80]    Michael Sipser. "Halting space-bounded computations." In *Theoretical Computer Science* 10.3 (1980), pp. 335–338 (Cited on p. 225).
DOI: 10.1016/0304-3975(80)90053-5

[Sta84]    Ryan Stansifer. *Presburger's article on integer arithmetic: Remarks and translation*. Tech. rep. Consulted version: https://ecommons.cornell.edu/server/api/core/bitstreams/a1a7a505-bb98-4d5b-abd5-e2ee527db229/content. Cornell University, 1984 (Cited on p. 239)

[Ste67]    Richard Edwin Stearns. "A Regularity Test for Pushdown Machines." In *Information and Control* 11.3 (1967), pp. 323–340 (Cited on p. 223).
DOI: 10.1016/S0019-9958(67)90591-8

[Str72]    Ross Street. "The formal theory of monads." In *Journal of Pure and Applied Algebra* 2.2 (1972), pp. 149–168 (Cited on p. 327).
DOI: 10.1016/0022-4049(72)90019-9

[SW21]    Howard Straubing and Pascal Weil. "Varieties." In *Handbook of Automata Theory, volume I: Theoretical Foundations*. Ed. by Jean-Éric Pin. 2021, Chapter 16, pp. 569–614 (Cited on p. 302).
DOI: 10.4171/Automata.
URL: https://hal.science/hal-03434221

[TG99]    Alfred Tarski and Steven Givant. "Tarski's System of Geometry." In *Bulletin of Symbolic Logic* 5.2 (1999), pp. 175–214 (Cited on p. 44).
DOI: 10.2307/421089

[Til87]    Bret Tilson. "Categories as algebra: An essential ingredient in the theory of monoids." In *Journal of Pure and Applied Algebra* 48.1 (1987), pp. 83–198 (Cited on p. 309).
DOI: 10.1016/0022-4049(87)90108-3

[UACM17]    Henning Urbat, Jiří Adámek, Liang-Ting Chen, and Stefan Milius. "Eilenberg Theorems for Free." In *International Symposium on Mathematical Foundations of Computer Science (MFCS)*. Vol. 83. Consulted version: https://arxiv.org/abs/1602.05831v3. 2017, 43:1–43:15 (Cited on p. 330).
DOI: 10.4230/LIPIcs.MFCS.2017.43

[UD54]    Peter Ungar and Blanche Descartes. "*k*-Chromatic graphs without triangles." In *The American Mathematical Monthly* 61.5 (1954), pp. 352–353 (Cited on p. 292).
DOI: 10.2307/2307489

[Val75]    Leslie G. Valiant. "Regularity and Related Problems for Deterministic Pushdown Automata." In *Journal of the ACM* 22.1 (1975), pp. 1–10 (Cited on p. 223).
DOI: 10.1145/321864.321865

**[Vrg+24]** Domagoj Vrgoč, Carlos Rojas, Renzo Angles, Marcelo Arenas, Vicente Calisto, Benjamín Farías, Sebastián Ferrada, Tristan Heuer, Aidan Hogan, Gonzalo Navarro, Alexander Pinto, Juan Reutter, Henry Rosales, and Etienne Toussiant. "MillenniumDB: A Multi-modal, Multi-model Graph Database." In *ACM Symposium on Principles of Database Systems (PODS)*. 2024, pp. 496–499 (Cited on pp. 107, 135).
DOI: 10.1145/3626246.3654757

**[Wik24]** Wikipedia contributors. *De Bruijn–Erdős theorem (graph theory)*. Date of last revision: 6 May 2024; date retrieved: 29 January 2025. 2024 (Cited on p. 256).
URL: https://en.wikipedia.org/wiki/De_Bruijn%E2%80%93Erd%C5%91s_theorem_(graph_theory)

**[Wil10]** Ross Willard. "Testing Expressibility Is Hard." In *Principles and Practice of Constraint Programming*. 2010, pp. 9–23 (Cited on p. 81)

**[WR10]** Alfred North Whitehead and Bertrand Russell. *Principia mathematica. Volume I*. 1910 (Cited on p. 6)

**[WR12]** Alfred North Whitehead and Bertrand Russell. *Principia mathematica. Volume II*. 1912 (Cited on p. 6)

**[Zhu20]** Dmitriy Zhuk. "A Proof of the CSP Dichotomy Conjecture." In *Journal of the ACM* 67.5 (2020) (Cited on pp. 42, 248).
DOI: 10.1145/3402029

**[Zie87]** Wieslaw Zielonka. "Notes on finite asynchronous automata." en. In *RAIRO Theoretical Informatics and Applications* 21.2 (1987), pp. 99–135 (Cited on p. 220).
URL: http://www.numdam.org/item/ITA%5C_1987%5C_%5C_21%5C_2%5C_99%5C_0/

# Index

## Abstract

This thesis investigates the central role of homomorphism problems—structure-preserving maps—in two complementary domains: database querying over finite, graph-shaped data, and constraint solving over (potentially infinite) structures.

Building on the well-known equivalence between conjunctive query evaluation and homomorphism existence, the first part focuses on conjunctive regular path queries, a standard extension of conjunctive queries that incorporates regular-path predicates. We study the fundamental problem of query minimization under two measures: the number of atoms (constraints) and the tree-width of the query graph. In both cases, we prove the problem to be decidable, and provide efficient algorithms for a large fragment of queries used in practice. The second part of the thesis lifts homomorphism problems to automatic structures, which are infinite structures describable by finite automata. We highlight a dichotomy, between homomorphism problems over automatic structures that are decidable in non-deterministic logarithmic space, and those that are undecidable—proving to be the more common case. In contrast to this prevalence of undecidability, we then focus on the language-theoretic properties of these structures, and show, relying on a novel algebraic language theory, that for any well-behaved logic (a pseudovariety), whether an automatic structure can be described in this logic is decidable.

## Résumé

Cette thèse étudie le rôle central joué par les problèmes d'homomorphismes—c'est-à-dire des fonctions préservant les structures relationnelles—dans deux domaines complémentaires : le requêtage de données sous forme de graphes, et la résolution de contraintes sur des structures potentiellement infinies.

En se fondant sur l'équivalence entre l'évaluation des requêtes conjonctives et l'existence d'homomorphismes, la première partie étudie les requêtes conjonctives à chemins réguliers, une extension du langage précédent incorporant des prédicats exprimant l'existence de chemins réguliers. Nous étudions le problème fondamental de la minimisation de ces requêtes selon deux métriques : le nombre d'atomes (contraintes) et la largeur arborescente du graphe sous-jacent. Dans les deux cas, nous montrons que le problème est décidable et proposons des algorithmes efficaces pour un fragment substantiel des requêtes utilisées en pratique.

La seconde partie de la thèse généralise les problèmes d'homomorphismes aux structures automatiques, structures infinies décrites par des automates finis. Nous dessinons une dichotomie entre les problèmes résolubles en espace logarithmique non-déterministe et ceux qui sont indécidables—ces derniers étant malheureusement majoritaires. *A contrario*, nous mettons en avant une large classe de problèmes plus syntaxiques : à l'aide d'une nouvelle théorie algébrique des langages, nous montrons que, pour toute logique raisonnable (une pseudovariété), il est décidable de savoir si une structure automatique peut être spécifiée dans cette logique.